\documentclass[PhD, binding=0.6cm, twoside]{packages/sapthesis}

\usepackage{microtype}

\usepackage{hyperref}
\hypersetup{colorlinks=true,
	linkcolor=blue,
	anchorcolor=blue,
	citecolor=blue,
	urlcolor=blue,
	pdftitle={Low dimensional interacting bosons},
	pdfauthor={Serena Cenatiempo}}

\title{Low dimensional interacting bosons}
\author{Serena Cenatiempo}
\IDnumber{\textcolor{black}{1226137}}  
\course[Physics]{Fisica} % only one argument for Laurea and Laurea Magistrale
\courseorganizer{Scuola di dottorato in Scienze MM.FF.NN.}
\cycle{XXIV}
\submitdate{December 2012}  % \today
\copyyear{2012}  
\advisor{Prof. Vincenzo Marinari }         
\advisor{Dr. Alessandro Giuliani}
\authoremail{serena.cenatiempo@roma1.infn.it}

\examdate{}
\examiner{}
\examiner{}
\examiner{}
\versiondate{December 7, 2012}  % 
\website{}
\ISBN{}

\usepackage{packages/enumerate}   %\usepackage{enumitem} 

\usepackage{amsbsy}
\usepackage{amsthm}
\usepackage{amsfonts}
\usepackage{amssymb}
\usepackage{dsfont}
\usepackage{tikz}

\usepackage{packages/feynmp}
\usepackage{prettyref}   
\usepackage{float}
\usepackage{units}

\newcommand{\Ball}[1]{\fmfv{decor.shape=circle,decor.filled=empty, decor.size=3thick}{#1}}
\newcommand{\bBall}[1]{\fmfv{decor.shape=circle,decor.filled=empty, decor.size=5thick}{#1}}
\newcommand{\BBall}[1]{\fmfv{decor.shape=circle,decor.filled=empty, decor.size=9thick}{#1}}
\newcommand{\SBall}[1]{\fmfv{decor.shape=circle, decor.filled=shaded, decor.size=4thick}{#1}}
\newcommand{\Square}[1]{\fmfv{decor.shape=diamond,decor.filled=empty, decor.size=4thick}{#1}}

\newcommand{\sSquare}[1]{\fmfv{decor.shape=diamond,decor.filled=shaded, decor.size=4thick}{#1}}

\usetikzlibrary{fit}
\usetikzlibrary{shapes.geometric}
\usetikzlibrary{decorations.pathmorphing}

\tikzset{
point/.style={circle,fill=black,inner sep=1pt},
vertex/.style={circle,fill=black,inner sep=1.5pt},   % VERTICI
bvertex/.style={circle,fill=black,inner sep=2.8pt},
Bvertex/.style={circle,fill=black,inner sep=4pt}, % VERTICE GRANDE
specialEP/.style={rectangle,fill=white,draw,inner sep=3pt},  % RETTANGOLO
whitevex/.style={circle,fill=white,draw, inner sep=2pt},
linelabel/.style={sloped,above,very near start, inner sep=1pt,execute at begin node=$\scriptstyle,execute at end node=$},
baseline=(current  bounding  box.center),doubled/.style={double distance= 1pt,line width=1.5pt},
th/.style={line width=0.5 pt, gray},  %linea thin GRIGIA
med/.style={line width=1 pt}  %linea con medio spessore
}

\newtheorem{lemma}{Lemma}[chapter]

\newtheorem{bound}{Result}
\newtheorem{result}{Main Result}

\numberwithin{equation}{chapter}

\newcount\driver
\newcount\bozza

scaled\magstep1                  \font\euftw=eufm10
scaled\magstep1 \font\msytw=msbm10 scaled\magstep1
 
scaled\magstep1                 \font\indbf=cmbx10 scaled\magstep2

scaled \magstep2

{\count255=\time\divide\count255 by 60
\xdef\hourmin{\number\count255}
        \multiply\count255 by-60\advance\count255 by\time
   \xdef\hourmin{\hourmin:\ifnum\count255<10 0\fi\the\count255}}

\let\a=\alpha \let\b=\beta    \let\g=\gamma     \let\d=\delta     \let\e=\varepsilon
\let\z=\zeta  \let\h=\eta     \let\th=\vartheta      \let\l=\lambda
\let\m=\mu    \let\n=\nu      \let\x=\xi        \let\p=\pi        \let\r=\rho
\let\s=\sigma \let\t=\tau     \let\f=\phi    \let\ph=\varphi   \let\c=\chi
\let\ps=\psi   \let\o=\omega     
\let\G=\Gamma \let\D=\Delta       \let\L=\Lambda    
             
\let\O=\Omega

\def\PP{{\cal P}}\def\EE{{\cal E}}\def\VV{{\cal V}}
\def\CC{{\cal C}}\def\HHH{{\cal H}}\def\WW{{\cal W}}
\def\TT{{\cal T}}\def\NN{{\cal N}}
\def\RR{{\cal R}}\def\LL{{\cal L}}
\def\DD{{\cal D}}\def\AA{{\cal A}}\def\GG{{\cal G}}\def\SS{{\cal S}}
\def\KK{{\cal K}}

\def\bT{{\bf T}}

\def\kk{{\bf k}}
\def\pp{{\bf p}}\def\qq{{\bf q}}\def\xx{{\bf x}}
\def\yy{{\bf y}}\def\nn{{\bf n}}
\def\rr{{\bf r}}

\def\bz{{\bf 0}}

       \def\oo{{\underline \omega}}
\def\ee{{\underline \varepsilon}}  

\def\un{{\underline \nu}}

\def\ux{{\underline x}}
\def\uk{{\underline k}}

\def\uy{{\underline y}}
\def\uz{{\underline z}}
\def\uw{{\underline w}}

\def\uQ{{\underline Q}}

  \def\hW{{\hat W}}

\def\hV{{\hat V}}
\def\hv{{\hat v}}
\def \HV{\hat \VV}

\def\bh{{\bar h}}

\def \BV{\bar\VV}

\def\bm{{\bar \m}}
\def\bl{{\bar \l}}
\def\bn{{\bar \n}}

\def\tl#1{{\tilde{#1}}}

\def\qed{\raise1pt\hbox{\vrule height5pt width5pt depth0pt}}

\def\MMM{\hbox{\euftw M}}          
\def\RRR{\hbox{\msytw R}}

        \def\ZZZ{\mathbb{Z}}

\def\Val{{\rm Val}}
\def\indic{\hbox{\raise-2pt \hbox{\indbf 1}}}

\let\dpr=\partial

\let\==\equiv

\let\io=\infty
\let\0=\noindent

\def\pagina{{\vfill\eject}}
\def\*{{\hfill\break\null\hfill\break}}
\def\bra#1{{\langle#1|}}
\def\ket#1{{|#1\rangle}}
\def\media#1{{\left\langle#1\right\rangle}}
\def\bmedia#1{{\bigl\langle#1\bigr\rangle}}
\def\bmed#1#2{{\bigl\langle#1\bigr\rangle_{#2}}}
\def\ie{\hbox{\it i.e.\ }}
\def\eg{\hbox{\it e.g.\ }}

\def\unit{\mathds{1}}   %IDENTITA'

\let\arr=\rightarrow
\def\PS#1{\ps^{(#1)}}
\def\TPS#1#2{\tl{\ps}^{(#1)}_{#2}}

\def\txe{\text{e}}
\def\txi{\text{in}}
\def\cst{(\text{const.})}

\def\TL#1{{\widetilde #1}}

\def\lft{\left}
\def\rgt{\right}

\def\tende#1{\,\vtop{\ialign{##\crcr\rightarrowfill\crcr
             \noalign{\kern-1pt\nointerlineskip}
             \hskip3.pt${\scriptstyle #1}$\hskip3.pt\crcr}}\,}
\def\otto{\,{\kern-1.truept\leftarrow\kern-5.truept\to\kern-1.truept}\,}

\def\Tr{\rm Tr}

\def\sqt[#1]#2{\root #1\of {#2}}

\def\T#1{{#1_{\kern-3pt\lower7pt\hbox{$\widetilde{}$}}\kern3pt}}
\def\VVV#1{{\underline #1}_{\kern-3pt
\lower7pt\hbox{$\widetilde{}$}}\kern3pt\,}
\def\W#1{#1_{\kern-3pt\lower7.5pt\hbox{$\widetilde{}$}}\kern2pt\,}

\def\Re{{\rm Re}\,}\def\Im{{\rm Im}\,}

\def\indica{\leaders \hbox to 0.5cm{\hss.\hss}\hfill}
\def\guida{\leaders\hbox to 1em{\hss.\hss}\hfill}
\mathchardef\oo= "0521

\def\fig#1#2#3#4#5{
\begin{figure}[#1]  %h=here t=top
\centering{
\includegraphics[width=#2\textwidth]{#3} %height=#1\textwidth  height=#1cm
\caption{ #4}  \label{#5}
}
\end{figure}
}

\def\feyn#1#2#3{
\vskip 0.5cm
\begin{figure}[t] %[htbp]
\centering{
{#1}
\caption{\small #2}  \label{#3}
}
\end{figure}
}

\def\feynH#1{
\vskip 0.5cm
\begin{figure}[h] %[htbp]
\centering{
{#1}
}
\end{figure}
}

\def\tik#1#2#3{
\vskip 0.2cm
\begin{figure}[t] %[htbp]
\centering{
{#1}
\caption{\small #2}  \label{#3}
}
\end{figure}
\vskip 0.2cm
}

\def\treelabelsize#1{{\footnotesize #1}}    %\small
\def\ins#1#2#3{\vbox to0pt{\kern-#2 \hbox{\kern#1 #3}\vss}\nointerlineskip}

\newdimen\xshift \newdimen\xwidth \newdimen\yshift
\newcount\griglia   

\def\insertplot#1#2#3#4#5#6{%
\xwidth=#1pt \xshift=\hsize \advance\xshift by-\xwidth \divide\xshift by 2%
\begin{figure}[ht]
\vspace{#2pt} \hspace{\xshift}
\begin{minipage}{#1pt}
#3 \ifnum\driver=1 \griglia=#6
\ifnum\griglia=1 \openout13=griglia.ps \write13{gsave .2
setlinewidth} \write13{0 10 #1 {dup 0 moveto #2 lineto } for}
\write13{0 10 #2 {dup 0 exch moveto #1 exch lineto } for}
\write13{stroke} \write13{.5 setlinewidth} \write13{0 50 #1 {dup 0
moveto #2 lineto } for} \write13{0 50 #2 {dup 0 exch moveto #1
exch lineto } for} \write13{stroke grestore} \closeout13
\includegraphics{griglia.ps} \fi
\includegraphics{#4.ps}\fi%
\ifnum\driver=2 \fi
\end{minipage}
\caption{#5}
\end{figure}
}
\newdimen\shift \shift=-1.5truecm
\def\lb#1{%

\ifnum\bozza=1
\label{#1}\rlap{\hbox{\hskip\shift$\scriptstyle#1$}}
\else\label{#1} \fi}

\def\be{\begin{equation}}
\def\ee{\end{equation}}
\def\bea{\begin{eqnarray}}\def\eea{\end{eqnarray}}
\def\bean{\begin{eqnarray*}}\def\eean{\end{eqnarray*}}
\def\bfr{\begin{flushright}}\def\efr{\end{flushright}}
\def\bc{\begin{center}}\def\ec{\end{center}}
\def\bal{\begin{align}} \def\eal{\end{align}}

\def\[#1\]{\begin{align}#1\end{align}}

\def\ba#1{\begin{array}{#1}} \def\ea{\end{array}}
\def\bd{\begin{description}}\def\ed{\end{description}}
\def\bv{\begin{verbatim}}\def\ev{\end{verbatim}}
\def\non{\nonumber}
\def\Halmos{\hfill\vrule height10pt width4pt depth2pt \par\hbox to \hsize{}}

\pgfdeclarelayer{background}
\pgfsetlayers{background,main}

\begin{document}

\frontmatter

\maketitle

\dedication{
To Luca, my bridegroom, \\[3pt] 
for his constant encouragement %\\[3pt]  
and invaluable support \\[3pt] 
during all the stages of my Ph.D. studies\\[3pt]
and for the wonderful days %
we have shared \\[3pt] 
since last August 29. 
 }

%\begin{abstract}
%\end{abstract}

\begin{acknowledgments}[Acknowledgments]

\vskip 0.5cm

The research leading to these results has received funding from the European Research Council under the European Union's Seventh Framework Programme ERC Starting Grant CoMBoS (grant agreement n$^o$ 239694). 

\vskip 0.2cm

I acknowledge gratefully the Ph.D. program of the Physics Department of ``La Sapienza'' University which provided ideal conditions to make my Ph.D. experience productive and stimulating. In particular I would like to thank Prof. E. Marinari, who helped me to find a direction for my research interests in the early months of my Ph.D. and has always supported my project. I am also thankful with the Physics Department of ``La Sapienza'' and the Mathematics Department of ``Roma Tre'' for the stimulating research environment I experienced. In the past four years I could meet leading researchers in the fields I am interested in and be involved in the current frontiers of physics and mathematical physics. 

\vskip 0.2cm

I wish to express my gratitude to Prof. G. Gallavotti. He is not only an outstanding researcher and inspiring professor, but also a great teacher, with an extraordinary dedication to his students. It has been a great opportunity and pleasure to be a Ph.D. student in his exceptional research team. I will never forget his hearty welcome in his group and his continuous encouragement and interest in my research. 

\vskip 0.2cm

My sincere thanks also go to Prof. G. Benfatto and Prof. V. Mastropietro, which have crucially contributed to this work. Prof. Benfatto introduced me to the problem of Bose condensation from a mathematical physics point of view and was a fundamental and generous guide, always willing for a discussion or some suggestions. The contribution of  Prof. Mastropietro, both in teaching me some of the techniques I used in this thesis and in discussing the physical implications of my work, was precious. I am also indebted to him for having always challenged me throughout these years, never accepting less than my best efforts.

\vskip 0.2cm

It is a pleasure to thank my advisor, Dr. A. Giuliani, for his essential guidance and valuable help in every phase of my Ph.D. Not only without his contributions of ideas, skill and experience this work would not have been possible, but I have learned a lot from him about how to carry out research in mathematical physics. I would like to warmly thank him for all the time he dedicated to our project and my questions, for our enlightening discussions on open problems in mathematical physics and for his wealth of useful advice.

\vskip 0.2cm

Let me conclude with a special thank to prof. R. Figari, my general physics professor and bachelor and master thesis advisor. His teaching skills and research attitude have been crucial for my education and later choices. Thanks to him I have discovered how the clearness of the mathematical rigour and the insight of physical intuition can go together and I have been fascinated by the aims and methods of mathematical physics since the early years of my university studies.

\vskip 0.2 cm

%My time in Rome was also enriched by the friendship of my PhD colleagues at ``La Sapienza'' university. I appreciated a lot the welcoming and stimulating atmosphere of the ``Pi room'' and  the time we spent together, also discussing on our different research fields and interests. Alessandra, Alessandro, Andrea, Antonella, Chiara, Dario, Francesco C., Francesco S., Natascia, Mariana, Matthieu, Paulina, Ulisse: it was great!  

%Finally my thanks to Dr. M. Porta, Dr. R. Greenblatt and Dr. M. Correggi, for many constructive discussions, and to Alessia, Chiara, Federica and Mikaela for their support during the final stages of this Ph.D.

%I thank all my Ph.D. colleagues and friends for the stimulating environment created. (Natascia)

%great friends which have made these years unique. It is impossible to mention all them, but I know 

%\end{document}

\end{acknowledgments}

\tableofcontents

% \chapter{Introduction}
%\input{intro-senza-sapclass} \input{intestazione-sap}  \begin{document} %\usepackage{showkeys}

{\setcounter{equation}{1}
\renewcommand{\theequation}{\roman{equation}.}
\setcounter{figure}{0}
\renewcommand{\thefigure}{\roman{figure}}

\chapter{Introduction}

\vskip 0.2cm

It may seem a long time ago, but it was only in 1995 that
%It may seem a long time ago when, in 1995, 
over the space of few months, three independent and different approaches succeeded in realising Bose--Einstein condensation (BEC) in dilute atomic vapor~\cite{Anderson1995, Bradley1995, Davis1995}, establishing the first clear experimental evidence of the prediction by Einstein\footnote{\,Evidence for superfluidity in liquid He was already obtained in 1938, and although superfluidity was almost immediately connected by London to Einstein's theory of BEC, in these systems the bosons are so closely packed that they can be only understood as strongly interacting systems, miles away from the non interacting description by Einstein. 
}~\cite{Einstein}.

%It sounds a long time ago, when, in 1995, within the space of few months, three independent and different approaches succeeded in realizing Bose--Einstein condensation (BEC) in dilute atomic vapor~\cite{Anderson1995, Bradley1995, Davis1995}  establishing the first clear experimental evidence of the prediction by Einstein\footnote{Evidence for superfluidity in liquid He was already obtained in 1938, and although superfluidity was almost immediately connected by London to Einstein’s theory of BEC, in these systems the bosons are so closely packed that they can be understood only as strongly interacting systems, miles away from the non interacting description by Einstein. }~\cite{Einstein}.

In less than two decades the field of Bose--Einstein condensation of atomic gases has
grown explosively, driven by the combination of new experimental techniques and theoretical advances.  Condensate states have emerged as quantum systems unique in the precision and flexibility wherewith they can be manipulated.  
The ultracold vapour has become an ultralow-temperature laboratory for the communities of atomic physics, quantum optics and condensed matter physics with a great variety of applications: quantum fluids~\cite{vortices},  qubits~\cite{Atom-chip1,Atom-chip2},  Josephson junctions~\cite{BEC_optical}, atom lasers~\cite{AtomLaser}, and the very recent experiments on BEC of photons~\cite{BEC_phonons}, just to mention a few.  Experiments on thin films~\cite{He_films} such as on gases in highly elongated magnetic and pancake-shaped optical traps (see \cite{BEC_2d, BEC_low_dim} and ref. therein), have also pushed forward the study of BEC in low dimensional systems. 

\vskip 0.2cm

However, in spite of its widespread experimental successes, BEC still  challenges us from a theoretical point of view. In fact, thus far, there are very few, also quite special, models in which we are able to prove BEC for interacting bosons. In particular BEC for an homogeneous system has been proved only in the special case of hard core bosons on a lattice at half-filling in three or more dimensions~\cite{hard-sphere-bosons}. 

More recently Bose condensation and superfluidity  have been proved for three and two dimensional bosons in a trap~\cite{Lse,LSeY5}, but only in the Gross-Pitaevskii (GP) scaling limit, a special case of a dilute limit where the density goes to zero as the particle number $N$ goes to infinity\footnote{\,With ``dilute'' we mean that the scattering length $a$ of the interacting potential is much smaller than the mean particle distance, \ie  $\r a^d \ll 1$, with $d$ the dimension of the system. In three dimensions the GP scaling is obtained by requiring $N/L$ to stay constant while $N \arr \io$; this implies that the density of the system goes to zero as $N^{-2}$ with increasing $N$. }.  However, even if the GP limit well reproduces the actual experimental data, its results only apply to certain finite length scales. As well known no claims of whatever phase transition can be made without performing the {\it thermodynamic limit}, where the density is kept fixed while the size of the periodic box where the system is defined tends to infinity. 

\vskip 0.2cm

The longstanding problem one would like to understand is the occurrence of BEC for an homogeneous system of bosons interacting with a repulsive short range potential, in the thermodynamic limit.  This problem has an obvious interest from a theoretical point of view, but it also has a great importance for the experimental physics, since the increasing ability in creating larger and larger condensates is making it necessary to better understand the behavior of bosonic systems beyond the GP limit.

\vskip 0.2cm

%------------------------------------------------FIG

\fig{t}{0.6}{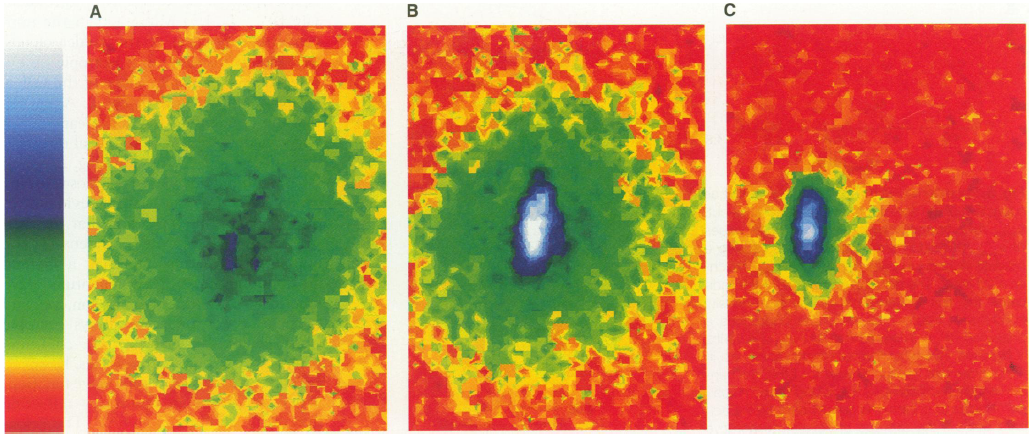}
{\small{{\bf BEC in a vapor of rubidium-87 atoms~\cite{Anderson1995}.} False--color images ($200 \m m $ by $270 \m m$) display the velocity distribution of the cloud. ({\bf A}) just before the appearance of the condensate, ({\bf B}) just after the appearance of the condensate, and ({\bf C}) after further evaporation has left a sample of nearly pure condensate. The circulate pattern of the not condensate fraction (mostly yellow and green) is an indication that the velocity distribution is isotropic, consistent with thermal equilibrium. The condensate fraction (mostly blue and white) is elliptical, indicative that it is a highly non termal distribution. The elliptical pattern is in fact an image of a single, macroscopically occupied quantum wave function.} }{BEC}

%-----------------------------------------------------------------------------------------------------

The usual picture of Bose condensation in the homogeneous interacting case is based on the approximate exactly soluble Bogoliubov model~\cite{Bogoliubov}, which predicts a linear spectrum of excitations for small momenta. This property is considered typical of superfluid behavior, according to Landau's argument~\cite{Landau}. However Bogoliubov's approximation is a quite rough truncation based on assumptions
that are not a priori justified. 
Although Bogoliubov predictions are believed to be correct in the weak coupling limit, a control of  these approximations is to date  beyond reach of rigorous analysis. The problem is that in the attempt of developing a perturbation theory around Bogoliubov solution, is faced with a theory plagued by ultraviolet and infrared divergences, 
%which may a priori totally change the physical properties of the system with respect to the ones predicted by Bogoliubov's. 
whose meaning could be that the interacting system has completely different physical properties with respect to the ones predicted by Bogoliubov's. 

\vskip 0.2cm

Since the early '60, the problem of studying the corrections to Bogoliubov's theory has attracted the attention of the theoretical physics community. The literature on perturbation theory (PT) for Bose condensation at zero or small temperatures is huge; here we can just outline some of the main contributions in this line.  

The first results date back to the works of Beliaev (1958,~\cite{Beliaev}) and Hugenholtz--Pines (1959,~\cite{Beliaev, Pines}). They both performed PT for zero temperature interacting bosons in the low density limit and obtained series expansion for the ground state energy and the phonon spectrum. The work of Hugenholts--Pines differs from Beliaev's for the fact that the zero--momentum state has been removed, 
in the same spirit of Bogoliubov's work. The choice of performing perturbation theory around Bogoliubov's solution rather than the non interacting one was then acquired in all the later papers on the subject. Hugenholts and Pines also showed, again with perturbative arguments, that the condition of minimum free energy can be equivalently posed as a condition on the structure of the perturbation theory, in particular as a condition on the one particle irreducible diagrams with two external lines.

In the same years, in the last paper of a series of five works~\cite{LeeYang5}, Lee and Yang showed that the thermodynamic functions of an interacting system of bosons in the condensed phase can be expressed in terms of the average occupation number. They  developed a variational principle which enabled them to compute the thermodynamic functions and the occupation number of the single particle state with momentum $k$
%the average of the particles with momentum $k$ 
in the condensed phase. As an example, the method was applied to a dilute system of Bose hard spheres, obtaining a low--density expansion for the free energy, with an explicit evaluation of the first few terms. 
%in particular a second order expression of the free energy, which had already been obtained by the same authors together with 
A similar result was due to the same authors with Huang by using a different method~\cite{Lee-Huang-Yang}.

%was mainly concerned with showing that 
In the work of Gavoret and Nozi\'eres~\cite{Gavoret} the excitation spectra for the interacting Bose gas for low momenta was obtained by combining perturbation theory and the use of Ward identities (WIs). The authors also showed that the spectra of quasi--particles and of the density--fluctuations excitations are identical for low momenta, \ie they both correspond to phonons with the usual macroscopic sound velocity. This is mathematically expressed by the identity of the poles of the one particle Green's function and of the correlation function of two elementary excitations. Other results on the superfluid behavior at zero temperature were obtained in the later works by Nepomnyashchii and Nepomnyashchii~\cite{Nepom} and  Popov and Seredniakov~\cite{Popov}. 

All the works we have just mentioned are based on diagrammatic techniques borrowed from Quantum Field Theory. However the results which have been achieved are obtained by summations over {\it special classes of diagrams} selected from the divergent perturbative series, and do not provide a systematic study of the divergences affecting the theory. 

\vskip 0.2cm

The first to obtain a fully consistent study of the infrared divergences for the  three dimensional weak interacting system at zero temperature, using exact  Renormalization Group (RG) techniques, was Benfatto in 1994~\cite{benfatto}. With the aim of studying the occurrence of BEC, which only depends on the long--distance behavior of the system, Benfatto considered a model with an ultraviolet momentum cutoff. For this model he proved that the theory is order by order finite in the running coupling constants, with explicit bounds on the coefficient of order $n$. The importance of Benfatto's work lies in the fact that it represents a strong justification of the generally accepted picture of BEC in a scheme which is the only one allowing in some cases to perform a full non perturbative construction, i.e. to control the convergence of the series defining the generating functions. 
On the other side, Benfatto strategy requires the study of the flow equations of six effective couplings and does not have any chance to be extended to the two dimensional case, where the thee and four body interactions are relevant in the RG sense.

\vskip 0.2cm

After Benfatto's work, Pistolesi et. al.~\cite{CaDiC1,CaDiC2} showed that the study of the flow equations for the effective parameters of the interacting boson problem can be drastically simplified by implementing local Ward identities. These reduce the number of independent running couplings to only two, allowing the authors to attack the two dimensional problem, for which they have found a non trivial fixed point and no anomalous dimensions. 
Pistolesi's work is bases on a non--rigorous RG scheme, which uses a {\it dimensional regularization} and does not allow, not even in principle, to fully construct the theory.
%The idea of exploiting the symmetries of the problem, even if applied within a non rigorous RG scheme, based on {\it dimensional regularization}, which do not allow even in principle to fully construct the theory, represents a crucial benchmark for our work. 
Still, the idea of exploiting the symmetries of the problem represents a crucial benchmark for our work. 

\vskip 0.2cm  

The inspiration of this thesis is to work up the ideas of Pistolesi et al. within the same {\it exact Renormalization Group} approach used by Benfatto, based on a {momentum regularization} scheme, in the same spirit of the RG approach by Wilson~\cite{Wilson}. Here by ``exact RG''  we mean that our method allows to obtain a construction {\it at all orders} of the thermodynamic functions and correlations, with explicit bounds on the $n$--th order coefficients. This result is obtained without neglecting the effects of the irrelevant terms, \ie the terms that become dimensionally smaller under the iterations of the RG transformation, but can still give {\it finite} contributions to the thermodynamic and correlation functions. 

\vskip 0.2cm

The technique we use is borrowed from the methods of {\it rigorous} or {\it constructive Renormalization Group}, in the form developed by Gallavotti in the 80's~\cite{GalReview} to study the ultraviolet stability of scalar fields.
In the context of fermionic systems the perturbative methods that we employ also provide, in certain cases, a way to {\it fully construct} the ground state of the interacting system in the weak coupling regime, \ie to prove the convergence of the resummed perturbation theory. 
Examples in this line,
%Some examples in this line  %Some examples in which this result was achieved
 in the context of low dimensional condensed matter systems, are the one--dimensional interacting fermions~\cite{BM-luttinger}, the 2d Hubbard model on the square lattice at positive temperature~\cite{BGM-Hubbard} and the short range half--filled 2d Hubbard model on the honeycomb lattice \cite{SRGraph1,SRGraph2}. 

In the context of bosonic theories the situation is quite different. In fact a {\it non perturbative construction} of the model cannot be obtained without combining perturbation theory with complementary methods. For example, in the case of the well known $\f^4_2$ and $\f^4_3$ real quantum field theories, the rigorous construction of the theory is obtained by combining perturbation theory with ``large fields'' estimates (see~\cite{GalReview, RV, Brydges,GK} for reviews).  Estimates of this sort are not available in the current context, since the reference Gaussian measure, which we are perturbing around, is a complex one, with semi--positive definite covariance; no method is currently available for the construction of perturbations of such complex measures. See however~\cite{BTFK_ultraviolet} for attempts in this direction. 

In this thesis we do not seek to solve the problem of construct the bosonic theory, but we aim to ``just'' construct the theory at all orders. Still in the latter case it is crucial to understand the effect of a momentum regularization, being the momentum regularization scheme the only one allowing in principle to perform a full not perturbative construction of the theory.  In fact if one wants in perspective to built the model it is essential to introduce momentum cutoffs, and then prove the convergence of the series defining the generating functions as long as the cutoffs are removed. 

It is also likely that WIs will play an important role in a future full construction of the theory, and therefore it is important to understand how they are implemented in a Wilsonian RG scheme. 

\vskip 0.2cm

%Coming back to our purpose, 
Regarding this issue, \ie to combine local WIs within an exact RG scheme based on a multiscale momentum decomposition, one should stress that this is not a trivial task at all. In fact 
%Wilsonian RG methods are based on a multiscale 
the momentum decomposition breaks the local gauge invariance, which Ward identities are based on. The presence of cutoffs then produces corrections to the ``naive'' (formal) WIs. Even if formally these corrections go to zero when we remove the cutoffs, they cannot naively be neglected, since they are dimensionally marginal in a RG sense. In some low dimensional systems of interacting fermions they result to be crucial for establishing the infrared behavior of the system. For example in Luttinger liquids if one uses the formal WIs, without their corrections, the anomalous dimensions are not found. Then one may be worried that the presence of these corrections may substantially change the structure of WIs, particularly in the two dimensional case, as well as the results obtained by Pistolesi et al. 

\vskip 0.2cm

A control of the corrections to WIs coming from the presence of cutoffs may be obtained
%This goal is achievable 
thanks to a remarkable technique developed by Benfatto and Mastropietro in~\cite{BM-luttinger}, which allows to exactly implement WIs within constructive RG scheme. 
%
%Using the technique developed in~\cite{BM-luttinger} for the rigorous analysis of Luttinger liquids it is possible get a \blue{rigorous and complete treatment of the effects of the cutoffs} within the constructive RG scheme. 
%
Pursuing this analysis to our system, the correction terms to the formal WIs appear as new marginal (in 3d) or relevant (in 2d) terms, which can be in turn written as series in the effective parameters appearing in the generating functions and, again, explicitly bounded at all orders. We remark that, since the corrections to WIs turn out to have the same ``dimensions'', in the RG language, of the other terms appearing in the formal WIs, they may possibly be responsible for anomalous dimensions as in the Luttinger liquid case. 
On the contrary, quite unexpectedly, one finds that they are of higher order in the small parameter $\l$ with respect to the terms already present in the formal WIs, and do not change qualitatively 
%the way in which the local WIs are used to control the flow equations
the conclusions obtained by Pistolesi et al., 
even in the two dimensional case.

\vskip 0.2cm

It has to be stressed that from a quantitative point of view, the corrections to local WIs, even if subleading, are possibly observable in the relations among the thermodynamical and response functions which can be derived from local WIs. 

\vskip 0.2cm

In this thesis we have studied with exact RG techniques a simplified model for a zero temperature three and two dimensional system of bosons interacting with a weak repulsive short range potential, obtained by introducing an ultraviolet momentum cutoff. In order to formulate our main result, 
we need to describe in some more detail  the strategy that we follow. For the purposes of this introduction we will outline the model in a quite informal way, referring to chap. \ref{model} for a more detailed description.

\pagina

\subsubsection{The model}

We consider a system of $N$ interacting bosons with mass $m$ in a $d$--dimensional box $\O$ of volume $|\O|=L^d$ with periodic boundary conditions. The Hamiltonian of the system is 
\[
H_{\O,N}=-\frac{\hbar}{2m}\sum_{i=1}^{N}\D_{\xx_{i}} +\lambda\sum_{1\leq i<j \leq N} v\left({\xx}_{i}-{\xx}_{j}\right)
\]
where $\xx_i$ is the position of the $i$-th particle and $\l\,v(\xx)$ is a weak repulsive short range potential, with $0<\l \ll 1$ representing the intensity of the interaction. We work in the grand canonical ensemble with  chemical potential $\m$ fixed in such a way that the system has fixed density $\r$ as $L \arr \io$. We choose units in such a way that $\hbar =2m=1$.

A formal but convenient way to calculate the partition function and every correlation functions of the previous system consists in writing their {\it coherent state path integral representation}~\cite{NO}.  The latter allows one to express in a compact way the perturbative series in $\l$ for the partition function and the correlations in terms of the correlation functions of the free system. If $Z_\L^0$ is the non--interacting partition function, the partition function $Z_\L$ of the interacting system at $\b^{-1}$ temperature in terms of the coherent states, here denoted by $\ph_{\xx,t}^{+}=(\ph_{\xx,t}^{-})^{*}$, is  given by
\[
\frac{Z_\L}{Z^0_\L} = \int P^0_\L(d\ph)\,e^{-V_\L(\ph)} 
\]
with $\L=[-\b/2, \b/2] \times [-L/2, L/2]^d$ and potential
\[
V_\L(\ph) =&\, \frac{\l}{2} \int_{\O \times \O} d^d \xx \;d^d \yy \int_{-\b/2}^{\b/2} dt \, |\ph_{\xx,t}|^2 \,v(\xx-\yy)\, |\ph_{\yy,t}|^2   \non \\
			& \qquad -\m \int_\O d^d\xx \, \int_{-\b/2}^{\b/2} dt\,|\ph_{\xx,t}|^2 
\]
The measure $P^0_\L(d\ph)$ is a complex Gaussian measure whose covariance in the zero temperature thermodynamic limit is given by
\[ \label{S}
 S^0(x,y) = \lim_{\b, L \arr \io} \int P^0_\L(d\varphi) \varphi_{x}^{-}\varphi_{y}^{+} =
\rho_0 \,+\,\frac{1}{(2\pi)^{d+1}} 
\int_{\RRR^{d+1}}  
d^d\kk \,dk_0 \,\frac{e^{-ik x}}{-ik_{0}+\kk^{2}}  
\]
We remark that the function $S^0(x,y)$ is a correlation function generalized to imaginary time -- the so called {\it Schwinger function} -- and the corresponding correlation function for the free system is obtained by taking in \eqref{S} the limit  $t \arr 0^-$. The first part of the Schwinger function \eqref{S} has the interpretation of the density of the condensate; the second term is the slowly decaying part of the correlation function in the non interacting case. \\

In order to study the occurrence of condensation in the interacting case, we assume a spontaneous symmetry breaking of the $U(1)$ symmetry of the system, by fixing a priori the condensate density. Then we try to fix the chemical potential $\m$ in order to generate a model whose condensate physical density is the prescribed one. This means that we require  the interacting $2$--point Schwinger function to converge to $\r_0$ in the zero temperature thermodynamic limit. If under this requirement we manage to prove that the perturbation theory around Bogoliubov model can be expressed in terms of series in the effective parameters with finite coefficients, this can be interpreted by saying that the correlation function thus obtained describes a Bose condensate state with condensate density $\r_0$ and chemical potential $\m$. In this approach we are regarding the condensate density $\r_0$ as a physical constant and $\m$ as a bare constant to be fixed to generate a model whose physical density is the prescribed one.  \\

With this aim, the steps leading to Bogoliubov approximation are reinterpreted within the functional integral scheme. First, inspired by \eqref{S}, we write the bosonic field $\ph^-_x$ as the sum of two independent Gaussian fields: the first, translational invariant, corresponds to the $k=0$ component of the original fields, having average $\r_0$; the second field, whose covariance is given by the second term in the r.h.s. of \eqref{S},  represents the fluctuations with respect to the condensed state. Then the replacement of the bosonic operators associated to the condensate state by c--numbers, the so called {\it c--number substitution} (see \eg~\cite{Book-Lieb}),  corresponds to writing 
 $\ph^-_x=\sqrt{\r_0}\,+\, \ps^-_x$, as explained with more details in sec. \ref{reinterpret}. 

With this substitution the potential $V_\L(\ph)$ can be rewritten as the sum of three terms: a first term depending only on $\r_0$, the term $Q_{\L}(\ps)$ which is  quadratic in the $\ps$ fields and a third term $\VV_\L(\ps)$ containing the cubic and quartic terms in the fluctuation fields.  Bogoliubov model is obtained by neglecting the latter term. 

At this point we include the quadratic potential $Q_\L(\ps)$  into the free measure of the $\ps$ fields. The covariance $g_{-+}^{B}(x)$ of the new measure $P_{B,\L}(d\ps)$ thus obtained represents the two point Schwinger function in the Bogoliubov approximation. It has a very different large distance behavior, with respect to the free correlation in \eqref{S}:
\[ \label{gB}
& g_{\a  \a'}^{B}(x)=\frac{1}{(2\pi)^{d+1}}\int d^{d}\kk dk_{0}\, e^{-ikx}\,
g^B_{\a \a'}(k) 
\]
with
\[
\lft(g^B_{\a \a'}(k) \rgt)^{-1}
=  \left(\begin{array}{cc}
-ik_0+\kk^2 +\l \hv(\kk) \r_0   & \l \hv(\kk) \r_0  \\[3pt]
\l \hv(\kk) \r_0  \quad &  ik_0 +\kk^2  + \l \hv(\kk) \r_0 
\end{array}\right)   
\]
 with $\hv(\kk)$ the Fourier transform of $v(\xx-\yy)$. Since the condensation problem depends only on the long--distance behavior of the system we consider a simplified model, obtained by modifing \eqref{gB} in
\[ \label{gB2}
& g_{\a  \a'}^{B\,(\leq0)}(x)=\frac{1}{(2\pi)^{d+1}}\int d^{d}\kk dk_{0}\,\chi_{0}(k) \, e^{-ikx}\,g^B_{\a \a'}(k) 
\]
with the function $\c_0(k)=\c_0(|k)|$  plaing the role of a prefixed ultraviolet cutoff. \\

With this formulation the calculation of the corrections to Bogoliubov model for a system which exhibits condensation with condensate density $\r_0$  is effectively equivalent to the study of the effective potential $\WW_{\L}(\r_0)$, defined by 
\[ \label{introW}
e^{-|\L|\,\WW_\L(\r_0)}= \int P^{(\leq 0)}_{B,\L}(d\ps)\,e^{-\VV_{\L}(\ps)}
\]
with $P^{(\leq 0)}_{B,\L}(d\ps)$ the measure with covariance \eqref{gB2} and
\[ \label{V}
\VV_\L(\ps)  = &\, \frac{\l}{2}\,\hv(0) \int_{\L}  \left( \psi^+_x \ps^-_x \right)^{2}dx
+ \l \hv(0) \sqrt{\r_0} \int_{\L} \psi^+_x \ps^-_x \lft( \psi^+_x +\ps^-_x \rgt) dx \non \\
& \qquad + (\l \hv(0) \r_0 -\m) \int_\L \psi^+_x \ps^-_x 
\]
the part of the interacting potential neglected in Bogoliubov approximation, after the c--number substitution. Note  that we have substituted everywhere $\hv(\kk)$ with $\hv(\bz)$ since the cutoff function is chosen in such a way that the potential appears local on the energy scales we consider. 
% (very different from the behavior of the free system, compare with the second term in the r.h.s. of \eqref{S}). 

%The calculation of the corrections to \eqref{gB2} due to the interaction \eqref{V} is obtained by studying the potential $\WW_{\L}(\r_0)$, in \eqref{introW}. As already

\vskip 0.2cm

The integral in \eqref{introW} is studied by a multiscale analysis, that is we  iteratively integrate  the fields of decreasing energy, starting from the momenta ``close'' to the ultraviolet momentum cutoff moving towards smaller momentum scales.  At each step the integral is rewritten as an integral involving only the momenta smaller than a certain value proportional to $2^{-h}$. At each step $h$ of the integration the effective potential may be written as the sum of a part containing the ``relevant'' and ``marginal'' terms (\ie terms which dimensionally tend to grow along the iterative integrations and are responsible for the divergence of the theory)  and a second part containing the ``irrelevant'' terms in the RG language.

The marginal and relevant terms  have the same structure of the initial potential and then, modulo the irrelevant terms, we get an effective theory very similar to the original one, except for the presence of new effective parameters and of  a ``dressed propagator'', obtained by including the quadratic marginal terms in the measure. {\it The ``effective'' parameters differ from their ``bare'' counterparts because the physical parameters appearing in their definitions are renormalized by the integration of the momenta on higher scales. }

\vskip 0.2cm

Using estimates based on the Gallavotti--Nicol\`o tree expansion one may prove that the irrelevant terms are all bounded if the effective couplings are bounded. The problem we are left with is to study how the effective couplings evolve under the multiscale integration. Each effective coupling at a fixed scale  is exactly written as a series in the effective couplings at larger scales. Even if the theory is renormalizable (\ie  the number of effective parameters is finite), to prove that there exist some initial values of the effective couplings such that the flow remains finite is quite complicate in $3d$ and impossible in $2d$, without the use of symmetries. The latter point needs some attention. While in the 3d case the use of WIs allows us to solve problem with a more satisfactory method than the one followed by Benfatto, but is not crucial, the use of WIs in the two dimensional case is even more crucial than what already pointed by Pistolesi et al.   

In fact our RG scheme allowed us to identify in the $2d$ case three new effectively marginal terms, which have not been identified before. As discussed in details in sec. \ref{cap2_ren_prop}, when one takes into account the non local terms which are present in the flow equations for the running coupling constants and does not neglect the interactions among coupling constants at different momentum scales, one finds that the {\it effective scaling dimensions} of the terms arising in the $2d$ perturbation theory are different from the {\it na\"ive scaling dimensions}. The two dimensional problem then appears incredibly complicate: not only two of the seven coupling constants already present in the $3d$ case become relevant in $2d$, but there are four additional marginal couplings. The latter are the three--particle effective interaction $\l_{6,h}$ -- already recognized by Pistolesi et al. -- and then three {\it effectively marginal} couplings arising from the interaction among different scales. The goal of solving the flows of these eleven couplings may seem hopeless. Amazingly, one finds that the additional four marginal couplings present in the $2d$ case are related among them by three global WIs, which again allow to reduce the number of  independent couplings to one, namely $\l_{6,h}$. In the two dimensional case, thanks to WIs, the eleven flows equations are traced back to the study of only two independent, coupled, flows equations. That is why the use of  WIs represents a key point of our analysis. 

\vskip 0.5cm

\subsubsection{Main results}

Using multiscale RG methods, we construct a renormalized expansion, allowing us to express the partition function of the system 
%and the Schwinger functions, from which the physical observables can be computed,
 as series in the effective couplings with {\it finite} coefficients at all orders, admitting explicit $n!$ bounds (see results \eqref{thm_nfactorial0}, \eqref{thm_nfactorial2} and \eqref{thm_nfactorial3} sec.~\ref{multiscale}). Now, if these effective couplings remain small in the infrared, the informations obtained from our expansion by lowest order truncations are reliable at weak coupling. Using the combination of multiscale methods and global and local WIs we succeed in proving that: \\

%Equazioni di flusso ridotte allo studio di un'unica equazione di flusso/unico parametri effettivo cosa non banale perche' il numero di.. e' 7 in 3d e 8 in 2d. Nota. non ci serve che $\l_h$ tenda a zero ma solo che resti piccolo. La prova del main result 1 usa in maniera non banale le WI. 

{ {\bf Main Result (1).} \it \label{R1} The flow equations for the effective couplings -- which are seven in the 3d case and eleven in the 2d case -- are reduced to the study of a unique effective parameter in $3d$ and two effective parameters in $2d$. 

Then all the running coupling constants stay small in the infrared, and  the interacting theory is well defined at all orders, provided that: in $3d$ the effective parameter $\l_h$ related to the intensity of the two--particles interaction stays small; in $2d$ the effective parameters $\l_h$ and $\l_{6,h}$
are such that $\l \l_h$  and $\l_{6,h}/(\l \l^2_{h})$ stay small. Here $\l_{6,h}$ is the effective parameter related to the intensity of the three--particles interaction and $\l$ in the small parameter giving the intensity of the interacting potential.\\
}

Moreover, in 3d we  prove that the flow of $\l_h$ has an asymptotically free flow in the infrared limit, as already established by Benfatto. In the 2d case a one--loop calculation shows that both $\l \l_h$ and $\l_{6,h}/(\l \l_h^2)$ admit fixed points of order one in the infrared limit $h \arr -\io$. This means that the perturbative scheme in $2d$ is not completely consistent, unless the fixed points are numerically so small that the perturbation theory makes sense. Of course, proving such a statement is beyond reach of the rigorous methods that we employ here. It may in principle be possible to play with other parameters, such as the condensate density or the range of the interacting potential, to make those fixed points smaller, see discussion on sec.~\ref{lambda6}.  

%Besides the quantity $\l_{6,h}/(\l \l^2_{h})$ stays small, while the product $\l \l_h$ is of order one, so the assumption of the main result (1) appear to be not unreliable. However the problem of proving that $\l_h$ has a fixed point of order one is beyond the possibility of a perturbative theory and then remains an open problem. \\

\vskip 0.2cm

We stress here that in the work by  Pistolesi et al.  the study of the marginal coupling $\l_{6,h}$ was completely neglected. Still the presence of $\l_{6,h}$ and of the other effectively marginal terms modify the leading order flow equation for $\l_h$, changing the value of its fixed point. Remarkably, the presence of these new couplings do not affect the conclusions on the behavior of the propagator, listed in the following, which in the $2d$ case are only based on the fact that $\l_h$ admits a fixed point.  \\

Let us denote by $g^{-+}_\l (x) = (2\pi)^{-d-1}\int d^{d+1}k\, e^{ikx}g^{-+}_\l (x) $ the renormalized propagator for the decaying fields $\ps^\pm_x$ in the infinite volume limit; under the previous hypothesis on the effective couplings our main result on the asymptotic behavior of the propagator can be informally stated as follows. \\

%Let us denote by $\media{\ldots}= \lim_{|\L|\arr \io} \media{\ldots}_\L$ the expectation value with respect to the interaction \blue{?} in the infinite volume limit; our main result on the asymptotic behavior of the propagator can be informally stated as follows.  \\

{ {\bf Main Result (2).} \it  There exist a choice of the counterterm $\m$ such that, both in three and two dimensions, for $k=(k_0, \kk)$ small, the expression of the renormalized propagator is
\[
g_\l^{-+}(k) \simeq \frac{\AA(\l)}{k_0^2 + c^2(\l)\,\kk^2}
\]
where ``$\simeq$'' means that we are considering the dominant singularity in $k$ as $k \arr 0$ and $\AA(\l)$ and $c(\l)$ are expressed by a series in the effective couplings with finite coefficients that admit $n!$--bounds at all orders. The first non trivial contribution to $\AA(\l)$ is $\l \r_0 \hv(\bz)$.   Regarding  the singularity of the propagator $k_0=\,\pm c(\l) |\kk|$, which has the physical interpretation of the dispersion relation of quasi--particles, we find that in our effective model
\[
c^2(\l)= c^2_B(1+ \cal{B}(\l))
\]
with $c_B=\sqrt{2\l \r_0 \hv(0)}$ the speed of sound predicted by Bogoliubov approximation and $\cal{B}(\l)$ given by a series whose construction is defined at all orders. In particular $\cal{B}(\l)$ goes to zero as $\l$ approaches zero.
%With `` $O(\l)$'' we mean that the corrections to Bogoliubov sound speed in the non approximate model can be expressed by series in the effective couplings with finite coefficients at all orders, whose first non--trivial contribution is of order $\l$.
} \\

It is an interesting feature that Bogoliubov linear spectrum is found to be independent of the dimension of the system, being exactly constrained by Ward identities.  In particular the correlations do not exhibit anomalous dimensions, i.e. the model is in the same universality class of the exactly soluble Bogoliubov model. The latter result is absolutely not trivial since one would expect such a situation in a super renormalizable and asintotically free theory, but not in a case in which we have, as in 2d, two not trivial fixed points.

\subsubsection{Summary}

The plan of the work we have just outlined is detailed along the thesis according to the following scheme. 

\vskip 0.1cm

In chap.~\ref{model} we review the concept of BEC for interacting bosons and get the exact solution of Bogoliubov by using a coherent state path integral representation for the partition function of the system.  The predictions of Bogoliubov approximation for the ground state energy and the chemical potential are calculated, both in $3d$ and $2d$. Then we state the effective model we are interested in. This will represent the starting point for the subsequent perturbative analysis. 

\vskip 0.1cm

In chap.~\ref{chap_multiscale}  we describe the multiscale analysis applied to the partition function. This chapter includes detailed analysis of the divergences affecting the ``naive'' perturbation theory, the definition of the renormalized expansions and the assumptions on the effective couplings which make the expansion meaningful.
We also describe the expansion for the generating functional of the density and current correlation functions, whose gauge invariance is used to derive global and local WIs. 

Due to the fact that our unperturbed reference model is Bogoliubov's two different regimes arise, in which the RG procedure must be defined differently. A ``high'' momenta regime, just below the ultraviolet momentum cutoff, in which Bogoliubov potential is negligible and the reference unperturbed model is the non interacting system; a ``low'' momenta regime, where Bogoliubov contribution dominates, which shows the most interesting features. In particular in the two dimensional case, in order to control the theory in the low momenta regime, it is necessary to introduce an ``effective scaling dimension'' and three new effective marginal parameters, as described in sec.~\ref{cap2_ren_prop}. 

\vskip 0.1cm

In chap.~\ref{flows} we study the flow of the running coupling constants in both  regimes, using the bounds previously derived in chap.~\ref{chap_multiscale} and some global and local WIs, which reduce the number of independent running couplings, this fact being crucial for the control of the two dimensional theory.  In this chapter the main results are given in a more detailed way than in this introduction.

\vskip 0.1cm

In chap.~\ref{WI} we derive the global and local WIs which have been used to control the flow of the running coupling constants in the low momenta regime. The global WIs also allow to classify the terms that can possibly appear in the theory by symmetry reasons and to state the renormalization condition in terms of properties of the flow of the effective chemical potential. The effect of the corrections to WIs due to the presence of cut--offs is also analyzed.

\vskip 0.1cm

Finally in chap.~\ref{conclusions} we draw the conclusions and briefly overview some of the perspectives which our work points to.

\vskip 0.1cm

In the remaining Appendices we collect a number of technical lemmas needed for the proof of the main result. Some second order computations have been also reported, in particular to make clearer some of the statements that thanks to the use of WIs are proved at all orders. \\

So, let's start. 

} %fine set counter e numerazione romana

\mainmatter

%\chapter{The model}
%\input{intro-senza-sapclass} \input{intestazione-sap} \begin{document} 
%\tableofcontents

\chapter{The model} \label{model}

\section{Definition of the problem}
  % Questo capitolo ha come obiettivo la formulazione del MAIN RESULT in maniera precisa

We are interested in the study of the properties of a gas of bosons of mass $m$ in a $d$ dimensional box $\O$ of side $L$ interacting via a weak repulsive two-body potential. The corresponding Hamiltonian is 
	\be \label{eq:Hamiltonian}
 H_{\O,N}=-\frac{\hbar^2}{2m}\sum_{i=1}^{N} \Delta_{\xx_i}+\lambda\sum_{1\leq i<j\leq N}v\left(\mathbf{x}_{i}-\mathbf{x}_{j}\right)
	\ee
where $\xx_{j}$ is the position of the $j$-th particle and the potential is assumed to be
non-negative, decreasing faster than $1/|\xx|^{d}$ at infinity (such potentials are simply called {\it repulsive potentials}, in quantum mechanics literature) and rotationally invariant. The parameter $\l \geq 0 $ gives the strength of the interaction. We  take periodic boundary conditions and assume  $v$ can  be periodically extended to $\RRR^{3}$.  The operator $H_{\O,N}$ acts on the symmetric subspace of the Hilbert space $\bigotimes^N L^2(\O)$, which we denote with $\HHH_{N,s}$.
We will work in the grand canonical ensemble considering the following Hamiltonian
\be 
H_\O=\bigoplus_{N\geq0}(H_{\O,N}-\m_{\O,\b} N) 
\ee
acting on the symmetric Hilbert space $\HHH_s\equiv \bigoplus_{N\geq0} \HHH_{N,s}$, with  $\m_\O$  the chemical potential, fixed in such a way that the total density of the system $\rho=\left\langle N\right\rangle /L^{d}$ is fixed. The thermodynamic properties of the system are obtained by averaging with respect to the Gibbs measure at inverse temperature $\b$. In particular we are interested in the study of the ground state and low temperature properties of the system, that is in computing:
\begin{itemize}
	\item the specific free energy at low temperatures (i.e. $\beta \gg 1$):
\be
f(\beta,\rho)=-\lim_{L\arr+\infty}\,\frac{1}{\beta L^d}\log\Tr_{\HHH_{s}}e^{-\beta(H_{\O,N}-\m_\O N)}
\ee
%with $\m$ fixed in such a way that the system has fixed density $\rho$; 	
	\item the specific ground state energy $e_{o}(\rho)=\lim_{\b\arr +\io}f(\b,\rho)$;
	\item the one-particle reduced density matrix 
%$$\left\langle a_{x}^{+}a_{y}\right\rangle =
$$S(\xx,\yy)= \sum_{N \geq 0} N\int d\xx_{2}\ldots d\xx_{N}\Gamma_N(\xx,\, \xx_{2},\,\ldots,\, \xx_{N}\,;\, \yy,\, \xx_{2},\,\ldots,\, \xx_{N})$$ 
and higher order correlation functions. Here $\Gamma_N$ is the density matrix in the sector with fixed number of particles $ N $. In particular $S(\xx,\yy)$ is related to the occurrence of BEC for interacting systems. In fact, according to the definition by Penrose and Onsager~\cite{Penrose-Onsager}, BEC is said to occur if $S(\xx,\yy)$ has an eigenvalue of the order of $\langle N \rangle$ in the thermodynamic limit. This is in particular true if the one--particle density matrix shows a long range order, i.e. tends to a constant in the thermodynamic limit~\cite{Yang-ODLRO}. 
	\item the density--density and current--current response functions, respectively the response of the system density to an infinitesimal perturbation proportional to the density of particles or to the probability flux. 
\end{itemize}

It is convenient to rewrite the Hamiltonian \eqref{eq:Hamiltonian} in the second quantization formalism, by introducing the creation and annihilitation operators $a^{+}_{\kk}$ and  $a_{\kk}$: 
%We define the number operators $N_{\mathbf{k}}$ and $N$ by 
%\begin{equation}\label{numberop}
%N_{\mathbf{k}} =  a^{\dagger}_{\mathbf{k}}a_{\mathbf{k}},\quad 
%N =   \sum_{\mathbf{k}} N_{\mathbf{k}} .
%\end{equation}
\begin{equation}\label{Ha}
H_{\O}  =  \sum_{\mathbf{k} } \KK (\mathbf{k}) \ 
a^{+}_{\mathbf{k}} a_{\mathbf{k}} \ 
+ \  \tfrac{\lambda }{2 |\O |}   \sum_{\mathbf{k},\mathbf{q},\mathbf{p} } 
\hat{v} (\mathbf{k}) \ 
a^{+}_{\mathbf{q} +\mathbf{k} }
a^{+}_{\mathbf{p}-\mathbf{k}} a_{\mathbf{q}}
 a_{\mathbf{p}}
\end{equation}
where $\KK(\kk) = \tfrac{\hbar^{2}}{2m} \,\kk^{2}$, $\hv(\kk)=\sum_\xx e^{i \kk \cdot \xx} v(\xx)$ and
%Here the subscript $\O$ in $\KK_\O(\kk)$ has been introduced to remind that, in the finite volume case, the variable $\kk$ is given by 
$\kk = 2\pi \,\nn/L$, with $\nn \in \ZZZ^d$. The Hamiltonian in~\eqref{Ha} acts on the Hilbert space of $N$-particles symmetric wave functions $\HHH_s$.
%The total number operator is given by $N =\sum_\kk  a^{+}_{\kk} a_{\kk}$. 
In what follows we will choose units in such a way that $\hbar=2m=1$\footnote{With this choice, the dimensions of the physical quantities speed ($c$), momentum ($\kk$), frequency ($k_0$) and energy ($E$) are respectively: $ [c] = [\kk] = [L]^{-1}$ and $[E] =[k_0]= [L]^{-2}$.}.  

Let now indicate with $a^\pm_\xx$ the bosonic field operators, related to the annihilation and creation operators as follows:
\be \label{a}
a^\pm_\xx= \frac{1}{L} \sum_{\kk \in \DD_L} a^\pm_\kk e^{\pm i \kk \cdot \xx} 
\ee
The equilibrium properties of the bosons system at $\b^{-1}$ temperature and in the grand canonical ensemble with chemical potential $\m_{\b}$, i.e. the average of any observable on the system, can be obtained once the following functions, the $2n$--point correlation functions or $n$--particle density matrices, are known:
\be \label{eq:corr_func}
S(\xx_1\,\ldots\,\xx_{2n})=\lim_{L\arr\io} \frac{ \Tr_{\HHH_s} \left[e^{-\b (H_{\O}-\m_{\O,\b} N)} a^+_{\xx_1}\ldots a^+_{\xx_n} a^-_{\xx_{n+1}}\ldots a_{\xx_{2n}}^-\right] }{\Tr_{\HHH_s} e^{-\b (H_{\O}-\m_{\O,\b} N)}}
\ee
In particular each single particle observable can be calculated once the $2$--point correlation function is known. The first example one has in mind is the occupation number $N_\kk=\media{a^+_\kk a^-_\kk} $, i.e. the average number of particles with momentum $\kk$, which corresponds to the Fourier transform of the two--point correlation function:
\be
S_{+-}(\xx, \yy)= \frac{ \Tr_{\HHH_s} e^{-\b (H_{\O}-\m_{\O,\b} N)} a^+_\xx a^-_\yy}{ \Tr_{\HHH_s} e^{-\b (H_{\O}-\m_{\O,\b} N)}} 
\ee
The linear response of a physical observable to an infinitesimal external perturbation is also connected with the  correlation functions. For instance the density--density response
%measures the response of the system density to a perturbation proportional to the density of particles and 
can be computed from the density-density correlation function. In fact, let consider the system described by the Hamiltonian $H-\int_\O h(\yy) \r(\yy)d \yy$, with $\r(\xx)=a^+_\xx a^-_\xx$ the density of particle operator and $h(\xx)$ an external field. The average density at the point $\xx$ is given by:
\be
\media{\r_\xx} =  \frac{ \Tr_{\HHH_s} e^{-\b (H_{\O}-\m_{\O,\b} N) +\b \int_\O d\yy h(\yy) \r(\yy)} a^+_\xx a^-_\xx}{ \Tr_{\HHH_s} e^{-\b (H_{\O}-\m_{\O,\b} N)+\b \int_\O d\yy h(\yy) \r(\yy)}} 
\ee
If one is interested in studying how the average density changes in $\xx$ when the external field is switched on, then can calculate the linear response $\c(\xx,\yy)$, given by the functional derivative of $\media{\r(\xx)}$ with respect to the external field in $\yy$, calculated at equilibrium (i.e. with $h(\xx)\equiv 0$). An explicit calculation shows that the linear response can be written in terms of the density--density correlation function:
\be
\c(\xx,\yy) = \media{\r(\xx)\r(\yy)} - \media{\r(\xx)}\media{\r(\yy)}
\ee
where the symbol $\media{\cdot}$ indicates $\frac{ \Tr_{\HHH_s} e^{-\b (H_{\O}-\m_{\O,\b} N)} \cdot}{ \Tr_{\HHH_s} e^{-\b (H_{\O}-\m_{\O,\b} N)}}$.

In order to study the interacting theory it will result more convenient to study,  in place of the correlation functions defined by \eqref{eq:corr_func}, the $2n$-point Schwinger function, \ie the correlation functions generalized to imaginary times,
defined as follows:
\be \label{eq:schw_funct}
 S_{\s_1,\ldots,\s_{2n}}(x_1\,\ldots\,x_{2n})=\lim_{L\arr\io} \frac{  \Tr_{\HHH_s} \left[e^{-\b (H_{\O}-\m_{\O,\b} N)} T \lbrace a^{\s_1}_{x_1}\ldots a_{x_{2n}}^{\s_{2n}} \rbrace \right] }{ \Tr_{\HHH_s} e^{-\b  (H_{\O}-\m_{\O,\b} N)}}
\ee
where $x_i \equiv (\xx_i, t_i)$, $0 \leq t_i \leq \b$ and the operator $T$ denotes the time-ordered product of the operators in the brackets, i.e. arranges the times in decreasing chronological order. In the \eqref{eq:schw_funct} $\s_i=\pm$, $\sum_{i=1}^{2n}\s_i=0$ and $a^\pm_{\xx,t}$ are the bosonic operators in the imaginary time Heisenberg representation, i.e. $ a^\pm_{\xx,t}= e^{(H-\m N)t}\, a^\pm_\xx \,e^{-(H-\m N)t} $. The equilibrium properties of the systems, i.e. the elements of  the $n$--particle density matrices, are obtained from the $2n$--point Schwinger functions by taking the limit $t\arr 0^-$. Then the Schwinger functions contain more information then what we really need. On the other hand they are a natural object in perturbation theory, since admit an useful perturbative expansion in terms of the Schwinger functions of the free system. Moreover, they can be usefully represented in terms of functional integrals,  as we will see in more details in section \eqref{Interacting}. 

Our goal will be the study of the equilibrium properties of the interacting Bose gas in presence of the phenomenon of Bose-Einstein condensation. As a first step, with the aim of stressing some ideas which will result useful for the interacting case treatment,  we will recall in the following section the concept of Bose condensation for free bosons both following the standard presentation and the Schwinger functions point of view.

\section{The non interacting case}

The properties of the ideal Bose gas  at $\b^{-1}$ temperature can be obtained from the grand--canonical partition function:
\be  %usare Z oppure \X ?
Z^0_{\O,\b} = \sum_{N\geq0} e^{-\b(H^0_{\O}-\m_{\O,\b}^0 N)}
 \ee
where $H^0_{\O}$ is obtained by \eqref{eq:Hamiltonian} taking $\l=0$ and the chemical potential $\m_{\O,\b}^0$ is determined by fixing the density of the system $\r_{\O,\b}$:
\be \label{eq:density}
\r_{\O,\b} =\frac{\media{N}_{\O,\b}}{|\O|} = \frac{1}{|\O|\b} \frac{\dpr}{\dpr \m^0_{\O,\b}} \ln Z^0_{\O,\b}
\ee
One finds (see for example~\cite{Book-Lieb}) that $Z^0_{\O,\b}$ factorizes into the contributions from the single particle energy levels $\KK(\kk)$; the result is
\be
Z^0_{\O,\b} = \prod_{\kk}\frac{1}{1-e^{-\b (\KK(\kk)-\m^0_{\O,\b})}}
\ee 
Note that in the non interacting bosonic case it is necessary that $\m^0_{\O,\b} <  0 $. For fixed $\m^0_{\O,\b} <0$ the density of the finite system is given by:
\be
\r_{\O,\b} = \frac{1}{|\O|} \sum_{\kk}\frac{1}{e^{\b (\KK(\kk)-\m_{\O,\b}^0)}-1}
\ee
In the thermodynamic limit the latter formula gives:
\be
\r_\b = \lim_{|\O| \arr \io} \r_{\O,\b}  = \frac{1}{(2\pi)^d} \int_{\RRR^d} d^d \kk \frac{1}{e^{\b (\KK(\kk)-\m_\b^0)}-1}
\ee
which is a monotonously increasing function of $\m^0_\b$, which for $\b$ finite is bounded as $\m^0_\b \arr 0$ by a critical density $\r_{\b}^{cr}$ in the three dimensional case. This phenomenon was interpreted by Einstein~\cite{Einstein} by saying that the particles exceeding the critical number all go into the lowest energy state, \ie in order to fix the system density at some number greater than $\r_\b^{cr}$ we have to let $\m^0_{\O,\b}$ to zero simultaneously with the increasing volume of the system $|\O|\arr \io$ . Then for $\r_\b>\r_\b^{cr}$ the chemical potential is null and the total density of the free system is 
\be \label{eq:density2}
\r_\b=\r_0 + \r_\b^{cr}
\ee
with $\r_0$ the density contribution from the lowest energy level, given by
\be\label{eq:condensate}
\r_{0}= \lim_{|\O|\arr \io} \frac{1}{|\O|} \frac{1}{e^{\b \left(\KK(\bz) -\m_{\O,\b}^0 \right) }-1}
\ee
The phenomenon that a single particle level of the non interacting system has a non zero density in the thermodynamic limit, \ie a macroscopic occupation, is called Bose--Einstein condensation (BEC). It is simple to see that there is no condensation into the excited energy levels, since $\KK(\kk)-\m_{\O}^0\geq \KK(\kk)-\KK(\bz)=\text{const.}\,L^{-2}$. 
In the zero temperature case, \ie the ground state, one finds $\r^{cr}_{+\io}$=0 in all dimensions $d$, which means that all the particles are in the condensate state. In a sector of fixed particle number, the ground state wave function is simply a product of single particle wave--functions in the lowest energy state. This will not be true, as one may expect, in the interacting ground state. 

Once the total density $\r_\b$ is fixed we can determine the values of the condensate density $\r_{0}=\r_{0}(\r_\b,\b)$ and chemical potential $\m^0_\b=\m^0_\b(\r_\b,\b)$ in the thermodynamic limit. In particular:
\begin{itemize}
 \item if $\b=+\io$ then $\r_0=\r_{+\io}$ and $\m^0_{+\io}=0$; 
 \item if $\b<+\io$ then $\r_{0}=\max(\r_\b-\r_\b^{cr}, 0)$ and $\m^0_\b$ is fixed using \eqref{eq:density}.
\end{itemize}
Note that for $\r_{0}\neq0$ one can equivalently fix the condensate density $\r_{0}$ and then calculate $\r_\b=\r_\b(\r_{0},\b)$ through \eqref{eq:density2}, while the dependence of $\m_{\O,\b}^0$ on $|\O|$ can be determined using  \eqref{eq:condensate}. \\

Let us consider now the Schwinger functions approach. Since the condensation phenomena is related to the density of the system, the interesting object would be the $2$--point Schwinger function: %\nota{questione $t_1=t_2$}
\be
S^{-+}_{\b,\O}(x_1,x_2)= 
\begin{cases}
 \media{ a_{x_1} a_{x_2}^{+} }_{\b,L} \quad t_1>t_2  \\
\media{a_{x_2}^{+} a_{x_1} }_{\b,L} \quad t_1 \leq t_2
\end{cases}
\ee  
The $2$--point correlation function $\media{a^+_{\xx_2} a_{\xx_1}}$ is obtained by $S^{-+}_{\b,L}(x_1,x_2)$ by taking the limit $t_1-t_2\arr0^-$. A standard calculation (see \eg~\cite{NO} 
%or the section \ref{A.free_calculations}
) gives:
\be \label{propagator}
S_{\b,\O}^{-+}(x)=\frac{1}{L^{d}}\sum_{\kk}e^{-i\kk \cdot \xx}\,
e^{-f_{\O,\b}(\kk)t}\left(\frac{\th(t>0)}{1-e^{-\b f_{\O,\b}(\kk)}}+\frac{\th(t\leq0)e^{-\b f_{\O,\b}(\kk)}}{1-e^{-\b f_{\O,\b}(\kk)}}\right)
\ee
with $x=x_1-x_2$ and $f_{\O,\b}(\kk)=\KK(\kk) -\m_{\O,\b}^0$. In this formalism the presence of the condensate state corresponds to the fact that for $\m^0_{\O,\b}=0$ the term in the sum with $k_{0}=|\kk|=0$ involves a division by zero.
%The condensation phenomena occurs for $\m=0$; however in such a case the term in the sum with $k_{0}=|k|=0$ involves a division by zero. This corresponds to the presence of a condensate state in the non interacting system, i.e. a single particle level (the ground state indeed) with a macroscopic occupation in the thermodynamic limit $L\arr+\io$. 
However taking $\m^0_{\O,\b}$ going to zero with the increasing volume $|\O|$ in such a way that the condensate density $\r_0$ is fixed, \ie
\be \label{muzero}
\frac{e^{\b \m^0_{\O,\b}}}{1-e^{\b \m^0_{\O\b}}}=|\O|\,\r_{0} + O(1) \qquad |\O| \gg 1
\ee
we can extract from $\lim_{L\arr+\io} S_{\b,L}$ the contribution coming from the condensate state, which is
\be
S^{-+}_0(x)=\lim_{\substack{L \arr \io \\[1pt] \r_{0}\,\text{fixed} }} \,\frac{1}{L^{d}}\, 
   e^{\m^0_{\O,\b} t}\;\frac{\th(t>0)+e^{\b \m^0_{\O,\b}}\,\th(t\leq0)}{1-e^{\b\m^0_{\O,\b}}} = \r_{0}
\ee
At this point we can take $\m^0_{\O,\b}=0$ in the remaining part of the integral, obtaining
\be \label{eq:S1}
S^{-+}_{\b}(x)=\r_0 + \frac{1}{(2\pi)^{d}} \int d^{d}\kk \, e^{-i \kk \cdot \xx}e^{-\KK(\kk) |t|}\left(\frac{\th(t>0)}{1-e^{-\b\KK(\kk)}}  +\frac{\th(t\leq0)e^{-\b\KK(\kk)}}{1-e^{-\b\KK(\kk)}}   \right)
\ee
with $S^{-+}_{\b}(x)=\lim_{|\O|\arr\io}S^{-+}_{\b,\O}(x)$. Note that, using the definition
\be 
N_\kk=\int_{\RRR^d} d^d\xx\, e^{i\kk \cdot \xx} S_\b(\xx)
\ee
with $S_\b(\xx)$ obtained  taking the limit $t\arr0^-$ in \eqref{eq:S1} one finds the expression \eqref{eq:density2} for the density of a free Bose gas in the thermodynamic limit. \\

In order to investigate the ground state properties of the system we take the  $\b \arr +\io$  limit of \eqref{eq:S1}:  
\be
S^{-+}(x)=  \lim_{\b\arr+\io}S^{-+}_{\b}(x) = \r_0 + \frac{1}{(2\pi)^{d}} \int_{\RRR^d} d^{d}\kk \, e^{-i \kk \cdot \xx}\,
e^{-\KK(\kk) t}\th(t>0) 
\ee
Introducing the Matsubara frequency $k_0$ we get a more covariant expression for $S(x)$: 
\be \label{free_propagator}
S^{-+}(x)= \r_0 + \frac{1}{(2\p)^{d+1}}\int_{\RRR^{d+1}} d^d \kk dk_0\,\frac{e^{-ikx}}{-ik_{0}+\kk^{2}}
\ee
where $k=(k_{0},\kk)$. Then the first part of the function $S(x)$ in $\eqref{free_propagator}$ has the interpretation of the density of the condensate; the second part is a slowly decaying term which vanishes in the  limit $t\arr 0$. As already stressed the knowledge of the whole expression for $S^{-+}(x)$ is superfluous if we are interested in the equilibrium properties of the free gas, but will result essential for the interacting case treatment. 
For what concern the  $2n$--point Schwinger functions with $n>1$ they can can be obtained by the $2$--point function by the Wick rule (as showed in~\cite{NO}).

\section{The interacting case}  \label{Interacting}

We want to approach the interacting case $\l \neq 0$. The first idea might be to express both the interacting partition function $Z_{\b,\O}(\l)$
%\be \frac{Z_{\b,\O}(\l)}{Z_{\b,\O}(0)} = e^{-|\Lambda| \left(E_{\b,\O}(\l)-E_{\b,\O}(0)\right)}  %= \int P_\L(d\ph) e^{-V_\L(\ph)}  \ee
and the interacting Schwinger functions 
\be 
 S_{\b, \O}^{\,\s_1,\ldots,\s_{2n}}(x_1\,\ldots\,x_{2n}; \l)=\frac{  \Tr_{\HHH_s} \left[e^{-\b (H_{\O}-\m_{\b, \O} N)} T \lbrace a^{\s_1}_{x_1}\ldots a_{x_{2n}}^{\s_{2n}} \rbrace \right] }{ \Tr_{\HHH_s} e^{-\b  (H_{\O}-\m_{\b,\O} N)}}
\ee
as formal series in $\l$ and $\m_{\O}$ with 
\[
 H_{\O} & = H_{\O}^{0} +V^{\l}_{\O} \non \\
\m_{\b,\O} & = \m_{\b,\O}^0 N + \m_{\b,\O}
\]
Here $H_{\O}^0$ is the Hamiltonian of the non interacting case, $V^{\l}_{\O}$ the interacting potential  (the second term in the r.h.s. of \eqref{Ha}) and $\m_{\b,\O}$ the correction to the chemical potential due to the interaction.
A useful tool to develop a systematic perturbation expansion in power of $\l$ is provided by functional integrals. A functional integral representation for the partition function and the Schwinger functions of our system may be obtained using the coherent states $\ket{\ph}$, i.e. the eigenstates  $\ket{\ph}$ of the annihilation operator 
\[
a_{\kk} \ket{\ph } =  \ph^-_{\kk} \ket{\ph} \; , \; \bra{\ph }  a^{+}_{\kk} &=  \ph^+_{\kk} \bra{\ph },
\]
with $\ph^+_{\kk} = (\ph^-_\kk)^*$, see \cite{NO} for a reference. From \eqref{a} follows:
\[
\ph^-_\xx = \frac{1}{L^d} \sum_\kk e^{- i\kk \cdot \xx} \ph^-_\kk
\]
With these definitions the expression for the partition function of the interacting system defined by \eqref{eq:Hamiltonian} in the coherent state representation is given by:
\[ 
Z_{\b,\O} & = \Tr_{\HHH_s} \ e^{-\b (H_\O-\m_{\b,\O} N)} =\lim_{M\arr \io} Z_{\b,\O}^{M} \label{Z} \\
Z_{\b,\O}^{M}& =  \int  \Biggl[ \prod_{j=1 }^{M} \prod_{\kk}   
\frac{d \ph^+_{\kk \tau_{j} } d\ph^-_{\kk \tau_{j} }  }{2\pi i }  
\Biggr]\   e^{- \SS_{\b,\O}^{M}}   \label{Z_M}
\]
where $\tau_{j}= \frac{j\beta }{M}$ with index $j=0,\cdots ,M$, the fields are periodic in the $\t$ index $\ph^-_{\kk,0}=\ph^-_{\kk,\b}$ and 
\[
\SS_{\b,\O}^{M} =  \tfrac{\b }{M}  \sum_{j=1 }^{M} 
\bigl(H_\O-\m_{\b,\O} N\bigr) (\ph^+_{\kk,\tau_{j-1} }, \ph^-_{\kk\tau_{j} }) 
+
\sum_{\kk}[ \sum_{j =1}^{M}  (\ph^+_{\kk \tau_{j} }-
\ph^+_{\kk \tau_{j-1} }) \ph^-_{\kk \tau_{j} }] 
\]
with
\[
& (H_\O-\mu_{\b,\O} N) (\ph^+_{\kk \t_{j-1} }, \ph^-_{\kk \t_{j} }) =  \non \\
& \sum_{\kk} (\KK(\kk)-\m_{\b,\O} ) \ph^+_{\kk \t_{j-1} } \ph^-_{\kk \t_{j} } \ 
+ \  \tfrac{\l }{2 |\O |}   \sum_{\kk,\qq,\pp } 
\hv (\kk) \ 
\ph^+_{\qq +\kk,\t_{j-1} } \ph^+_{\pp-\kk,\tau_{j-1}} \ph^-_{\qq,\t_{j} }
 \ph^-_{\pp ,\t_{j} }
\] 
 We define the Fourier transform in $\t $ 
\[  \label{tauFT}
\ph^-_{\tau } = & \frac{1}{M } \sum_{k_{0}} e^{-ik_{0}\tau } \ph^-_{k_{0}}  \non \\ 
\ph^-_{k_{0}} = &  \sum_{\tau } e^{ ik_{0}\tau } \ph^-_{\tau }    
\]
where $k_{0}= (2\pi m)/\beta $ with $m=0,\dotsc ,M-1$. With this convention the Fourier transform does not preserve the norm and $\sum_\t |\ph_\t|^2 =\tfrac{1}{M} \sum_{k_0} |\ph_{k_0}|^2$; the integral  \eqref{Z_M} becomes
\be
Z_{\b,\O}^{M} =
%\int  \prod_{j=1}^M \prod_{\kk} \frac{d \ph^+_{\kk,\t_j } d\ph^-_{\kk,\t_j} }{2\pi i}  =
\int   \prod_{\kk,k_0} \frac{d \ph^+_{\kk,k_0 } d\ph^-_{\kk,k_0} }{2\pi i M} e^{- \SS_{\b,\O}^{M}(\ph_{\kk,k_0})}
\ee
with
\[   \label{LM}
\SS_{\b,\O}^{M}(\ph_k) =  
 \sum_{k} & \tfrac{1}{M} \left[  \tfrac{\b }{M}\,f_{\kk,\L} \  e^{i \frac{k_{0}\b }{M}} + 
\left(1- e^{i \frac{k_{0}\b }{M}} \right)  \right] |\varphi_{k}|^{2}  \non \\ 
& \quad +\  \tfrac{\l \b}{2 |\O | M^4 }   \sum_{k,q,p} \hv (\kk) \ 
e^{i \frac{(p_{0}+q_{0})\b }{M}} \ph^+_{q+k }\ph^+_{p-k}\ph^-_{q }\ph^-_{p } 
\]
where 
\[ \label{fk_def}
f_{\kk,\L} = \KK(\kk) - \m_{\L}
\] 
and we replaced $k= (k_{0},\kk)$ and  $\L=[{-\b}/2,\,{\b}/2]\times[-L/2,\,L/2]^{d}$ to make notations compact. For $\l=0$ the \eqref{Z_M} is a gaussian integral 
\[ 
 Z^0_{\L}  =  \lim_{M\arr \io} \int   \prod_{k} \frac{d \ph^+_{k } d\ph^-_{k} }{2\pi i M}\;   
e^{-\frac{1}{2} \sum_k  
	 \begin{pmatrix}  \ph^+_{k} & \ph^-_{-k}  \end{pmatrix} 
       C^{0,k}_{M,\L}
      \begin{pmatrix}  \ph^-_{k}  \\  \ph^+_{-k} \end{pmatrix}}  \label{Z0}
\]
with
\[ \label{free-covariance}
C^{0,k}_{M,\L}= \tfrac{1}{M} \begin{pmatrix}
 1-  \left( 1- \tfrac{\b}{M} f_{\kk,\L}  \right) e^{i \frac{k_{0}\b }{M}}  &  0    \\
0      &      1-  \left( 1- \tfrac{\b}{M} f_{\kk,\L}  \right) e^{-i \frac{k_{0}\b }{M}} \end{pmatrix}
\]
If the real part of the eigenvalues of $C^{0,k}_{M,\L}$ is positive, \ie  provided that $\m_{\L}<0$, the integral is absolutely convergent\footnote{There is a further condition to be fulfilled in order to have positive eigenvalues, that is $\b f_\kk / M < 1$ for each $\kk$. Note that for $M$ finite and large $\kk$ this condition may be not satisfied. However if we put an ultraviolet cutoff on $\kk$, as we will do in the following, for $M$ sufficiently large the theory is well defined.} 
and we have
\be
Z^0_{\L} = \lim_{M\arr \io}  \prod_k \frac{1}{M} \frac{1}{\sqrt{\det C^{0,k}_{M,\L}}}= \prod_\kk \frac{1}{1- e^{-\b f_{\kk,\L}} }
\ee
where we have used the identity
\[ \label{identity_ik0}
\prod_{n=0}^{M-1} \left(z- e^{i\frac{2\pi n}{M}}  \right) = z^{M}-1
\]
Note that in order to get the correct expression for the partition function, one has to carry out the calculation with $M$ finite and only at the end  take the limit $M\arr \io$. On the contrary, if we are interested in the calculation of the 2--point Schwinger functions we may consider from the beginning only the dominant terms in $\b /M$. For example, in the non interacting case, we recover \eqref{free_propagator} approximating the exponential factor in \eqref{free-covariance} as $e^{\pm i\,k_0\,\b/M}\simeq 1\pm i k_0 \b/M$:
\[ \label{Swinger_libera}
S^{-+}_{\L}(x-y) & =  \bmedia{\ph^-_{\xx, x_0} \ph^+_{\yy,y_0}}_{\b,\O} \non \\
%= \frac{1}{L} \sum_\kk e^{-i \kk \cdot \xx} \bmedia{\ph^-_{\kk, x_0} \ph^+_{\kk,0}} 
& = \lim_{M\arr \io} \frac{1}{M^2\,|\O|} \sum_{\kk, k_0, p_0} e^{-i \kk \cdot (\xx-\yy)} e^{-i k_0 x_0+ i p_0 y_0 } \bmedia{\ph^-_{\kk, k_0} \ph^+_{\kk,p_0}} 
% \non \\ & = \r_0 + \frac{1}{\b |\O|} \sum_{\kk,k_0} e^{-i \bigl(\kk \cdot (\xx-\yy) + k_0 (x_0 - y_0) \bigr)}\,\frac{1}{f_\kk -i k_0}
\]
where 
\[ \label{Swinger_libera2}
 \bmedia{\ph^-_{\kk, k_0} \ph^+_{\kk,p_0}} & =   \frac{1}{Z^{0,M}_{\L}}  \int   \prod_{k} \frac{d \ph^+_{k } d\ph^-_{k} }{2\pi i M}\;   e^{-\frac{1}{2} \sum_{k} \sum_{\s,\s'}   \ph^\s_{\s k}   (C^{0,k}_{M,\L})^{\s \s'}  \ph^{\s'}_{-\s' k} } \ph^-_{\kk, k_0} \ph^+_{\kk,p_0}  \non \\[6pt]
& =  (C^{0,k}_{M,\L})^{-1}_{-+} \,\d_{k_0,p_0} %= \frac{M^2}{\b}\, \frac{1}{f_\kk -i k_0+O(1/M)}
\]
and 
\[
(C^{0,k}_{M,\L})^{-1}_{-+}  \d_{k_0,p_0} = \begin{cases}
-\frac{M^2}{\b \m^0_{\b,\O}} = M^2 |\O| \r_0 \lft(1 + O\lft(1/|\O|\rgt) \rgt)  &  k=0 \\[9pt]
\; \frac{M^2}{\b}\, [f_{\kk,\L} -i k_0+O\lft(1/M\rgt)]^{-1}  & k \neq 0
\end{cases}
\]
where we used \eqref{muzero}. Then finally, by taking the limit $\b,|\O|\arr +\io$, eq.~\eqref{Swinger_libera} becomes
\[
S^{-+}(x-y)  = \r_0 + \frac{1}{(2\pi)^{d+1}} \int d^d \kk d k_0  e^{-i \bigl(\kk \cdot (\xx-\yy) + k_0 (x_0 - y_0) \bigr)}\,\frac{1}{f_\kk -i k_0}
\]
In the following we will use the formal notation 
\[ \label{free_measure}
 P^0_{\L}(d\ph) & \,= 
 \lim_{M \arr +\io }\frac{  % espressione in spazio k
 \prod_{k} \frac{d \ph^+_{k } d\ph^-_{k}}{2\pi i}\; e^{-\frac{1}{2} \sum_{k} \sum_{\s,\s'}   \ph^\s_{\s k}   (C_{M,\L}^{0,k})^{\s \s'}  \ph^{\s'}_{-\s' k} }
    }
{ \int \prod_{k} \frac{d \ph^+_{k} d\ph^-_{k} }{2\pi i}\;   e^{-\frac{1}{2} \sum_{k} \sum_{\s,\s'}   \ph^\s_{\s k}   (C_{M,\L}^{0,k})^{\s \s'} \ph^{\s'}_{-\s' k}}  
} \non \\[6pt]
& \text{``=''} \, \frac{
 \prod_{x} \frac{d \ph^+_{x } d\ph^-_{x}}{2\pi i}\; e^{- \int_\L dx   \ph^+_{x} (-\dpr_0 +\D)   \ph^-_y }
    }
{ \int \prod_{x} \frac{d \ph^+_{x} d\ph^-_{x} }{2\pi i}\;   e^{- \int_\L dx   \ph^+_{x} (-\dpr_0 +\D)   \ph^-_y}  
}
\]
with $\s,\s'=\pm 1$. 
%and $\Lambda=[{-\beta}/2,\,{\beta}/2]\times[-L/2,\,L/2]^{d}$.
%and the integral over $dx$ formal since for $L$ and $\b$ finite one has to substitute it with a sum over $x$.
%, $k_0=\frac{2\pi}{\b} m$, $m \in \ZZZ$
The complex gaussian measure $P^0_\L(d\ph)$  satisfies the following properties:  
\bea
\int P^0_\L(d\ph) &=&\unit \\
\int P^0_\L(d\ph) \ph^-_x\ph^+_y &=& S^{-+}_{\L}(x-y) \\
\int P^0_\L(d\ph) \ph^-_x\ph^-_y &=& \int P^0_\L(d\ph) \ph^+_x\ph^+_y =0
\eea
 The interacting partition function $Z_\L$ can be formally written as:
\be \label{Z_interacting}
\frac{Z_\L}{Z^0_\L} = %e^{-|\L| \left(E_0(\l)-E_0(0)\right)} = 
\int P^0_\L(d\ph) e^{-V_\L(\ph)} 
\ee
with 
\be \label{eq:V}
V_\L(\ph) = 
%\frac{\b}{M} \left[ \tfrac{\l}{2 |\O |M }   \sum_{k,q,p} \hv (\kk) \ e^{i \frac{(p_{0}+q_{0})\b }{M}} \ph^+_{q+k }\ph^+_{p-k}\ph^-_{q }\ph^-_{p } - \m_I \sum_{k} e^{i \frac{k_{0}\b }{M}} |\varphi_{k}|^{2} \right]   
  \frac{\l}{2}\int_{\L \times \L} dxdy\,|\ph_{x}|^{2}v(\xx-\yy)\d(x_0 - y_0)|\ph_{y}|^{2} - \m_\L \int_\L dx|\ph_{x}|^{2}
\ee  
The expression for the Schwinger functions in terms of functional integrals is:   
\be
 S_\L^{\s_1,\ldots,\s_n}(x_1\,\ldots\,x_n)= \frac{1}{Z_\L} \int P^0_\L(d\ph) e^{-V_\L(\ph)} \ph^{\s_1}_{x_1}\ldots \ph^{\s_n}_{x_n}
\ee
The strategy so far depicted, i.e. to study the interacting system as a perturbation of the free case, is not the most convenient. Based on Bogoliubov model~\cite{Bogoliubov}, the well--known approximate exactly solvable model for interacting bosons, one expects the interacting Schwinger functions to have a large distance behavior very different from the free ones. Then a natural approach is to use as starting point for the perturbative expansion just Bogoliubov Hamiltonian. 
This idea was immediately recognized and dates back to the work by Hugenholtz and Pines~\cite{Pines}. All the later attempts of developing a perturbation theory free of infrared divergences and calculating the corrections to Bogoliubov theory
(as in the papers by Gavoret and Nozi\'eres~\cite{Gavoret}, Nepomnyashchii~\cite{Nepom}, Popov~\cite{Popov}, Yang~\cite{LeeYangGROUP},
%\cite{LeeYang1},\cite{ LeeYang2},\cite{ LeeYang3},\cite{ LeeYang4},\cite{ LeeYang5}
where partial summations are used to remove the divergences), as the very recent works by Pistolesi et al.~\cite{CaDiC1,CaDiC2} and Benfatto~\cite{benfatto}, are based on this idea.

%who first obtained a systematic control of the perturbative theory around Bogoliubov model in the three dimensional case.  \nota{ARTICOLI LEE-YANG: inserire referenza, calcolano le correzioni a bogoliubov. Specificare risommazioni parziali.}

Since we are following the same strategy, the next section is devoted to the description of Bogoliubov theory in the grand canonical ensemble and in the functional integral representation. In the section \ref{eff_mod} we finally define the effective model we are dealing with.

\section{Bogoliubov approximation}  \label{reinterpret}

Bogoliubov approximate model~\cite{Bogoliubov} was the first attempt to explain BEC for interacting bosons and predicts a linear spectrum for small momenta at zero temperature, property which is considered typical of superfluid behavior, according to  Landau argument~\cite{Landau}. 
In this section we will reinterpret the steps leading to Bogoliubov approximation within the functional integral scheme. 
%A more detailed discussion of Bogoliubov approximation in the grand canonical ensemble, but in the Fock space representation, can be found in appendix \ref{Bog}.

The first step of Bogoliubov approximation consists in keeping in $\SS^M_{\b,\O}$ defined in \eqref{LM} only the terms that are at most quadratic in  the fields $\ph_{k}$, with $k\neq (0, \bz)$, and write explicitly the terms containing $\ph_{0}$. 
 The result is
\[ \label{step1}
\SS_{\L,M}^{B} = & \frac{\b}{M^2} \left(-  \mu^B_\L  |\ph_{0}|^{2}   +\  \tfrac{\l \hv (0)  }{2 |\O |M^2 }  |\ph_{0}|^{4} \right)\non \\
& + \frac{1}{2M} \sum_{k\neq 0} \left\{ 2\tilde{F}_{k} |\varphi_{k}|^{2} \ 
 +\  \tfrac{\b}{M} \tfrac{\lambda  \hat{v} (\mathbf{k}) }{|\O |M^2 }   \ 
\left[
\ph^+_{k} \ph^+_{-k } (\ph^-_{0})^{2} +
(\ph^+_{0})^{2} \ph^-_{k}  \ph^-_{-k }
 \right]\right\} 
\]
where
\begin{align}
\tilde{F}_{k} &= \tfrac{\b}{M} \left(\KK (\kk)-\m^B_\L +\l   (\hv (0) + \hv (\kk) )  \tfrac{ |\ph_{0}|^{2}   }{|\O |M^2 }   
\right) \ 
 e^{i \frac{k_{0}\beta }{M}}  + 
\left(1- e^{i \frac{k_{0}\beta }{M}} \right)  %\frac{}{\frac{\beta }{M} }
\notag\\
&=  1-   e^{i \frac{k_{0}\beta }{M}} 
\left( 1- \frac{\beta }{M} F_{\kk,M}  \right)
\end{align}
and
\[
F_{\kk,M} = \KK (\kk)-\m^B_\L +\l (\hv(0) + \hat{v} (\kk) )
 \tfrac{ |\ph_0|^{2}   }{|\O |M^2 }   
\]
with $\hv(\kk)$ real, being $v(\xx)$ symmetric. Note that $\ph^{\pm}_0 =M \bmedia{\ph^{\pm}_{\bz, \t}}_\t$. Then the separation of the $\ph^\pm_0$ fields corresponds in writing the zero spatial momentum field as $\ph^\s_{\bz,\t}= \x^\s + \eta^\s_{\bz,\t}$ with $\xi^\s =\bmedia{\ph^{\s}_{\bz, \t}}_\t $ the average over the temporal index and  $\eta^\s_{\bz,\t}=\sum_{k_0 \neq 0} e^{i \s k_0 \t} \ph^\s_{k_0}$ the fluctuation with respect to it. Note that in \eqref{step1} also the cubic and quartic term in the fields $\ph^\s_{\bz, k_0}$ are neglected. Using the rescaling $ \ph^\pm_{0}= \xi^\pm M $ we obtain:
\begin{equation}
\SS_{\O, M}^{B} = \b  \Bigl\{ 
-  \m_\L^B  |\xi|^{2}  +\  \tfrac{\l  \hv (0)  }{2 |\O| }  |\xi|^{4} 
 +\frac{1 }{2} \sum_{k\neq 0}
\begin{pmatrix}
\ph^+_{k} & \ph^-_{-k}  
\end{pmatrix}
C^{B,k}_{M,\L}
 \begin{pmatrix}
 \ph^-_{k}  \\
 \ph^+_{-k}
\end{pmatrix}
 \Bigr\}
 \end{equation}
where
\[  \label{C_Bogoliubov}
C^{B,k}_{M,\L}= \tfrac{1}{M}
\begin{pmatrix}
 1-  \left( 1- \tfrac{\beta }{M} F^\x_{\kk}  \right) 
e^{i \frac{k_{0}\b }{M}}  & \tfrac{\b }{M} g^{\x,+}_{\kk}\\
\tfrac{\b }{M} g^{\x,-}_{\kk} &  1-  \left( 1- \tfrac{\b }{M} F^\x_{\kk}  \right) 
e^{-i \frac{k_{0}\b }{M}} 
\end{pmatrix}
\]
and
\[ \label{Feg}
& F^\x_{\kk} = \KK (\kk)-\m^B_\L +\l (\hv(0) + \hat{v} (\kk) )
 \tfrac{\,|\x|^{2}   }{|\O | }  
& g^{\x,\pm}_{\kk} =  \tfrac{\l  \hv (\kk)}{|\O|} (\xi^\pm)^2
\]
The partition function becomes
\be \label{Z_Bog_1}
Z_{\L,M}^{B} = M  \int
\frac{d\xi^+  d\xi^-   }{2\pi i}
\ e^{-\b \bigl\{ -  \m^B_\L  |\xi|^{2}  +\  \tfrac{\l \hv (0)  }{2 |\O| }  |\xi|^{4} \bigr\} }
\prod_{k>0 } I_{M,k} (\xi)
\ee
where  $k>0$ means the first non zero coordinate in the four component vector $k$ is positive and  
\[
I_{M,k} (\xi)  & =  \frac{1}{M^2} \int  \frac{d\ph^+_{k} d\ph^+_{-k} d\ph^-_{k }  d\ph^-_{-k } }{( 2\pi i )^{2}}  
\  e^{- %\tfrac{1}{2}
\begin{pmatrix} \ph^+_{k} & \ph^-_{-k}   \end{pmatrix}
C^{B,k}_{M,\O}
 \begin{pmatrix}  \ph^-_{k}  \\  \ph^+_{-k} \end{pmatrix}} 
  \non \\[6pt] 
& = (M^2 \det C^{B,k}_{M,\L})^{-1} 
\]   
as long as the eigenvalues of $C^{B,k}_{M,\L}$ have a real positive part.  The determinant of $C^{B,k}_{M,\L}$ can written as
\[ \label{det_C_B}
\det C^{B,k}_{M,\L}= \tfrac{1}{M^2} (A_{\mathbf{k}}-B_{\mathbf{k}} e^{i \frac{k_{0}\beta }{M}})
(A_{\mathbf{k}}-B_{\mathbf{k}} e^{-i \frac{k_{0}\beta }{M}})
%= (A_{\mathbf{k}}-B_{\mathbf{k}} e^{i \frac{k_{0}\beta }{M}})(A_{-\mathbf{k}}-B_{-\mathbf{k}} e^{-i \frac{k_{0}\beta }{M}})
\]
with 
\[
A_{\kk} - B_{\kk} &= \frac{\b }{M} \sqrt{ (F_{\kk}^{\x})^2-| g^\x_{\kk} |^{2} } \non \\[6pt]
A_{\kk}+B_{\kk} &= 2 \,\sqrt{  1- \frac{\b }{M}  F^\x_{\kk} + \frac{1}{4} \left(\frac{\b }{M} \right)^{2} 
( (F^\x_{\kk})^{2}  -| g^\x_{\kk} |^{2} ) } %\notag\\
%&= 2 - \frac{\beta }{M}  F_{\kk} +  O \left(\frac{\b^{2} }{M^{2}} \right)  
\]
The eigenvalues of $C^{B,k}_{M,\L} $ are given by
\[
\l_\pm = \frac{1}{2} \lft( \Tr\,C^{B,k}_{M,\L} \pm \sqrt{(\Tr\,C^{B,k}_{M,\L})^2 -4 \det C^{B,k}_{M,\L}} \rgt)
\]
where both the trace and the determinant of $C^{B,k}_{M,\L} $ are real.  The conditions assuring the eigenvalues to be positive result result in ${\Tr}\,C^{B,k}_{M,\L} >0$ and $\det C^{B,k}_{M,\L} >0$. The trace is positive as long as  $\b F_\kk /M < 1$, which holds for $M$ sufficiently big\footnote{Note that for large $|\x|$ the trace might be negative; however the complete action $\SS_{\L,M}$, \ie the action before Bogoliubov approximation was made, is well defined. In the following we first integrate over the $|\x|$ variable, by fixing its value at the critical point, and then we perform Bogoliubov's approximation.}.  By rewriting \eqref{det_C_B} as 
\[ 
\det C^{B,k}_{M,\L}= \tfrac{1}{M^2} \lft( (A_\kk - B_\kk)^2 +2A_\kk B_\kk \bigl(1 - \cos(k_0 \b /M) \bigr) \rgt)
\]
we see that the determinant is positive for each $k$ provided that
%results
 \be \label{mu-require}
(F^\x_\kk)^2 - |g^\x_\kk|^2 > 0 \quad \forall \kk  \quad \Leftrightarrow \quad  \m^B_\O < \l \hv(\bz) {|\x|^2}/{|\O|}
\ee
Coming back to the computation of the partition function in \eqref{Z_Bog_1}, by using \eqref{det_C_B} and the fact that $A_{\kk}$, $B_{\kk}$ are symmetric in $\kk$ and do not depend on $k_{0}$, we get 
%the sum over $k$ in  as $\sum_{k}= 2 \sum_{k>0}$, getting
\be
\prod_{k>0} I_{kM} (\xi) = \prod_{k \neq 0} \frac{1}{ 
(A_{\kk}-B_{\kk} e^{i \frac{k_{0}\b }{M}}) } 
= \left(A_\bz - B_\bz \right)  \prod_{\kk} \frac{1}{A_{\kk}^{M}- B_{\kk}^{M}}
\ee
where we used the identity \eqref{identity_ik0}. The factor $ \left(A_\bz - B_\bz \right)$ comes from the product over $k_0 \neq 0$ of the $\kk=0$ mode:
\be
\prod_{k_0 \neq 0} \frac{1}{A_\bz- B_\bz e^{i \frac{k_{0}\b }{M}} } = 
(A_\bz - B_\bz) \prod_{k_0 } \frac{1}{A_\bz- B_\bz e^{i \frac{k_{0}\b }{M}} } = \frac{A_\bz - B_\bz}{A_{\bz}^{M}- B_{\bz}^{M}}
\ee
Inserting all these results in the partition function
we have
\begin{equation}
Z_{M,\L}^{B}= \b  \int
\frac{d\xi^+  d\xi^- }{2\pi i}\, 
 e^{-\beta |\O |\Bigl\{ -  \m^B_\L  \tfrac{\,|\xi|^{2}}{|\O|} +\  \tfrac{1}{2}\l \hv(0)   \frac{|\xi|^{4}}{|\O|^2}  - \frac{1}{\b |\O|} C_\bz(|\x|) \Bigr\} }
\prod_{\kk} \frac{1}{A_{\kk}^{M}- B_{\kk}^{M}}
\end{equation}
with $C_\bz(\xi) = \tfrac{1}{2}\log \left(F^2_\bz -|g_\bz|^2 \right)$. 
%positive being $\m^B_\O < \l \hv (0) |\xi|^2/ |\O|$. 
By taking the limit $M$ which goes to infinity one finds
\begin{align}
\lim_{M\arr \io} A_{\kk}^{M}& =  e^{- \tfrac{\b}{2} \left(  F^\x_{\kk} 
- \sqrt{ (F^\x_{\kk})^{2}-| g^\x_{\kk} |^{2}} \right) }\\
\lim_{M\arr \io} B_{\kk}^{M}&=  e^{-  \tfrac{\b}{2} \left(  F^\x_{\kk} 
+ \sqrt{ (F^\x_{\kk})^{2}-| g^\x_{\kk} |^{2}} \right) }
\end{align}
and 
\[ \label{Bog-1step}
Z_{\L}^{B}= \b  \int \frac{d\xi^+  d\xi^-  }{2\pi i}\, e^{-\beta |\O |\, \WW^B_{\b,\O} (\x)} 
\]
with $\WW^B_{\b,\O} (\x)$ a specific free energy equal to
\[ \label{WW_xi}
\WW^B_{\b,\O} (\x)= & -  \m^B_\L  \tfrac{\,|\xi|^{2}}{|\O|} +\  \tfrac{1}{2  }\l \hv(0)   \frac{|\xi|^{4}}{|\O|^2}  - \frac{1}{2 |\O|} \sum_\kk \left(  F^\x_{\kk} - \sqrt{ (F^\x_{\kk})^{2}-| g^\x_{\kk} |^{2}} \right)  \non \\
   & -\tfrac{1}{\b |\O|}\log \prod_\kk  \frac{1}{1- e^{- \b \sqrt{ (F^\x_{\kk})^{2}-| g^\x_{\kk} |^{2} }}} - \frac{1}{\b |\O|} C_\bz(|\x|) 
\]
%\be
%\z_\kk(|\xi|) =\frac{e^{ \b \left(  F_{\kk} - \sqrt{ F_{\kk}^{2}-| g_{\kk} |^{2}} \right) }}{1- e^{-2 \b \sqrt{ F_{\kk}^{2}-| g_{\kk} |^{2} }} }
%\ee
Let us denote  $\WW^B_{\O} (\x)= \lim_{\b \arr +\io}\WW^B_{\b,\O} (\x) $, then
\[ \label{WW_xi_0}
\WW^B_{\O} (\x)= & -  \m^B_\O  \tfrac{\,|\xi|^{2}}{|\O|} +\  \tfrac{1}{2  }\l \hv(0)   \frac{|\xi|^{4}}{|\O|^2}  - \frac{1}{2 |\O|} \sum_\kk \left(  F^\x_{\kk} - \sqrt{ (F^\x_{\kk})^{2}-| g^\x_{\kk} |^{2}} \right)
\]
with $F^\x_\kk$ and $g^{\x,\pm}_\kk$ defined in \eqref{Feg} and the chemical potential  $\m^B_\O$ fixed by the density of the system.  The first two terms in the r.h.s. of \eqref{WW_xi_0} define a ``Mexican hat'' potential in the $\x^\pm/\sqrt{|\O|}$ variables. Being the the last term in the r.h.s. of \eqref{WW_xi_0} negative, one may be worried that this correction may destroy the double well shape.  In order to investigate this point, we first assume that $\WW^B_{\O} (\x)$ has a double well shape, an hypothesis whose consistency can be checked as follows. Using a saddle point approximation -- as we will see below -- we find the expression \eqref{murelation} for the chemical potential in the thermodynamic limit, with $F_\kk$ and $g_\kk$ defined in \eqref{Fg_after_sub}. 
%obtained by substituting $|\x^\pm|=\sqrt{\r_0 |\O|}$, with $\r_0$ the condensate density, in the definitions for $F^\x_\kk$ and $g^{\x,\pm}_\kk$. 
Then we verify that by substituting in \eqref{WW_xi_0} the approximate value for $\m^B_\O$ that has been found,  the potential \eqref{WW_xi_0} shows actually a double well shape. 
%This control can be performed at each order in $\l$, being $\m=\m(\l,\r_0)$ in \eqref{murelation} invertible. 
A straightforward asymptotic analysis of \eqref{WW_xi_0} in the $\x$ variable, with $\m^B_\O$ given by \eqref{leadingorder} and $\l$ fixed and small, shows that the last term in \eqref{WW_xi_0} is subleading both in $\l$ and $\x$ with respect to the first two terms. This can also see in fig.~\ref{double-well}, where the plain lines represent the first two terms in \eqref{WW_xi_0} for different values of $\l$ and the dotted lines a numerical integration of \eqref{WW_xi_0}, after the substitution \eqref{leadingorder}. \\

%Preciso che $\m$ e' fissato dalla formula sotto dove sia gli F che i g sono calcolati in $\r_0$, mentre nella formula per $\WW$ dipendono da $\x$.

\fig{t}{0.6}{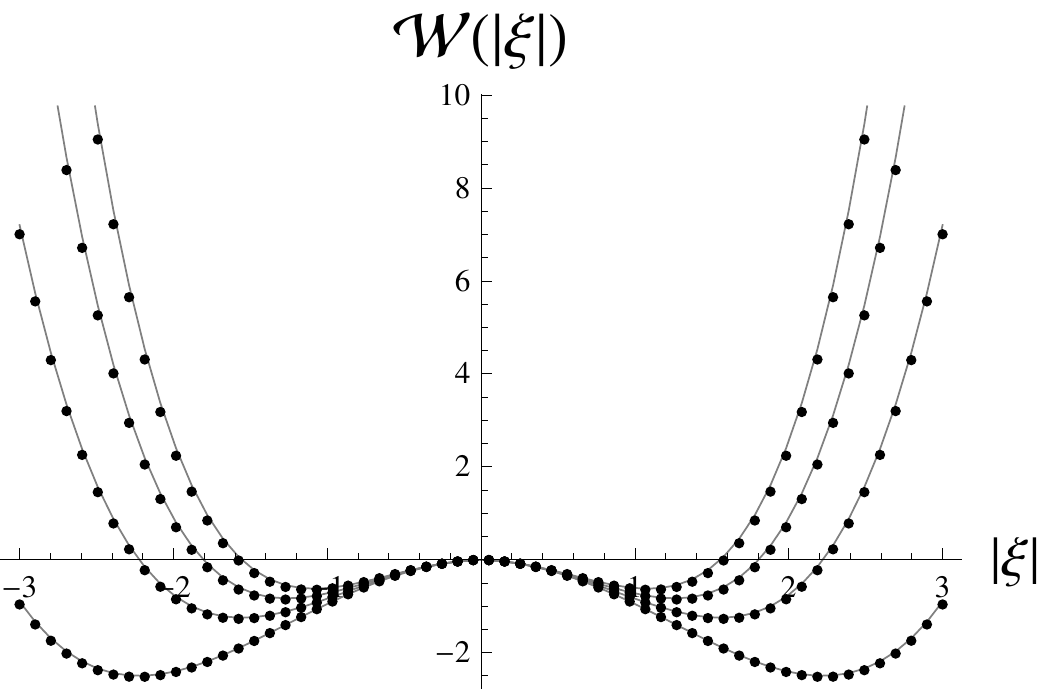}{Double well potential in the variable $|\x|$ in the three dimensional case for different values of $\l$. Moving from the lowest to the highest curve we have $\l=0.1,0.2,0.3,0.4$. The plain line represents the first two terms in the r.h.s. of \eqref{WW_xi_0}; the dotted line is obtained by numerical integration of \eqref{WW_xi_0}. We see that the shape of the double well is not affected by the last term in the r.h.s. of \eqref{WW_xi_0}. }{double-well}

The second step of Bogoliubov approximation consists in interpreting $|\x|^2$ as the average number of particles at $\kk=0$, that is
\be \label{step2}
|\x|^2 =\r_0 |\O|
\ee
This substitution corresponds in the Fock space representation to the fact that  one can replace the operators $a_\bz$ and $a^+_\bz$ everywhere in the Bogoliubov Hamiltonian (or in the total bosonic Hamiltonian) by the c--number $\sqrt{N_0}$ without making an error in the ground state energy for particle in the thermodynamic limit, as proved in appendix D of~\cite{Book-Lieb}.
Then \eqref{Bog-1step} becomes
\be  \label{Bog-2step}
Z^{B}_{\L} = \b |\O | \int_0^{+\io} d\r_0 \, e^{-\b |\O|\,\WW^B_{\b,\O} (\r_0)}
%e^{-\b |\O|  \bigl\{ -  \m \r_0  +\  \tfrac{ 1 }{2 } \l \hv (0) \r_0^{2}  \bigr\}} \prod_\kk  \z_\kk (\r_0)
%\sum_{\kk} \left(  F_{\kk} - \sqrt{ F_{\kk}^{2}-| g_{\kk} |^{2}} \right)  }
\ee
with $\WW^B_{\b,\O}(\r_0)$ obtained from \eqref{WW_xi} with the substitution \eqref{step2}. Denoting by  $\WW^B_{\O} (\r_0)= \lim_{\b \arr +\io}\WW^B_{\b,\O} (\r_0) $ we have
\be \label{1.WW}
\WW^B_{\O}(\r_0) = -  \m^B_\O \r_0  + \tfrac{ 1 }{2 } \l \hv (0) \r_0^{2}  - \frac{1}{2 |\O|} \sum_\kk \left(  F_{\kk} - \sqrt{ F_{\kk}^{2}- g_{\kk}^{2}} \right)
\ee
where we are here denoting with
\[ \label{Fg_after_sub}
& F_{\kk} = \KK (\kk)-\mu +\l (\hv(0) + \hat{v} (\kk) )\r_0 \non \\
&  g_\kk = \l \hv(\kk) \r_0
\]
the quantities in \eqref{Feg} after the substitution \eqref{step2}. The condensate density $\r_0$ is fixed via a self--consistent equation, which will be written below.

%We stress that the contribution in \eqref{Bog-2step} due to the $\ph^\s_{\bz, k_0}$ fields, namely $C_\bz(\r_0)$, will give no contribution to the energy in the thermodynamic limit, being a finite quantity. 

\subsection{The ground state energy }
We are interested in calculating the ground state energy per particle in the Bogoliubov approximation  and in the thermodynamic limit when the average total number of particles $ \media{N}$ and $L$ tend to infinity with density 
$\rho =\media{N}/ |\O|$ fixed:
\be \label{gse}
e^B_{0}(\r)= % \lim_{\bar N\rightarrow \infty} \frac{E_{0, \O} (\rho )}{ \langle N\rangle} = 
\lim_ {\b \arr \io} \lim_{L \arr \io} \frac{E^B_{\b, \O} (\rho )}{|\O | \rho }
\ee 
For  fixed $\O $ the ground state energy is obtained by
\be \label{ground_state}
E^B_{\b, \O} -E^0_{\b, \O}= 
- \lim_{\b \arr \io}\frac{1}{\b |\O|} \frac{\dpr}{\dpr \b} \log \frac{Z^B_{\b,\O}}{Z^0_{\b,\O}} + \left(\m^B_{\b,\O} -\m^0_{\b, \O} \right)\rho |\O | 
\ee 
In the limit $\b \arr \io$  \eqref{1.WW} the integral in \eqref{Bog-2step} will concentrate around the saddle defined by
\be \label{saddle}
\frac{\dpr}{\dpr \r_0} \WW^B_{\O}=0
\ee
In the following we will skip the subscript $\b$ whenever $\b = +\io$. The derivative of  \eqref{1.WW} with respect to $\r_0$ can be computed explicitly and we obtain 
\be \label{murelation}
\mu^B_\O =  \l \hv (0)\r_0 +  \tfrac{1}{2|\O |}  \sum_{\kk} \bigl(\l \hv (0) + \l \hv (\kk)\bigr)
\tfrac{F_\kk - \sqrt{ F_\kk^2- g_\kk^2}}{ \sqrt{ F_\kk^2- g_\kk^2}} - \tfrac{1}{2|\O |}
 \sum_{\kk} \tfrac{ \l  \hat{v} (\kk)g_\kk}{ \sqrt{ F_\kk^2- g_\kk^2}}
 \ee
At the leading order in $\lambda $ we have
\begin{equation}\label{leadingorder}
\mu^B_{\O} =  \lambda \hat{v} (0)\rho_{0} 
\end{equation}
To obtain the first corrections we insert this value in $F_\kk$. In the regime $\l \hv(0) \r_0 R_0^2 \leq 1$, with $R_0$ the range of the interacting potential, we obtain the following corrections in the three and two dimensional cases: 
\[
\mu^B_{3d} & = \l \hv(0) \r_0 \left(1 - c^2_{1}\, \l   +  c_{2} \l \sqrt{\l \r_0 \hv(0)^3} +o (\lambda^{5/2} ) \right)  \label{mu3d} \\
\mu^B_{2d} & = \l \hv(0) \r_0 \bigl(1 +O(\l\, |\log \l \r_0 \hv(0)|) \bigr)  \label{mu2d}
\]  
%\nota{referring to \eqref{mu2d} the $o(\l)$ is $\l $}
with $c_1$ and $c_2$ explicit and explicitly computable integrals; however for the aim of this work it is not necessary to write them explicitly.
%One can evaluate explicitely an asintotic series of $\l$, the first few terms look as follows. $c_1=\int dk \ldots$, $c_2=\int dk \ldots$.
Note that  the second order correction to $\m^B$ must be negative, in order to satisfy the condition \eqref{mu-require}. We also stress that, differently from the free case, the chemical potential in the interacting case depends on $\l$ and is different from zero also in presence of condensation. \\

We can now calculate Bogoliubov prediction for the ground state energy. Inserting in  \eqref{1.WW} the leading order term for the chemical potential $\m^B=\l \hv(0) \r_0$ and taking $\r=\r_0$ (as we will see in the next paragraph $\r-\r_0 =O(\l^{d/2})$) one obtains, up to an error term of order $O(\l^{(d+2)/2})$ :
%since we will find a result correct up to order $o(\l^{5/2})$ we can neglect the corrections to the chemical potential  
\be \label{Ecanonical}
e^B_ {0,\O} = 
\frac{1}{2} \l \hv(\bz) \r_{0}
- \frac{1}{2\r_0 |\O |} \sum_{\kk \neq \bz}
(\kk^2 +\l \hv(\kk) \r_0 - \sqrt{\kk^4 +2\kk^2 \l \hv(\kk) \r_0}) 
\ee
where $e^B_ {0,\O}=\lim_{\b \arr \io} e^B_{\b, \O}$. The  \eqref{Ecanonical} corresponds to the expression for Bogoliubov ground state energy calculated in the canonical ensemble (as showed for example in appendix A of \cite{Book-Lieb}).  
In the thermodynamic limit  \eqref{Ecanonical} becomes:
\be \label{Ecanonical-limit}
e^{B}_{0}= \frac{1}{2}\l \hv(\bz)\r_0 -\frac{1}{2\r_0 (2 \pi)^d } \int d^d\kk \left(\kk^2 +\l \hv(\kk) \r_0 - \sqrt{\kk^4 +2\kk^2 \l \hv(\kk) \r_0} \right) 
\ee
In the following we will review the predictions of Bogoliubov model for the ground state energy in the regime $\l \hv(\bz) \r_0 R_0^2 \leq 1$, both for two and three dimensions. 
We will choose the interaction potential as
\be
 v_{R_0}(x)=R_{0}^{-2} \,e^{-\frac{|x|}{R_0}} 
\ee 
then the condition becomes $\l \r_0 R_0^d \leq 1$. This particular choice of $v_{R_0}(x)$ is not relevant and has been done only for simplicity reasons.  

\subsubsection{Three dimensional prediction of the ground state energy for  $\e \ll 1$}
In the $3d$ case, the regime  $\e= \l \r_0R_0^3\leq 1$ corresponds to $a_0 / R_0 \geq \sqrt{ \r_0 a_0^3}$, with $a_0= \l \hv(\bz)/ 8\pi$ the first term in the Born series for the scattering length $a$ of the interacting potential. Under the conditions
\be
1\gg\frac{a_0}{R_{0}}\gg\sqrt{\rho_{0}a_{0}^{3}}\gg\left(\frac{a_0}{R_{0}}\right)^{2}\label{eq:Bogoliubov_regime}
\ee
it is convenient to rewrite the integral in \eqref{Ecanonical-limit} as $I=I_0+I_1$ where
\be
I_0 =\frac{1}{16\pi^3 \r_0} \int d^3\kk \left(\kk^2 +\l \hv(\kk) \r_0 - \sqrt{\kk^4 +2\kk^2 \l \hv(\kk) \r_0} -\frac{\l^2  \hv^2(\kk) \r_0^2}{2\kk^2} \right)  
\ee
and
\[
I_1 =  \frac{\l^2   \r_0}{32 \pi^3} \int d^3\kk \frac{\hv^2(\kk)}{\kk^2}
\]
The integral $I_1$ converges since $\hv(\kk)$ goes to zero faster then $|\kk|^{-1}$. In $I_0$ the integrand goes to zero faster then $|\kk|^3$ and is absolutely convergent. If we replace $\hv(\kk)$ by $\hv(\bz)$  (this substitution leads to errors of order $o(\l^{5/2})$)  the integral can be calculated exactly, giving:
\be
I_0=4\pi\rho_{0} a_0 \frac{128}{15\sqrt{\pi}}\sqrt{\rho_{0}a_0^3}
\ee 
For what concern $I_1$, this is proportional to second term $a_1$ of the Born series, in particular $I_1= 4\pi \r_0 a_1$. Then we get the following expression for the ground state energy in the Bogoliubov approximation:
\be \label{LHYformula}
e_{0,B}^{3d}=4\pi\rho_{0} \left(a_0+a_1 +\frac{128}{15\sqrt{\pi}}a_0\sqrt{\rho_{0}a_0^{3}}+o(a_0\sqrt{\rho_{0}a_0^{3}})\right)
\ee
where the error term comes from the substitution $\hv(\kk)\simeq \hv(\bz)$ and from considering only the leading order expressions for the chemical potential and the total density.  The right side condition of \eqref{eq:Bogoliubov_regime} allows to write $a\simeq a_{0}+a_{1}$ to the desired accuracy (i.e.up to error terms that are much smaller than $a_{0}\sqrt{\rho_{0}a_{0}^{3}}$, \ie $o(\l^{5/2})$ ) and to write the $\eqref{LHYformula}$ as
\be \label{LHYformula2}
e_{0,3d}^{B} =4\pi\rho_{0}a \left(1 +\frac{128}{15\sqrt{\pi}}\sqrt{\rho_{0}a^{3}}+o(\sqrt{\rho_{0}a^{3}})\right)
\ee
The leading term of \eqref{LHYformula2} was proved to be correct for the non approximate Hamiltonian \eqref{eq:Hamiltonian}  by Dyson (upper bound)~\cite{Dyson-upper-bound} and Lieb-Yngvason (lower bound)~\cite{LY-lower-bound}. The second order correction was first derived by Lee, Huang and Yang~\cite{LeeYang0, Lee-Huang-Yang}, this is why eq.\eqref{LHYformula2} is known as the Lee-Huang-Yang formula.  The proof of the latter formula is, so far, an open problem, even if a few recent papers present partial results~\cite{Erdos-Schlein-Yau, YauYin, Giuliani-Seiringer,Seiringer_mean_field}.

Note that the condition $a_0/R_{0}\gg\sqrt{\rho_{0}a_0^{3}}$ is necessary to have $e_{0}^B \leq4\pi\rho_{0}a_{0}$, i.e. the right side of \eqref{LHYformula}
must be equal to $4\pi\rho_{0}a_{0}$ plus a negative correction,
which requires $|a_{1}|\gg a_{0}\sqrt{\rho_{0}a_{0}^{3}}$. 
The condition $\r_0 a_0^3 \ll 1$ also guarantees that Bogoliubov theory is asymptotically correct
up to term of order $\rho_{0}a_{0}\sqrt{\rho_{0}a_{0}^{3}}$, \ie that the contributions coming from the cubic and quartic terms of the interaction  (neglected in Bogoliubov theory)  are smaller. 

\subsubsection{Two dimensional prediction of the ground state energy for  $\e \ll 1$}
For what concerns the $2d$ case, in the regime $\l \r_0 R^2_0 \ll 1$, Bogoliubov theory predicts the following leading order for the ground state energy
\be \label{E_Bog_2d}
e^{2d}_{0,B} = \frac{1}{2} \l \r_0 \hv(\bz) \left[1+ O\left( \l \log(\l \r_0 R^2_0)\right)\right]
\ee 
while the corrections due to the quartic terms are of order $\lambda^{2}$. In the $2d$ case the relation between the first order of the scattering length and the strength of the interaction is $a_0=R_0 e^{-4\pi /\l}$. Then, in the case $\r_0 R_0^2=1$ we find the prediction for the ground state energy to be
\be \label{E_Bog_2d2}
e^{2d}_{0,B} =\frac{4\pi \r_0}{|\log \r_0 a_0^2|} \left[1+O\left(\log(|\log \r_0 a_0^2|)\right) \right]
\ee
The leading term of the latter formula, first calculated by Schick~\cite{Schick},  was rigorously proved to be correct for nonnegative finite range two--body potential  in~\cite{LY2D}.  The (negative) correction to the leading term in \eqref{E_Bog_2d2} is of the same order of what has been found for a two dimensional hard core gas of bosons~\cite{Hines}. 
% The first derivation of the correct asymptotic formula for the ground state energy per particle of a dilute, homogeneous, two-dimensional Bose gas in the thermodynamic limit was postulated as late as 1971 by Schick [12] and was rigorously proved to be correct in~\cite{LY2D}.

\subsection{Choice of the free parameter}
As we have already stressed in the non interacting case the three parameters $\r_{\b,\O} $, $\rho_{0}$ and $\mu_{\b,\O}$ are dependent and their relation depends on the strength of interaction $\l$ and on the temperature. We have two different options to choose the free parameter.

\paragraph{Option 1} The natural choice, even if not the most convenient from a technical point of view, is to fix the total density of the system $\r$ and then choose $\r_{0}(\b, \O)$ as a function of $\m_{\b,\O}$ via  \eqref{murelation}  and $\m_{\b, \O} $ as a function of $\r $ via \eqref{rhorel}. 

\paragraph{Option 2} A second possible choice consists in keeping the density of the condensate $\r_0$ fixed. Then $\m_{\b, \O} $ is adjusted as a function of $\r_0$ so that \eqref{murelation} is satisfied. In this case the total density particle $\r_{\b,\O} $ is computed as a function of $\m_{\b,\O} $ by the formula
\begin{equation}\label{rho}
\rho_{\b, \O} = \frac{1}{|\O |}\media{N}_{B}= 
\frac{1}{|\O |\b } \frac{\dpr}{\dpr \m }
\ln  Z^{B}_{\b,\O}.
\end{equation}
This second choice is the one we will use in our treatment of the full interacting problem, \ie we will fix the density of the condensate and then prove that we can choose the correction to the chemical potential coming from the cubic and quartic terms of the interaction (neglected in Bogoliubov theory) in such a way that \eqref{saddle} is satisfied.\\

In order to calculate the relation between the total and the condensate densities in the Bogoliubov approximation, we use the relation \eqref{rho} with $Z^B_{\b,\O}$ defined in \eqref{Bog-2step}, by taking into account that the integral over $\r_0$ is concentrated around the saddle point we fixed by the choice of $\m_{\b,\O}$.
We obtain 
\[  \label{rhoBog}
\rho_{\b, \O} = & - \frac{\dpr}{\dpr \mu } \WW^B_{\b,\O} %\non \\
=   - \frac{\dpr}{\dpr \mu } \WW^B_{\O}+ \frac{1}{|\O |}
  \sum_{\kk } \frac{e^{-\beta E_\kk }}{1-e^{-\beta E_\kk }}
\frac{F_\kk}{E_\kk}
%\frac{\dpr}{\dpr \mu } E_\kk
\]  
with $ \WW^B_{\b,\O}(\r_0)$ and $ \WW^B_{\O}(\r_0)$ defined in \eqref{WW_xi} and \eqref{1.WW} respectively and $E_\kk= \sqrt{F^2_\kk - g^2_\kk}$. For $\b \arr \io$ the contributions from the last term in \eqref{rhoBog} can be neglected
(due to the $e^{-\beta E_{\mathbf{k}} } $ factor) and in the limit $\b, |\O| \arr \io$ we get
\[  \label{rhorel}
\r = \rho_{0} + \int  \frac{d^d \kk}{(2\pi)^d}
%\frac{1}{2|\O |} \sum_{\mathbf{k} \neq 0} 
\left( \frac{ F_{\mathbf{k}} }{ E_{\mathbf{k}} }-1  \right) 
=  \rho_{0}\left( 1+ O (\l^{\frac{d}{2}} ) \right),
\]
where $d=2,3$. We see as the total density of the system in Bogoliubov model is not equal to $\r_0$ even at zero temperature. For $d=1$ the integral over $\kk$ in \eqref{rhorel} diverges for small momenta, \ie Bogoliubov approximation fails. This is due to the fact that in the one--dimensional case, in the presence of repulsive interaction, no condensation is expected, not even at zero temperature. \\
%there is no BEC in a repulsive 1D Bose gas even at zero temperature in the thermodynamic limit

The expression of the total density in the Bogoliubov approximation for $\b$ finite and in the thermodynamic limit is obtained by taking the limit $|\O|\arr \io$ of \eqref{rhoBog}:
\[
\lim_{|\O|\arr \io}\r_{\b, \O}= \r +  \int  \frac{d^d \kk}{(2\pi)^d}
% \frac{1}{2|\O |} \sum_{\mathbf{k} \neq 0} 
\,\frac{ F_{\kk} }{ E_{\kk} }\,\frac{1}{e^{\b E_\kk}-1}  
\]
The latter expression diverges for small momenta in the two dimensional case. This is related to the fact that in the two dimensions condensation does not occur at non zero temperature, due to the very general Mermin-Wagner-Hohenberg theorem~\cite{Hohenberg, Mermin-Wagner}. In particular  Hohenberg~\cite{Hohenberg} proved  a rigorous inequality that can be used to rule out the existence of long--range order for bosonic and fermionic systems in one or two dimensions and $T\neq 0$.

\subsection{Schwinger functions for Bogoliubov approximation}

The two point Schwinger functions for Bogoliubov model is calculated by using \eqref{Swinger_libera} and \eqref{Swinger_libera2}, once the matrix $C^{0,k}_{M,\L}$ is substituted with $C^{B,k}_{M,\L}$ defined in \eqref{C_Bogoliubov}. We find:
\[ \label{prop_Bogoliubov}
S_{\s \s'}^B (x) = \r_0 + g_{\s \s'}^B (x) 
\]
where 
\[
g^B_{\s \s'}(x)=\frac{1}{(2\pi)^{d+1}}\int d^{d+1}k\,g^B_{\s \s'}(k)
\] and
\[ \label{prop_psi}
\lft(g_{\s \s'}^B(k)\rgt)^{-1} & = \lim_{M\arr \io} \frac{1}{M^2 |\O|} \lft(C^{B,k\neq0}_{M,\L}\rgt)_{\s \s'}  \non \\[6pt] 
& = \left( \begin{array}{cc}
-ik_{0}+\kk^2 +\l \r_0 \hv(\kk) & \l \r_0 \hv(\kk) \\[3pt]
\l \r_0 \hv(\kk)    &  ik_0 +\kk^2  +\l \r_0 \hv(\kk)
\end{array}\right)
\]
where the first row and the second column correspond to $\s=+$, while the second row and the first column to $\s=-$. 
The matrix $g^B_{\s \s'}(k)$ is the propagator of the fluctuations fields $\ph^\pm_k$ with $k\neq 0$, which in the following will be denoted as $\ps^\pm_k$ to distinguish them from the fields $\x^\pm$. 

We will denote the Gaussian measure with propagator \eqref{prop_psi} as $P_{B,\L}(d\ps)$. The latter measure can be also thought as obtained by adding to the free measure $P_\L^0(d\ps)$, with propagator $g_{\a \a'}^{0}(x) $ obtained by setting $\l=0$ in \eqref{prop_psi}, the quadratic potential in the $\ps^\pm$ fields given by Bogoliubov approximation, that is
\[
 P_{B,\L}(d\ps) =  P_\L^0(d\ps)\,e^{-V_{B,\L}(\ps)} 
\]
with $V_{B,\L}(\ps)$ obtained by \eqref{eq:V} by writing $\ph^\pm_x= |\O|^{-1/2}\x^\pm + \ps^\pm_\x$, then by substituting $|\O|^{-1}|\x|^2=\r_0$ and finally by neglecting the cubic and quartic terms in the fields $\ps^\pm_x$:
\[  \label{VB1}
 V_{B,\L}(\ps)  = & \frac{\l}{2} \r_0 \int_{\L \times \L} (\ps^+_x + \ps^-_x) v(x-y) (\ps^+_y + \ps^-_y) dx dy  \non \\ 
& + (\l \r_0 \hv(\bz) - \m^B_{\O}) \int_\L |\ps_x|^2 dx
\]
Here $(\l \r_0 \hv(\bz) -\m^B_\O)$ is zero at leading order, see \eqref{leadingorder}. Note that the two--point Schwinger functions obtained by \eqref{prop_psi} have a singularity of the type $(k_0^2 +c_B^2 \kk^2)^{-1}$, with $c_B^2=2 \l \r_0 \hv(0)$, to be compared with the singularity $(-ik_0 + \kk^2)^{-1}$ of the free Bose gas, see \eqref{free_propagator}. 
The anomalous behavior of $S^{-+}_B(x)$ is related to the emergence  in the Bogoliubov model of a linear spectrum for small momenta, as one can see in the basis of the creator and annihilator operators $(b_{k}^{+},\, b_{k})$ which diagonalizes Bogoliubov Hamiltonian. In this basis we have:
\be \label{eq:bogol_diagonal}
H^{B}_{\b,\O}=\sum_{\kk \neq0}E'_\kk b_{\kk}^{+}b_{\kk}+\WW^B_{\b,\O}(\r_0)
\ee
with $\WW^B_{\b,\O}$ defined in \eqref{WW_xi} and $E'_\kk=\sqrt{\kk^{4}+2\l \r_0\hv(k) \kk^2}$ , \ie $E'_\kk \simeq c_{B}|\kk|$ for small momenta. Then $c_{B}$ has the physical interpretation of  the velocity of the Bogoliubov quasi-particles. \\

%\blue{The spectra of the quasi--particles and of the density--fluctuations excitations are shown to be identical for $\kk \arr 0$. Both corresponds to phonons with the usual macroscopic sound velocity. This is expressed mathemathically by the indentity of the poles of the one particle Green's function and the correlation function of two elementary excitations.~\cite{Gavoret}}

{\it Remark.} Introducing a functional representation of \eqref{eq:bogol_diagonal} in terms of the eigenstates $\lbrace \tl{\ph}_{k}^{-} \rbrace$ of the $b_k$ operators we get a diagonal propagator for the $\tl{\ps}_k$ fields: 
\[
\lft(\tl{g}^{B}_{\s \s'}(k)\rgt)^{-1}=\left(\begin{array}{cc}
ik_{0}+c_{B}|\kk| & 0\\
0 & -ik_{0}+c_{B}|\kk|
\end{array}\right)
\]
Since we intend to make perturbation theory around Bogoliubov model, it might seem more intuitive to work in the $\tl{\ps}_k$ representation. However in such a basis the vertices of the interaction would depend on the momenta, even at the bare level, which is not very convenient.  In addition, the RG treatment is greatly simplified by introducing an appropriate linear combination of the eigenvectors of the free Hamiltonian, see \eqref{psi_lt}. \\ 

The form \eqref{prop_Bogoliubov} of the covariance of Bogoliubov approximate model shows that Bogoliubov measure may be written as the product of two independent measures over the fields $\tl{\xi}=\xi/\sqrt{|\O|}$ and $\ps^\pm_x$. As a consequence, the partition function of Bogoliubov model can be written as
\[  \label{ZB}
\frac{Z^B_\L}{Z^0_\L} = \int P_\L \bigl(d \tl{\xi} \bigr) P_{B,\L} \bigl(d\ps \bigr) 
\]
with $P_\L(d\tl{\xi})$ and $P_{B,\L}(d\ps) $ the measures with covariances $\r_0$  and  $g^B_{\s \s'}(x)$ in the thermodynamic limit. In particular
\[
P(d \tl{\x}) = \lim_{|\L| \arr +\io } P_\L(d \tl{\x}) =\frac{d \tl{\x}^+ d \tl{\x}^-}{2\pi i \r_0}\, e^{- \frac{|\tl{\x}|^2}{\r_0}} 
\]
so that in the thermodynamic limit
\[
\lim_{|\L| \arr +\io}  \bmedia{\tl{\x}^+ \tl{\x}^-}_\L = \int P(d \tl{\x}) \,\tl{\x}^+\,\tl{\x}^-\,  = \r_0 
\]
The expression for $Z^B_\L$ in \eqref{ZB} has an exact correspondence with \eqref{Bog-1step} with the identification
\[
e^{-|\L|\, \lft(\WW^B_\L(\tl{\x})+  \m^0_\O|\tl{\x}|^2\rgt)} = \int P_{\L}^{0}(d\ps)\,e^{-V_{B,\L}(\tl{\x},\ps)}
\]
with $V_{B,\L}(\tl{\x},\ps)$ Bogoliubov potential before the substitution $|\tl{\x}|^2=\r_0$ and $ \m_0$ satisfying \eqref{muzero}. 

\fig{t}{0.38}{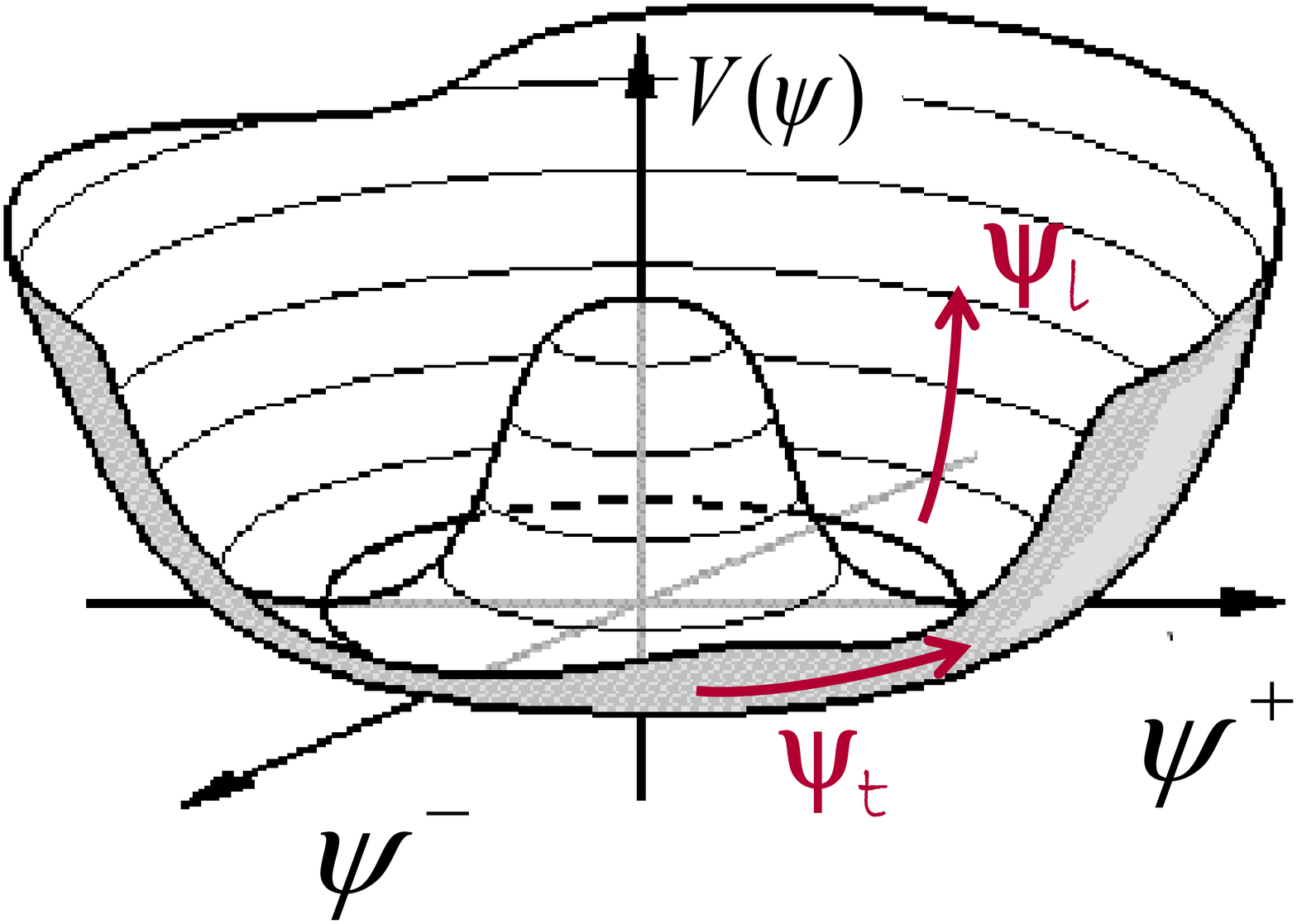}{A pictorial representation of the fields $\ps^l_x$ and $\ps^t_x$ defined by \eqref{psi_lt}; the latter represent the longitudinal and transverse component of the fields $\ps^\pm_x$ with respect to the direction of the broken symmetry.}{mexican}

Note that since the specific free energy $\WW^B_{\L}(\xi)$ has a ``Mexican hat'' structure, see \eqref{WW_xi}, in the thermodynamic limit the partition function of the system and each correlation function is given by the value of the integral in the saddle, that is $|\tl{\x}|^2=\r_0$ with $\r_0$ minimizing $\WW^B_{\L}(\xi)$. Fixing $\tl{\x}^\pm=\sqrt{\r_0}$, as we do, corresponds to choose a particular minumum and then in forcing a symmetry breaking of the gauge symmetry of the theory, defined by the one--parameter group of unitary transformations of the bosonic fields $\ph_x^\pm \arr e^{i\vartheta} \ph_x^\pm$.\footnote{The gauge symmetry of the bosonic system described by \eqref{Ha} can be explicitly broken by adding a term $\tl{\l} \sqrt{|\O|} (a^+_\bz + a_\bz)$ to the Hamiltonian. In the grand canonical state defined by the so modified Hamiltonian the operator $a_\bz$ has a non--zero expectation value, which goes to zero as $\tl{\l}\arr 0$ for any fixed volume $|\O|$. Gauge symmetry breaking means that this expectation value, divided by $\sqrt{|\O|}$ remains non--zero even as $\tl{\l} \arr 0$, after the thermodynamic limit has been taken.} When this is done we are left with the integral
\be 
e^{-|\L|\,\lft( \WW^B_\L(\r_0) +\m^0_\L \r_0\rgt)} = \int P_{\L}^{0}(d\ps)\,e^{-V_{B,\L}(\ps)}
\ee
%In the limit $\beta,L \rightarrow \infty$ we expect just the values of the fields that minimize $\WW_B(\xi)$ to be relevant, i.e. $\xi^+\xi^-=\rho_0$, as already seen; moreover in this picture.
This picture suggests the introduction of the adimensional fields
\[ \label{psi_lt}
\ps^l_{x}=\frac{1}{\sqrt{2\r_0}} \,\left( \ps_{x}^{+} + \ps_{x}^- \right)  \quad , \quad
\ps^t_{x}=\frac{1}{\sqrt{2i \r_0}} \,\left( \ps_{x}^{+} - \ps_{x}^- \right) 
\]
which represent the longitudinal and transverse component of the field $\ps^-_x$ to the direction of the broken symmetry, see fig.~\ref{mexican}. As we will see in chap.~\ref{multiscale}, the $\ps_x^{l,t}$ fields have a very convenient scaling, as was first noticed in \cite{benfatto}. Their propagator is given by
\be \label{eq:Bog_fields_lt}
g^B_{\a \a'}(x)= \frac{1}{(2\p)^{d+1}}\int_{\RRR^{d+1}} d^d \kk dk_0\,p^B_{\a_1 \a_2}(k)\frac{e^{-ikx}}{k_0^2 + \kk^2(\kk^2 + 2\l \hv(\kk) \r_0) }
\ee
with 
\[ \label{p}
p^B_{\a_1 \a_2}(k) 
= \frac{1}{\r_0} \left(\begin{array}{cc}
\kk^2  & k_{0} \\
-k_{0} \quad &  \kk^2  + \l \hv(\kk) \r_0 
\end{array}\right)
\]
Here the first row and column correspond to $\a=l$ and the second row and column to $\a=t$. The expression of Bogoliubov potential \eqref{VB1} in the basis of the longitudinal and transverse fields $\ps^l_x$ and $\ps^t_x$ is: 
\be
V_{B,\L}(\ps) = \l \r_0^2 \int_{\L \times \L} \ps^l_x\, v(x-y)\,  \ps^l_y dx dy
\ee
while the part of the potential containing the cubic and quartic terms of the interaction, neglected in Bogoliubov approximation, is given by:
\[  \label{V_not_local}
V_{0,\L}(\ps)  = \r_0^2\, \biggl[& \, \frac{\l}{8} \int_{\L\times \L}
\left(\bigl(\psi^l_x \bigr)^{2}+\bigl(\psi^t_x \bigr)^{2}\right) v(x-y) \left(\bigl(\psi^l_y \bigr)^{2}+\bigl(\psi^t_y \bigr)^{2}\right) dx dy \non \\[6pt]
& + \frac{\l}{2} \,\sqrt{2} \int_{\L \times \L}\psi^l_y v(x-y) \lft(\bigl(\psi^l_x \bigr)^2+\bigl(\psi^t_x \bigr)^{2}\rgt)dx dy   \, \biggr] \non \\[6pt] &
+ \frac{ \n}{2}\,\r_0 \int_{\L} \lft(\bigl(\psi^l_x \bigr)^{2}+\bigl(\psi^t_x \bigr)^{2} \rgt) dx
\]
with $\n=\l\r_0 \hv(0)-\m_B$ the correction to the chemical potential in the Bogoliubov approximation, as in \eqref{leadingorder}. 

%QUALI CORRELAZIONI CI INTERESSA CALCOLARE? Vogliamo calcolare le correlazioni in presenza di rottura spontante della simmetria di gauge della teoria, ovvero in presenza di condensato. Per questo e' conveniente scrivere $a_0= \sqrt{\r_0}$ piu' fluttuazioni e in questa scrittura e' imposta la rottura spontanea di simmetria. I campi di fluttuazione possono essere scritti in modo naturale come una parte reale e immaginaria che scalano in modo opportuno}

%\pagina

\section{The effective model}   \label{eff_mod}

Our goal is the exact calculation of the zero temperature properties of the interacting system described by \eqref{Z_interacting}, in the presence of a condensate state. In particular we want to investigate if the correlations of such model exhibit or not anomalous dimensions with respect to Bogoliubov model. In fact, the infrared divergences affecting the perturbation theory of the interacting problem may a priori completely change the nature of the propagator at low momenta. \\

%With the aim of studying the condensation problem I will consider a simplified model for a zero temperature two dimensional system of bosons interacting with a repulsive short range potential, obtained introducing an ultraviolet momentum cutoff. I will prove that the interacting theory is well defined at all orders in terms of an effective parameter related to the intensity of the interaction and that  the correlations do not exhibit anomalous dimensions, i.e. the model is in the same universality class of the exactly soluble Bogoliubov model. 

Our approach consists in assuming a spontaneous symmetry breaking of the $U(1)$ symmetry of the system, which means to fix a priori the condensate density $\r_0$, and then in showing that the correction to the chemical potential with respect to Bogoliubov's one, $\n=\m - \m_B$,  can be fixed in such a way that in the thermodynamic limit the interacting $2$--point Schwinger function converges to $\r_0$ as $L,\b \arr \io$.  
If this program succeeds, in the sense that we manage to prove that the correlation function of the interacting system can be expressed as a series in some effective parameters, with finite coefficients that admits explicit bounds at all orders, this will be interpreted by saying that the correlation function thus obtained describes a Bose condensate state with condensate density $\r_0$ and chemical potential $\m$. In this approach we regards $\r_0$ as a physical constant and $\n$ as a bare constant to be fixed to generate a model whose physical condensate density is the prescribed one. \\
%then we have proven the usual claim that the zero temperature Bose gas has a linear dispersion for small momenta. 

In order to find an expression for the partition function (and then for the correlation functions) of the interacting system as a perturbation around the exactly solvable Bogoliubov approximation, we will be interested in studying the potential $\WW_\L(\tl{\x})$, defined by the following functional integral
\[
e^{-|\L|\, \WW_\L(\tl{\x}) } = \int P_{B,\L}^{0}(d\ps)\,e^{-V_{0,\L}(\tl{\x},\ps)}
\]
In developing the announced program, we will assume the structure of the minima of the potential $\WW^B_{\L}(\tl{\x})$ to be qualitatively unchanged by the addition of the cubic and quartic terms of the interaction. As a consequence the partition function and the correlation functions for $\b, L \arr +\io$ will be given by the value of the functional integral representing them in the saddle point
$ \dpr_\tl{\xi} \WW_{\L}(\tl{\x}) =0 $.
This means that $\n$ will be chosen so that the free energy $\WW_\L(\r_0)$ is minimum for the fixed value of $\r_0$. With this assumption the study of the properties of the interacting system turn in studying the specific free energy $\WW_\L(\r_0)$, defined by:
	\be \label{eq:W_xi}
e^{-|\L|\, \WW_\L(\r_0)} = \int P_{B,\L}(d\ps)\, e^{-V_0(\ps)}
	\ee
with $P_{B,\L}$ the measure with propagator \eqref{eq:Bog_fields_lt} and $V_0(\ps)$ the potential containing the cubic and quartic terms of the interaction, neglected in Bogoliubov approximation, defined in \eqref{V_not_local}. 

Since the condensation problem depends only on the long-distance behavior of the system, we shall consider a simplified model obtained by modifying the decaying part of the propagator of the free measure $P^0_{\L}(d\ps)$ into
\[ \label{free_prop_c0}
g^{\c_0}_{\a \a'}(x) =\frac{1}{(2\pi)^{d+1}} \int d^d\kk dk_0\,\c_0(|k|)\,g_{\a \a'}(k)
\]
where
\be   \label{k_cutoff}
 |k|^2:= k_0^2 + \kk^2(\kk^2 + 2\l \hv(0) \r_0 \kk^2) 
\ee
and  $\chi_0(|k|)$ is a regularization of the characteristic function of the set \mbox{$R_0^{-4}|k|^2\leq1$}, playing the role of an ultraviolet cutoff at the scale of the inverse range $R_0^{-1}$ of the interaction potential. We will denote with  $P_{\c_0}(d\psi)$ the measure with covariance \eqref{free_prop_c0}. The choice of \eqref{k_cutoff} as argument of the cutoff function will be clear in a while. When we add to $P_{\c_0}(d\psi)$ the quadratic Bogoliubov potential, we get a new measure $P_{B,\c_0}(d\psi)$ with propagator
\[ \label{eq:prop_R0}
g_{\a  \a'}^{B,\c_0}(x)=\frac{1}{(2\pi)^{d+1}}\int d^{d}\kk dk_{0}\,\chi_{0}(|k|) \, e^{-ikx}\,
\frac{p_{\a \a'}^{B,\c_0}(k)}{|k|^2}
\]
The matrix $p^{B,\c_0}_{\a \a'}(k)$ has a slight different expression with respect  \eqref{p} due to the presence of the ultraviolet cutoff. In fact:  
\[
p_{\a \a'}^{B,\c_0}(k)
= \frac{1}{\r_0} \left(\begin{array}{cc}
\kk^2   & k_{0} \\
-k_{0} \quad &  \kk^2  + 2\l \hv(\kk) \r_0 \c_0(k)
\end{array}\right)   
\]
To get \eqref{eq:prop_R0} we have used the fact that the product between the measure $P_{\c_0}(d\psi)$ and the exponential of a quadratic term in the $\ps^\pm$ fields gives a new gaussian measure
\[
P_{Q_0,\c_0}(d\psi) =P_{\c_0}(d\psi)\,e^{-\frac{1}{2}\sum_{\a\a'} 
\int\psi_{k}^{\a}Q^0_{\a\a'}\psi_{-k}^{\a'}dk}
\]
with propagator given by 
\[ \label{mult_gauss_measures}
g^{\c_0, Q_0}_{\a \a'}(x) & =\frac{1}{(2\pi)^{d+1} }\int d^d\kk dk_0\,\c_0(k)\,g^{Q_0}_{\a \a'}(k)  \non \\[6pt]
\lft(g^{Q_0}_{\a \a'}(k)\rgt)^{-1} & =\lft(g_{\a \a'}(k)\rgt)^{-1}+\c_{0}(k)Q_{\a \a'}(k)
\]
see \eqref{change_integration}.  Note that:
\begin{itemize} 
\item The choice of $R_0^{-1}$ as scale of the ultraviolet cutoff has been done since we are interested in the weak coupling regime $\l \r_0 R_0^d \leq1$, which corresponds to the regimes $\sqrt{\l\r_0}\leq R_0^{-1}$ for $d=2$ and 
$\sqrt{\l a_0}\leq R_0^{-1}$ for $d=3$, $a_0$ being the first order Born approximation to the  three dimensional scattering length $a$ of the potential $v_{R_0}(x)$.
\item The model we are studying is ``morally'' a system of interacting bosons on a lattice with size equal to the range of the potential; however we choose to put a rotational invariant ultraviolet momentum cutoff rather then consider a lattice since this choice preserves the original symmetries of the model. This will greatly simplify the infrared problem. Two interesting open problems are: a) to remove the ultraviolet momentum cutoff; b) to extend the results obtained to a non rotational symmetric theory of bosons on a lattice.  
\end{itemize}
Due to the presence of the ultraviolet cutoff, the short range potential appears as a local potential, up to errors of order $O\left((kR_{0})^{2}\right)$; from now on we will then consider the local potential
\[  \label{V}
\VV_I(\ps)  = \r_0^2 \hv(\bz) & \, \biggl[\frac{\l}{8} \int_{\L}
\left(\bigl(\psi^l_x \bigr)^{2}+\bigl(\psi^t_x \bigr)^{2}\right)^{2}dx
+ \frac{\l}{2} \,\sqrt{2} \int_{\L}\lft(\bigl(\psi^l_x \bigr)^{3}+\bigl(\psi^t_x \bigr)^{2} \ps^{l}_x\rgt)dx  \non \\ &
+ \l  \int_{\L} \bigl(\psi^l_x \bigr)^{2}dx \, \biggr]
+ \frac{ \n}{2}\,\r_0 \int_{\L} (\bigl(\psi^l_x \bigr)^{2}+\bigl(\psi^t_x \bigr)^{2}) dx
\]
%Here and in the following the calligraphic ``$\VV$'' will be used to indicate local potentials and the subscript ``$I$'' to indicate the full interacting potential, including Bogoliubov quadratic term. 
Here and in the following the subscript ``$I$'' will be used to indicate the full interacting potential, including Bogoliubov quadratic term. Introducing the adimensional parameter $\e=2\l \r_0 \hv(\bz) R_0^2$ 
we get:
\[
\VV_I(\ps) =   \r_0 R_0^{-2} & \, \biggl[\frac{\e}{16} \int_{\L}
\left(\bigl(\psi^l_x \bigr)^{2}+\bigl(\psi^t_x \bigr)^{2} \right)^{2}dx
+ \frac{\e}{4} \,\sqrt{2} \int_{\L}\lft(\bigl(\psi^l_x \bigr)^{3}+\bigl(\psi^t_x \bigr)^{2}\ps_x^l\rgt)dx  \non \\[6pt]
& 
+\frac{\e}{2}  \int_{\L} \bigl(\psi^l_x \bigr)^{2}dx + \frac{ \n}{2}\,R_0^2 \int_{\L} \lft(\bigl(\psi^l_x \bigr)^{2}+\bigl(\psi^t_x \bigr)^{2}\rgt) dx \, \biggr]
\]
where $\r_0 R_0^{-2}$ is a dimension fixing factor introduced to keep track of the dimension of the various quantities and has the physical dimension of an action density in space time. Bogoliubov potential corresponds to the term $ \r_0 R_0^{-2} \frac{\e}{2}  \int_{\L} (\psi^l_{x})^{2}dx$ and the correction to Bogoliubov potential is:
\[
\VV_0(\ps)  = \r_0 R_0^{-2} &\, \biggl[\frac{\e}{16} \int_{\L}
\left(\bigl(\psi^l_x \bigr)^{2} +\bigl(\psi^t_x \bigr)^{2} \right)^{2}dx
+ \frac{\e}{4} \,\sqrt{2} \int_{\L}\lft (\bigl(\psi^l_x \bigr)^{3}+\bigl(\psi^t_x \bigr)^{2}\ps^l_x \rgt)dx  \non \\[6pt]
& + \frac{ \n}{2}\,R_0^2 \int_{\L} \lft(\bigl(\psi^l_x \bigr)^{2}+\bigl(\psi^t_x \bigr)^{2}\rgt) dx \,\biggr]
\]
With this new notations the inverse propagator of Bogoliubov measure is given by
\[
g^{-1}_{\a \a'}(k)
= \r_0 \left(\begin{array}{cc}
\kk^2 + \e \c_0(k) R_0^{-2}   & k_{0} \\
-k_{0} \quad &  \kk^2 
\end{array}\right)
\]

\vskip 0.5cm

%In fact $ \hv(k)=8 \pi \hv(0) / [1+(|k|R_0)^2]^2$.

\subsubsection{Effect of the ultraviolet cutoff}

The effective model we have introduced is representative of the long--distance behavior of the system and is useful to approach the condensation problem. However in order to obtain a prediction for the ground state energy or chemical potential, one have to study also the ultraviolet problem, i.e. to integrate the momenta excluded by the cutoff function. In fact the prediction of the physical quantities are affected by the presence of the cutoff function, as one may already see at the level of Bogoliubov approximation. 

Let consider first the effect of the cutoff on the $\kk$ variable. We consider the expressions \eqref{murelation}, \eqref{Ecanonical-limit} and  \eqref{rhorel}  for the chemical potential, the ground state energy and the density of the system in Bogoliubov approximation. Putting an ultraviolet momentum cutoff corresponds to neglect the integration over momenta such that $|\kk| \geq R_0^{-1}$. One finds that:
\begin{itemize}
\item for $d=3$ the ground state energy and the chemical potential are changed at order $O(\l^2 \r_0 R_0)$; in particular in the Lee--Huang--Yang formula the contribution coming from the ultraviolet is the one which permits of reconstruct the series for the scattering length at the correct order;
\item for $d=2$ the ground state energy and the chemical potential are changed at order $O(\r_0 \l^2)$; for what concerns the ground state energy the second order correction is of order $\l \log(\l \r_0 R_0^2)$ and then the presence of the cutoff functions does not affect it. 
\item for $d=2,3$ the total density of the system is changed by $O\left( \r_0 (\l^2 \r_0 R_0^d) \right)$. 
\end{itemize}
For what concerns the ultraviolet cutoff on the $k_0$ variable, it corresponds to approximate $e^{ik_0}$ with $ik_0$ in the Schwinger functions calculations. When this is done Bogoliubov predictions change of order $\l$, \eg the ground state energy (in the canonical ensemble) becomes:
\be
\tl{e}^B_0 = \frac{1}{2} \l \hv(\bz) \r_{0} - \frac{1}{2 \r_0 |\O |} \sum_{\kk \neq \bz}
(\kk^2 - \sqrt{\kk^4 +\e \kk^2 }\,)
\ee
to be compared with \eqref{Ecanonical}. This means that the parameters $\l$ and $\m$ appearing in our effective model have to be considered effective parameter, different from the physical ones. To get the correct values of the physical observables one might first integrate the ultraviolet momenta to get the exact values of $\l$ and $\m$ at the scale $R_0^{-1}$. The temporal ultraviolet integration is achievable with no much effort and one can see that effectively the $O(\l)$ term is restored. The integration on the spatial ultraviolet also seems feasible but for the time being we have not make much efforts in this direction; for the aims of the actual work we are only interested in the qualitative properties of the correlations, in order to get information on the occurrence of BEC.  

%\bibliographystyle{unsrt}  \bibliography{biblioBOSE} \end{document}

%\chapter{Multiscale analysis}
%\input{intro-senza-sapclass} \input{intestazione-sap} \begin{document} %\tableofcontents

\chapter{Multiscale analysis} \label{chap_multiscale}
The present chapter is devoted to the description of the 
%iterative integration 
scheme we shall follow in order to compute the functional integral defining the effective potential $\WW_{\L}(\r_0)$, introduced in the previous chapter, which contains all the  informations on the large distance behavior of the interacting bosonic system in presence of condensation. For convenience we report here the definition of $\WW_{\L}(\r_0)$ given in sec. \ref{eff_mod}:
\be \label{free-energy}
e^{-|\L|\, \WW_{\L}(\r_0)} = \frac{1}{\NN} \int P^\L_{Q_0,\c_0}(d\psi) e^{-\VV_0(\ps)}
\ee 
with $\L$ the volume in the space time, $\VV_0(\ps)$ the correction to Bogoliubov potential  
\[ \label{BV_I}
\VV_0(\ps)  = \r_0 R_0^{-2} &\, \biggl[\frac{\e}{16} \int_{\L}
\left(\bigl(\psi^l_x \bigr)^{2} +\bigl(\psi^t_x \bigr)^{2} \right)^{2}dx
+ \frac{\e}{4} \,\sqrt{2} \int_{\L}\lft (\bigl(\psi^l_x \bigr)^{3}+\bigl(\psi^t_x \bigr)^{2}\ps^l_x \rgt)dx  \non \\[6pt]
& + \frac{ \n}{2}\,R_0^2 \int_{\L} \lft(\bigl(\psi^l_x \bigr)^{2}+\bigl(\psi^t_x \bigr)^{2}\rgt) dx \,\biggr]
\]
and $P^\L_{Q_0,\c_0}(d\psi)$ the measure with propagator 
%\eqref{eq:prop_R0} in the thermodynamic limit and for zero temperature
\be \label{eq:prop_R0_cap3}
g_{\a \a'}^{\c_0,Q_0}(x)=\frac{1}{(2\pi)^4}\int d^{d}\kk dk_{0}\,\chi_{0}(k) g^{Q_0}_{\a \a'}(k)\, e^{-ikx}
\ee
with
\[ \label{inverse_prop}
\lft(g_{\a \a'}^{Q_0}(k)\rgt)^{-1} = \rho_{0}\left(\begin{array}{cc}
\kk^{2} + \e R_0^{-2}  \c_0(k)  & -k_{0}\\
k_{0} & \quad \kk^{2}
\end{array}\right)
\] 
where the first row and column correspond to $\a=l$, the second row and column to $\a=t$ and $\e=2\l \r_0 \hv_\bz R_0^{2}$. Note that $\e=c^2_B R_0^2$, with $c_B$ the speed of sound of Bogoliubov quasi particles. The function $\c_0(k)\=\chi_0(|k|^2)$, with
\be
 |k|^2= k_0^2 + \kk^2(\kk^2 + \e R_0^{-2} \kk^2) 
\ee
 is a $C^\io$ regularization of the characteristic function of the set \mbox{$R_0^{-4}|k|^2\leq1$} playing the role of an ultraviolet cutoff at the scale of the inverse range $R_0^{-1}$ of the interaction potential.

Given the formal functional expression for $\WW_{\L}(\r_0)$, for any finite $\L$ (\ie for each finite temperature $\b^{-1}$ and volume $|\O|=L^d$ ) the theory is order by order finite, with the integrals in \eqref{eq:prop_R0_cap3} representing sums over $\DD_\b \equiv \{k_0=2\pi n /\b, n\in \ZZZ \}$ and $\DD_L \equiv \{\kk =2\pi \nn /L, \nn \in \ZZZ^d \}$ and  $\b$ and $|\O|$ playing the role of an infrared cutoff. The problem is to perform suitable resummations that allows to re--express the specific free energy in terms of a modified expansion, whose $n$--th order is uniformly convergent as $\b$ and $L$ go to infinity.
%
%make the resummed series finite at all orders uniformly in $|\L|$, in order to send $\b$ and $L$ to infinity. 
%Once this is done
% effective potential  \eqref{free-energy}  it is know, 
%we can also derive from the effective potential  \eqref{free-energy} every Schwinger functions, as we will see in section \ref{Schwinger-functions}. 
%

The goal just described can be achieved by rigorous Renormalization Group (RG) method, in the form presented in~\cite{GalReview} (see also~\cite{Mastropietro} for an updated introduction) and already applied to a certain number of infrared problems in condensed matter systems~\cite{BM-luttinger, BGM-Hubbard, SRGraph1,SRGraph2,EMGraph1,EMGraph2} as detailed in the introduction. 

The Renormalization Group scheme consists in performing the integration in  \eqref{free-energy} in an iterative way, starting from the momenta ``close'' to the ultraviolet cutoff and moving towards smaller momentum scales. At the $n$--th step of the iteration the functional integral \eqref{free-energy} is rewritten as an integral involving only the momenta smaller than a certain value, proportional to $\g^{-n}$ where $\g >1$ is a constant. 
%the same constant (to be chosen sufficiently close to 1) appearing in the definition of the cutoff function.   
Both the propagators and the interaction will be replaced by ``effective'' ones; they differ from their ``bare'' counterparts because the parameters appearing in their definitions are ``renormalized'' by the integration of the momenta on higher scales. In the following will be convenient to introduce the {\it scale label} $h \leq 0$ as $h:=-n$.  The definition of the effective parameters is given in such a way that the new expansion which is obtained, the so called {\it renormalized expansion}, may be bonded at all orders uniformly in the infrared cutoff, under certain assumptions on the size of the effective parameters.

The aim of this chapter is to describe the main steps of the Renormalization Group scheme, as applied to our problem, to a reader which is not familiar with this technique. For this reason instead of immediately state the final result, we will proceed by progressive steps, first identifying the ``dangerous'' terms  making the perturbation theory divergent, then introducing a first {\it renormalized expansion} without renormalization of the covariance, finally defining the effective potential we are dealing with. This may be found needlessly long by the expert reader; however we are persuaded that he could easily skip the introductory parts and select the setup of the analysis and the final results, the latter contained in sect. \ref{effective} and in the $n!$ bounds described by results \eqref{thm_nfactorial0} pag. \pageref{thm_nfactorial0},  \eqref{thm_nfactorial2} pag. \pageref{thm_nfactorial2} and \eqref{thm_nfactorial3} pag. \pageref{thm_nfactorial3}.  \\

The chapter is organized as follows:
\begin{enumerate}[\it Sec. 2.1 \:]
\item the multiscale decomposition scheme is introduced. 
\item The ``naive'' expansion in terms of ``Gallavotti--Nicol\`o'' trees is presented; this part is not original but essentially taken by~\cite{GalReview, BM,GM}. 
\item The renormalized expansion is discussed, with a particular remark on the ideas leading to the definitions of the effective parameters.
\item A different renormalized expansion, in which also the covariance is renormalized by the integration over the higher momentum scales, is introduced; $n!$ bounds on the kernels of the effective interaction are stated, under some assumptions on the size of the effective couplings, which will be proven in the next chapter.
\item  We describe how the same integration scheme can be applied to the generating functional of the density and current correlations; the latter object, from which the density and current response function can be calculated, in this thesis is a crucial technical object, for which local Ward identities will be derived.
\end{enumerate}

%The dangerous terms for perturbation theory and the renormalized expansion are discussed respectively in section \ref{dimensional} and \ref{Localization}. In section \ref{ren_bound} we then describe the result of the iteration, that is the bounds the kernels of the effective interaction satisfy at each step, {\it under the assumption that the size of one of the effective coupling (the local quartic term) remains small}. The latter property will be proven in next chapter. 

\section[Multiscale decomposition]{Multiscale decomposition and trees expansion} \label{multiscale}

Given a fixed positive number $\g$ larger than 1, let us fix the following form for the cutoff function $\c_0(k)$:
\be
\chi_0(|k|^2)=  
\begin{cases}
\;1 \: & \text{if} \; |k|^2 \leq \frac{2}{\g^2+1}\, R_0^{-4}\\[6pt]
\;0 \: & \text{if} \; |k|^2 > \frac{2\g^2}{\g^2+1}\, R_0^{-4}
\end{cases}
\ee
This particular choice of $\c_0$ is not relevant; it is done only to simplify some calculations in the following. The multiscale decomposition consists in integrating iteratively the fields of decreasing energy scale, i.e. such that \mbox{$|k|^2\simeq\g^{2h}R_0^{-4}$} with $h$ a scale index $h \leq 0$. This can be done splitting the fields $\ps^{\a}_x$ in a sum of independent bosonic fields 
\be
\ps^{\a}_x=\sum_{h=-\io}^0 \ps_x^{\a,(h)}
\ee 
with propagators obtained by substituting $\c_0(k)$ in  \eqref{eq:prop_R0_cap3} with
\be \label{f_h} 
f_h(k)= \c_h(k)-\c_{h-1}(k)
\ee
 being $\c_h(k)= \c (\g^{-h}k)$. The latter decomposition, called ``decomposition in scales'' is done in order to introduce, in spite of a propagator which decades too slowly at infinity, a sum of propagators, each one with good decay properties at infinity, which are described in the lemma \ref{lemma}.

Due to the form \eqref{inverse_prop} of the propagator, two different behavior appear, depending on the value of $\kk^{2}$ with respect to $\e R_0^{-2}$:
\begin{description}
\item  [$\kk^{2}> \e R_0^{-2}$\,:] the contribution of the interaction in the measure is negligible and the propagator behaves as the one of the free gas; 
\item [$\kk^{2} < \e R_0^{-2}$\,:] the contribution of Bogoliubov interaction dominates with respect to $\kk^2$ and the propagator becomes very different from the free one. 
\end{description}
Let denote with $\bh$  the scale index which divides the two region, that is   
\be
\kk^2 =\g^{\bh} R_0^{-2}= \e R_0^{-2} \quad \Rightarrow \quad \g^\bh = \e
\ee
In the following it will be convenient to rewrite the propagator in terms of dimensionless momenta $k'=(k'_0, \kk')$, defined in such a way that $|k'|^2=R_0^4 |k|^2$, that is
\[ \label{scalings}
& k'_0  = R_0^{2} \,k_0 \non \\[6pt]
&  \kk'^2 = R_0^{2} \,\kk^2
%\begin{cases}
%R_0^{2}\, \kk^2 & \quad \text{for } \kk^{2}> \e R_0^{-2} \non\\
%\e\, R_0^{2}\, \kk^2 & \quad \text{for } \kk^{2}\leq \e R_0^{-2} 
%\end{cases}
\]
In terms of the adimensional variable $k'$ the propagator at scale $h$ becomes:
\be \label{adimensional_prop}
g_{\a \a'}^{(h)}(x)=\frac{1}{(2\pi)^{d+1}}\int d^{d+1}k' \,f_{h}(|k'|)\, 
\frac{ p^{h}_{\a \a'}(k')}{|k'|^2}\, e^{-ik'x}
\ee
where
\[
p_{\a \a'}^{h}(k') & \simeq \r_0^{-1}R_0^2 \left(\begin{array}{cc}
\kk'^{2}   & -k'_{0}\\
k'_{0} & \quad \kk'^{2}
\end{array}\right)  &&
|k'|^2 \simeq k'^2_0 + \kk'^4  \quad \text{for } h > \bh \label{prop_up} \\[6pt]
p_{\a \a'}^{h} (k') & \simeq \r_0^{-1}R_0^2 \left(\begin{array}{cc}
\kk'^{2}    & -k'_{0}\\
k'_{0} &   \e
\end{array}\right)  &&
|k'|^2\simeq k'^2_0 + \e \kk'^2  \quad \text{for } h \leq \bh  \label{prop_down}
\]  
and  the symbol ``$\simeq$'' means that we are considering the dominant behavior in $k$. Note that there is a natural connection between the propagators in two regions, since for $h=\bh$  \eqref{prop_up} and  \eqref{prop_down} coincide. 
The propagator at scale $h$ defined in \eqref{adimensional_prop} satisfies the following lemma, which is proved in appendix \ref{App-lemma}. 

{\lemma  \label{lemma}
Let $\bh$ the scale index such that $\g^{\bh}= \e$. Here $\e =2\l \r_0 R_0^{d}$ is an adimensional parameter related to the intensity $\l $ and the range $R_0$ of the interacting potential and to the density $\r_0$ of the condensed state. Then, the propagator $g_{\a\a'}^{(h)}(x)$ defined in \eqref{adimensional_prop} satisfies the following bounds:  
\[ 
& \lft|g_{\a \a'}^{(h)}(x) \rgt|  \leq \lft(\r_0 R_0^{-2} \rgt)^{-1} \g^{\frac{d}{2}h}\, \frac{C_N}{1+\lft[(\g^h x_0)^{2}  + (\g^{\frac{h}{2}} \xx )^{2} \rgt]^N} & h>\bh    \label{prop_above} \\[6pt] 
 & \lft|g_{\a \a'}^{(h)}(x) \rgt|  \leq \lft(\r_0 R_0^{-2} \rgt)^{-1} \e^{\frac{d}{2}}\,(\e^{-1}\g^h)^{\lft(d-1 + \d_{\a l}+ \d_{\a' l} \rgt)} \, \frac{C_N }{1+\lft[(\g^h x_0)^{2}  + (\g^h \sqrt{\e}\,\xx )^{2} \rgt]^N}  & h \leq \bh \label{prop_below}
\]
with $\a,\a'=l,t$, $d$ the spatial dimension of the system and $C_N$ a constant depending only on the integer $N$. The dimensional factor $(\r_0 R_0^{-2})$ has the dimensions of an action density in space time. \\}

Note that in the region $h\geq \bh$ the behavior of the propagator is independent on $\a, \a'$; in fact the propagator is exactly equal to the free one, being he contribution to coming from Bogoliubov interaction is smaller than $\kk^2$.  On the contrary in the region $h < \bh$ , due to Bogoliubov contribution, the fields $\ps^{l, (h)}_x$ and $\ps^{t, ( h)}_x$ have different scalings dimension. A dependence on $\e$ also appears in the propagator due to the fact that at each scale $h$ we have $\e\kk^2 \simeq \g^{2h}$. 
%
%In particular we can define dimensionless fields $\tl{\ps}^{\a,(h)}$ such that 
%\[  \label{dimensions_fields}
%& \ps^{\a,(h)}_x= \g^{\frac{d}{4}h} \,\tl{\ps}^{\a,(h)}_x \quad  \a=l,t & \text{for } h > \bh  \non \\
%& \ps^{l,(h)}_x= \g^{\frac{d+1}{2}h} \,\tl{\ps}^{l,(h)}_x \; ,  \; \ps^{t,(h)}_x= \g^{\frac{d-1}{2}h} \,\tl{\ps}^{t,(h)}_x \qquad & \text{for } h\leq \bh
%\]
%

Once the original field is decomposed into the $\ps^{\a,(h)}$ fields, using the {\it addition principle} for gaussian measures (see \eqref{addition_principle}) we can rewrite \eqref{free-energy} as
\[ \label{free2}
e^{-|\L|\, \WW_{\L}(\r_0)} = \int \prod_{h=-\io}^0 P^\L_{Q_0,f_h}(d\psi) e^{-\VV_0(\ps)}
\]
with $P^\L_{Q_0,f_h}(d\psi)$ denoting the measure with covariance $g^{(h)}(x)$. Since we will get uniform estimates in the volume $\L$ from now on we will skip the dependence on it in the free energy and the measure. 

In order to calculate \eqref{free2} we will consider a new potential  $\WW_{h^*}(\r_0)$ with an infrared cutoff at scale $h^*$, \ie defined by
\[ \label{free3}
e^{-|\L|\, \WW_{h^*}(\r_0)} =  \int P_{Q_0,\c_{[h^*,0]}}(d\psi) \,e^{-\VV_0(\ps)} 
%\int \prod_{h=h^*}^0 P^\L_{Q_0,f_h}(d\psi) e^{-\VV_0(\ps)}
\]
where we have introduced the notation
\[
 P_{Q_0,\c_{[h,0]}}(d\psi):= \prod_{j=h}^0 P_{Q_0,f_j}(d\psi)
\] 
with $\c_{[p,q]}=\sum_{i=p}^q f_i(k)$. With this definition
\be
\WW(\r_0)= \lim_{h^*\arr -\io} \WW^{(h^*)}(\r_0)
\ee
and our result will consist in proving uniform bounds in $h^*$, allowing us to take the limit $h^* \arr -\io$. 

The functional integral in \eqref{free3} is evaluated by integrating one by one each of the bosonic fields at a certain scale, starting from that with greater energy.  After the integration of the fields $\PS{0}, \ldots, \PS{h+1}$ we can rewrite the effective potential in a way similar to the \eqref{free3} but with a new effective interaction $\BV_h$: 
\[  \label{free4}
e^{-|\L|\, \WW_{h^*}(\r_0)} =  e^{-\bar{E}_h(\r_0)} \int P_{Q_0,\c_{[h^*,h]}}(d\PS{\leq h}) \,e^{-\BV_h(\PS{\leq h})} 
\]
Here $E_h(\r_0)$ is the contribution to $|\L|\WW_{h^*}(\r_0)$ coming from the integration over the first $|h|$ energy scales. We define the {\it effective potential at scale $(h-1)$} the potential $\BV_{h-1}(\PS{\leq h-1})$ such that:
\[ \label{Vbar}
e^{-\BV_{h-1}(\PS{\leq h-1})-\tl{E}_h(\r_0)} & =  
% \int P_{Q_0,\c{[h,0]}} (d\PS{h}) \,e^{-\BV_0(\PS{0}+\ldots +\PS{h}+\PS{\leq h-1})}   \non \\ & =
 \int P_{Q_0,f_{h}} (d\PS{h}) \,e^{-\BV_{h}(\PS{h}+\PS{\leq h-1})} 
\]
Using \eqref{Vbar} one can see that the identity \eqref{free4} is reproduced at scale $(h-1)$ with $\bar{E}_{h-1}=\bar{E}_h + \tl{E}_h$. The reason for using the symbol ``$\BV$''  to indicate the effective potentials $\{ \BV_h \}_{h \in [h^*+1, 0]}$ is the fact that towards this chapter we will introduce a second family of effective potentials, $\{ \VV_h \}_{h \in [h^*+1, 0]}$, obtained as a slight modification of the first one.  

\subsubsection{Gallavotti--Nicol\`o trees expansion}
Let now consider the first step of the iterative integration in \eqref{free2}, \ie the integration on the fields $\psi^{(0)}$. We get:
\be
e^{-\BV_{-1}(\PS{\leq -1})}=\int P_{Q_0, f_0}(d\ps)e^{-\BV_0 (\PS{0}+ \PS{\leq -1})}
\ee
It turns convenient to rewrite the integration over the $\ps^{(0)}$ fields in terms of the {\it truncated expectation} $\EE^T(X;n)$ defined by
\[
\EE^T(X;n)= \frac{\dpr^n}{\dpr \l^n} \log \int P(d\ps) e^{\l X(\ps)} \Bigl|_{\l=0}
\]
with $X$ a function defined on the fields $\ps$ and $n$ a positive integer number. By using the properties of the truncated expectations, see appendix~\ref{A1}, we have
\[ \label{ch2_inv_exp}
\int P(d \ps) e^{X(\ps + \f)} = \exp \lft[\, \sum_{n =0}^{\io} \frac{1}{n!} \EE^T \lft( X(\cdot+\f) ;n \rgt) \,\rgt] 
\]
with $\f$ a non integrated field. By using \eqref{ch2_inv_exp} we can rewrite $-\BV_{(-1)}$ as a sum over truncated expectations:
\be \label{expansion1}
-\BV_{-1}(\PS{\leq -1})= \sum_{n=0}^{\io} \frac{1}{n!}\, {\EE_{0}}^T\left(-\BV_0(\,\cdot + \PS{\leq -1});\,n \right)
\ee
The latter sum can be graphically represented as a sum over ``trees'', as depicted in the picture \ref{t1}. For each of the trees showed in fig. \ref{t1}
\begin{enumerate}[-- i --]
\item the left index is called ``root'' of the tree; we can associate to it the label $-1$ to take in mind that we are calculating $-\BV_{-1}$;
\item each of the points of the tree with label $+1$ (end points) represents a term $-\BV_0(\PS{\leq 0})$; 
\item the vertices with label $0$ represent truncated expectations with respect the field $\PS{0}$, \ie $\EE_0^T$;
\item to each of the trees is associated a combinatorial factor $\frac{1}{n!}$ with $n$ the number of end points. 
\end{enumerate}
%----------------------------------
%
%
\tik{
\begin{tikzpicture}[scale=0.7]
\draw (-1.4,0) node{$-\BV_{-1}=$};
\foreach \x in {0,...,2}   
    {
        \draw[th] (\x , -1) -- (\x , 1);   
        \node[vertex] at (\x,0){};   
    }
\draw[med] (0,0) -- (2,0);   
\draw (3,0) node{$+$};
%%%%
\foreach \x in {4,...,6}   
    {
        \draw[th] (\x , -1) -- (\x , 1);    
    } 
\draw[med] (4,0) node[vertex]{} --   (5,0) node [vertex] {} -- (6,0.8) node[vertex] {};  
\draw[med] (5,0) node [vertex] {} -- (6,-0.8) node[vertex] {};  
\draw (7,0) node{$+$};
%%%
\foreach \x in {8,...,10}   
    {
        \draw[th] (\x , -1) -- (\x , 1);      
    }
\draw[med] (8,0) node[vertex]{} --   (9,0) node [vertex] {} -- (10,0) node[vertex] {};  
\draw[med] (9,0) -- (10,0.8) node[vertex] {};  
\draw[med] (9,0) -- (10,-0.8) node[vertex] {};  
\draw (11,0) node{$+$};
\draw (12,0) node{$\cdots$};
%\foreach \x in {12,...,14}   
   % {
      %  \draw[th] (\x , -1) -- (\x , 1);      
    %}
%\draw[med] (12,0) node[vertex]{} --   (13,0) node [vertex] {} -- (14,0.8) node[vertex] {};  
%\draw[med] (13,0) -- (14,0.25) node[vertex] {};  
%\draw[med] (13,0) -- (14,-0.25) node[vertex] {};  
%\draw[med] (13,0) -- (14,-0.8) node[vertex] {};  
\draw (13,0) node{$=$};
\foreach \x in {14,...,16}   
    {
\draw[th] (\x , -1) -- (\x , 1);   
     }   
\draw[med] (14,0) node[vertex]{} --   (15,0) node [bvertex] {} ;  
\treelabelsize{
\draw (0,-1.5) node {$-1$};  
\draw (1,-1.5) node {$0$};
\draw (2,-1.5) node {$1$};
\draw (6,-1.5) node {$-1$};  
\draw (5,-1.5) node {$0$};
\draw (4,-1.5) node {$1$};
\draw (8,-1.5) node {$-1$};  
\draw (9,-1.5) node {$0$};
\draw (10,-1.5) node {$1$};
\draw (14,-1.5) node {$-1$};
\draw (15,-1.5) node {$0$};
\draw (16,-1.5) node {$1$}; }
\end{tikzpicture}
}{Graphical representation of the expansion \eqref{expansion1}. We will refer to the elements of the graphical sum as ``trees''. We have explicitly indicated the  label associated to the vertical lines only for first tree, but the same labels hold for each of the trees in the sum. The thick vertex corresponding to the vertical line with label $0$ is a graphical way to represent all the trees contributing to $-\BV_{-1}$.  }{t1}
%
%
% ----------------------------------------------------------
The graphical representation of the effective potentials in terms of trees results very useful in the iteration of the integration. In fact, if we consider the second step of the integration, we get:
\be 
e^{-\BV_{-2}(\PS{\leq -2})}=\int P_{-1}(d\ps)e^{-\BV_{-1} (\PS{-1}+ \PS{\leq -2})}
\ee
which becomes
\[ \label{expansion2}
-\BV_{-2}(\PS{\leq -2})= & \sum_{n=0}^{\io} \frac{1}{n!} \EE^T_{-1}\left(-\BV_{-1}(\,\cdot + \PS{\leq -2});\,n \right) \non \\
= & \sum_{n=0}^{\io} \frac{1}{n!}{\EE_{-1}}^T\left( 
\sum_{n'=0}^{\io} \frac{1}{n'!} \EE^T_{0} \left(-\BV_{0};\,n'\right);\,n \right)
\]
%----------------------------------
%
%
\tik{
\[
&
\begin{tikzpicture}[scale=0.7]
\draw (-1.4,0) node{$-\BV_{-2}=$}; 
\foreach \x in {0,...,2}   
    {
        \draw[th] (\x , -1) -- (\x , 1);    
    }
\node[vertex] at (0,0){};
\node[vertex] at (1,0){};
\node[bvertex] at (2,0){};
\draw[med] (0,0) -- (2,0);   
\draw (3,0) node{$+$};
%%%%
\foreach \x in {4,...,6}   
    {
        \draw[th] (\x , -1) -- (\x , 1);    
    } 
\draw[med] (4,0) node[vertex]{} --   (5,0) node [vertex] {} -- (6,0.8) node[bvertex] {};  
\draw[med] (5,0) node [vertex] {} -- (6,-0.8) node[bvertex] {};  
\draw (7,0) node{$+$};
%%%
\foreach \x in {8,...,10}   
    {
        \draw[th] (\x , -1) -- (\x , 1);      
    }
\draw[med] (8,0) node[vertex]{} --   (9,0) node [vertex] {} -- (10,0) node[bvertex] {};  
\draw[med] (9,0) -- (10,0.8) node[bvertex] {};  
\draw[med] (9,0) -- (10,-0.8) node[bvertex] {};  
\draw (11,0) node{$+$};
\draw (12,0) node{$\cdots$};
\treelabelsize{
\draw (0,-1.5) node {$-2$};  
\draw (1,-1.5) node {$-1$};
\draw (2,-1.5) node {$0$};
\draw (4,-1.5) node {$-2$};  
\draw (5,-1.5) node {$-1$};
\draw (6,-1.5) node {$0$};
\draw (8,-1.5) node {$-2$};  
\draw (9,-1.5) node {$-1$};
\draw (10,-1.5) node {$0$};
}
\end{tikzpicture} \non \\[12pt]   %------------------------------------------------------------
&  \hskip 1cm
\begin{tikzpicture}[scale =0.7]
\draw (-1,0) node{$=$};
\foreach \x in {0,...,3}   
    {
\draw[th] (\x , -1) -- (\x , 1);   
\node[vertex] at (\x,0){};  
     }   
\draw[med] (0,0) -- (3,0);
%%%%
\draw (4,0) node{$+$};
%%%%
\foreach \x in {5,...,8}   
    {
        \draw[th] (\x , -1) -- (\x , 1);    
    } 
\draw[med] (5,0) node[vertex]{} -- (6,0) node[vertex]{} --   (7,0) node [vertex] {} -- (8,0.8) node[vertex] {};  
\draw[med] (7,0) node [vertex] {} -- (8,-0.8) node[vertex] {};  
\draw (9,0) node{$+$};
%%%
\foreach \x in {10,...,13}   
    {
        \draw[th] (\x , -1) -- (\x , 1);      
    }
\draw[med] (10,0) node[vertex]{} --   (11,0) node [vertex] {} -- (12,0) node[vertex] {} -- (13,0) node[vertex] {};  
\draw[med] (12,0) -- (13,0.8) node[vertex] {};  
\draw[med] (12,0) -- (13,-0.8) node[vertex] {}; 
\treelabelsize{
\draw (0,-1.5) node {$-2$};
\draw (1,-1.5) node {$-1$};
\draw (2,-1.5) node {$0$};
\draw (3,-1.5) node {$1$};
\draw (5,-1.5) node {$-2$};
\draw (6,-1.5) node {$-1$};
\draw (7,-1.5) node {$0$};
\draw (8,-1.5) node {$1$};
\draw (10,-1.5) node {$-2$};
\draw (11,-1.5) node {$-1$};
\draw (12,-1.5) node {$0$};
\draw (13,-1.5) node {$1$};
} 
\end{tikzpicture}  
\non \\[12pt]  %------------------------------------------------------------
& \hskip 1.2cm \begin{tikzpicture}[scale =0.7]
\draw (-1,0) node{$+$};
\foreach \x in {0,...,3}   
    {
\draw[th] (\x , -1) -- (\x , 1);    
     }   
\draw[med] (0,0) node[vertex]{} -- (1,0) node[vertex]{} --   (2,0.4) node [vertex] {} -- (3,0.8) node[vertex] {}; 
\draw[med] (1,0) -- (2,-0.4) node [vertex] {} -- (3,-0.8) node[vertex] {}; 
%%%%
\draw (4,0) node{$+$};
%%%%
\foreach \x in {5,...,8}   
    {
        \draw[th] (\x , -1) -- (\x , 1);    
    } 
\draw[med] (5,0) node[vertex]{} -- (6,0) node[vertex]{} --   (7,0) node [vertex] {} -- (8,0) node[vertex] {};  
\draw[med] (6,0) node [vertex] {} -- (7,0.4) node[vertex] {} -- (8,0.8) node[vertex] {}; 
\draw[med] (6,0) node [vertex] {} -- (7,-0.4) node[vertex] {} -- (8,-0.8) node[vertex] {};  
\draw (9,0) node{$+$};
%%%
\foreach \x in {10,...,13}   
    {
        \draw[th] (\x , -1) -- (\x , 1);      
    }
\draw[med] (10,0) node[vertex]{} --   (11,0) node [vertex] {} -- (12,0.4) node[vertex] {} -- (13,0.8) node[vertex] {};  
\draw[med] (12,0.4) -- (13,0) node[vertex] {};  
\draw[med] (11,0) -- (12,-0.4) node[vertex] {}-- (13,-0.8) node[vertex] {};  
\draw (14,0) node{$+$};
\draw (15,0) node{$\cdots$};
\treelabelsize{
\draw (0,-1.5) node {$-2$};
\draw (1,-1.5) node {$-1$};
\draw (2,-1.5) node {$0$};
\draw (3,-1.5) node {$1$};
\draw (5,-1.5) node {$-2$};
\draw (6,-1.5) node {$-1$};
\draw (7,-1.5) node {$0$};
\draw (8,-1.5) node {$1$};
\draw (10,-1.5) node {$-2$};
\draw (11,-1.5) node {$-1$};
\draw (12,-1.5) node {$0$};
\draw (13,-1.5) node {$1$};
} 
\end{tikzpicture} \non
\]
}{Graphical representation of the expansion \eqref{expansion2}. The first line represents the expansion of $-\BV_{-2}$ in terms of $-\BV_{-1}$, while the second and third lines is the graphical representation of $-\BV_{-2}$ in terms of the element of the interacting potential $\BV_{I}$. The latter expansion is obtained using the definition of the thick vertex at scale $0$ given in fig. \ref{t1}.}{t2}
%
%
% ----------------------------------------------------------

The graphical representation of the equation \eqref{expansion2} is given in picture \ref{t2}. Iterating the strategy so far discussed $|h|$ times, we obtain a representation for the effective potential at scale $h$, $\BV_h$ in terms of {\it Gallavotti--Nicol\`o trees}, that is:
\be \label{sviluppo_h}
-\BV_h(\PS{\leq h}) = \sum_{n=1}^{\io} \sum_{\t \in \TT_{h,n}} \BV_h(\t, \PS{\leq h})
\ee
where the sum is over the set $\TT_{h,n}$ of the possible trees with $n$ end points at scale $1$ and a root with scale index $h$.  A possible tree $\t \in \TT_{h,n}$ is depicted in picture \ref{t3}.
\fig{ht}{0.8}{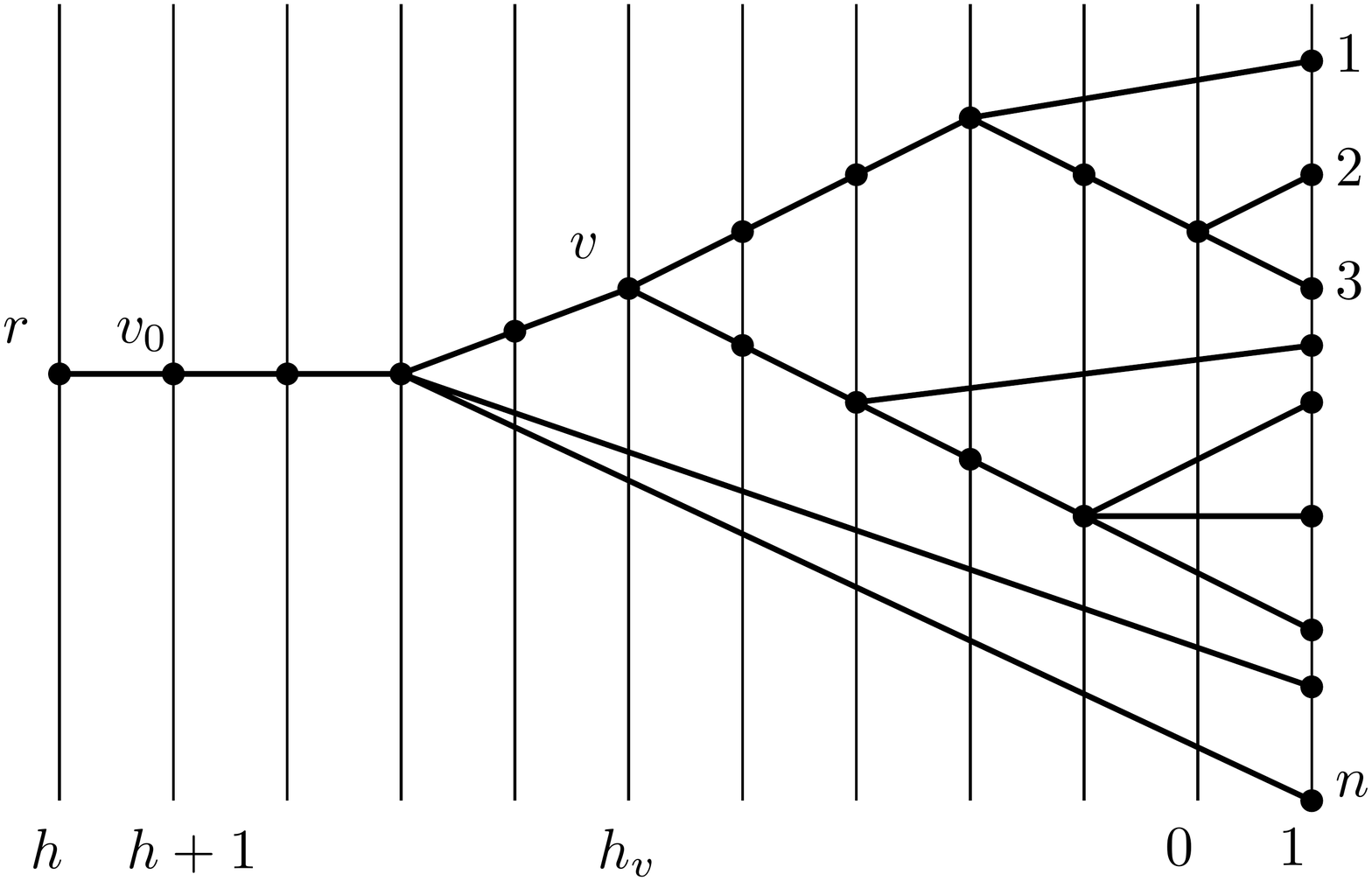}{Example of tree of order $n$ contributing to the graphical representation of $\BV_h$. All the endpoints are on the vertical line with label $1$. }{t3}
It is useful to introduce some definitions and notations regarding the structure of the trees in \eqref{expansion2}: 
%\begin{enumerate}[-- i --]
%\item Let consider the family of trees which can be constructed by joining a point 
%\end{enumerate}
\begin{description}
\item[\it Scale or frequency:] the integer index taking values in $[h,1]$ associated to the vertical lines. 
\item[\it End--points:] points of the tree associated to the vertical line with scale $1$; the number $n$ of end points is called {\it order of the tree}. With each end point $v$ we associate a factor $\BV_0(\PS{\leq 0})$ and a  space--time point $x_v \in \L$, corresponding to the integration variable in the $x$--space representation in the terms contributing to $\BV_0(\PS{\leq 0})$ . 
\item[\it Vertices:] the $n\geq 1$ points on the vertical lines labeled by a scale taking values in $[h+1,1]$. Since the structure of the tree induces a partial ordering between the vertices, we will say that $v_1<v_2$ if $v_1$ is on the path which joins $r$ with $v_2$. Given a vertex $v$ we will denote with $v'$ the vertex immediately preceding $v$ on $\t$. 
\item[\it Root of the tree:] the point $r$ in the tree with scale index $h \leq 0$; note that it is not considered a vertex according to the previous definition.
\item[\it Trivial vertices:] denoting with $s_v$ the number of vertices which immediately follow the vertex $v$, $v$ is {\it non trivial} if $s_v >1$ or $s_v=0$, while is called {\it trivial} if $s_v=1$. The non trivial vertices  with $s_v>1$ are the branching point of the trees; those with $s_v=0$ are the endpoints.
\item[\it Subtrees:] given a tree $\t$ and a vertex $v$, we will call $\t_v$ the subtree of $\t$ whose root is $v$. 
\item[\it Labeled tree:] a tree is said labeled if a scale index is associated to each of its vertices, as in picture \ref{t3}.
\item[\it Unlabeled tree:] for each labeled tree $\t$ it exists an unlabeled tree which has the same topological structure of $\t$, but don't have either trivial vertices, or frequency indices with the exception of the index $r$ to distinguish the root. The unlabeled tree corresponding to the tree of picture \ref{t3} is showed in picture \ref{t4}. Two unlabeled trees are identified if they can be superposed by a suitable continuous deformation, so that the endpoints with the same index coincide. The combinatorial factor associated to an unlabeled tree is the same of one of the corresponding labeled trees, \ie $1/n(\t)!$
\end{description}
\fig{ht}{0.55}{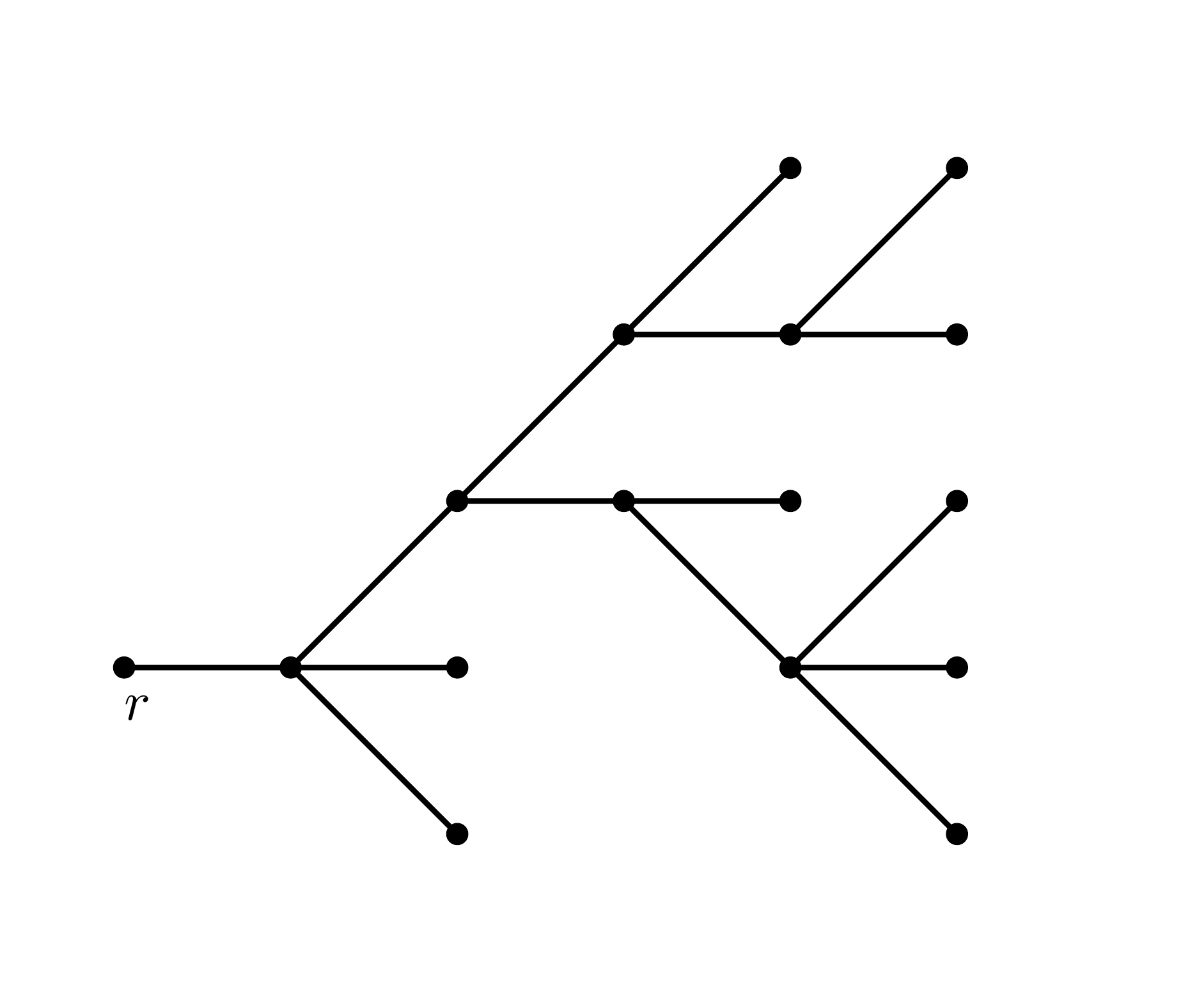}{The unlabeled tree corresponding to the tree in fig. \ref{t3}. The topological structure of the unlabeled tree is the same of the labeled one, but the trivial vertices are absent. The only label is the letter $r$ associated to the root.}{t4}
Once introduced this definitions we see that it is convenient to expand the sum over the labeled trees in \eqref{sviluppo_h} as
\be
\sum_{\t \in \TT_{h,n}}= \sum_{\t \in \TT^*_{n}} \sum_{\{h_v\}_{v\in \t}}
\ee
where $\TT^*_{n}$ is the set of the rooted unlabeled trees with $n$ final points. The set of indices $\{h_v\}$ satisfies the following properties: the labels take all values in $[h+1,0]$ and are consistent with the vertices ordering induced by the tree structure; the label of the end points is always equal to $1$. Some technical results on the counting of trees are resummed in appendix \ref{App-counting}.

%In the following we resum some technical results on the counting of trees (see~\cite{GalReview, GM} for a proof):
%\begin{enumerate}[i --]
%\ item The number of rooted unlabeled trees with $n$ end--points is bounded by $C_1^n$ with $C_1$ constant. 
%\item The number of rooted unlabeled trees with $m$ vertices is bounded by $C_2^m$ with $C_2$ constant. 
%\item Let $\TT_{h,n}$ to be the set of labeled trees with $n$ end points and root label $h$. For $\g >1$ and $\d >0$ we have
%\be \label{sum_over_scale_indices}
%\sum_{\TT_{h,n}} \prod_{v\, \text{not e.p.}} \g^{-\d(h_v -h_{v'})} \leq C_2^n
%\ee
%where e.p. is a shortcut for ``endpoints''. % is the set of the end points of the tree $\t$. 
%\item Given $n$ points, the number of \blue{unlabeled} trees connecting them is bounded by $C^n\, n!$
%\end{enumerate}

\subsubsection{Contribution to $\BV_h(\t, \PS{\leq h})$ from a single tree}

In the following we will describe the rules to calculate the contribution to $\BV_h(\PS{\leq h})$ coming from a particular tree $\t$. First of all we notice that the factor $-\BV_0(\PS{\leq0})$ associated to the end points is a sum of different terms proportional to $\l_0$, $\m_0$ and $\n_0$; then it is convenient to associate to each end point a second label $\s$, with values $\l, \m, \n$ denoting the term of the potential associated to that particular end point. 
The terms  $\BV_h(\PS{\leq h})$ can be defined recursively as follows:
\begin{enumerate}[1)]
\item Let us denote with $\t_0$ the trivial tree, that is the tree of order one given by a single line connecting the root with a unique end point with identity index $\s$; then the root index is $h=0$ and 
\be
\BV_0(\t_0, \PS{\leq 0})= \BV^\s_0(\PS{\leq 0})
\ee
with $\BV^\s_0$ representing one of the contributions of the terms of the interacting potential in \eqref{BV_I}, \ie 
\[
& \BV^\l_0(\PS{\leq 0})= \r_0 R_0^{-2}\, \l_0 \int dx \bigl(\ps^t_x \bigr)^4 
&& \BV^{\l'}_0(\PS{\leq 0})= \r_0 R_0^{-2}\, \l'_0 \int dx \bigl(\ps^t_x \bigr)^2 \bigl(\ps^l_x \bigr)^2 \non \\
& \BV^{\l''}_0(\PS{\leq 0})= \r_0 R_0^{-2}\, \l''_0 \int dx \bigl(\ps^l_x \bigr)^4
&& \BV^\m_0(\PS{\leq 0})= \r_0 R_0^{-2}\, \m_0 \int dx\,  \psi_x^l \bigl(\ps^t_x \bigr)^2  \non \\
& \BV^{\m'}_0(\PS{\leq 0})= \r_0 R_0^{-2}\, \m'_0 \int dx  \bigl(\ps^l_x \bigr)^3
&& \BV^\n_0(\PS{\leq 0})= \r_0 R_0^{-2}\, \n_0 \int dx  \bigl(\ps^t_x \bigr)^2
\] 
with $2 \l_0 =2\l''_0= \l'_0= \e/8$, $\m_0=\m'_0= \e \sqrt{2}/4$ and $\n_0= \n R_0^2 /2$. 
\item Let us consider a not trivial tree $\t$ with scale $h$; if $v_0$ is the first non trivial vertex following the root in $\t$ and $\t_1, \ldots, \t_{s_{v_0}}$ are the subtrees of $\t$ having $v_0$ as root, then
\be \label{valore_albero}
\BV_h(\t, \PS{\leq h})= \frac{1}{s_{v_0}!} \EE^T_{h+1} \left( 
\BV_{h+1}(\t_1, \PS{\leq h+1}), \ldots, \BV_{h+1}(\t_{s_{v_0}}, \PS{\leq h+1})
\right)
\ee 
\item The \eqref{valore_albero} can be iterated, obtaining
\[ \label{valore_albero2}
\BV_h(\t, \PS{\leq h}) & = \prod_{v \in V(\t)} \frac{1}{s_v!} \non \\
& \EE^T_{h+1} \left( \EE^T_{h+2} \biggl( \ldots \EE^T_{-1} \Bigl(
\EE^T_0 \bigl(\, \BV(\t_0, \PS{\leq 0}),\ldots \bigr),\ldots \Bigr)
, \ldots \biggr), \ldots\right)
\]
with $V(\t)$ the set of vertices of $\t$. Note that the expectations in \eqref{valore_albero2} have to be calculated from the end points back to the root.
\end{enumerate}

Each time we encounter a vertex $v$ which is not an end point we have then to calculate an expression similar to \eqref{valore_albero}. The latter is a sum of several contributions, differing for the choices of the fields contracted under the action of the truncated expectation $\EE^T_{h_v}$ associated to each not endpoint vertex $v$.
In order to further expand the expression \eqref{valore_albero}, we introduce more definitions, allowing us to distinguish the fields that are contracted or not in the expectations.
\begin{description}
\item[\it Field label $f$:] it labels the field variables appearing in the monomials associated with the endpoints, \ie the lines going out of the vertices. For each field $f$ we denote with $x(f)$ and $\a(f)$  respectively the space--point time and the $\a$ index $(\a=l,t)$ associated to the bosonic field variable with label $f$. Then we may have (or not) a further label $\dpr_0(f)$ or $\dpr_\xx(f)$ which indicates the presence (or absence) of a derivative with respect the $x_0$ or $\xx$ variables, which will act on the propagator as the field $f$ will be contracted. Even if our initial potential $\BV_0$ does not contain any field with derivative, we will include this case since this kind of lines will be appear in the bounds of sections \ref{Localization} and \ref{ext_fields}.
\item[\it Set of field labels $I_v$.] Given an end point $v$, then $I_v$ is the set of field labels associated to $v$; if $v$ is not an end point then we denote with $I_v$ the set of field label associated with the end points following the vertex $v$. 
\item[\it Set of external fields $P_v$. ] We denote with $P_v \subseteq I_v$ the set of the external fields coming out from the vertex $v$. In particular, if $v$ is an endpoint, then $P_v = I_v$. If $v$ is not an end point and $v_1, \ldots, v_s $ are the $s_v\geq1$ vertices immediately following it, then $P_v \subseteq \cup_i P_{v_i}$; then $P_v$ is the subset of $I_v$ containing the labels of the fields which result not contracted after  the action of all the expectations $\EE^T_{h_{v_i}}$ with $i=1,\ldots,s$. 
\item[\it Set of the internal fields ${\cal I}_v$. ] If $v$ is not an end point we define $Q_{v_i}=P_v \cap P_{v_i}$ with $v_i$ one of the $s$ vertices immediately following $v$ on the tree. This definition implies  $P_v =\cup_i Q_{v_i}$ The union of the subsets $P_{v_i} \setminus Q_{v_i}$  is by definition the set of the internal fields of $v$, and is not empty if $s_v>1$.
\end{description}
We will use the symbols $|I_v|$, $|P_v|$ and $|{\cal I}_v|$ in order to indicate the number of fields respectively in the sets $I_v$, $P_v$ and ${\cal I}_v$.
Using the previous definitions one can prove (see \cite{GalReview, GM}) that each time we encounter a vertex $v$ which is not an end point we have to calculate an expression of the kind
\be \label{expectations}
\frac{1}{s_{v}!} \EE^T_{h_v} \left( 
\TPS{\leq h_v}{P_{v_1}}, \ldots,\TPS{\leq h_v}{P_{v_{s_v}}})
\right)
\ee
with
\be \label{field_product}
\TPS{\leq h_{v}}{P_{v}}= \prod_{f \in P_{v}} \ps^{\a(f),\,(\leq h_{v})}_{x(f)}
\ee
\ie a product of $|P_{v}|$ fields on scale $\leq h_v$. Therefore the effect of the truncated expectation $\EE^T_{h_v}$ is to contract the fields on scale $h_v$ appearing  in \eqref{expectations} in all the possible ways. 

Given $\t \in \TT_{h,n}$, there are many possible choices of the subsets $P_v$, with $v \in \t$, compatible with all the constraints. We shall denote by $\PP_\t$ the family of all these choices and by $\{P_v\}_{v \in \t}$ 
%$\bP_\t=\{P_v, v \in \t\}$ 
the elements of $\PP_\t$. We can then rewrite $\BV_h(\t, \PS{\leq h})$ as
\[ \label{sviluppo4}
\BV_h(\t, \PS{\leq h}) =\sum_{\{P_v\}_{v \in \t}} \int dx_{v_0} \, K_{\t, P_{v_0}}^{h+1}(x_{v_0})\,\ps_{P_{v_0}}^{(\leq h)}
\] 
where $x_{v_0}=\cup_{f\in I_{v_0}}\{ x(f)\}$ is the set of integration variables associated with $\t$ and $\TPS{\leq h}{P_{v_0}}$ is the product of all the external fields of the tree $\t$, \ie the fields which result not contracted after the last expectation $\EE^T_{h+1}$.
%is defined in \eqref{field_product}.
%\be
%\TPS{\leq h}{P_{v_0}} = \prod_{f \in P_{v_0}} \ps^{\a(f),\,(\leq h)}_{x(f)}
%\ee
Using \eqref{valore_albero} and \eqref{sviluppo4} we see that the kernel $K_{\t, P_{v_0}}^{h+1}(x_{v_0})$ is defined inductively by the equation, valid for any $v\in \t$ that is not an endpoint
\be \label{kernel}
K_{\t, P_{v}}^{h_v}(x_{v})= \sum_{P_{v_1}, \ldots, P_{v_{s_v}}}  \frac{1}{s_v!} \prod_{i=1}^{s_v}  \, K_{\t, P_{v_i}}^{h_v+1}(x_{v_i}) \, \EE^T_{h_v}\left(\TPS{h_v}{P_{v_1} \setminus Q_{v_1} }, \ldots, \TPS{h_v}{P_{v_{s_v}} \setminus Q_{v_{s_v}} } \right)
\ee
while if $v$ is an endpoint, then we have
\be
K_{\t, I_{v}}^{(1)}(x_{v})= r_v
\ee
with $r_v$ the term in $\BV_0$ associated to the endpoint $v$, \ie $r_v= \l_0, \m_0, \n_0$. Iterating \eqref{kernel} we get
\[ \label{kernel2}
K_{\t, P_{v_0}}^{h}&(x_{v_0}) = \non \\ 
& \sum_{\{P_v\}_{v \in \t}} 
\left( 
 \prod_{v \notin V_{\text{e.p.}}(\t)} \,\frac{1}{s_v!} \, \EE^T_{h_v}\Bigl(\TPS{h_v}{P_{v_1} \setminus Q_{v_1} }, \ldots, \TPS{h_v}{P_{v_{s_v}} \setminus Q_{v_{s_v}} } \Bigr) \rgt)
\lft( \prod_{v \in V_{\text{e.p.}}(\t)} r_v \rgt)
\]

\subsubsection{Feynman diagrams representation}

The truncated expectations in \eqref{kernel2} can be conveniently expressed into a sum of Feynman diagrams. Let's start to describe which are the diagrams contributing to the truncated expectation in \eqref{expectations}:
\begin{enumerate}
\item Given a vertex $v$, for each of its subvertex $v_1, \ldots, v_{v_s}$ we draw an element of a diagram and some  ``half--lines'' emerging from these elements representing the fields in the sets $P_{v_1}, \ldots, P_{v_{s_v}}$.  
\item For each vertex $v_i$ we draw a box $G_{v_i}$ containing all the endpoints following $v$ in the tree $\t$. These boxes, which we will call {\it clusters}, are such that the external lines to the vertex $v_i$, labeled by the indices in $P_{v_i}$, are also the external lines of the box $G_{v_i}$. Note that, by construction, there will be an inclusion relation by clusters such that $G_v \supset G_w$ if $v \prec w$.
\item We consider the half--lines representing the fields belonging to the sets $P_{v_i} \setminus Q_{v_i}$ (\ie the internal lines respect to the vertex $v$) and we contract them in pairs, in such a way that the subclusters $G_{v_1}, \ldots, G_{v_{v_s}}$, enclosing the sets $P_{v_1}, \ldots, P_{v_s}$ are all connected, see fig. \ref{clusters}. The latter property is required by the truncated expectation $\EE^T_{h_v}$; on the other hand it does not forbid the external lines of the same cluster to be contracted between themselves. 
\item To each line $l$ obtained joining the half--line representing $\ps^{\a, (\leq h_v)}_{x_i}$ with the half--line representing $\ps^{\a', (\leq h_v)}_{x_j}$ we associate a propagator $$g^{(h_v)}_l \= g_{\a \a'}^{(h_v)}(x_i - x_j)$$ If $l$ is a line contained in a diagram $\G$, we shall write $l \in \G$. 
\end{enumerate}

\tik{
\begin{tikzpicture}
\tikzstyle{every node}= [draw, shape= circle, inner sep=5pt, line width=1.3 pt]
%\draw[step=0.5cm, color=gray] (-2, -2) grid (4,2);
\draw (-2,0) node(v1){$G_{v_1}$};
\draw (0,-1.5) node(v2){$G_{v_2}$};
\draw (3,-0.5) node(v3){$G_{v_3}$};
\draw (0.5,1) node(v4){$G_{v_4}$};
\draw[line width=1.3pt]  (v1)--(v2);
\draw[line width=1.3pt, dashed]  (v2)--(1.5,-1);
\draw[line width=1.3pt] (1.5,-1)--(v3)--(v4);
\draw[line width=1.3pt,dashed]  (v4)--(v1);
\draw[line width=1.3pt]  (v4)--(v2);
\draw[line width=1.3pt, dashed]  (v4)--(1.7,2);
\draw[line width=1.3pt]  (v1)--(-1.7,-1.5); %(0, -3);
\end{tikzpicture}
}{A Feynman diagram obtained by joining some of the half lines $P_{v_1}, \ldots, P_{v_n}$ emerging from the clusters $G_{v_1}, \ldots, G_{v_n}$. The two different types of line, the plain and the dashed one, are used to denote respectively the fields $\ps^t$ and $\ps^l$.}{clusters}

We denote by $\GG(P_v)$ the set of all the Feynman diagrams which can be obtained by following the given prescriptions. To each diagram $\G \in \GG(P_v)$ corresponds a number, which will be called the {\it value} of the graph, given by the product of the propagators of the lines $l \in \G$:
\be
\Val(\G)= \prod_{l \in \G} g^{(h_v)}_l
\ee
Then the following equality holds:
\be \label{tru}
\EE^T_{h_v}\left(\TPS{h_v}{P_{v_1} \setminus Q_{v_1} }, \ldots, \TPS{h_v}{P_{v_{s_v}} \setminus Q_{v_{s_v}} } \right) = \sum_{\G \in \GG(P_v)} \Val(\G)
\ee 
If we repeat the latter construction for each of the truncated expectations in \eqref{kernel2}  we see that the kernel $K_{\t, P_{v_0}}^{h}(x_{v_0})$   can be written as a sum on the set $\GG(P_{v_0})$ of Feynman diagrams with the following properties
\begin{enumerate}[1)]
\item all the elements of the diagram are connected by the lines of $\G$; 
\item all the clusters $G_v$ are connected by the lines of $\G$; for each cluster $G_v$ the set $P_v$ determine the external lines of any diagram $\G_v$ which can be obtained by contracting the fields corresponding to the labels $f \in P_w$ with $v \preceq w$.  
\item the propagators contained in the cluster $G_v$ but not in some smaller cluster have scale $h_v$; 
\item the diagram $\G$ has $|P_{v_0}|$ external lines, with labels in $P_{v_0}$.
\end{enumerate}
An element $\G \in \GG(P_{v_0})$ is constructed moving along the tree $\t$ from the endpoints to the root; when a vertex $v$ is reached, we construct a diagram $\G_v$ formed by lines $l$ on scales $h_l \geq h_v$.   With these definitions we get:
\be
K^{(h)}_{\t, P_{v_0}}(x_{v_0}) = \sum_{\G \in \GG(P_{v_0})} \Val(\G)
\ee
Note that once a structure of a cluster has been fixed, there will be a lot of diagrams compatible with it: in fact we have a lot of different ways to contract between themselves the lines $P_v$ external to the each cluster $G_v$, once $P_{v_0}$ has been fixed. 
We stress some of the properties of the cluster structure we introduced: 
\begin{enumerate} 
\item The hierarchy structure of clusters provides an arrangement of endpoints which is the same underlying the tree structure in \eqref{valore_albero2}. So, given a tree, we can represent it as a set of clusters and viceversa, see fig. \ref{t4}, where only the clusters associated to nontrivial vertices are drawn.
\item Given a cluster $G_v$, if all the maximal subclusters $G_{v_1}, \ldots, G_{v_{v_s}}$ contained inside $G_v$ are thought as points, then the set of points so obtained is connected: so it is possible to single out a set of $s_v -1$ lines connecting them. Such a set will be called an {\it anchored tree}: it realizes a minimal connection between the maximal subclusters of $G_v$. 
\item We can associate a scale label to clusters. A cluster is said to be on scale $h$ if contains endpoints which are contracted by lines on scale $h' \geq h$ such that there is at least one line on scale $h$.  By extension we can consider also the endpoints as (trivial) clusters on scale $h=1$.  
\item We can associate a scale label to a line $l \in \G$; this is defined as the label of the smaller cluster which encloses the line. 
%\item  
%Each truncated expectation as \eqref{expectations} sees the clusters $P_{v_1}, \ldots,P_{v_{s_v}} $ as points: by this we mean that its action is independent on the internal structures of the subclusters $G_{v_1}, \ldots, G_{v_{v_s}}$ and depends only on the external lines of such clusters. The crucial properties is that, 
\end{enumerate}
An example of Feynman diagram, with the tree and the cluster structure associated to it, is given in fig.~\ref{t5}
%
%-----------------------------------------
\tik{
\[  \t = 
\parbox{4cm}{\centering{ \vskip 0.5cm
\begin{tikzpicture}%[scale=0.7]
%\draw (-0.5,0) node{$\t=$};
\foreach \x in {0,...,3}   
    {
        \draw[very thin] (\x , -1.5) -- (\x , 1.5);      
    }
\draw (0,-2) node {$-2$};  
\draw (1,-2) node {$-1$};
\draw (2,-2) node {$0$};
\draw (3,-2) node {$1$};
\draw[med] (0,0) node[vertex] {} -- (1,0) node[vertex] {} -- (2,0.5)node[vertex] {} -- (3,1)node[vertex, label=0:$1$] {};  
\draw[med] (2,0.5) -- (3,0)node[vertex, label=0:$2$]{};
\draw[med] (1,0) node[vertex] {} -- (2,-0.5) node[vertex] {} -- (3,-1)node[vertex,, label=0:$3$] {};
\end{tikzpicture}
}} 
\; \Leftrightarrow \;
\parbox{3.5cm}{\centering{
\begin{tikzpicture} [scale=0.9]
\draw[med]  (-0.4,0) circle(1cm);
\draw[med]  (0,0) circle(2cm);
\draw (-0.4, 1.2) node[label=right:$0$]{};
\draw (0, 2.2) node[label=right:$-1$]{};
\draw (-0.8, -0.1) node[vertex, label=$1$]{};
\draw (-0.1, -0.1) node[vertex, label=$2$]{};
\draw (1.2, -0.1) node[vertex, label=$3$]{};
%\draw[line width=1.3pt]
\end{tikzpicture}
}}  
\quad \Leftarrow \: \: \G=
\parbox{3.2cm}{\centering{ 
\begin{tikzpicture} [scale=0.9]
\draw[med]  (0,0.75) circle(0.75cm);
\draw[med]  (0,-0.75) circle(0.75cm);
\draw[med] (-1,-0.25) rectangle(1,1.75);
\draw (1.2,1.5) node[label=left:$0$]{};
\draw[med] (-1.5,-1.75) rectangle(1.5,2);
\draw (1.7,-1.5) node[label=left:$-1$]{};
\draw (0,0) node[vertex, label=$2$]{};
\draw (0,1.5) node[vertex, label=270:$1$]{};
\draw (0,-1.5) node[vertex, label=$3$]{};
\draw[med] (-1, -2.5)-- (0, -1.5) -- (1, -2.5);
\draw[med] (-1, 2.5)-- (0, 1.5) -- (1, 2.5);
%\draw[line width=1.3pt]
\end{tikzpicture}
}}  \non
\]
}
{An example of tree $\t$ of order 3 with the corresponding cluster structure, where only the non trivial vertices are depicted. The cluster structure uniquely identifies a tree $\t$ and viceversa.
%To each cluster is associated a scale label, corresponding to the frequency of the first non trivial vertex immediately preceding the vertices contained in the cluster. 
Down left an element of the class of Feynman diagrams compatible with $\t$; in the example all the elements of the diagram are assumed of type $\l$, in general one has to consider all the different vertices in $\BV_0$. }{t5}
%
%---------------------------------------------------------
The final expression for the effective potential in terms of the trees and Feynman diagrams  described in this section
\[ \label{final}
\BV_h(\PS{\leq h})& = \sum_{n=1}^\io \; \sum_{\t \in \TT^*_{h, n}} 
\sum_{\{h_v\}_{v\in \t}} \sum_{\{P_{v}\}_{v\in \t}}
 \int dx_{v_0} \PS{\leq h}_{P_{v_0}} \sum_{\G \in \GG(\{P_v\})} \Val(\G) %\\
%& =\sum_{n=1}^\io  \sum_{P_{v_0}}  \int dx_{v_0} \PS{\leq h}_{P_{v_0}} || V^{(h)}_{n_l, n_t}||
\] 
is called {\it non renormalized expansion}. This name comes from the fact that, as we will see in the next section, it will be necessary to introduce a different expansion, that will be called {\it renormalized}, in order to control the divergences emerging in \eqref{final}.
In \eqref{final}  $\GG(\{P_{v}\})$ is the set of the connected Feynman diagrams compatible with the cluster structure described by the set $\{P_{v}\}$.

We can write the \eqref{final} as a sum of kernels with fixed numbers $n^\txe_t$ and  $n^\txe_l$ of external legs of type $t$ and $l$:
\[ \label{potential_xspace}
\BV_{h}(\psi^{(\leq h)}) = \sum_{n^\txe_{l}+n^\txe_{t}\geq2}  \int dx_{P^*_{v_0}} V_{n^\txe_{l}n^\txe_{t}}^{(h)}(x_{P^*_{v_0}}) \, \PS{\leq h}_{P^*_{v_0}} 
%= &\sum_{n_{l}+n_{t}\geq2}  \int dx_{1}\ldots dx_{n_{l}}dy_{1}\ldots dy_{n_{t}} \non \\ 
%& W_{n_{l}n_{t}}^{(h)}(x_{1},\ldots x_{n_{l}};\, y_{1},\ldots,y_{n_{t}}) \psi_{x_{1}}^{l}\ldots\psi_{x_{n_{l}}}^{l}\psi_{y_{1}}^{t}\ldots\psi_{y_{n_{t}}}^{t}
\]
where the star in $P^*_{v_0}$ represents that the set of $|P^*_{v_0}|$ external fields is composed by $n^\txe_l$ external fields are of type $l$ and $n^\txe_t$ of type $t$, that is 
\[
& x_{P^*_{v_0}}=\{x_{1},\ldots x_{n^\txe_{l}};\, y_{1},\ldots,y_{n^\txe_{t}}\} \\
& \PS{\leq h}_{P^*_{v_0}}=\psi_{x_{1}}^{l} \ldots\psi_{x^\txe_{n_{l}}}^{l}\psi_{y_{1}}^{t}\ldots\psi_{y_{n^\txe_{t}}}^{t}
\]
and  the kernels $V_{n^\txe_{l}n^\txe_{t}}^{(h)}$ are defined by:
\[ \label{final_kernel}
V_{n^\txe_{l},\,n^\txe_{t}}^{(h)}(x_{P^*_{v_0}}) = 
\sum_{n=1}^\io \sum_{\t \in \TT_{h, n}} 
\sum_{ \substack{\{P_{v}\} \\ n^\txe_l, n^\txe_t\, \text{fixed}}} \sum_{\G \in \GG(\{P_{v}\})} \int dx_{I_{v_0}\setminus P^*_{v_0}  }  \Val(\G)
\]

\section{Dimensional bounds} \label{dimensional}
Let $V_{n^\txe_{l}n^\txe_{t}}^{(h);\,n}(x_{P^*_{v_0}})$ be the contribution due to the trees of order $n$ to the kernel $V_{n^\txe_{l}n^\txe_{t}}^{(h)}(x_{P_{v_0}})$ defined in \eqref{final_kernel}. We are interested in bounding the following object:
\[  \label{sum1}
|| V_{n^\txe_l, \,n^\txe_t}^{(h);\, n} || & \, = \frac{1}{\b L^d}\,\int dx_{P_{v_0}} | V_{n^\txe_{l},\,n^\txe_{t}}^{(h);\, n}(x_{P_{v_0}})|  \non \\
& = \sum_{\t \in \TT_{h,n}} 
\sum_{ \substack{\{P_{v}\} \\ n^\txe_l, n^\txe_t \text{fixed}}} 
\sum_{\G \in \GG(\{P_{v}\})} \lft| \frac{1}{\b L^d} \int dx_{v_0} \Val(\G) \rgt|
\]
Since the bounds we will derive are based only on dimensional arguments, such bounds are called {\it dimensional bounds}. 
%with
%\be
%\BV_h(\PS{\leq h}) =\sum_{n=1}^\io  \sum_{P_{v_0}}  \PS{\leq h}_{P_{v_0}} || V^{(h)}_{n_l, n_t}||
%\ee
In the latter sum $|P^*_{v_0}|= n^\txe_l + n^\txe_t$
is the number of external lines of the diagram, $n^\txe_l$ and $n^\txe_t$  are the numbers of external lines of type $l$ or $t$. In the following we will use also the symbols $n^\txe_{\dpr_0}$ and  $n^\txe_{\dpr_\xx}$ to indicate the number of external lines which have a label $\dpr_0$ or $\dpr_\xx$ respectively. 

In the following we will describe how to get a dimensional bound for the contribution in the sum \eqref{sum1} coming from a generic Feynman diagram $\G$. In the multiscale integration of the effective potential, we need to consider two main sets of trees: 
\begin{enumerate}[(1)]
\item the set of trees $\TT^>_{k,n}$ contributing to the calculation of the potential $\BV_k$ with $k\geq \bh$, \ie the trees with root at scale $k$ and $n$ endpoints at scale $1$. The Feynman diagrams compatible with these trees have propagators at scales $h_v > \bh$, whose behavior is described by  \eqref{prop_above}.
\item the set of trees $\TT^<_{h,n}$ contributing to the calculation of the potential $\BV_h$ with $h < \bh$. This may be conveniently seen as the set of trees with root at scale $h$ and $n$ endpoints at scale $(\bh+1)$, whose values correspond to the terms in $\BV_\bh$ calculated as in step (1). The Feynman diagrams compatible with these trees have propagators at scales $h_v \leq \bh$, whose behavior is described by  \eqref{prop_below}. It has to be stressed that the integration over the scale labels $[\bh+1,0]$ gives rise to vertices at scale $\bh$ with every possible number of external legs. In order to simplify the discussion, in the rest of the chapter we will consider only the case in which the vertices at scale $\bh$ are of the {\it same type} of the vertices of the original potential $\VV_0$. In the two dimensional case also the vertex with six plain legs will be considered, for reasons that will become clear at the end of this section. In appendix \eqref{app:hbar} we show that the discussion is not changed when we consider more general vertices. 

%The corresponding Feynman diagrams have lines contracted both above and below $\bh$, then we have to use both the estimates \eqref{prop_above} and \eqref{prop_below}. In place of considering the whole set of trees in $\TT_{h,n}$, for the moment we will consider the set $\TT^{\,<}_{h,n} \subset \TT_{h,n}$ of trees with endpoints.
% The Feynman diagrams compatible with the trees in $\TT^{\,<}_{h,n}$ have contracted lines only at scales $h_v \leq \bh$, whose behavior is described by  \eqref{prop_below}. The problem of considering the remaining trees in  $\TT_{h,n}$, \ie the trees  with branches connecting vertices above and below $\bh$, will be trivial once the contributions from the set $\TT^{\,<}_{h,n}$ and $\TT^{\,>}_{\bh,n}$ have been investigated. 
\end{enumerate}
From now on we will use the word ``half--line'' to refers to the fields $\ps^{l,t}$ and the word ``line'' when two of these fields are contracted. Referring to $\int dx_{v_0} |\Val(\G)|$, let's consider the following steps.
\begin{enumerate}[a)]
\item We perform the integration over the $x_{v_0}$ variables along the lines of an anchored tree which realizes a minimal connection inside each subclaster $G_v$ with $v>v_0$. The anchored tree is made by $\sum_{v > v_0} (s_v -1)$ lines and the following property may be proved by induction:
\[ 
\sum_{v\, \text{not e.p.}} (s_v -1)= m_{v_0} -1 
\]
with $m_{v_0}=|x_{v_0}|$ the number of endpoints following $v_0$ in $\t$. Then, using the lemma \ref{lemma} we can use the decaying part of the $g^{h_v}(x)$ propagator to bound each of the $(s_v -1)$ integrations $\int d^d x $  with their dimensional estimate $\g^{-\d_I h_v}$, where
\[
& \d_I = \frac{d}{2}+1  & \text{for } h> \bh \non \\
&  \d_I = d+1  & \text{for } h\leq \bh 
\]
being 
$\g^{2h} \simeq  k^2_0 + \kk^4$  for  $h> \bh $ and $\g^{2h} \simeq  k^2_0 + \kk^2 $ for $h\leq \bh$. Note that, at the end, all the integrations have been performed up to one, corresponding to a single endpoint of the tree: such an integration gives a factor $(\b L^d)$. 
\item We bound each contracted half--line in $\G$ with its dimensional estimate, using lemma \ref{lemma}. If we indicate with $\gamma^{\d_j h}$ with $j=l,t$ the dimensional estimate for the fields $\ps_x^{l,t}$ on scale $h$ we have
\[
& \d_l =\d_t =\frac{d}{4}  &\text{for } h> \bh \non \\
& \d_l =\frac{d+1}{2} \; , \quad \d_t =\frac{d-1}{2}&  \text{for } h\leq \bh 
\]
\item For each derivative $\dpr_0$ or $\dpr_\xx$ acting on a propagator $g^{(h_v)}(x)$ we get an extra contribution $\g^{\d_0 h_v}$ or $\g^{\d_1 h_v}$ to the dimensional estimates with
\[
& \d_0 =1 \; , \quad \d_1 =\frac{1}{2} &\text{for } h> \bh \non \\
& \d_0 =\d_1 = 1  &  \text{for } h\leq \bh 
\]
Then, for each contracted half--line bringing a label $\dpr_0$ or $\dpr_\xx$ we have a $\g^{\d_0}$ or $\g^{\d_1 h_v}$ factor in the bound.
\item We assume each initial coupling to be bounded by a constant $\h$. With ``initial coupling'' we mean the couplings $\{|\e|, \,|\n_0|\}$ at scale $h=0$ for the $\t^>$ trees and the couplings at scale $\bh$ for the trees $\t^<$ .  
\end{enumerate}
Putting together the previous estimate and taking into account the fact that the number of lines of a diagram of order $n$ is bounded by $C^n$, we get:
\[ \label{power1}
\bigl\|\Val(\G)\bigr\| & = \frac{1}{\b L^d} \int dx_{v_0} |\Val(\G)|   \non \\
& \leq  \, C^n \,C_d(P_v;\e,\r_0, R_0) %\non \\ &
 \prod_{v\, \text{not e.p.} }  \g^{- \d_I h_v (s_v -1)}  
\g^{(\d_l \tl{n}^\txi_{l,v} +\d_t \tl{n}^\txi_{t,v} + \d_0 \tl{n}^\txi_{\dpr_0,v} + \d_1 \tl{n}^\txi_{\dpr_\xx,v} ) h_v}
\]
where 
$\tl{n}_{j,v}^\txi$  is the number of half--lines %(\ie lines before the contraction) 
of $j$-type contained in the  cluster corresponding to the vertex $v$ but in none of the more inner clusters, the latter corresponding to  the vertices which follow $v$.  Then $ \tfrac{1}{2} \sum_{j=l,t} {\tl{n}}_{j,v}^{\txi}$ is the number of contractions occurring on scale $h_{v}$. 

The factor $C_d(P_v;\e,\r_0, R_0)$ contains the dependence of a generic Feynman graph with labels $\{P_v\}$ on the parameters $\e$, $\r_0$ and $R_0$; it also depends on the number $n$ of vertices, on the spatial dimension and on the region of momenta we are considering. In order not to overwhelm the discussion, its calculation is postponed in appendix \eqref{order_e}. In the following we will only report its value in the different cases.

Now, for each factor $\g^{h_v \d_v}$ in \eqref{power1} we can extract a factor $\g^{h \d_v}$ (with $h$ the scale label of the vertex $v_0$) and then rewrite the remaining $(h_v-h)$ factors using the following properties, which can be proved by induction:
\[ \label{prop1}
& \sum_{v\, \text{not e.p.}} \tl{n}^\txi_{j,v}= n^\txi_{j,v} \non \\
& \sum_{v\, \text{not e.p.}} (h_v -h)\,(s_v -1)= \sum_{v\, \text{not e.p.}} (h_v -h_{v'})\, (m_v -1) \non \\
& \sum_{v\, \text{not e.p.}} (h_v -h)\, \tl{n}^\txi_{j,v}= \sum_{v\, \text{not e.p.}} (h_v -h_{v'}) \, n^\txi_{j,v} 
\] 
where
\begin{description}
\item[$m_{v}$] is the total number of vertices contained in the cluster $G_v$. In the folloing we shall consider only the case in which the vertices $v \in \t$ are of same type of the vertices of the original potential $\BV_0$. However the extension of the result will be trivial. We denote with $m_{4,v}, m'_{4,v}, m_{4,v}'', m_{3,v}, m'_{3,v}$ and $m_{2,v}$ the number of vertices of type $\l, \l', \l'', \m, \m'$and $\n$ contained in the cluster $G_v$ and with $m_{6,v}$ the number of vertices with six plain legs, which will be considered only in the two dimensional case, for $h\leq \bh$. 
\be   \label{prop2}
m_{v} = m_{6,v}\c(d=2, h\leq \bh)+m_{4,v}+ m'_{4,v} + m''_{4,v} + m_{3,v} +m'_{3,v} + m_{2,v}
\ee
\item[\it $n^\txi_{j,v}$] is the total number of contracted half--lines contained in the cluster $G_v$, \ie lines which are contracted at scale equal or greater than $h_v$. In particular we have
\[ \label{prop3}
& n^\txi_{t,v}= 6m_{6,v}\c(d=2, h \leq \bh)+ 4m_{4,v} + 2m'_{4,v}+2m_{3,v} + 2m_{2,v}-n^\txe_{t,v} \non \\
& n^\txi_{l,v}= 2m'_{4,v} + 4m''_{4,v}+ m_{3,v}+3m'_{3,v} -n^\txe_{l,v}
 \non \\
& n^\txi_{\dpr,v}= - n^\txe_{\dpr,v}
\]
\end{description}
Note that for $h \geq \bh$, since the propagator does not distinguish the lines $l$ from $t$'s, the vertices with same number of external legs have the same dimensional behavior. Then it is sufficient to consider the total number of vertices with fixed number of external legs, \eg $ \bar{m}_{4,v} =  m_{4,v}+m'_{4,v}+m''_{4,v}$. 
Using the \eqref{prop1}--\eqref{prop3} we can rewrite the product in \eqref{power1} as 
\[ \label{power2}
\g^{h\, \tl{\d}_{v_0}}\prod_{v\, \text{not e.p.} }  \g^{(h_v -h _{v'})\,\tl{\d}_v } 
\]
with $\g^{h\tl{\d}_{v_0}}$ coming from
\[
\g^{h \sum_{v} \lft(- \d_I (s_v -1) + \d_l \tl{n}^\txi_{l,v} +\d_t \tl{n}^\txi_{t,v} + \d_0 \tl{n}^\txi_{\dpr_0,v} + \d_1 \tl{n}^\txi_{\dpr_\xx,v}\rgt)}  
\] 
and $\tl{\d}_v$ depending on the region of momenta we are considering. In particular for $\bh < h \leq 0$ and $h\leq \bh$ we have respectively
\[
\tl{\d}^{\,>}_{v} =& \; \frac{d}{2}+1 -\frac{d}{4}n_{v}^\txe-\frac{1}{2}n_{\dpr_\xx,v}^\txe-n_{\partial_{0},v}^\txe%  \non \\[6pt] &
+\left(\frac{d}{2}-1\right)\bar{m}_{4,v}+\left(\frac{d}{4}-1\right)\bar{m}_{3,v}-\bar{m}_{2,v}  \non \\[9pt]
\tl{\d}^{\,<}_{v} =&\; d+1 -\left(\frac{d+1}{2}\right)n_{l,v}^\txe-\left(\frac{d-1}{2}\right)n_{t,v}^\txe-n_{\dpr,v}^\txe  + \lft(d-3\rgt)m_{4,v}\non \\[6pt] 
& \quad +\lft(d-1\rgt)m_{4,v}^{'} +\lft(d+1\rgt)m_{4,v}^{''} 
  +\left(\frac{d-3}{2}\right)m_{3,v}+\left(\frac{d+1}{2}\right)m_{3,v}^{'}  -2m_{2,v} 
\]
with $n_{\dpr,v}^\txe=n_{\dpr_0,v}^\txe+n_{\dpr_\xx,v}^\txe$. In the last expression it turns to be convenient to recompose the factors depending on the number of vertices, as in the following example:
\be \label{endpoint}
  h\,m_{2, v_0} + \sum_{v\, \text{not e.p.}}(h_v - h_{v'})m_{2,v} = \sum_{v \text{\,e.p.}} h_{v}\, \c(m_{2,v})
\ee
where $\c(m_{2,v})$ is equal to $1$ if $v$ is of type $m_2$, otherwise is zero. For each endpoint $h_v=1$, but we prefer to maintain the writing $h_v$ in \eqref{endpoint} for reasons that will become clear in the following section. Moving the contribution of the vertices to the endpoint as done in \eqref{endpoint} and writing explicitly the factor dependence on $\e$, $\r_0$ and $R_0$ (see appendix \ref{order_e} for details) the \eqref{power1} becomes:

\subsubsection{\it Region $ \bh <h \leq 0$}
\[  \label{power_fin_up}
& \bigl\|\Val(\G)\bigr\|\leq  C^n  \lft(\r_0 R_0^{-2}\rgt) \lft(\l \e^{-1}\rgt)^{1-\frac{n^\txe_l+n^\txe_t}{2}} \,  
\g^{h\, \d^{\,>}_{v_0}} \prod_{v\, \text{not e.p.} } \g^{(h_v -h _{v'})\,\d^{\,>}_v }  \non \\[6pt]
& \; \prod_{v \text{\,e.p.}} 
\lft[\l \,\g^{h_v \lft(\frac{d}{2}-1\rgt)}\rgt]^{\c(m_{4,v})}  
\lft[ \sqrt{\l \e\,} \,\g^{h_v \lft(\frac{d}{4}-1\rgt)}\rgt]^{\c(m_{3,v})} \,\g^{h_v \lft(\frac{d}{4}-1\rgt) \c(m_{2,v})}  
\]
with 
\[
\d^{\,>}_{v}=\frac{d}{2}+1-\frac{d}{4} n_{v}^\txe - \frac{1}{2} n_{\dpr_{\xx},v}^\txe-n_{\dpr_{0},v}^\txe
\]
where the expression of $C(P_v; \e,\r_0, R_0)$ in the region  $ \bh <h \leq 0$ is calculated in \eqref{bar_C0}. In \eqref{power_fin_up} $\r_0R_0^{-2}$ is the dimension fixing factor (it is an action density in space time), while $\l \e^{-1}=(\r_0 R_0^d)^{-1}$ is an adimensional factor that for the purposes of this chapter may be though equal to one. The reason why we are keeping this factor is related to the study of the two dimensional case and will be sufficiently highlighted in the course of the work, see in particular sec. \ref{lambda6}.
\subsubsection{\it Region $h \leq \bh$}
In the three dimensional case the following bound holds:
\[  
\bigl\|\Val(\G)\bigr\|_{3d} & \leq  C^n \lft(\r_0 R_0^{-2}\rgt) \bar{C}_{3d}(P_v;\e,\l)\,
\,\g^{h\, \d^{3d,<}_{v_0}} \prod_{v\, \text{not e.p.} }  \g^{(h_v -h _{v'})\,\d^{3d,<}_v }  \non \\[6pt]
&  \prod_{v \text{\,e.p.}} \lft(\l \e^{\frac{3}{2}} \g^{-2h_v} \rgt)^{\c(m_{2,v})}
\g^{ h_v \left[2\c(m'_{4,v}) + 4\c(m''_{4,v})  + 2 \c(m'_{3,v}) \rgt]}  \label{fin_down3d}
\]
with
\[
& \bar{C}_{3d}(P_v;\e,\l)= \lft(\l \e^{-\frac{1}{2}}\rgt)^L  \e^{-3 +2n_l^\txe + n_t^\txe +\frac{1}{2}n_{\dpr_\xx}}    \\[6pt]
& \d^{3d,<}_{v}  = 4-2n_{l,v}^{\text{e}}-n_{t,v}^{\text{e}}-n_{\partial,v}^{\text{e}}
\label{dim_min_3d}
\]
see \eqref{3d_bh} for a calculation of $\bar{C}_{3d}(P_v;\e,\l)$. 
Here $L$ is the loop number, equal to $L=1+ m_4 +\frac{1}{2}m_3 -\frac{1}{2}n_l^\txe -\frac{1}{2}n_t^\txe$. With respect to the bound in the higher momentum region, the order in the small parameter $\e$ for a certain Feynman diagram with fixed $\{P_v\}$ depends on the number of loops, instead than the number of vertices. We again remark that in this chapter $\e=\l \r_0 R_0^d$ and $\l$ may be identified, since we can consider $\r_0R_0^2$ of order one.  In the two dimensional case the following bound holds:
\[  
& \bigl\|\Val(\G)\bigr\|_{2d}  \leq  C^n \lft(\r_0 R_0^{-2}\rgt) \bar{C}_{2d}(P_v;\e,\l)\,
\,\g^{h\, \d^{2d,<}_{v_0}} \prod_{v\, \text{not e.p.} }  \g^{(h_v -h _{v'})\,\d^{2d,<}_v }  \non \\[6pt]
& \qquad  \prod_{v \text{\,e.p.}} \lft( \e \g^{-2h_v} \rgt)^{\c(m_{2,v})} \, \l^{\c(m_{6,v})}
\g^{ h_v \left[  -\c(m_{4,v})  -\frac{1}{2} \c(m_{3,v})  + \c(m'_{4,v}) + 3\c(m''_{4,v})  + \frac{3}{2}\c(m'_{3,v})\rgt]}      \label{fin_down2d}
\]
with
\[
& \bar{C}_{2d}(P_v;\e,\l)= \l^L  \e^{-2 +\frac{3}{2}n_l^\txe + \frac{1}{2}n_t^\txe +\frac{1}{2}n_{\dpr_\xx}}    \\[6pt]
& \d^{2d,<}_{v}  = 3 -\frac{3}{2}n_{l,v}^{\text{e}}-\frac{1}{2}n_{t,v}^{\text{e}}-n_{\partial,v}^{\text{e}} \label{dim_min_2d}
\]
see \eqref{2d_bh} for a calculation of $\bar{C}_{2d}(P_v;\e,\l)$. 
The factors $\d^{\,>}_v$ and $\d^{\,<}_v$ are referred to as the {\it scaling dimensions} of the kernels $G_v$ with $n^\txe_v= n^\txe_{l,v} + n^\txe_{t,v}$ external fields. The factors associated to the endpoints are not important in the previous estimates and give constant factors, however notice that the dimension appearing in front to the $\c(m_v)$ functions depend on the dimension $d$, apart for the vertex $m_{2,v}$.  

The above estimates are of course finite, but the problems come out if one wants to perform the sum over the scales $\{h_v\}_{v \in \t}$ of $\t \in \TT_{h,n}$ in the limit $h \arr -\io$. In fact in order to get this sum we need the {\it unrenormalized dimension} $\d_v$ to be negative, being $h_v - h_{v'}\geq 1$ (see the lemma \ref{lem:labeled_trees}). This is not true for each diagram or cluster, in particular

\begin{description}
\item[\it Region $\bh<h \leq 0$.] The ``dangerous'' diagrams giving $\d^{\,>}_v\geq 0$, neglecting for the moment the contributions coming from the derivatives, are those for which
\[
n_{v}^\txe\leq2+\frac{4}{d} \qquad 
\begin{cases}
n_{v}^\txe \leq4  &  \text{for }d=2  \\ 
n_{v}^\txe\leq 3 & \text{for }d=3
\end{cases}
\]
\item [\it Region $h \leq \bh$.]
If $n_{t,v}^\txe=0$ the dangerous diagrams are those for which
\[
n_{l,v}^\txe\leq2  
\]
for all the spatial dimensions. If $n_{l,v}^\txe=0$ then the ``dangerous''  diagrams for $d=2,3$ satisfies
\[
n_{t,v}^\txe \leq \,2\;\frac{d+1}{d-1} \qquad 
\begin{cases}
 n_{t,v}^\txe \leq 6 &    \text{for }d=2  \\ 
 n_{t,v}^\txe \leq 4  & \text{for }d=3
\end{cases}
\]
\end{description}
In the renormalization group approach, the diagrams with zero dimension are called {\it marginal}, while the diagrams with positive dimension are called {\it relevant}. 

In the following we will be interested in the local part of the clusters $G_v$. By parity reasons some of the local diagrams are vanishing; in particular, indicating with $\hV^{(h_v)}_{n^\txe_l, n^\txe_t}(0,\ldots,0)=\int dx_1 \ldots dx_{n_l^\txe + n_t^\txe} V^{(h_v)}_{n^\txe_l, n^\txe_t}(x_1,\ldots,x_{n_l^\txe + n_t^\txe})$ the local part of a cluster $G_v$ with $n^{\text{e}}_{l,v}$ external legs of type $\ps^l$ and $n^{\text{e}}_{t,v}$ external legs of type $\ps^t$ we have 
\[
\hV^{(h_v)}_{01}(0)  &=  0  \non \\
\hV^{(h_v)}_{11}(0,0) & =  0 \non \\
\hV^{(h_v)}_{21}(0,0,0)=\hV^{(h_v)}_{03}(0,0,0) & =  0    \\
\hV^{(h_v)}_{31}(0,\ldots,0)=\hV^{(h_v)}_{13}(0,\ldots,0) & =  0 \non \\
\hV^{(h_v)}_{41}(0,\ldots,0)=\hV^{(h_v)}_{23}(0,\ldots,0)=\hV^{(h_v)}_{05}(0,\ldots,0) & = \non 0
\]

%---------------------------------------------------------
\feyn{
\begin{fmffile}{feyn-TESI/rel3_a}
\unitlength = 1cm  
\def\myl#1{2.5cm} %larghezza parbox
\[
 & \parbox{\myl}{\centering{
		\begin{fmfgraph*}(2,1.25)
			\fmfright{i1,i2}
			\fmfleft{o1}
			\fmf{plain}{i1,v,i2}
			\fmf{dashes, tension=1.5}{o1,v}
			\fmfdot{v}
			\fmfv{label={$V_{12}^{(h)}$},label.angle=100, label.dist=10pt}{v}
		\end{fmfgraph*}
            \\ $\frac{1}{4}$
	}} 
	\quad \parbox{\myl}{\centering{
		\begin{fmfgraph*}(2,1.25)
			\fmfright{i1,i2}
			\fmfleft{o1}
			\fmf{dashes}{i1,v,i2}
			\fmf{dashes, tension=1.5}{o1,v}
			\fmfdot{v}
			\fmfv{label={$V_{30}^{(h)}$},label.angle=100, label.dist=10pt}{v}
		\end{fmfgraph*}
            \\ $\frac{1}{4}$
	}}  
	 \parbox{\myl}{\centering{	
		\begin{fmfgraph*}(2 ,1.25)    
			\fmfleft{i1}
			\fmfright{o1}
			\fmf{plain}{i1,v1,o1}
			\fmfv{label={$V_{02}^{(h)}$},label.angle=90, label.dist=10pt}{v1}   
			%decor.shape=circle,decor.filled=full,decor.size=2thick}{v3}
			\fmfdot{v1}
		\end{fmfgraph*}  
		\\ $1$ 
           }} 
\quad \parbox{\myl}{\centering{	
		\begin{fmfgraph*}(2 ,1.25)
			\fmfleft{i1}
			\fmfright{o1}
			\fmf{dashes}{i1,v1,o1}
			\fmfv{label={$V_{20}^{(h)}$},label.angle=90, label.dist=10pt}{v1}   
			\fmfdot{v1}
		\end{fmfgraph*}  
		\\ $1$   
	}}  \non \\[12pt]
& \hskip 1.5cm \parbox{\myl}{\centering{
		\begin{fmfgraph*}(2,1.25)
			\fmfright{i1,i2}
			\fmfleft{o1}
			\fmf{plain}{i1,v,i2}
			\fmf{plain, tension=1.5}{o1,v}
			\fmfdot{v}
			\fmfv{label={$V_{03}^{(h)}$},label.angle=100, label.dist=10pt}{v}
		\end{fmfgraph*}
            \\ $\frac{1}{4}$
	}} \quad
\parbox{\myl}{\centering{
		\begin{fmfgraph*}(2,1.25)
			\fmfright{i1,i2}
			\fmfleft{o1}
			\fmf{dashes}{i1,v,i2}
			\fmf{plain, tension=1.5}{o1,v}
			\fmfdot{v}
			\fmfv{label={$V_{21}^{(h)}$},label.angle=100, label.dist=10pt}{v}
		\end{fmfgraph*}
            \\ $\frac{1}{4}$
	}} 
\; \parbox{\myl}{\centering{	
		\begin{fmfgraph*}(2 ,1.25)
			\fmfleft{i1}
			\fmfright{o1}
                   \fmf{plain}{i1,v1}
			\fmf{dashes}{v1,o1}
			\fmfv{label={$V_{11}^{(h)}$},label.angle=90, label.dist=10pt}{v1}   
			\fmfdot{v1}
		\end{fmfgraph*}  
		\\ $1$   
	}}  \non
\]
\end{fmffile}
}
{{\bf Relevant and marginal clusters for $d=3$ and $\bh <h \leq 0$}. The solid lines represent the $\ps^t$ fields, the dashed lines the $\ps^l$ fields. The number under each diagram is its scaling dimension; for each external derivatives with respect to $x_0$ or $\xx$ the scaling dimension is decreased by one or one--half respectively. The local part of the diagrams on the second line is zero.}{rel3d_first}

%---------------------------------------------------------
\feyn{
\begin{fmffile}{feyn-TESI/rel3_b}
\unitlength = 1cm  
\def\myl#1{2.5cm} %larghezza parbox
\[
 \parbox{\myl}{\centering{	 	   
		\begin{fmfgraph*}(2,1.25)
			\fmfleft{i1,i2}
			\fmfright{o1,o2}
			\fmf{plain}{i1,v,o2}
			\fmf{plain}{i2,v,o1}
			\fmfv{label=$V_{04}^{(h)}$,label.angle=90, label.dist=10pt}{v}
			%\fmflabel{$\lambda_h$}{v}
			\fmfdot{v}
			%\fmf{photon}{v1,v2}
		\end{fmfgraph*} \\
 		$0$
	}} &
	\quad \parbox{\myl}{\centering{
		\begin{fmfgraph*}(2,1.25)
			\fmfright{i1,i2}
			\fmfleft{o1}
			\fmf{plain}{i1,v,i2}
			\fmf{dashes, tension=1.5}{o1,v}
			\fmfdot{v}
			\fmfv{label=$V_{12}^{(h)}$,label.angle=100, label.dist=10pt}{v}
		\end{fmfgraph*}
            \\ $0$
	}} 
	 \quad \parbox{\myl}{\centering{	
		\begin{fmfgraph*}(2 ,1.25)
			\fmfleft{i1}
			\fmfright{o1}
			\fmf{plain}{i1,v1,o1}
			\fmfv{label={$V_{02}^{(h)}$},label.angle=90, label.dist=10pt}{v1}   
			%decor.shape=circle,decor.filled=full,decor.size=2thick}{v3}
			\fmfdot{v1}
		\end{fmfgraph*}  
		\\ $2$ \non
	}}  
\quad \parbox{\myl}{\centering{	
		\begin{fmfgraph*}(2 ,1.25)      %Z_h
			\fmfleft{i1}
			\fmfright{o1}
			\fmf{dashes}{i1,v1,o1}
			\fmfv{label={$V_{20}^{(h)}$},label.angle=90, label.dist=10pt}{v1}   
			\fmfdot{v1}
		\end{fmfgraph*}  
		\\ $0$ 
           }} \\[12pt]
& \quad \parbox{\myl}{\centering{	
		\begin{fmfgraph*}(2 ,1.25)    
			\fmfleft{i1}
			\fmfright{o1,o2}
			\fmf{plain}{i1,v1}
			\fmf{plain}{o2,v1,o1}	
			\fmfv{label={$V_{03}^{(h)}$},label.angle=90, label.dist=10pt}{v1}   
			\fmfdot{v1}
		\end{fmfgraph*}  
		\\ $1$ 
           }} 
 \quad \parbox{\myl}{\centering{	
		\begin{fmfgraph*}(2 ,1.25)    
			\fmfleft{i1}
			\fmfright{o1}
			\fmf{plain}{i1,v1}
			\fmf{dashes}{v1,o1}	
			\fmfv{label={$V_{11}^{(h)}$},label.angle=90, label.dist=10pt}{v1}   
			\fmfdot{v1}
		\end{fmfgraph*}  
		\\ $1$ 
           }}  \non
\]
\end{fmffile}
}{{\bf Relevant and marginal clusters for $d=3$ and $h\leq \bh$}. The scaling dimension for each diagram is reported. For each external derivatives with respect to $x_0$ or $\xx$ the scaling dimension is decreased by one. The diagrams on the second line have vanishing local part.}{rel3d_second}

%----------------------------------------------------------
\feyn{
\begin{fmffile}{feyn-TESI/rel2_a}
\unitlength = 1cm  
\def\myl#1{2.3cm} %larghezza parbox
\[
    &  \parbox{\myl}{\centering{	 	   
		\begin{fmfgraph*}(2,1.25)
			\fmfleft{i1,i2}
			\fmfright{o1,o2}
			\fmf{plain}{i1,v,o2}
			\fmf{plain}{i2,v,o1}
			\fmfv{label=$V_{04}^{(h)}$,label.angle=90, label.dist=10pt}{v}
			\fmfdot{v}
		\end{fmfgraph*} \\
 		$0$
	}} 
	\quad \parbox{\myl}{\centering{	 	   
		\begin{fmfgraph*}(2,1.25)
			\fmfleft{i1,i2}
			\fmfright{o1,o2}
			\fmf{plain}{i1,v,o2}
			\fmf{dashes}{i2,v,o1}
			\fmfv{label=$V_{22}^{(h)}$,label.angle=90, label.dist=10pt}{v}
			\fmfdot{v}
		\end{fmfgraph*} \\
 		$0$
	}}
	\quad \parbox{\myl}{\centering{	 	   
		\begin{fmfgraph*}(2,1.25)
			\fmfleft{i1,i2}
			\fmfright{o1,o2}
			\fmf{dashes}{i1,v,o2}
			\fmf{dashes}{i2,v,o1}
			\fmfv{label=$V_{40}^{(h)}$,label.angle=90, label.dist=10pt}{v}
			\fmfdot{v}
		\end{fmfgraph*} \\
 		$0$
	}}   
	 \quad \parbox{\myl}{\centering{
		\begin{fmfgraph*}(2,1.25)
			\fmfright{i1,i2}
			\fmfleft{o1}
			\fmf{plain}{i1,v,i2}
			\fmf{dashes, tension=1.5}{o1,v}
			\fmfdot{v}
			\fmfv{label=$V_{12}^{(h)}$,label.angle=100, label.dist=10pt}{v}
		\end{fmfgraph*}
            \\ $\frac{1}{2}$
	}} 
	\quad \parbox{\myl}{\centering{
		\begin{fmfgraph*}(2,1.25)
			\fmfright{i1,i2}
			\fmfleft{o1}
			\fmf{dashes}{i1,v,i2}
			\fmf{dashes, tension=1.5}{o1,v}
			\fmfdot{v}
			\fmfv{label=$V_{30}^{(h)}$,label.angle=100, label.dist=10pt}{v}
		\end{fmfgraph*}
            \\ $\frac{1}{2}$
	}}  \non \\[12pt]
	 & 
 \parbox{\myl}{\centering{	
		\begin{fmfgraph*}(2 ,1.25)
			\fmfleft{i1}
			\fmfright{o1}
			\fmf{plain}{i1,v1,o1}
			\fmfv{label={$V_{02}^{(h)}$},label.angle=90, label.dist=10pt}{v1}   
			%decor.shape=circle,decor.filled=full,decor.size=2thick}{v3}
			\fmfdot{v1}
		\end{fmfgraph*}  
		\\[-3pt] $1$ 
           }} 
\quad \parbox{\myl}{\centering{	
		\begin{fmfgraph*}(2 ,1.25)
			\fmfleft{i1}
			\fmfright{o1}
			\fmf{dashes}{i1,v1,o1}
			\fmfv{label={$V_{20}^{(h)}$},label.angle=90, label.dist=10pt}{v1}   
			\fmfdot{v1}
		\end{fmfgraph*}  
		\\[-3pt] $1$   
	}}   \non \\[12pt]
&   \parbox{\myl}{\centering{	 	   
		\begin{fmfgraph*}(2,1.25)
			\fmfleft{i1,i2}
			\fmfright{o1,o2}
			\fmf{plain}{i1,v}
                   \fmf{dashes}{i2,v}
			\fmf{plain}{o2,v,o1}
			\fmfv{label=$V_{13}^{(h)}$,label.angle=90, label.dist=10pt}{v}
			\fmfdot{v}
		\end{fmfgraph*} \\
 		$0$
	}}  \quad
\parbox{\myl}{\centering{	 	   
		\begin{fmfgraph*}(2,1.25)
			\fmfleft{i1,i2}
			\fmfright{o1,o2}
			\fmf{dashes}{i1,v}
                   \fmf{plain}{i2,v}
			\fmf{dashes}{o2,v,o1}
			\fmfv{label=$V_{31}^{(h)}$,label.angle=90, label.dist=10pt}{v}
			\fmfdot{v}
		\end{fmfgraph*} \\
 		$0$
	}} \quad \parbox{\myl}{\centering{	
		\begin{fmfgraph*}(2 ,1.25)    
			\fmfleft{i1}
			\fmfright{o1,o2}
			\fmf{plain, tension=1.5}{i1,v1}
			\fmf{plain}{o2,v1,o1}	
			\fmfv{label={$V_{03}^{(h)}$},label.angle=90, label.dist=10pt}{v1}   
			\fmfdot{v1}
		\end{fmfgraph*}  
		\\ $\frac{1}{2}$ 
           }} \quad \parbox{\myl}{\centering{	
		\begin{fmfgraph*}(2 ,1.25)    
			\fmfleft{i1}
			\fmfright{o1,o2}
			\fmf{plain, tension=1.5}{i1,v1}
			\fmf{dashes}{o2,v1,o1}	
			\fmfv{label={$V_{21}^{(h)}$},label.angle=90, label.dist=10pt}{v1}   
			\fmfdot{v1}
		\end{fmfgraph*}  
		\\ $\frac{1}{2}$ 
           }} \quad \parbox{\myl}{\centering{	
		\begin{fmfgraph*}(2 ,1.25)    
			\fmfleft{i1}
			\fmfright{o1}
			\fmf{plain}{i1,v1}
			\fmf{dashes}{v1,o1}	
			\fmfv{label={$V_{11}^{(h)}$},label.angle=90, label.dist=10pt}{v1}   
			%decor.shape=circle,decor.filled=full,decor.size=2thick}{v3}
			\fmfdot{v1}
		\end{fmfgraph*}  
		\\ $1$ 
           }}  \non
\]
\end{fmffile}
}{{\bf Relevant and marginal diagrams for $d=2$ and $\bh <h \leq 0$}, with their scaling dimensions. Note that in the region $h > \bh$ there is no distinction between the $\ps^l$ and $\ps^t$ fields. For each external derivatives with respect to $x_0$ or $\xx$ the scaling dimension is decreased by one or one--half respectively. The local part of the diagrams on the third line is zero.}{rel2d_first}

%-------------------------------------------------------
%
\feyn{
\begin{fmffile}{feyn-TESI/rel2_b}
\unitlength = 1cm  
\def\myl#1{2.5cm} %larghezza parbox
\[
& \parbox{\myl}{\centering{	 	   
		\begin{fmfgraph*}(2,1.25)
			\fmfleft{i1,i2,i3}
			\fmfright{o1,o2,o3}
			\fmf{plain}{i1,v,o2}
			\fmf{plain}{i2,v,o1}
			\fmf{plain}{i3,v,o3}
			\fmfv{label=$V_{06}^{(h)}$,label.angle=90, label.dist=10pt}{v}
			\fmfdot{v}
		\end{fmfgraph*} \\
 		$0$
	}}
	\quad \parbox{\myl}{\centering{	 	   
		\begin{fmfgraph*}(2,1.25)
			\fmfleft{i1,i2}
			\fmfright{o1,o2}
			\fmf{plain}{i1,v,o2}
			\fmf{plain}{i2,v,o1}
			\fmfv{label=$V_{04}^{(h)}$,label.angle=90, label.dist=10pt}{v}
			%\fmflabel{$\lambda_h$}{v}
			\fmfdot{v}
			%\fmf{photon}{v1,v2}
		\end{fmfgraph*} \\
 		$1$
	}}
	\quad \parbox{\myl}{\centering{
		\begin{fmfgraph*}(2,1.25)
			\fmfright{i1,i2}
			\fmfleft{o1}
			\fmf{plain}{i1,v,i2}
			\fmf{dashes, tension=1.5}{o1,v}
			\fmfdot{v}
			\fmfv{label=$V_{12}^{(h)}$,label.angle=100, label.dist=10pt}{v}
		\end{fmfgraph*}
            \\ $\frac{1}{2}$
	}}  
	 \quad \parbox{\myl}{\centering{	
		\begin{fmfgraph*}(2 ,1.25)
			\fmfleft{i1}
			\fmfright{o1}
			\fmf{plain}{i1,v1,o1}
			\fmfv{label={$V_{02}^{(h)}$},label.angle=90, label.dist=10pt}{v1}   
			%decor.shape=circle,decor.filled=full,decor.size=2thick}{v3}
			\fmfdot{v1}
		\end{fmfgraph*}  
		\\ $2$ \non
	}} 
	 \quad \parbox{\myl}{\centering{	
		\begin{fmfgraph*}(2 ,1.25)
			\fmfleft{i1}
			\fmfright{o1}
			\fmf{dashes}{i1,v1,o1}
			\fmfv{label={$V_{20}^{(h)}$},label.angle=90, label.dist=10pt}{v1}   
			%decor.shape=circle,decor.filled=full,decor.size=2thick}{v3}
			\fmfdot{v1}
		\end{fmfgraph*}  
		\\ $0$ \non
	}}   \non \\[12pt] &
 \hskip 1.5cm \parbox{\myl}{\centering{	 	   
		\begin{fmfgraph*}(2,1.25)
			\fmfleft{i1,i2}
			\fmfright{o1,o2,o3}
			\fmf{plain, tension=1.5}{i1,v,i2}
			\fmf{plain}{o2,v,o1}
			\fmf{plain}{v,o3}
			\fmfv{label=$V_{05}^{(h)}$,label.angle=90, label.dist=10pt}{v}
			\fmfdot{v}
		\end{fmfgraph*} \\
 		$\frac{1}{2}$
	}} \quad
 \parbox{\myl}{\centering{	 	   
		\begin{fmfgraph*}(2,1.25)
			\fmfleft{i1}
			\fmfright{o1,o2,o3}
			\fmf{dashes, tension=2}{i1,v}
			\fmf{plain}{o2,v,o1}
			\fmf{plain}{v,o3}
			\fmfv{label=$V_{13}^{(h)}$,label.angle=90, label.dist=10pt}{v}
			\fmfdot{v}
		\end{fmfgraph*} \\
 		$0$
	}} \quad
\parbox{\myl}{\centering{	 	   
		\begin{fmfgraph*}(2,1.25)
			\fmfleft{i1}
			\fmfright{o1,o2}
			\fmf{plain, tension=1.5}{i1,v}
			\fmf{plain}{o2,v,o1}
			\fmfv{label=$V_{03}^{(h)}$,label.angle=90, label.dist=10pt}{v}
			\fmfdot{v}
		\end{fmfgraph*} \\
 		$\frac{3}{2}$
	}} \quad
 \parbox{\myl}{\centering{	
		\begin{fmfgraph*}(2 ,1.25)    
			\fmfleft{i1}
			\fmfright{o1}
			\fmf{plain}{i1,v1}
			\fmf{dashes}{v1,o1}	
			\fmfv{label={$V_{11}^{(h)}$},label.angle=90, label.dist=10pt}{v1}   
			%decor.shape=circle,decor.filled=full,decor.size=2thick}{v3}
			\fmfdot{v1}
		\end{fmfgraph*}  
		\\ $1$ 
           }} \non
\]
\end{fmffile}
}{{\bf Relevant and marginal clusters for $d=2$ and $h\leq \bh$.} The local part of the diagrams on the second line is zero. With respect to the $3d$ case there is an additional marginal diagram ($V^{(h)}_{04}$) and the scaling dimensions of $V^{(h)}_{04}$ and $V^{(h)}_{12}$ are higher, compare with fig. \ref{rel3d_second}.}{rel2d_second}
%------------------------------------------------

\vskip -2cm
The previous identities guarantee that the only diagrams with two, three or four external legs which can be generated at each scale are equal to the ones in the original potential. 
They are some of the consequences of a more general Ward identity, see 
in chapter \ref{GWI} eq. \eqref{GWI_fourier}.
 The marginal and relevant diagrams (with non vanishing local part) with their dimensions, for both regions $h >\bh$ and $h \leq \bh$ and for two and three spatial dimensions, are showed in figures \ref{rel3d_first}  -- \ref{rel2d_second}, where the solid lines represent the $\ps^t$ fields and the dashed lines the $\ps^l$ fields. 

Since the diagrams for which $\d_v \leq 0$ give ``dangerous'' contribution to the effective potential, they need to be ``renormalized'', that means that we need to introduce a different expansion such that the contributions coming from them are treated in a special way. 
 Before we describe the renormalization procedure, we may wonder what happens if we impose the requirement $\d^>_v, \d^<_v <0$ for every diagram $\G$, which corresponds to remove ``by hand'' the dangerous subgraphs.  If we make this assumption, the contribution of order $n$ to the kernel  $V^{(h);\, n}_{n_l,\,n_t}$ of effective potential at scale $h$ with fixed external legs $(n_l,\,n_t)$ can be bounded as follows:
\[ \label{nfactorial_bound}
|| V^{(h);\, n}_{n_l,\,n_t}||  
& \leq \sum_{\t \in \TT_{h,n}} \,
\sum_{\substack{ \{P_{v}\} \\ n_l,\,n_t\,\text{fixed} }} 
\, \sum_{G \in \GG(\{P_{v}\})}  C^n \,C_d(P_v;\e,\r_0, R_0)\, \g^{h \d_{v_0}} \prod_{v\,\text{not e.p.}} 
\frac{1}{s_v!}\, \g^{(h_v - h_{v'})\d_v} \non \\
& \leq  \sum_{\t \in \TT_{h,n}} \, \sum_{ \{P_{v}\}}  C^n \,C_d(P_v;\e,\r_0, R_0)\, \g^{h \d_{v_0}} \prod_{v\,\text{not e.p.}} 
\frac{(s_v !)^2}{s_v!}\, \g^{-\frac{1}{d+3} (h_v - h_{v'}) - \frac{d}{4(d+3)} n^\txe_v } \non  \\[6pt]
& 
%\leq C^n \h^n \g^{h \d_{v_0}} \frac{(2n)!}{n!} 
\leq C^n \,C_d(P_v;\e,\r_0, R_0)\, \g^{h \d_{v_0}} n!
\]
where in the first line 
\begin{enumerate}[i)]
\item for each vertex $v$ we bounded the number of Feynman diagrams formed by $s_v$ elements and with $P_v$ external lines by $ C_1^n (2 s_v !) \leq C_2^n (s_v !)^2$, see lemma \ref{lem:Feyn_diagrams}. We see as the factor $1/s_v!$ arising from the tree expansion is not enough to compensate the number of Feynman diagrams and a factor $s_v!$ survives in the estimate.
\item the following bound, holding both for $\d_v^>$ and $\d_v^<$,
\be
 \d_v (h_v - h_{v'}) \leq -\frac{1}{d+3}(h_v - h_{v'}) - \frac{d}{4(d+3)} n^\txe_v
\ee
comes from the conditions $\d_v \leq -1$, $n^\txe_v \geq 2$ and $h_v - h_{v'} \geq 1$.
\end{enumerate}
Then, in the second line
\begin{enumerate}[i)]
\item we have bounded the sum on the labeled trees of the factors $\g^{-\frac{1}{d+3}\,(h_v - h_{v'})\,}$ with $C^n$, see lemma \ref{lem:labeled_trees};
\item we have bounded  the sum of the factors $\g^{- \frac{d}{4(d+3)}  n^\txe_v}$ over the choices of labels $\{P_v\}$  with $C^n$, see lemma \ref{lem:label_choice};
\item we have used the bound $\prod_v s_v! \leq n!\,$;
\item for simplicity of notation we have used always the same constant C, even if it is changing in each of the previous estimates.
\end{enumerate}
The bound \eqref{nfactorial_bound} shows that if we manage to find an expansion such that the scaling dimension is negative for each cluster, we can take the limit $h \arr -\io$ and prove that the effective potential is {\it order by order finite}, with the coefficient of arbitrary order $n$ bounded by $\text{(const.)}^n n!$. These are the so called {\it ``n!-bounds''}. 

On the other hand, due to the presence of the $n!$ in \eqref{nfactorial_bound}, the estimate for the contribution of order $n$ to the effective potential at scale $h$ is not summable with respect to $n$, and we cannot prove the theory to be well defined, even in the case with $\d_v <0$, without the help of independent methods with respect to the ones of perturbative theory\footnote{The problem of proving the convergence of the series defining the quantities of interest is a problem intrinsic at all the bosonic theories. In fact, since these theories are not analytic in $\l$ (they only have sense for $\l$ positive) there is no hope to find a perturbative expansion with coefficients without the $n!$ term.}.
%\footnote{ In the cases in which a full non perturbative construction of a bosonic theory has been achieved, as in the cases of the $\f^4_{3}$ and $\f^4_2$ quantum field theories,  }

\section{Localization and renormalized expansion}\label{Localization}

In section \ref{multiscale} we introduced a perturbative expansion for the effective potential at scale $h$ and found estimates on the terms contributing to it which are finite for each finite $h$ but not uniform in the limit $h\arr -\io$, due to the presence of marginal and relevant couplings.  In this section we will describe how to construct a new perturbative expansion, the {\it renormalized expansion}, which will allow us to overcame the latter problem and prove $n!$ bounds on the effective potentials. 

%The size of the marginal and relevant couplings and of the new effective interaction must to be controlled, and in particular it must be proven that the weight of the local quartic term in $\BV_h$ remains smaller under iterations. Then we have to find a suitable definition of the new parameters after each integration step, so that the flow of the effective coupling constants can be controlled. In this section we describe the definition of localization, crucial for the definition of the effective coupling constants.

The first step in the costruction of the renormalized expansion is the definition of the {\it localization operator} $\LL$ that acts on the effective potential such that, if $\RR = \unit - \LL$ then
\be
\BV^{(h)} = \LL \BV^{(h)} + \RR \BV^{(h)}
\ee
In the following we will call $\LL \BV^{(h)}$ the {\it relevant} or {\it local} part of the effective potential and $\RR \BV^{(h)}$ the {\it irrelevant} part. The action of $\LL$ on the effective potential gives
\be
\LL \BV^{(h)}(\PS{\leq h}) = \sum_i \, r_{i,h} F^i(\PS{\leq h})
\ee
where $F^i(\PS{\leq h})$ is the integral of a monomial in the $\ps$ fields and $r_{i,h}$ are called the {\it running coupling constants} at scale $h$. The explicit form of the functions $F^i(\PS{\leq h})$ will depend on the dimension $d$ and on the value of $h$ with respect to $\bh$, then we will discuss it later. For the moment we see that the definition of the localization operator leads to a new tree expansion, in which a new label, with value $\LL$ or $\RR$ appears. The renormalized tree of lower orders are shown in figure \ref{ren_tree1} and \ref{ren_tree2}. For what concern figure \ref{ren_tree1} note that we on the second line an endpoint at scale $h=0$ is defined, representing the sum of all the trees with an index $\LL$ on the non trivial vertex at scale $h=0$. In figure \ref{ren_tree2} the latter construction is carried on  the second step of the perturbative expansion. As a consequence the trees giving $-\LL \BV_{-2}$ have not only endpoints at scale $h=1$ as usual, but also endpoints at scale $h=0$, representing the local terms $-\LL \BV_{-1}$. In the same way we introduce the definition of endpoints at scale $h=-2$, representing $-\LL \BV_{-2}$. 
%
%----------------------------------
%
%
\tik{
\[ &
\begin{tikzpicture}[scale=0.7]
\draw (-1.6,0) node{$-\RR \BV_{-1}=$};
\foreach \x in {0,...,2}   
    {
        \draw[th] (\x , -1.2) -- (\x , 1.2);   
        \node[vertex] at (\x,0){};   
    }
\draw[med] (0,0) -- (2,0);   
\draw (3,0) node{$+$};
\node[anchor=north east, label=$\RR$] at (1,0){};
%%%%
\foreach \x in {4,...,6}   
    {
        \draw[th] (\x , -1.2) -- (\x , 1.2);    
    } 
\draw[med] (4,0) node[vertex]{} --   (5,0) node [vertex] {} -- (6,0.8) node[vertex] {};  
\node[anchor=north east, label=$\RR$] at (5,0){};
\draw[med] (5,0)  -- (6,-0.8) node[vertex] {};  
\draw (7,0) node{$+$};
%%%
\foreach \x in {8,...,10}   
    {
        \draw[th] (\x , -1.2) -- (\x , 1.2);      
    }
\draw[med] (8,0) node[vertex]{} --   (9,0) node [vertex] {} -- (10,0) node[vertex] {};  
\draw[med] (9,0) -- (10,0.8) node[vertex] {};  
\draw[med] (9,0) -- (10,-0.8) node[vertex] {};  
\node[anchor=north east, label=$\RR$] at (9,0){};
\draw (11,0) node{$+$};
\foreach \x in {12,...,14}   
    {
        \draw[th] (\x , -1.2) -- (\x , 1.2);      
    }
\draw[med] (12,0) node[vertex]{} --   (13,0) node [vertex] {} -- (14,0.8) node[vertex] {};  
\draw[med] (13,0) -- (14,0.25) node[vertex] {};  
\draw[med] (13,0) -- (14,-0.25) node[vertex] {};  
\draw[med] (13,0) -- (14,-0.8) node[vertex] {};  
\node[anchor=north east, label=$\RR$] at (13,0){};
\draw (15,0) node{$+ \cdots$};
\treelabelsize{
\draw (0,-1.5) node {$-1$};  
\draw (1,-1.5) node {$0$};
\draw (2,-1.5) node {$1$};
\draw (4,-1.5) node {$-1$};  
\draw (5,-1.5) node {$0$};
\draw (6,-1.5) node {$1$};
\draw (8,-1.5) node {$-1$};  
\draw (9,-1.5) node {$0$};
\draw (10,-1.5) node {$1$};
\draw (12,-1.5) node {$-1$};  
\draw (13,-1.5) node {$0$};
\draw (14,-1.5) node {$1$};
}
\end{tikzpicture} 
%---------------------------------------------------------------------------------------------------
\non \\[6pt] &
\begin{tikzpicture}[scale=0.7]
\draw (-1.6,0) node{$-\LL \BV_{-1}=$};
\foreach \x in {0,...,2}   
    {
        \draw[th] (\x , -0.8) -- (\x , 0.8);   
    }
\draw[med] (0,0) -- (1,0);  
\node[vertex] at (0,0){};  
\node[vertex] at (1,0){}; 
\treelabelsize{   
\draw (0,-1.5) node {$-1$};  
\draw (1,-1.5) node {$0$};
\draw (2,-1.5) node {$1$};}
\end{tikzpicture} 
%---------------------------------------------------------------------- 
\non \\[12pt] & \hskip 1.25cm = 
\begin{tikzpicture}[scale=0.7]
\foreach \x in {0,...,2}   
    {
        \draw[th] (\x , -1.2) -- (\x , 1.2);   
        \node[vertex] at (\x,0){};   
    }
\draw[med] (0,0) -- (2,0);   
\draw (3,0) node{$+$};
\node[anchor=north east, label=$\LL$] at (1,0){};
%%%%
\foreach \x in {4,...,6}   
    {
        \draw[th] (\x , -1.2) -- (\x , 1.2);    
    } 
\draw[med] (4,0) node[vertex]{} --   (5,0) node [vertex] {} -- (6,0.8) node[vertex] {};  
\node[anchor=north east, label=$\LL$] at (5,0){};
\draw[med] (5,0)  -- (6,-0.8) node[vertex] {};  
\draw (7,0) node{$+$};
%%%
\foreach \x in {8,...,10}   
    {
        \draw[th] (\x , -1.2) -- (\x , 1.2);      
    }
\draw[med] (8,0) node[vertex]{} --   (9,0) node [vertex] {} -- (10,0) node[vertex] {};  
\draw[med] (9,0) -- (10,0.8) node[vertex] {};  
\draw[med] (9,0) -- (10,-0.8) node[vertex] {};  
\node[anchor=north east, label=$\LL$] at (9,0){};
\draw (11,0) node{$+$};
\foreach \x in {12,...,14}   
    {
        \draw[th] (\x , -1.2) -- (\x , 1.2);      
    }
\draw[med] (12,0) node[vertex]{} --   (13,0) node [vertex] {} -- (14,0.8) node[vertex]{};  
\draw[med] (13,0) -- (14,0.25) node[vertex] {};  
\draw[med] (13,0) -- (14,-0.25) node[vertex] {};  
\draw[med] (13,0) -- (14,-0.8) node[vertex] {};  
\node[anchor=north east, label=$\LL$] at (13,0){};
\draw (15,0) node{$+ \cdots$};
\treelabelsize{
\draw (0,-1.5) node {$-1$};  
\draw (1,-1.5) node {$0$};
\draw (2,-1.5) node {$1$};
\draw (4,-1.5) node {$-1$};  
\draw (5,-1.5) node {$0$};
\draw (6,-1.5) node {$1$};
\draw (8,-1.5) node {$-1$};  
\draw (9,-1.5) node {$0$};
\draw (10,-1.5) node {$1$};
\draw (12,-1.5) node {$-1$};  
\draw (13,-1.5) node {$0$};
\draw (14,-1.5) node {$1$};
}
\end{tikzpicture} 
\non
\]
}{The renormalized expansion at lowest order. Note the definition of the endpoint at scale $h=0$ given on the second line: it corresponds to the sum of all the possible trees whose root has indices $h=0$ and $\LL$.}{ren_tree1}
%
%
%
%------------------------------------------------------------------- REN_TREE2
\tik{
\vskip 0.5cm
\[
& \begin{tikzpicture}[scale=0.7]
\draw (-1.6,0) node{$-\LL \BV_{-2}=$};
\foreach \x in {0,...,3}   
    {
        \draw[th] (\x , -1.2) -- (\x , 1.2);     
    }
\draw[med] (0,0) node[vertex]{} -- (1,0) node[vertex] {} ; 
\node[anchor=north east, label=$\LL$] at (1,0){};
\draw (4,0) node{$=$};
\foreach \x in {5,...,8}   
    {
        \draw[th] (\x , -1.2) -- (\x , 1.2);     
    }
\draw[med] (5,0) node[vertex]{} -- (6,0) node[vertex] {}  -- (7,0) node[vertex] {}; 
\node[anchor=north east, label=$\LL$] at (6,0){};
%\draw (9,0) node{$+$};
\treelabelsize{
\draw (0, -1.7) node{$-2$};
\draw (1, -1.7) node{$-1$};
\draw (2, -1.7) node{$0$};
\draw (3, -1.7) node{$1$};
\draw (5, -1.7) node{$-2$};
\draw (6, -1.7) node{$-1$};
\draw (7, -1.7) node{$0$};
\draw (8, -1.7) node{$1$};
}
\end{tikzpicture}  \non \\[6pt]
%--------------------------------------------------2nd line
&  \hskip 1.3cm  %\; \;
\begin{tikzpicture}[scale=0.7]
\draw (0.3,0) node{$+$};
\foreach \x in {1,...,4}   
    {
        \draw[th] (\x , -1.2) -- (\x , 1.2);   
    } 
\draw[med] (1,0) node[vertex]{} -- (2,0) node[vertex] {}  -- (3,0) node[vertex] {}; 
\draw[med] (3,0) node[vertex]{} -- +(1,1) node[vertex] {};
\draw[med] (3,0) node[vertex]{} -- +(1,-1) node[vertex] {};
\node[anchor=north east, label=$\LL$] at (2,0){};
\node[anchor=north east, label=$\RR$] at (3,0){};
%%%%
\draw (5,0) node{$+$};
%%%%
\foreach \x in {6,...,9}   
    {
        \draw[th] (\x , -1.2) -- (\x , 1.2);    
    } 
\draw[med] (6,0) node[vertex]{} -- ++  (1,0) node [vertex] {} --++ (1,0.5) node[vertex] {} --++ (1,0.5) node[vertex] {};   
\draw[med] (7,0)  -- ++ (1,-0.5) node[vertex] {}-- ++ (1,-0.5) node[vertex] {};  
\node[anchor=north east, label=$\LL$] at (7,0){};
\node[anchor=north east, label=$\RR$] at (8,0.5){};
\node[anchor=north east, label=$\RR$] at (8,-0.5){};
%%%
\draw (10,0) node{$+$};
%%%
\foreach \x in {11,...,14}   
    {
        \draw[th] (\x , -1.2) -- (\x , 1.2);      
    }
\draw[med] (11,0) node[vertex]{} --   (12,0) node [vertex] {};  
\draw[med] (12,0) -- ++(1,0.5) node[vertex] {} -- ++(1,0.5) node[vertex] {};  
\draw[med] (12,0)  -- ++(1,-0.5) node[vertex] {};  
\node[anchor=north east, label=$\LL$] at (12,0){};
\node[anchor=north east, label=$\RR$] at (13,0.5){};
\treelabelsize{
\draw (1, -1.7) node{$-2$};
\draw (2, -1.7) node{$-1$};
\draw (3, -1.7) node{$0$};
\draw (4, -1.7) node{$1$};
\draw (6, -1.7) node{$-2$};
\draw (7, -1.7) node{$-1$};
\draw (8, -1.7) node{$0$};
\draw (9, -1.7) node{$1$};
\draw (11, -1.7) node{$-2$};
\draw (12, -1.7) node{$-1$};
\draw (13, -1.7) node{$0$};
\draw (14, -1.7) node{$1$};
}
\end{tikzpicture}  \non \\[6pt]
%--------------------------------------------------3rd line
& \hskip 1.3cm 
\begin{tikzpicture}[scale=0.7]
\draw (0.3,0) node{$+$};
\foreach \x in {1,...,4}   
    {
        \draw[th] (\x , -1.2) -- (\x , 1.2);   
    } 
\draw[med] (1,0) node[vertex]{} -- (2,0) node[vertex] {} -- (3,0) node[vertex] {} - ++(1,1) node[vertex] {}; 
\draw[med] (3,0) node[vertex]{} -- ++(1,-1) node[vertex] {};
\draw[med] (3,0) node[vertex]{} -- ++(1,0) node[vertex] {};
\node[anchor=north east, label=$\LL$] at (2,0){};
\node[anchor=north east, label=$\RR$] at (3,0){};
%%%%
\draw (4.5,0) node{$+$};
%%%%
\foreach \x in {5,...,8}   
    {
        \draw[th] (\x , -1.2) -- (\x , 1.2);    
    } 
\draw[med] (5,0) node[vertex]{} -- ++  (1,0) node [vertex] {} --++ (1,0.5) node[vertex] {} --++ (1,0.5) node[vertex] {};  
\draw[med] (6,0)  -- ++ (1,0) node[vertex] {} -- ++ (1,0) node[vertex] {};  
\draw[med] (6,0)  -- ++ (1,-0.5) node[vertex] {}-- ++ (1,-0.5) node[vertex] {};  
\node[anchor=north east, label=$\LL$] at (6,0){};
\node[anchor=north east, label=$\RR$] at (7,0.5){};
\node[ label=$\RR$] at (6.7,-0.5){};
\node[anchor=north east, label=$\RR$] at (7,-1.3){};
%%%
\draw (8.5,0) node{$+$};
%%%
\foreach \x in {9,...,12}   
    {
        \draw[th] (\x , -1.2) -- (\x , 1.2);      
    }
\draw[med] (9,0) node[vertex]{} --   (10,0) node [vertex] {};  
\draw[med] (10,0) -- ++(1,0.5) node[vertex] {} -- ++(1,0.5) node[vertex] {};  
\draw[med] (10,0)  -- ++(1,-0.5) node[vertex] {} -- ++(1,-0.5) node[vertex] {};  
\node[anchor=north east, label=$\LL$] at (10,0){};
\node[anchor=north east, label=$\RR$] at (11,0.5){};
\draw[med] (11,0.5)  -- ++(1,-0.5) node[vertex] {};
\node[anchor=north east, label=$\RR$] at (11,-0.5){};
%%%%
\draw (12.5,0) node{$+$};
%%%
\foreach \x in {13,...,16}   
    {
        \draw[th] (\x , -1.2) -- (\x , 1.2);      
    }
\draw[med] (13,0) node[vertex]{} --   (14,0) node [vertex] {};  
\draw[med] (14,0) -- ++(1,0.5) node[vertex] {} -- ++(1,0.5) node[vertex] {};  
\draw[med] (14,0)  -- ++(1,0) node[vertex] {}; 
\draw[med] (14,0)  -- ++(1,-0.5) node[vertex] {}; 
\node[anchor=north east, label=$\LL$] at (14,0){};
\node[anchor=north east, label=$\RR$] at (15,0.5){};
\draw (17,0) node{$+\,\ldots$};
\treelabelsize{
\draw (1, -1.7) node{$-2$};
\draw (2, -1.7) node{$-1$};
\draw (3, -1.7) node{$0$};
\draw (4, -1.7) node{$1$};
\draw (5, -1.7) node{$-2$};
\draw (6, -1.7) node{$-1$};
\draw (7, -1.7) node{$0$};
\draw (8, -1.7) node{$1$};
\draw (9, -1.7) node{$-2$};
\draw (10, -1.7) node{$-1$};
\draw (11, -1.7) node{$0$};
\draw (12, -1.7) node{$1$};
\draw (13, -1.7) node{$-2$};
\draw (14, -1.7) node{$-1$};
\draw (15, -1.7) node{$0$};
\draw (16, -1.7) node{$1$};
}
\end{tikzpicture}  \non
\]
}{The second iterative step of the renormalized expansion. The roots of the trees in the figure are at scale $h=-2$, while the endpoints are both at scale $h=0$ and $h=1$.The latter correspond to $-\LL \BV_{-1}$, see fig. \ref{ren_tree1}. }{ren_tree2}
\fig{!h}{0.85}{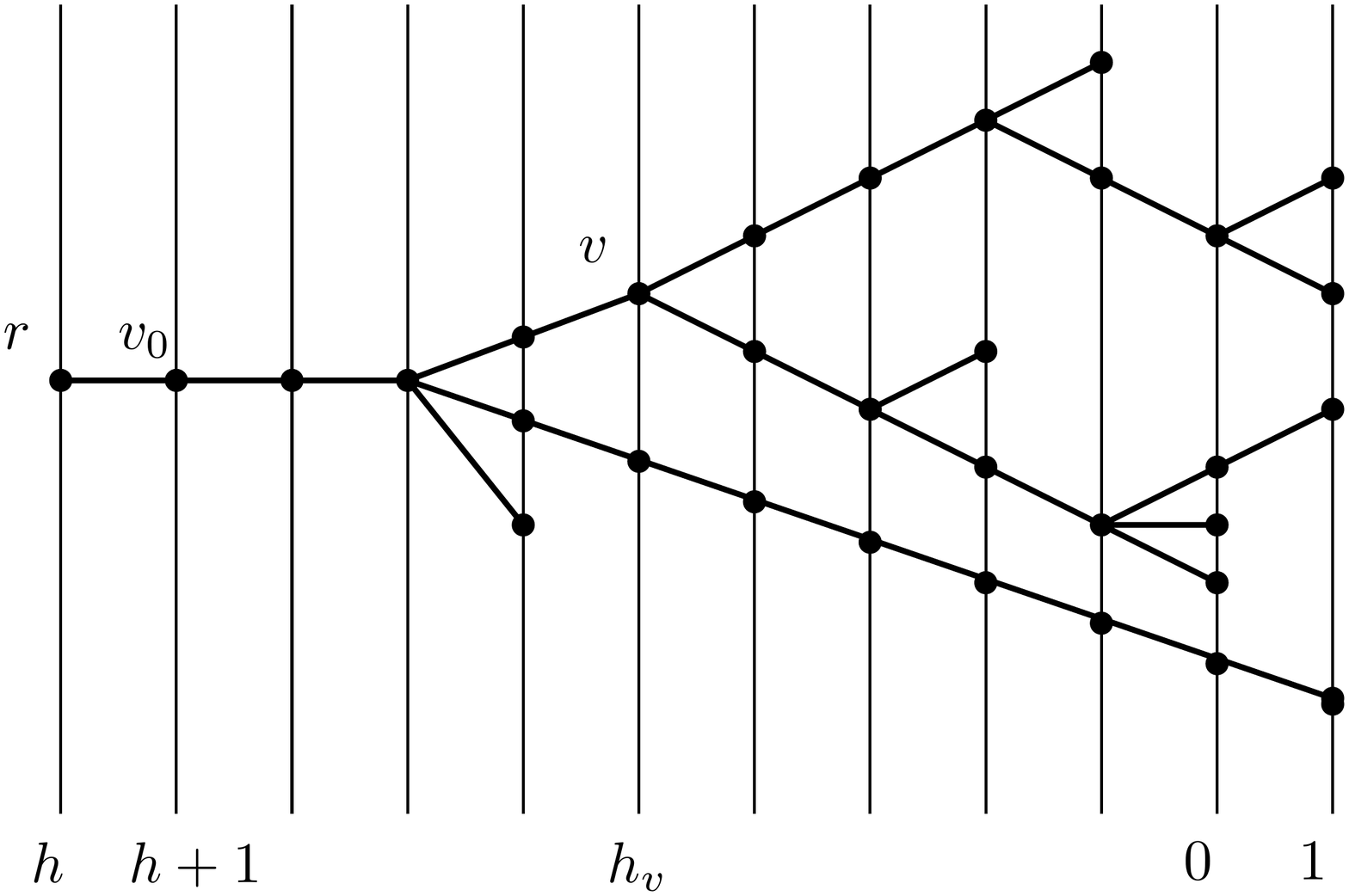}{An example of renormalized tree contributing to the effective potential at scale $h$. With each vertex different from $v_0$ and from the endpoints a label $\RR$ is associated, which is not reported in the picture; to the vertex $v_0$ is associated the index $\LL$ or $\RR$. The endpoints with scale $h^* \in [h+2,0]$ represent $-\LL \BV_{(h^*-1)}$; the endpoints at scale $1$ corresponds as before to the elements of $\BV_0$.}{ren_tree_h}
%
%\end{minipage}

The iteration of the latter definitions and constructions leads to the renormalized expansion for the effective potential at scale $h$, which is obtained with some slight modifications of the unrenormalized expansion described in section \ref{multiscale}:
\begin{enumerate}[1)]
\item with each vertex different from $v_0$ and from the endpoints a label $\RR$ is associated to; 
\item with the vertex $v_0$ is associated the index $\LL$ or $\RR$ according as the tree contributes to the relevant or irrelevant part of $\BV_h$;
\item there exist endpoints with scale labels in $[h+2,0]$. An endpoint with scale label $h^* <1$ corresponds to $-\LL \BV_{(h^*-1)}$. An endpoint with scale $h=1$ corresponds to a term $\LL \BV_I$ or $\RR \BV_I$, with $\BV_I$ the correction to Bogoliubov potential. 
\end{enumerate}
A possible tree contributing to the renormalized expansion for $\BV_h$ is depicted in picture \ref{ren_tree_h}.

Introducing a localization procedure corresponds in defining iteratively some new effective couplings. Let denote with $\rr_h=\{r^i_h\}$ the vector of the endpoints at scale $h$ of the renormalized tree expansion. Here the index $i$ refers to the different contribution to $\LL \BV_h$. Let us consider the set of the trees with root at scale $(h-1)$ and label $\LL$  associated to the vertex $v_0$, at scale $h$; then the renormalized expansion provides a method to calculate the coupling $r^i_{h-1}$ as a function of the couplings at scales greater than $(h-1)$. In particular the sum of the trees with index $\LL$ on $v_0$ provides the form of the {\it beta function} defined by the following relation:
\be \label{beta_funct}
\rr_{h-1} = \G\, \rr_h +\b_h \left( \rr_h, \rr_{h+1}, \ldots, \rr_0 \rgt)
\ee
where $\G$ is a suitable linear transformation we shall define later on. The effective couplings $\rr_h$ are called the {\it running coupling constants at scale $h$}. Referring to fig. \ref{ren_tree2} the beta function $\b_{-1}$ is given by the sum of the trees on the second and third lines, with the index $\LL$ on the vertex $v_0$ at scale $-1$ and $n> 1$ endpoints; the two diagrams on the first line of fig.  \ref{ren_tree2} corresponds respectively to $\rr_{-1}$ and $\rr_0$. \\

%then we have to find a suitable definition of the new parameters after each integration step, so that the flow of the effective coupling constants can be controlled. In this section we describe the definition of localization, crucial for the definition of the effective coupling constants.

Once this definitions have been introduced we are faced with two problems: the first one is to prove that, under some assumption on the running coupling constants, the beta function is well defined order by order (this will be done using the Feynman diagrams expansion in way similar to the one used in the previous section). The second problem is to control the flow of the running coupling constants generated by the beta function, in order to prove that there exist some initial values $\rr_0$ such that the flow remains finite and the assumptions on $\rr_h$, which have been used at first, are verified. 
%If , then the theory is well defined, in the sense that the interacting partition function would be expressed by a series with explicit estimates on the coefficient of order $n$. 
The rest of the chapter is devoted to prove the existence of the beta function; then in chapter \ref{flows} we will discuss how to control the flow of the running coupling constants. 
In order to estimate the contributions to the beta function we need to define explicitly the action of the localization operator for the two and three dimensional cases and in the two regions $h \geq \bh$ and $h < \bh$. The discussion of the section \ref{dimensional} suggests that the action of $\LL$  would be not trivial only on the contribution to $\BV_h$ coming from the marginal and relevant clusters shown in fig. \ref{rel3d_first}  -- \ref{rel2d_second}.  Let consider the kernels \eqref{final_kernel}. The action of $\LL$ on a marginal kernel will correspond to extract the local part of that kernel, \eg given the kernel $W_{20}^{(h)}(x,y)$ which is marginal for $h \leq \bh$ we define
\[ \label{loc_marginal}
\LL \, \int dx  dy V_{20}^{(h)}(x,\,y) \ps^{l,(h)}_{x}\ps^{l,(h)}_{y}:= \int dx dy V_{04}^{(h)}(x,y) \ps^{l,(h)}_{x}\ps^{l,(h)}_{x} 
\]
%\ie $\LL W_{n^\txe_{l}n^\txe_{t}}^{(h)}(x_1, \ldots, x_{n_l + n_t})= W_{n^\txe_{l}n^\txe_{t}}^{(h)}(x)$, 
Then, the relevant kernel will be localized trough a suitable Taylor expansion. For example, given the kernel $V_{02}^{(h)}(x,y)$ which is relevant with scaling dimension 2 for $h \leq \bh$ we define
\[  \label{loc_relevant}
 \LL \, \int dx  dy V_{20}^{(h)}(x,\,y) \ps^{t,(h)}_{x}\ps^{t,(h)}_{y} :=  &
\non \\ 
  \, \int dx  dy V_{02}^{(h)}(x,y) \ps^{t,(h)}_{x} \Bigl(\ps^{t,(h)}_{x}
& + \sum_{i=0}^3 (y_i - x_i) \dpr_i \ps^{t,(h)}_x   \non \\ 
 & +\frac{1}{2} \, \sum_{i,j=0}^3 (y_i - x_i)(y_j - x_j)  \dpr_i \dpr_j \ps^{t,(h)}_x  \Bigr)  
\]
The action of the $\LL$ operator results even plainer in momentum space. Let us rewrite the expression for the effective potential \eqref{potential_xspace} in the momentum space:
\[
\BV^{(h)}(\psi^{\leq h}) = &\sum_{n^\txe_{l},\,n^\txe_{t}} \int \frac{d^{d+1} k_1}{(2\pi)^{d+1}} \ldots \frac{d^{d+1} k_{n_l}}{(2\pi)^{d+1}} \frac{d^{d+1} p_1}{(2\pi)^{d+1}} \ldots \frac{d^{d+1} p_{n_t}}{(2\pi)^{d+1}}  \non \\
& \qquad \hV^{h}_{n^\txe_l, n^\txe_t}\bigl(\{k_i\}, \{p_j\}\bigr)\, 
\d\Bigl(\sum_{i=1}^{n^\txe_l} k_i + \sum_{j=1}^{n^\txe_t} p_i  \Bigr) \lft( \prod_{i=1}^{n^\txe_l} \ps^l_{k_i} \rgt) \lft( \prod_{j=1}^{n^\txe_t} \ps^t_{p_j} \rgt)
\]
Then \eqref{loc_relevant} and \eqref{loc_marginal} correspond to the following actions on the kernels of the potential in momentum space:
\[
\LL\, \hV_{02}^{(h)}(k,p) & :=  \hV_{20}^{(h)}(0,0) \non \\
\LL\, \hV_{02}^{(h)}(k,p)  & :=  \hV_{02}^{(h)}(0,0) +k \dpr_k \hV_{02}^{(h)}(0,0) + \frac{1}{2} k^2 \dpr_{k}^2 \hV_{02}^{(h)}(0,0) 
\]
Following this plan we define the localization operator in the regions $\bh < h \leq0$ and $h \leq \bh$ as follows.

\subsection{Localization for $ \bh< h \leq 0$}  

We remind that in this region the fields $\ps^t_x$ and $\ps^l_x$ have the same dimensional behavior, then the dimensional estimate of the kernels of the effective potentials $W_{n^\txe_l n^\txe_t}^{(h)}$ depends only on the total number of external legs $n_\txe = n^\txe_l + n^\txe_t $, independently on their type. That's why the localization procedure for kernels with the same number of external legs will be the same. 

{\centering \subsubsection{\it Three dimensions}}
In the three dimensional case we get:
\[  \label{loc_3d_a}
& \LL^{3d}_{\,>}\, \hV_{12}^{(h)}(k_1,k_2,k_3) :=   \, \hV_{12}^{(h)}(0,0,0)  \non \\
& \LL^{3d}_{\,>}\, \hV_{30}^{(h)}(k_1,k_2,k_3)  := \,  \hV_{30}^{(h)}(0,0,0)  \non \\
& \LL^{3d}_{\,>}\, \hV_{02}^{(h)}(k,p)  :=   \, \hV_{02}^{(h)}(0,0)+ \frac{1}{2} \, \sum_{i,j=1}^3 k_i \, k_j \,\dpr_{k_i} \dpr_{k_j} \hV_{02}^{(h)}(0,0)  \non \\
& \LL^{3d}_{\,>}\, \hV_{20}^{(h)}(k,p)  :=   \, \hV_{20}^{(h)}(0,0) +\frac{1}{2} \, \sum_{i,j=1}^3 k_i \, k_j \, \dpr_{k_i} \dpr_{k_j} \hV_{02}^{(h)}(0,0)   \non \\
&\LL^{3d}_{\,>}\, \hV_{11}^{(h)}(k,p)  :=   k_0\,\dpr_0  \hV_{11}^{(h)}(0,0)  \non \\[6pt]
&\LL^{3d}_{\,>}\, \hV_{n^\txe_{l}n^\txe_{t}}^{(h)}(k_{1},\ldots,\, k_{n^\txe_l+n^\txe_t})  :=  0 \quad \text{otherwise}
\]
We have not included in \eqref{loc_3d_a} some of the possibly local terms since they are indentically zero for symmetry reasons. In particular the term
\[
  \int dx \ps^{t,(h)}_x \dpr_j \ps^{t,(h)}_x & = \int dk\, \ps^{t,(h)}_k i k_j\, \ps^{t,(h)}_{-k}   %\non \\[6pt]
 %\int dx \lft(\ps^{t,(h)}_x\rgt)^2 \dpr_j \ps^{t,(h)}_x & = \int dk\, \lft(\ps^{t,(h)}_k\rgt)^2 i k_j\, \ps^{t,(h)}_{-k}  \qquad \forall j=0,\ldots,3 
\] 
is the integral of a total derivative with the fields satisfying periodic boundary conditions.  Besides the terms $\int dk \ps^{l,(h)}_{-k} \ps^{t,(h)}_{k}$ and $ \int dk  \ps^{l,(h)}_{-k} i \,\kk \,\ps^{t,(h)}_k  $ are identically zero for parity reasons. We will use the following symbols to indicate the integrals of monomials of the fields given by the localization procedure:
\[ \label{monomials}
 &  F^{(h)}_{n_l^\txe, n_t^\txe} = \int dk_1 \ldots dk_{n_l^\txe}\, dp_1 \ldots dp_{n_l^\txe}
\ps^{l,\,(h)}_{k_1} \ldots \ps^{l,\,(h)}_{k_{n_l^\txe}}\, \ps^{t,\,(h)}_{p_1} \ldots \ps^{t,\,(h)}_{p_{n_t^\txe}}
\d \lft(k_1 + \ldots +p_{n_t^\txe} \rgt) \non \\
& F^{(h)}_{\a,\, \dpr_0 \a'} = \int dk dp\,  \ps^{\a,(h)}_{k}  i p_0 \, \ps^{\a',(h)}_p \,\d(k+p) \non \\
& F^{(h)}_{\dpr_0 \a,\, \dpr_0 \a'}  = \int dk dp\,  \ps^{\a,(h)}_k (- p_0^2) \ps^{\a',(h)}_p \,\d(k+p) \non \\
& F^{(h)}_{\dpr_\xx \a,\, \dpr_\xx \a'} = \int dk dp\,  \cdot  \ps^{\a,(h)}_k (-\pp^2)\, \ps^{\a',(h)}_p  \,\d(k+p) 
\]
The previous discussion implies that the local potential for $d=3$ and $\bh <h < 0$ is of the form:
\[ \label{local_pot_3d_a}
\LL^{3d}_{\,>}\, \BV^{(h)}(\ps) =  \, \r_0 R_0^{-2} & \, \biggl( \g^{\frac{h}{4}} \bar{\m}_h \, F_{12} + \g^{\frac{h}{4}}\bar{\m}'_h F_{30}+ \g^{h}\bar{\n}_h \,F_{02}   + \tfrac{1}{2}\,\bar{z}_h\,F_{20}   \non \\[6pt] 
&  \;   +  \tfrac{1}{2}\,\bar{a}_h\, F_{\dpr_\xx t,\, \dpr_\xx t}  +  \tfrac{1}{2}\,\bar{a}'_h\, F_{\dpr_\xx l,\, \dpr_\xx l} +  \bar{e}_h\,F_{t,\, \dpr_0 t} \biggr)
\]
as graphically depicted in fig. \eqref{local3d_a}, where the terms with derivatives on the external legs represent the monomials $F^{(h)}_{\dpr_i \a,\, \dpr_i \a'}$ defined in \eqref{monomials}. The running coupling constants $\rr_h$ in \eqref{local_pot_3d_a} are defined with in front their dimensional estimate, \eg the running coupling constant corresponding to the local term with two external $t$ fields is defined as $\g^{h}\n_h$.

%
%---------------------------------------
\feyn{
\begin{fmffile}{feyn-TESI/loc3_a}
\unitlength = 1cm  
\def\myl#1{3cm} %larghezza parbox
\def\myll#1{3cm} 
\[
& \LL^{3d}_{\,>}\, \BV_h \quad  = 
  \parbox{\myll}{\centering{
		\begin{fmfgraph*}(2,1.25)
			\fmfright{i1,i2}
			\fmfleft{o1}
			\fmf{plain}{i1,v,i2}
			\fmf{dashes, tension=1.5}{o1,v}
			\fmfdot{v}
			\fmfv{label=$\g^{\frac{h}{4}}\bar{\m}_h$,label.angle=120}{v}
		\end{fmfgraph*}
	}} +
\parbox{\myll}{\centering{
		\begin{fmfgraph*}(2,1.25)
			\fmfright{i1,i2}
			\fmfleft{o1}
			\fmf{dashes}{i1,v,i2}
			\fmf{dashes, tension=1.5}{o1,v}
			\fmfdot{v}
			\fmfv{label=$\g^{\frac{h}{4}} \bar{\m}'_h$,label.angle=120}{v}
		\end{fmfgraph*}
	}} 
	+\parbox{\myll}{\centering{	
		\begin{fmfgraph*}(2 ,1.25)
			\fmfleft{i1}
			\fmfright{o1}
			\fmf{plain}{i1,v1,o1}
			\fmfv{label={$\g^{h}\bar{\n}_h$},label.angle=90}{v1}   
			%decor.shape=circle,decor.filled=full,decor.size=2thick}{v3}
			\fmfdot{v1}
		\end{fmfgraph*}
	}}  \non \\[12pt]  & \hskip 0.5cm
+  \parbox{\myl}{\centering{
		\begin{fmfgraph*}(2,1.25)
			\fmfleft{i1}
			\fmfright{o1}
			\fmf{dashes}{i1,v1,o1}
			\fmfv{label={$\g^h \bar{z}_h$},label.angle=90}{v1} 
			\fmfdot{v1}
		\end{fmfgraph*}
	}}  
+ \parbox{\myl}{\centering{	
		\begin{fmfgraph*}(2,1.25)
			\fmfleft{i}
			\fmfright{o}
			\fmf{plain,label=$\dpr_\xx$,label.dist=-0.2w}{i,v}
			\fmf{plain,label=$\dpr_\xx$,label.dist=-0.2w}{v,o}
			\fmfv{label=$\bar{a}_h$,label.angle=-90}{v}   
			\fmfdot{v}
		\end{fmfgraph*}	
}} 
 + \parbox{\myl}{\centering{	
		\begin{fmfgraph*}(2,1.25)
			\fmfleft{i}
			\fmfright{o}
			\fmf{dashes}{i,v}
			\fmf{plain,label=$\partial_0$,label.dist=-0.2w}{v,o}
			\fmfv{label=$\bar{e}_h$,label.angle=-90}{v}   
			\fmfdot{v}
		\end{fmfgraph*} 
}} 
+ \parbox{\myl}{\centering{	
		\begin{fmfgraph*}(2,1.25)
			\fmfleft{i}
			\fmfright{o}
			\fmf{dashes,label=$\dpr_\xx$,label.dist=-0.2w}{i,v}
			\fmf{dashes,label=$\dpr_\xx$,label.dist=-0.2w}{v,o}
			\fmfv{label=$\bar{a}'_h$,label.angle=-90}{v}   
			\fmfdot{v}
		\end{fmfgraph*}
}}   \non
\]
\end{fmffile}
}{{\bf Local potential for $d=3$ and $\bh<h\leq 0$}. The running coupling constants are defined with their dimensions, \eg being the local term with two external $t$ fields relevant with dimension $\d^{\,>}=1$ the corresponding running coupling  is defined as $\g^{h}\n_h$.The diagram with two external derivatives with respect to $k_0$ has negative scaling dimension, however it is convenient to localize it, as will result clear in the next chapter.}{local3d_a}

%----------------------------------------------------------------------------------------

{\centering \subsubsection{\it Two dimensions}}
Since the two legged vertices have the same scaling dimensions than in the three dimensional case, the localization procedure for $\hV_{20}^{(h)}$, $\hV_{02}^{(h)}$ and $\hV_{11}^{(h)}$ is the same defined in \eqref{loc_3d_a}. For what concern the diagrams with more than two legs we have:
\[ \label{loc_2d_a}
&\LL^{2d}_{\,>}\, \hV_{04}^{(h)}(k_1,\ldots, k_4) :=   \, \hV_{04}^{(h)}(0,\ldots,0)  \non \\
&\LL^{2d}_{\,>}\, \hV_{22}^{(h)}(k_1,\ldots, k_4) := \,  \hV_{22}^{(h)}(0,\ldots,0)  \non \\
&\LL^{2d}_{\,>}\, \hV_{40}^{(h)}(k_1,\ldots, k_4)  := \,  \hV_{40}^{(h)}(0,\ldots,0)  \non \\[6pt]
&\LL^{2d}_{\,>}\, \hV_{12}^{(h)}(k_1, k_2, k_3)  :=   \, \hV_{12}^{(h)}(0,0,0)  \non \\
& \LL^{2d}_{\,>}\, \hV_{30}^{(h)}(k_1, k_2, k_3)  := \,  \hV_{30}^{(h)}(0,0,0) \non \\[6pt]
&   \LL^{2d}_{\,>}\, \hV_{n^\txe_{l}n^\txe_{t}}^{(h)}(k_{1},\ldots,\, k_{n^\txe_l+n^\txe_t }) :=  0
\]
In the previous definition we have taken into account the fact that the terms
\[
 \int dx \lft(\ps^{\a,(h)}_x\rgt)^2 \dpr_\a \ps^{l,(h)}_x & = \int dk\, \lft(\ps^{t,(h)}_k\rgt)^2 i k_j\, \ps^{\a,(h)}_{-k}  \qquad \forall j=0,\ldots,d
\] 
with $\a=l,t$ are zero being the integral of a total derivative with the fields satisfying periodic boundary conditions. On the other side the term $ \int dk   \bigl(\ps^{t,(h)}_k\bigr)^2 i \,k_i \,\ps^{l,(h)}_k  $ for each $i=0,\dots,d$ is identically zero by parity reasons. The local potential for $d=2$ and $h > \bh$ is of the form
\[ \label{local_pot_2d_a}
\LL^{2d}_{\,>}\, \BV_{h}(\ps) =  \, \r_0 R_0^{-2} & \,  
\Bigl( \bar{\l}_h \,F_{04} +\bar{\l}'_h \, F_{22}+\bar{\l}''_h \,F_{40}  + \g^{\frac{h}{2}} \bar{\m}_h \,F_{12}  + \g^{\frac{h}{2}}\bar{\m}'_h \,F_{30} + \g^{h}\bar{\n}_h \,F_{02}  \non \\[6pt]
&
+  \tfrac{1}{2}\,\bar{z}_h \,F_{20}+   \tfrac{1}{2}\,\bar{a}_h \, F_{\dpr_\xx t,\, \dpr_\xx t}  +  \tfrac{1}{2}\,\bar{a}'_h  \,F_{\dpr_\xx l,\, \dpr_\xx l} +  \bar{e}_h \, F_{ t,\, \dpr_0 t}   \Bigr)
\]
as graphically depicted in fig. \eqref{local2d_a}.

%---------------------------------------
\feyn{
\begin{fmffile}{feyn-TESI/loc2_a}
\unitlength = 1cm  
\def\myl#1{3 cm} %larghezza parbox
\def\myll#1{2.5cm} 
\[
&  \LL^{2d}_{\,>}\, \BV_h \quad = 
 \parbox{\myl}{\centering{	 	   
		\begin{fmfgraph*}(2,1.25)
			\fmfleft{i1,i2}
			\fmfright{o1,o2}
			\fmf{plain}{i1,v,o2}
			\fmf{plain}{i2,v,o1}
			\fmfv{label=$\bar{\l}_h$,label.angle=90}{v}
			\fmfdot{v}
		\end{fmfgraph*}
	}}+
 \parbox{\myl}{\centering{	 	   
		\begin{fmfgraph*}(2,1.25)
			\fmfleft{i1,i2}
			\fmfright{o1,o2}
			\fmf{dashes}{i1,v,o2}
			\fmf{plain}{i2,v,o1}
			\fmfv{label=$\bar{\l}'_h$,label.angle=90}{v}
			\fmfdot{v}
		\end{fmfgraph*}
	}} +
 \parbox{\myl}{\centering{	 	   
		\begin{fmfgraph*}(2,1.25)
			\fmfleft{i1,i2}
			\fmfright{o1,o2}
			\fmf{dashes}{i1,v,o2}
			\fmf{dashes}{i2,v,o1}
			\fmfv{label=$\bar{\l}''_h$,label.angle=90}{v}
			\fmfdot{v}
		\end{fmfgraph*}
	}} \non \\[15pt] & \hskip 1.5 cm
%------------------------------------------------------------
	 + \parbox{\myl}{\centering{
		\begin{fmfgraph*}(2,1.25)
			\fmfright{i1,i2}
			\fmfleft{o1}
			\fmf{plain}{i1,v,i2}
			\fmf{dashes, tension=1.5}{o1,v}
			\fmfdot{v}
			\fmfv{label=$\g^{\frac{h}{2}}\bar{\m}_h$,label.angle=120}{v}
		\end{fmfgraph*}
	}} +
\parbox{\myl}{\centering{
		\begin{fmfgraph*}(2,1.25)
			\fmfright{i1,i2}
			\fmfleft{o1}
			\fmf{dashes}{i1,v,i2}
			\fmf{dashes, tension=1.5}{o1,v}
			\fmfdot{v}
			\fmfv{label=$\g^{\frac{h}{2}} \bar{\m}'_h$,label.angle=120}{v}
		\end{fmfgraph*}
	}} 
	+\parbox{\myl}{\centering{	
		\begin{fmfgraph*}(2 ,1.25)
			\fmfleft{i1}
			\fmfright{o1}
			\fmf{plain}{i1,v1,o1}
			\fmfv{label={$\g^{h}\bar{\n}_h$},label.angle=90}{v1}   
			%decor.shape=circle,decor.filled=full,decor.size=2thick}{v3}
			\fmfdot{v1}
		\end{fmfgraph*}
	}}   \non \\[12pt] & \hskip 1.5cm +  
%-----------------------------------------------------------------
\parbox{\myll}{\centering{
		\begin{fmfgraph*}(2,1.25)
			\fmfleft{i1}
			\fmfright{o1}
			\fmf{dashes}{i1,v1,o1}
			\fmfv{label={$\g^h \bar{z}_h$},label.angle=90}{v1} 
			\fmfdot{v1}
		\end{fmfgraph*}
	}}
+ \parbox{\myll}{\centering{	
		\begin{fmfgraph*}(2,1.25)
			\fmfleft{i}
			\fmfright{o}
			\fmf{plain,label=$\dpr_\xx$,label.dist=-0.2w}{i,v}
			\fmf{plain,label=$\dpr_\xx$,label.dist=-0.2w}{v,o}
			\fmfv{label=$\bar{a}_h$,label.angle=-90}{v}   
			\fmfdot{v}
		\end{fmfgraph*}	
}}  
+ \parbox{\myll}{\centering{	
		\begin{fmfgraph*}(2,1.25)
			\fmfleft{i}
			\fmfright{o}
			\fmf{dashes}{i,v}
			\fmf{plain,label=$\partial_0$,label.dist=-0.2w}{v,o}
			\fmfv{label=$\bar{e}_h$,label.angle=-90}{v}   
			\fmfdot{v}
		\end{fmfgraph*} 
}} 
+ \parbox{\myll}{\centering{	
		\begin{fmfgraph*}(2,1.25)
			\fmfleft{i}
			\fmfright{o}
			\fmf{dashes,label=$\dpr_\xx$,label.dist=-0.2w}{i,v}
			\fmf{dashes,label=$\dpr_\xx$,label.dist=-0.2w}{v,o}
			\fmfv{label=$\bar{a}'_h$,label.angle=-90}{v}   
			\fmfdot{v}
		\end{fmfgraph*}
}} \non
\]
\end{fmffile}
}{{\bf Local potential for $d=2$ and $\bh<h\leq 0$}. Note that the scaling dimension of the two--legged diagrams is independent on the spatial dimension.}{local2d_a}

The values of the running coupling constants at $h=0$, both in the three and two dimensional case, are
\[ \label{bare_couplings_3d}
& \bar{\l}_0 = 2 \bar{\l}'_0 = \bar{\l}''_0 = \frac{\e}{16}  \non \\   
& \bar{\m}_0=\bar{\m}'_0= \frac{\e}{4}\sqrt{2}  \non \\
&  \bar{\n}_0 = \frac{\n}{2} R_0^2  \non \\[6pt]
& \bar{z}_0 = \e  \qquad \bar{a}_0 = \bar{a}'_0 = \bar{e}_0 = 0
\]
with the correction to the chemical potential $\n$ fixed in such a way that the flow of effective chemical potential is bounded for each $h$, see sec.~\ref{nu}.
%calculated starting from the value of $\n_\bh$; the latter depends on the renormalization condition \eqref{nu_system} as showed in chapter~\ref{flows}. 

%----------------------------------------------------------------

\subsection{Localization for $h \leq \bh$}  
%We define the action of the localization operator for $h > \bh$ on the kernels with non negative scaling dimensions following the same strategy showed in defining $\LL_{\,<}$.

{\centering \subsubsection{\it Three dimensions}}
 The action of the localization operator for $h\leq\bh$ and in the three dimensional case, which we denote with $\LL^{3d}_{\,<}$ is defined as follows. On the marginal kernels
\[ 
& \LL^{3d}_{\,<}\, \hV_{04}^{(h)}(k_1,\ldots, k_4) := \hV_{04}^{(h)}(0, \ldots, 0) \non \\
& \LL^{3d}_{\,<}\, \hV_{12}^{(h)}(k_1,k_2,k_3)  := \hV_{12}^{(h)}(0,0,0)  \non \\
& \LL^{3d}_{\,<}\, \hV_{20}^{(h)}(k,p)  :=  \hV_{20}^{(h)}(0,0)   \non \\
& \LL^{3d}_{\,<}\, \hV_{02}^{(h)}(k,p)  :=   \, \hV_{02}^{(h)}(0,0)  +\frac{1}{2} \, \sum_{i,j=0}^3 k_i k_j \dpr_{k_i} \dpr_{k_j} \hV_{02}^{(h)}(0,0) \non \\
& \LL^{3d}_{\,<}\, \hV_{11}^{(h)}(k,p)  :=   k_0 \dpr_0 \hV_{11}^{(h)}(0,0) \label{loc_3d_b}
\]
For the remaining kernels
\[
\LL^{3d}_{\,<}\, \hV_{n^\txe_{l}n^\txe_{t}}^{(h)}(k_{1},\ldots,\, k_{n^\txe_l+n^\txe_t}) :=  0
\]
In the definition of the localization procedure it has been taken into account the fact that some of the diagrams one would have to include in the localization procedure are zero by symmetry reasons (as also detailed in the discussion of the localization procedure in the region $\bh<h\leq 0$). 
%We did not include in \eqref{loc_3d_b} some local term since they are indentically zero for symmetry reasons. For example the terms
%\[
 % \int dx \ps^{t,(h)}_x \dpr_j \ps^{t,(h)}_x & = \int dk\, \ps^{t,(h)}_k i k_j\, \ps^{t,(h)}_{-k}   \non \\[6pt]
 %\int dx \lft(\ps^{t,(h)}_x\rgt)^2 \dpr_j \ps^{t,(h)}_x & = \int dk\, \lft(\ps^{t,(h)}_k\rgt)^2 i k_j\, \ps^{t,(h)}_{-k}  \qquad \forall j=0,\ldots,3 
%\] 
%are integrals of a total derivative with the fields satisfying periodic boundary conditions.  Then the term $\int dk \ps^{l,(h)}_{-k} \ps^{t,(h)}_{k}$ and $ \int dk  \ps^{l,(h)}_{-k} i \,\kk \,\ps^{t,(h)}_k  $ are identically zero for parity reasons. 
The previous discussion implies that the local potential for $d=3$ and $h < \bh$ is of the form:
\[ \label{local_pot_3d_b}
\LL^{3d}_{\,<}\, \BV^{(h)}(\ps)=\,  \r_0 R_0^{-2} &\, \biggl( \l_h  F^{(h)}_{04} + \m_h F^{(h)}_{12} + \g^{2h}\n_h F^{(h)}_{02} + \tfrac{1}{2}\,z_h\, F^{(h)}_{20}  \non \\[6pt] 
& +   \tfrac{1}{2}\,a_h\, F^{(h)}_{\dpr_\xx t,\, \dpr_\xx t} 
 + \tfrac{1}{2}\, b_h\, F^{(h)}_{\dpr_0 t,\, \dpr_0 t}+  e_h\, F^{(h)}_{l, \dpr_0 t} \biggr)
\]
as graphically depicted in fig. \ref{local3d_b}, where 
%the derivative on one of the external legs represents the action of the derivative on the kernel $W_{n_l^\txe, n_t^\txe}(k)$.  
the terms with derivatives on the external legs represent the monomials $F^{(h)}_{\dpr_i \a,\, \dpr_i \a'}$ defined in \eqref{monomials}.

The initial values $\rr_\bh$ of the running coupling constants at the beginning of the second region are calculated studing the beta function in the region $h > \bh$, as showed in section \ref{flows}. One finds that:
\be \label{couplings_hbar_3d}
\begin{array}{llll}
\l_\bh =\l_0 \left(1+ O\lft(\l \e^{1/2}\rgt)\right)    & \m_\bh =\m_0 \left(1+ O\lft(\l \e^{1/2}\rgt)\right) \\[6pt]
z_\bh= z_0 \left(1+ O\lft(\l \e^{1/2}\rgt)\right) \quad  & \n_\bh= O\lft(\l \e^{1/2}\rgt)  \\[6pt]
a_\bh= O\lft(\l \e^{1/2}\rgt)  &    e_\bh= O\lft(\l \e^{1/2}\rgt)  \qquad b_\bh =  O\lft(\l \e^{-1/2}\rgt)
\end{array}
\ee 
%with $\n_{\bh}$ to be fixed in such a way that  $\g^{2h^*}\n_{h^*} \arr 0$ as $h^* \arr -\io$. The latter condition is equivalent to the saddle point requirement $\dpr_\x \WW_\L(\x)=0$ as showed in section \ref{ren_condition}.  
%We can hope to have a perturbative control of the model only if we can show that all the running coupling constants appearing in \eqref{local_pot_3d_b} stay small for $h \arr -\io$, for a suitable choice of $\n_{\bh}$.
Comparing \ref{couplings_hbar_3d} with \ref{bare_couplings_3d} one notices that the integration on the fields in the region $h > \bh$ has only effect on the higher order correction to the initial values of the coupling constants at scale $0$.

%
%------------------------------------
\feyn{
\begin{fmffile}{feyn-TESI/loc3_b}
\unitlength = 1cm  
\def\myl#1{2.5cm} %larghezza parbox
\def\myll#1{3cm} 
\[
& \LL^{3d}_{\,<}\, \BV^{3d}_h \quad = 
 \parbox{\myll}{\centering{	 	   
		\begin{fmfgraph*}(2,1.25)
			\fmfleft{i1,i2}
			\fmfright{o1,o2}
			\fmf{plain}{i1,v,o2}
			\fmf{plain}{i2,v,o1}
			\fmfv{label=$\lambda_h$,label.angle=90}{v}
			%\fmflabel{$\lambda_h$}{v}
			\fmfdot{v}
			%\fmf{photon}{v1,v2}
		\end{fmfgraph*}
	}}
	+ \parbox{\myll}{\centering{
		\begin{fmfgraph*}(2,1.25)
			\fmfright{i1,i2}
			\fmfleft{o1}
			\fmf{plain}{i1,v,i2}
			\fmf{dashes, tension=1.5}{o1,v}
			\fmfdot{v}
			\fmfv{label=$\mu_h$,label.angle=100}{v}
		\end{fmfgraph*}
	}} 
	+ \quad \parbox{\myll}{\centering{	
		\begin{fmfgraph*}(2 ,1.25)
			\fmfleft{i1}
			\fmfright{o1}
			\fmf{plain}{i1,v1,o1}
			\fmfv{label={$\gamma^{2h}\nu_h$},label.angle=90}{v1}   
			%decor.shape=circle,decor.filled=full,decor.size=2thick}{v3}
			\fmfdot{v1}
		\end{fmfgraph*}
	}} \non \\[12pt] & \hskip 0.5cm
 +  \parbox{\myl}{\centering{
		\begin{fmfgraph*}(2,1.25)
			\fmfleft{i1}
			\fmfright{o1}
			\fmf{dashes}{i1,v1,o1}
			\fmfv{label={$z_h$},label.angle=-90}{v1} 
			\fmfdot{v1}
		\end{fmfgraph*}
	}}
+\parbox{\myl}{\centering{	
		\begin{fmfgraph*}(2,1.25)
			\fmfleft{i}
			\fmfright{o}
			\fmf{plain,label=$\dpr_0$, label.dist=-0.2w}{i,v}
			\fmf{plain,label=$\dpr_0$,label.dist=-0.2w}{v,o}
			\fmfv{label=$b_h$,label.angle=-90}{v}   
			\fmfdot{v}
		\end{fmfgraph*}
	}}+
 \parbox{\myl}{\centering{	
		\begin{fmfgraph*}(2,1.25)
			\fmfleft{i}
			\fmfright{o}
			\fmf{plain,label=$\dpr_\xx$,label.dist=-0.2w}{i,v}
			\fmf{plain,label=$\dpr_\xx$,label.dist=-0.2w}{v,o}
			\fmfv{label=$a_h$,label.angle=-90}{v}   
			\fmfdot{v}
		\end{fmfgraph*}	
}} +
 \parbox{\myl}{\centering{	
		\begin{fmfgraph*}(2,1.25)
			\fmfleft{i}
			\fmfright{o}
			\fmf{dashes}{i,v}
			\fmf{plain,label=$\dpr_0$,label.dist=-0.2w}{v,o}
			\fmfv{label=$e_h$,label.angle=-90}{v}   
			\fmfdot{v}
		\end{fmfgraph*} }} \non
\]
\end{fmffile}
}{{\bf Local potential for $d=3$ and $h\leq\bh$.} The running coupling constants are defined with their dimensions, \eg being the local term with two external $t$ fields relevant with dimension $\d=2$ the corresponding running coupling  is defined as $\g^{2h}\n_h$.}{local3d_b}
%---------------------------------------------------
%
{\centering \subsubsection{\it  Two dimensions}}

Since the two legged vertices have the same scaling dimensions than in the three dimensional case, the localization procedure for $\hV_{20}^{(h)}$, $\hV_{02}^{(h)}$ and $\hV_{11}^{(h)}$ will be the same defined in \eqref{loc_3d_b}. The action of the $\LL$ operator on the marginal and relevant kernels with more than two legs is defined as follows:
\[ \label{loc_2d_b}
& \LL^{2d}_{\,<}\, \hV_{06}^{(h)}(k_1,\ldots, k_5) := \hV_{06}^{(h)}(0,\ldots,0) \non \\
& \LL^{2d}_{\,<}\, \hV_{04}^{(h)}(k_1,\,k_2,\,k_3)  :=  \hV_{04}^{(h)}(0,\ldots,0) \non \\
& \LL^{2d}_{\,<}\, \hV_{12}^{(h)}(k,\,p)   :=   \, \hV_{12}^{(h)}(0,\,0)  
\]
where $\int dx (\ps^{t,(h)}_x)^3\, \dpr_i \ps^{t,(h)}_x=0$ since it is a total derivative of the fields. The localization gives zero on the irrelevant kernels. Then the local potential for $d=2$ and $h < \bh$ is of the form:
\[ \label{local_pot_2d_b}
\LL^{2d}_{\,<} \, \BV_h(\ps)=\, \r_0 R_0^{-2} & \,  \biggl( \l_{6,h} \, F_{06} +\g^h \l_h \, F_{04} + \g^{\frac{h}{2}}\m_h \,F_{12}
 + \g^{2h}\n_h \,F_{02}   \non \\[6pt]
 & + \tfrac{1}{2}\,z_h\, F_{20} +   \tfrac{1}{2}\,a_h\,  F_{\dpr_\xx t,\, \dpr_\xx t}  + \tfrac{1}{2}\,b_h\, F_{\dpr_0 t,\, \dpr_0 t}+  e_h\,F_{t,\, \dpr_0 t} \biggr)
\]
as graphically depicted in fig. \ref{local2d_b}.  The initial values of the running coupling constants at the beginning of the second region are calculated studing the beta function in the region $h \geq \bh$, as showed in section \ref{flows}. One finds: 
\be
\begin{array}{lll}
 \l_\bh =\frac{1}{16}\left(1+ O\lft(\l \rgt)\right)   & \m_\bh =\frac{\sqrt{2}}{4}\,\sqrt{\e} \left(1+ O\lft(\l \rgt)\right)    \\[6pt]
z_\bh= z_0\left(1+ O\lft(\l \rgt)\right)   &  \n_\bh = O\lft(\l \rgt)   \\[6pt]
a_\bh= O\lft(\l \rgt)  &   e_\bh= O\lft(\l\rgt)  \qquad b_\bh = O\lft(\l \e^{-1}\rgt)    
\end{array}
\ee 

\feyn{
\begin{fmffile}{feyn-TESI/loc2_b}
\unitlength = 1cm  
\def\myl#1{2.5cm} %larghezza parbox
\def\myll#1{2.2cm} 
\[
& \LL^{2d}_{\,<} \BV_h    \;= 
 \parbox{\myll}{\centering{	 	   
		\begin{fmfgraph*}(2,1.25)
	  		\fmfleft{i1,i2,i3}
			\fmfright{o1,o2,o3}
			\fmf{plain}{i1,v,o1}
			\fmf{plain}{i2,v,o2}
                  \fmf{plain}{i3,v,o3}
			\fmfv{label=$\l^h_{6}$,label.angle=90}{v}
			%\fmflabel{$\lambda_h$}{v}
			\fmfdot{v}
			%\fmf{photon}{v1,v2}
		\end{fmfgraph*}
	}}
  \;+ \parbox{\myll}{\centering{	 	   
		\begin{fmfgraph*}(2,1.25)
			\fmfleft{i1,i2}
			\fmfright{o1,o2}
			\fmf{plain}{i1,v,o2}
			\fmf{plain}{i2,v,o1}
			\fmfv{label=$\g^h \l_h$,label.angle=90, label.dist=0.15w}{v}
			%\fmflabel{$\lambda_h$}{v}
			\fmfdot{v}
			%\fmf{photon}{v1,v2}
		\end{fmfgraph*}
	}}
	+\; \parbox{\myll}{\centering{
		\begin{fmfgraph*}(2,1.25)
			\fmfright{i1,i2}
			\fmfleft{o1}
			\fmf{plain}{i1,v,i2}
			\fmf{dashes, tension=1.5}{o1,v}
			\fmfdot{v}
			\fmfv{label=$\g^{\frac{h}{2}}\mu_h$,label.angle=115}{v}
		\end{fmfgraph*}
	}} 
	+ \quad \parbox{\myll}{\centering{	
		\begin{fmfgraph*}(2 ,1.25)
			\fmfleft{i1}
			\fmfright{o1}
			\fmf{plain}{i1,v1,o1}
			\fmfv{label={$\gamma^{2h}\nu_h$},label.angle=90}{v1}   
			%decor.shape=circle,decor.filled=full,decor.size=2thick}{v3}
			\fmfdot{v1}
		\end{fmfgraph*}
	}} \non \\[12pt]  & \hskip 0.5cm
  +  \parbox{\myl}{\centering{
		\begin{fmfgraph*}(2,1.25)
			\fmfleft{i1}
			\fmfright{o1}
			\fmf{dashes}{i1,v1,o1}
			\fmfv{label={$z_h$},label.angle=-90}{v1} 
			\fmfdot{v1}
		\end{fmfgraph*}
	}}
+\parbox{\myl}{\centering{	
		\begin{fmfgraph*}(2,1.25)
			\fmfleft{i}
			\fmfright{o}
			\fmf{plain,label=$\dpr_0$, label.dist=-0.2w}{i,v}
			\fmf{plain,label=$\dpr_0$,label.dist=-0.2w}{v,o}
			\fmfv{label=$b_h$,label.angle=-90}{v}   
			\fmfdot{v}
		\end{fmfgraph*}
	}}+
 \parbox{\myl}{\centering{	
		\begin{fmfgraph*}(2,1.25)
			\fmfleft{i}
			\fmfright{o}
			\fmf{plain,label=$\dpr_\xx$,label.dist=-0.2w}{i,v}
			\fmf{plain,label=$\dpr_\xx$,label.dist=-0.2w}{v,o}
			\fmfv{label=$a_h$,label.angle=-90}{v}   
			\fmfdot{v}
		\end{fmfgraph*}	
}} +
 \parbox{\myl}{\centering{	
		\begin{fmfgraph*}(2,1.25)
			\fmfleft{i}
			\fmfright{o}
			\fmf{dashes}{i,v}
			\fmf{plain,label=$\dpr_0$,label.dist=-0.2w}{v,o}
			\fmfv{label=$e_h$,label.angle=-90}{v}   
			\fmfdot{v}
		\end{fmfgraph*} }} \non
\]
\end{fmffile}
}{{\bf Local potential for $d=2$ and $h\leq \bh $.} The running coupling constants are defined with their scaling dimensions. With respect to the three dimensional case the three and four legged terms are relevant and there is a new six--legged marginal coupling.}{local2d_b}

%----------------------------------------------------------------------------------------

\subsection{Effect of the renormalization procedure}   \label{ren_bound}
The renormalization procedure defined in the previous section is introduced precisely to guarantee that each kernel of the renormalized expansion by construction has a negative scaling dimension. We will prove this statement only in the case $h\leq\bh$ and $d=3$ and in this section we will use simply $\LL$ to refer to $\LL^{3d}_{\,<}$. The extension of the same ideas to the region $h > \bh$ and to $d=2$ is trivial.

Let consider the kernel $\hV_{20}^{(h)} (k_1,k_2)$ and its local part, see \eqref{loc_3d_b}. The action of the $\RR$ operator on this kernel corresponds to subtract from it the first term of its Taylor expansion around $k_1=k_2 =0$: 
\[ \label{R}
 \RR \hV_{20}^{(h)} (k_1,k_2) & \,= \hV_{20}^{(h)}(k_1,k_2)- \hV_{20}^{(h)}(0,0) \non \\
& =\, \hV_{20}^{(h)}(k_1,k_2) - \hV_{20}^{(h)}(k_1,0) + \hV_{20}^{(h)}(k_1,0)  -W_{20}^{(h)}(0,0) \non \\
& = \int_0^1 dt \frac{d}{dt} \hV_{20}^{(h)}(k_1,t k_2) + \int_0^1 dt \frac{d}{dt} \hV_{20}^{(h)}(t k_1,0) \non \\
& = \int_0^1 dt k_2 \dpr_{k_2} \hV_{20}^{(h)}(k_1,t k_2) + \int_0^1 dt k_1 \dpr_{k_1} \hV_{20}^{(h)}(tk_1,0) \non \\
& ``='' \int_0^1 dt \uk \dpr_\uk \hV_{20}^{(h)}(\uk) 
\]
The expression in the last line of \eqref{R} is only formal, but it is useful in order to derive a dimensional estimate for the action of the renormalization operator. In fact we will bound separately  the external momentum $k$ which multiplies the kernel $\hV_{20}^{(h)}(\uk)$ and the derivative which acts on it. For what regards $k$ it is associated to one of the lines going out from the cluster $\hV_{20}^{(h)}(\uk)$ and it will be integrated with the propagator $g^{(h_{v_1})}$ corresponding to this line. Since the scale $h_{v_1}$ of a line external to a cluster $G_v$ is necessarily smaller than $h_v$ the momentum $k$ gives a dimensional gain $\g^{h_{v_1}}\leq \g^{h_{v'}}$, with $h_{v'}$ the scale of the first non trivial vertex preceding $v$ on $\t$. On the contrary the derivative $\dpr_k$ acts on one of the lines contained in the cluster $G_v$, with scale $h_{v_2}\geq h_v$. Then a dimensional estimate for the derivative is the factor $\g^{-h_{v_2}}\leq \g^{-h_v}$. Putting together the two estimates we get that the action of $\RR$ on $\hV_{20}^{(h)}(\uk)$ gives an extra dimensional factor
\be
\g^{-(h_v - h_{v'})}
\ee 
with respect to the not renormalized extimate. In the same way we can prove that the action of $\RR$ on $\hV_{11}^{(h)}(k_1, k_2)$ corresponds in subtracting to it the first two terms of its Taylor expansion and that we can write
\[ \label{defR2}
\RR \hV_{11}^{(h)}(k_1, k_2)& = \hV_{11}^{(h)}(k_1, k_2)- \hV_{11}^{(h)}(0, 0) \non \\
&  - (k_1 \,\dpr_{k_1} +k_2 \,\dpr_{k_2} ) \hV_{11}^{(h)}(k_1, k_2)\bigl|_{k_1, k_2=0} \non \\
& ``='' \int_0^1 dt \uk^2 \dpr^2_\uk \hV_{11}^{(h)}(t \uk)
\]
Then $\RR \hV_{11}^{(h)}(k_1, k_2)$ has a dimensional gain $\g^{-2(h_v -h_{v'})}$. 
The same discussion can be repeated for the remaining kernels; one finds that the action of the $\RR$ operator on a generic vertex with $(n^\txe_{l,v}\,, n^\txe_{t,v})$ external legs is dimensionally equivalent to a factor
\[ \label{improvement}
 \g^{-z_v (h_v - h_{v'})}
\]
where $z_v$ is the {\it improvement of the scaling dimension due to renormalization}. In the  $d=3$ case and in the region $h \leq \bh$ the improvement of the scaling dimension is the following
\be \label{zRen_3d}
z^{3d,<}_v =
\begin{cases}
\;4  &  \;(n^\txe_{l,v}\,, n^\txe_{t,v}) =(0,2) \\
\;3 &  \;(n^\txe_{l,v}\,, n^\txe_{t,v}) =(0,3),\,(1,1) \\
\;2  &  \;(n^\txe_{l,v}\,, n^\txe_{t,v}) =(0,4),\,(1,2),\,(2,0) \\
\; 0  &  \;\text{otherwise}
\end{cases}
\ee
where we have also taken into account that some of the diagrams are null for symmetry reasons.

Apart from the latter dimensional gain, it is important to check if there are other factors coming from the definition of the renormalization procedure, which may destroy the $n!$ bounds. Even if we will not enter in the detail of this discussion we list the problems which may arise and the respective solutions:
\begin{enumerate}[i)]   %\nota{ref. tesi alessandro}
\item  In the case of more renormalizations acting in clusters one inside the other we could worry about to have $m=O(n)$ derivatives acting on the same propagator of some internal cluster. In fact in this case the bound for $|\dpr^m g^h(x)|$ has a $(m!)^\a$ factor, $\a>1$, due to the fact that the propagator is not analytic. The  $(m!)^\a$ will make the $n!$ bounds lost. By studying in detail the action of the $\RR$ operator over a cluster which has already been renormalized, by using the complete expression for the action of $\RR$ rather than the formal expressions on the last lines of  \eqref{R} and \eqref{defR2}, one sees that the derivatives do not accumulate on the same propagator (see \eg~\cite[sec. 3.3]{XYZ} or~\cite{THGiuliani}).
\item  In the renormalization operation different contributions arise, since, given a derivative, we are free of choosing on which of the internal propagator make the derivative to act. We need to bound the number of these contributions. This require a more precise definition of the renormalization procedure, consisting in fixing for each cluster $G_v$ two of the $s_v$ possible subclusters of $G_v$, on which eventually make the two derivatives coming from the renormalization of $v$ to act. With the latter definition one can prove that the number of terms generated by the renormalization operation is bounded by $C^n$.
\item Another problem may arise if the momenta $k$ coming from the renormalization accumulate on the same external line. However still this phenomenon can be controlled, thanks to some freedom intrinsic in our definition of the renormalization procedure. We refer to~\cite[sec. 3.3]{XYZ} for the discussion of this point.
%We can avoid this situation, since we have freedom of choosing the external line on which make $k$ to fall. In fact we can decide to associate the $k$ coming from the renormalization of the cluster $v$ to one of the internal line of the cluster $v'$ immediately preceding $v$ in $\t$. This choice is feasible being the diagram connected. 
\end{enumerate}

We can now get the bound for a generic Feynman graph $\G_\LL$  which contributes to $\LL \BV_h$ with $h\leq \bh$ and in three dimensions. It is sufficient to add to the not renormalized estimate  \eqref{fin_down3d}  the extra dimensional factors \eqref{improvement} coming from the action of $\RR$ on the vertices of $\t$. Besides, with respect to the unrenormalized trees, the trees contributing to the $\LL \BV_h$  have endpoints at each scales $h \leq \bh$. Then we also have a product over the values of the running coupling constants $\rr_{h_v}$. One gets:
\[ \label{ren_bound}
\bigl\|\Val(\G_\LL)\bigr\|  
\leq C^n  \,& \lft(\r_0 R_0^{-2}\rgt) \bar{C}_{3d}(P_v;\e,\r_0, R_0)\,  \g^{h\, \d^{3d,<}_{v_0}}\hskip -0.3cm\prod_{v\, \text{not e.p.} }  \g^{(h_v -h _{v'})\,(\d^{3d,<}_v -z_v) } \non \\
& \prod_{v \text{\,e.p.}} \lft(\l_{h_v}\rgt) \bigl(\e^{-1/2}\m_{h_v}\bigr) \lft(\e \n_{h_v}\rgt) (a_{h_v}) (b_{h_v})  (e_{h_v} z_{h_v}) 
\]
In \eqref{ren_bound} $h$ the frequency of the root, 
$ \d^{3d,<}_{v}  = 4-2n_{l,v}^\txe- n_{t,v}^\txe - n_{\dpr,v}^\txe $ the unrenormalized scaling dimension and $\bar{C}_{3d}(P_v;\e,\l)= \bigl(\l \e^{-\frac{1}{2}}\bigr)^L  \e^{-2 +\frac{3}{4}n_l^\txe + \frac{1}{2}n_t^\txe +\frac{1}{2}n_{\dpr_\xx}}$, see \eqref{3d_h}. 
Note that, due to the choice of defining the running coupling constants with their dimension, the bound \eqref{ren_bound} does not contains the dimensional factors associated to the endpoints. In fact, for each endpoint representing a local term at scale $h_v$ the bad dimensional factor appearing in the product over the endpoints of type $m_2$ in \eqref {fin_down3d} 
\[
\prod_{v \text{\,e.p.}} \g^{- 2 h_v \c(m_{2,v}) } 
\]
is exactly compensated by the dimension in front of the running coupling constant, \ie by the fact that for each endpoint of type $m_{2}$ we have a factor $\g^{2h_v}\n_{h_v}$. For what concerns the product over the endpoints corresponding to irrelevant terms (\ie with negative scaling dimension) they can  only have scale label $h=1$ and then the products on this points give a constant factor. 
Assuming that
\be \label{assumption3d}
\h^{3d}_*= \sup_{h^*<h \leq \bh} \max \{ |\l_h|, |\e^{-1/2}\m_h|,| \e \n_h|,|a_h|, ,|b_h|,|z_h|,|e_h|\} 
\ee
with $\h_*$ a small constant we get
\[ \label{ren_bound2}
\bigl\|\Val(\G_\LL)\bigr\|  
\leq C^n \,(\h_*^{3d})^n \,& \lft(\r_0 R_0^{-2}\rgt) \bar{C}_{3d}(P_v;\e,\r_0, R_0) \non \\[3pt] 
&  \g^{h\, \d^{3d,<}_{v_0}} \prod_{v\, \text{not e.p.} }  \g^{(h_v -h _{v'})\,(\d^{3d,<}_v -z^{3d,<}_v) } 
\]
with  $z^{3d,<}_v$ defined in \eqref{zRen_3d} and
\[
\d^{3d,<}_v = 4 -2n_{l,v}^\txe -n_{t,v}^\txe -n_{\dpr,v}^\txe
\]
The factor $\g^{h \d_{v_0}}$ in \eqref{ren_bound2} it is absorbed in the definition of the beta function. Assume for example that we are calculating $\LL V^{h}_{02}(x-y)=\g^{2h} \n_h$: the beta function $\b^\n_{h+1}$ is given by the sum of all the renormalized trees with $n>1$ vertices, root at scale $h$ and a label $\LL$ on the first vertex following the root $v_0$. The dimensional estimate of each of these trees has in front a factor $\g^{2h}$; then, defining the beta function without this factor we find the following flow equation for $\n_h$:
\[
& \g^{2h}\n_h = \g^{2(h+1)} \n_{h+1} + \g^{2h}\b^\n_{h+1} \non \\
& \n_h = \g^2 \n_{h+1} + \b_{h+1}^\n
\]
to be compared with \eqref{beta_funct}. Note that the renormalization procedure is defined in such a way that $\d^{3d,<}_v -z^{3d,<}_v <0$ for every cluster. Since the {\it renormalized scaling dimension}
\be
D^{3d,<}_v = \d^{3d,<}_v - z^{3d,<}_v
\ee
is always negative, using the estimates discussed at the end of the section \ref{multiscale}, we can prove $n!$ bounds for the renormalized expansion, under the assumptions \eqref{assumption3d}.  \\

A very analogous discussion can be carry forward also for $h\geq \bh$ and in two dimensions. We will not belabor on this part since in the next section we will introduce a new family of effective potentials, for which we will give every detail.  We only report here for completeness the improvement of the scaling dimension for the $2d$ case
\[ \label{zRen_2d}
z_v^{2d,\,<}  = 
	\begin{cases}
       \;4  & \;(n^\txe_{l,v},n^\txe_{t,v}) = (0,2) \\
	\;3  & \;(n^\txe_{l,v},n^\txe_{t,v}) = (0,5),\,(0,3),\,(1,1) \\
	\;2 &    \; (n^\txe_{l,v},n^\txe_{t,v})= (0,6),\,(0,4),\,(1,2),\,(2,0) \\
	\;1 &    \; (n^\txe_{l,v},n^\txe_{t,v}) = (1,3)\\
	\;0 & \;\text{otherwise}
	\end{cases}  
\]
with respect to the scaling dimension 
\[
\d^{2d,<}_v = 3 -\frac{3}{2}n_{l,v}^\txe -\frac{1}{2}n_{t,v}^\txe -n_{\dpr,v}^\txe
\]
The analogous of the assumption \eqref{assumption3d} in the two dimensional case is 
\be \label{assumption2d}
\h^{2d}_*=\sup_{h^*<h \leq \bh} \max \{ |\l \l_{6,h}|,|\l \l_h|, |\sqrt{\l \e^{-1}\,}\m_h|,| \e \n_h|,|a_h|, ,|b_h|,|z_h|,|e_h|\}  \ll 1
\ee
For what concerns the renormalized dimension for $h >\bh$ we remind that
\[
& \d^{3d,>}_v = \frac{5}{2} -\frac{3}{4}n_{v}^\txe -\frac{1}{2}n_{\dpr_\xx,v}^\txe -n_{\dpr_0,v}^\txe \non \\
& \d^{2d,>}_v = 2 -\frac{1}{2}n_{v}^\txe -\frac{1}{2}n_{\dpr_\xx,v}^\txe -n_{\dpr_0,v}^\txe
\]
while the improvements of the scaling dimensions are
\[  \label{zRen_up}
  z_v^{3d,\,>}  = 
\begin{cases}
\;2 &   \;  n^\txe_v=2 \\
\;1 &   \; n^\txe_v=3  \\
\;0 & \;\text{otherwise}
\end{cases}  \qquad 
  z_v^{2d,\,>}  = 
	\begin{cases}
	\;2 &    \;n^\txe_v=2 \\
	\;1 &   \;n^\txe_v=3,4 \\
	\;0 & \;\text{otherwise}
	\end{cases}  
\]

%Note that each step of the multiscale integration corresponds to a very large resummation of Feynman graphs, then the cancellations making the theory finite may be very complicate to recognize also at the first orders of the perturbative expansion. 

%At each step $h$ of the integration $V_{h}$ has a part that is growing under the RG transformation (the so called ``relevant'' and ``marginal'' terms) and a part that is smaller and smaller, called ``irrelevant''. Performing a suitable resummation, the result of the integration at each scale is exactly written as a series in the relevant and marginal terms, the running coupling constants $ \vec{r}_h = \vec{r}_h\,(\vec{r}_{h+1},\dots,\vec{r}_0)$. Each step of this multiscale integration corresponds to a very large resummation of Feynman graphs.

\vskip 1cm

{\bf Short memory property.} An immediate collorary of the dimensional estimates \eqref{ren_bound} for the three dimensional case is that contributions from trees in $\TT_{h,n}$ with a vertex $v$ on scale $h_v=k >h$ admit an improved bound with respect to the $n!$ bound in \ref{ren_bound2}, of the form
$\leq C^n \h^n n!\, \g^{h\,D^{3d,<}_{v_0}} \g^{\th (h-k)}$, for any $0<\th <2$. In fact, in the  basic power counting we have the product
\[
\prod_{\substack{v\, \text{not e.p.} \\ v>v_0}} \g^{-(h_{v}-h_{v'})\,D^{3d,<}_{v}}
\]
with $D^{3d,<}_v < 2$. However  $D^{3d,<}_v<0$ is a sufficient condition to garantee the convergence. Then, for any $0<\th<2$ we can rewrite the previous product as
\[
%& \prod_{\substack{v\, \text{not e.p.} \\ v>v_0}} \g^{-(h_{v}-h_{v'})\,\th \d_{v}} \Big[\prod_{\substack{v\, \text{not e.p.} \\ v>v_0}} \g^{-(h_{v}-h_{v'})\th \d_{v}}\Big] \non \\
%& \qquad \leq 
\prod_{\substack{v\, \text{not e.p.} \\ v>v_0}} \g^{-(h_{v}-h_{v'}) \,(D^{3d,<}_{v}-\th)} \,\g^{-\th(h_{v}-h_{v'})}   \label{short_mem1}
\]
For a tree with a branch $\mathcal{B}$ going from scale $h$ to scale $k \gg h$ the product of the factors $\g^{-\th(h_{v}-h_{v'})}$ in \eqref{short_mem1} along the branch gives the factor
\[
\prod_{v \in \mathcal{B}} \g^{-(h_{v}-h_{v'})\,\th}= \g^{-(k - h)\th}
\]
which can be thought as a dimensional gain with respect to the ``basic'' dimensional bound. This improved bound is usually referred to as the short memory property, since indicates that long trees (\ie trees with non trivial interactions at scale more and more distant from $h$) are exponentially suppressed. 

%This means that the trees have smaller and smaller values if . In other words the result does not depend on the detail of the ultraviolet region but only on the scales near $h$. This fact motivates the name ``short memory property'' for this result.

The short memory factors that can be extracted from the dimensional estimate of a generic  diagram in three and two dimensions, above and below $\bh$ are:
\[
& 0<\th^{3d}_> < \frac{1}{2}\quad ; \quad  0<\th^{3d}_< < 2   \non \\
& 0<\th^{2d}_> < \frac{1}{2} \quad ; \quad   0<\th^{2d}_< < \frac{1}{2}
\]

\pagina

\section{Effective potentials and renormalized measure} \label{effective}

In the previous section we have seen that the effect of the localization operator $\LL$ is to identify at each step of the multiscale integration some new effective couplings. 
%Among these, there are some local quadratic terms in the fluctuation fields, so that we can always write the action of the localization operator as the sum of two pieces
For each dimension and each value of $h$ we may write
\[
\LL \BV_{h}(\ps) = \LL_0 \BV_{h}(\ps) + \LL_Q \BV_{h}(\ps)
\]
with $\LL_Q \BV_{h}(\ps)$ defined by 
\[
& \LL_Q \BV_{h}(\ps)  = \r_0 R^{-2}_0\,  \biggl( \tfrac{1}{2}\,\bar{z}_{h} \int(\psi_{x}^{l})^{2}dx  \biggl) &  \bh<h\leq 0 \non \\[6pt]
& \LL_Q \BV_{h}(\ps)  = \r_0 R^{-2}_0  \biggl( \tfrac{1}{2}\,z_{h} \int(\psi_{x}^{l})^{2}dx+ \tfrac{1}{2}\, a_{h}\int(\dpr_\xx \psi_{x}^{t})^{2}dx  \non \\
& \hskip 3cm + \tfrac{1}{2}\,b_{h}\,\int(\partial_{0}\psi_{x}^{t})^{2}dx   +e_{h}\int\psi_{x}^{l}\partial_{0}\psi_{x}^{t}dx \biggr) & h \leq \bh
\]
In the high momentum region $\LL_Q \BV_{h}(\ps) $ contains only the local quadratic term proportional to $\bar{z}_h$, with $\bar{z}_0=\e$ Bogoliubov contribution to the measure. In the low momentum region, on the contrary, we define $\LL_Q \BV_{h}(\ps) $ so that it contains all the local quadratic terms in the bosonic fields. 

Now we can iteratively define a new family of effective potential $\VV_h$, as follows.
%by adding to the  measure at each step of the multiscale iteration the quadratic local part $\LL_Q \BV_{h}$.  
Let consider the functional integral in \eqref{free4} after the first step of the integration:
\[   \label{BV_h}
e^{-|\L|\, \WW_{h^*}(\r_0)} =  e^{-\EE_{-1}(\r_0)} \int P_{Q_0,\c_{[h^*,-1]}}(d\PS{\leq -1}) \,e^{-\BV_{-1}(\PS{\leq -1})} 
\]
with $P_{Q_0,\c_{[h^*,-1]}}(d\PS{\leq -1})$ the measure with covariance
\[ \label{g0}
g^{[h^*,-1]}_{\a \a'}(x)=\int\frac{d^{d+1}k}{(2\pi)^{d+1}}\,e^{-ikx}\chi_{[h^*,-1]}(k)\,g^{\,(0)}_{\a \a'}(k)
\]
 with $\a,\a'=l,t$ and 
\[
g^{\,(0)}_{\a \a'}(k)^{-1} = Q_{\a \a'}^{(0)}(k)= \r_0 R_0^{-2} \left(\begin{array}{cc}
\kk^{2}+\e \chi_{[h^*,0]}(k)\quad &  k_{0}\\
- k_{0} & \e^{-1}\kk^{2}
\end{array}\right)
\]
We can {\it renormalize} the measure $P_{Q_0,\c_{[h^*,-1]}}(d\PS{\leq -1})$ by adding to the exponent of its gaussian weight the local quadratic terms in $\LL_Q \BV_{-1}(\ps) $, by using the property \eqref{mult_gauss_measures}. In this way we get:
\[   \label{BV_2}
e^{-|\L|\, \WW_{h^*}(\r_0)} =  e^{-\EE_{-1}(\r_0)} \, e^{-t_{-1}}\,\int P_{Q_{-1},\c_{[h^*,-1]}}(d\PS{\leq -1}) \,e^{-\VV_{-1}(\PS{\leq -1})} 
\]
where $t_{-1}$ takes into account the different normalization of the functional integrals. The measure $P_{Q_{-1},\c_{[h^*,h]}}(d\PS{\leq h})$ is given by
\[
\int P_{Q_{-1}, \c_{[h^*,h]}}(d\PS{\leq h}) :=  \,e^{-t_{-1}}\int P_{Q_0, \c_{[h^*,h]}}(d\PS{\leq h})\, e^{-\LL_Q \HV_{h}(\PS{\leq h})}  
\]
with the matrix $Q_{-1}$ defined as
\[
Q_{\a \a'}^{(-1)}(k) = Q_{\a\a'}^{(0)}(k)  + q_{\a \a'}^{(-1)} \, \c_{[h^*,-1]}(k) 
\]
with $Q_{ll}^{(0)}(k)=z_0\,\c_{[h^*,0]}(k)$, $Q_{tt}^{(0)}=Q_{ll}^{(0)}=Q_{b}^{(0)}=0$ and $q_{\a \a'}^{(-1)}$ the matrix of the local terms $\LL\BV_{-1}$ generated by the integration over the $\PS{-1}$ fields:
\[
q_{\a \a'}^{(-1)} &  =  \left(\begin{array}{cc}
 z_{-1} & 0\\
0 & 0
\end{array}\right)
\]
Its propagator has the same form than \eqref{g0} but with $g^{(0)}_{\a \a'}$ substituted by
\[
g^{\,(-1)}_{\a \a'}(k)^{-1} & = g^{\,(0)}_{\a \a'}(k)^{-1} + Q_{\a \a'}^{(-1)}(k) 
 \non \\[6pt]
& = \left(\begin{array}{cc}
\kk^{2}+\e \chi_{[h^*,0]}(k)+z_{-1} \chi_{[h^*,-1]}(k)\quad &  k_{0}\\
- k_{0} & \e^{-1}\kk^{2}
\end{array}\right)
\]
At this point we can integrate \eqref{BV_2} at scale $-1$; defining
\[ \label{VV_2}
e^{-\VV_{-2}(\PS{\leq -2}) - \tl{\EE}_{-1} } = \int P_{Q_{-1},f_{-1}}(d\PS{\leq -1}) \,e^{-\VV_{-1}(\PS{\leq -1})}
\]
we get
\[
& e^{-|\L|\, \WW_{h^*}(\r_0)} =  e^{-\EE_{-2}(\r_0)} \,\int P_{Q_{-1},\c_{[h^*,-2]}}(d\PS{\leq -2}) \,e^{-\VV_{-2}(\PS{\leq -2})} 
\]
with $\EE_{-2}= \EE_{-1}+t_{-1}+\tl{\EE}_{-1}$. Iterating the previous scheme we may write \eqref{BV_h} as follows:
\[ 
e^{-|\L|\, \WW_{h^*}(\r_0)} & =  e^{-\EE_{-2}(\r_0)} \int P_{Q_{-2},\c_{[h^*,-2]}}(d\PS{\leq -2}) \,e^{-\VV_{-2}(\PS{\leq -2})}  \non \\[6pt]
= \ldots \, & = e^{-\EE_{h}(\r_0)} \int P_{Q_{h},\c_{[h^*,h]}}(d\PS{\leq h}) \,e^{-\VV_{h}(\PS{\leq -2})} 
\]
with $E_{h-1}= E_{h}+t_{h}+\tl{E}_{h}$ at each step. The procedure so far described leads to the definition of new families of measures  $P_{Q_{h},\c}(\PS{\leq h})$ and effective potentials $\VV_{h}(\PS{\leq h})$. The effective potentials are defined as in \eqref{VV_2}
\[
 e^{-\VV_{h}(\PS{\leq h})- \tl{\EE}_h}  =  \int  P_{Q_{h+1}, f_{h+1}}(\PS{ h+1})\,e^{-\VV_{h+1}(\PS{\leq h}+ \PS{h+1})} 
\]
For $\bh<h <0 $ the difference between the potentials $\BV_h$ and $\VV_h$ is only that the latter does not contain the local term proportional to $z_h$, which has been included in the measure. For $h\leq \bh$ the effective potential $\VV_{h}(\PS{\leq j}$ does not contain any of local quadratic terms generated by the integration on scale $h+1$, that is:
\[
\LL \VV_{h}(\ps)   = \r_0 R^{-2}_0 & 
\biggl( \l_{6,h}\,\c(d=2)\,\int \left(\psi_{x}^{t}\right)^{6}dx + \g^{(d-3)h}\l_{h} \int \left (\psi_{x}^{t}\right)^{4}dx  \non \\[6pt]
& + \g^{\frac{d-3}{2}\,h} \m_{h}\int(\psi_{x}^{t})^{2}\psi_{x}^{l}dx + \g^{2h}\nu_{h} \int(\psi_{x}^{t})^{2}dx  \biggr) 
\]
For what concerns the family of measures $P_{Q_{h}, f_h}(d\PS{\leq h})$, they have the following propagator
\be \label{g_h}
g^{\,(h)}_{\a \a'}(x)= \int\frac{d^{d+1}k}{(2\pi)^{d+1}}e^{-ikx}f_h(k)\,g^{\,(h)}_{\a \a'}(k)
\ee
with 
\[
\lft( g^{\,(h)}_{\a \a'}(k) \rgt)^{-1}  & =
 \lft(g^{\,(0)}_{\a \a'}(k) \rgt)^{-1} + \sum_{j=h}^{0}q^{\,(j)}_{\a \a'}\, \chi_{[h^*,j]}(k)   \non \\[6pt] 
&= \left(\begin{array}{cc}
\kk^2 + Z_{h}(k) & E_{h}(k)\,k_{0}\\
E_{h}(k)\,k_{0} & \qquad A_{h}(k)\,\kk^{2}+B_{h}(k)\,k_{0}^{2}
\end{array}\right)  
\]
where for each scale $h$
\[
Z_{h}(k) & =  Z_{h-1}+ z_{h}\c_{[h^*,h]}(k) 
\]
while we have 
\[
E_h(k)=A_{h}(k)=1 \quad  B_h(k)=0 \qquad \bh<h\leq0
\]
and
\[
E_{h}(k) & = E_{h+1}(k) + e_{h+1}\,\c_{[h^*,h]}(k) \non \\
A_{h}(k) & =  A_{h+1}(k) + a_{h+1}\,\c_{[h^*,h]}(k) \non \\
B_{h}(k) &=  B_{h+1}(k) + b_{h+1}\,\c_{[h^*,h]}(k) \qquad  h \leq \bh
\]
with initial conditions
\[
Z_{0}(k)=\e \chi_{[h,0]}(k);\quad A_{0}= E_{0}=1;\quad B_{0}=0
\]
Then it turns to be
\[ \label{wave_funct}
Z_{h}(k) & =  \sum_{j=h}^{0}\,z_{j}\,\c_{[h^*,\,j]}(k) \non \\
A_{h}(k) -1 & =  \sum_{j=h}^{\bh}\,a_j \,\c_{[h^*,\,j]}(k)\non \\
B_{h}(k) & =  \sum_{j=h}^{\bh}\,b_j \,\c_{[h^*,\,j]}(k) \non \\
E_{h}(k) -1 & =  \sum_{j=h}^{\bh}\,e_j\,\c_{[h^*,\,j]}(k) 
\]
The functions  $\{A_h(k), B_h(k), E_h(k),Z_h(k)\}$ are called  {\it wave function renormalization functions}, since they represent the renormalization of the propagator due to the integration on the fields living on momentum scales greater then $\g^{h}$.

%With these definitions the integration of the fields living on momentum scales greater then $\g^{h}$ produce an effective theory very similar to the original one, modulo the presence of a new effective potential $\VV_h$ and also a new propagator, which is renormalized by the local quadratic terms generated on scales equal and greater than $j$. 

\subsubsection{Wave function renormalization constants}

The wave function renormalization functions depends on $|k|$ through the cutoff function, see \eqref{wave_funct}. When we evaluate them at a certain value of $|k|$ we get the  {\it wave function renormalization constants}. Given a generic wave function renormalization function $Q_h(k)$ with $h \geq h^*$ we choose as scale of the localization $\tl{k}$ such that  $|\tl{k}|^2= \frac{2\,\g^2}{\g^2 +1} \,\g^{2h}$. For this value of $|k|$ we have that all the cutoff functions $\chi_{[h^*,j]}(|k|)$, with $j \geq h$ appearing in the definition \eqref{wave_funct} are equal to one; then the renormalization constants take their maximum value in $|\tl{k}|$. We define:    
\[
Z_{h}  := & \max_{k} Z_{h}(k)=  \sum_{j=h}^{0} z_{j} \non \\
A_{h}  := & \max_{k} A_{h}(k)=A_{0}+\sum_{j=h}^{\bh} a_{j}\non \\
B_{h}  := & \max_{k} B_{h}(k)=\sum_{j=h}^{\bh} b_j\non \\
E_{h}  := & \max_{k} E_{h}(k)=E_{0}+\sum_{j=h}^{\bh} e_{j}
\]
The definition of the renormalization constants is well defined if it is independent on the particular choice of the cutoff function $\c_{[h^*,0]}(k)$ appearing in the effective potential, that is
\[ \label{def_Z}
Z_{j}^{[h^*]}\equiv Z_j  \qquad \forall j> h 
\]
with the label $[h^*]$ in $Z_{j}^{[h^*]}$ referring to the lower scale of the cutoff function. This is of course an essential condition to have a definition which makes sense in the limit $h \arr -\io$ and is easily verified. \\

{\it Proof of \eqref{def_Z}. } For simplicity of notation in the following we will take $|\tl{k}|=\g^h$ for a certain scale $h$. Let consider two different models with cutoff functions $\c_{[\hat{h},0]}(k)$ and $\c_{[h^*,0]}(k)$ with $\hat{h}<h^*$;  the renormalized wave functions at scale $h$ are respectively:
\[
Z_{h}^{[\hat{h}]}(k) = & Z_{0}+\sum_{j=h}^{0} z_j \chi_{[\hat{h},j]}(k) \non \\
Z_{h}^{[h^*]}(k) = & Z_{0}+\sum_{j=h}^{0} z_j\chi_{[h^*,j]}(k)
\]
Since $\chi_{[\hat{h},j]}(\g^h)=\chi_{[h^*,j]}(\g^h)$ for each $h>h^*$ then $Z_{h}^{[\hat{h}]}=Z_{h}^{[h^*]}$ for each $h > h^*$. 
Then, if  $h=h^*$, the definition of $Z_{h^*}^{[\hat{h}]}$ and $Z_{h^*}^{[h^*]}$ are not equivalent but their difference is small, in fact:
\[
Z_{h^*}^{[h^*]} & =  Z_{0}+\sum_{j=h^*+1}^{0} z_j \chi_{[h^*,j]}(\g^{h^*})+z_{h^*} f_{h^*}(\g^{h^*}) \non \\
% & =  Z_{0}+\sum_{j=h^*+1}^{0} z_{j}\chi_{[\hat{h},j]}(\g^{h^*})+z_{h^*}f_{h^*}(\g^{h^*}) \non \\
 & =  Z_{0}+\sum_{j=h^*}^{0} z_j\chi_{[\hat{h},j]}(\g^{h^*})+z_{h^*}\left(f_{h^*}(\g^{h^*})-\chi_{[\hat{h},h^*]}(\g^{h^*})\right) \non \\
 & =  Z_{h^*}^{[\hat{h}]} - z_{h^*}\chi_{[\hat{h},h^*-1]}(\g^{h^*})
\]
where $z_{h^*}\chi_{[\hat{h},h^*-1]}(\g^{h^*})$ is a very small quantity, as discussed just below; note also that $\chi_{[\hat{h},h^*-1]}$ which would equal to zero in case of sharp cutoff functions.  \\

Let now consider the renormalized propagator \eqref{g_h}. It is simple to see that the dependence on $k$ of the renormalized wave functions is in general very weak. In fact, if we consider a propagator on scale $h>h^*$, with $h^*$ the infrared cutoff 
%\[
%g_{\a \a'}^{(j)}(x)=\int\frac{d^{4}k}{(2\pi)^{4}}e^{-ikx}\,f_{j}(k)\,g_{j}^{\a \a'}(k)
%\]
the wave function renormalization constants have to be evaluated on
the support of $f_{h}(k)$. On this support $\chi_{[h^*,j]}(k)=1$ for all $j>h$ and then we have, for example
\[ \label{Z_k}
f_{h}(k)\, Z_{h}(k) & =  f_{h}(k) \,\Bigl( \sum_{j=h+1}^{0} z_j+ z_{h}f_{h}(k)  \Bigr)\non \\[6pt]
 & = f_{h}(k) \,\Bigl(  Z_{h+1} + z_h f_{h}(k)\Bigr)  = f_{h}(k) \,\Bigl(  Z_h + z_h (f_{h}(k)-1)\Bigr)
\]
where $z_h$ is the last scale contribution to the beta function of $\hV^{(h)}_{20}(k)$ and is a small quantity with respect to $Z_{h}$: 
\[
& \frac{z_h}{Z_h}= O(\l \e^{-1/2} \l_h)  & d=3  \non \\[6pt]
& \frac{z_h}{Z_h}= O(\l \l_h)  & d=2 
\]
where $\e^{-1}\l_h \ll1$ and $\l \l_h \ll 1$ are the assumption for the $n!$--bounds, see \eqref{assumption3d} and \eqref{assumption2d}. Under the same assumptions we can prove that also $a_h$, $b_h$ and $e_h$ are small quantities with respect to $A_h$, $B_h$ and $E_h$. Then we may conclude that {\it on the support of $f_h(k)$} 
%\[
%&  Z_h(k) = Z_h \lft(1+ (f_h(k)-1)\,O\bigl(\l \e^{-1/2} \l_h)\, \rgt)  & d=3  \non \\[6pt]
%&  Z_h(k) = Z_h \lft(1+ (f_h(k)-1)\,O\bigl(\l \l_h)\, \rgt)  & d=2
%\]
we can  
\begin{enumerate}[a)]
\item  neglect  the dependence on $k$ of $Z_{h}$, $A_{h}$, $B_{h}$ and $E_{h}$ in the renormalized propagator 
\item approximate the wave function renormalization constants at scale $h+1$ by the corresponding constants at scale $h$, since $Z_{h}=Z_{h+1}+ z_h$. 
\end{enumerate}

\pagina

\subsection{Bounds for the renormalized propagator}  \label{cap2_ren_prop}

We are now ready to derive the dimensional bound for the Feynman graphs produced by the multiscale integration with the renormalized measure we have just introduced. In order to do that we need  it is crucial to find bounds for the renormalized propagator defined by \eqref{g_h}.

\subsubsection{\bf High momenta region $\bh < h \leq 0$ }
The discussion for the high momentum region $\bh < h \leq 0$ is trivial, since only the local quadratic term has been included in the free measure. Then the trees contributing to the expansion of the effective potentials $\VV_h$ with $\bh <h\leq 0$ are the same than the ones in the multiscale expansion for $\BV_h$, except for the fact that there are no endponts with two external dashed legs and that the propagator at scale $h$ is given by
\[
g^{\,(h)}_{\a \a'}(k)   & = \frac{\left(\begin{array}{cc}
\kk^{2}&  k_{0}\\
\, k_{0} &  \kk^{2} + Z_{h}(k)
\end{array} \right)} {  k_{0}^{2} +\lft( \kk^{2}+ Z_{h}(k)\rgt)\,\kk^{2} }   
\]
with $Z_h(k) \leq \kk^2$ for each $h \geq \bh$, by definition of $\bh$. We remind that $g^{\,(h)}_{\a \a'}(x) $ is obtained by integrating $g^{\,(h)}_{\a \a'}(k) $ over the support of $f_h(k)$, see \eqref{g_h}. The dimensional estimate of the renormalized propagator in this region is not changed with respect the non--renormalized one. The following bounds hold. \\

\begin{bound} \label{thm_nfactorial0} 
{\bf ($n!$--bounds $\bh<h \leq 0$)} 
Let consider the quantity $|| V^{(h);n}_{n^\txe_l,n^\txe_t}||$ defined by the expression \eqref{sum1} with $\bh < h \leq 0$, but with the sum over the trees running over the renormalized trees 
obtained by the localization procedure defined in~\eqref{loc_3d_a} and \eqref{loc_2d_a}, respectively for the $3d$ and $2d$ cases.
 Here  $||V^{(h);n}_{n^\txe_l,n^\txe_t}||$ is an estimate for the coefficient of order $n$ of the renormalized expansion for the kernel of the effective potential at scale $\bh<h \leq 0$ with fixed number of external legs. Let $\bar{\h}$ defined by 
\[
& \bar{\h}= \sup_{\bh<h\leq 0} \{ \l \e^{-1}\bl_{h},\, \sqrt{\l \e^{-1}\,} \bm_{h},\, \bn_{h},\, \bar{a}_h,\, \bar{e}_h \}  
\]
be small enough (where in $3d$ the endpoint of type $\bl_h$ can be present only at scale $0$). Then
\[
||V^{(h);n}_{n^\txe_l,n^\txe_t}||  \leq C^n\, \bar{\h}^{n} \,n! \, \lft(\r_0 R_0^{-2}\rgt) C_0(P_{v_0};\e,\r_0, R_0)\,  \g^{h\, \d^{>}_{v_0}}
\]
for some constant $C$. Here $\e=\l \r_0 R_0^d$, the scaling dimensions in three and two dimensions are
\[
& \d^{3d,>}_v = \frac{5}{2} -\frac{3}{4}n_{v}^\txe -\frac{1}{2}n_{\dpr_\xx,v}^\txe -n_{\dpr_0,v}^\txe \non \\
& \d^{2d,>}_v = 2 -\frac{1}{2}n_{v}^\txe -\frac{1}{2}n_{\dpr_\xx,v}^\txe -n_{\dpr_0,v}^\txe
\]
and
\[
C_0(P_{v_0};\e,\r_0, R_0)= \lft(\l \e^{-1}\rgt)^{1-\frac{1}{2}(n^\txe_l+n^\txe_t)}=\lft(\r_0 R_0^d\rgt)^{-1+\frac{1}{2}(n^\txe_l+n^\txe_t)}
\] 
\end{bound}

\vskip 1cm

\subsubsection{\bf Low momenta region $h \leq \bh$ }

Note that with respect to the multiscale expansion for $\BV_h$, the trees contributing to the expansion of the effective potentials $\VV_h$ with $h\leq \bh$ do not have any endpoint with two external legs, since the local quadratic terms have all been included in the measure. Let assume that 
\[ \label{ass_prop}
&  1  \leq \, A_h  \leq 1+ o(1)  &   1-E_h  &\leq \, \e\,B_h \leq 1  \non \\
&   0 \leq E_h \leq 1  & 
 1 &\leq \frac{Z_h}{\e\,E_h}  \leq 1+ o(1)
\]
These bounds are proven in chapter \ref{flows} at all orders, using the beta function equation. Using \eqref{ass_prop} we can prove that $\DD_h(k) \geq \e Z_h \lft(k_0^2 + \e \kk^2 \rgt)$ and then
\[  \label{prop_last_scale}
&\bigl|\,g^{(h)}_{tt}(k)\,\bigr| \leq  \cst \, \e \g^{-2h} \non \\[6pt]
&\bigl|\,g^{(h)}_{lt}(k)\,\bigr| \leq  \cst \,\g^{-h} \non \\[6pt]
&\bigl|\,g^{(h)}_{ll}(k)\,\bigr| \leq \cst \, Z_h^{-1} 
\]
to be compared with the bounds for the unrenormalized propagator \eqref{prop_below}. We see that the only difference is on the factor $Z_h^{-1} $ appearing in the longitudinal--longitudinal correlation, instead of $\e^{-1}$. 

\subsubsection{\it Bounds for the ``last scale'' propagator}  

We have proved that the renormalized propagator $g^{(h)}(k)$, under the assumptions \eqref{ass_prop}, has the same behavior of the unrenormalized one (apart for an extra $Z_h^{-1}$ for the $g^{(h)}_{ll}$ propagator) for each $h > h^*$, with $h^*$ the lower boundary of the cutoff function in the effective potential $\WW_{h^*}(\r_0)$. The behavior of the ``last scale'' propagator, \ie $g^{(h^*)}(x)$ is a bit more subtle to analyze; in fact let consider the momenta $k$ such that $|k| \leq 2\g^{2h^*}/(\g^2+1)$ with $f_{h^*}(k) \neq 0$. In this region, which is the very left tail of the cutoff function $\c_{[h^*,0]}(k)$, all the cutoff functions $\c_{[h^*,j]}(k)$  contained in the definition of the wave function renormalization functions are right equal to $f_{h^*}(k)$.  Then we have:
\[ \label{Z_last_scale}
& Z_{h^*}(k) = Z_{h^*} f_{h^*}(k)  \non \\
& A_{h^*}(k) = 1 + \lft(A_{h^*} -1 \rgt)f_h(k) \non \\
& B_{h^*}(k) = B_{h^*} f_{h^*}(k)  \non \\
& E_{h^*}(k) = 1 + \lft(E_{h^*} -1 \rgt)f_h(k) 
\]
with $Z_{h^*}$, $A_{h^*}$, $B_{h^*}$ and $E_{h^*}$ satisfying \eqref{ass_prop}. If we are really around $|k|=\g^{2h^*}$ then $f_{h^*}(k)=1$ and we get $Q_{h^*}(k) = Q_{h^*}$ for each of the wave functions renormalization functions. The problem appears for $f_h(k)$ going to zero, since the propagator in momentum space $g^{(h)}_{ll}(k)$ and $g^{(h)}_{lt}(k)$ are not longer bounded. In fact we have 
\[
\frac{1}{\DD_{h^*}(k)} \leq \frac{1}{E_{h^*}\,f_{h^*}(k)\g^{2{h^*}}}
\]
while the numerators $p^{(h^*)}_{ll}(k)$ and  $p^{(h^*)}_{lt}(k)$ are finite. However, if we consider the behavior of the product of the propagator in the momenta space with the cutoff function, \ie
\[
\tl{g}^{(h^*)}_{\a \a'}(k) =f_{h^*}(k)\, g^{(h^*)}_{\a \a'}(k)
\]
then we may prove the following uniform bounds in $k$ for $\tl{g}^{(h^*)}_{\a \a'}(k)$:
 %\nota{Inserire prova in appendice}
\[  \label{prop_last_scale}
&\bigl|\,\tl{g}^{(h^*)}_{tt}(k)\,\bigr| \leq  \cst \, \e \g^{-2h^*} \non \\[6pt]
&\bigl|\,\tl{g}^{(h^*)}_{lt}(k)\,\bigr| \leq  \cst \,\g^{-h^*} \non \\[6pt]
&\bigl|\,\tl{g}^{(h^*)}_{ll}(k)\,\bigr| \leq \cst \, Z_{h^*}^{-1} 
\]
We see as the last scale renormalized propagators satisfy the same estimates than the renormalized propagators at scale $h >h^*$ on the support of $f_{h^*}(k)$. The fact that, in order to prove bounds which are also valid on the left tail of the cutoff function $\c_{[h^*,h]}(k)$, we need {\it one cutoff function for each propagator} will be crucial in the analysis of the local WIs, see sec.~\ref{WI_zero_momentum}. This is the reason why we have devoted so much attention on this point.  \\

\subsubsection{\it Statement of the $n!$ bounds for $\VV_h(\ps)$}

We can now write the counterpart of  \eqref{ren_bound} for the kernels of the effective potentials $\VV_h$. With respect to \eqref{ren_bound} we have only endpoints with at least three external legs and an extra factor $Z_{h_v}^{-1}$ must be added, for each longitudinal propagator at scale $h_v$. If $\G$ is a generic Feynman diagram contributing to $\LL V^{(h);n}_{n^\txe_l,n^\txe_t}$ the following bounds hold:
\[ \label{ren_bound2}
\bigl\|\Val(\G)\bigr\|  
 \, \leq & \,C^n   \lft(\r_0 R_0^{-2}\rgt) C_{3d}(P_v;\e,\r_0, R_0)\,  \g^{h\, \d^{3d,<}_{v_0}}\hskip -0.3cm\prod_{v\, \text{not e.p.} }  \g^{(h_v -h _{v'})\,(\d^{3d,<}_v -z^{3d,<}_v) } \non \\
& \prod_{v \text{\,e.p.}} \lft(\l_{h_v}\rgt) \lft(\e^{-1/2}\m_{h_v}\rgt) \lft( \e \n_{h_v}\rgt)  \,\prod_{n^{h_v}_{ll}} \lft(\e\,Z^{-1}_{h_v} \rgt)  \qquad  \qquad d=3  \non \\[6pt]
\bigl\|\Val(\G)\bigr\|    \,
\leq & \, C^n  \lft(\r_0 R_0^{-2}\rgt) C_{2d}(P_v;\e,\r_0, R_0)\,  \g^{h\, \d^{2d,<}_{v_0}}\hskip -0.3cm\prod_{v\, \text{not e.p.} }  \g^{(h_v -h _{v'})\,(\d^{2d,<}_v -z^{2d,<}_v) } \non \\
& \prod_{v \text{\,e.p.}} \lft(\l \l_{h_v}\rgt) \lft(\sqrt{\l \e^{-1}\,}\m_{h_v}\rgt) \lft( \e \n_{h_v}\rgt)\lft(  \l\,\l_{6,h_v}\rgt)  \,\prod_{n^{h_v}_{ll}} \lft(\e\,Z^{-1}_{h_v} \rgt)  \qquad d=2
\]
with $z^{3d,<}_v$ and $z^{2d,<}_v$ defined respectively in \eqref{zRen_3d} and \eqref{zRen_2d} and scaling dimensions equal to
\[
& \d^{3d,<}_v = 4 -2n_{l,v}^\txe -n_{t,v}^\txe -n_{\dpr,v}^\txe \non \\
& \d^{2d,<}_v = 3 -\frac{3}{2}n_{l,v}^\txe -\frac{1}{2}n_{t,v}^\txe -n_{\dpr,v}^\txe
\]
The effect of the renormalization of the measure on the bound for a generic Feynman diagram, under the assumptions \eqref{ass_prop}, consists only in the product of $Z_h^{-1}$'s over the longitudinal--longitudinal propagators. In order to bound this factor we consider some assumptions on $Z_h$ and $\m_h$, which are a posteriori justified by two global Ward identities relating $\l_h$, $\m_h$ and $Z_h$ among them, as described in sect.~\ref{role}. We assume that
\[ 
& Z_h \geq \cst \,\m_h   &  d=3  \label{Z_ass3} \\
& Z_h \geq \cst (\m_h)^2/\l_h    &  d=2  \label{Z_ass2}
\]
and also that in the two dimensional case
\[ \label{m_ass}
& c_1 \g^{\frac{h}{2}}\l_h \leq \m_h \leq c_2 \g^{\frac{h}{2}}\l_h & d=2
\]
We now note that to each propagator $g^{(h)}_{ll}(k)$ we can associate the two vertices of type $\m$ where the two dashed ``half--lines'' come from. In fact the vertices of type $\m$ are the only vertices carrying dashed lines. Due to the presence of the cutoff function $f_h(k)$ in $g^{(h)}_{ll}(k)$, at least one of these two $\m$ vertices must be at scale $h$. Then in the three dimensional case, using \eqref{Z_ass3}, we can always bound $Z_h^{-1}$ with the corresponding vertex $\m_h$, \ie 
\[  \label{bound_ll3}
& \frac{\e^{-1/2} \m_h}{ \e^{-1}Z_h}   \leq  \cst\,  \e^{1/2}   & d=3 
\]

\vskip 0.2cm 
In the three dimensional case the following bounds hold:

\begin{bound}  \label{thm_nfactorial2}
{\bf ($n!$--bounds $h \leq \bh$, $d=3$)} Let consider the quantity $||  V^{(h);n}_{n^\txe_l,n^\txe_t}||$ defined by the expression \eqref{sum1}, but with the sum over the trees running over the renormalized trees 
obtained by the localization procedure defined in~\eqref{loc_3d_b}.  Here   $||V^{(h);n}_{n^\txe_l,n^\txe_t}||$ is an estimate for the coefficient of order $n$ of the renormalized expansion for the kernel of the effective potential at scale $h \leq \bh$ with fixed number of external legs. Let $\h^{3d}_*$ defined by 
\[
& \h^{3d}_*= \sup_{h^*<h\leq \bh} \{ \l_{h},\, \e^{-1/2}\m_{h},\,  \e \n_{h} \}  
\]
be small enough. If \,  $A_h -1 \leq o(1)$, $0  \leq  B_h \leq \e^{-1}$ , $
  \frac{Z_h}{E_h} = \e(1+o(1))$ and $Z_h \geq \cst \m_h$, then 
\[
||V^{(h);n}_{n^\txe_l,n^\txe_t}||  \leq C^n\, (\h^{3d}_*)^{n-n_{ll}} \,n! \, \lft(\r_0 R_0^{-2}\rgt) C_{3d}(P_{v_0};\e,\r_0, R_0)\,  \g^{h\, \d^{<}_{v_0}}
\]
for some constant $C$. Here $\e=\l \r_0 R_0^d$ and
\[
C_{3d}(P_{v_0};\e,\r_0, R_0) & = \lft(\l \e^{-\frac{1}{2}}\rgt)^L  \e^{-2 +\frac{3}{2}n_l^\txe + \frac{1}{2}n_t^\txe +\frac{1}{2}n_{\dpr_\xx} +\frac{1}{2}n_{ll}} 
\]
depends only on the loop number, once the labels $\{P_{v_0}\}$ of the external legs are fixed.
\end{bound}   

\vskip 0.2cm

{\it The two dimensional case is much more subtle.} In fact, as we can see from \eqref{Z_ass2} in order to bound $Z^{-1}_h$ both the two $\m$ vertices must be  at scale $h$. In this case we have
\[ \label{assZ_2}
& \frac{\l \e^{-1}\,\m^2_h}{\e^{-1} Z_h} \leq \cst \l \l_h   & d=2
\]
where $\l \l_h$ is supposed to be bounded. On the opposite, if one of the two $\m$ vertices is at scale $k >h$, using \eqref{Z_ass2} and \eqref{m_ass}, we find
\[ \label{div}
\frac{\l \e^{-1}\,\m_h \m_k}{\e^{-1} Z_h} \leq \cst \l \l_h\,\frac{\m_k}{\m_h} \leq \cst \l \l_k \,\g^{-\frac{1}{2}(h-k)}
\]
where $\g^{-\frac{1}{2}(h-k)}=\g^{\frac{1}{2}|h-k|}$ tends to infinity as $|h-k|\arr +\io$. If one tries to compensate this bad factor by the short memory property, one immediately notes that there are some kernels with external dashed lines for which the short memory factor is not sufficient: for example the kernel $V^{(k)}_{14}$ has renormalized scaling dimension $-1/2$, so that, if the dashed line of this kernel is contracted with a dashed line at scale $h$ we can only extract a factor $\g^{\th(h-k)}$ with $\th <1/2$, which is not sufficient to compensate the divergence in \eqref{div}.

%-------------------------------------------------------
%
\feyn{
\begin{fmffile}{feyn-TESI/effective}
\unitlength = 1cm  
\def\myl#1{2.4cm} %larghezza parbox
\[
& \parbox{\myl}{\centering{	 	   
		\begin{fmfgraph*}(2,1.25)
			\fmfleft{i1,i2}
			\fmfright{o1,o2}
			\fmf{dashes}{i1,v}
                    \fmf{plain}{v,o2}
			\fmf{plain}{i2,v,o1}
			\fmfv{label=$V_{13}^{(h)}$,label.angle=90, label.dist=10pt}{v}
			\fmfdot{v}
		\end{fmfgraph*} \\
 		%$-1+\frac{1}{2}=-\frac{1}{2}$
$-\frac{1}{2}$
	}}
\quad \parbox{\myl}{\centering{
		\begin{fmfgraph*}(2,1.25)
			\fmfright{i1,i2}
			\fmfleft{o1}
			\fmf{dashes}{i1,v,i2}
			\fmf{plain, tension=1.5}{o1,v}
			\fmfdot{v}
			\fmfv{label=$V_{12}^{(h)}$,label.angle=100, label.dist=10pt}{v}
		\end{fmfgraph*}
            \\ %$-\frac{3}{2}+1=-\frac{1}{2}$
		$-\frac{1}{2}$
	}} \quad
	 \parbox{\myl}{\centering{
		\begin{fmfgraph*}(2,1.25)
			\fmfright{i1,i2}
			\fmfleft{o1}
			\fmf{plain}{i1,v,i2}
			\fmf{dashes, tension=1.5}{o1,v}
			\fmfdot{v}
			\fmfv{label=$V_{12}^{(h)}$,label.angle=100, label.dist=10pt}{v}
		\end{fmfgraph*}
            \\ $-1$ % $-\frac{3}{2}+\frac{1}{2}=-1$
	}}   \quad 
%\non \\[15pt] 	 & \hskip 1.5cm
 \parbox{\myl}{\centering{	
		\begin{fmfgraph*}(2 ,1.25)
			\fmfleft{i1}
			\fmfright{o1}
			\fmf{dashes}{i1,v1,o1}
			\fmfv{label={$V_{20}^{(h)}$},label.angle=90, label.dist=10pt}{v1}   
			\fmfdot{v1}
		\end{fmfgraph*}  
		\\ $-1$   %$-2+1=-1$ 
	}}  \quad
 \parbox{\myl}{\centering{	
		\begin{fmfgraph*}(2 ,1.25)    
			\fmfleft{i1}
			\fmfright{o1}
			\fmf{plain}{i1,v1}
			\fmf{dashes}{v1,o1}	
			\fmfv{label={$V_{11}^{(h)}$},label.angle=90, label.dist=10pt}{v1}   
			\fmfdot{v1}
		\end{fmfgraph*}  
		\\ $-\frac{3}{2}$     %$-\frac{3}{2}$ 
           }} 
 \non \\[18pt] & \hskip 2.8cm
 \parbox{\myl}{\centering{	 	   
		\begin{fmfgraph*}(2,1.25)
			\fmfleft{i1}
			\fmfright{o1,o2,o3,o4}
			\fmf{dashes, tension=2.5}{i1,v}
			\fmf{plain}{o2,v,o1}
			\fmf{plain}{o3,v,o4}
			\fmfv{label=$V_{14}^{(h)}$,label.angle=90, label.dist=10pt}{v}
			\fmfdot{v}
		\end{fmfgraph*} \\
 		$0$   %$-\frac{1}{2}+\frac{1}{2}=0$
	}} \quad
 \parbox{\myl}{\centering{	 	   
		\begin{fmfgraph*}(2,1.25)
			\fmfleft{i1,i2}
			\fmfright{o1,o2}
			\fmf{dashes}{i1,v,i2}
			\fmf{plain}{o2,v,o1}
			\fmfv{label=$V_{22}^{(h)}$,label.angle=90, label.dist=10pt}{v}
			\fmfdot{v}
		\end{fmfgraph*} \\
 		$0$   %$-1+1=0$
	}} \quad
\parbox{\myl}{\centering{	 	   
		\begin{fmfgraph*}(2,1.25)
			\fmfleft{i1}
			\fmfright{o1,o2}
			\fmf{dashes, tension=1.5}{i1,v}
			\fmf{dashes}{o2,v,o1}
			\fmfv{label=$V_{03}^{(h)}$,label.angle=90, label.dist=10pt}{v}
			\fmfdot{v}
		\end{fmfgraph*} \\
 		$0$    %$-\frac{3}{2}+\frac{3}{2}=0$
	}} \non
\]
\end{fmffile}
}{{\bf Effective renormalized scaling dimensions of the kernels such that $\widehat{\d}_v^{2d,<}>0$.} For each kernel the effective renormalized dimension is reported, which is equal to the renormalized scaling dimension $\widehat{D}_v^{\,2d,<}=\widehat{\d}_v^{\,2d,<}-z_v^{2d,<}$, see \eqref{zRen_2d} for a definition of $z_v^{2d,<}$. The kernels on the last line are effectively marginal.}{effective}
%------------------------------------------------

To take into account the presence of the ``dangerous'' diagrams where all the external dashed lines are contracted with other dashed lines at lower scales, we define an {\it effective scaling dimension $\widehat{\d}_{v}^{\,2d,<}$}  obtained from the scaling dimension $\d_v^{2d,<}=3 - \frac{3}{2}n^\txe_{l,v} -\frac{1}{2}n^\txe_{t,v} -n^\txe_{\dpr,v}$ by increasing of $1/2$ the dimension of each dashed line:
\[ \label{effective_dim}
\widehat{\d}_{v}^{\,2d,<}= 3 - n^\txe_{l,v} -\frac{1}{2}n^\txe_{t,v} -n^\txe_{\dpr,v}
\]
The effective dimension \eqref{effective_dim} is negative in the following cases: 
 $n_l^\txe \geq 4$;
 $n_l^\txe =3$ and $n_t^\txe >0$;
$n_l^\txe =2$ and $n_t^\txe >2$;
 $n_l^\txe =1$ and $n_t^\txe >4$. 
In the remaining cases, when  $\widehat{\d}_{v}^{\,2d,<}$ is not negative, we can wonder if  the {effective renormalized dimension}
$\widehat{D}_v^{\,2d,<}=\widehat{\d}_{v}^{\,2d,<} - z_v^{2d,<}$, with $z_v^{2d,<}$ given by \eqref{zRen_2d}, is negative, \ie if the bad factor coming from the contraction of the dashed lines is compensated by the renormalization procedure.

The renormalized effective dimensions $\widehat{D}_v^{\,2d,<}$  for the diagrams with positive effective dimension $\widehat{\d}_v^{\,2d,<}$  are shown in fig. \ref{effective}. There are five kernels for which the effect of renormalization (the factor $z_v^{2d,<}$) is sufficient to compensate the increasing of the dimensions due to the contraction of the external dashed legs. However three new  {\it effective marginal kernels} appears. One is then worried that  the renormalized perturbation theory is again affected by logarithmic divergences.  

The solution of this apparent problem stays in defining a localization procedure also on the new marginal kernels, even if they appear as irrelevant in a ``naive'' dimensional analysis. As usual, we define
\[  \label{L_extra}
 \LL^{2d}_{\,<}\, \hV_{14}^{(h)}(k_1,\ldots, k_4)  := \hV_{14}^{(h)}(0,\ldots,0) & = \r_0 R_0^{-2} \,\o_h \,F_{14} \non \\
 \LL^{2d}_{\,<}\, \hV_{22}^{(h)}(k_1,\,k_2,\,k_3)   :=  \hV_{22}^{(h)}(0,\ldots,0) & =\r_0 R_0^{-2} \g^{-h} \, \l'_h \,F_{22} \non \\
 \LL^{2d}_{\,<}\, \hV_{30}^{(h)}(k,\,p)    :=   \, \hV_{30}^{(h)}(0,\,0)   & = \r_0 R_0^{-2} \g^{-\frac{3}{2}h}\, \m'_h \,F_{30}
\]
with $F_{n_l^\txe, n_t^\txe}$ defined in \eqref{monomials} and $\o_h$, $\l'_h$ and $\m'_h$ defined with in front their ``naive'' scaling dimensions. In this way the renormalized part of the kernels $\hV_{14}^{(h)}$, $\hV_{22}^{(h)}$ and $\hV_{30}^{(h)}$ has negative effective dimension. Then for each irrelevant kernel we can extract a short memory factor equal to $1/2$ times the number of dashed legs outgoing from the kernel itself and control the bad factor  $Z_h^{-1}$ coming from the propagators $g^{(h)}_{ll}$.

The problem of controlling the bad factor $\g^{-\frac{1}{2}(h-k)}$, arising in the contraction of two external legs coming from different scales, ends in controlling the flow equations of the new couplings $\o_h$, $\l'_h$ and $\m'_h$. This may seem an hopeless perspective, since in the two dimensional case we already have eight running couplings to be controlled. The latter goal will be achievable thanks to the use of three additional global WIs (with respect the WIs which we will also use in the three dimensional case) allowing us to prove the following bounds on the new marginal couplings:
\[ \label{ass_new}
& \o_h \leq \cst \,\l_{6,h} \g^{\frac{h}{2}} \non \\
& \l'_h \leq \cst \,\l_{6,h} \g^{h} \non \\
& \m'_h \leq \cst \,\l_{6,h} \g^{\frac{3}{2}h} 
\]
The behavior in \eqref{ass_new} is just what we need in order to control the ``bad'' factors which may be generated by the contraction of the dashed external legs of the new vertices $\l'_h$, $\m'_h$ and $\o_h$. Note that, due to the property of the cutoff functions $f_j(k)$ the outgoing lines of the running coupling constants at scale $h$ may be only contracted at scale $h-1$ or $h+1$; then the dangerous situation expressed by eq. \eqref{div} never occurs.

By using \eqref{ass_new} and \eqref{m_ass}  we get
\[  \label{asym_2d_2}
   \bigl\|\Val(\G)\bigr\| \,&\leq \cst \r_0 R_0^{-2}\,  \l^{1-\frac{1}{2}(n_l^\txe+n_t^\txe)}\,\e^{-2 +\frac{3}{2}n_l^\txe+ \frac{1}{2}n_t^\txe + \frac{1}{2}n_\dpr^\txe}  \non \\
& \quad \g^{h\,\d_{v_0}^{2d,<}} \prod_{v \text{not e.p.}} \g^{(h_v - h_{v'})(\,\widehat{\d}_{v_0}^{\,2d,<} - z_v^{2d,<}\,)}
\non \\[6pt]
&  \prod_{v \text{e.p.}}\, \lft(\,\l \l_{h_v} \,\rgt)^{m_{4,v}+ m'_{4,v}+\frac{1}{2}(m_{3,v}+m'_{3,v}) +\frac{3}{2}m_{5,v} +2m_{6,v}}\,(\e \n_{h_v})^{m_{2,v}}\, \non \\[6pt]
&\,\lft(\,\frac{\l_{6,{h_v}}}{\e \l_{h_v}^2} \,\rgt)^{m_{6,v}+ m_{5,v} + m'_{4,v} +m'_{3,v}} \, \lft(\e^{-1}\g^{h_v} \l_{h_v}\rgt)^{\frac{1}{2}(m_{3,v} + m_{5,v} + 2m'_{4,v} +3m'_{3,v})- n^v_{ll}}
\]
where $\widehat{\d}_v^{\,2d,<}$ is defined in \eqref{effective_dim} and  $m_{5,v}$, $m'_{4,v}$,  $m'_{3,v}$ and $n^v_{ll}$ are respectively the number of vertices $\o_{h_v}$, $\l'_{h_v}$, $\m'_{h_v}$ or longitudinal propagator at scale $h_v$. Note that the exponent of $\l \l_h$ exactly reconstructs the number $L_v$ of loop at scale $h_v$, since
\[
L_v -1 +\frac{1}{2}(n_{l,v}^\txe +n_{t,v}^\txe)=m_{4,v} + m'_{4,v} + \frac{1}{2}(m_{3,v}+m'_{3,v}) +\frac{3}{2}m_{5,v} +2m_{6,v} 
\]
Besides, the factor on the last line of \eqref{asym_2d_2} may be bounded by one, under the assumption that $\l_h$ is finite; in fact this factor  is one when all the internal dashed lines at scale $v$ are contracted among them and goes to zero with $h \arr -\io$ if there are internal dashed lines at a certain scale $v$ which are not all contracted among them. 
Let now assume that 
\[
\frac{\l_{6,h}}{\e \l_h^2} \leq \cst 
\]
and let 
\[
& \h^{2d}_*= \sup_{h^*<h\leq \bh} \{ \l \l_h,\, \e\,\n_h \}  
\] 
The following bounds hold.

\begin{bound}  \label{thm_nfactorial3}
{\bf ($n!$--bounds $h \leq \bh$, d=2)} Let consider the quantity $||  V^{(h);n}_{n^\txe_l,n^\txe_t}||$, with $h \leq \bh$, defined by the expression \eqref{sum1} but with the sum over the trees running over the renormalized trees obtained by the localization procedure defined in~\eqref{loc_2d_b} and \eqref{L_extra}.  Here  $||V^{(h);n}_{n^\txe_l,n^\txe_t}||$ is an estimate for the coefficient of order $n$ of the renormalized expansion for the kernel of the effective potential at scale $h \leq \bh$ with fixed number of external legs. Let $\h^{2d}_*$ defined by 
\[
& \h^{2d}_*= \sup_{h^*<h\leq \bh} \{ \l \l_h,\,\l_{6,h}/ (\e \l_h^2),\, \e\,\n_h\}   \non
\]
be small enough. If \,  $A_h -1 \leq o(1)$, $0  \leq  B_h \leq \e^{-1}$, $
  \frac{Z_h}{E_h}  = \e(1+o(1))$ and the assumptions \eqref{Z_ass2}, \eqref{m_ass} and \eqref{ass_new} hold, then
\[
|| V^{(h);n}_{n^\txe_l,n^\txe_t}||  \leq C^n\, (\h^{2d}_*)^{n} \,n! \, \lft(\r_0 R_0^{-2}\rgt) C_{2d}(P_{v_0};\e,\r_0, R_0)\,  \g^{h\, \d^{2d,<}_{v_0}}
\]
for some constant $C$. Here $\e=\l \r_0 R_0^d$ and
\[ \label{nfactorial3}
C_{2d}(P_{v_0};\r_0, R_0) & =  \l^{1-\frac{1}{2}(n_l^\txe + n_t^\txe)}\,\e^{-2 +\frac{3}{2}n_l^\txe + \frac{1}{2}n_t^\txe +\frac{1}{2}n_{\dpr_\xx}}
\]
depends only on the loop number, once the labels $\{P_{v_0}\}$ of the external legs are fixed.
\end{bound}   

\hskip 0.5cm

Note that the factor $\d^{2d,<}_{v_0}$ in the bound \eqref{nfactorial3} is the ``naive'' {\it scaling dimension} of a kernel with $n_l$ external leg of type $l$ and $n_t$ external legs of type $t$. The effective scaling dimension we have discussed in the last pages only enters in the product over the branches of the trees, as shown in \eqref{asym_2d}. Since the effective renormalized scaling dimension $\widehat{D}_v^{\,2d,<}$ is at least $-1/2$, we still can extract a short memory factor $\g^{\th(h/k)}$, provided that $0<\th <1/2$.   
%Kernels with positive, vanishing or negative scaling dimensions are called {\it relevant}, {\it marginal} or {\it irrelevant} operators, respectively. 
 \\

{\it Remark.} The assumption  on $\bar{\h}$, $\h^{3d}_*$ and $\h^{2d}_*$ in the bounds \ref{thm_nfactorial0}, \ref{thm_nfactorial2} and \ref{thm_nfactorial3}, so as the assumptions on the behavior of the wave function renormalization constants, will be verified in the next chapter. Note that the action of the $\RR$ operator corresponds to extract the non divergent part from a divergent diagram. The remaining divergent parts of each kernel are enclosed in the definition of the running coupling constants, which become the unknown parameters of the problem. To prove that they exist some initial values $\rr_0$ such that the flows of the running coupling constants stays small corresponds to prove that there are some cancellations between the divergent parts of the perturbative expansion, such that the resummed theory is finite at the end. Note that all the analysis of this chapter is based on the fact that the scaling dimension $\d_{v_0}$, both in three and two dimensions, is independent on the number of endpoints of the tree $\t$ and there is only a finite number of diagrams with non negative dimensions, that is the model is {\it renormalizable}.  \\

\pagina

\feyn{
\begin{fmffile}{feyn-TESI/currents}
\unitlength = 1cm  
\def\myl#1{3cm} %larghezza parbox
\[
 \parbox{\myl}{\centering{	 	   
		\begin{fmfgraph*}(2,1.25)
			\fmfleft{i1}
			\fmfright{o1,o2}
			\fmf{wiggly,label=${J_0}$, tension=1.5}{i1,v}
			\fmf{plain, left=0.3}{v,o2}
                    \fmf{plain, right=0.3}{v,o1}
			\fmfdot{v}
		\end{fmfgraph*} \\
 		$m_{J_0}$
	}}
	\quad \parbox{\myl}{\centering{
		\begin{fmfgraph*}(2,1.25)
			\fmfleft{i1}
			\fmfright{o1,o2}
			\fmf{wiggly,label=${J_0}$, tension=1.5}{i1,v}
			\fmf{dashes, left=0.3}{v,o2}
                    \fmf{dashes, right=0.3}{v,o1}
			\fmfdot{v}
		\end{fmfgraph*}
            \\ $m'_{J_0}$
	}} 
	 \quad \parbox{\myl}{\centering{	
		\begin{fmfgraph*}(2 ,1.25)
			\fmfleft{i1}
                 	\fmfright{o1,o2}
                   \fmf{wiggly, label=${J_i}$, tension=1.5}{i1,v}
			\fmf{dashes, right=0.3}{v,o1}
                    \fmf{plain, left=0.3, label= $\dpr_i$, label.dist=0.1cm}{v,o2}   
			\fmfdot{v}
		\end{fmfgraph*}  
		\\ $m_{J_1}$ \non
	}} 
\]
\end{fmffile}
}{The new vertices appearing in the non renormalized tree expansion for the generating functional of density and current correlations. The external fields $J_\n$ are depicted with a wiggly line. The index $i$ assumes values between one and $d$, so the last picture have to be imagined as repeated $d$ times. Here $m_{J_1}=\sum_i m_{J_i}$. }{ext_fields}

\section{Generating functional of correlations} \label{sec:gen_fun}

In order to study the flow of the running coupling constants we need to introduce a new effective potential  $\WW(J_\n)$, depending on $(d+1)$ external fields $J_\n$ with \mbox{$\n=0,\ldots,d$}. The potential  $\WW(J_\n)$ is defined by the following functional integral
\[ \label{WW}
e^{|\L|\WW(\r_0,J_\n)}=\int P^\L_{Q_0,\c_0} (d\ps) e^{-\VV_0(\ps)
\,+\,\r_0\int dx\left[J_{x}^{0} \left( (\psi_{x}^{t})^2 +(\psi_{x}^{l})^2\right)
+J_i^{1} \cdot \left(\ps^{l} i\dpr_{x_i} \ps_{x}^{t} - \psi^{t} i\dpr_{x_i} \ps_{x}^{l} \right)\right]}   
\]
where repeated indexes are summed. Here $\VV_0(\ps)$ and  $P^\L_{Q_0,\c_0} (d\PS{\leq0})$ are defined in \eqref{BV_I} and \eqref{eq:prop_R0_cap3} respectively.  
The functional \eqref{WW} corresponds to the generating functional of density and current correlations, which are obtained by deriving $\WW(\r_0,J_{\n})$ twice with respect to one of the external fields. For example, by deriving twice $\WW(\r_0,J_\n)$ with respect to the external fields $J_0$, and then setting the external fields equal to zero, we obtain the density response function, as defined at the beginning of chapter \ref{model}.

%The multiscale integration uses to compute the partition function can be suitably modified in order to compute the the kernels of the effective potentials produced by the multiscale integration of $\WW_{[h^*,0]}(\f,J)$

As for the partition function, we introduce a reference potential $\WW_{h^*}(\r_0,J_\n)$ with an infrared cutoff at scale $h^*$. Then we proceed in a way analogous to the one described in the previous sections and iteratively integrate the fields $\PS{0},\ldots,\PS{h+1}$. After the integration of the first $|h|$ scales we are left with a functional integral similar to \eqref{free4} but now involving new terms depending on $J$.  

 The kernels of the effective potentials produced by the multiscale integration of $\WW_{[h^*,0]}(\r_0,J)$ can be represented as sums over trees, suitable modified with respect to the ones used for the partition function, which in turn can be evaluated as sums over Feynman diagrams. 

The non renormalized trees contributing to $\WW_{[h^*,0]}(\r_0,J)$ turn to have three new type of endpoints, $m_{J_0,v}$ $m'_{J_0,v}$ and $m_{J_1,v}$, as depicted in the picture \ref{ext_fields} and that we will be an extra label identifying an {\it external fields} of type $J_\n$ (here depicted as a wiggly line) from a bosonic line. With respect to the bosonic fields, the external fields are never contracted, so they always correspond to external lines. Analogously to \eqref{potential_xspace} the effective generating functional on
scale $h$ reads:
\[
 \WW_h(J_{\n},\psi^{\leq h}) = & \sum_{n_{l}+n_{t}+m_0+m_1\geq2}\int dx_{1}\ldots dx_{n_{l}}dy_{1}\ldots dy_{n_{t}}dz_{1}\ldots dz_{m_{0}}dw_{1}\ldots dw_{m_{1}} \non \\[6pt]
 &  \; V_{n_{l}n_{t};m_{0}m_{1}}^{(h)}(x_{1},\ldots x_{n_{l}};\, y_{1},\ldots,y_{n_{t}};\, z_{1}\ldots z_{m_{0}};\, w_{1}\ldots w_{m_{1}}) \non \\[6pt]
 & \qquad  \psi_{x_{1}}^{l}\ldots\psi_{x_{n_{l}}}^{l} \psi_{y_{1}}^{t} \ldots \psi_{y_{n_{t}}}^{t}J_{z_{1}}^{0} \ldots J_{z_{m_{0}}}^{0}J_{w_{1}}^{1}\ldots J_{w_{m}}^{1}
\]
where the subscript $m_0$ represent the total number of external field of type $J_0$ and the subscript $m_1$ the total number of external field of type $i$, with $i$ chosen between $1,\ldots,d$.
Let $W_{n_{l}n_{t};m_{0}m_{1}}^{(h);n }$ be the contribution to the kernel $W_{n_{l}n_{t};m_{0}m_{1}}^{(h)}$  due to the trees of order $n$. In the following we will get a bound for the quantity
\[ \label{kernel_ext_fields}
|| V_{n_{l}n_{t};m_{0}m_{1}}^{(h);n } || & = \frac{1}{\b L^d}\int d \ux\, d \uy\, d \uz \,d \uw\; |V_{n_{l}n_{t};m_{0}m_{1}}^{(h);n }(\ux, \uy, \uz,\uw)| \non \\
& = \sum_{\t \in \TT_{h,n}} 
\sum_{ \substack{\{P_{v}\} \\ n_l, n_t \text{fixed}}} 
\sum_{\G \in \GG(\t)} \lft|  \frac{1}{\b L^d} \int dx_{v_0} \Val(\G) \rgt|
\]
where
\begin{enumerate}[--]
\item $\ux=\{x_1, \ldots, x_{n_l} \}$, $\uy=\{y_1, \ldots, y_{n_t} \}$, $\uz=\{z_1, \ldots, z_{m_0} \}$, $\uw=\{w_1, \ldots, w_{m_1} \}$;
\item $\TT_{h,n}$ is the set of the trees described in section \ref{multiscale} but with endpoints which may also be of type $m_{J_0}$, $m'_{J_0}$ or $m_{J_1}$;
\item $\GG(\t)$ is the set of the connected Feynman diagrams compatible
with the tree $\t$, having $n_{l}$ and $n_{t}$ external lines of type $l$ or $t$ and $m_{0}$ external lines of type $J_{0}$ and $m_{1}$ external lines of type $J_{1}$. 
\end{enumerate}

\feyn{
\begin{fmffile}{feyn-TESI/loc_ext_a}
\unitlength = 1cm  
\def\myl#1{3cm}
\[
\parbox{\myl}{\centering{       %\m_h^{J_0}
 	\begin{fmfgraph*}(2,1)   
			\fmfright{i1,i2}
			\fmfleft{o1}
			\fmf{plain,left=0.3}{v,i2}
                    \fmf{plain,right=0.3}{v,i1}
			\fmf{wiggly,label=${J_0}$, tension=1.5}{o1,v}
			%\fmfv{label=$\mu_h^{J_0}$,label.angle=-90,label.dist=-0.4w}{v}  
                    \fmfdot{v}
		\end{fmfgraph*} \\[6pt]
              $\bm_h^{J_0}$  
		}} 
\parbox{\myl}{\centering{	
		 \begin{fmfgraph*}(2,1.25)    %Z_h^{J_0}
			\fmfright{i1}
			\fmfleft{o1}
			\fmf{dashes}{i1,v}
			\fmf{wiggly,label=${J_0}$}{o1,v}
			%\fmfv{label=$Z^{J_0}_h$,label.angle=-90, label.dist=-0.4w}{v}	
			\fmfdot{v}		
		\end{fmfgraph*} \\[6pt]
              $\g^{\frac{d}{4}}\bar{Z}_h^{J_0}$
		}} 
\parbox{\myl}{\centering{	
		 \begin{fmfgraph*}(2,1)    %m_h^{J_1}
			\fmfleft{i1}
			\fmfright{o1,o2}
			\fmf{plain, label=$\partial_i$, label.dist=0.005cm,right=0.3}{o2,v}
			\fmf{dashes, left=0.3}{o1,v}
			\fmf{wiggly,label=${J_1}$, tension=1.5}{i1,v}
			\fmfdot{v}		
		\end{fmfgraph*} \\[6pt]
             $\bm_h^{J_1}$ 
		}} 
 \parbox{\myl}{\centering{	
		 \begin{fmfgraph*}(2,1.25)    %E_h^{J_1}
			\fmfright{i1}
			\fmfleft{o1}
			\fmf{plain, label=$\dpr_i $,label.dist=0.05w}{i1,v}
			\fmf{wiggly,label=${J_1}$}{o1,v}
			%\fmfv{label=$E^{J_1}_h$,label.angle=-90, label.dist=-0.4w}{v}
			\fmfdot{v}			
		\end{fmfgraph*} \\[6pt]
		 $\g^{\frac{d}{4}}\bar{E}_h^{J_1}$ 
		}}  \non 
\]
\end{fmffile}
}{{\bf Vertices of the renormalized expansion for $h\geq \bh$ and $d=2,3$ with external fields.} The other diagrams with external fields are irrelevant due to our choice of the localization procedure. The fields $J_0$ and $J_1$ counts respectively as a plain line with derivative with respect to $x_0$ and a plain line with derivative with respect to $\xx$ in the dimensional estimate.}{loc_ext_a}

\subsection{Localization for  $\bh < h \leq 0$}

Introducing also the new vertices $m_{J_0,v}$, $m'_{J_0,v}$  and $m_{J_1,v}$ we get the following dimensional bound for the kernels of the unrenormalized theory:
\[ \label{unrJ}
||V_{n_{l}n_{t};m_{0}m_{1}}^{(h); n}|| & \leq 
\sum_{\t \in \TT_{h,n}} \,
\sum_{\substack{ \{P_{v}\} \\ n_{j}, m_i \:\text{fixed} }}\, \sum_{G \in \GG( P_v) }  
C^n (\l \e^{-1})^{(1-\frac{1}{2}(n_l^\txe+\n_t^\txe))}\,\e^{m_{J_0}-\frac{1}{2}m_{J_1}} \non \\
 & \quad \g^{h \d^{J,>}_{v_0}} \prod_{v\,\text{not e.p.}} 
\frac{1}{s_v!}\, \g^{(h_v - h_{v'})\d^{J,>}_v}  \non \\
& \quad \prod_{v\, \text{e.p.}} \bigl(\l \g^{\left(\frac{d}{2}-1\right) \c(m_{4,v}) } \bigr) \bigl(\sqrt{\l \e\,}\g^{\left(\frac{d}{4}-1 \right) \c(m_{3,v})} \bigr) \bigl(\g^{(\frac{d}{4}-1)\c(m_{2,v})} \bigr)
\]
where the scaling dimension -- with respect to \eqref{power_fin_up} -- is 
\[ \label{dimension_current_a}
\d^{J,>}_{v} = \frac{d}{2}+1-\frac{d}{4} n^\txe_{v} - \frac{1}{2} n^\txe_{\dpr_\xx, v} -n^\txe_{\dpr_0,v} -m_{J_{0},v} - \frac{1}{2} m_{J_1,v}
\]
and we are using the same definitions than section \ref{multiscale}. We note that the scaling dimension $\d^{<}_{v}$ in \eqref{dimension_current_a} depends on the number of vertices containing  $J_{0}$ and $J_{1}$ external fields. We can as usual move this contribution to the endpoints; we will get the factor
\[
\prod_{v\, \text{e.p.}}\g^{- \c(m_{J_{0},v}) -\frac{1}{2}\c(m_{J_{1},v})}
\]
However, due to the fact that the $J_0$ and $J_1$ fields are not contracted with other fields, there is an arbitrariness in the choice of their dimensions, in the sense that we can arbitrarily improve the internal dimension of the clusters containing these fields, at the expense of the external dimension. Let see in details what we mean in the case of the $J_0$ field. First of all we observe that we can choose to move to the endpoints only a part of the contribution $- \c(m_{J_{0},v})$, let say $ (\e_0 -1) \c(m_{J_{0},v})$, as follows:
\[
& \g^{-h \,m_{J_{0},v_0}}  \prod_{\substack{v\,\text{not e.p.}\\ v>v_{0}}} 
\g ^{- (h_{v}-h_{v'}) m_{J_{0},v}}  \non \\
& = \g^{-\e_0 h \,m_{J_{0},v_0}} \prod_{\substack{v\,\text{not e.p.}\\v>v_{0} }
} \g^{-\e_0 (h_{v} -h_{v'})m_{J_{0},v}} 
\prod_{v\,\text{e.p.}}\g^{\left(\e_0-1\right) h_{v} \c(m_{J_{0},v})}  \label{endpoint_J}
\]
Using the  identity
\be
n^\txe_{J_{0},v}  =  m_{J_{0},v} +m'_{J_{0},v}
\ee
we can also write
\[
\eqref{endpoint_J} & =  \g^{-\e_0 h\, (n^\txe_{J_{0},v} -m'_{J_{0},v})} \prod_{\substack{v\,\text{not e.p.}\\v>v_{0}}} 
\g^{-\e_0\,(h_{v}-h_{v'})(n^\txe_{J_{0},v} -m_{J_{0},v}^{'})} 
\prod_{v\,\text{ e.p.}} \g^{\left(\e_0 -1 \right) h_{v} \c(m_{J_{0},v})} \non \\
& =
\g^{- \e_0 h\, n^\txe_{J_{0},v} } \prod_{\substack{v\,\text{not e.p.}\\ v>v_0 }} 
\g^{-\e_0 (h_{v}-h_{v'})\, n^\txe_{J_{0},v} } 
\prod_{v\,\text{ e.p.}} \g^{\e_0 h_{v} \c(m'_{J_{0},v})} \, 
\g^{\left(\e_0 -1\right)\, h_{v} \c(m_{J_{0},v})}
\]
In this latter identity $\e_{0}$ appears as the dimension of each external $J_{0}$-field. As a consequence of this choice the vertex $m'_{J_{0}}$ has dimension
$-\e_{0}$, while the vertex $m_{J_{0}}$ has dimension $1-\e_0$. Then $m_{J_{0}}$ is relevant for $0\leq\e_{0}<1$, marginal for $\e_{0}=1$ and irrelevant
for $\e_{0}>1$. The arbitrariness in the choice of $\e_{0}$ does not gives any ambiguity in the calculation of the physical observables: in fact when we improve the external dimension, the gains from the short memory factors get worse; on the contrary if the external dimension gets worse, the gains from the short memory factor will improve. 

Following the same ideas we can choose the dimension of the $J_{1}$ external field; in this case we denote as $\e_1$ the part of the $-1/2$ factor we decide to move to the endpoints and use the identity
\[
n^\txe_{J_{1},v}=m_{J_{1},v}
\]
If we choose $\e_{0},\e_{1}\neq0$ we get an expression equal to \eqref{unrJ} with
a new scaling dimension
\[ \label{dimension_current_a}
\bar{\d}^{J,>}_{v} = \frac{d}{2}+1-\frac{d}{4} n^\txe_{v} - \frac{1}{2} n^\txe_{\dpr_\xx, v} -n^\txe_{\dpr_0,v}  - \e_{0}n^\txe_{J_{0},v} -\e_{1}n^\txe_{J_{1},v}
\]
and the following term in the product over the endpoints 
\[
%||V_{n_{l}n_{t};m_{0}m_{1}}^{(h); n}|| & \leq 
%\sum_{\t \in \TT_{h,n}} \,
%\sum_{\substack{ \{P_{v}\} \\ n^{j} \:\text{fixed} }}\, \sum_{G \in \GG( \t) }  
%\red{C^n \h^n} \g^{h \bar{\d}^>_{v_0}} \prod_{v\,\text{not e.p.}} 
%\frac{1}{s_v!}\, \g^{(h_v - h_{v'})\bar{\d}^{>}_v}  \non \\
%
%& \quad \prod_{v\, \text{e.p.}}\g^{h_{v} \left[\left(\frac{d}{2}-1\right) \c(m_{4,v}) + \left(\frac{d}{4}-1 \right) \c(m_{3,v})\right]} \non \\
%& \quad 
\prod_{v\, \text{e.p.}}\g^{h_{v} \lft[(\e_0-1)\c(m_{J_{0},v}) + \e_0 \c(m'_{J_{0},v}) + (\e_1-1/2) \c(m_{J_1,v}) \rgt]}
\]
We choose the localization procedure in such a way that the dimension of the field $J_0$ is equal to dimension of the derivative with respect to $\dpr_0$ and the dimension of the field $J_1$ equal to dimension of a spatial derivative $\dpr_i$, that is:
\[ \label{e0_e1_a}
 \e_0 = 1   \qquad  \e_1= \frac{1}{2} 
\]
both in three and two dimensions. This corresponds to the following localization procedure.  As far as the kernels with $m=0$ are concerned, we use the same definitions introduced for the effective potential, which will not be repeated here. For the terms with $m_{0},m_{1}\geq1$, we choose the following localization procedure, both for the three and two dimensional case:
\[ \label{loc_ext_fields_a}
&\LL_>\,\hV_{02;10}^{(h)}(k,p)  :=  \hV_{02;10}^{(h)}(0,0)\non \\
&\LL_>\,\hV_{11;01}^{(h)}(k,p)  :=  \sum_{i=1}^3 k_i \dpr_{k_i} \hV_{11;10}^{(h)}(k,p) \big|_{k=p=0}\non \\
&\LL_>\,\hV_{10;10}^{(h)}(k)  :=  \hV_{10;10}^{(h)}(0) \non \\
&\LL_>\, \hV_{01;01}^{(h)}(k) :=  %W_{01;01}^{(h) }(x) +
 \sum_{i=1}^3 k_i \, \dpr_{k_i} \hV_{01 ;01}^{(h)}(0)\big|_{k=0} \non \\[6pt]
&\LL_>\,\hV_{n_{l}n_{t};m_{0}m_{1}}^{(h)}(k_{1},\ldots,\, k_{n+m}) :=  0 \text{\qquad otherwise}
\]
The corresponding scaling dimension is:
\[ \label{bar_up}
\bar{\d}^{J,>}_{v}= &\, \frac{d}{2}+1-\frac{d}{4}n_{v} -\frac{1}{2}n_{\underline{\partial},v} -n_{\partial_{0},v}  -n_{J_{0},v} -\frac{1}{2} n_{J_1,v} 
\]
With this choice of $\LL_>$ all vertices with two or more external $J$-fields and a non trivial dependence on $\psi$ are irrelevant. Note that all the terms with one external $J_\n$ field different from $J_{0}\psi_{l}$, $J_{0}\dpr_0\psi_{t}$, $J_{0}(\psi_{t})^2$, $J_{1}\dpr_\xx \psi_{t}$ and  $J_{1}\dpr_{\xx}\psi_{t}\ps_l$ are zero by parity. With the previous definitions we obtain an expansion for $V_{n_{l}n_{t};m_{0}m_{1}}^{(h);n}$ in terms of renormalized \textit{Gallavotti-Nicol\`o} trees,  whose vertices are either on scale $0$, or  of the form $\bl_{h}$, $\bm_{h}$, $\bn_{h}$ and $\bm_{h}^{J_{0}}$, $\bm_{h}^{J_1}$, $\bar{Z}_{h}^{J_{0}}$ or $\bar{E}_{h}^{J_{1}}$, see fig. \ref{loc_ext_a} for a representation of the vertices with external fields.

%-----------------------------------------

\subsection{Localization for $h\leq\bh$}

%Introducing also the new vertices $m_{J_0,v}$ and $m_{J_1,v}$ we get the following dimensional bound for the unrenormalized theory. \\
The unrenormalized dimensional bound for the kernel 
$||V_{n_{l}n_{t};m_{0}m_{1}}^{(h); n}|| $ in the region $h \leq \bh$ is the 
%following
%\[ \label{unrJ_low}
%& ||V_{n_{l}n_{t};m_{0}m_{1}}^{(h); n}||  \leq 
%\sum_{\t \in \TT_{h,n}} \,
%\sum_{\substack{ \{P_{v}\} \\ n^{j} \:\text{fixed} }}\, \sum_{G \in \GG( \t) }  
%C^n C_d(P_v;\e,\r_0,R_0) \g^{h \d^{J,<}_{v_0}} \prod_{v\,\text{not e.p.}} 
%\frac{1}{s_v!}\, \g^{(h_v - h_{v'})\d^{J,<}_v}  \non \\
%& \quad \prod_{v\, \text{e.p.}}\g^{h_{v} \left[\left(d-3\right) \c(m_{4,v}) + \left(d-1\right)\c(m'_{4,v}) + \left(d+1\right) \c(m''_{4,v}) + \left(\frac{d-3}{2} \right) \c(m_{3,v}) + \left(\frac{d+1}{2} \right)\c(m'_{3,v})\right]}
%\]
same obtained for the free energy, see \eqref{fin_down3d} and \eqref{fin_down2d}, apart for a new scaling dimension, given by
\[ \label{dimension_current_b}
\d^{J,<}_{v} =  d+1-\left(\frac{d+1}{2}\right)n^\txe_{l,v}-\left(\frac{d-1}{2}\right)n^\txe_{t,v}-n^\txe_{\partial,v}-2m_{J_{0},v}
\]
and a different order in $\e$,  due to the the new vertices with external fields, as discussed at the end of appendix \ref{order_e}. 

As discussed in the previous section we can arbitrarily fix the dimension of the external field $J_0$ and $J_1$. In this case  the scaling dimension $\d^{J,<}_{v}$ in \eqref{dimension_current_b} does not depend explicitly on the number of vertices containing a $J_{1}$-external field. However using the identity
\[
n^\txe_{J_{1},v}=m_{J_{1},v}
\]
we can always multiply the external and internal dimensions by 
\[
1 & =  \g^{\e_{1}(m_{J_{1}}-n^\txe_{J_{1}}) } 
\prod_{\substack{v\,\text{not e.p.}\\ v>v_{0}}}
\g^{\e_{1}(m_{J_{1,v}}-n^\txe_{J_{1},v})} \\
 & = \g^{-\e_{1}n^\txe_{J_{1}}} \prod_{\substack{v\,\text{not e.p.}\\v>v_{0}}}
\g^{-\e_{1}n^\txe_{J_{1},v}} \prod_{m_{J_{1},v}}\g^{\e_{1} h_{v}}
\]
so that $m_{J_{1},v}$ becomes irrelevant once $\e_{1}\neq 0$.  As we choose $\e_{0},\e_{1}\neq0$ the scaling dimension becomes
\[ \label{dimension_current_b}
\bar{d}^{J,<}_{v} & = & d+1-\left(\frac{d+1}{2}\right)n^\txe_{l,v}-\left(\frac{d-1}{2}\right)n^\txe_{t,v} - n^\txe_{\partial,v} - \e_0 n^\txe_{J_{0},v} - \e_1 n^\txe_{J_1,v}
\]
and we get a new contribution coming from the product over the endpoints with one external field, that is
\[ 
%& ||V_{n_{l}n_{t};m_{0}m_{1}}^{(h); n}||  \leq 
%\sum_{\t \in \TT_{h,n}} \,
%\sum_{\substack{ \{P_{v}\} \\ n^{j} \:\text{fixed} }}\, \sum_{G \in \GG( \t) }  
%\red{C^n \h^n} \g^{h \bar{\d}_{v_0}} \prod_{v\,\text{not e.p.}} 
%\frac{1}{s_v!}\, \g^{(h_v - h_{v'})\bar{\d}^<_v}  \non \\
%& \quad \prod_{v\, \text{e.p.}}\g^{h_{v} \left[\left(d-3\right) \c(m_{4,v}) + \left(d-1\right)\c(m'_{4,v}) + \left(d+1\right) \c(m''_{4,v}) + \left(\frac{d-3}{2} \right) \c(m_{3,v}) + \left(\frac{d+1}{2} \right)\c(m'_{3,v})\right]} \non \\
%& 
\quad \prod_{v\, \text{e.p.}}\g^{h_{v} \lft[(\e_0-2)\c(m_{J_{0},v}) + \e_0 \c(m'_{J_{0},v}) + \e_1 \c(m_{J_1,v}) \rgt]}
\]
As a guide to fix the dimensions $\e_0$ and $\e_1$ in the lower region we will use some Ward identities (WIs) relating the vertices with external $J$ fields with the vertices without external fields, derived in section \ref{WI} and depicted in fig. \ref{formal_local_WI}. We choose the dimension of the external fields in such a way that the vertices related by the WIs have the same dimensions. This correspond to fix the dimension of the field $J_0$ equal to dimension of a $\ps^l$ field or, which is the same, of a $\ps^t$ field plus a derivative and the dimension of the field $J_1$ equal to dimension of a $\ps^t$ field plus a derivative, that is
\[ \label{e0_e1_b}
 \e_0 = \e_1= \frac{d+1}{2} = 
\begin{cases}
2 & d=3 \\[3pt]
\frac{3}{2} & d=2
\end{cases}
\]
The corresponding scaling dimension results to be:
\[  \label{bar_down}
 \bar{\d}^{\,<}_{v} = &\,  d+1-\left(\frac{d+1}{2}\right) \lft(n^\txe_{l,v}+ n^\txe_{J_{0},v} +n^\txe_{J_{1},v} \rgt) -\left(\frac{d-1}{2}\right)n^\txe_{t,v} -n^\txe_{\partial,v} 
\]
\feyn{
\begin{fmffile}{feyn-TESI/WI}
\unitlength = 1cm  
\def\myl#1{3cm}
	\begin{align*}
	\parbox{\myl}{\centering{	
		\begin{fmfgraph*}(2,1.25)
			\fmfleft{i}
			\fmfright{o}
			\fmf{dashes}{i,v}
			\fmf{plain,label=$\dpr_0$,label.dist=-0.2w}{v,o}
			\fmfv{label=$E_h$,label.angle=90,label.dist=-0.4w}{v}   
                   \bBall{v}
		\end{fmfgraph*}	
		}}
	& \simeq \quad
\parbox{\myl}{\centering{       %\m_h^{J_0}
 	\begin{fmfgraph*}(2,1)   
			\fmfright{i1,i2}
			\fmfleft{o1}
			\fmf{plain}{i1,v,i2}
			\fmf{wiggly,label=${J_0}$, tension=1.5}{o1,v}
			\fmfv{label=$\mu_h^{J_0}$,label.angle=-90,label.dist=-0.4w}{v}  
             \bBall{v}
		\end{fmfgraph*}
		}} 
		\\[12pt]
		\sqrt{2} \parbox{\myl}{\centering{	
		\begin{fmfgraph*}(2,1.25)
			\fmfleft{i}
			\fmfright{o}
			\fmf{plain,label=$\partial_0$, label.dist=-0.2w}{i,v}
			\fmf{plain,label=$\partial_0$,label.dist=-0.2w}{v,o}
			\fmfv{label=$B_h$,label.angle=90, label.dist=-0.4w}{v}   
                \bBall{v}
		\end{fmfgraph*}
	}} 
	& \simeq \quad
	\parbox{\myl}{\centering{	
		 \begin{fmfgraph*}(2,1.25)    %E_h^{J_0}
			\fmfright{i1}
			\fmfleft{o1}
			\fmf{plain, label=$\partial_0$, label.dist=0.051w}{i1,v}
			\fmf{wiggly,label=${J_0}$}{o1,v}
			\fmfv{label=$E^{J_0}_h$,label.angle=-90, label.dist=-0.4w}{v}	
                   \bBall{v}		
		\end{fmfgraph*}
		}} 
	\\[12pt]
	%----------------------------------------------------------------
 \sqrt{2} \parbox{\myl}{\centering{	
		\begin{fmfgraph*}(2,1.25)
			\fmfleft{i}
			\fmfright{o}
			\fmf{plain,label=$\dpr_i$,label.dist=-0.2w}{i,v}
			\fmf{plain,label=$\dpr_i$,label.dist=-0.2w}{v,o}
			\fmfv{label=$A_h$,label.angle=90,label.dist=-0.4w}{v}   
                   \bBall{v}
		\end{fmfgraph*}
		}} 
		 & \simeq \quad
		 \parbox{\myl}{\centering{	
		 \begin{fmfgraph*}(2,1.25)    %E_h^{J_1}
			\fmfright{i1}
			\fmfleft{o1}
			\fmf{plain, label=$\dpr_i$, label.dist=0.05w}{i1,v}
			\fmf{wiggly, label=$J_i$}{o1,v}
			\fmfv{label=$E^{J_i}_h$,label.angle=-90, label.dist=-0.4w}{v}		
                  \bBall{v}	
		\end{fmfgraph*}
		}} 
	\end{align*}
\end{fmffile}
}{{\bf Local Word identities derived in section \ref{WI}}. The ``$\simeq$'' symbol is used since the identities are not complete as depicted: the extra term coming from the presence of the cutoff function $\c_0$ in our effective model has been neglected in this picture.}{formal_local_WI}
The localization procedure for the terms with $m_{0},m_{1}\geq1$ which corresponds to this choice is the following, valid both for the three and two dimensional case:
\[ \label{loc_ext_fields_b}
&\LL_<\,\hV_{02;10}^{(h)}(k,p)  :=  \hV_{02;10}^{(h)}(0,0)\non \\
&\LL_<\,\hV_{10;10}^{(h)}(k)  :=  \hV_{10;10}^{(h)}(0) \non \\
&\LL_<\,\hV_{01;10}^{(h)}(k)  := % W_{10;10}^{(h)}(x) +
k_0 \, \dpr_{k_0} \hV_{01;10}^{(h)}(k)\big|_{k=0}\non \\
&\LL_<\, \hV_{01;01}^{(h)}(k) :=  %W_{01;01}^{(h) }(x) +
 \sum_{i=1}^3 k_i \,\dpr_{k_i}\hV_{01;01}^{(h)}(k)\big|_{k=0} \non \\
&\LL_<\,\hV_{n_{l}n_{t};m_{0}m_{1}}^{(h)}(k_{1},\ldots,\, k_{n+m}) :=  0 \text{\qquad otherwise}
\]
%with $W_{10;10}^{(h)}(x)$, $W_{01;01}^{(h)}(x)$, $\dpr_{\xx}W_{10;10}^{(h)}(x)$ and $\dpr_0 W_{01;01}^{(h)}(x)$  zero by parity. 
With this choice of $\LL_<$ the fields $J_0$ and $J_1$ turn to have just the dimensions \eqref{e0_e1_b}. Then all vertices with two or more external $J$-fields and a non trivial dependence on $\psi$ are irrelevant, the terms  $J_{0}\ps^{l}$, $J_{0}\dpr_{0}\ps^{t}$ and $J_{1}\dpr_\xx \ps^{t}$ are marginal and the term $J_{0}(\ps^{t})^{2}$ is marginal for $d=3$ and relevant for $d=2$, but with the same dimension than $\m_h$ by construction. The terms $J_{0}\psi_{t}$, $J_{0}\dpr_\xx \psi_{t}$, $J_{1}\psi_{t}$, $J_{1}\dpr_{0}\psi_{t}$, $J_{1}\psi_{l}$ and $J_{1}(\psi_{t})^2$
are zero by parity.  With these definitions  we obtain an expansion for $V_{n_{l}n_{t};m_{0}m_{1}}^{(h);n}$ in terms of renormalized \textit{Gallavotti-Nicol\`o} trees, whose vertices are either on scale $\bar{h}$, or are of the form $\l_{h}$,$\m_{h}$ , $ \n_{h}$, $\l_{6,h}$ in the two dimensional case, $\m_{h}^{J_{0}}$, $Z_{h}^{J_{0}}$, $E_{h}^{J_{0}}$ or $E_{h}^{J_{1}}$, see fig. \ref{loc_ext_3d_b}. \\

{\bf Additional running coupling constants for $d=2$.} Apart for the previous running coupling constants, in the two dimensional case it is also crucial to localize the following additional kernels, which correspond to $\m'_h$, $\l'_h$ and $\o_h$ once the $J_0$ field is substituted with a dashed line:
\[ \label{new_J_kernels}
&\LL_<\,\hV_{20;10}^{(h)}(k_1,k_2)  :=  \hV_{02;10}^{(h)}(0,0)\non \\
&\LL_<\,\hV_{12;10}^{(h)}(k_1,k_2,k_3)  :=  \hV_{10;10}^{(h)}(0,0,0) \non \\
&\LL_<\,\hV_{04;10}^{(h)}(k_1,k_2,k_3)  :=  \hV_{04;10}^{(h)}(0,0,0)
\]
At this stage of our analysis the localization seems not necessary, since the effective scaling dimensions of the last kernels are all negative, this property depending on the fact that the $J_0$ lines is never contracted. However one finds that the localization of these additional terms is necessary in order to study the flow of $\m_h^{J_0}$, as stressed in sec.~\ref{2dJ_flow}. The running coupling constants defined in \eqref{new_J_kernels} are graphically represented in fig.~\ref{ext_2d} pag.~\ref{ext_2d}. \\ 

{\it Remark.} The reader may wonder why we have not used the WIs in fig. \ref{formal_local_WI} to fix the dimensions of the external field also in the higher momentum region $\bh<h\leq 0$. A first reason is that in this case the dimension of the $\ps^l$ field and that of $\dpr_0 \ps^t$ are different and we can not satisfy both the requirements that the vertices appearing in the first and the second WIs in fig. \ref{formal_local_WI} have the same dimensions. A second more important reason is that we will use WIs are used only in the region $h \leq \bh$ and the choice to have WIs relating running coupling constants with the same dimensions will simplify the discussion.
\feyn{
\begin{fmffile}{feyn-TESI/loc_ext_3d_b}
\unitlength = 1cm  
\def\myl#1{2.6cm}
\[
\parbox{1.5cm}{\centering{
{\textcolor{white}{c}} \\[0.9cm] % \vskip{1.5 cm}
$d=3$  \\[12pt]
$d=2$
}}
\parbox{\myl}{\centering{       %\m_h^{J_0}
 	\begin{fmfgraph*}(2,1)   
			\fmfright{i1,i2}
			\fmfleft{o1}
			\fmf{plain, left=0.3}{v,i2}
                   \fmf{plain, right=0.3}{v,i1}
			\fmf{wiggly,label=${J_0}$, tension=1.5}{o1,v}
			%\fmfv{label=$\mu_h^{J_0}$,label.angle=-90,label.dist=-0.4w}{v}  
                    \fmfdot{v}
		\end{fmfgraph*} \\[6pt]
              $\mu_h^{J_0}$ \\[12pt]
              $\g^{\frac{h}{2}}\mu_h^{J_0}$ 
		}} 
\parbox{\myl}{\centering{	
		 \begin{fmfgraph*}(2,1.25)    %Z_h^{J_0}
			\fmfright{i1}
			\fmfleft{o1}
			\fmf{dashes}{i1,v}
			\fmf{wiggly,label=${J_0}$}{o1,v}
			%\fmfv{label=$Z^{J_0}_h$,label.angle=-90, label.dist=-0.4w}{v}	
			\fmfdot{v}		
		\end{fmfgraph*} \\[6pt]
              $Z_h^{J_0}$ \\[12pt]
              $Z_h^{J_0}$ 
		}} 
\parbox{\myl}{\centering{	
		 \begin{fmfgraph*}(2,1.25)    %E_h^{J_0}
			\fmfright{i1}
			\fmfleft{o1}
			\fmf{plain, label=$\partial_0$, label.dist=0.051w}{i1,v}
			\fmf{wiggly,label=${J_0}$}{o1,v}
			%\fmfv{label=$E^{J_0}_h$,label.angle=-90, label.dist=-0.4w}{v}	
			\fmfdot{v}		
		\end{fmfgraph*} \\[6pt]
             $E_h^{J_0}$ \\[12pt]
              $E_h^{J_0}$ 
		}} 
 \parbox{\myl}{\centering{	
		 \begin{fmfgraph*}(2,1.25)    %E_h^{J_1}
			\fmfright{i1}
			\fmfleft{o1}
			\fmf{plain, label=$\dpr_i$,label.dist=0.05w}{i1,v}
			\fmf{wiggly,label=${J_1}$}{o1,v}
			%\fmfv{label=$E^{J_1}_h$,label.angle=-90, label.dist=-0.4w}{v}
			\fmfdot{v}			
		\end{fmfgraph*} \\[6pt]
		 $E_h^{J_1}$ \\[12pt]
              $E_h^{J_1}$ 
		}}  \non 
\]
\end{fmffile}
}{{\bf Vertices of the renormalized expansion for $h<\bh$ and $d=2,3$ with external fields. } The only difference between the three and two dimensional case is in the different dimension of the $\m_h^{J_0}$ vertex, which reflects the fact that $\m_h$ is marginal for $d=3$ and relevant with dimension $1/2$ in two dimensions. The fields $J_0$ and $J_1$ counts as a dashed line or a plain line with derivative in the dimensional estimate.}{loc_ext_3d_b}

%-----------------------------------------------------------------------------

\pagina 

\subsubsection{On the $n!$~bounds}

The $n!$ bounds as the ones obtained for the kernels of the effective potentials and described by results \eqref{thm_nfactorial0}, \eqref{thm_nfactorial2} and \eqref{thm_nfactorial3} can be also proved for the kernels $|| V^{(h);n}_{n_l,n_t; m_0,m_1}||$ of the functional $\WW(\r_0,J)$. In particular we get exactly the same result, which will not repeated here, but with scaling dimensions $\bar{\d}^{J,>}_v$ and $\bar{\d}^{J,<}_v$, given by \eqref{bar_up} and \eqref{bar_down}, provided that also the product over the new vertices at each scale $h_v$ can be bounded.
% Moreover the dependence on $\e$ is slightly changed. 
\vskip 0.2cm

In particular, as discussed in appendix \ref{order_e}, in the region $\bh < h \leq 0$ the product over the vertices $\bm_h^{J_0}$ and $\bm_h^{J_1}$ gives an extra factor to the estimate for $|| V^{(h);n}_{n_l,n_t; m_0,m_1}||$ with respect to the one for the kernels of the effective potentials, which is
\[
\prod_{v \,\text{e.p.}}  \lft(\e \bm_{h_v}^{J_0} \rgt)^{m_{J_0,v}} \, \lft(\e^{-\frac{1}{2}} \bm_{h_v}^{J_1} \rgt)^{m_{J_1,v}}
\] 
We stress that $\bm_h^{J_0}$ and $\bm_h^{J_1}$ are the unique running coupling constants with external $J$ fields which can contribute to local diagrams, since $\bar{Z}_h^{J_0}$ and $\bar{E}_h^{J_1}$ have only one bosonic line. The prove that $\e \bm_{h_v}^{J_0}$ and $\e^{-\frac{1}{2}} \bm_{h_v}^{J_1} $ are bounded will given in sec.~\ref{transient}. 

\vskip 0.2cm
Regarding the region $h \leq \bh$, here the kernel $J_1 \ps^l\dpr_i \ps^t$ becomes irrelevant and we are left with an unique running coupling constant, $\m_h^{J_0}$. Then we only have to prove that 
\[
& \prod_{v \,\text{e.p.}}  \lft(\e \m_{h_v}^{J_0} \rgt)^{m_{J_0,v}} 
\] 
can be bounded. We advance here that by studying the flow equation of $\m_h^{J_0}$ we will find $\m_h^{J_0}\leq \e^{-1}\m_h$ both in $3d$ and $2d$, and then the product over endpoints of type $\m_h^{J_0}$ endpoints is equivalent to product over endpoints of type $\m_h$, apart for a different contribution to the order in $\e$, as we will see in more detail in the next chapter.  
%The contribution to the order in $\e$ coming from this coupling is $\e$ for $d=3$ and $\e^{\frac{1}{2}}$ for $d=2$. In particular, when we substitute a $\m_h$ vertex with external dashed leg with a $\m_h^{J_0}$ vertex, the order in $\e$ is changed by $\e^{-1}.$ 
\\

%\[
%& \bar{\h} =\max_{\bh<h\leq 0} \Bigl\{ \l \e^{-1}\bl_h,\, \sqrt{\l \e^{-1}\,}\bm_h,\, \bn_h,\, \bar{a}_h,\, \bar{e}_h,\, \e\bm^{J_0}_h \Bigr\}  \non \\
%& \h_*=\max_{h^*<h\leq \bh} \left\{ |\lambda_{\bh}|,\,|\mu_{\bh}|,|\mu_{\bh}^{' J_{0}}|,\,|\nu_{\bh}|,\,|\mu_{\bh}^{J_{0}}|,\,|\mu_{\bh}^{J_{1}}|\right\} 
%\]
%and
%\[
%C(\e,\r_0,R_0)= 
%\]

A  summary of the running coupling constants defined with the localization procedure can be found in table \ref{table:dimensions}. 

\pagina

\vskip 5cm
\begin{minipage}{11.5cm}
\begin{table}[H] 
\renewcommand{\arraystretch}{1.5}  
\begin{centering}
\begin{tabular}{|c||c|c||c|c|}  
\hline 
& \multicolumn{2}{|c||}{$\bh< h \leq 0$} &  \multicolumn{2}{|c|}{$h \leq \bar{h}$} \\
\hline 
 &  $d=3$ & $d=2$ & $d=3$ & $d=2$\\
\hline 
\hline 
$V^{(h)}_{06;00}$ & \multicolumn{2}{|c||}{ irrelevant}   & irr. & $\l_{6,h}$ \\
\hline 
$V^{(h)}_{04;00}$ & irr. & $\l_h$  & $\bl_h$ &  $\g^h \l_h$ \\
\hline 
$V^{(h)}_{12;00}$ & $\g^{h/4}\bm_h$ & $\g^{h/2}\bm_h$  & $\m_h$ & $\g^{h/2}\m_h$\\
\hline 
\hline 
$V^{(h)}_{02;00}$ &  \multicolumn{2}{|c||}{$\g^h \bn_h$}  & \multicolumn{2}{c|}{$\g^{2h} \n_h$}\\
\hline 
$V^{(h)}_{20;00}$ &  \multicolumn{2}{|c||}{$\g^h \bar{Z}_h$} & \multicolumn{2}{c|}{$Z_h$} \\
\hline 
$ V^{(h)}_{l\, \dpr_0 t;00}$ &  \multicolumn{2}{|c||}{ $\bar{A}_h$} & \multicolumn{2}{c|}{$A_h$} \\
\hline 
$V^{(h)}_{\dpr_0 t, \dpr_0 t;00}$ &  \multicolumn{2}{|c||}{ irrelevant} & \multicolumn{2}{c|}{$B_h$} \\
\hline 
$V^{(h)}_{\dpr_\xx t, \dpr_\xx t;00}$ &  \multicolumn{2}{|c||}{ $\bar{E}_h$} & \multicolumn{2}{c|}{$E_h$} \\
\hline 
\hline 
$V^{(h)}_{02;10}$ &   \multicolumn{2}{|c||}{ $\bm_h^{J_0}$} & $\m_h^{J_0}$ &    $\g^{h/2}\m^{J_0}_h$ \tabularnewline
\hline 
$V^{(h)}_{02;01}$ &  \multicolumn{2}{|c||}{ $\bm_h^{J_1}$} &   \multicolumn{2}{|c|}{irrelevant} \tabularnewline
\hline 
$W^{(h)}_{10;10}$ &  $ \g^{3/4}\bar{Z}^{J_0}_h$ & $ \g^{1/2}\bar{Z}^{J_0}_h$ & \multicolumn{2}{|c|}{$ Z^{J_0}_h$} \\
\hline 
$ W^{(h)}_{\dpr_0 t;10}$ & \multicolumn{2}{|c||}{irrelevant}  & \multicolumn{2}{|c|}{$ E^{J_0}_h$} \\
\hline 
$ W^{(h)}_{\dpr_\xx t;01}$ &  $ \g^{3/4}\bar{E}^{J_1}_h$ & $ \g^{1/2}\bar{E}^{J_1}_h$ & \multicolumn{2}{|c|}{$ E^{J_1}_h$} \\
\hline 
\end{tabular}
\par\end{centering} 
\vskip 1cm
\caption{Comparison between the dimension of the vertices of the renormalized expansion in the two regions $\bh<h\leq0$ and $h \leq \bh$ and for two and three spatial dimensions.}
 \label{table:dimensions}
\end{table}
\end{minipage}

%\end{document}

%\chapter{Flows of the running couplings}
%\input{intro-senza-sapclass} \input{intestazione-sap}\usepackage{showkeys} \begin{document} \tableofcontents 

\chapter{The flow of the running coupling constants} \label{flows}

In chap. \ref{multiscale} we proved that the effective model we introduced is order by order finite in the renormalized coupling constants -- $\a^{3d}_{i,h}=\{\l_h, \e^{-1/2}\m_h, \e \n_h\}$ in the three dimensional case and $\a^{2d}_{i,h}=\{\l \l_h, \l_{6,h}/(\l \l_h^2), \e\n_h\}$ in the two dimensional case -- provided these are bounded for all $h \leq0$ and that  
%\nota{La cosa che ha significato fisico e' $c^2(\l)=A_h/B_h$ per cui basta che $A_h$ sia costante e $B_h\neq0$}
\[ \label{ass1}
&  A_h -1 \leq o(1) \non \\
& \e \leq B_h  \leq \e^{-1} \non \\
& \frac{Z_h}{E_h} = \e \, (1+o(1)) 
\]
Under the previous assumptions the dominant behavior of the renormalized propagator as $k \arr 0$ is not changed with respect to the Bogoliubov propagator, except for the effective longitudinal propagator, which is renormalized by $Z_h^{-1}$, with $Z_h$ the longitudinal wave function renormalization constant. In order to control the effect of the renormalization of the longitudinal propagator in the three dimensional case we must have
\[  \label{ass2}
 |Z_h| \geq  \cst |\m_h|
\]
while in the two dimensional case the following assumptions are needed
\[   \label{ass3}
& Z_h \geq \cst \m^2_h/\l_h >0  \non \\
& 0 < c_1 \g^{\frac{h}{2}} \l_h \leq \m_h \leq c_2 \g^{\frac{h}{2}}\l_h \non \\[6pt]
& |\o_h| \leq \cst |\l_{6,h}| \g^{\frac{h}{2}} \non \\
& |\l'_h| \leq \cst |\l_{6,h}| \g^{h} \non \\
& |\m'_h| \leq \cst |\l_{6,h}| \g^{\frac{3}{2}h} 
\]
%
%This behavior depends on the fact that the two point functions $Z_h$ and $E_h$ tends to zero, with the ratio $Z_h/E_h$ close to $\e$, and that the effective transverse wave function renormalization constants $A_h$ and $B_h$ approach to constant values in the infrared limit $h \arr -\io$. 
%
Even if we do not prove the convergence of the series but only $n!$ bounds we expect that our series gives meaningful information only as long as the running coupling constants satisfy these conditions. 

In this chapter we will describe how to control the flow of the running coupling constants under the iteration of the RG transformation and to prove the previous assumptions. We will refer in particular to the asymptotic region $h\leq \bh$, while the discussion of the transient region is postponed in appendix \ref{app_transient}. \\

We will prove the following result.

{\result {\bf (Coupling constants flow $h \leq \bh$)} Both in $3d$ and $2d$ there are three exact relations (see eq.~\eqref{GWI_m}, \eqref{GWI_Z}, \eqref{LWI_E}) relating $\m_h$, $Z_h$ and $E_h$ to $\l_h$ and two  relations allowing to control the flows of $A_h$ and $B_h$, see eq.~\eqref{LWI_A} and \eqref{LWI_B}. Moreover in $2d$ there are three additional exact relations (see eq.~\eqref{WI_new1} ) relating $\o_h$, $\l'_h$ and $\m'_h$ to $\l_{6,h}$.

So the a priori seven running coupling constants of the three dimensional case are reduced to the study of two independent running couplings, $\l_h$ and $\n_h$, with $\n_h$ controlled by the choice of the chemical potential. Similarly the a priori eleven couplings of the two dimensional case are reduced to the study of three independent running couplings, $\l_h$, $\l_{6,h}$ and $\n_h$, with $\n_h$ fixed by the choice of the chemical potential.

Then, the flows of all the renormalized coupling constants are bounded (with explicit bounds) provided that the following conditions are verified:
\begin{description} 
\item[\quad \it (i) $3d$ case:] $\l_h$ is smaller than one for each $h$;
\item[\quad \it (ii) $2d$ case:]  the effective parameters $\l_h$ and $\l_{6,h}$ are such that  $\l \l_h$ and $\l_{6,h}/(\e \l^2_h)$ are smaller than one for each $h$. 
\end{description}

In the three dimensional case we can prove that $\l_h$ vanishes for $h \arr -\infty$, i.e the flow is asymptotically free, see sec.~\ref{lambda3}. In the two dimensional case a one--loop calculation suggests $\l \l_h$ and $\l_{6,h}/(\e \l^2_h)$ to be of order one in the infrared limit, as showed in sec. \ref{lambda6}. 
} 

\vskip 0.5cm

%The prove of this statement is of course beyond the possibility of a perturbative theory and then remains an open problem. \\

The second result that we will prove in this chapter is concerned with the renormalized propagator, which in the region $h \leq \bh$ is given by
\[ \label{RENg}
& g_{\a \a'}^{(h)}(x) = \frac{1}{(2\pi)^{d+1}} \int dk_0 d^{d}\kk\, f_h(k_0^2 +\e \kk^2) \,e^{-ikx} \, g_{\a \a'}^{(h)}(k)  \non \\[9pt]
& g_{\a \a'}^{(h)}(k)= (\r_0 R_0^{-2})^{-1}\, \frac{
\left(\begin{array}{cc}
 A_h \kk^2 + B_h k_0^2   \; & E_h k_{0}\\
- E_h k_{0}\; &  \kk^2 + \e Z_h 
\end{array}\right)
}{\lft(E_h^2 +B_h Z_h \rgt)k_0^2 + A_h Z_h \kk^2 + \kk^2 \lft(A_h \kk^2 + B_h k_0^2  \rgt) }
\]
The behavior of $A_h$, $B_h$ and $E_h$ is constrained by three local Ward Identities.  In particular we will prove that both in $3d$ and $2d$
\[ \label{wavefunc}
 &  E_h = \e^{-1}Z_h \lft(1 + O\bigl(\l \bigr)  \rgt) \non \\[3pt]
 & A_h = 1+  O\bigl( \l \,\e^{(d-2)/2} \bigr) \non \\[3pt]
 & B_h = \e^{-1} \lft( 1-E_h + O(\l) \rgt)   
\]
Moreover, an additional local WI allows to prove that for each $h \leq \bh$
\[  \label{propWI}
E^2_h+ Z_h B_h  = Z_h \e^{-1}\,\lft(1 + O(\l) \rgt)
\]
The identity \eqref{propWI} implies that, if we denote with $\DD_h(k)$ the denominator of $g_{\a\a'}^{(h)}(k)$ in \eqref{RENg}
\[
\DD_h(k) & \simeq Z_h \e^{-1} \lft( k_0^2 + \e \kk^2 \rgt) 
\]
with the symbol ``$\simeq$'' here and below referring to the dominant contribution both in the small parameter $\l$ and as $k \arr 0$. 
Using the relation between $Z_h$ and $\l_h$ we also have
\[
& Z_h = \frac{\e}{1+ \l \e^{\frac{1}{2}}|h -\bh|}  & d=3 \non \\[6pt]
& Z_h = 16\,\g^h \l_h \qquad 1 \leq \l_h < \l_* & d=2 
\]
with $\l_*=O(\l^{-1})$ the fixed point for $\l_h$.
%then the two point functions $Z_h$ and $E_h$ vanish in the deep infrared limit.  
%\blue{ With respect to the renormalized propagator, we can prove that Bogoliubov linear spectrum is exactly constrained by Ward identities and in particular that the two point functions $Z_h$ and $E_h$ vanish in the deep infrared limit with their ratio approaching to $\e$, while $A_h$ and $B_h$ tend to constants values. } \\
%
Let us now consider only the dominant behavior of the propagator as $k \arr 0$. As a consequence of the fact that $Z_h/E_h = \e$ at leading order in $\l$ and $E^2_h +B_h Z_h = Z_h B_{-\io}$
%that $A_h$ and $B_h$ tends to constants in the deep infrared 
we have
\[
g_{\a \a'}^{(h)}(k)  \simeq (\r_0 R^{2}_0)^{-1}\, \frac{
\left(\begin{array}{cc}
B_h (k_0^2 + \frac{A_h}{B_h} \kk^2)   \; & E_h k_{0}\\
-E_h k_{0}\; &  Z_h
\end{array}\right)
}{Z_h B_{-\io} \lft(k_0^2 + \frac{A_h}{B_{-\io}} \kk^2\rgt)} 
\]
In the deep infrared limit $h \arr -\io$, which corresponds to $k \arr 0$, both the longitudinal--longitudinal and the transverse--longitudinal propagator vanish, and
%\[
%g_{\a \a'}^{(-\io)}(k) \simeq  (\r_0 R^{2}_0)^{-1}\, 
 %\frac{
%\left(\begin{array}{cc}
%0  \; & 0\\
%0\; &  (B_{-\io})^{-1}
%\end{array}\right)}
%{k_0^2 + \frac{A_{-\io}}{B_{-\io}}\,\kk^2}
%\]
the interacting pair Schwinger function of the fields $\ps^{\pm}$, here denoted with $g_\e^{\s \s'}(k)$,  behaves as:
\[
g_{\l}^{-+}(k)\simeq-g_\l^{--}(k)\simeq -g_\l^{++}(k) \simeq \frac{\r_0}{2}\, g^{(-\io)}_{tt}(k) = \frac{R_0^2}{2}\,\frac{(B_{-\io})^{-1}}{k_0^2 +\tfrac{A_{-\io}}{B_{-\io}}\,\kk^2}
\]
Using $A_{-\io}=1+\cal{A}(\l)$ and $B_{-\io}=\e^{-1}(1+\cal{B}(\l))$ and coming back to the dimensional variables $k'_0=R_0^{-2}k_0$ and $\kk'^2=R_0^{-2}\kk^2$ we obtain
\[
g_{\l}^{-+}(k)\simeq -g_\l^{--}(k)\simeq -g_\l^{++}(k) \simeq \frac{\l \r_0 \hv(\bz)\,(1+{\cal{B}}(\l))^{-1}}{k_0^2 +c^2(\l)\,\kk^2}
\]
with $c(\l)$ the renormalized speed of sound, given by
%\[
%c^{2}(\e)=\frac{A_{-\io}}{B_{-\io}}\,R_0^{-2}=c_{B}^{2} \lft(1+\CC(\l)\rgt)
%\]
%and \[ \CC(\l) =\frac{1+\AA(\l)}{1+\cal{B}(\l)} \]
\[ \label{clambda}
c^2(\l) = R_0^{-2}\,\frac{A_{-\io}}{B_{-\io}} = c_B^2 \lft( 1 +O\bigl( \l \bigr) \rgt)
\]
with  $c_{B}=\sqrt{2\l \r_0 \hv(\bz)}$ the speed of sound predicted by Bogoliubov
approximation. It remains to be seen if the ultraviolet integration changes the magnitude in $\l$ of the correction term in \eqref{clambda}. The previous discussion is enclosed in the following result. 

\vskip 0.5cm

{ \result {\bf (Renormalized propagator)} Let us consider the  fields $\ph^\pm_{x}$ defined in sec. \ref{Interacting}, with $x=(x_0,\xx)$. Let us write $\ph^\pm_{x_0,\xx}=\x^\pm + \ps^\pm_{x_0,\xx}$, with 
%$\x^\pm = |\O|^{-1}\sum_\kk \ph^\pm_{0,\kk}$ 
$\x^\pm=\langle \ph^\pm_{x_0,\xx}\rangle_{x_0,\xx}$ the fields with covariance equal to the condensate density in the thermodynamic limit 
and $\ps^\pm_{x_0,\xx}$ the fluctuation fields with respect to the average $\x^\pm$. Order by order in perturbation theory  the interacting propagator of the fields $\ps^\pm_x$ for spatial dimensions $d=2,3$ is equal to
\[
g^{-+}_\e(x) \simeq \int d^d \kk dk_0 \, e^{-ikx} \frac{\GG(\l)}{k_0^2 + c^2(\l)\,\kk^2}
\]  
where ``$\simeq$'' means that we are considering the dominant singularity as $k \arr 0$ and 
\[
c(\l)= c^2_B \lft(1+\CC(\l)\rgt)
\]
with $c_B$ the speed of sound in the Bogoliubov approximation. 
Here $\GG(\l)$ and $\CC(\l)$ are expressed by series in the effective couplings with finite coefficients that admit $n!$--bounds at all orders. Moreover: (i) the first non trivial contribution to $\GG(\l)$ is $\l \r_0 \hv(\bz)$; (ii)  $\CC(\l)$ goes to zero as $\l \arr 0$. 

The interacting propagator shows the linear dispersion relation of quasi--particles predicted by Bogoliubov theory  with a renormalized speed of sound $c(\l)$.}

\vskip 0.5cm

The scheme of the present chapter is the following:
\begin{enumerate}[\it Sec. 3.1 \:]
\item In this section we list the global and local Ward identities necessary to control the flow in the region $h \leq \bh$. In order to conceptually separate how to prove and how to use these identities, their derivation has been postponed to the next chapter.  
%For what concerns the local WI in chap.~\ref{WI} we also study the correction terms to the formal WIs coming from the presence of cutoffs. Since these turn to be of higher order in the small parameter $\l$, so as we are interested in obtaining relations between the coupling constants we can use the formal WIs. This result is not trivial at all and is proven in chap.~\ref{WI}, where the complete expression of the local WIs is reported. 
The expression of the global and local WIs in the transient region can be found in appendix \ref{app_transient}.
\item We show how the flow of the relevant coupling $\n_h$ is controlled by the choice of the chemical potential, under some assumptions on the coupling constants that will be verified in the next sections.
\item The values of the running coupling constants at $h=\bh$ are reported. The detailed study of the flow of the coupling constants in the first region $\bh \leq h \leq 0$ can be found in appendix \ref{app_transient}. 
%The latter is easily obtained by dimensional estimates, see section \ref{transient}.  
%\item The flow in the lower region $h \leq \bh$ is obtained by using some global and local Ward identities, which reduce the number of independent running coupling to one, namely $\l_h$. 
\item In this section we discuss how to prove the results \eqref{wavefunc} for the renormalized wave functions $A_h$, $B_h$ and $Z_h$. 
The proof is based on the use of local WIs, which relate $A_h$, $B_h$ and $Z_h$ with the coupling constants with external fields $\m_h^{J_0}$, $E^{J_0}_h$ and $E_h^{J_1}$. 
%We describe how to control  (in the region $h \leq \bh$) the flow of the running coupling constants with external fields which appear in the local Ward identities. In particular 
The flows of $\m_h^{J_0}$ and $E^{J_0}_h$  are studied by a comparison with the flows equation for $\m_h$ and $E_h$ respectively, while $E_h^{J_1}$ is studied using a dimensional argument. The derivation of \eqref{propWI} is also discussed.

\item  We present the one--loop computations for the two particles effective interaction $\l_h$ in the $3d$ case and for the three and two particles effective interactions $\l_h$ and $\l_{6,h}$ in the $2d$ case. The first computation shows that in the three dimensional case $\l_h$  is asymptotically free in the infrared limit. In two dimensions, the leading order computations suggests $\l \l_h$ and $\l \l_{6,h}$ to have non trivial fixed points of order one. 
\end{enumerate}

%may be related to the flow of the two point function $Z_h$ through some Ward identities. 
% thanks to remarkable cancellations following from some Ward identities. 

%Assuming that the coupling constants to be bounded, the theory is renormalizable.

%{\it Remark.} In order to get estimates on the beta function and taking advantage from the short memory properties we will consider only ``short trees'', \ie trees where all the vertices and propagator are at the same scale. Two particularly interesting cases are those in which all the vertex are at scale $\bh$ or at scale $h\ll \bh$. The order in $\{ \e, \r_0, R_0\}$ in both these case are discussed in appendix \ref{order_e} and used many times along the chapter.  \\

We find convenient to remind here the scaling dimensions for the kernels of the effective potential both in the transient region than for $h \leq \bh$:
\[
d=3 \quad
\begin{cases}
\d^>_{v} & =  \frac{5}{2}-\frac{3}{4}n_{v}^\txe-n_{J_{0},v}^\txe -\frac{1}{2}n_{J_{1},v}^\txe  - n_{\dpr_{0},v}^\txe -\frac{1}{2}n_{\dpr_\xx,v}^\txe \\[6pt]
\d^<_{v} & =  4-2n_{l,v}^\txe -n_{t,v}^\txe -n_{\dpr,v}^\txe - 2n_{J_{0}}^\txe-2n_{J_{1}}^\txe
\end{cases} \\[12pt]
d=2 \quad
\begin{cases}
\d^>_{v} & =  2 -\frac{1}{2}n_{v}^\txe -n_{J_{0},v}^\txe -\frac{1}{2}n_{J_{1},v}^\txe  - n_{\dpr_{0},v}^\txe -\frac{1}{2}n_{\dpr_\xx,v}^\txe \\[6pt]
\d^<_{v} & =  3 -\frac{3}{2} n_{l,v}^\txe - \frac{1}{2}n_{t,v}^\txe -n_{\dpr,v}^\txe - \frac{3}{2}n_{J_{0}}^\txe-\frac{3}{2}n_{J_{1}}^\txe
\end{cases} 
\]

%The effect of the renormalization in the region $h \leq \bh$ is described by: \\

%\blue{AGGIUNGERE $z_v$}

\pagina

%However the task of combining our renormalization group approach with Ward identities is not trivial, since 

%
%
%
%--------------------------------------------------------------
\feyn{
\begin{fmffile}{feyn-TESI/flow}
 \unitlength = 0.8cm
\def\myl#1{2.2cm}
\begin{align*}
\parbox{\myl}{\centering{	 	   
		\begin{fmfgraph*}(2,1.5)
			\fmfleft{i1,i2}
			\fmfright{o1,o2}
			\fmf{plain}{i1,v,o2}
			\fmf{plain}{i2,v,o1}
			\fmfv{label=$\lambda_{h-1}$,label.angle=90, label.dist=-0.4w}{v}
			\bBall{v}
		\end{fmfgraph*}
	}} 
	& =
	\parbox{\myl}{\centering{	 	   
		\begin{fmfgraph*}(1.8,1.4)
			\fmfleft{i1,i2}
			\fmfright{o1,o2}
			\fmf{plain}{i1,v,o2}
			\fmf{plain}{i2,v,o1}
		\fmfv{label=$\l_{h}$, label.angle=90, label.dist=-0.5w}{v}
                \Ball{v}
		\end{fmfgraph*}
	}} 
	+
	\parbox{\myl}{\centering{
			\begin{fmfgraph*}(2.5,1.4) 
			\fmfleft{i1,i2}
			\fmfright{o1,o2}
			\fmf{plain}{i1,v1,i2}
			\fmf{plain,left=0.7, tension=0.4}{v1,v2}
			\fmf{plain,right=0.7, tension=0.4}{v1,v2}
			\fmf{plain}{o1,v2,o2}
			\Ball{v1,v2}
			\end{fmfgraph*}  
			}} 
			+
	\parbox{\myl}{\centering{
			\begin{fmfgraph*}(2.5,1.4) 
			\fmfleft{i1,i2}
			\fmfright{o1,o2}
			\fmf{plain}{i1,v1,i2}
			\fmf{plain,right=0.4, tension=0.6}{v1,v2}
			\fmf{plain, left=0.4, tension=0.6}{v1,v3}
			\fmf{dashes, left=0.4, tension=0.2}{v3,v2}
			\fmf{plain}{o1,v2}
			\fmf{plain}{o2,v3}
			\Ball{v1,v2,v3}
			\end{fmfgraph*}  
			}} 
			+
	\parbox{\myl}{\centering{
			\begin{fmfgraph*}(2.5,1.4) 
			\fmfleft{i1,i2}
			\fmfright{o1,o2}
			\fmf{plain}{i1,v1}
			\fmf{plain}{i2,v2}
                   \fmf{plain}{v3,o1}
			\fmf{plain}{v4,o2}
                   %linee centrali
                   \fmf{plain, tension=0.6}{v1,v3}
	             \fmf{plain, tension=0.6}{v2,v4}
			\fmf{dashes, tension=0.4}{v1,v2}
			\fmf{dashes, tension=0.4}{v3,v4}
			\Ball{v1,v2,v3,v4}
			\end{fmfgraph*}  
			}} + \; \ldots \\[12pt]
			%-------------------------------\mu_h----------------------------
	\parbox{\myl}{\centering{
		\begin{fmfgraph*}(2,1.5)
			\fmfright{i1,i2}
			\fmfleft{o1}
			\fmf{plain}{i1,v,i2}
			\fmf{dashes, tension=1.5}{o1,v}
			\fmfv{label=$\mu_{h-1}$,label.angle=100,label.dist=-0.4w}{v}    %shaded
                   \bBall{v}
		\end{fmfgraph*}
	}} 
	& = \parbox{\myl}{\centering{
		\begin{fmfgraph*}(2,1.4)
			\fmfright{i1,i2}
			\fmfleft{o1}
			\fmf{plain}{i1,v,i2}
			\fmf{dashes, tension=1.5}{o1,v}
			\Ball{v}
			\fmfv{label=$\mu_h$,label.angle=-100}{v}
		\end{fmfgraph*}
	}} +\;
	\parbox{\myl}{\centering{
			\begin{fmfgraph*}(2.8,1.4) 
			\fmfleft{i1}
			\fmfright{o1,o2}
			\fmf{dashes, tension=1.5}{i1,v1}
			\fmf{plain,left=0.7, tension=0.4}{v1,v2}
			\fmf{plain,right=0.7, tension=0.4}{v1,v2}
			\fmf{plain}{o1,v2,o2}
			\Ball{v1,v2}
			\end{fmfgraph*}  
			}} 
			+\;
	\parbox{\myl}{\centering{
			\begin{fmfgraph*}(2.8,1.4) 
			\fmfleft{i1}
			\fmfright{o1,o2}
			\fmf{dashes,tension=1.5}{i1,v1}
			\fmf{plain,right=0.4, tension=0.6}{v1,v2}
			\fmf{plain, left=0.4, tension=0.6}{v1,v3}
			\fmf{dashes, left=0.4, tension=0.2}{v3,v2}
			\fmf{plain}{o1,v2}
			\fmf{plain}{o2,v3}
			\Ball{v1,v2,v3}
			\end{fmfgraph*}  
			}} + \; \ldots \\[12pt]
			%----------------------------------------Zh--------------------
			\parbox{\myl}{\centering{
		\begin{fmfgraph*}(2,1.1)
			\fmfleft{i1}
			\fmfright{o1}
			\fmf{dashes}{i1,v1,o1}
			\fmfv{label={$Z_{h-1}$},label.angle=90,label.dist=-0.4w}{v1} 
			\bBall{v1}
		\end{fmfgraph*}
		}}
		& =
		\parbox{\myl}{\centering{
		\begin{fmfgraph*}(2,1)
			\fmfleft{i1}
			\fmfright{o1}
			\fmf{dashes}{i1,v1,o1}
			\fmfv{label={$Z_h$},label.angle=-90}{v1} 
			\Ball{v1}
		\end{fmfgraph*}
	}}			+\;
	\parbox{\myl}{\centering{
			\begin{fmfgraph*}(2.8,1.4) 
			\fmfleft{i1}
			\fmfright{o1}
			\fmf{dashes, tension=1.5}{i1,v1}
			\fmf{plain,left=0.7, tension=0.4}{v1,v2}
			\fmf{plain,right=0.7, tension=0.4}{v1,v2}
			\fmf{dashes}{o1,v2}
			\Ball{v1,v2}
			\end{fmfgraph*}  
			}}   \\[12pt]
			%----------------------------------------Eh--------------------
			\parbox{\myl}{\centering{
		\begin{fmfgraph*}(2,1.1)
			\fmfleft{i1}
			\fmfright{o1}
                   \fmf{plain}{i1,v1}
			\fmf{dashes, label=$\dpr_0$, label.dist=-0.25w}{v1,o1}
			\fmfv{label=$E_{h-1}$, label.angle=-90}{v1} 
			\bBall{v1}
		\end{fmfgraph*}
		}}
		& =
		\parbox{\myl}{\centering{
		\begin{fmfgraph*}(2,1)
			\fmfleft{i1}
			\fmfright{o1}
                   \fmf{plain}{i1,v1}
			\fmf{dashes}{v1,o1}
			\fmfv{label={$E_h$},label.angle=-90}{v1} 
			\Ball{v1}
		\end{fmfgraph*}
	}}			+\;
	\parbox{\myl}{\centering{
            \begin{fmfgraph*}(2.8,1.4) 
                   \fmfleft{i}
			\fmfright{o}
			\fmftop{t}
			\fmfbottom{b}
			\fmf{phantom, tension=4}{t,v3}    
			\fmf{phantom, tension=4}{b,v4}   
                   \fmf{plain, tension=1.5}{i,v1}			
			\fmf{plain,left=0.4, tension=1.2}{v1,v3}
			\fmf{plain,left=0.4}{v3,v2}			
			\fmf{dashes,right=0.4}{v1,v4}
			\fmf{plain,right=0.4}{v4,v2}
			\fmf{dashes, tension=1.5, label=$\dpr_0$, label.dist=0.05w}{o,v2}
			\Ball{v1,v2}			
			\end{fmfgraph*}  
			}}   \\[6pt]
	%----------------------------------------nu_h--------------------
			\parbox{\myl}{\centering{
		\begin{fmfgraph*}(2,1.1)
			\fmfleft{i1}
			\fmfright{o1}
			\fmf{plain}{i1,v1,o1}
			\fmfv{label={$\nu_{h-1}$},label.angle=90,label.dist=-0.4w}{v1} 
			\bBall{v1}
		\end{fmfgraph*}
		}}
		& =
		\parbox{\myl}{\centering{
		\begin{fmfgraph*}(2,1)
			\fmfleft{i1}
			\fmfright{o1}
			\fmf{plain}{i1,v,o1}
			\fmfv{label={$\nu_h$},label.angle=-90}{v} 
			\Ball{v}
		\end{fmfgraph*}
	}}			+\;
	\parbox{\myl}{\centering{
		\begin{fmfgraph*}(2,2)
			\fmfleft{i1}
			\fmfright{o1}
			\fmf{plain, tension=0.8}{i1,v}
			\fmf{plain}{v,v}
			\fmf{plain, tension=0.8}{v,o1} 
			\Ball{v}
		\end{fmfgraph*}
	}}			\;+\;
	\parbox{\myl}{\centering{
			\begin{fmfgraph*}(2.8,1.4) 
			\fmfleft{i1}
			\fmfright{o1}
			\fmf{plain, tension=1.5}{i1,v1}
			\fmf{plain,left=0.7, tension=0.4}{v1,v2}
			\fmf{dashes,right=0.7, tension=0.4}{v1,v2}
			\fmf{plain, tension=1.5}{o1,v2}
			\Ball{v1,v2}
			\end{fmfgraph*}  
			}}  	\;+\;
	\parbox{\myl}{\centering{
			\begin{fmfgraph*}(2.8,1.4) 
            \fmfleft{i}
			\fmfright{o}
			\fmftop{t}
			\fmfbottom{b}
			\fmf{phantom, tension=4}{t,v3}    
			\fmf{phantom, tension=4}{b,v4}   
            \fmf{plain, tension=1.5}{i,v1}			
			\fmf{plain,left=0.4, tension=1.2}{v1,v3}
			\fmf{dashes,left=0.4}{v3,v2}			
			\fmf{dashes,right=0.4}{v1,v4}
			\fmf{plain,right=0.4}{v4,v2}
			\fmf{plain, tension=1.5}{o,v2}
			\Ball{v1,v2}			
			\end{fmfgraph*}  
			}} 
\end{align*}
\end{fmffile}	
}{{\bf Flow equations for $d=3$, $h \leq \bh$, at leading order in $\e$.} The dots on the first two lines denote the fact that some of the leading order diagram have not been depicted; these are the three and four vertex diagrams obtained by a different contraction of the plain and dashed legs. The flow equation of $A_h$ and $B_h$ are obtained by deriving twice with respect to $\pp$ or $p_0$ the diagrams on the last line, with $p$ the external momentum. In the two dimensional case the six $t$ legged flow equation must to be added in addition to those depicted. }{flow_leading}

\section{Role of the symmetries} \label{role}

A crucial aspect of our approach is the subtle use of  Ward identities (WI), which reduce the number of independent running couplings. In fact, in order to prove the theory to be renormalizable, one need to control one relevant and six marginal couplings in three dimensions and three relevant and eight marginal couplings in two dimensions. Moreover the flow equations of the latter couplings are not trivial even at leading order, as one can already  see in the ``simpler'' three dimensional case, see in fig. \ref{flow_leading}. The Ward identities simplify (at least from a methodological point of view) the three dimensional case treatment and turn to be crucial for the control of the two dimensional theory.

The derivation of the WIs will be discussed in more details in the next chapter. These are related to the gauge invariance of the generating functional $\WW(\f, J_0, J_1)$, defined in section \ref{sec:gen_fun}, with $\f$, $J_0$ and $J_1$ external fields, under the transformation of the bosonic fields \mbox{$\psi_{x}^{\pm}\arr e^{i\th}\psi_{x}^{\pm}\,$}, with $\th$ depending on $x$ (local gauge invariance) or not (global gauge invariance). 

It is important to stress that while global WI are exact identities, the multiscale momentum decomposition breaks the local gauge invariance, which local WI are based on. However the corrections to the formal WI  may be studied within the RG approach, following a strategy proposed and developed in \cite{BM-luttinger}. 
%Since the corrections to local Ward identities turn to be of higher order in the small parameter $\e$ and do not change the conclusions we are interested in this chapter, we will neglect them and deal with the formal local WIs only.  
The discussion of the correction terms to the local WIs and of the techniques to control them has been postponed to the next chapter. 

A last remark regards the use of the symbol ``$\simeq$'' which in this chapter, as along all the thesis, refers to identity which are true at leading order in $\g^h$, as $h \arr -\io$.  \\

{\bf Global WIs.}
%The Global Ward identities encode relation between the kernels of the effective potential, in particular stating that the non vanishing vertices are the same of the original potential. In order to control the flows equation, the following identities are used:
The following two global Ward identitie give relations between the effective three and four body interactions $\l_h$ and $\m_h$ and the effective longitudinal wave function renormalization $Z_h$:

\begin{center}
\[  
 \hline \non \\[-12pt]
 h\leq\bh \qquad \qquad  d & =3  & d &=2  \non \\ \hline & \non \\[-9pt]
\m_h  & \simeq 4\sqrt{2} \,\l_h      &  \m_h  & \simeq4\sqrt{2}\,\g^{\frac{h}{2}}\l_h     
 \label{GWI_m}\\[3pt]
Z_h & \simeq 2\,\sqrt{2}\,\m_h + 2\,\g^{2h}\n_h     &  Z_h & \simeq 2\,\sqrt{2}\,\g^{\frac{h}{2}} \m_h + 2\,\g^{2h}\n_h    \label{GWI_Z}
\]
\vskip 0.3cm 
\end{center} 
The meaning of these global WI's at the lowest order in perturbation theory is shown in appendix \ref{B.WI-leading}.  Regarding the term $\g^{2h}\n_h$ it is subdominant in the small parameter $\e$ at the beginning of the lower region (\ie for $h \lesssim \bh$,see sec.~\ref{nu}) and becomes subdominant in $h$ as $|h|$ grows.
%For what concerns the identity \eqref{GWI_Z}, being
%\[
%& 0\leq \g^{2h}\n_h \leq  O(\l\,\e^\frac{3}{2}) & d=3 \non \\[6pt]
%& 0\leq \g^{2h}\n_h \leq  O(\l\,\e)   & d=2 
%\] 
%, the identity \eqref{GWI_m} implies that 
%\[
%&  0 \leq Z_h -2\,\sqrt{2}\m_h \leq O(\l\,\e^\frac{3}{2}) & d=3 \non \\[6pt]
%& 0 \leq Z_h -2\,\sqrt{2}\,\g^{\frac{h}{2}}\,\m_h \leq O(\l\,\e) & d=2
%\]
%\blue{ For example, for $d=3$ and at scale $h=\bh$ we have $Z_\bh = 2\sqrt{2}\m_\bh=\e$, while $\g^{2\bh}\n_\bh=O(\l \e^\frac{3}{2})$, which has the same order in $\l$ adn $\e$ of the one loop diagrams contributing to  $Z_\bh$ and $\m_\bh$.} Moreover as $h \arr \io$, since $\n_h$ is bounded as we will prove in the next section, the term $\g^{2h}\n_h$ goes to zero and the identity between $Z_h$ and $\m_h$ becomes exact.
Then in the asymptotic region $h \arr -\io$ we have 
\[
 16\,\l_h & \simeq 2\,\sqrt{2}\,\m_h \simeq Z_h & d=3 \non \\[6pt]
 16\,\g^h \,\l_h& \simeq 2\,\sqrt{2}\,\g^{\frac{h}{2}}\m_h \simeq Z_h & d=2
\]
Note that in the two dimensional case the identity relating $Z_h$ with $\l_h$ implies that if  $\l_{h}\arr\l_{*}$  then $Z_{h}\g^{-h}$  must be finite, \ie $Z_{h}= \cst\,\g^{h}$  where only the constant may depend on the value of the interaction.  The statements $Z_{h}\sim\g^{h}$  or $\l_{h}\arr \l_{*}$  are equivalent, and in fact we will study the simpler RG equation for $Z_h$ in spite of the one for $\l_h$. \\

In the two dimensional case three additional global WIs must been used to control the flow of the marginal couplings $\o_h$, $\l'_h$ and $\m'_h$, \ie
\[ \label{WI_new1}
 6\sqrt{2}\,\l_{6,h}  - \g^{-\frac{h}{2}}\o_h &\simeq  0 \non \\
\g^{-h}\l'_h-   24\,\l_{6,h} & \simeq 2 \g^{h}\l_h \non \\
 \g^{-\frac{3}{2}h}\m'_h -16\sqrt{2}\,\l_{6,h} & \simeq 4\sqrt{2}\,\g^h \l_h 
\]
By using these identities and being $0<\g^h\l_h\leq \e$ we see that the last three assumptions in \eqref{ass3} are satisfied. 
In the asymptotic region, assuming $\l_*$ to have a fixed point, the terms on the r.h.s. of the last two lines in \eqref{WI_new1} vanish and we get
\[
\o_h & \simeq  6\sqrt{2}\,\g^{\frac{h}{2}}\,\l_{6,h} \non \\
\l'_h & \simeq  24\,\g^{h}\,\l_{6,h} \non \\
\m'_h & \simeq  16\sqrt{2}\,\g^{\frac{3}{2}h}\,\l_{6,h} 
\]
Note that the behavior in $h$ of the coupling constants $\o_h$, $\l'_h$ and $\m'_h$ is exactly what occurs to compensate the divergences arising by the contraction of all the dashed legs of these kernels with other dashed legs at lower scale.  

\vskip 1cm

{\bf Local WIs.} Through the following local Ward identities the flows of the coupling constants with an external field $J_0$ or $J_1$ are used to get information on the behavior of the wave function renormalization constants $E_h$, $A_h$ and $B_h$ in the lower region $h \leq\bh$. 
\begin{center}
\[  
 \hline \non \\[-12pt] \hskip -4cm
 \text{Local WIs} \quad  h \leq \bh \hskip 3.5cm
d & =2,3    \non \\ \hline & \non \\[-9pt]
\g^{\frac{3-d}{2}\,h}\,2\,\m_h^{J_0} (1+O(\l)) & \simeq  E_h       \label{LWI_E}   \\[3pt]
E_h^{J_1} (1+O(\l\,\e)) & \simeq \sqrt{2}\,\lft(1-A_h  \rgt)     \label{LWI_A} \\[3pt]
E_h^{J_0}(1+O(\l)) & \simeq -   \sqrt{2}\,B_h          \label{LWI_B}  
%Z_h^{J_0}& = 2\sqrt{2}\lft(\e^{-1}Z_h -1\rgt)      \label{LWI_Z}
\]
\vskip 0.3cm
\end{center}
The three identities \eqref{LWI_A}, \eqref{LWI_A} and \eqref{LWI_E}, together with the global WIs allows to control the flow. 

In the three dimensional case the flow of the three coupling constants $\l_h$, $\m_h$ and $\n_h$ and the four single scale renormalization constants $A_{h},B_{h},E_{h},Z_{h}$ can be also studied by using only the two global WIs and by a leading order computation, as done by Benfatto in~\cite{benfatto}. On the contrary in the two dimensional case the use of the local WIs is essential to solve the flow.

%\blue{spostare al capitolo 3!} In the bound of a generic graph contributing to the beta function, two $\m_h$ vertices are essentially equivalent to one $\l_h$ vertex. Nella seconda regione l'ordine in $\e$ \`e indipendente dal numero di vertici e dipende solo dal numero di loop. nella prima regione, invece, dipende dal numero di vertici.

\pagina

\section{Choice of the chemical potential} \label{nu}

%In order to prove the main results 1. and 2. we shall proceed by induction: we will first assume that $\h^*= \max_{k \leq 0} \{ |\a_{i,k}|\}$ is small , that $E_h/Z_h = \e (1+o(1))$, $ 0 \leq A_h -1 \leq  \e $ and $0 \leq B_h \leq \e^{-1} $  for all $h \leq0$ 
%and a suitable constant $C>0$, and we will show that, by properly choosing the values of the chemical potential $\n_0$  we can control the flow of $\n_h$. Once this is done, we will show that the flows of the remaining coupling constants constants $\a_{i,h}$  effectively remain bounded and small for all $h \leq \bh$ and that the all the assumptions \eqref{ass1}, \eqref{ass2} and \eqref{ass3} are satisfied. This will be possible thanks to some remarkable relations among the coupling constants themselves and between coupling constants and wave function renormalization constants, following from some Ward identities.  \\

In this section we prove that, under the assumptions \eqref{ass1}, \eqref{ass2} and \eqref{ass3}, we can control the flow of $\n_h$ by properly choosing the values of the chemical potential $\n_0$ at the scale of the interacting potential. In particular we will prove that 
\[ \label{bound_nu}
\sup_{k\leq \bh}|\n_h| < \tl{\n}
\]
with $\tl{\n}$ a suitable constant and then the factor $\e \n_h$ appearing in the dimensional bounds \eqref{thm_nfactorial2} pag.~\pageref{thm_nfactorial2} and \eqref{thm_nfactorial3} pag.~\pageref{thm_nfactorial3} is small for each $h$. One can wonder if the choice $|\n_h| \leq \tl{\n}$ is compatible with the fact that the chemical potential must be fixed in such a way that the effective potential $\WW(\x)$ reaches its minimum for the fixed density $\r_0$. However in sec. \ref{ren_condition} we will show that the condition $\dpr_\x W(\x)=0$ corresponds to the requirement 
\[
\lim_{h \arr -\io} \g^{2h}\n_h
\] 
which is of course satisfied by \eqref{bound_nu}. 

We will proceed as follows: first we will show that we can fix $\n_\bh$ is such a way that $\n_{h}$ is bounded for each $h \leq \bh$. The value of $\n_\bh$ is then related to  $\bn_\bh$ by a continuity condition at $h=\bh$. Finally we show that $\n_\bh$ fixes the initial value $\bn_0$ of the chemical potential in our effective model.  Then the couplings $\bn_h$ and $\n_h$ satisfies the following system:
\[ \label{nu_system}
\begin{cases}
\; |\n_h| \leq \cst  & \text{boundary condition } \\[6pt]
\; \nu_{h-1}=\g^{2}\nu_{h}+\b_{h}^{\nu} & h\leq\bar{h}\\[6pt]
\; \g^{\bh}\bn_\bh=\g^{2\bh}\nu_{\bh} & \text{continuity condition at $\bh$ }\\[6pt]
\; \bn_{h-1}=\g\,\bn_{h}+\bar{\b}_{h}^{\n}  & h>\bh 
\end{cases}
\]
In the following we will also prove that  $\lim_{h\arr-\io}\nu_{h}= \n_*$ with $\n_*$ a constant. 

The beta functions $\b_{h}^{\n}$ and $\bar{\b}_{h}^{\n}$ can be expressed as sums over Gallavotti--Nicol\`o (GN) trees with at least two endpoints and with at least one endpoint on scale $h$, the reason being that the local part of the trees with only endpoints at scale lower than $h-1$ is zero by the support properties of the single--scale propagators that enter the definition of $\b^\#_h$.  Iterating the first equation in \eqref{nu_system} we obtain that for each $h\leq\bh$:
\[ \label{nu}
\g^{2h}\n_{h} = \g^{2\bh}\, \n_\bh + \sum_{j=h+1}^{\bh}\g^{2(j-1)}\, \b_{j}^{\nu}
\]
with $\nu_{\bar{h}}$ fixed by the condition $\n_{-\io}=\n_*$ 
\[ \label{nu_hbar}
\n_{\bh} = - \sum_{j\leq \bh} \g^{2(j-\bh-1)}\,\b_{j}^{\n}
\]
the beta function $\b^\n_j$ appearing in \eqref{nu_hbar} is a function of all the running coupling constants, included $\n_k$ with $k \geq j$. Substituting \eqref{nu_hbar} in \eqref{nu} we obtain the flow equation for $\n_h$ under the condition on $\n_{-\io}$:
\[   \label{nu_h}
\nu_{h}=- \sum_{j\leq h} \g^{2( j -h -1 )} \b^\n_j(\z_j,\n_j; \ldots; \z_\bh, \n_\bh) 
\]
where we have denoted with $\{\z_{k}\}$ the set of all the running coupling constants at scale $k$ except $\n_k$. We note that the solution to the flow equation induced by the beta function $\b^\n_j$ with initial condition $\lim_{h \arr -\io}\n_h=\n_*$ is a fixed point of the map $\bT: \MMM_K \arr \MMM_K$ defined as
\[ \label{mapT}
(\bT \un)_h = -\sum_{j\leq h} \g^{2( j -h -1 )} \b^\n_j(\z_j,\n_j; \ldots; \z_\bh, \n_\bh) 
\]
with $\MMM_{K,d}$ with $d=2,3$  the space of sequences $\un = \{\n_h\}_{h\leq\bh}$ such that 
\[
& |\n_h| \leq K |\l \e^{-\frac{1}{2}}|  & d=3 \non \\
& |\n_h| \leq K  & d=2
\]
 We shall think $\MMM_{K,d} $ as a Banach space with norm $||\cdot||$ where $||\un||=\sup_{k \leq \bh} |\n_h|$. The fact that, for $K$ sufficiently large, $\bT$ is a map from $\MMM_{K,d}$ to itself is a simple consequence of the bound $|\b^\n_h| \leq c_0 |\l \e^{-1/2}|$ in $3d$ and $|\b^\n_h| \leq c_0 $ in $2d$. For example, in the $3d$ case
\[
|(\bT \un)_h | \leq \sum_{j \leq h} \g^{2(j-h)} c_0 |\l \e^{-\frac{1}{2}}| \leq c'_0 |\l \e^{-\frac{1}{2}}|
\]
In order to prove that the map $\bT$ admits a fixed point it is sufficient to show that $\bT$ is a contraction on $\MMM_{K,d}$, \ie if $\un, \un' \in \MMM_{K,d}$
\[ \label{contraction}
|| T \un - T\un' || \leq L || \un - \un'||
\]
with $L<1$. First note that, we can always split the beta function $\b^\n_h$ as the sum of two parts: a first part $\b^\n_{h,1}$ containing the diagrams which not contain any vertex $\n_k$ and a second part containing at least a vertex $\n_k$:
\[ \label{Beta2}
\b^\n_h(\z_h,\n_h; \ldots; \z_\bh, \n_\bh) = \b^\n_{h,1}(\z_h; \ldots; \z_\bh) + \b^\n_{h,2}(\z_h,\n_h; \ldots; \z_\bh, \n_\bh)
\]
The one--loop diagrams contributing to $\b^\n_{h,1}$ are shown in the last line of fig.~\ref{flow_leading}; the one--loop diagrams contributing to $\b^\n_{h,2}$ are shown in fig.~\ref{chemical}.
Using the expansion in GN trees, the short memory factor and the fact that the trees contributing to $\b^\n_{h,2}$ have at least two endpoints, we find that
\[ \label{beta2_bound}
& |\b^\n_{h,2}(\n) - \b^\n_{h,2}(\n')| \leq C \, \l \e^{-\frac{1}{2}}|\z^*| |\n - \n' | & d=3  \non \\[6pt]
& |\b^\n_{h,2}(\n) - \b^\n_{h,2}(\n')| \leq C \,|\z^*| |\n - \n' | & d=2
\]
with
\[
|\z^*| = 
\begin{cases}
\max_{ h \leq \bh} \{ |\l_h|\} & d=3 \\[6pt]
\max_{ h \leq \bh} \{|\l \l_h|, |\l_{6,h}/(\l \l_h^2)| \} & d=2
\end{cases}
\]
The bounds \eqref{beta2_bound}, plugged back into eq.~\eqref{mapT}, imply \eqref{contraction}  provided that  $\l \e^{-\frac{1}{2}}|\z^*|$ is sufficiently small in three dimensions and $|\z^*|$ is sufficiently small in two dimension. Under the previous conditions $\bT$ has a unique fixed point $\un_*$ in $\MMM_{K,d}$ such that 
\[ \label{fixed_nu}
\n_{\bh} = -\sum_{j\leq \bh} \g^{2(j-\bh-1)}\,\b_{j}^{\n}(\un_*)
\]
In three dimensions we will prove that $0<\l \e^{-\frac{1}{2}}|\z^*|<\l \e^\frac{1}{2}$ and the discussion is completely consistent. In two dimensions the chemical potential is controlled provided that $\l \l_h$ and $\l_{6,h}/(\l \l_h^2)$ are sufficiently small, which is the same condition we need to have a consistent perturbation theory.  However, as shown in sec.~\ref{lambda6} a one--loop computation shows that $\l \l_h$ and $\l_{6,h}/(\l \l_h^2)$ admit fixed points of order one; the possibility of proving that these fixed points are sufficiently small that the perturbation theory makes sense is beyond reach of the methods that we employ in this thesis. \\

{\it Remark.} In \eqref{Beta2} we have neglected the fact that also the couplings $\{\z_k\}$ depends on $\n_k$. In fact this dependence is weak, due to the fact that 
$\n_k$ does not appear in the flow equation of $\z_k$ at leading order. However one can take into account this dependence, fixing $\z_h=\z_h(\un)$ as a function of $\un$,  see \eg~\cite[Appendix A5]{AshkinTeller}. \\

Let us consider the flow equation  \eqref{fixed_nu}.  The dominant diagrams contributing to $\b^n_j$ are the one--loop diagrams without $\n$ vertices. 
%The dominant diagrams contributing to $\n_\bh$ are the ones contributing to $\b_\bh^\n$, \ie the diagrams where all the vertices are at scale $\bh$. Using \eqref{3d_bh} and \eqref{2d_bh} we see that the leading order diagrams contributing to $\b^\#_\bh$ are the one loop diagrams without vertices $\n_\bh$ in $3d$ and without vertices $\n_\bh$ and $\l_{6,\bh}$ in $2d$. The leading order diagrams for $\n_\bh$ are shown in the last line of fig. \ref{flow_leading}. 
By using the short memory property, one can prove that it exists a constant $C$ such that:
\[
 & |\b^\n_j |\, \leq \begin{cases}
C\,\l\,\e^{-\frac{1}{2}}  & d=3  \\[3pt]
C\,\l\,\e^{-1}   & d=2  
\end{cases}
%\leq C\, \l\,\e^{-\frac{1}{2}}  & d=3 \non \\[3pt]
% & |\b^\n_\bh | \leq C\,\l\,\e^{-1}   & d=2  
\]
and then
\[ 
 |\g^{2\bh}\,\n_{\bh} | \leq 
\begin{cases}
\cst \,\l\,\e^{\frac{3}{2}} \qquad & d=3 \\[6pt]    \label{nu_bar}
\cst \,\l\, \e \qquad  & d=2
\end{cases}  
\]

% if one wants to keep in the flow equations of $\l_h$ and $\n_h$ only the leading terms, then one has to consider the one loop graphs {\it without $\n_h$} vertices. Then ; hence the growth of $\n_h$ is controlled in a simple way by the right choice of $\bn_0$, if one can control the flow of $\l_h$. }

%where $\l_j(\un)$ is the solution of
%\[
%\l_{h-1}=\l_h + \b^\l_h(\l_h,\n_h; \ldots; \l_\bh, \n_\bh)
%\]
%obtained as a function of the {\it parameter} $\un$. 
%If we find a fixed point $\un_*$ of \eqref{mapT}, the beta function for $\l_h$ and $\n_h$ will be simultaneously solved by $\ul(\un_*)$ and $\un_*$ respectively and the solution will have the desired smallness properties for $\l_h$ and $\n_h$.  

\feyn{
\begin{fmffile}{feyn-TESI/chemical}
 \unitlength = 0.8cm
\def\myl#1{2.5cm}
\begin{align*}
	\b^\n_{h,2} \;= \parbox{\myl}{\centering{
		\begin{fmfgraph*}(2.2,3)
			\fmfleft{i1}
			\fmfright{o1}
                    \fmftop{t}
                     \fmfbottom{b}
                    \fmf{phantom, tension=1}{t,v2}
                    \fmf{phantom, tension=0.3}{b,v}
			\fmf{plain, tension=1}{i1,v,o1}
			\fmf{plain, right=0.8, tension=0.2}{v,v2}
                   \fmf{plain, right=0.8, tension=0.2}{v2,v}
			\Ball{v,v2}
		\end{fmfgraph*}
	}}			\;+\;
	\parbox{\myl}{\centering{
			\begin{fmfgraph*}(2.8,3) 
            \fmfleft{i}
			\fmfright{o}
			\fmftop{t}
			\fmfbottom{b}
			\fmf{phantom, tension=1}{t,v3}    
			\fmf{phantom, tension=1}{b,v4}   
            \fmf{plain, tension=2}{i,v1}			
			\fmf{plain,left=0.4, tension=1.2}{v1,v3}
			\fmf{plain,left=0.4}{v3,v2}			
			\fmf{dashes,right=0.4}{v1,v4}
			\fmf{dashes,right=0.4}{v4,v2}
			\fmf{plain, tension=2}{o,v2}
			\Ball{v1,v2,v3}			
			\end{fmfgraph*}  
			}}  	\;+\;
	\parbox{\myl}{\centering{
			\begin{fmfgraph*}(2.8,1.4) 
            \fmfleft{i}
			\fmfright{o}
			\fmftop{t1,t2}
			\fmfbottom{b}
			\fmf{phantom, tension=4}{t1,v3}    
                   \fmf{phantom, tension=4}{t2,v33}  
			\fmf{phantom, tension=4}{b,v4}   
            \fmf{plain, tension=2}{i,v1}			
			\fmf{plain,left=0.4, tension=1}{v1,v3}
			\fmf{plain, tension=10, left=0.2}{v3,v33}			
                   \fmf{dashes,left=0.3, tension=0.8}{v33,v2}
			\fmf{dashes,right=0.4}{v1,v4}
			\fmf{plain,right=0.3}{v4,v2}
			\fmf{plain, tension=2}{o,v2}
			\Ball{v1,v2,v3}		
			\end{fmfgraph*}  
			}}
\end{align*}   \vskip -0.5cm
\end{fmffile}
}{Leading order diagrams contributing to the beta function $\b^\n_{h,2}$ defined in \eqref{Beta2}. For each additional $\n_k$ vertex the estimate \eqref{beta2_bound} is improved of $\e|\n_k|$.}{chemical}	

\subsubsection{Initial value of the chemical potential}

The flow in the higher momenta region is obtained by iterating the first equation in \eqref{nu_system}:
\[
\bn_{h}  =\g^{-h}\,\n_{0}+\sum_{j=h+1}^{0}\g^{\,j-h-1}\,\bar{\b}_{j}^{\n}
\]
with $\n_0$ fixed by the value of $\bn_\bh$:
\[ \label{bar_nu_0}
\n_0=\g^{\bh}\,\bn_\bh - \sum_{j = \bh +1}^0 \g^{\,j-1}\,\bar{\b}_{j}^{\n} 
\]
The leading order contributions to the beta function $\bar{\b}^\n_j$ are the tadpole diagram obtained by the contraction of two legs of $\bl_h$ and the diagrams with two $\bm_h$ vertices, see \eqref{bar_C}, since two $\bm_h$ vertices count as a $\l_h$ vertex. In three dimension the tadpole exists only at scale $h=0$, being the four--legged diagram irrelevant; moreover $\bm_h \leq \e\g^{-\frac{h}{4}}$. In two dimensions both the tadpole and the diagrams with two vertices $\m_h$ are bounded by $\cst \l$, since $\bl_{h}=\e$  and $\m_h\leq \sqrt{\e}$. We then have
\[ \label{beta_bar_nu_2d}
|\bar{\b}^\n_j | \leq 
\begin{cases}
c_1 \l \d_{j0} + c_2\l \e \g^{-\frac{j}{2}}  &  d=3  \\
c_3 \, \l & d=3  
\end{cases}
\]
with $c_1$, $c_2$ and $c_3$ suitable constants. Using the latter estimates on the beta function and the estimates \eqref{nu_bar} on $\n_\bh$ we get
\[ 
|\bn_0| & \leq C \l\, \bigl(1 + c'_1\,\e +c'_2\,\e^{\frac{3}{2}} \bigr) & d=3 \non \\
|\bn_0| & \leq C \l\, \bigl(1 + O(\e)  \bigr) & d=2
\]

\vskip 0.2cm

{\it Remark.} We remind that the leading order value of the chemical potential in the Bogoliubov approximation is $\m_B=\l \r_0 \hv(\bz)=\e R_0^{-2}/2$. The correction to this value in our effective model is given by $\n=\n_0\,R_0^{-2}/2$, where $R_0^{-2}$ restores the exact physical dimensions, being $\n_0$ adimensional. In the three dimensional case, the terms of order $\l \e$ and $\l \e^{\frac{3}{2}}$ we find are of the same order of the expected corrections to the leading order value, see \eqref{mu3d}.
%the term of order $\l\,\e$ both in the three and two dimensional case is the expected corrections to the leading order value, \ie $O(\l^2 \r_0 R_0)$ in $d=3$ and $O(\l^2 \r_0)$ in $d=2$.  
However as already stressed in chap.~\ref{model} the coefficients of these terms do not represent the values of the first corrections to $\m_B$ in the Hamiltonian model, due to the presence of the ultraviolet momentum cutoff on the $\kk$ variable. As an example of this fact, the term of order $\l$ in $\bn_0$, which is not compatible with Bogoliubov prediction, comes from the ultraviolet cutoff in the $k_0$ variable, as discussed at the end of chap.~\ref{model} . \\

\pagina

\section{Values of the running coupling constants at $h=\bh$}

We report below the initial values of the running coupling constants at scale $\bh$, for the three and two dimensional case, with their lower scaling dimensions. The estimates in the following table are discussed in appendix \ref{app_transient}. 

\vskip 0.5cm

\begin{center}

\begin{table}[H]
\renewcommand{\arraystretch}{1.5}
\noindent \begin{centering} 
\begin{tabular}{|c||c|c||c|c|}
\hline 
\multicolumn{5}{|c|}{RCC at $h =\bh$}\\ 
\hline 
\multicolumn{5}{c}{}\\
\hline
 &  $\d^{3d}_<$ & $r_\bh^{3d}$ & $\d^{2d}_<$  &  $r_\bh^{2d}$ \\
\hline 
\hline 
$\l_{6,\bh}$ &   & & $0$ & $O\bigl(\l \e \bigr)$ \\
\hline 
$\l_{\bh}$ &  $0$ & $\tfrac{1}{16}\,\e\,(1 + O\bigl(\l \e^{\frac{1}{2}})\bigr) $ & $1$ & $\tfrac{1}{16}\lft(1 + O\bigl(\l \bigr)\rgt)$ \\
\hline 
$\m_{\bh}$ &   $0$ & $\tfrac{\sqrt{2}}{4}\,\e\,(1 + O\bigl(\l \e^{\frac{1}{2}})\bigr)$ & $ 1/2$ & $\frac{\sqrt{2}}{4}\,\sqrt{\e}\,\left(1+O\bigl(\l \bigr) \rgt)$ \\
\hline 
%$\n_{h}$ & $\e \lft(1+o(1) \rgt)$ & $1$ &  & $2$ & $\e^{\frac{1}{2}}$\tabularnewline
%\hline
\multicolumn{5}{c}{}\\[-3pt] 
\hline
\multicolumn{5}{|c|}{Wave function renormalization constants}\\
\hline 
$Z_{\bh}$ & 0 & $\e \,(1 + O\bigl(\l \e^{\frac{1}{2}})\bigr)$ & $0$ & $\e\,\left(1+O\bigl(\l \bigr) \rgt)$ \\
\hline 
$E_{\bh}$  & $0$ & $1+O(\l\,\e^{\frac{1}{2}})$ & $0$ & $1+O(\l )$ \\
\hline
$A_{\bh}$ & $0$ & $1+O(\l\,\e^{\frac{1}{2}})$ & 0 & $1+O(\l )$ \\
\hline 
$B_{\bh}$   & $0$ & $O(\l\,\e^{-\frac{1}{2}})$ & 0 & $ O \lft( \l\, \e^{-1} \rgt) $\\
\hline 
\multicolumn{5}{c}{}\\[-3pt] 
\hline
\multicolumn{5}{|c|}{Coupling constants with external fields}\\
\hline 
$\m_{\bh}^{J_{0}}$  & $0$ & $1+O(\l \, \e^{\frac{1}{2}})$ & 1/2 & $ \e^{-\frac{1}{2}}\lft(1+O\bigl(\l \bigr) \rgt) $ \\
\hline 
$\m_{\bh}^{J_{1}}$ & $-2$ & $\e^2 \bigl(1+O(\l \, \e^{\frac{1}{2}})\bigr)$ & $-3/2$ & $\e^{\frac{3}{2}}\,\lft(1+O\bigl(\l \bigr)  \rgt)$ \\
\hline 
$Z_{\bh}^{J_{0}}$  & $0$ & $O(\l)$ & $0$ & $O\lft(\l^2 \rgt)$ \\
\hline 
$E_{\bh}^{J_{1}}$  & $0$ & $O(\l\,\e^{\frac{1}{2}})$ & $0$ &  $ O\bigl(\l^2 \bigr)$ \\
\hline 
$E_{\bh}^{J_{0}}$  & $0$ & $O(\l \, \e^{-\frac{1}{2}})$ & $0$ &  $O\bigl(\l^2\,\e^{-1} \bigr)$ \\
\hline 
\multicolumn{5}{c}{}\\[-9pt] 
%\hline
%\multicolumn{5}{|c|}{Kernels with two external fields}\\
%\hline
%$J_{\bh}^{J_{0}}$  & $0$ & \blue{$O\lft(\l^2\e^{-1}\rgt)$}  & $0$ &  $O\lft(\l^2\e^{-1}\rgt)$ \\
%\hline
\end{tabular}
\par\end{centering}
\vskip 0.2cm
\noindent \centering{} 
\caption{Values of running coupling constants at $h=\bar{h}$ for $d=3$ and $d=2$. Here $\e= \l \r_0 R_0^d $. Note that, while $\bar{A}_\bh$ and $\bar{E}_\bh$ are smaller than one, $A_\bh$ and $E_\bh$ starts from $1$, since we have included the local quadratic terms in the free measure.
The values of the other endpoints at scale $\bh$ are $\l'_\bh=\e^2\l_\bh$, $\l''_\bh=\e^4\l_\bh$, $\m'_\bh=\e^2\m_\bh$.}  \label{valueBAR} 
\end{table}
\end{center}

\vskip -1cm
It will be also useful in the following the value at $\bh$ of the kernel with two $J_0$ external fields, $J_{\bh}^{J_{0}}$, which is 
\[
J_{\bh}^{J_{0}}= \begin{cases}
O\lft(\l \e^{-1}\rgt)  & d=3 \\
O\lft(\l^2\e^{-1}\rgt) & d=2
\end{cases}
\]

\pagina

\section{Coupling constants with external fields for $h\leq \bh$} \label{flow_J0}

In this section we will discuss how to control the running coupling constant with external fields $J_0$ and $J_1$. The discussion of the three and two dimensional cases are quite different, since in the two dimensional case, as described in chap.~\ref{multiscale} we need to localize some extra effectively marginal kernels. 

As already discussed in the previous sections in the region $h \leq \bh$ the localization scheme is defined including at each step the local quadratic terms in the measure. Including the local quadratic terms at scale $h =\bh$ in the free measure, we obtain the following renormalized propagator: 
\[
g_{\a \a'}^{(\bh)}(k) & = (\r_0 R_0^{-2})^{-1}\,
\frac{\left(\begin{array}{cc}
A_\bh \kk^{2} + B_\bh k_0^2 \; & E_\bh k_{0}\\
-E_\bh k_{0}\; & \kk^2 +Z_\bh 
\end{array}\right)}
{(E^2_\bh + B_\bh Z_\bh)k_{0}^{2} + Z_\bh A_\bh \kk^{2}+\kk^2 (A_\bh \kk^{2} + B_\bh k_0^2 )} \non \\[6pt]
&  \simeq (\r_0 R_0^{-2})^{-1}\,
\frac{\left(\begin{array}{cc}
\kk^{2}  \; &  k_{0}\\
- k_{0}\; & \e
\end{array}\right)}
{k_{0}^{2} + \e \kk^{2}}
\]
where ``$\simeq$'' means leading order in $k$, as $k$ goes to zero, and in $\e$, see the values of $A_\bh$, $B_\bh$, $E_\bh$ and $Z_\bh$ in table~\ref{valueBAR} in three and two dimensions. The structure $k_0^2 + \e \kk^2$ in the denominator is preserved along the flow in the second region (this is also a consequence of WIs, as discussed up ahead). This property justifies the choice of the cutoff function in the second region as:  $\c_h(k):=\c_h(k_{0}^{2}+\e\,\kk^{2}) $. 

\subsection{Three dimensions}

Aim of this subsection is to prove the results \eqref{wavefunc} for the $3d$ case. With this purpose we study the flow of the coupling constants with an external field $J_\n$, which are related to $E_h$, $A_h$ and $B_h$ through the local WIs. \\

Below we will use many times the dimensional estimate \eqref{thm_nfactorial2} and the short memory property \eqref{short_mem1}. We remind the reader that the short memory factor we can extract along each branch of a tree in the lower momenta region $h\leq \bh$ is $\g^{\th(h-\bh)}$ with $0<\th < 2$. 

%We remind that the trees contributing to the flow equations in the region $h \leq \bh$ have end points on scale lower or equal than $\bh$. In particular the endpoints at scale $\bh$ are all the possible diagrams emerging by the integration over the transient region $\bh < h \leq 0$, with arbitrary number of external legs. In the following we will consider only endpoints at scale $\bh$ which have the same structure of the initial potential at scale $h=0$. However one can easily proof that nothing is changed if we consider all the possible irrelevant diagrams at scale $\bh$. The discussion of the latter point can be found in appendix~\ref{app-counting}. 

%In table \ref{values_bar_h} the initial values of the running coupling constants at scale $\bh$ are reported, whose derivation can be found in appendix~\ref{app-transient}. 

Let us start with the discussion of the flows of the three coupling constants with the external field $J_0$. These are studied by analogy with the flows of the ``corresponding'' couplings without external fields, where the correspondence is given by the substitution of the vertex giving the external field $J_0$ with a vertex $\m_h$, as better explained below. 

{\centering \subsubsection{I. Flow for $\mu_{h}^{J_{0}}$, d=3}}

The flow of $\mu_{h}^{J_{0}}$ is studied by analogy with the flow of $\m_h$. In fact it is sufficient to note that the dashed external line in $\m_h$ may only belong to a vertex $\m_k$ for $ h\leq k < \bh$ or to an irrelevant vertex at scale $\bh$. Similarly, the external $J_0$ field in $\m_h^{J_0}$ can only come from  a vertex $\m_k^{J_0}$, for $h \leq k < \bh$ or from an irrelevant vertex at scale $\bh$. The two legged relevant vertices with one dashed or $J_0$ line, which are $Z_h$, $Z_{h}^{J_{0}}$ and $E_{h}^{J_{0}}$, cannot contribute to the beta function of $\m_h$ or $\m_h^{J_0}$ since they only have two external legs and then the local part of the diagrams made with them is zero. Then, apart for diagrams containing irrelevant vertices, the beta function $\b_h^{J_0, \m}$ for $\mu_{h}^{J_{0}}$ is equal to the beta function of $\m_{h}$, once the vertex $\m_{k}$, $k \geq h$, giving the external dashed line is replaced by a vertex $\m_{k}^{J_{0}}$. \\

In the following we will denote with $\b_{h}^{\m}$ the part of the beta function of $\m_h$ such that the external dashed lines comes from a vertex $\m_k$, and with $\b_{h}^{*\,\m}$ the diagrams contributing to the same beta function where the external dashed line comes from an irrelevant vertex at scale $\bh$. In three dimensions $\m_h$ and $\m_h^{J_0}$ are marginal. Their flow equations  are
\[ 
& \m_{h-1}-\m_{h}  =  \b_{h}^{\m}+\b_{h}^{*\,\mu}  \label{flow_mu} \\
& \m_{h-1}^{J_{0}}-\m_{h}^{J_{0}}  =  \b_{h}^{J_{0},\mu} \label{flow_mu_J}
\]
where $\b_{h}^{\m}$ and $\b_{h}^{J_{0},\m}$ differ only for a vertex, that is we can write
\[
 \b_{h}^{\m} & = \sum_{h+1 \leq k\leq \bh}\m_{k} \, \b_{h,k}^{\m} \non \\
 \b_{h}^{J_{0},\m} & = \sum_{h+1 \leq k\leq \bh} \mu_{k}^{J_{0}}\, \b_{h,k}^{\mu} 
\]
with $\m_k\,\b_{h,k}^\m$ the sum of all the diagrams contributing to the beta function for $\m_h$ where the external dashed line comes from a vertex $\m_k$ and equivalently $\m_k^{J_0}\,\b_{h,k}^\m$ is  the sum of all the diagrams contributing to the beta function for $\m_h^{J_0}$ where the external wiggly line $J_0$ comes from a vertex $\m^{J_0}_k$. Then, assuming
\[
\frac{\m_k^{J_0}}{\mu_{k}} = \frac{\m_h^{J_0}}{\m_h} \lft(1 + O(\l \e^{\frac{1}{2}}) \rgt) \label{eq:condition_mu}
\]
we can rewrite $\b_{h}^{J_{0},\m}$ as follows
\[
\b_{h}^{J_{0},\m} = \sum_{h+1 \leq k\leq \bh}\frac{\mu_{k}^{J_{0}}}{\mu_{k}}\, \m_{k}\,\b_{h,k}^{\mu}=\frac{\mu_{h}^{J_{0}}}{\mu_{h}}\,\beta_{h}^{\mu} \lft(1 + O(\l \e^{\frac{1}{2}}) \rgt)
\]
%
%
%--------------------------------------------------------------------------------------------
\feyn{
\begin{fmffile}{feyn-TESI/beta_mu_star}
\unitlength = 0.8 cm  
\def\myl#1{2.5cm}
\def\myll#1{2.5cm}
\[
%-------------------------------\mu_h----------------------------
	\b_h^\m & = \;
	\parbox{\myll}{\centering{
			\begin{fmfgraph*}(2.8,2) 
			\fmfleft{i1}
			\fmfright{o1,o2}
			\fmftop{t}
			\fmfbottom{b}
			\fmf{phantom, tension=1.8}{t,v3}    %---
			\fmf{phantom, tension=1.8}{b,v4}    %---
            \fmf{dashes, tension=1.5}{i1,v1}			
			\fmf{plain,left=0.3}{v1,v3}
			\fmf{plain,left=0.3}{v3,v2}			
			\fmf{plain,right=0.3}{v1,v4}
			\fmf{plain,right=0.3}{v4,v2}
			\fmf{plain, tension=1.5}{o1,v2,o2}
		\Ball{v1,v2}
		\end{fmfgraph*}
			}} 
			+\;
	\parbox{\myll}{\centering{
			\begin{fmfgraph*}(2.8,2) 
			\fmfleft{i1}
			\fmfright{o1,R,o2}
                   \fmftop{T}
                   \fmfbottom{B}
			\fmf{phantom, tension=1.8}{B,vB}  %-------
                    \fmf{phantom, tension=1.8}{T,vT}  %-------
			\fmf{dashes,tension=1.5}{i1,v1}
			\fmf{plain, left=0.3, tension=1.2}{v1,vT}
                   \fmf{plain, left=0.2, tension=0.4}{vT,v3}
                    \fmf{phantom, tension=1}{R,vR}  %------
			\fmf{dashes, left=0.3, tension=1}{v3,vR}
                   \fmf{dashes, left=0.3, tension=0.4}{vR,v2}			
                  \fmf{plain, right=0.3, tension =1.4}{v1,vB}
	            \fmf{plain, right=0.2, tension=0.4}{vB,v2}           
		     \fmf{plain, tension=0.8}{o1,v2}
			\fmf{plain,tension=0.8}{o2,v3}
                   \Ball{v1,v2,v3}
			\end{fmfgraph*}  
			}}  
+\;
	\parbox{\myll}{\centering{
			\begin{fmfgraph*}(2.8,2) 
			\fmfleft{i1}
			\fmfright{o1,R,o2}
                   \fmftop{T}
                   \fmfbottom{B}
			\fmf{phantom, tension=1.8}{B,vB}  %-------
                    \fmf{phantom, tension=1.8}{T,vT}  %-------
			\fmf{dashes,tension=1.5}{i1,v1}
			\fmf{plain, left=0.3, tension=1.2}{v1,vT}
                   \fmf{dashes, left=0.2, tension=0.4}{vT,v3}
                    \fmf{phantom, tension=1}{R,vR}  %------
			\fmf{plain, left=0.3, tension=1}{v3,vR}
                   \fmf{plain, left=0.3, tension=0.4}{vR,v2}		
                  \fmf{plain, right=0.3, tension =1.4}{v1,vB}
	            \fmf{dashes, right=0.2, tension=0.4}{vB,v2}           
	             \fmf{plain, tension=0.8}{o1,v2}
			\fmf{plain,tension=0.8}{o2,v3}
                   \Ball{v1,v2,v3}
			\end{fmfgraph*}  
			}}  +\;
	\parbox{\myll}{\centering{
			\begin{fmfgraph*}(2.8,2) 
			\fmfleft{i1}
			\fmfright{o1,R,o2}
                   \fmftop{T}
                   \fmfbottom{B}
			\fmf{phantom, tension=1.8}{B,vB}  %-------
                    \fmf{phantom, tension=1.8}{T,vT}  %-------
			\fmf{dashes,tension=1.5}{i1,v1}
			\fmf{plain, left=0.3, tension=1.2}{v1,vT}
                   \fmf{dashes, left=0.2, tension=0.4}{vT,v3}
                    \fmf{phantom, tension=1}{R,vR}  %------
			\fmf{plain, left=0.3, tension=1}{v3,vR}
                   \fmf{dashes, left=0.3, tension=0.4}{vR,v2}			
                  \fmf{plain, right=0.3, tension =1.4}{v1,vB}
	            \fmf{plain, right=0.2, tension=0.4}{vB,v2}           
			\fmf{plain, tension=0.8}{o1,v2}
			\fmf{plain,tension=0.8}{o2,v3}
                   \Ball{v1,v2,v3}
			\end{fmfgraph*}  
			}}  
\non \\[20pt]
%-------------------------------------------------------- beta *
   \beta^{*\mu}_h & = \; 
 \parbox{\myl}{\centering{
	\begin{fmfgraph*}(2.8,2)
			\fmfleft{i1}
			\fmfright{o1,o2}
			\fmftop{t}
			\fmfbottom{b}
			\fmf{phantom, tension=1.8}{t,v3}    %---
			\fmf{phantom, tension=1.8}{b,v4}    %---
                   \fmf{plain, tension=1.5}{i1,v1}			
			\fmf{plain,left=0.3}{v1,v3}
			\fmf{plain,left=0.3}{v3,v2}			
			\fmf{dashes,right=0.3}{v1,v4}
			\fmf{dashes,right=0.3}{v4,v2}
		       \fmf{dashes, tension=1.5}{o1,v2}
                   \fmf{plain, tension=1.5}{v2,o2}
			\fmfdot{v2}
			\fmfv{label=$\l'_\bh$}{v2}
	\Ball{v1}	
		\end{fmfgraph*}
		}}  + \; 
 \parbox{\myl}{\centering{
	\begin{fmfgraph*}(2.8,2)
			\fmfleft{i1}
			\fmfright{o1,o2}
			\fmftop{t}
			\fmfbottom{b}
			\fmf{phantom, tension=1.8}{t,v3}    %---
			\fmf{phantom, tension=1.8}{b,v4}    %---
                   \fmf{plain, tension=1.5}{i1,v1}			
			\fmf{plain,left=0.3}{v1,v3}
			\fmf{dashes,left=0.3}{v3,v2}			
			\fmf{dashes,right=0.3}{v1,v4}
			\fmf{plain,right=0.3}{v4,v2}
			\fmf{dashes, tension=1.5}{o1,v2}
                   \fmf{plain, tension=1.5}{v2,o2}
			\fmfdot{v2}
			\fmfv{label=$\l'_\bh$}{v2}
	\Ball{v1}	
		\end{fmfgraph*}
		}} \non \\[12pt] 
& + \;
\parbox{\myll}{\centering{
			\begin{fmfgraph*}(2.8,2) 
			\fmfleft{i1}
			\fmfright{o1,R,o2}
                   \fmftop{T}
                   \fmfbottom{B}
			\fmf{phantom, tension=1.8}{B,vB}  %-------
                    \fmf{phantom, tension=1.8}{T,vT}  %-------
			\fmf{dashes,tension=1.5}{i1,v1}
			\fmf{dashes, left=0.3, tension=1.2}{v1,vT}
                   \fmf{plain, left=0.2, tension=0.4}{vT,v3}
                    \fmf{phantom, tension=1}{R,vR}  %------
			\fmf{plain, left=0.3, tension=1}{v3,vR}
                   \fmf{plain, left=0.3, tension=0.4}{vR,v2}		
                  \fmf{dashes, right=0.3, tension =1.4}{v1,vB}
	            \fmf{plain, right=0.2, tension=0.4}{vB,v2}           
	             \fmf{plain, tension=0.8}{o1,v2}
			\fmf{plain,tension=0.8}{o2,v3}
                   \fmfv{label=$\m_\bh$, label.angle=-120}{v1}
 			\Ball{v1,v2,v3}
			\end{fmfgraph*}  
			}}  + \;
\parbox{\myll}{\centering{
			\begin{fmfgraph*}(2.8,2) 
			\fmfleft{i1}
			\fmfright{o1,R,o2}
                   \fmftop{T}
                   \fmfbottom{B}
			\fmf{phantom, tension=1.8}{B,vB}  %-------
                    \fmf{phantom, tension=1.8}{T,vT}  %-------
			\fmf{dashes,tension=1.5}{i1,v1}
			\fmf{dashes, left=0.3, tension=1.2}{v1,vT}
                   \fmf{dashes, left=0.2, tension=0.4}{vT,v3}
                    \fmf{phantom, tension=1}{R,vR}  %------
			\fmf{plain, left=0.3, tension=1}{v3,vR}
                   \fmf{plain, left=0.3, tension=0.4}{vR,v2}		
                  \fmf{dashes, right=0.3, tension =1.4}{v1,vB}
	            \fmf{dashes, right=0.2, tension=0.4}{vB,v2}           
	             \fmf{plain, tension=0.8}{o1,v2}
			\fmf{plain,tension=0.8}{o2,v3}
                   \fmfv{label=$\m_\bh$, label.angle=-120}{v1}
 			\Ball{v1,v2,v3}
			\end{fmfgraph*}  
			}}+ \;
\parbox{\myll}{\centering{
			\begin{fmfgraph*}(2.8,2) 
			\fmfleft{i1}
			\fmfright{o1,R,o2}
                   \fmftop{T}
                   \fmfbottom{B}
			\fmf{phantom, tension=1.8}{B,vB}  %-------
                    \fmf{phantom, tension=1.8}{T,vT}  %-------
			\fmf{dashes,tension=1.5}{i1,v1}
			\fmf{dashes, left=0.3, tension=1.2}{v1,vT}
                   \fmf{dashes, left=0.2, tension=0.4}{vT,v3}
                    \fmf{phantom, tension=1}{R,vR}  %------
			\fmf{plain, left=0.3, tension=1}{v3,vR}
                   \fmf{dashes, left=0.3, tension=0.4}{vR,v2}		
                  \fmf{dashes, right=0.3, tension =1.4}{v1,vB}
	            \fmf{plain, right=0.2, tension=0.4}{vB,v2}           
	             \fmf{plain, tension=0.8}{o1,v2}
			\fmf{plain,tension=0.8}{o2,v3}
                   \fmfv{label=$\m_\bh$, label.angle=-120}{v1}
 			\Ball{v1,v2,v3}
			\end{fmfgraph*}  
			}}
\non
\]
\end{fmffile}
}{{\bf Flow of $\m_h$, $d=3$.} On the first line the diagrams contributing to $\b_h^\m$ at the main order in $\e$, which corresponds to the one loop diagrams without irrelevant vertices. On the second line the leading order diagrams contributing to $\b^{* \m}_h$, \ie the diagrams with only one irrelevant vertex at scale $\bh$; these are obtained by the first line by substituing the vertex $\l_h$ with $\l'_\bh$ or the vertex $\m_h$ giving the external $l$ line an irrelevant vertex $\m'_\bh.$   }{beta_mu_star}
%
%
%
%---------------------------------------------------------------------------------------
Then, by using \eqref{flow_mu} and \eqref{flow_mu_J}, we can express the flow of $\m_h^{J_0}$ in terms of the flow of $\m_h$ as follows
%\[
%\mu_{h-1}^{J_0}-\mu_{h}^{J_0} =  \frac{\mu_{h}^{J_0}}{\mu_{h}}\,\lft( \m_{h-1}-\m_{h}-\b_{h}^{*\mu}\rgt)
%\]  
\[
\frac{\mu_{h}^{J_{0}}}{\mu_\bh^{J_0}}  & =  \prod_{k=h+1}^{\bh} \lft(1+\frac{\b_{k}^{\m}}{\m_{k}}\,\bigl(1 + O(\l \e^{\frac{1}{2}}) \bigr) \rgt)
%=   \prod_{k=h+1}^{\bh}\left(\frac{\m_{k-1}-\b_{k}^{*\m}}{\mu_{k}}\right) 
\non \\
& =\prod_{k=h+1}^{\bh}\frac{\mu_{k-1}}{\mu_{k}}\left(1-\frac{\beta_{k}^{*\mu}}{\mu_{k-1}}  + O(\l \e^{\frac{1}{2}}) \right) \non \\
& =\frac{\mu_{h}}{\mu_{\bh}}\prod_{k=h+1}^{\bh}\left(1-\frac{\beta_{k}^{*\mu}}{\mu_{k-1}} +O(\l \e^{\frac{1}{2}}) \right)
\]
Here $\b_{h}^{*\,\mu}$ contains the diagrams where the external dashed line of $\mu_{h-1}$ comes from one of the irrelevant vertices $\l'_\bh$ or $\m'_\bh$, as shown in fig. \ref{beta_mu_star}. 
With respect to the dimensional estimate for $\b_h^\m$, the beta function $\b_{h}^{*\,\mu}$ has a factor $\e^2 \l_h /Z_h$ coming from the substitution of a vertex at scale $h$ with an irrelevant vertex. The $Z_h^{-1}$ factor comes from the fact that when we substitute $\l_h$ with $\l'_h$ or $\m_h$ with $\m'_h$  we get an additional $g^{(h)}_{ll}$ propagator\footnote{ Regarding the substitution of  one vertex $\m_h$ with $\m'_h$, one can convince immediately that there is a unique additional $g^{(h)}_{ll}(k)$ propagator, even if we are changing two internal plain legs with dashed legs, looking at the dominant diagrams in the asymptotic limit. These are the second diagrams on the first and third lines of fig. \ref{beta_mu_star}. For $h \leq \bh$ but not ``close'' to the infrared limit if one sums all the diagrams with three three-legged vertices contributing to $\b_h^\m$ and  $\b_h^{*\m}$ one finds the same result.}. 

Moreover, by the short memory property \ref{short_mem1}, these diagrams are exponentially suppressed as $h \arr -\io$. In fact, due to the long branch connecting the irrelevant vertex to the vertex at scale $h+1$ in the tree expansion for $\b_{h}^{*\,\mu}$, we can always extract a factor $\g^{-(\bh -h)}$ from the usual estimate.  By a leading order calculation one finds 
%$\b_h^{*\m}$ to be positive and 
\[ \label{bound:b_star_mu}
 \b_{h}^{*\,\m} = O\lft(\l\,\e^{\frac{1}{2}} Z_h \, \g^{- (\bh -h)}\rgt) \qquad \qquad d=3 
\]
Using \eqref{bound:b_star_mu} one get the following estimate: 
\[ \label{flow_mJ0}
\frac{\m_h^{J_0}}{\m_\bh^{J_0}}  = \frac{\m_h}{\m_\bh}  \lft(1+{\cal M}_{1,h}(\l) + {\cal M}_{2,h}(\l) \rgt)
\]
with ${\cal M}_{1,h}(\l) =O(\l \e^{\frac{1}{2}})$, ${\cal M}_{2,h}(\l) =O(\l \e^{-\frac{1}{2}}\g^h)$ and ${\cal M}_{1,\bh}(\l)=-{\cal M}_{2,\bh}(\l)$.
Asymptotically, being $\m_h=\m_\bh \,\bigl(1+c\,\l\,\e^{\frac{1}{2}}|h-\bh|\bigr)^{-1}$ we have
\[
\m_h^{J_0} = \frac{1}{1+c\,\l \e^{\frac{1}{2}}|h-\bh|}\,\lft(1+O(\l \e^{\frac{1}{2}}) \rgt)
\]
The \eqref{flow_mJ0} also shows that the assumption \eqref{eq:condition_mu} is  verified. 
By using the WI \eqref{LWI_E} and the flow equation \eqref{flow_mJ0} we get
\[
E_h & \simeq \e^{-1}\,\m_{h}\,\lft(1 + O(\l)\rgt) & d=3  \non
\]

%\[
%E_h & \simeq \e^{-1}\,\m_{h}\,\lft(1 + O(\l \e^{\frac{1}{2}})\rgt) & d=3  \non
%\]

\feyn{
\begin{fmffile}{feyn-TESI/beta_E_star}
\unitlength = 1 cm  
\def\myl#1{2.5cm}
\[
 &  \beta^{\m}_h  = \; 
 \parbox{\myl}{\centering{
	\begin{fmfgraph*}(2.8,2)
			\fmfleft{i1}
			\fmfright{o1}
			\fmftop{t}
			\fmfbottom{b}
			\fmf{phantom, tension=1.8}{t,v3}    %---
			\fmf{phantom, tension=1.8}{b,v4}    %---
                   \fmf{dashes, tension=1.5}{i1,v1}			
			\fmf{plain,left=0.3}{v1,v3}
			\fmf{plain,left=0.3}{v3,v2}			
			\fmf{plain,right=0.3}{v1,v4}
			\fmf{dashes,right=0.3}{v4,v2}
		       \fmf{plain, tension=1.5, label= $\dpr_0$}{o1,v2}
			\fmfdot{v2}
	\fmfv{decor.shape=circle,decor.filled=empty,decor.size=.08w}{v1,v2}	%, label=$\m_h$, label.angle=-120
	%\fmfv{decor.shape=circle,decor.filled=empty,decor.size=.08w, label=$\m_h$, label.angle=-60}{v2}	
		\end{fmfgraph*}
		}}  \hskip 2cm
 \beta^{\m, J_0}_h  = \; 
 \parbox{\myl}{\centering{
	\begin{fmfgraph*}(2.8,2)
			\fmfleft{i1}
			\fmfright{o1}
			\fmftop{t}
			\fmfbottom{b}
			\fmf{phantom, tension=1.8}{t,v3}    %---
			\fmf{phantom, tension=1.8}{b,v4}    %---
                   \fmf{wiggly, tension=1.5, label=$J_0$}{i1,v1}			
			\fmf{plain,left=0.3}{v1,v3}
			\fmf{plain,left=0.3}{v3,v2}			
			\fmf{plain,right=0.3}{v1,v4}
			\fmf{dashes,right=0.3}{v4,v2}
		       \fmf{plain, tension=1.5, label= $\dpr_0$}{o1,v2}
			\fmfdot{v2}
	\fmfv{decor.shape=circle,decor.filled=empty,decor.size=.08w}{v1}	%, label=$\m_h$, label.angle=+120
	\fmfv{decor.shape=circle,decor.filled=empty,decor.size=.08w}{v2}	%, label=$\m_h$, label.angle=-60
		\end{fmfgraph*}
		}}  \non \\
& \hskip 2cm  \beta^{E,*}_h  = \;
 \parbox{\myl}{\centering{
	\begin{fmfgraph*}(2.8,2)
			\fmfleft{i1}
			\fmfright{o1}
			\fmftop{t}
			\fmfbottom{b}
			\fmf{phantom, tension=1.8}{t,v3}    %---
			\fmf{phantom, tension=1.8}{b,v4}    %---
                   \fmf{dashes, tension=1.5}{i1,v1}			
			\fmf{dashes,left=0.3}{v1,v3}
			\fmf{plain,left=0.3}{v3,v2}			
			\fmf{dashes,right=0.3}{v1,v4}
			\fmf{dashes,right=0.3}{v4,v2}
		       \fmf{plain, tension=1.5, label= $\dpr_0$}{o1,v2}
			\fmfdot{v1}
	\fmfv{label=$\m'_\bh$, label.angle=-120}{v1}	
	\fmfv{decor.shape=circle,decor.filled=empty,decor.size=.08w
%label=$\m_h$, label.angle=-60
}{v2}	
		\end{fmfgraph*}
		}} %\qquad +\quad
\non 
\]
\vskip -0.5cm
\end{fmffile}
}{{\bf Flow of $E_h^{J_0}$, $d=3$.} On the first line the beta function of $E_h$ and $E_h^{J_0}$ at leading order; the beta function of $E_h^{J_0}$ at leading order is obtained by substituting the vertex $\m_h$ with external dashed line in $\b^\m_h$ with a vertex $\m^{J_0}_h$. On the second line the leading order diagram contributing to $\b^{E,*}_h$.
}{beta_E_star}

\vskip 0.5cm

{\centering \subsubsection{II. Flow for $E_{h}^{J_{0}}$, d=3}}

The flow for $E_{h}^{J_{0}}$ is obtained by comparing the beta function for $E_{h}^{J_{0}}$ with the one for $E_{h}$, following the same ideas used in the study of the flow of $\m^{J_0}_h$. Let denote with $\b^E_h$ the part of the beta function of $E_h$ where the external dashed line comes from a relevant vertex $\m_k$ and with  $\b_{h}^{E,*}$  the graphs for which the external dashed line is given by an irrelevant vertex at scale $\bh$. The beta function of $E_h^{J_0}$ is obtained by substituting  $\m_k$ with $\m_k^{J_0}$, see fig.~\ref{beta_E_star}. We have:
\[
E_{h-1} & =  E_{h}+\beta_{h}^{E}+\beta_{h}^{E,*} \non \\[3pt]
E_{h-1}^{J_{0}} & =  E_{h}^{J_{0}}+\beta_{h}^{E,{J_{0}}}
\]
and then
\[
E_{h}^{J_{0}} - E_{\bh}^{J_{0}} = \sum_{k=h+1}^{\bh}  \b_{k}^{E,J_0 } 
\]
Denoting with $\b_{h,k}^{E}$ the sum of all the diagrams contributing to $\b_{h}^{E}$ where the external dashed line comes from a vertex $\m_k$ at scale $h<k\leq \bh$ we can write:
\[
& \b_{h}^{E}=\sum_{k=h+1}^\bh \m_{k}\,\b_{h,k}^{E} \non \\
& \b_{h}^{E, J_0}=\sum_{k=h+1}^\bh \m_{k}^{J_{0}}\b_{h,k}^{E} = \e^{-1} \b_{h}^{E} \, \lft(1+O(\l \e^{\frac{1}{2}})\rgt)
\]
where we have used  \eqref{flow_mJ0} to estimate $\m_k^{J_0}/\m_k$ .
Then 
\[
E_{h}^{J_{0}} - E_{\bh}^{J_{0}}  & =  \e^{-1} \bigl(1 + O(\l \,\e^{\frac{1}{2}}) \bigr) \Big(E_h - E_\bh +\sum_{k=h+1}^{\bh} |\b^{E,*}_k|\Big)  
\]
with 
%is dimensionally bounded by $\l\, \e^{\frac{1}{2}}\g^{-(h-\bar{h})}$. 
%By a leading order calculation one find $\b_h^{*\m}$ to be positive and 
\[ \label{bound:b_star_E}
\b_{k}^{E,*} = O\lft(\l\,\e^{\frac{1}{2}}  \, \g^{- (\bh -k)}\rgt) 
\]
Being 
\[
 E_{\bh}^{J_{0}}=O\lft(\l\,\e^{-\frac{1}{2}}\rgt) \hskip 1cm
  E_{\bh}=1+ O\lft(\l\e^{\frac{1}{2}}\rgt)\quad
\]
we find
\[ \label{E0_3d}
 E_{h}^{J_0}=-\e^{-1} \lft( 1- E_h \rgt)  +O(\l\,\e^{-\frac{1}{2}}) 
\]
with
\[
 E_h= \lft(1+\l\,\e^{\frac{1}{2}}|h-\bh| \rgt)^{-1} 
\]
In the infrared limit $h \arr -\io$ we get
\[ 
 E_{h}^{J_0}=-\e^{-1} \lft(1  +O(\l\,\e^{\frac{1}{2}})  \rgt)
\]
The local WI \eqref{LWI_B} together with  \eqref{E0_3d} give
\[
\sqrt{2}\,B_h & \simeq \e^{-1} \lft( 1- E_h +O\bigl(\l \bigr) \rgt)   & d=3
%\sqrt{2}\,B_h & = \e^{-1} \lft( 1- E_h +O\bigl(\l\,\e^{\frac{1}{2}}\bigr) \rgt)   & d=3
\]
The cancellation making $B_{h}$ finite  (instead than logarithmically divergent as one could expect by the na\"ive dimensional estimate) can be easily seen at the one--loop level, see appendix~\ref{appB.WI_B}. However it has to be stressed that the one--loop calculations do not say much: if we could not exclude that a similar cancellation takes place at all orders there would always be the possibility that higher orders produce a completely different behavior. That is why the WIs results to be crucial. \\
%corresponding to completely different physical properties of the system. 

\vskip 0.5cm
{\centering \subsubsection{III. Flow for $E_{h}^{J_{1}}$ , d=3}}

The argument allowing us to control the flow of $E_{h}^{J_{1}}$ is based on a dimensional estimate. In fact, the  beta function of the marginal coupling $E_{h}^{J_{1}}$ can be easily estimated, noting that the external $J_{1}$ line can only be originated by an irrelevant vertex  $\mu_{\bh}^{J_{1}}$ or from a marginal vertex of type $E_{k}^{J_{1}}$, as shown in figure \eqref{flow_E_J1}. However, the local part of the diagrams in which the external $J_{1}$-line comes from a vertex $E_{k}^{J_{1}}$ are zero due to the presence of the cutoff function $f_h(k)$, which is zero for vanishing external momentum. 

For what regards the remaining diagrams, \ie the ones where the wiggly line comes from  $\mu_{\bh}^{J_{1}}$, we can extract from then a short memory factor. Denoting with $\b_h^{E, J_1}$ the beta function of $E^{J_1}_h$ we have
\[
\b_h^{E,J_1}= O\lft(  \l\,\e^{\frac{1}{2}}\g^{-(\bh-h)} \rgt)
\]
This implies that $E_{h}$ is not log-divergent as a priori expected by the dimensional estimate, but tends to a constant. In particular, being $E_{\bh}^{J_{1}}=O(\l \e^{\frac{1}{2}})$  we have
\[ \label{E3d}
E_{h}^{J_{1}}  =  O(\l \e^{\frac{1}{2}}) 
\]
Using the local WI \eqref{LWI_A} and \eqref{E3d} we find:
\[
 A_{h} -1& \simeq O(\l \,\e^{\frac{1}{2}})    & d=3
\]
Note that  $A_{h}$ is marginal and then a priori logarithmic divergent. 
The cancellation making $A_{h}$ finite order by order in perturbation theory can be easily seen at level of the one--loop computation (see appendix \ref{B.WI-leading}) and it is a consequence of the symmetry of the integrals under the change of variable $(\pp+\kk,-\pp)\arr(-\pp,\kk+\pp)$.

\feyn{
\begin{fmffile}{feyn-TESI/beta_E1}
\unitlength = 1 cm  
\def\myl#1{3cm}
\[
    \parbox{\myl}{\centering{
	\begin{fmfgraph*}(2.5,1.5) 
			\fmfleft{i1}
			\fmfright{o1}
			\fmf{wiggly,label=${J_1}$}{i1,v}
			\fmf{plain, label=$\dpr_\pp$, label.dist=-0.15w}{v,o1}
			\fmfv{decor.shape=circle,decor.filled=empty,decor.size=.2w}{v}		
		\end{fmfgraph*}
		}} \quad = & \quad 
\parbox{\myl}{\centering{
	\begin{fmfgraph*}(2.5,1.5)
			\fmfleft{i}
			\fmfright{o}
			\fmf{wiggly,label=${J_1}$, tension=1.2}{i,v1}
			\fmf{plain,left=0.7, tension=0.4}{v1,v2}
			\fmf{dashes,right=0.7, tension=0.4}{v1,v2}
			\fmf{plain, tension=1.2, label=$\dpr_\pp$, label.dist=-0.15w}{v2,o}
			\fmfdot{v1}
                    \fmfv{label=$\dpr_\xx$, label.angle=90}{v1}
	\fmfv{decor.shape=circle,decor.filled=empty,decor.size=.08w}{v2}	
			\end{fmfgraph*}  
			}} \quad + \quad 
\parbox{\myl}{\centering{
\begin{fmfgraph*}(2.5,1.5)
			\fmfleft{i}
			\fmfright{o}
			\fmftop{t}
			\fmfbottom{b}
			\fmf{wiggly,label=${J_1}$, tension=1.4}{i,v1}   
			\fmf{plain, tension=1.4, label=$\dpr_\pp$, label.dist=0.03w}{v2,o}
			\fmf{phantom, tension=1.5}{t,v3}    %----
			\fmf{phantom, tension=1.5}{b,v4}    %----
			\fmf{plain,left=0.3, label=$\dpr_\xx$, label.dist=0.02w}{v1,v3}
			\fmf{dashes,left=0.3}{v3,v2}
			\fmf{dashes,right=0.3}{v1,v4}
			\fmf{plain,right=0.3}{v4,v2}
			\fmfdot{v1}
	\fmfv{decor.shape=circle,decor.filled=empty,decor.size=.08w}{v2}	
			\end{fmfgraph*}  
			}}  \quad + \quad \ldots \non \\
	& \quad \parbox{\myl}{\centering{		
	\begin{fmfgraph*}(3,2.5)
			\fmfleft{i1}
			\fmfright{o1}
			\fmf{wiggly,label=${J_1}$}{i1,v1}
			\fmf{plain, tension=0.7, label=$\dpr_\xx$, label.dist=-0.12w}{v1,v2}
                    \fmf{plain}{v2,v2}
			\fmf{plain, label=$\dpr_\pp$, label.dist=-0.12w, tension=0.8}{v2,o1}
		\fmfv{decor.shape=circle,decor.filled=empty,decor.size=.08w}{v1}
		\fmfv{decor.shape=circle,decor.filled=empty,decor.size=.08w}{v2}		
		\end{fmfgraph*}
		}} \quad +\quad \ldots  \non 
\]   \vskip -1cm
\end{fmffile}
}{{\bf Flow for $E_{h}^{J_{1}}$.} Diagrams contributing to the flow of $E_{h-1}^{J_{1}}$  at the main order in $\e$. The symbol $\dpr_\pp$ on the external leg denotes the derivation of the diagrams with respect to the external momentum $\pp$. The symbol $\dpr_\xx$ over a propagator means derivation of the propagator with respect the variable $\xx$.}{flow_E_J1}  
%\nota{vanno bene simboli in fig. \ref{flow_E_J1}?}
%

\pagina

\subsection{Two dimensions} \label{2dJ_flow}

In this subsection we prove the results \eqref{wavefunc} in the $2d$ case. With this purpose we study the flow of the coupling constants with an external field $J_\n$ in the two dimensional case. 

In $2d$ there are four running coupling constants with external dashed legs, as shown in the first line of fig.~\ref{ext_2d}. However for each of this coupling there is a corresponding coupling with external $J_0$ field, which can be obtained by substituting one of the dashed legs with the $J_0$ fields, see the second line of fig.~\ref{ext_2d}. 

The reason why we have chosen to localize also the vertices $\m'^{J_0}_h$ , $\l'^{J_0}_h$ and $\o^{J_0}_h$  is exactly to use the analogy between the coupling constants with dashed legs and $J_0$ fields to control the flows of the latter. In fact, if one do not localize the kernels $V^{(h)}_{20;10}$, $V^{(h)}_{12;10}$ and $V^{(h)}_{04;10}$ and try to repeat an argument similar to the one pursued in three dimensional case, immediately realize that the short memory is not sufficient to extract the short memory factor we need to prove that the sum over the scale labels is summable, as happens in \eqref{flow_mJ0}. 

We also remark that the couplings $\m^{J_0}_h$, $\m'^{J_0}_h$ , $\l'^{J_0}_h$ and $\o^{J_0}_h$ satisfies global WIs similar to the ones holding for the vertices $\m_h$, $\m'_h$, $\l'_h$ and $\o_h$, as proved in chap.~\ref{WI} and represented in fig.~\ref{extWI}. 

%As shown in appendix \ref{order_e} the leading order diagrams in the small parameter $\e$ in the two dimensional case are the one loop diagrams without $\n_h$ and $\l_{6,h}$ vertices close to $\bh$; asymptotically the dominant diagrams are the one loop diagrams without $\n_h$ vertices where all the internal dashed legs are contracted among them. 

\feyn{
\begin{fmffile}{feyn-TESI/ext_2d}
\unitlength = 1 cm  
\def\myl#1{2.3cm}
\[
 & \parbox{\myl}{\centering{
	\begin{fmfgraph*}(1.4,1)
			\fmfleft{i1}
			\fmfright{o1,o2}	
			\fmf{dashes, tension=1.2}{i1,v}
			\fmf{plain}{o1,v,o2}
			\Ball{v}
		\end{fmfgraph*}  \\[3pt]
            $\m_h$
		}}  
\parbox{\myl}{\centering{
	\begin{fmfgraph*}(1.4,1)
			\fmfleft{i1}
			\fmfright{o1,o2}	
			\fmf{dashes, tension=1.2}{i1,v}
			\fmf{dashes}{o1,v,o2}
			\Ball{v}
		\end{fmfgraph*} \\[3pt]
            $\m'_h$
		}}  
\parbox{\myl}{\centering{
	\begin{fmfgraph*}(1.4,1)
			\fmfleft{i1}
			\fmfright{o1,o2,o3}	
			\fmf{dashes, tension=1.8}{i1,v}
			\fmf{plain}{o1,v,o2}
                   \fmf{dashes}{o3,v}
			\Ball{v}
		\end{fmfgraph*} \\[3pt]
            $\l'_h$
		}}  
\parbox{\myl}{\centering{
	\begin{fmfgraph*}(1.4,1)
			\fmfleft{i1}
			\fmfright{o1,o2,o3,o4}	
			\fmf{dashes, tension=2.5}{i1,v}
			\fmf{plain}{o1,v,o2}
                   \fmf{plain}{o3,v,o4}
			\Ball{v}
		\end{fmfgraph*} \\[3pt]
            $\o_h$
		}}  \non \\[12pt]
%%%%%%%%%%%%%%%
&
 \parbox{\myl}{\centering{
	\begin{fmfgraph*}(1.4,1)
			\fmfleft{i1}
			\fmfright{o1,o2}	
			\fmf{wiggly, tension=1.2, label=$J_0$}{i1,v}
			\fmf{plain}{o1,v,o2}
			\Ball{v}
		\end{fmfgraph*} \\[3pt]
            $\m^{J_0}_h$
		}}  
\parbox{\myl}{\centering{
	\begin{fmfgraph*}(1.4,1)
			\fmfleft{i1}
			\fmfright{o1,o2}	
			\fmf{wiggly, tension=1.2, label=$J_0$}{i1,v}
			\fmf{dashes}{o1,v,o2}
			\Ball{v}
		\end{fmfgraph*} \\[3pt]
             $\m'^{J_0}_h$
		}}  
\parbox{\myl}{\centering{
	\begin{fmfgraph*}(1.4,1)
			\fmfleft{i1}
			\fmfright{o1,o2,o3}	
			\fmf{wiggly, tension=1.8, label=$J_0$}{i1,v}
			\fmf{plain}{o1,v,o2}
                   \fmf{dashes}{o3,v}
			\Ball{v}
		\end{fmfgraph*}\\[3pt]
               $\l'^{J_0}_h$
		}}  
\parbox{\myl}{\centering{
	\begin{fmfgraph*}(1.4,1)
			\fmfleft{i1}
			\fmfright{o1,o2,o3,o4}	
			\fmf{wiggly, tension=2.5, label=$J_0$, label.dist=-0.6cm}{i1,v}
			\fmf{plain}{o1,v,o2}
                   \fmf{plain}{o3,v,o4}
			\Ball{v}
		\end{fmfgraph*} \\[3pt]
             $\o^{J_0}_h$
		}}  
 \non
\]
\end{fmffile}
}{{\bf Local terms, $d=2$.} Correspondence between the coupling constants with at least one dashed leg and the coupling constants with one external $J_0$ field.
}{ext_2d}

\feyn{
\begin{fmffile}{feyn-TESI/extWI}
\unitlength = 1 cm  
\def\myl#1{2.3cm}
\[
\sqrt{2}\,\parbox{\myl}{\centering{
	\begin{fmfgraph*}(1.4,1)
			\fmfleft{i1}
			\fmfright{o1,o2,o3}	
			\fmf{wiggly, tension=1.8, label=$J_0$}{i1,v}
			\fmf{plain}{o1,v,o2}
                   \fmf{dashes}{o3,v}
			\Ball{v}
		\end{fmfgraph*}
		}}  + 2
 \parbox{\myl}{\centering{
	\begin{fmfgraph*}(1.4,1)
			\fmfleft{i1}
			\fmfright{o1,o2}	
			\fmf{wiggly, tension=1.2, label=$J_0$}{i1,v}
			\fmf{plain}{o1,v,o2}
			\Ball{v}
		\end{fmfgraph*} 
		}}  =
\parbox{\myl}{\centering{
	\begin{fmfgraph*}(1.4,1)
			\fmfleft{i1}
			\fmfright{o1,o2}	
			\fmf{wiggly, tension=1.2, label=$J_0$}{i1,v}
			\fmf{dashes}{o1,v,o2}
			\Ball{v}
		\end{fmfgraph*} 
		}}  
\non \\[12pt]
4\sqrt{2} \parbox{\myl}{\centering{
	\begin{fmfgraph*}(1.4,1)
			\fmfleft{i1}
			\fmfright{o1,o2,o3,o4}	
			\fmf{wiggly, tension=2.5, label=$J_0$, label.dist=-0.6cm}{i1,v}
			\fmf{plain}{o1,v,o2}
                   \fmf{plain}{o3,v,o4}
			\Ball{v}
		\end{fmfgraph*} 
		}}   =
\parbox{\myl}{\centering{
	\begin{fmfgraph*}(1.4,1)
			\fmfleft{i1}
			\fmfright{o1,o2,o3}	
			\fmf{wiggly, tension=1.8, label=$J_0$}{i1,v}
			\fmf{plain}{o1,v,o2}
                   \fmf{dashes}{o3,v}
			\Ball{v}
		\end{fmfgraph*}
		}}  \non
\]
\end{fmffile}
}{Global WIs relating the coupling constant with one external $J_0$ field.
}{extWI}

{\centering \subsubsection{I. Flow for $\m_{h}^{J_{0}}$, d=2}}

In the same spirit of what already done in the three dimensional case we want to control the flow of $\m_h^{J_0}$ by the comparison with the one of $\m_h$.
In two dimensions both $\m_h$ and $\m_h^{J_0}$ are relevant with dimension $1/2$. Their flow equations are
\[ 
\m_{h-1}-\g^{\frac{1}{2}}\m_{h} & =  \tl{\b}_{h}^{\m}+\b_{h}^{*\,\mu}  \label{flow_mu2} \\[3pt]
\m_{h-1}^{J_{0}}-\g^{\frac{1}{2}}\m_{h}^{J_{0}} & =  \tl{\b}_{h}^{J_{0},\mu} +\b_{h}^{* J_0} \label{flow_mu_J2}
\]
where $\tl{\b}_{h}^{\m}$ and $\tl{\b}_{h}^{J_{0},\mu}$ are the part of the beta function with the external dashed line or the $J_0$ line comes from one of the local terms at scale $h \leq k <\bh$. The remaining part of the two beta functions,  $\b_{h}^{*}$ and $\b_{h}^{* J_0}$ respectively, contains the diagrams where the external dashed or wiggly line comes from an irrelevant vertex at scale $\bh$. The main point here is that $\tl{\b}_{h}^{\m}$ and $\tl{\b}_{h}^{J_{0},\mu}$ are identical once one has changed the relevant vertex giving the external dashed line (\ie one of the vertices $\m_k$, $\m'_k$, $\l'_k$ or $\o_k$) with the corresponding vertex with the wiggly external line.\\

Let us consider the trees contributing to $\b_{h}^{*\,\mu}$. 
Having some attention one can check that we can always extract from the branch connecting the irrelevant vertex giving the external dashed line a short memory factor   $\g^{\th(h -\bh)}$ with $\th <1$. It is sufficient to remember that the effective internal dimension in two dimensions and in the lower region is equal to the renormalized scaling dimension $\d^{2d,<}_v - z_v^{2d,<}$ plus $1/2$ for each external dashed leg which may be contracted at lower scale. There are two vertices with effective scaling dimension lower than $1$, as reported in fig.~\ref{danger}. However in our case the external dashed leg coming from scale $\bh$ remains external along all the path connecting the irrelevant vertex to the root. Then the vertices along this path have effective dimension $-1$ or lower. Then, using \eqref{2d_hstar} and the short memory property we get  
\[  \label{mu_star2}
|\b_{h}^{*\,\mu}| \leq  \l \l_* \,\g^{\frac{3}{4}h}
\]  
In the previous bound the factor $\g^{-\frac{3}{4}\bh}$ which one had expected in the short memory factor is compensated by the initial value of the irrelevant endpoint at $\bh$.  The choice of taking $\th=3/4$ depends on the fact that we need a short memory larger than $1/2$ to get a summable contribution, as discussed below. 
%%
%\nota{in \eqref{mu_star2} ho usato un solo loop, ma ne ho almeno due giusto?}
%%
Regarding $\b_{h}^{*\,J_0}$ we have in hand the same short memory factor than for $\b_{h}^{*\,\mu}$, the discussion being very similar, since a $J_0$ line dimensionally behaves as a non contracted dashed line. Using \eqref{2d_hstar} we get
\[
|\b_{h}^{*\,J_0}| \leq \e^{-1} \l \,\g^{\frac{1}{2}(h-\bh)}
\]
%---------------------------------------------------------------
\feyn{
\begin{fmffile}{feyn-TESI/danger}
\unitlength = 1 cm  
\def\myl#1{2.3cm}
\[
 & \parbox{\myl}{\centering{
	\begin{fmfgraph*}(1.4,1)
			\fmfleft{i1}
			\fmfright{o1,o2}	
			\fmf{dashes, tension=1.2}{i1,v}
			\fmf{plain}{o1,v}
			\fmf{dashes, tension=0.8}{v,o2}
			\fmfdot{v}
		\end{fmfgraph*}  \\
              $-\frac{3}{2}$
		}}  
 \parbox{\myl}{\centering{
	\begin{fmfgraph*}(1.4,1)
			\fmfleft{i1}
			\fmfright{o1,o2,o3}	
			\fmf{dashes, tension=1.8}{i1,v}
			\fmf{plain}{o1,v}
			\fmf{plain}{o3,v,o2}
			\fmfdot{v}
		\end{fmfgraph*}  \\
              $-1$
		}}  \non
\]
\end{fmffile}
}{{\bf Irrelevant kernels, $d=2$.} We report here the two irrelevant vertices with internal effective scaling dimension $1/2$. For each kernel the renormalized scaling dimension $\d^{2d,<}_v -z^{2d,<}_v$ is reported; the effective scaling dimension is obtained by adding $1/2$ for each dashed line which can be contracted at some lower scale. 
}{danger}

Now we are ready to discuss how to bound the flow equation of $\m_h^{J_0}$. Using \eqref{flow_mu2} and \eqref{flow_mu_J2}, assuming that
\[
\frac{\m_k^{J_0}}{\mu_{k}} = \frac{\m_h^{J_0}}{\m_h} \lft(1 + O(\l ) \rgt) \label{cond_mu2}
\]
and following the same strategy used for the three dimensional case, one finds
\[
\frac{\m^{J_0}_{h}}{\m^{J_0}_{\bh}} =\frac{\m_{h}}{\mu_{\bh}}\prod_{k=h+1}^{\bh}\left(1-\frac{\b_{k}^{*\m}}{\m_{k-1}}\right) \lft(1 + O(\l ) \rgt)
+ \sum_{k=h+1}^{\bh} \b^{*J_0}_k
\]
Using \eqref{mu_star2} we get
\[
\frac{|\b_{k}^{*\m}|}{\m_{k-1}} \leq O\lft(\l\,\g^{\frac{k}{4}}\rgt) 
\]
which is summable over $k$ thanks to the fact that we could extract from $\b_{k}^{*\m}$  a short memory factor greater than $1/2$. Finally we obtain
\[ \label{flow_mJ0_2d}
\frac{\m_h^{J_0}}{\m_\bh^{J_0}}  = \frac{\m_h}{\m_\bh}  \lft(1+O(\l)   \rgt) 
\] 
with
\[
\m_\bh=\,\frac{\sqrt{2}}{4}\e^{-\frac{1}{2}}\lft(1+O(\l) \rgt) \hskip 1cm \m^{J_0}_\bh=\e^{-\frac{1}{2}}\lft(1+O(\l) \rgt)
\]
Then
\[
2\sqrt{2}\,\m_h^{J_0} = \e^{-1}\m_h \lft(1+O(\l ) \rgt)  
\]
which also verifies \eqref{cond_mu2}. With similar discussions we can also see that at leading order $\m'^{J_0}_h =\e^{-1}\m'_h$,  $\l'^{J_0}_h =\e^{-1}\l'_h$ and $\o^{J_0}_h =\e^{-1}\o_h$.

By using the WIs \eqref{GWI_Z} and \eqref{LWI_E} and the flow equation  \eqref{flow_mJ0_2d} we get
\[
E_h &  \simeq \e^{-1}\,Z_h \lft(1+O(\l ) \rgt)   & d=2
\]

\vskip 1cm

{\centering \subsubsection{II. Flow for $E_{h}^{J_{0}}$, d=2}}

Once the flow of $\m_h^{J_0}$, $\m'^{J_0}_h$, $\l'^{J_0}_h$ and $\o^{J_0}_h$ are controlled, the flow equations of the two--legged vertices with external fields are studied in complete analogy with the three dimensional case, apart from the fact that now there are four vertices giving the external $J_0$ line. Denoting with $\b^{E,*}_k$ and $\b^{E_{J_0},*}_k$ the parts of the beta function for $E_h$ and $E_h^{J_0}$ with the external dashed or wiggly line coming from an irrelevant diagram at scale $\bh$ we have:
\[
E_{h}^{J_{0}} - E_{\bh}^{J_{0}} & = \e^{-1} \bigl(1 + O(\l) \bigr) \Big(E_h - E_\bh +\sum_{k=h+1}^{\bh} |\b^{E,*}_k|\Big)  +\sum_{k=h+1}^{\bh} \b^{E_{J_0},*}_k
\]
Being
\[
& \b_{k}^{E,*} = O( \l  \g^{\th k})   \hskip 1cm
 \b_{k}^{E_{J_0},*} =  O(\l \e^{-1}  \g^{\th k}) 
\]
with $0<\th<1/2$ and 
\[
 E_{\bh}^{J_{0}}=O\lft(\l^2\e^{-1}\rgt) \hskip1cm  E_{\bh}=1+O\lft(\l \rgt)
\]
we can conclude that
\[   \label{E0_2d}
 E_{h}^{J_0}=-\e^{-1} \lft( 1- E_h  +O(\l) \rgt)  
\]
Since 
\[
& E_h = \e^{-1}\,\l_*\,\g^h  
\]
in the infrared limit
\[
 E_{-\io}^{J_0}=-\e^{-1} \lft(1 +O(\l)\rgt)
\]
Using the local WI \eqref{LWI_B} one obtains
\[ 
\sqrt{2}\,B_h & \simeq  \e^{-1} \lft( 1- E_h +O(\l) \rgt)    & d=2
\]

\feyn{
\begin{fmffile}{feyn-TESI/Estar2}
\unitlength = 1 cm  
\def\myl#1{3.8cm}
\[
\parbox{\myl}{\centering{
	\begin{fmfgraph*}(3,2)
			\fmfleft{i1}
			\fmfright{o1}
			\fmftop{t}
			\fmfbottom{b}
			\fmf{phantom, tension=1.8}{t,v3}    %---
			\fmf{phantom, tension=1.8}{b,v4}    %---
                  \fmf{dashes, tension=2.2}{i1,v1}	
			\fmf{dashes,right=0.3}{v1,v4}
			\fmf{plain, right=0.3}{v4,v2}		
			\fmf{dashes,tension=1}{v1,vc}
			\fmf{plain,tension=1}{vc,v2}
			\fmf{dashes, left=0.3}{v1,v3}			
			\fmf{plain,left=0.3}{v3,v2}			
			\fmf{plain, tension=2.2, label= $\dpr_0$, label.dist=-0.4cm}{o1,v2}
			\fmfdot{v1}
			\Ball{v1,v2}	
		\end{fmfgraph*}
		}}
\parbox{\myl}{\centering{
	\begin{fmfgraph*}(3,2)
			\fmfleft{i1}
			\fmfright{o1}
			\fmftop{t}
			\fmfbottom{b}
			\fmf{phantom, tension=1.8}{t,v3}    %---
			\fmf{phantom, tension=1.8}{b,v4}    %---
                  \fmf{dashes, tension=1.5}{i1,v1}			
			\fmf{dashes,tension=0.5, left=0.5}{v1,v3}
			\fmf{dashes,left=0.3}{v3,v2}			
			\fmf{dashes,right=0.3}{v1,v4}
			\fmf{plain,right=0.3}{v4,v2}
			\fmf{dashes, tension=0.5, right=1}{v1,v3}
		       \fmf{plain, tension=1.5, label= $\dpr_0$, label.dist=0.05cm}{o1,v2}
			\fmfdot{v1}
			\Ball{v1,v2,v3}	
		\end{fmfgraph*}
		}}
\] \vskip -0.5cm
\end{fmffile} 
}{Example of diagrams contributing to $\b^{E,*}_k$.}{Estar2}

\vskip 0.5cm

{\centering \subsubsection{III. Flow for $E_{h}^{J_{1}}$, d=2}}

The flow of $E_{h}^{J_{1}}$ is controlled with the same dimensional argument used in the three dimensional case. From each diagram contributing to the beta function of $E_h^{J_1}$ we can extract a short memory factor due to the presence of a long branch connecting   $\mu_{\bh}^{J_{1}}$ to the root. We have
\[
\b_h^{E,J_1}= O\lft(  \l\,\,\g^{-\frac{1}{2}(\bh-h)} \rgt)
\]
Being $E_{\bh}^{J_{1}}=O(\l^2)$ in two dimensions we conclude that
\[ \label{E2d}
& E_{h}^{J_{1}}  =  O(\l ) 
\]
By the  local WI \eqref{LWI_A} we get
\[
&  A_{h}-1  \simeq 1+ O(\l ) & d=2 
\]

\vskip 0.5cm

%\pagina

\subsection{A local WI for the propagator}  \label{local_for_prop}

Using the local Ward identity 
\[ \label{extra}
B_h = J_h \lft(1 +O(\l) \rgt)
\] 
see \eqref{WI_J_complete} for a derivation, it is possible to proof an identity useful to control the behavior of the propagator, \ie
\[ \label{WI_prop}
E^2_h+ Z_h B_h   \simeq Z_h \e^{-1}\,\lft(1 + O(\l) \rgt)
\]
holding both in three and two dimensions.
The starting point of the proof is the flow equation for $J_h$, which we can study through the analogy with the flow of $Z_h$. In fact we can image of substituting the two vertices giving the two external dashed lines in $Z_h$ with identical vertices with wiggly dashed lines. In the three dimensional case, denoting with $\m_k \m_k' \b_{h;k,k'}^{Z} $  the diagrams contributing to the beta function of $Z$ where the external dashed lines comes from $\m_k$ and $\m_k'$ we get
\[
J_{h-1} - J_h & =  \sum_{k,k'=h+1}^\bh \frac{\m_k^{J_0}}{\m_k}\,\frac{\m_{k'}^{J_0}}{\m_{k'}} \,\b_{h;k,k'}^{Z}  + \b^{*J_{J_0}}_h
\]
where $\b^{J_{J_0}*}_h$ contains the terms with the two $J_0$ lines coming from irrelevant vertices. Using that
\[
& \frac{\m_k^{J_0}}{\mu_{k}} = \frac{\m_h^{J_0}}{\m_h} \lft(1 + O(\l ) \rgt) & d=2,3 
\]
%\[
%\frac{\m_h^{J_0}}{\m_h} \lft(1-c_1 \l \e^{\frac{3}{2}}\g^k \rgt) & \leq \frac{\m_k^{J_0}}{\m_k} \leq \frac{\m_h^{J_0}}{\m_h} & d=3  \non \\
 %\frac{\m_h^{J_0}}{\m_h} \lft(1-c_2 \l \l_*\e^{\frac{1}{4}}\g^\frac{k}{4} \rgt) & \leq \frac{\m_k^{J_0}}{\m_k} \leq \frac{\m_h^{J_0}}{\m_h} & d=2
%\]
we obtain
\[
J_h - J_\bh & = \lft(\frac{\m_h^{J_0}}{\m_h}\rgt)^2 \Bigl( \frac{Z_\bh}{2} -\frac{Z_h}{2} -\sum_{k=h+1}^\bh \,\b_{k}^{* Z}  \Bigr) +\sum_{k=h+1}^\bh \,\b_{k}^{*J_{J_0}}
\]
Using the estimates %\eqref{b_star_Z} and \eqref{b_star_Z2} on the beta function $\b^{Z,*}_h$ in the three and two dimensional cases 
\[ \label{b_star_Z}
& \b_{h}^{Z,*} = O\lft(\l\,\e^{\frac{3}{2}}\, \g^{-(\bh -h)}\rgt) & d=3 \non \\
& \b_{h}^{Z,*} = O\bigl(\l^2\, \g^{\th(h-\bh)}\bigr) \quad 0<\th<1/2  & d=2
\]
and the analogous estimate for $\b^{J_{J_0} *}_h$ one gets
\[
J_h + \lft(\frac{\m_h^{J_0}}{\m_h}\rgt)^2 \frac{Z_h}{2} & = \e^{-1}\bigl(1+O(\l) \bigr)  & d=3 \non \\
J_h + \lft(\frac{\m_h^{J_0}}{\m_h}\rgt)^2 \frac{Z_h }{2}& = \e^{-1}\bigl(1+O(\l^2\e^{-1}) \bigr)  & d=2 
\]
 Finally, using the Ward identities \eqref{GWI_Z}, \eqref{LWI_E} and \eqref{extra} we get \eqref{WI_prop}. 

%E_h=2\sqrt{2}\e^{-1}\m_h}
\pagina 

\section{One-loop computations}

As shown in appendix \ref{order_e}  in order to keep in each flow equation only the leading terms in the small parameter $\e$ it is sufficient to consider the one loop graphs without $\n_h$ vertices. This property holds both at the beginning of the region $h \leq \bh$ and in the asymptotic region $h \arr -\io$, but in the latter case the dominant diagrams for $h \arr -\io$ are those with all the internal dashed lines contracted among them, see \eqref{3d_hstar} and \eqref{2d_hstar}. We shall study the flow equations for the effective parameters $\l_h$ in $3d$ and $\{\l_h, \l_{6,h}\}$ in $2d$, by keeping only these contributions; the properties of the corresponding solutions will be used to justify the approximation. 

\subsection{3d case: one--loop computation for $\l_h$} \label{lambda3}
We have seen how the use of WIs allows to reduce the flow of the running coupling constants $\l_h$, $\m_h$, $E_h$ and $Z_h$ to one unique independent coupling. In particular the leading order calculation of the beta function for $Z_h$ is very simple, since only one diagram contributes to it, see fig. \ref{flow_leading}. An explicit calculation (see appendix \ref{lowest_order}) gives
\[
Z_{h-1}= Z_h - \frac{1}{8} \l\,\e^{-\frac{1}{2}} \b^{3d}_{2}\,Z_h^2
\]
with
\[
\b_{2}^{3d} = \int \frac{d^{4}k}{(2\pi)^{4}} \frac{f_0(k)}{\lft(k_0^2 +\kk^2\rgt)^2} 
\]
If we approximate in the integral for $\b_{2}^{3d}$ the cutoff function $f_0(k)$ with the characteristic function $\th_\g(k_0^2 +\kk^2)$ of the set $\{2\g^{-2}/(1+\g^2) \leq k_0^2 +\kk^2 \leq 2\g^2/(1+\g^2)\}$ we obtain
\[
\b_{2,\th_\g}^{3d}=\frac{\log \g}{8\pi^2 }
\]
Then, being $Z_\bh=\e$, we find 
\[
Z_h = \frac{\e}{1+ \bar{c} \l\,\e^{\frac{1}{2}}|h-\bh|}
\]
with $\bar{c}=1/(64\pi^2)$. Since all the neglected terms in the beta function are at least of order $1/|h|^3$ (see \eqref{3d_hstar}, appendix \ref{order_e} ) they cannot change in a substantial way the asymptotic properties of the flow: only the value of the constant $\bar{c}$ depends on them. Then the effective interactions $\l_h$ and $\m_h$ and the wave functions renormalization constants $Z_h$ and $E_h$ are asymptotically free in the infrared limit.

\subsection{2d case: one loop computation for $\l_h$ and $\l_{6,h}$}   \label{lambda6}

In the two dimensional case the asymptotic flow of the running coupling constants is reduced to the flows of the two independent effective parameters $\l \l_h$ and $\l \l_{6,h}$. Let us consider the 
\[
 x_h = \l \l_h  \hskip 1cm y_h = \l \l_{6,h}
\]
whose initial values are
\[
x_\bh = \l  \hskip 1cm y_\bh=0
\]
A leading order explicit calculation in the case $\r_0 R_0^2=1$ (see \eqref{leading_2d} for the details) gives
\[ 
x_{h-1} - \g\,x_h &= - (\g-1)\,x_h^2\,\lft[ 4\,\tl{\b}^{2d}_2\;-\; 3\,\tl{\b}_1^{2d}\,\lft(\frac{y_h}{x_h^2}\rgt) \;+\; \frac{9}{4}\,\tl{\b}^{2d}_0 \lft(\frac{y_h}{x^2_h}\rgt)^2 \rgt]  \label{FP1}
 \\[12pt]
y_{h-1} - y_{h} &=  -(1-\g^{-1})\,\lft[ \, 8\tl{\b}^{2d}_3\,x_h^3 -12\,\tl{\b}^{2d}_2 y_h x_h  + \frac{27}{8}\,\tl{\b}^{2d}_0\, \frac{y^3_h}{x^3_h} \,\rgt]    \label{FP2}
\]
with
\[  \label{beta_def}
& \b_{n}^{2d} = \int \frac{d^{4}k}{(2\pi)^{4}} \frac{f_0(k)}{\lft(k_0^2 +\kk^2\rgt)^{n}}= (1-\g^{-1}) \,\tl{\b}_{n}^{2d}
\]
and  $\lim_{\g \arr 1} \tl{\b}_{n}^{2d}=1/\pi^2$ for $n=0,1,2,3$. The qualitative behavior of  \eqref{FP1} and \eqref{FP2} is equivalent to the differential equations obtained by taking the limit $\g \arr 1$, which are:
\[
\pi^2 \frac{dx}{dt} & = \pi^2 x - x^2\lft[ 4 - 3 \,\frac{y}{x^2} + \frac{9}{4} \lft(\frac{y}{x^2}\rgt)^2 \rgt]  \label{FP1b} \\[6pt]
\pi^2 \frac{dy}{dt} &= x^3 \lft[8 -12 \,\frac{y}{x^2} +  \frac{27}{8} \lft(\frac{y}{x^2}\rgt)^3 \rgt] \label{FP2b}
\]
with the limit $h \arr -\io$ corresponding to $t \arr \io$.  It is simple to see that the r.h.s. of  \eqref{FP1b} is always positive; than $x(t)$ grows starting from its initial value $\l$. Since $x(t)\neq 0$ for each $t$, also the r.h.s. of \eqref{FP2b} is always positive and $y(t)$, starting from zero, grows up to a fixed point, identified by the solution of the equation  
\[ \label{FP_z}
8\,-12\, z_*  + \frac{27}{8}\,z^2_* =0
\]
with $z(t)=y(t)/x^2(t)$.  As a consequence of that also $x(t)$ reaches a fixed point 
whose expression at leading order is
\[
x_* = \frac{1}{4-\; 3\,z_*\, + \frac{9}{4}\,z^2_* }
\]
We see that the presence of the effective marginal terms changes the fixed point for $x$, even at leading order, which otherwise would be $x_*=1/4$. Even if the qualitative behavior is not changed by the addition of the new terms and it turns to be of order one, as already stated in literature~\cite{CaDiC1}, this result is not trivial. In fact the computation of \eqref{FP1} and \eqref{FP2} in presence of the new terms is quite subtle, as shown in details in appendix~\ref{leading_2d}. A numerical analysis of the flow equations \eqref{FP1} and  \eqref{FP2} is shown in fig.\ref{fixedpoint}.
%--------------------
\fig{t}{0.6}{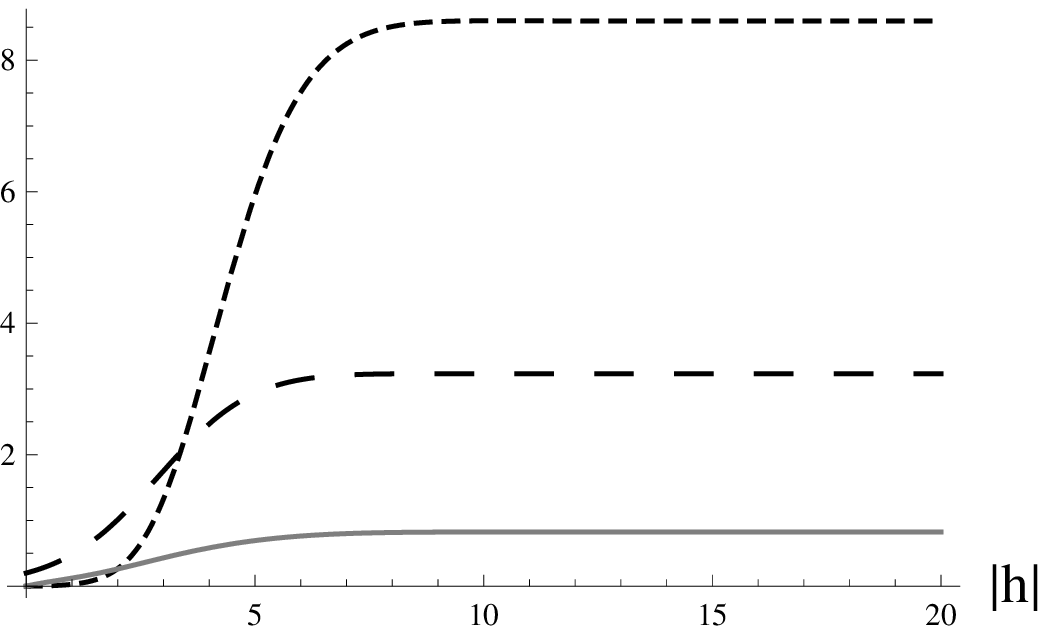}{{\bf Effective parameters, $d=2$.} Numerical solution of the leading order flow equations for $\l \l_h$ (long--dashed line) and $\l \l_{6,h}$ (dashed line). The plain line represents $\l_{6,h}/(\l \l^2_h)$ which is the additional factor to be added in the renormalized bounds for each vertex of type $\l_6$, $\o_h$, $\l'_h$ and $\l'_h$, see \eqref{asym_2d}.}{fixedpoint}
%------------------------------

We conclude with a remark about the role of the parameter $\r_0 R_0^2$, which we have carefully carried along the calculation, the reason being that one could hope to reduce the fixed point values by playing on the values of the condensate density and range of the potential. This is not the case.  In fact if $\r_0 R_0^2 \neq 1$ the flow equations for $x_h$ and $y_h$ are the same that \eqref{FP1} and  \eqref{FP1} apart for changing $y_h=\l \l_{6,h}$ with $y'_h=(\r_0 R_0^2)^{-1} y_h$.  So, while the fixed point for $y_h$ depends on $\r_0 R_0^2$ and in particular
\[
\frac{y_h}{x_h^2} = \r_0 R_0^2  \,z_* 
\]
the fixed point for $x_h$ only depends on the solution of \eqref{FP_z} which is unchanged. 

%\blue{ The problem to prove that $\l^*$ is of order one is beyond the possibility of a pertubative theory and then remains an open problem.} \\

%---------------APPENDICI

%\appendix
%\input{APP-transient}
%\input{APP-second-order.tex}

%\end{document}

%---- WI
%\input{intro-senza-sapclass} \input{intestazione-sap} \usepackage{showkeys}   \begin{document}% \tableofcontents

\chapter{Ward Identities} \label{WI}

In this chapter we discuss how to derive the global and local Ward identities and how to control the corrections to the formal Ward identities coming from the cutoffs. Referring to the latter purpose we prove here the identities widely used in chap.~\ref{multiscale}:
\[
A_h -1 & \simeq O(\l \,\e^{\frac{d-2}{2}}) \non \\[3pt]
\sqrt{2}\,B_h & \simeq \e^{-1}\lft(1-E_h + O(\l)\rgt)  \non \\[3pt]
E_h & \simeq \e^{-1} \m_h  \bigl(1+ O(\l)\bigr) \non \\[3pt]
E^2_h + Z_h B_h & \simeq \e^{-1} Z_h (1 +O(\l))
\]
where ``$\simeq$'' means that the identity holds at the leading order in $k$ as $k\arr 0$. 

We will also compare these identities with the ``formal'' ones obtained by neglecting the corrections coming from the ultraviolet cutoff. We will see that the correction terms affects the local WIs at the second non trivial order in the small parameter $\l$. Even if this does not change the way in which the local WIs are used to control the flow equations, the prediction on the physical observables are quantitatively changed. For example, the correction to the speed of sound with respect to Bogoliubov prediction is changed at the first non trivial order in $\l$ if one consider the complete WIs instead of the formal ones. Regarding the magnitude in $\l$ of the prediction on the renormalized speed of sound it remains to understood if it depends on the presence of the ultraviolet cutoff; however this is an example of how the correction terms to the formal WIs may quantitatively change the relations between physical observables.

%\blue{ The correction to the local WI's coming to the presence of the cutoff function in our effective model do not change the qualitative behavior of the flow, giving higher order corrections in the small parameter $\l$ to the formal WIs. On the other side the correction terms are different from zero in the infrared limit $h^* \arr -\io$ and then quantitatively change the relations between physical observables which can be derived starting from the WI's.  }

% NOTA: in 3d le formal WI sembrano cambiate gia' al secondo ordine perche' in E_h passo da \l \e^{1/2} a ordine \l.

The Ward identities we are going to derive are relations between the kernels of the generating functional $\WW_{h^*}(\f, J_0, {\bf J}_1)$ defined in section \ref{sec:gen_fun} derived by performing a phase transformation of the fluctuation fields $\ps^\pm$. 
The expression of the functional integral in terms of the original fluctuations fields
\[ \label{G0}
e^{|\L|\WW_{h^*}(\f,J_{0},{\bf J}_{1})}= &\, \int P^{\,0}_{\c_{[h^*,0]}} (d\ps)
 e^{-\VV_I(\ps+\f) 
+ \int_\L dx \left[\,J_{x}^{0} \,\psi_x^{+}\psi_x^{-}
+\,{\bf J}_{x}^{1} \cdot \left(\ps_x^- \dpr_\xx \ps_x^{+} 
- \psi_x^+ \dpr_\xx \ps_x^- \right)\right] }  
\]
where $P^{\,0}_{\c_{[h^*,0]}} (d\ps)$ is the measure with covariance
\[
g^{0,[h^*]}_{\s \s'}(x) &= \frac{1}{(2\pi)^{d+1}} \int d^{d+1}k\, \c_{[h^*,0]}(k)\, g^{0}_{\s \s'}(k) \non \\[6pt]
(g^{0}_{\s \s'}(k))^{-1} &=  \left(\begin{array}{cc}
-i k_0 +\kk^{2} & 0\\
0 & i k_0 +\kk^{2}
\end{array}\right)
\]
and
\[ \label{4.V}
\VV_I(\ps+\f)  = &\,  \frac{\l}{2}\, \hv(\bz) \int_{\L}
\left[ \lft(\sqrt{\r_0}+\ps_x^+ +\f_x^+  \rgt) \lft(\sqrt{\r_0}+\ps_x^- +\f_x^-  \rgt) \right]^{2}dx \non \\
 & - \m_I \int_{\L} \lft(\sqrt{\r_0}+\ps_x^+ +\f_x^+  \rgt) \lft(\sqrt{\r_0}+\ps_x^- +\f_x^-  \rgt) dx 
\]
with $\m_I=\m_B +\n$ the interacting chemical potential. We remark that the relations between the kernels of the potential \eqref{G0}, which we will refer to as ``one--step'' potential, are not the ones we are interested in. In order to control the flow equations we need relations between the kernels of the effective multiscale potentials obtained including at each step of the integration the local quadratic terms in the measure. With this aim we will consider a sequence of ``one--step'' potentials, or {\it reference models}, with infrared cutoff on scale $h$, which have trivial invariance properties under the phase transformation of the $\ps$ fields. For each $h$ it is simple to proof that the difference between the kernels of the effective potential and the kernels of the ``one--step'' potential is subdominant both in $h$ as $h \arr -\io$ and in the small parameter of the perturbation theory, see appendix \ref{one-step} for the discussion of this point. Since we are interested in the behavior of the system in the infrared limit $h \arr -\io$, for the purposes of our work it is equivalent to use the reference model or the effective model. \\

%The local Ward identities we are going to derived hold for the kernels $\hW^{(h^*)}_{n_l n_t}(k_1,\ldots,k_{n_l +n_t})$ of the ``one--step potentials'' obtained integrating the fields at scales higher than $h^*$. 

%We stress here that the effective potentials introduced in our multiscale scheme do not have trivial invariance properties under the phase transformation of the $\ps$ fields. However we can also obtain relations for the kernel of the effective potentials, by using the fact that the difference between the kernels of the effective potential and the kernels of the one--step potential is small, as shown in appendix~\ref{one-step}. \\

The scheme of the chapter is the following:
\begin{enumerate}[\it Sec. 4.1 \:]
\item We derive the global WIs for the one--step potential, whose interpretation at the one--loop level is also reported. By the use of a global WI, the condition of minimum of the effective potential $\WW(\x)$ is shown to be equivalent to a condition on the effective chemical potential $\n_h$.
\item This section is devoted to the derivation of the local WIs, which is obtained through the following steps:
\begin{enumerate}[a.]
\item We derive the formal WIs where ``formal'' means that we are neglecting the corrections terms coming from the cutoff functions. 
\item We derive the complete (\ie ``non--formal'') local WIs, taking into account the contributions coming from the cutoff functions. 
\item We describe the multiscale integration of the correction terms, which correspond to new marginal (also relevant in $2d$) kernels whose beta functions can be again rewritten as a series in the running coupling constants. 
\item  We study  the flow of the new marginal running coupling constants corresponding to the correction terms and prove that they give corrections of higher order in $\l$ to the formal WI's.
\end{enumerate}
\end{enumerate}

\vskip 1cm

{\it Remark.} The reader may notice that with respect to the works \cite{BM-luttinger, GRanom} where the Ward identities are derived for the Schwinger functions, \ie by differentiation of the potential
\[ \label{Schwing}
& {\cal U}_{\L, h^*}(\f,J_{0},{\bf J}_{1})= \non \\[6pt]
& \quad \frac{1}{|\L|} \log \int P^{0}_{\c_{[h^*,0]}} (d\ps)
 e^{-\VV_I(\ps) 
+ \int_\L dx \left[ \ps^+_x\f^-_x + \ps^-_x \f^+_x + \,J_{x}^{0} \,\psi_x^{+}\psi_x^{-}
+\,{\bf J}_{x}^{1} \cdot \left(\ps_x^- \dpr_\xx \ps_x^{+} 
- \psi_x^+ \dpr_\xx \ps_x^- \right)\right] } 
\]
here we analyze directly the kernels of the effective potentials, which can be obtained by differentiation from eq.~\eqref{G0}. In our context, to the purpose of deriving relations between the running coupling constants, it is equivalent to study this slightly modified potential rather than the one in eq.~\eqref{Schwing}.
%Our choice is due to to the fact that we are interested in finding relations between the running coupling constants and the wave function renormalization constants, in order to control their flows. 

\pagina
%-----------------------------------------------------------------------
\section{Global WIs}  \label{GWI}

The global WI's are derived by the gauge invariance of the generating functional $\WW(\f, J_0, J_1)$ under the transformation of the bosonic fields \mbox{$\psi_{x}^{\pm}\arr e^{\pm i\th}\psi_{x}^{\pm}\,$}, with $\th$ constant. Note that the Jacobian of the latter transformation is one and the density and current terms are invariant under a global phase transformation of the $\ps^\pm_x$ fields, then only $\VV_I(\ps)$ is effected by the rotation. We have:
\[ \label{G0_f}
& \WW_{h^*}(\f,J)= \non \\
& \quad \frac{1}{|\L|} \log \int P_{\c_{[h^*,0]}} (d\ps)
 e^{-\VV_I(e^{i\th}\ps+\f) 
+ \int_\L dx \left[\,J_{x}^{0} \,\psi_x^{+}\psi_x^{-}
+{\bf J}_{x}^{1} \cdot \left(\ps_x^- \dpr_\xx \ps_x^{+} 
- \psi_x^+ \dpr_\xx \ps_x^- \right)\right] }  
\]
Deriving the latter expression with respect to $\th$ and then setting $\th=0$ we obtain the equality:
\[ \label{G1}
\langle \int dx\left[\frac{\dpr \VV_I}{\dpr \f_{x}^{+}} \left(\phi_{x}^{+}+\sqrt{\r_0}\right) - \frac{\dpr \VV_I}{\dpr \f_{x}^{-}} \left(\f_{x}^{-}+\sqrt{\rho_{0}}\right)\right] \rangle_{h^*} =0
\]
with
\[
\langle \,\cdot\, \rangle_{h^*}=\frac{1}{Z}\int P_{\c_{[h^*,0]}} (d\ps)\,(\,\cdot\,) \,
e^{-\VV_I(\ps_x +\f_x)
+\int_\L dx\left[\,J_{x}^{0} \,\psi_x^{+}\psi_x^{-}
+\,{\bf J}_{x}^{1} \cdot \left(\ps_x^- \dpr_\xx \ps_x^{+} - \psi_x^+ \dpr_\xx \ps_x^- \right)\right]}  
\]
Introducing the fields $\psi_{x}^{l}$ and $\psi_{x}^{t}$ such that \mbox{$\psi_{x}^{\pm}=\sqrt{\frac{\r_0}{2}} \lft(\psi_x^{l}\pm i \psi_x^{t} \rgt)$}  and $\f_{x}^{l}$ and $\f_{x}^{t}$ such that \mbox{$\f_{x}^{\pm}=\sqrt{\frac{\r_0}{2}} \lft(\f_x^{l}\pm i \f_x^{t} \rgt)$} the generating functional becomes
\[ \label{G0_long}
& \WW_{h^*}(\f,J)= \non \\
& \quad \frac{1}{|\L|} \log \int P_{\c_{[h^*,0]}} (d\ps)
 e^{-\VV_I(\ps+\f) 
+ \r_0 \int_\L dx \left[\,J_{x}^{0} \,\frac{1}{2}\,\lft((\psi_x^{l})^2 + (\psi_x^{t})^2 \rgt)
+ {\bf J}_{x}^{1} \cdot \left(\ps_x^l i\,\dpr_\xx \ps_x^{t} 
- \psi_x^t i\,\dpr_\xx \ps_x^l \right)\right] }  
\]
and the relation \eqref{G1} may be written as
\[  \label{G2}
 \langle \int dx\left[ \,
\frac{\dpr \VV_I}{\dpr \phi_{x}^{t}}\left(\phi_{x}^{l}+\sqrt{2}\right) -
\frac{\dpr \VV_I}{\dpr \phi_{x}^{l}}\left(\phi_{x}^{t}\right) 
\, \right]\rangle_{h^*}  =0
\]
where
\[
\big\langle\frac{\dpr \VV_I(\psi+\phi)}{\dpr \phi_{x}^{\a}}\big\rangle_{h^*}= 
-\frac{\d \WW_{h^*}(\f) }{\d \phi_{x}^{\a}} \quad\qquad\a=l,t
\]
and we may rewrite \eqref{G2} as
\[ \label{GWI}
\int dx\left[\frac{\delta\WW_{h^*}(\f)}{\delta\phi_{x}^{t}}\,(\phi_{x}^{l}+\sqrt{2})-\frac{\d \WW_{h^*}(\f)}{\d \phi_{x}^{l}}\,(\phi_{x}^{t})\right] 
=0 
\] 
We remind to the reader the expression of the interacting potential in terms of the $\ps^l$ and $\ps^t$ fields:
\[ \label{4.VI}
\VV_I(\ps) & = \r_0 R_0^{-2} \, \Biggl(\frac{\e}{16} \int_{\L}
\left(\bigl(\psi_x^l \bigr)^2+\bigl(\psi_x^t\bigr)^2 \right)^{2}dx
+ \frac{\e}{4} \,\sqrt{2} \int_{\L}\bigl(\psi_x^l \bigr)^2 \bigl(\psi_x^l+\psi_x^t\bigr)dx  \non \\
& \qquad + \frac{\e}{2} \int_\L \bigl(\psi_x^l \bigr)^2\,dx + \frac{ \n}{2}\,R_0^2 \int_{\L} \left(\bigl(\psi_x^l \bigr)^2+\bigl(\psi_x^t\bigr)^2 \right) dx \,\Biggr)
\]
with $\n=\l \r_0 \hv(\bz)-\m_I$ the correction to the interacting chemical potential with respect Bogoliubov chemical potential $\m_B=\l \r_0 \hv(\bz)$. 

In order to obtain relations among the kernels of the one-step effective potential $\WW_{h*}(\f)$, defined by
\[
W_{mn}^{(h^*)}(x_{1},\ldots,\, x_{m};\, y_{1},\ldots,\: y_{n})=
\frac{1}{m!n!}\,
\frac{\delta^{m+n}\,\WW_{h^*}(\f)}{\delta\phi_{x_{1}}^{l}\ldots\delta\phi_{x_{m}}^{l}\delta\phi_{y_{1}}^{t}\ldots\delta\phi_{y_{n}}^{t}}\Big|_{\phi=J_0=0,\,{\bf J}=\bz}
\]
all we have to do is apply to \eqref{GWI} an arbitrary functional derivative with respect to the $\f^{\a}$ fields and then set all the external fields equal to zero. By deriving \eqref{GWI}  $m$ times with respect to the field $\f^l$ and $n$ times with respect to $\f^t$  we obtain the following relations for the kernels $W_{mn}^{(h^*)}(x_{1},\ldots,\, x_{m};\, y_{1},\ldots,\: y_{n})$:
\[
\sqrt{2} &\,m!\, (n+1)! \int dx  W^{(h^*)}_{m,n+1}(x_1,\ldots,x_{m};\,y_1,\ldots,y_n,x)  \non \\
& +(m-1)!\,(n+1)! \sum_{j=1}^m W^{(h^*)}_{m-1,n+1}(x_1,\ldots,x_{j-1},x_{j+1},\ldots,x_m;\,y_1,\ldots,y_n,x_j) \non \\
& \quad -(m+1)!\, (n-1)! \sum_{j=1}^n W^{(h^*)}_{m+1,n-1}(x_1,\ldots,x_m, y_j;\,y_1,\ldots,y_{j-1},y_{j+1},\ldots,y_n) =0
\]
This implies the following relation for the Fourier tranforms of the kernels evaluated at zero external momentum:
\[  \label{GWI_fourier}
&  \sqrt{2}\,(n+1) \hW_{0,n+1}^{(h^*)}(0,\ldots,0)- \hW_{1,n-1}^{(h^*)}(0,\ldots,0)=0  & \text{for } m =0 \non \\[6pt]
& \sqrt{2}\,(m+1) \hW_{m,1}^{(h^*)}(0,\ldots,0)+ \hW_{m-1,1}^{(h^*)}(0,\ldots,0) =0 &  \text{for } n=0 \non \\[6pt]
& \sqrt{2}\,(n+1) \hW_{m,n+1}^{(h^*)}(0,\ldots,0)+\,\,(n+1) \hW_{m-1,n+1}^{(h^*)}(0,\ldots,0) \non \\[6pt]
& \qquad \qquad \qquad -\,(m+1)\hW_{m+1,n-1}^{(h^*)}(0,\ldots,0)=0  & \text{for } m,n \neq 0
\]

\vskip 0.2cm

%{\bf WI's for RRC's and wave function renormalization constants.} 
As shown in appendix~\ref{one-step} the difference between the zero momentum kernels $\hW_{m n}^{(h)}$ of the one--step effective potential $\WW_{h}(\f,J)$ and the running coupling constant at scale $h$, obtained by the multiscale integration, is small, since it comes only from the integration on the last scale $h$. Then we can use the identities \eqref{GWI_fourier} to get relations between the running coupling constants. 

The identities relating $\l_h$, $\m_h$ and $Z_h$ are obtained choosing $m=n=1$ or $m=0$ and $n=3$; in these two cases we get:
\[
\,\sqrt{2}\,\hW_{12}^{(h)}(0,0,0) +\,\hW_{02}^{(h)}(0,0)-\hW_{20}^{(h)}(0,0)=0 \non \\[6pt]
4\,\sqrt{2}\,\hW_{04}^{(h)}(0,0,0,0) -\hW_{12}^{(h)}(0,0,0) =0
\]
Since the difference between the kernels $\hW_{n_l n_t}$ of the one--step potential and the kernels $\hV_{n_l n_t}$ of the multiscale potential is subdominant in $h$ as $h\arr -\io$, see appendix~\ref{one-step}, the previous relation correspond to 
\[ \label{GW_flow_3d}
2\,\sqrt{2}\,\m_{h} +2\g^{2h}\n_h -Z_h \simeq 0 \non \\[6pt]
4\,\sqrt{2}\,\l_h - \m_h\simeq 0
\]
in the three dimensional case
and 
\[ \label{GW_flow_2d}
2\,\sqrt{2}\,\g^{\frac{h}{2}}\m_{h} +2\g^{2h}\n_h -Z_h \simeq 0 \non \\[6pt]
4\,\sqrt{2}\,\g^{h}\l_h - \g^{\frac{h}{2}}\m_h\simeq 0
\]
in the two dimensional case, where the symbol ``$\simeq$'' denotes that the two sides are equal apart for subdominant terms in $h$. 
%
% the iterative inclusion in the measure of the local quadratic terms in the fields $\psi$. 

The global Ward identities necessary to control the flow of the three additional marginal couplings $\l'_h$, $\m'_h$ and $\o_h$ arising in the two dimensional case are:
\[
& 6\,\sqrt{2}\,\hW^{(h)}_{06} -\hW^{(h)}_{14}=0 \non \\[6pt]
& 2\,\sqrt{2}\,\hW^{(h)}_{14} -\hW^{(h)}_{22}=-2\hW^{(h)}_{04} \non \\[6pt]
& 2\,\sqrt{2}\,\hW^{(h)}_{22} -3\hW^{(h)}_{30}=-2\hW^{(h)}_{12} 
\]
which correspond to
\[
& 6\,\sqrt{2}\,\l_{6,h} \simeq \g^{-\frac{h}{2}}\o_h \non \\[6pt]
& 2\,\sqrt{2}\,\g^{-\frac{h}{2}}\o_h -\g^{-h} \l'_h\simeq-2\g^h\l_h \non \\[6pt]
& 2\,\sqrt{2}\,\g^{-h} \l'_h -3 \g^{-\frac{3}{2}h} \m'_h \simeq -2\g^{\frac{h}{2}}\m_h 
\]
Further global Ward identities are useful to individuate which are the null local kernels of the effective potentials; in particular 
\[
& \sqrt{2}\,\hW_{01}^{(h)}=0   & \Rightarrow   \quad \hW_{01}^{(h)}=0 \non \\
& 2\,\sqrt{2}\,\hW_{11}^{(h)}+\hW_{01}^{(h)}=0   & \Rightarrow \quad \hW_{11}^{(h)}=0 \non \\
& 3\,\sqrt{2}\,\hW_{03}^{(h)}-\hW_{11}^{(h)}=0 & \Rightarrow   \quad \hW_{03}^{(h)}=0 \non \\
& 3\,\sqrt{2}\,\hW_{21}^{(h)}+\hW_{11}^{(h)}=0 & \Rightarrow  \quad \hW_{21}^{(h)}=0 \non \\
& 3\,\sqrt{2}\,\hW_{13}^{(h)}+3\,\hW_{03}^{(h)}-2\,\hW_{21}^{(h)}=0 & \Rightarrow   \quad \hW_{13}^{(h)}=0 \non \\
& 3\,\sqrt{2}\,\hW_{31}^{(h)}+\hW_{21}^{(h)}=0 & \Rightarrow   \quad \hW_{31}^{(h)}=0
\] 
Comparing the previous result with the expression of the potential $\VV_I(\ps)$ in \eqref{4.VI} we note that the previous Ward Identities fix that the kernels with four, three or two legs that may be generated at each scale $h$ cannot be different from the ones already present in the original potential.

\subsection{Renormalization condition} \label{ren_condition}

The condition that the effective potential $\WW_\L(\x)$ reaches its minimum for $|\x|^2=\r_0$ corresponds to the request
\[
0=\frac{\d}{\d \x} \WW_{\L}(\x) = -\bmed{\frac{\d }{\d \ps^l_x} \,\VV_I(\ps +\x)}{-\io}
\]
where we have used \eqref{4.V} and that $\frac{\d }{\d \x } \,\ph^\pm_x= \frac{\d }{\d \ps^l_x} \ph^\pm_x $. Since $\WW(\r_0)=\lim_{h^* \arr -\io}W_{h^*}(\r_0)$, the minimum condition corresponds to
\[
\lim_{h^* \arr -\io} \hW_{10}^{(h^*)}=0
\] 
%
%Finally the identity allows to fix the behavior of the relevant coupling $\g^{2h}\n_h$ through the renormalization condition, which is a condition on $\hW^{[h^*]}_{10}$.
Then, using the global Ward identity
\[ \label{ren_WI}
2\,\sqrt{2}\,\hW^{(h^*)}_{02} -\hW^{(h^*)}_{10}=0
\]
we see that the minimum condition is equivalent to the requirement 
%the condition $\hW^{[-\io]}_{02}=0$, that is 
\be \label{nu_infinity}
\lim_{h^* \arr -\io}\g^{2h^*}\n_{h^*} =0  
\ee
which is certainly satisfied in our model, since we have fixed $\n_{-\io}$ so that $|\n_h|$ is bounded for each $h$, see sec.~\ref{nu}.

%
%
%-------------------------------------------------------------- PERTURBATIVE WIs
\feyn{
\begin{fmffile}{feyn-TESI/GWI}
 \unitlength = 0.8cm
\def\myl#1{2.8cm}
\begin{align*}
	&   \bar{\b}^\l_h \;=
	\parbox{\myl}{\centering{
			\begin{fmfgraph*}(2.5,1.4) 
			\fmfleft{i1,i2}
			\fmfright{o1,o2}
			\fmf{plain}{i1,v1,i2}
			\fmf{plain,left=0.8, tension=0.4}{v1,v2}
			\fmf{plain,right=0.8, tension=0.4}{v1,v2}
			\fmf{plain}{o1,v2,o2}
			\fmfv{decor.shape=circle,decor.filled=empty,decor.size=.08w}{v1,v2}			
\end{fmfgraph*}  
			}} 
			+  \parbox{\myl}{\centering{
			\begin{fmfgraph*}(2.5,1.4) 
			\fmfleft{i1,i2}
			\fmfright{o1,o2}
			\fmf{plain}{i1,v1,i2}
			\fmf{dashes,left=0.8, tension=0.4}{v1,v2}
			\fmf{dashes,right=0.8, tension=0.4}{v1,v2}
			\fmf{plain}{o1,v2,o2}
			\fmfv{decor.shape=circle,decor.filled=empty,decor.size=.08w}{v1,v2}
			\end{fmfgraph*}  
			}} +
\parbox{\myl}{\centering{
			\begin{fmfgraph*}(2.5,1.4) 
            \fmfleft{i1,i2}
			\fmfright{o1,o2}
			\fmftop{t}
			\fmfbottom{b}
			\fmf{phantom, tension=4}{t,v3}    
			\fmf{phantom, tension=4}{b,v4}   
            \fmf{plain, tension=1.5}{i1,v1,i2}			
			\fmf{plain,left=0.4, tension=1}{v1,v3}
			\fmf{dashes,left=0.4, tension=0.8}{v3,v2}			
			\fmf{plain,right=0.4, tension=1.2}{v1,v4}
			\fmf{dashes,right=0.4, tension=0.8}{v4,v2}
			\fmf{plain, tension=1.2}{o1,v2,o2}
			\fmfv{decor.shape=circle,decor.filled=empty,decor.size=.08w}{v1,v2}					\end{fmfgraph*}  
			}}  \\[12pt]
			%-------------------------------\mu_h----------------------------
		& \bar{\b}^\m_h \;=
	\parbox{\myl}{\centering{
			\begin{fmfgraph*}(2.5,1.4) 
			\fmfleft{i1}
			\fmfright{o1,o2}
			\fmf{dashes, tension=1.5}{i1,v1}
			\fmf{plain,left=0.8, tension=0.4}{v1,v2}
			\fmf{plain,right=0.8, tension=0.4}{v1,v2}
			\fmf{plain}{o1,v2,o2}
			\fmfv{decor.shape=circle,decor.filled=empty,decor.size=.08w}{v1,v2}
			\end{fmfgraph*}  
			}} 
			+\;
\parbox{\myl}{\centering{
			\begin{fmfgraph*}(2.5,1.4) 
			\fmfleft{i1}
			\fmfright{o1,o2}
			\fmf{dashes, tension=1.5}{i1,v1}
			\fmf{dashes,left=0.8, tension=0.4}{v1,v2}
			\fmf{dashes,right=0.8, tension=0.4}{v1,v2}
			\fmf{plain}{o1,v2,o2}
			\fmfv{decor.shape=circle,decor.filled=empty,decor.size=.08w}{v1,v2}
			\end{fmfgraph*}  
			}} 
			+\;
\parbox{\myl}{\centering{
			\begin{fmfgraph*}(2.5,1.4) 
            \fmfleft{i}
			\fmfright{o1,o2}
			\fmftop{t}
			\fmfbottom{b}
			\fmf{phantom, tension=4}{t,v3}    
			\fmf{phantom, tension=4}{b,v4}   
            \fmf{dashes, tension=1.5}{i,v1}			
			\fmf{plain,left=0.4, tension=1}{v1,v3}
			\fmf{dashes,left=0.4, tension=0.8}{v3,v2}			
			\fmf{plain,right=0.4, tension=1.2}{v1,v4}
			\fmf{dashes,right=0.4, tension=0.8}{v4,v2}
			\fmf{plain, tension=1.2}{o1,v2,o2}
			\fmfv{decor.shape=circle,decor.filled=empty,decor.size=.08w}{v1,v2}					\end{fmfgraph*}  
			}}  +
\parbox{2.5cm}{\centering{
			\begin{fmfgraph*}(2.5,1.4) 
                   \fmfleft{i}
			\fmfright{o1,o2}
			\fmftop{t}
			\fmfbottom{b}
			\fmf{phantom, tension=4}{t,v3}    
			\fmf{phantom, tension=4}{b,v4}   
            \fmf{dashes, tension=1.5}{i,v1}			
			\fmf{dashes,left=0.4, tension=1}{v1,v3}
			\fmf{plain,left=0.4, tension=0.8}{v3,v2}			
			\fmf{dashes,right=0.4, tension=1.2}{v1,v4}
			\fmf{plain,right=0.4, tension=0.8}{v4,v2}
			\fmf{plain, tension=1.2}{o1,v2,o2}
			\fmfv{decor.shape=circle,decor.filled=empty,decor.size=.08w}{v1,v2}					\end{fmfgraph*}  
			}} 
			\\[12pt]
			%----------------------------------------Zh--------------------
			&  
	\bar{\b}^Z_h \;= \parbox{\myl}{\centering{
			\begin{fmfgraph*}(2.5,1.4) 
			\fmfleft{i1}
			\fmfright{o1}
			\fmf{dashes, tension=1.5}{i1,v1}
			\fmf{plain,left=0.8, tension=0.4}{v1,v2}
			\fmf{plain,right=0.8, tension=0.4}{v1,v2}
			\fmf{dashes}{o1,v2}
			\fmfv{decor.shape=circle,decor.filled=empty,decor.size=.08w}{v1,v2}
			\end{fmfgraph*}  
			}} +\;  
\parbox{\myl}{\centering{
			\begin{fmfgraph*}(2.5,1.4) 
			\fmfleft{i1}
			\fmfright{o1}
			\fmf{dashes, tension=1.5}{i1,v1}
			\fmf{dashes,left=0.8, tension=0.4}{v1,v2}
			\fmf{dashes,right=0.8, tension=0.4}{v1,v2}
			\fmf{dashes}{o1,v2}
			\fmfv{decor.shape=circle,decor.filled=empty,decor.size=.08w}{v1,v2}
			\end{fmfgraph*}  
			}} +\;
\parbox{\myl}{\centering{
			\begin{fmfgraph*}(2.5,1.4) 
                   \fmfleft{i}
			\fmfright{o}
			\fmftop{t}
			\fmfbottom{b}
			\fmf{phantom, tension=4}{t,v3}    
			\fmf{phantom, tension=4}{b,v4}   
            \fmf{dashes, tension=1.5}{i,v1}			
			\fmf{plain,left=0.4, tension=1}{v1,v3}
			\fmf{dashes,left=0.4, tension=0.8}{v3,v2}			
			\fmf{plain,right=0.4, tension=1.2}{v1,v4}
			\fmf{dashes,right=0.4, tension=0.8}{v4,v2}
			\fmf{plain, tension=1.2}{o,v2}
			\fmfv{decor.shape=circle,decor.filled=empty,decor.size=.08w}{v1,v2}					\end{fmfgraph*}  
			}} 
\end{align*}
\end{fmffile}	
}{{\bf Flow equations $\bh <h \leq 0$.} Second order contributions to the beta function of $\bl_h$, $\bm_h$ and $\bn_h$. }{GWI1}

\subsection{Perturbative interpretation of the WIs}

The interpretation of the global Ward identities \eqref{GW_flow_3d} and \eqref{GW_flow_2d} is plain if we look at the leading order contributions to the beta function of the kernels $W^{(h)}_{04}$, $W^{(h)}_{12}$ and $W^{(h)}_{20}$. The interpretation is quite different in the higher momentum region  $\bh < h < 0$ and lower momentum region $h\leq \bh$. 

In the higher momentum region the scaling dimensions of the kernels do not depend on the type (plain or dashed) of the legs; then we have three four--legged running coupling constants, \ie $\l_h$, $\l'_h$ and $\l''_h$, and two three--legged running coupling constant, \ie $\m_h$ and $\m'_h$. Moreover the order  in the small parameter $\e$ increases with the number of vertices, so the leading order contributions to the beta function come from the second order diagrams, see fig.~\ref{GWI1}. The global WIs are an expression of the fact that at each order the diagrams contributing to the beta functions of $\bl_h$, $\bm_h$ and $\bar{Z}_h$ are exactly the same, the only difference staying in the external legs, see again fig.~\ref{GWI1}. More precisely it is sufficient to change one of the vertices $\l_k$ or $\l'_k$ on the first line of fig.~\ref{GWI1} with a vertex $\m_k$ or $\m'_k$ to get the beta function for $\bm_h$ instead of the one for $\bl_h$. In the same manner, by changing both  the two four--legged vertices in $\bar{\b}^\l_h$ we obtain the beta function of $\bar{Z}_h$.
With this exchanges the contracted legs remains the same, but the combinatorial factors associated at each diagram change, in such a way that the constants factor in \eqref{GW_flow_3d} and \eqref{GW_flow_2d} are obtained. \\

In the region  $h \leq \bh$ the dimensional scaling of the plain and dashed legs become different and the running coupling constants with more than two external legs are $\{\l_h, \m_h\}$ in $3d$ and $\{\l_h, \m_h, \l'_h, \m'_h, \o_h, \l_{6,h}\}$ in $2d$. The
 order in $\e$ in this region is given by the number of loops instead than the number of vertices. This depends on the identity
\[ \label{identity}
2\,\sqrt{2}\,\m_h  \lft[ g^{(h)}_{tt}(k)\,g^{(h)}_{ll}(k)+ \bigl(g^{(h)}_{tl}(k)\bigr)^2 \rgt] = g^{(h)}_{tt}(k)
\]
which implies that the sum of a certain combination of $n$--order diagrams gives the same contribution in the small parameter $\e$ than a $(n-1)$--order diagram (the order here refers to the number of vertices). For example, referring to the flow equation of $\m_h$, on the second line of fig.~\ref{flow_leading} pag.~\pageref{flow_leading}, the identity \eqref{identity} implies that the sum of the three diagrams of third order contributing to the beta function gives a contribution of the same order in $\e$ with respect to the second order diagram in the same figure. In this case the global WIs ensure that the sum of all the $n$--loop diagrams for each $n$ are proportional through \eqref{GW_flow_3d} and \eqref{GW_flow_2d}. The validity of the global WI  for $\l_h$, $\m_h$ and $Z_h$  at the one--loop level is easily checked, see appendix~\ref{B.flow} for an explicit computation.

\pagina
%-----------------------------------------------------------------------------------------------

\section{Local WIs}

In this section we will derive the local Ward identities, which have been used in chap.~\ref{flows} in order to control the flow of $A_h$, $B_h$, $E_h$ and to bound the renormalized propagator. Moreover one may also derive relations between the two point renormalization functions and the density--density and current--current kernels. As well known the latter identities are related to the physical relations between two points correlation functions and the response functions.

\subsection{Derivation of the formal WIs}

In this section we discuss the ``formal'' local WIs, \ie the local WIs obtained for a modification of the generating functional \eqref{G0}, obtained by considering the measure $P^{\,0}(d\ps)$, having the same propagator of $P^{\,0}_{\c_{[h^*,0]}}(d\ps)$ but without the cutoff function $\c_{[h^*,0]}(k)$. The corrections to the formal WI due to the presence of the cutoff function in \eqref{G0} will be discussed in the next section. 

Under a local gauge transformation of the fluctuation fields $\psi_{x}^{\pm} \arr e^{\pm i\th_x}\psi_{x}^{\pm}$ the measure $P(d\ps)$ is not invariant under the transformation. 
It is simple to see it using the formal expression for the measure  
\[
P(d\psi)= \frac{1}{\NN}\, D\psi \,e^{-\int dx [\psi_{x}^{+}(-\partial_{x_0}+\Delta)\psi_{x}^{-}]}
\]
with $\NN$ a normalization factor. By performing a local gauge transformation of $\WW_{h^*}(\f, J_0, {\bf J}_1)$, deriving
%\begin{eqnarray*}
%\psi_{x}^{+}(-\dpr_0+\D)\psi_{x}^{-} & \arr & e^{i\th_{x}}\psi_{x}^{+} (-\dpr_0+\Delta)(e^{-i\th_{x}}\psi_{x}^{-})
%\end{eqnarray*}
with respect to $\th_x$ and setting $\th_x$ to zero we get a term similar to the one in  \ref{GWI} (except for the absence of the integration over the $x$ variable) plus an extra term coming from the measure which is the media $\bmed{\cdot}{h^*}$ of the following derivative:
\[ \label{current}
&  \frac{\dpr}{\dpr \th_x} \lft\{-\int dx [e^{i\th_{x}}\psi_{x}^{+}(-\dpr_{x_0}+\D)(e^{-i\th_{x}}\psi_{x}^{-})] \rgt\}_{\th_x=0}  
\non \\[6pt]  
&\hskip 2cm
 = i\,\lft[ \dpr_{x_0} (\psi_{x}^{+}\psi_{x}^{-}) + \dpr_{\xx} (\psi_{x}^{-}\dpr_{\xx} \psi_{x}^{+}-\psi_{x}^{+}\dpr_{\xx} \psi_{x}^{-}) \rgt] \non \\[6pt]
& \hskip 2cm 
= i\,\r_0 \lft[ \,\frac{1}{2}\,\dpr_{x_0} \lft((\psi_{x}^{l})^2+(\psi_{x}^{t})^2\rgt)+ i\,\dpr_{\xx} (\psi_{x}^{l}\dpr_{\xx} \psi_{x}^{t}-\psi_{x}^{t}\dpr_{\xx} \psi_{x}^{l})
\right] \non \\[6pt]
& \hskip 2cm 
=  i \dpr_{x_\n} j^\n_x \qquad \n=0,1,2,3
\]
with $\dpr_{x_\nu}=(\dpr_{x_0},\dpr_{\xx})$ and
\[
j_{x}^{\nu}=\begin{cases}
\,\frac{\r_0}{2}\lft[(\psi_{x}^{l})^2+(\psi_{x}^{t})^2\rgt] \quad & \n=0 \\[6pt]
\,\r_0\,\lft[\psi_{x}^{l}\,i\dpr_{\xx} \psi_{x}^{t}-\psi_{x}^{t}\,i\dpr_{\xx} \psi_{x}^{l} \rgt] & i=1,2,3
\end{cases}
\]
Note that if we introduce the adimensional variables $(x'_0, x'_i)=(R_0^{-2}\,x_0, R^{-1}_0 x_i)$ we get  that the terms $\dpr_{x_\n} j_x^\n$ have dimensions $\r_0 R_0^{-2}$, as expected. Taking into account the definition in \eqref{current} we get the following local ward identity:
\[  \label{LWId}
\frac{\delta\WW_{h^*}}{\delta\phi_{x}^{t}}\,\lft(\phi_{x}^{l}+\sqrt{2}\rgt)-\frac{\delta \WW_{h^*} }{\delta\phi_{x}^{l}}\,\phi_{x}^{t}  -i \, \bmed{\dpr_{x_\nu}\, j_{x}^{\nu}}{h^*}= 0 
\]
In the Fourier space, using $\ps^\a_x = |\L|^{-1} \sum_k e^{ikx} \ps^\a_k$ we have
\[ \label{current_F}
-i\,\dpr_{x_\nu}\, j_{x}^{\nu} = |\L|^{-2} \sum_{k,p} e^{ikx} \lft[ \frac{1}{2}\,p_0 \lft(\ps^l_{k+p}\ps^l_{-k} + \ps^t_{k+p}\ps^t_{-k} \rgt) +\pp \lft(\pp + \kk\rgt) \ps^l_{k+p}\ps^t_{-k} \rgt]
\]
The term $\bmed{j_{x}^{\nu}}{h^*}$ can be written as functional derivative of the potential $\WW_{h^*}(\f, J_\n)$ with respect to the external fields $J_0$ and ${\bf J}_1$:
\[
\bmed{j_{x}^{\nu}}{h^*}=\frac{\d \WW_{h^*}(\f,J_\n)}{\d J_{\nu}^{x}} \bigg|_{\,\phi,J_\n=0}
\]
In the following we will indicate the kernels with an external leg of type $\n$ and one or two external plain (or dashed) legs as follows
\[ \label{kernel_nu}
W_{01;\nu}^{(h^*)}(y;\,x) & = \frac{\d^{2}\WW_{h^*}(\phi,J_\n)}{ \d\phi_{y}^{t} \d J_{x}^{\nu}} \Big|_{\phi,J=0} = - \,\bmed{\frac{\d \VV_I(\f)}{\d \phi_{y}^{t}};\, j_{x}^{\nu}}{h^*}
\non \\[6pt]
2!\,W_{02;\nu}^{(h^*)}(y,z;\, x) & =\frac{\d^{3}\WW_{h^*}(\phi,J_\n)}{ \d\phi_{y}^{t} \d \phi_{z}^{t} \d J_{x}^{\nu}} \Big|_{\phi,J=0} = \frac{\d}{\d J_{x}^{\n}} \,\bmed{\frac{\d^{2}\VV_I(\f)}{\d \phi_{y}^{t} \d\phi_{z}^{t}}}{h^*}
\non \\[6pt]
& =- \,\bmed{\frac{\d^{2}\VV_I(\f)}{\d \phi_{y}^{t}\d \phi_{z}^{l}};\, j_{x}^{\nu}}{h^*} +\, \bmed{\frac{\d\VV_I(\f)}{\d \phi_{z}^{t}}; \frac{\d\VV_I(\f)}{\d \phi_{y}^{t}};\, j_{x}^{\nu}}{h^*}
\]
where the semicolon indicates the connected expectation and the $2!$ factor on the second lines corresponds to the combinatorial factor $n_l!\,n_t!$.
%\ie $\bmedia{A\,;\,B}= \bmedia{A\,B}-\bmedia{A} \bmedia{B}$. 
The Fourier transform of  \eqref{kernel_nu} is defined by 
\[
W_{11;\nu}^{(h)}(x,y;\, z) & =  \frac{1}{(\b L^{d})^{2}}\sum_{k,p} e^{ipx} e^{iky}  e^{-i(k+p)z}\,\hW_{11;\nu}^{(h)}(p,k)
\]
when we have taken into account the fact that the kernels are translational invariant in the $x$ space. The local WI's are obtained by \ref{LWI} deriving with respect to $\f_{y}^{l}$ or $\phi_{y}^{t}$ an appropriate number of times. The local WIs which are used to control the flow are listed in the following.\\

\begin{enumerate}[\bf (1)]

\item{\bf Local WI for $A_{h}$.} This identity is obtained by deriving \eqref{LWId} once with respect to $\phi^{t}_y$ and then setting to zero the external sources:
\[
2\sqrt{2}\,W^{(h)}_{02}(x,y)-W^{(h)}_{10}(x) \d(x-y) - i \dpr_{x_\n} \,W^{(h)}_{01;\nu}(y;\,x) =0
\] 
In the momentum space the previous identity becomes
\[
2\sqrt{2}\, \hW_{02}^{(h)}(p)- \hW_{01}^{(h)}(0) +\,p_\n \hW_{01;\n}^{(h)}(p) =  0
\]
Using the global Ward identity $\hW_{01}^{(h)}(0)=2\sqrt{2}\,\hW_{02}^{(h)}(0)$ one gets
\[ \label{WI_AB}
2\sqrt{2}\, \lft(\hW_{02}^{(h)}(p)- \hW_{02}^{(h)}(0)\rgt)  = -\,p_\n \hW_{01;\n}^{(h)}(p) 
\]
Now, choosing as external momentum $p = (0,\pp)$ : 
\[
2\sqrt{2}\, \lft(\hW_{02}^{(h)}(\pp)- \hW_{02}^{(h)}(\bz)\rgt)  = -\,\pp \cdot \hW_{01;{ 1}}^{(h)}(\pp) 
\]
where $\pp \cdot \hW_{01;{1}}^{(h)}(\pp)$ is a shortcut for $\sum_i p_i \,\hW_{01;i}^{(h)}(\pp)$. We have 
\[
2\sqrt{2}\, \lft(\hV_{02}^{(h)}(\pp)- \hV_{02}^{(h)}(\bz)\rgt)  \simeq -\,\pp \cdot \hV_{01;{ 1}}^{(h)}(\pp) 
\]
The previous relation at the second order in $\pp$ gives
\[
\sqrt{2}\,(A_{h}-1)\simeq-E_{h}^{J_{1}}  \label{4.WI_A}
\]
both for the two and three dimensional case. The identity \eqref{4.WI_A} is immediately verified at $h=0$ being $E_{0}^{J_{1}}=0$ and $A_{0}=1$.  \\[6pt]

\item{\bf Local WI for $B_{h}$ . } This identity is obtained by choosing in \eqref{WI_AB} the external momentum as $p=(p_0; \bz)$:
\[
2\sqrt{2}\, \lft(\hW_{02}^{(h)}(p_0)- \hW_{02}^{(h)}(0)\rgt)  = -\,p_0 \hW_{01;0}^{(h)}(p_0) 
\]
The latter relation at the second order in $p_{0}$ gives:
\[
\sqrt{2}\, B_{h}\simeq -E_{h}^{J_{0}}        \label{4.WI_B}
\]
both for the two and three dimensional case. For $h=0$ the \eqref{4.WI_B} is immediately verified since $B_{0}=E_{0}^{J_{0}}=0$.  \\[6pt]

\item{ \bf Local WI for $E_h$ . } By deriving \eqref{LWId} twice with respect to the field $\phi^{t}$ we get the identity
\[
3\sqrt{2}\, W^{(h)}_{03}(x,y,z) -W^{(h)}_{11}(x,y) \d(x-z) & - W^{(h)}_{11}(x,z) \d(x-y) \non \\[6pt] 
& = 2\,i\, \dpr_\n W^{(h)}_{02;\n}(y,z;x) 
\]
that in momentum space corresponds to
\[
3\sqrt{2}\, \hW^{(h)}_{03}(p,k) - \hW^{(h)}_{11}(-k)  - \hW^{(h)}_{11}(k+p) = -2\,p_\n \hW^{(h)}_{02;\n}(k ;p) 
\]
Choosing as external momentum $p =(p_{0},\bz)$ we obtain the relation
\[
3\sqrt{2}\,\hW_{03}^{(h)}(k,p) - \lft(\hW_{11}^{(h)}(k+p)- \hW_{11}^{(h)}(k)\rgt) =-2\,p_0 \,\hW^{(h)}_{02;0}(k; p)
\]
that at the first order in $p_0$ gives
\[ \label{4.WI_E_h}
& E_{h}\simeq 2\,\m_{h}^{J_{0}}  & d=3 \non \\
& E_{h}\simeq \g^{\frac{h}{2}}2\,\m_{h}^{J_{0}}  & d=2
\]
in the lower momenta region. For $h=\bh$ in fact $\m_\bh=1/2$ and $E_\bh=1$. For $ \bh < h \leq 0 $ one has to take into account the fact that the kernels on the two sides of the equation have different scaling dimensions and that $\bar{E}_0$ is equal to zero, while $\bm_0$ to $1/2$. One obtains \eqref{LWI_E_min}.

\item{\bf Alternative identity for $E_{h}$.}

By deriving \eqref{LWId} once with respect to $\phi^{l}_y$  and then setting to zero the external sources:
\[
2\sqrt{2}\,W^{(h)}_{11}(x,y)+W^{(h)}_{10}(x) \d(x-y) - i \dpr_{x_\n} \,W^{(h)}_{10;\nu}(y;\,x) =0
\] 
which  in the momentum space corresponds to
\[
2\,\sqrt{2}\, \hW_{11}^{(h)}(p)+\hW_{01}^{(h)}(0) +\,p_\n \hW_{10;\n}^{(h)}(-p)  =  0 
\]
Using the global Ward identity $2\sqrt{2}\,\hW_{11}^{(h)}(0)= -\hW_{10}^{(h)}(0)$ and then choosing the external momentum to be $p=(p_0,\bz)$ one get
\[
2\,\sqrt{2}\, \lft(\hW_{11}^{(h)}(p_0)-\hW_{11}^{(h)}(0)\rgt) +\,p_0 \hW_{10;0}^{(h)}(-p_0)  =  0 
\]
The previous relation at the first order in $p_0$ gives 
\[
2\sqrt{2}\,\lft(E_{h}-1\rgt)\simeq Z_{h}^{J_{0}}
\] 
both for the two and three dimensional case and $h \leq \bh$. 

\item {\bf Local WI for the density--density kernel $J_{h}$.}  By deriving \eqref{LWId}  with respect to the external field $J_\n^y$ one gets the identity
\[
\sqrt{2}\,W^{(h)}_{01;\m}(x;y) - i \dpr_{x_\n} \,W^{(h)}_{00;\n \m}(x,y) =0
\] 
that is
\[   \label{WI_JK}
\sqrt{2}\,\hW^{(h)}_{01;\m}(p) + p_\n \,\hW^{(h)}_{00;\n \m}(p) =0
\] 
Choosing $p=(p_0,\bz)$ and $\m=0$ we get
\[
\sqrt{2}\,\hW^{(h)}_{01;0}(p_0) + p_0 \,\hW^{(h)}_{00;00}(p) =0
\]
which at the first order in $p_0$ gives
\[
\sqrt{2}\,E_{h}^{J_0}\simeq \,- J_{h}
\]
Combining the latter identity with \eqref{4.WI_B} 
\[  
B_{h}\simeq J_{h}    \label{4.WI_J}
\]
which is an useful identity in order to study the behavior of the propagator, as seen in section \ref{local_for_prop}.  

\item {\bf Local WI for the current--current kernel $K_{h}$.}  By using \eqref{WI_JK} and choosing $p=(0,\pp)$ and $\m=j$ we get
\[
\sqrt{2}\,\hW^{(h)]}_{01; j}(\pp) + \pp \cdot \,\hW^{(h)}_{00;{\bf 1} j}(\pp) =0
\]
Combining the latter identity with \eqref{4.WI_A} 
\[  
2\,\lft(A_{h}-1\rgt)\simeq K_{h}    \label{4.WI_K}
\]
where $K_h$ is the kernel with two $J_1$ external fields.
\end{enumerate}

\vskip 0.5cm
\subsubsection{Global WIs between kernels with external fields}

By applying to the identity \eqref{GWI} a functional derivative with respect one of the external fields $J_\n^y$ one obtains:
\[ \label{GWI_J}
\int dx\left[\frac{\d^2\WW_{h^*}(\f)}{\d J_y^\n\,\d\phi_{x}^{t}}\,(\phi_{x}^{l}+\sqrt{2})-\frac{\d^2 \WW_{h^*}(\f)}{\d J_y^\n\,\d \phi_{x}^{l}}\,(\phi_{x}^{t})\right] 
=0 
\] 
Starting from \eqref{GWI_J} and by applying an arbitrary functional derivative with respect to the $\f^\a$ fields we obtain a new series of global WIs involving the kernel with external $J_\n$ fields. Two particularly useful identities are obtained choosing $\n=0$ and $(m,n)=(1,1), (0,3)$ with $m$ and $n$ the number of derivatives with respect to the fields $\ps^l$ or $\ps^t$ respectively:
\[ \label{GWI_J2}
\sqrt{2}\, \hW_{12;0}(0,0,0) +2 \hW_{02;0}(0,0) & =\hW_{20;0}(0,0) \non \\[6PT]
4\sqrt{2}\, \hW_{04;0}(0,0,0,0) & = \hW_{12;0}(0,0,0) 
\]
where we are using the definitions \eqref{kernel_nu} for the kernels with external fields. The global WIs in \eqref{GWI_J2} are represented in fig.~\ref{extWI} pag.~\ref{extWI} and useful to control the flow of the running coupling constants with external fields in the two dimensional case. 
%-----------------------------------------------------------------------------------------------
\pagina

\subsection{Corrections to WIs due to cutoffs}

Now we are ready to include the corrections to local Ward Identities coming from the presence of a cutoff function $\c_{[h^*,0]}(k)$ in the generating functional \eqref{G0}. The formal expression of the measure in this case is the following
\[
P_{\chi_{[h^*,0]}}(d\psi)=\frac{1}{\mathcal{N}}\, D(\psi)\, e^{-\int dx \,\psi_{x}^{+}\,D_{[h^*,0]}\psi_{x}^{-}}
\]
with $D_{[h^*,0]}$ an operator on the $\ps$ fields whose action is defined by
\[
D_{[h^*,0]}\psi_{x}^{-}=\frac{1}{|\L|}\,\sum_{k}e^{-ikx}\chi_{[h^*,0]}^{-1}(k)(i k_{0} -\kk^{2})\psi_{k}^{-}
\]
As in the previous section, the local Ward identities are derived by performing a local gauge transformation of the fields $\ps^\pm_x$ in the generating functional $\WW_{h^*}(\f,J)$, by deriving it with respect to the phase variable $\th_x$ and then by setting $\th_x$ to zero. We obtain:
\[  
& \frac{\delta\WW_{h^*}}{\delta\phi_{x}^{t}}\,\lft(\phi_{x}^{l}+\sqrt{2}\rgt)-\frac{\delta \WW_{h^*} }{\delta\phi_{x}^{l}}\,\phi_{x}^{t}  \non \\
& \qquad +\bmed{\frac{\dpr}{\dpr \th_x} \lft\{\int dx [e^{i\th_{x}}\psi_{x}^{+}D_{[h^*,0]}(e^{-i\th_{x}}\psi_{x}^{-})] \rgt\}_{\th_x=0} }{h^*} = 0 
\]
%which differs from \eqref{LWI} for the term on the last line. 
where
\[ \label{local1}
\int dx & [e^{i\th_{x}}\psi_{x}^{+}D_{[h^*,0]}(e^{-i\th_{x}}\psi_{x}^{-})] = \int dx\psi_{x}^{+}D_{[h^*,0]}\psi_{x}^{-} \non \\ 
&+i\int dx\left(\th_{x}\psi_{x}^{+}D_{[h^*,0]}\psi_{x}^{-}-\psi_{x}^{+}D_{[h^*,0]}(\th_{x}\psi_{x}^{-})\right) + O(\th_x^2)
\]
The second term on the r.h.s. of \eqref{local1} can be conveniently written in the Fourier space, using $\psi_{x}^{\pm}=|\L|^{-1} \sum_{k}e^{\pm ikx}\psi_{k}^{\pm}$ and $\th_{x}= |\L|^{-1} \sum_{k}e^{-ikx}\th_{k}$:
\[
\frac{i}{|\Lambda|^{2}}\sum_{k,p} \th_k \left\{ \psi_{k+p}^{+} \chi^{-1}(k)(i k_{0}-\kk^{2})\psi_{k}^{-}-\psi_{k}^{+} \chi^{-1}(k) (i k_{0}-\kk^{2}) \psi_{k-p}^{-}\right\} 
\]
Once derived \eqref{local1} with respect to $\th_{x}$ and set $\th_x=0$ we obtain:
\[ \label{local2}
T_{x}(\psi^{\pm}) & = \frac{i}{|\L|^{2}}\sum_{k,p}e^{ipx}\psi_{k+p}^{+} \Bigl\{\chi_{[h^*,0]}^{-1}(k) \bigl[i k_{0}-\kk^{2}\bigr] \non \\ 
& \hskip 3cm - \chi_{[h^*,0]}^{-1}(k+p)\big[i(k_{0}+p_{0})-(\kk+\pp)^{2}\big]\Bigr\} \psi_{k}^{-} \non \\[6pt]
&=  \frac{1}{|\L|^{2}}\sum_{k,p}e^{ipx}\psi_{k+p}^{+}\,\lft[2\,T_0(k,p) +\,i T_1(k,p)\rgt]\,\psi_{k}^{-}
% & =  \frac{1}{|\L|^{2}}\sum_{k,p}e^{ikx}\psi_{k+p}^{+}\left[(k_{0}+p_{0})\chi_{[h^*,0]}^{-1}(k+p)-p_{0}\chi_{[h^*,0]}^{-1}(p)\right]\psi_{p}^{-} \non \\
 %&   -\frac{i}{|\L|^{2}}\sum_{k,p}e^{ikx}\psi_{k+p}^{+}\left[(\kk+\pp)^{2}\chi_{[h^*,0]}^{-1}(k+p)-\pp^{2}\chi_{[h^*,0]}^{-1}(p)\right]\psi_{p}^{-}
\]
with
\[
T_{0}(k,p) & =  \frac{1}{2}\left[(k_{0}+p_{0})\chi^{-1}(k+p)-k_{0}\chi^{-1}(k)\right]  \non \\[6pt]
T_{1}(k,p) & =   \lft[(\kk+\pp)^{2}\chi^{-1}(k+p)-\kk^{2}\chi^{-1}(k) \rgt]
\]
Changing the basic fields 
\[
\psi_{k}^{\pm} & = \sqrt{\frac{\r_0}{2}}\, \lft(\psi_{\pm k}^{l} \pm i\psi_{\pm k}^{t}\rgt)
\]
the first contribution to~\eqref{local2} becomes
\[ \label{T0}
T^0_x(\psi^{l},\psi^{t})=\frac{\r_0}{|\L|^{2}} \sum_{k,p} e^{ipx} \,T_0(k,p) \left[\psi_{k+p}^{l}\psi_{-k}^{l}+\psi_{k+p}^{t} \psi_{-k}^{t}\right]
\]
where we have used the fact that 
\[
\sum_{k,p}e^{ipx} T_{0}(k,p) \left[\psi_{k+p}^{t}\psi_{-k}^{l}-\psi_{k+p}^{l}\psi_{-k}^{t}\right]=0
\]
being $T_{0}(-(k+p),p)=T_{0}(k,p)$. Similarly
\[ \label{T1}
& T_{x}^{1}(\psi^{l},\psi^{t})=\frac{\r_0}{|\L|^{2}} \sum_{k,p}e^{ipx} \,T_{1}(k,p) \,\psi_{k+p}^{l}\psi_{-k}^{t}
\]
where we have used that
\[
\sum_{k,p}\psi_{k+p}^{\a}T_{1}(k,p)\psi_{-p}^{\a}=\frac{1}{2}\sum_{k,p}\left[\psi_{k+p}^{\a}T_{1}(k,p)\psi_{-k}^{\a}+\psi_{-k}^{\a}\left(-T_{1}(k,p)\right)\psi_{k+p}^{\a}\right]=0
\]
being $T_{1}(-(k+p),p)=-T_{1}(k,p)$. Taking into account the terms coming from the derivative of the free measure we get the following identity:
\[  \label{WI_complete}
& \frac{\delta\WW_{h^*}}{\delta\phi_{x}^{t}}\,\lft(\phi_{x}^{l}+\sqrt{2}\rgt)-\frac{\delta \WW_{h^*} }{\delta\phi_{x}^{l}}\,\phi_{x}^{t} + \bmed{T_{x}(\ps^l, \psi^t)}{h^*} =0
\]
where $T_{x}(\ps^l, \psi^t)=T_{x}^{0}(\ps^l, \psi^t)+T_{x}^{1}(\ps^l, \psi^t)$. The term $T_{x}(\ps^l, \psi^t)$ is the sum of two pieces: one piece already present in the formal local WI \eqref{LWI} plus a correction term which is different from zero only if the cutoff function is different from the identity. In order to extract the  correction term it is useful to rewrite $T^0_x$ and $T^1_x$ as follows:
\[
T_{0}(k,p) & =    \frac{1}{2}\,p_{0}+\frac{1}{2}\left[(k_{0}+p_{0})(\chi^{-1}(k,p)-1)-k_{0}(\chi^{-1}(k)-1)\right] \non \\[6pt]
 & =  \frac{1}{2}\,p_{0}+C_{0}(k,p) \non \\[6pt]
T_{1}(k,p) & =  \lft[(\kk+\pp)^{2}-\kk^{2}\rgt]+\left[(\kk+\pp)^{2}(\chi^{-1}(k+p)-1)-\kk^{2}(\chi^{-1}(k)-1)\right] \non \\[6pt]
 & =  \pp\cdot(\pp+2\kk)+C_{1}(k,p)
\]
to be compared with \eqref{current_F}. Then \eqref{WI_complete} may be rewritten as:
\[  \label{WI_complete2}
& \frac{\delta\WW_{h^*}}{\delta\phi_{x}^{t}}\,\lft(\phi_{x}^{l}+\sqrt{2}\rgt)-\frac{\delta \WW_{h^*} }{\delta\phi_{x}^{l}}\,\phi_{x}^{t} -i \, \bmed{\dpr_{x_\nu}\, j_{x}^{\nu}}{h^*} + \bmed{C_{x}(\ps^l, \psi^t)}{h^*} =0
\]
with $C_x(\ps^l, \ps^t)=C^0_x(\ps^l, \ps^t)+C^1_x(\ps^l, \ps^t)$ and $C^0_x$ and $C^1_x$ defined as  $T^0_x$ and $T^1_x$ in \eqref{T0} and \eqref{T1} with $T^\n(k,p)$ substituted by $C^\n(k,p)$.

\subsection{Multiscale integration of the correction terms to the WIs}

In order to perform the multiscale integration of the correction terms to the local WIs we introduce a new effective potential $\TL{\WW}_{[h,0]}(\f, \tl{J})$ defined by: 
\[ \label{W_tilde}
e^{|\L|\,\TL{\WW}_{[h,0]}(\f, \tl{J})} := \int P^{\,0}_{[h,0]}(d \ps) e^{-\VV_I(\ps + \f)  +\TL{B}(\tl{J},\f)}
\]
with
\[
%& B(J,\f) = \r_0 R_0^2 \int dx  \left[\,J_{x}^{0} \,\frac{1}{2}\,\lft((\psi_x^{l})^2 + (\psi_x^{t})^2 \rgt) + {\bf J}_{x}^{1} \cdot \left(\ps_x^l i\,\dpr_\xx \ps_x^{t} - \psi_x^t i\,\dpr_\xx \ps_x^l \right)\right]  \non \\[6pt]
& \TL{B}(\tl{J},\f) =\r_0 R_0^2  \int \frac{d^{d+1}p}{(2\pi)^{d+1}} \,
 \int \frac{d^{d+1}k}{(2\pi)^{d+1}} \biggl[\, \tilde{J}_{0}^{p}\, C_{0}(k,p)\left(\psi_{k+p}^{t}\psi_{-k}^{t} +\psi_{k+p}^{l}\psi_{-k}^{l}\right) \non \\
& \hskip 3cm +\tilde{J}_{1}^{p}\, C_{1}(k,p)\,\psi_{k+p}^{l}\psi_{-k}^{t}\,\biggr]
\]
and
\[ \label{C_nu}
C_{0}(k,p) & =  \frac{1}{2} \left[(k_{0}+p_{0})(\c_{[h,0]}^{-1}(k,p)-1)-k_{0}(\c_{[h,0]}^{-1}(k)-1)\right] \non \\[6pt]
C_{1}(k,p) & =  (\kk+\pp)^{2}(\c_{[h,0]}^{-1}(k+p)-1)-\kk^{2}(\c_{[h,0]}^{-1}(k)-1)
\]
With these definitions the correction terms arising in the Ward identities can be written as a convenient number of functional derivatives with respect to the fields $\f^l$ and $\f^t$ of
\[
C_{0}(x) & =  \frac{\d }{\d \tl{J}_{0}^{x}}\TL{\WW}_{[h,0]}(\tl{J},\f) \Big|_{\tl{J}=\f=0} \qquad C_{1}(x)=\frac{\d}{\d \tl{J}_{1}^{x}}\TL{\WW}_{[h,0]}(\tl{J},\f) \Big|_{\tl{J}=\f=0}
\]
In the following we will denote as $\TL{W}^{(h)}_{n_l,n_t;\n}(\{k_i\};p)$ the generic kernel with one external $\tl{J}_\n$ line,  $n_l$ external dashed lines and $n_t$ external plain lines. Here $\{k_i\}$ are the momenta of the $\ps^\a$ fields  $\a=l,t$  and $p$ is the momentum flowing throught $\tl{J}$. Then, for example
\[
\TL{W}^{(h)}_{11;\,\n}(k+p,-k;p) & =  \frac{\d^{3}}{\d \TL{J}_{\n}^{p}\, \d \f_{k+p}^{l} \d\f_{-k}^{t}}\,\TL{\WW}_{[h,0]}(\tl{J},\f) \Big|_{\f=\tl{J}=0}
\]
where $\n = (0, i)$ is the space-time index with $i=1,2,3$. The main difference with respect to the generating functional of the kernels with external fields $J_\n$ is the presence ot the correction term proportional to $C_\n(k,p)$. \\

The functional integral in \eqref{W_tilde} can be again studied by RG methods, see \cite{BM-luttinger}. A crucial role is played by the properties of the function $C_\n(k,p)$; it is easy to verify that, denoted with $\tl{g}^{(h)}(k)=f_h(k)\,g^{(h)}(k)$ the quantity
\[ \label{C_crucial}
 \tl{g}^{(i)}(k+p)\,C_\n(k,p)\,\tl{g}^{(j)}(k)
\] 
is non vanishing only if at least one of the indices $i$ or $j$ is equal to $0$ or both are equal to the last scale $h$; moreover, when it is nonvanishing it is dimensionally bounded from above by $$\cst |p_\n| \g^{-i -j}$$
see appendix~\ref{C_properties}. 
%This crucial property comes from the fact that, if not vanishing, the quantity
%\[ \label{lost}
%f_{i}(k+p)\,C_\n(k,p)\,f_{j}(k) \leq \cst |p_\n| \tl{f}_{i}(k) 
%\]
%with $\tl{f}_{l}(k)$ a cuoff function with the same support than $f_l(k)$, as showed in appendix~\ref{C_properties}.
%
If we start by integrating at scale $0$ the integral in \eqref{W_tilde} we find
\[
e^{|\L|\,\TL{\WW}_{[h,0]}(\f, \tl{J})} = e^{|\L| E_{-1} + \TL{S}(\tl{J})}  \int P_{[h,-1]}(d \ps^{(\leq -1)}) \,e^{-\VV_{-1}(\PS{\leq -1} + \f) + \TL{B}^{(-1)}(\ps)}
\]
where $\TL{S}(\tl{J},\f)$ collects the terms depending on $\tl{J}$ but independent on $\ps$ and 
\[
\TL{B}^{(-1)}(\ps) = \TL{B}_J^{(-1)}(\ps) + \TL{Q}_R^{(-1)}
\]with $\TL{B}_J^{(-1)}(\ps)$ linear in $\tl{J}$ and $\TL{Q}_R^{(-1)}$ the rest, \ie the kernels which are al least quadratic in $\tl{J}$. The kernels appearing in  $\TL{B}_J^{(-1)}(\ps)$, in particular, are the kernels $\TL{W}^{(-1)}_{n_l,n_t;\n}(\{k_i\};p)$ with any number of external dashed and plain legs.

\feyn{
\begin{fmffile}{feyn-TESI/squared}
 \unitlength = 0.8cm
\begin{align*}
\parbox{3.5cm}{\centering{	
		 \begin{fmfgraph*}(2,1.25)
			\fmfright{o1,o2}
			\fmfleft{i1}
                   \fmf{wiggly, label=$\tl{J}_0$, tension=1.5}{i1,v}
			\fmf{plain, right=0.3}{v,o1}
			\fmf{plain, left=0.3}{v,o2}
                   \bBall{v}								
\end{fmfgraph*}
		}}  
\parbox{3.5cm}{\centering{	
		 \begin{fmfgraph*}(2,1.25)
			\fmfright{o1,o2}
			\fmfleft{i1}
                   \fmf{wiggly, label=$\tl{J}_0$, tension=1.5}{i1,v}
			\fmf{plain, right=0.3}{v,o1}
			\fmf{plain, left=0.3}{v,o2}
                   \Square{v}								
\end{fmfgraph*}
		}}  
\end{align*}
\end{fmffile}
}{{\bf Example of vertices with an outgoing $\tl{J_0}$ field.} The kernel on the left is the monomial $\TL{W}^{(h)}_{02;0}(\{k_i\};p)$. The kernel on the right represents one of the correction terms coming from the cutoff function; in particular the squared vertex in the picture represents the matrix kernel $C_0(k,p)$. }{squared}

As usual the new kernels appearing in $\TL{B}^{(-1)}(\ps)$ can be represented as sums over Feynman graphs. By the properties of the $C_\n(k,p)$ function, see \ref{C_properties} for details, it follows that 
\[
\TL{W}^{(-1),C}_{n_l,n_t;\n}\bigl(\{k_i\};p \bigr):= p_\n\, \bar{W}^{(-1)}_{n_l,n_t;\n}\bigl(\{k_i\};p\bigr)
\]
with $\bar{W}^{(-1)}_{n_l,n_t;\n}$  dimensionally bounded uniformely in $p$ and  having the same scaling dimension of the kernels $\hV^{(-1)}_{n_l,n_t;\n}$ of the effective potential $\VV_{-1}(\ps)$. Then we define the action of the $\RR=1 - \LL$ operator exactly as done in section \ref{ext_fields} for the latter kernels. This define new marginal effective couplings on scale $(-1)$:
\[
& \LL \bar{W}^{(-1)}_{02;0}(k;p) = \bar{W}^{(-1)}_{02;0}(0;0):=\bm_{-1}^{\tl{J}_0} \non \\[6pt]
& \LL \bar{W}^{(-1)}_{11;1}(k;p) = \bar{W}^{(-1)}_{11;1}(0;0):=\bm_{-1}^{\tl{J}_1} \non \\[6pt]
& \LL \bar{W}^{(-1)}_{10;0}(p) =\LL \bar{W}^{(-1)}_{10;0}(0) := \g^{\frac{d}{4}}\bar{Z}_{-1}^{\tl{J}_0} \non \\[6pt]
& \LL \bar{W}^{(-1)}_{01;1}(p) = \sum_{i=1}^d p_i \dpr_{i} \bar{W}^{(-1)}_{01;1}(p_i)\big|_{p_i=0}:= \sum_{i=1}^d p_i \cdot \g^{\frac{d}{4}} \bar{E}_{i,-1}^{\tl{J}_1} 
\]
We now iterate the same procedure, and step by step the local parts of the kernels of type $\tl{J}_0^p \ps^t_{k+p} \ps^t_{-k}$, $\tl{J}_1^p \ps^l_{k+p} \ps^t_{-k}$, $\tl{J}^p_0 \ps^l_{p}$ and $\tl{J}_1^{p_i} \dpr_{p_i}\ps^t_{p_i}$
%$\bar{W}^{(-1)}_{02;0}$, $\bar{W}^{(-1)}_{11;1}$, $\bar{W}^{(-1)}_{10;0}$ and $\bar{W}^{(-1)}_{01;1}$ 
are collected together to form new running coupling constants, $\bm_{h}^{\tl{J}_{0}}$, $\bm_{h}^{\tl{J}_1}$, $\bar{Z}_{h}^{\tl{J}_{0}}$ and $\bar{E}_{h}^{\tl{J}_{1}}$ respectively. 

Note that starting from scale $(-1)$ the $\tl{J}_\n$ external line can be attached to a simple vertex, corresponding to one of the monomials $\TL{W}^{(h)}_{n_l,n_t;\n}\bigl(\{k_i\};p\bigr)$ or to a ``squared'' vertex, representing 
$\tl{J}^p_0 \,C_0(k,p)\bigl(\ps^t_{k+p} \ps^t_{-k} +\ps^l_{k+p} \ps^l_{-k}\bigr)$ or  
$\tilde{J}_{1}^{p}\, C_{1}(k,p)\,\psi_{k+p}^{l}\psi_{-k}^{t}$, see fig.~\ref{squared}.

Once arrived at scale $\bh$, as already seen for the multiscale analysis of the potential  $\VV_{h}(\ps, J)$, the scaling dimension of the kernels changes and we have to introduce a new localization procedure and new running coupling constants. Even in the lower momenta region the definition of localization is taken equal to the one described for the kernels $\hW^{(h)}_{n_l,n_t;\n}$, which we will repeat here. We only remind that  the running coupling constants with external field $\tl{J}$ in the region $h \leq \bh$ are 
\[
&\{ \m_{h}^{\tl{J}_{0}}, E_{h}^{\tl{J}_{0}}, Z_{h}^{\tl{J}_{0}}; E_{h}^{\tl{J}_{1}} \} & d=3 \non \\
&\{ \m_{h}^{\tl{J}_{0}},\m'^{\tl{J}_{0}}_{h}, \l'^{\tl{J}_{0}}_{h}, \o_{h}^{\tl{J}_{0}}, E_{h}^{\tl{J}_{0}}, Z_{h}^{\tl{J}_{0}}; E_{h}^{\tl{J}_{1}} \} & d=2
\] 
In three dimension all the running coupling constants with external field are marginal; in two dimensions $\m_{h}^{\tl{J}_{0}}$ is relevant with dimension $1/2$, $\m'^{\tl{J}_{0}}_{h}$, $\l'^{\tl{J}_{0}}_{h}$ and $\o'^{\tl{J}_{0}}_{h}$ are irrelevant with dimensions $-3/2$, $-1$ and $-1/2$ respectively and the others couplings are marginal.

\vskip 0.5cm

\subsubsection{Choice of the localization point}

So far, in the definition of the localization procedure we have chosen to localize the kernel of our effective potential at zero external momentum. However, as fare we are interested to the leading order in $\g^{h^*}$ as $h^* \arr -\io$, this choice is equivalent to localize at finite external momentum of order $\g^{h^*}$,  with $h$ the lowest scale in the multiscale integration of $\WW_{h^*}(\r_0,J_\n)$. 
%(then very small in the asymptotic regime, even if finite) 

In the next section we will see as this freedom in the choice of the localization point is crucial  to analyze the correction terms to the WIs coming from the cutoff. In fact the lowest scale contributions to the correction terms  is {\it not bounded} for zero external momentum. However it is sufficient to choose the external momentum of order $|p|=\g^{h^*}$ to make all these terms well defined. \\

%for technical reasons related to the corrections terms to the WI's coming from the cutoff function (reasons which are described in appendix \ref{WI_zero_momentum}) we cannot evaluate the local WI's for zero external momentum, as done for the formal Ward identities. In fact the last scale contributions coming from the cutoff function is not bounded for zero external momentum. 

%A subtle point of our analysis of the correction terms to the WIs coming from the cutoff is the choice of the localization point.

%\blue{Nelle definizioni di localizzazione che abbiamo introdotto abbiamo scelto di localizzare in zero ma abbiamo la liberta' di scegliere se localizzare in zero o a scala $\g^h$. Qualsiasi punto di localizzazione su scala $\g^h$ e' sufficiente. In effetti useremo questa liberta' perche'...}

%Abbiamo ricavato le WI per momento esterno nullo, ma possiamo mostrare che le stesse identit' valgono per le funzioni di rinormalizzazione della funzione d-onda perch\'e la differenza tra tali funzioni calcolate a momento nullo o a momento $\g^h$ \`e sottodominante, vedi dettagli in appendice 

%\subsection{WI's at small but finite external momentum}  

The differences arising when we localize $|p|=\g^{h^*}$ rather than in $|p|=0$ are all subdominant in $h$ as $h \arr -\io$. In particular, in appendix~\ref{APPzero} we describe how to prove the following properties:
\begin{enumerate}[i)]
\item the difference between the quadratic local terms of the one--step potential $\{\hat{A}_{h^*}, \hat{B}_{h^*},\hat{E}_{h^*},\hat{Z}_{h^*}\}$ evaluated for external momentum  $\g^{h^*}$ and the wave function renormalization constants $\{A_{h^*},B_{h^*},E_{h^*}Z_{h^*}\}$ localized in $\g^{h^*}$ is subleading both in the small parameter of the perturbation theory and $h^*$;
\item the difference between to localize the kernels $\hW^{(h^*)}_{12}(p)$, $\hW^{(h^*)}_{12;0}(p)$ and $\widehat{W}^{(h^*)}_{12;0}(p)$ for $|p|=\g^{h^*}$ or $p=0$ is subdominant in the small parameter of the perturbation theory;
%of the one step potential in zero or $|p|=\g^{h^*}$ is small with respect the other terms in the WIs. We are in particular interested in evaluating this difference for the kernel  $\hW^{(h^*)}_{12}(p)$ and the kernels with external $J_\n$ or $\tl{J}_\n$ fields appearing in the local WIs;
\item the kernels which by parity reasons vanish for zero external momentum, are of order $\g^{h^*}$ when we localize in $|p|=\g^{\h^*}$, and then dimensionally negligible with respect to the other contribution to the WIs;
\item from the dimensional point of view, the discrete derivative with respect to $p_\n$ evaluated for $p_\n=\g^{h^*}$ has the same behavior of the derivative with respect to $p_\n$, with $p_\n$ going to zero. 
\end{enumerate}

\vskip 1cm

\subsection{Lowest scale contributions and localization}  \label{WI_zero_momentum}

In this section we will describe the reason why it is not possible to evaluate the local WIs for external momentum equal to zero. The problem comes from the lowest scale term generated by the contraction of both the bosonic fields of the correction terms $C_{\n}(k,p)\ps^\a_{k+p}\ps^{\a'}_{-k}$ at scale $h^*$. In fact in this contraction one of the cutoff function associated to the two propagators is ``lost'', as described in appendix~\ref{C_properties}. On the other side we have already stressed in chap.~\ref{multiscale} how in order to bound the last scale propagators we need one cutoff function for each of the propagators, see \eqref{prop_last_scale}. One may wonder if is it still possible to get some bounds, even worst then \eqref{prop_last_scale}, in presence of a single cutoff function. The following example shows that this is not possible. 

\subsubsection{Bound for $g^{h^*}_{ll}(k+p)C_{0}(k,p)g^{h^*}_{ll}(k)$ }

%The problem that it's encountered if we try to choose $p=0$ in the local WI's it's clear  when we calculate the one--loop diagram 

Let consider a Feynman diagram containing a squared vertex $C_0(k,p)$ contracted with two $g^{(h^*)}_{ll}(k)$ propagators. We can find such a diagram in the lowest order part of the beta function for $\m_{h^*}^{\tl{J}_0}$ or in the higher order contributions to the flow of $\m_{h^*}^{\tl{J}_0}$ and $E_{h^*}^{\tl{J}_0}$. The integral we would like to bound is
\[ \label{I}
I= \int \frac{d^{d+1}k}{(2\pi)^{d+1}}\,f_{h^*}(k)\, g^{(h^*)}_{ll}(k+p)\,g^{(h^*)}_{ll}(k)
\]
with
\[
g^{(h^*)}_{ll}(k)=\frac{[A_{h^*}(k)\kk^2 + B_{h^*}(k)k_0^2]\,}{D_{h^*}(k)}
\] 
We remind that on the support of $f_{h^*}(k)$, we have $Z_{h^*}=Z_h\,f_{h^*}(k)$, $A_{h^*}=1+(A_h-1)\,f_{h^*}(k)$, $E_{h^*}=1+(E_h-1)\,f_{h^*}(k)$ and $B_{h^*}=B_h\,f_{h^*}(k)$, see \eqref{Z_last_scale}. Now let choose $p=(0,\bz)$ and consider the following contribution to \eqref{I}:
\[ \label{I_1}
I_1& =\int  \frac{d^{d+1}k}{(2\pi)^{d+1}}\,f_{h^*}(k)\, 
\frac{\kk^{4}}{\left(k_0^{2}+Z_{h^*}f_{h^*}(k) \kk^{2} \right)^{2}} \non \\ 
& \leq \frac{1}{(2\pi)^{d+1}}\int_{\substack{\kk^{2}\in\text{supp.}\\ f_{h^*}(k)}}
d^{d}\kk\, \kk^{4}\,f_{h^*}(k)\,\int_{0}^{+\io}dk_{0}\,\frac{1}{\left(k_{0}^{2}+Z_{h^*}f_{h^*}(k)\kk^{2}\right)^{2}}
\]
The integral in the $k_0$ variable can be calculated by the residua calculus (there are two poles of order two in $k_0=\pm i\sqrt{Z_{h^*}f_{h^*}(k)\,\kk^{2}}$), getting
\[ \label{I_0}
\int_{0}^{+\infty}dk_{0} \frac{1}{\left(k_{0}^{2}+Z_{h^*}f_{h^*}(k)\kk^{2}\right)^{2}}
=\frac{2\pi}{4\left(Z_{h^*}f_{h^*}(k)\kk^{2}\right)^{\frac{3}{2}}}
\]
Then, passing to spherical coordinates $d^d{\kk}=(2\r)^{d-1}\pi d\r$
\[
I_1& \leq \frac{1}{8\pi^{d-1}}\int_{\g^{h^*-1}}^{\g^{h^*+1}}
d\r\, \frac{\r^d}{\left(Z_{h^*}\right)^{\frac{3}{2}} \sqrt{f_{h^*}(\r)}}
\]
which is a singular integral since $f_h(\r)$ goes to zero as far $\r \arr \g^{h^*-1}$. \\ 

Let's now calculate the same integral than \eqref{I_1}, but for small, finite, external momentum, for example $p=(p_0,\bz)$:
\[
I_1(p_0) \leq& \frac{1}{(2\pi)^{d+1}}\int_{\substack{\kk^{2}\in\text{supp.}\\ f_{h^*}(k) }
}d^{d}\kk\,\kk^{4} f_{h^*}(k) \non \\
& \qquad \int_{0}^{+\infty} dk_{0}
\frac{1}{\left(k_{0}^{2}+Z_{h^*}f_{h^*}(k) \kk^{2}\right) \left((k_0+p_0)^{2}+Z_{h^*}f_{h^*}(k)\kk^{2}\right)}
\]
Again the integral in $k_{0}$, which we denote with $I_{k_{0}}(p_{0})$, can be calculated using the residua theorem; now the
the integrand function has four poles: $k_{0}=\pm i\sqrt{Z_{h}f_{h}\kk^{2}}$ and $ k_{0}=-p_0\pm i\sqrt{Z_{h}f_{h}\kk^{2}}$.
\[
I_{k_{0}}(p_{0})=\frac{2\pi}{\sqrt{Z_{h^*}f_{h^*}(k)\,\kk^{2}}\left(p_{0}^{2}+4Z_{h^*}f_{h^*}(k)\,\kk^{2}\right)}
\]
In the limit $p_{0}\arr 0$ of course $I_{k_{0}}(p_{0})$ gives \eqref{I_0}. The integral $I_1(p_0)$ turns to be
\[
I_1(p_0)=& \frac{1}{\pi^{d-2}}\int_{\g^{h^*-1}}^{\g^{h^*+1}} d\r\, \frac{\r^{d+2}\,f_{h^*}(k)}{\sqrt{Z_{h^*}f_{h^*}(k)} \left(p_0^{2}+4Z_{h^*}f_{h^*}(k)\r^2\right)} 
\]
which is bounded since for $f_{h^*}(k) \arr 0$ the integrand behaves as $\sqrt{f_{h^*}(k)}$. In particular using that $\r^2 =|\kk|^2\leq \e^{-1}\g^{2h^*}$, $Z_h \leq \e$ and $p_0=\g^{h^*}$ one find
\[
I_1(p_0) \leq \g^{(d+1)h^*} \e^{-\frac{d+3}{2}}Z_h^{-\frac{1}{2}}  \leq 
\g^{(d+1)h^*} \e^{-\frac{d}{2}}Z_{h^*}^{-2} 
\]
which is the same estimate holding for the integral of two longitudinal propagators with their cutoff functions:
\[
\frac{1}{(2\pi)^{d+1}}\int d^{d+1}k\, f_{h^*}(k+p)\,f_{h^*}(k)\,g^{h^*}_{ll}(k+p_{0})\, g^{h^*}_{ll}(k)
\]
Choosing $p$ such that $|p|=\g^h$ we are guaranteed that all the last scale diagrams coming from the correction to the WIs are bounded and in particular have the same bounds of the corresponding diagrams without the squared vertex $C_\n(k,p)$.

\pagina 

\subsection{Flow of the running coupling constants with $\tl{J}_\n$}

In this section we want to describe how to control the flow of the new running coupling constants appearing when we perform the multiscale integration of the correction terms to the local WIs. The localization procedure is chosen at external momentum of order $\g^{h^*}$ so that the lowest scale contributions are well defined. 

In the following we will denote generically  with $\a_{h}^{\tl{J}}$ one of the running coupling constants with a $\tl{J}$-field. We will describe only the evolution of the running coupling constants in the region $h \leq \bh$. With the same ideas one can see that  in the higher momentum region $\bh <h \leq -1$ the value of the couplings remains equal to their initial value $\a_{-1}^{\tl{J}}$. The coupling constant $\a^{\tl{J}}_h$ evolve according to the following flow equation
\[ \label{alpha}
\a_{h-1}^{\tl{J}_\n}- \g^\d\,\a_{h}^{\tilde{J}_{\n}}= \beta_{\a,h}^{\tl{J}_{\n}}+\b_{\a,h}^{\tl{J}_\n,\,I} + \b_{\a,h}^{\tl{J}_{\n},C}
\]
with $\d$ the scaling dimension of the coupling $\a^{\tl{J}}$.  The beta function in \eqref{alpha} has been split in three parts, which are:  
\begin{description}

\item [$\b_{h}^{\tl{J}_\n}$] which collects the terms coming from diagrams where the $\tl{J_\n}$ line comes from a vertex of type $\mu_{k}^{\tl{J}_{\n}}$ for some $h \leq k\leq \bh$. Note that for $\n \neq 0$ the vertex $\mu_k^{\tilde{J}_1}$ only exists at scale $\bh$, since it is irrelevant. 

\item [{$\b_{h}^{\tl{J}_\n,\, I}$}] collects the terms where the $\tl{J}_\n$ line comes from one of the irrelevant vertices at scale $h=\bh$. Due to the short memory property, these diagrams can be dimensionally bounded by $\cst \,\e^{c'}\,\g^{\th (h-\bh)}$ for any $0<\th<2$ in $3d$ and $0<\th<1$ in $2d$, with $c'$ an appropriate constant. 

\item [$\b_h^{\tl{J}_{\n},\,C}$]  is the contribution to $\a^{\tl{J}}_{h-1}$ coming from graphs where the $\tl{J}_\n$ line is attached to a squared vertex  representing $C_{\n}(k,p)$ with outgoing $\ps$ fields not both contracted at scale $0$\footnote{ In this case we get terms proportional to the vertex $\m_{-1}^{\tl{J}_\n}$, which have been already considered in the definition of $\b_{h}^{\tl{J}_\n}$.}. Due to the properties of the correction term $\b_h^{\tl{J}_{\n},2}$ is different from zero only if $h$ is equal to the lowest scale included in the cutoff function, \ie the cutoff function defining $\WW_{h^*}(\r_0,J_\n)$, which is $\c_{[h^*,0]}(k)$.  More precisely the non zero contributions to $\b_h^{\tl{J}_{\n},\,C}$ are given by the diagrams where both the $\ps$ fields outgoing from the squared vertex are contracted at scale $h$ or if one of them is contracted at scale $h^*$ and the remaining at scale $h=0$. However in the latter case we have at least a propagator on scale
$0$ or $-1$ and, by the short memory property, the diagram can be dimensionally bounded by $\cst \,\e^{c}\,\g^{\th h}$ for any $0<\th<1$ with $c$ an appropriate constant.

%Let denote by $\TL{W}^{(-1),C}_{n_l,n_t;\n}$ 

\end{description}

\vskip 0.3cm

%The leading contributions in $h$ to the flow equation for $\a_{h}^{\tl{J}_\n}$ are then the diagrams belonging to $\b_{h}^{\tl{J}_\n}$ and those obtained contracting the external $\ps$ fields of the squared vertex at scale $h^*$, with $h^*$ the lower scale of the cutoff function $\c_{[h^*,0]}(k)$. 

In order to study the flows of $\mu_{h}^{J_{0}}$, $E_{h}^{J_{0}}$, $Z_{h}^{J_{0}}$ and $E_{h}^{J_{1}}$, for each $h> h^*$ 
we may use exactly the same ideas used to study the couplings with external $J_\n$ field, see sec.~\ref{flow_J0}.
%
%The study of the flow equations of the running coupling constants with field $\tl{J}_0$ are studied by the analogy with the flows of the ``corresponding'' couplings with the same number and type of external legs but with $\tl{J_0}$ substituted by a dashed leg.
This allow us to prove that the diagrams contributing to $\b_{h}^{\a,\tl{J}}$  have always an extra $\l$ with respect the diagrams contributing to $\b_{h}^{\a,J}$. This simply depends on the fact that $\m_\bh^{\tl{J}_0}=\l\,\m_\bh^{J_0}$ and $\m_\bh^{\tl{J}_1}=\l\,\m_\bh^{J_1}$ as proved in appendix \ref{appB.initial_val}. 
When $h$ is equal to the lowest scale $h^*$ of the multiscale integration it is sufficient to add to the previous result the contribution coming from $\b_{h^*}^{\tl{J}_{\n},\,C}$.

\vskip 1cm

{ \center \subsubsection{ I. Flow of $\m_{h}^{\tl{J}_{0}}$}}

For each $h>h^*$, since  $\b_{\m,h}^{\tl{J}_{0},C}=0$, the flow equation of  $\m_{h}^{\tl{J}_0}$ can be studied with exactly the same strategy used to study the flow equation of $\m_h^{J_0}$; in fact the two beta functions are identical once the vertex with the outgoing $\tl{J}_0$ line is replaced with a the same vertex with outgoing  $J_0$ line. We do not repeat the details here (see sec.~\ref{flow_J0} for the analogous discussion of the flow of $\m_h^{J_0}$) but only present the conclusions. One can proof that:
\[
& \frac{\m_h^{\tl{J}_0}}{\m_\bh^{\tl{J}_0}}= \frac{\m_h}{\m_\bh} \lft(c_3+ O\bigl(\l\e^{\frac{1}{2}}\bigr) \rgt) & d=3 \non \\
& \frac{\m_h^{\tl{J}_0}}{\m_\bh^{\tl{J}_0}}= \frac{\m_h}{\m_\bh} \lft(c_2+ O\bigl(\l \bigr) \rgt) & d=2
\]
with $c_3$ and $c_2$ explicitly computable constants. Then 
\[  \label{m_tilde0}
 \m_h^{\tl{J_0}}= c_3\,\l \m_h^{J_0} & = c_3\l \e^{-1} \m_h\, \lft(1+ O\bigl(\l\e^{\frac{1}{2}}\bigr) \rgt)   & d=3 \non \\
 \m_h^{\tl{J_0}}= \,c_2\,\l \, \m_h^{J_0} & = \,c_2\,\l \e^{-1}\,\m_h\, \lft(1+ O\bigl(\l \bigr) \rgt) & d=2
\]
At scale $h^*$ we must add to \eqref{m_tilde0} the lowest scale contribution $\b_{\m,h^*}^{\tl{J}_{0},C}$ which is
\[
\b_{\m,h^*}^{\tl{J}_{0},C} = 
\begin{cases}
O \Bigl(\l \e^{-1}\m_{h^*} \,\l \e^{\frac{1}{2}} \bigl(1+ \l \e^{\frac{1}{2}}|h^* -\bh| \bigr)^{-1} \Bigr) & d=3 \\[6pt]
O \Bigl( \l \e^{-1} \m_{h^*}\, \z_* \Bigr)  & d=2
\end{cases}
\]
with $\z_*=\max_{h^* \leq h \leq \bh} \{\l \l_h, \l_{6,h}/(\e \l_h^2)\} $. We see that the lowest scale contribution do not change \eqref{m_tilde0} at leading order.

\vskip 0.5cm

{\center \subsubsection{ II. Flow of $E_{h}^{\tl{J}_0}$}}

For each $h>h^*$, since  $\b_{E,h}^{\tl{J}_{0},C}=0$, the flow equation of  $E_{h}^{\tl{J}_0}$ can be studied with exactly the same strategy used to study the flow equation of $E_h^{J_0}$, see sec.~\ref{flow_J0}. The only difference between the two contexts is the fact that
\[ 
\frac{\m_{h}^{\tilde{J}_{0}}}{\m_{h}^{\tilde{J}_{0}}} = c_d \,\l \, \frac{\mu_{h}^{J_{0}}}{\mu_{\bh}^{J_0}}
\]
with $c_d$ an explicitly computable constant dependent on the dimension $d$, whose second order expression can be found in appendix \ref{appB.initial_val}.
 One obtains:
\[ \label{E_tilde0}
& E_h^{\tl{J_0}} = -c_3 \l \,\e^{-1}\Bigl(1-E_h + O(\l \e^{\frac{1}{2}}) \Bigr)  & d=3 \non \\
& E_h^{\tl{J_0}}  = -c_2\l\, \e^{-1}\Bigl(1-E_h +O(\l) \Bigr) & d=2
\]
For what concerns the contribution $\beta_{E,h^*}^{\tilde{J}_{0},C}$ coming from the contractions of both external legs of the correction terms on scale $h^*$ we obtain
\[
\b_{E,h^*}^{\tl{J}_{0},C} = 
\begin{cases}
O \Bigl(\l \e^{-1}E_{h^*} \,\l \e^{\frac{1}{2}} \bigl(1+ \l \e^{\frac{1}{2}}|h^* -\bh| \bigr)^{-1} \Bigr) & d=3 \\[6pt]
O \Bigl( \l \e^{-1} E_{h^*}\, \z_* \Bigr)  & d=2
\end{cases}
\]
with $\z_*=\max_{h^* \leq h \leq \bh} \{\l \l_h, \l_{6,h}/(\e \l_h^2)\} $. We see that the lowest scale contribution do not change \eqref{E_tilde0} at leading order.

%Quando inseriamo i termini locali quadratici nella misura libera, compare una dipendenza da $k$ dovuta alla presenza della funzione di cutoff. Poiche' il propagatore e' valutato su scala $j$, cioe' c'e' $f_j(k)$ la dipendenza da $k$ appare solo sull'ultima scala. Tuttavia questa dipendenza non ci permette di valutare le correzioni alle WI a momento nullo, perche' per $p=0$ abbiamo delle divergenze. Il problema si risove valutando le WI per momento esterno $p=\g^h$; questa scelta non cambia le WI per le RRC all'ordine dominante. \\

\vskip 0.5cm

{\center \subsubsection{ III. Flow of $E_{h}^{\tl{J}_1}$}}

%For each $h>h^*$, since  $\b_{E_1,h}^{\tl{J}_{0},C}=0$,
Neglecting the contribution coming from the lowest scale, which again is subdominant in $|h|$, the flow of $E_{h}^{\tl{J}_{1}}$ is controlled with the same dimensional argument used to control $E_h^{J_1}$. The beta external $J_{1}$-line in the beta function $E_{h}^{\tilde{J}_{1}}$ may only come from an irrelevant vertex at scale $\bh$, as for example
\[
& \mu_{\bar{h}}^{\tilde{J}_{1}}=
\begin{cases}
O(\l\e^{3})  & d=3 \non \\
O(\l \e^\frac{5}{2}) & d=2
\end{cases}
\]
Then we can always extract from the beta function for $E_h^{J_1}$ a short memory factor and prove that
%\[
%& |\b_{k,1}^{E_{\tl{J}_{1}}} | \leq \cst\, \l \e^{-\frac{3}{2}} \m_\bh^{J_1} \g^{k-\bh} & d=3 \non \\
%& |\b_{k,1}^{E_{\tl{J}_{1}}} | \leq \cst\, \l \e^{-\frac{3}{2}} \m_\bh^{J_1} \g^{k-\bh} & d=2
%\]
%and then
\[
E_{h}^{\tilde{J}_{1}} & =  O(\l^2 \e^{\frac{3}{2}}) & d=2,3 
\]

\pagina

%---------------------------------------------------------- LWI with corrections
\feyn{
\begin{fmffile}{feyn-TESI/LWI}
 \unitlength = 0.9cm
	\begin{align*}
	\parbox{2.5cm}{\centering{	
		\begin{fmfgraph*}(2,1.25)
			\fmfleft{i}
			\fmfright{o}
			\fmf{dashes}{i,v}
			\fmf{plain,label=$\partial_0$,label.dist=-0.22w}{v,o}
			\fmfv{label=$E_h$,label.angle=90,label.dist=-0.4w}{v}   
			\Ball{v}
		\end{fmfgraph*}	
		}}
	& = \quad
\parbox{2cm}{\centering{
 	\begin{fmfgraph*}(2,1)
			\fmfright{i1,i2}
			\fmfleft{o1}
			\fmf{plain}{i1,v,i2}
			\fmf{wiggly,label=${J_0}$, tension=1.5}{o1,v}
                    \Ball{v}
			\fmfv{label=$\mu_h^{J_0}$,label.angle=-90,label.dist=-0.4w}{v}  
		\end{fmfgraph*}
		}} 
		\quad +\quad \parbox{2cm}{\centering{
 	\begin{fmfgraph*}(2,1)
			\fmfright{i1,i2}
			\fmfleft{o1}
			\fmf{plain}{i1,v}  %foreground=(0.6,,0.2,,0.4)
			\fmf{plain, label=$\partial_0$, label.dist=0.05w}{v,i2}
			\fmf{wiggly,label=${\tilde{J}_0}$, tension=1.5}{o1,v}
			\SBall{v}  
		\end{fmfgraph*}
		}} \quad +\quad \parbox{2cm}{\centering{
 	\begin{fmfgraph*}(2,1)
			\fmfright{i1,i2}
			\fmfleft{o1}
			\fmf{plain}{i1,v}  %foreground=(0.6,,0.2,,0.4)
			\fmf{plain, label=$\partial_0$, label.dist=0.05w}{v,i2}
			\fmf{wiggly,label=${\tl{J}_1}$, tension=1.5}{o1,v}
			\SBall{v}  
		\end{fmfgraph*}
		}} 
		\\[12pt]
%-------------------------------------------------------------------------
		\parbox{2.5cm}{\centering{	
		\begin{fmfgraph*}(2,1.25)
			\fmfleft{i}
			\fmfright{o}
			\fmf{plain,label=$\partial_0$, label.dist=-0.22w}{i,v}
			\fmf{plain,label=$\partial_0$,label.dist=-0.22w}{v,o}
			\fmfv{label=$B_h$,label.angle=90, label.dist=-0.4w}{v}   
                    \Ball{v}
		\end{fmfgraph*}
	}} 
	& = \quad
	\parbox{2cm}{\centering{	
		 \begin{fmfgraph*}(2,1.25)
			\fmfright{i1}
			\fmfleft{o1}
			\fmf{plain, label=$\partial_0$, label.dist=0.051w}{i1,v}
			\fmf{wiggly,label=${J_0}$}{o1,v}
			\fmfv{label=$E^{J_0}_h$,label.angle=-90, label.dist=-0.4w}{v}			
                    \Ball{v}
		\end{fmfgraph*}
		}} 
		\quad +\quad 
		\parbox{2cm}{\centering{	
		 \begin{fmfgraph*}(2,1.25)
			\fmfright{i1}
			\fmfleft{o1}
			\fmf{plain, label=$\partial_0^2$,label.dist=0.05w}{i1,v}
			\fmf{wiggly,label=$\tilde{J}_0$}{o1,v}
			\fmfv{label.angle=-90, label.dist=-0.4w}{v}		
			\SBall{v}			
		\end{fmfgraph*}
		}} \quad +\quad 
		\parbox{2cm}{\centering{	
		 \begin{fmfgraph*}(2,1.25)
			\fmfright{i1}
			\fmfleft{o1}
			\fmf{plain, label=$\partial_0^2$,label.dist=0.05w}{i1,v}
			\fmf{wiggly,label=$\tl{J}_1$}{o1,v}
			\fmfv{label.angle=-90, label.dist=-0.4w}{v}		
			\SBall{v}			
		\end{fmfgraph*}
		}} 
	\\[12pt]
	%----------------------------------------------------------------
 \parbox{2.5cm}{\centering{	
		\begin{fmfgraph*}(2,1.25)
			\fmfleft{i}
			\fmfright{o}
			\fmf{plain,label=$\dpr_\pp$,label.dist=-0.22w}{i,v}
			\fmf{plain,label=$\dpr_\pp$,label.dist=-0.22w}{v,o}
			\fmfv{label=$A_h$,label.angle=90,label.dist=-0.4w}{v}
                    \Ball{v}   
		\end{fmfgraph*}
		}} 
		 & = \quad
		 \parbox{2cm}{\centering{	
		 \begin{fmfgraph*}(2,1.25)
			\fmfright{i1}
			\fmfleft{o1}
			\fmf{plain, label=$\dpr_\pp$,label.dist=0.05w}{i1,v}
			\fmf{wiggly,label=$J_1$}{o1,v}
			\fmfv{label=$E^{J_1}_h$,label.angle=-90, label.dist=-0.4w}{v}	
                    \Ball{v}		
		\end{fmfgraph*}
		}} 
		\quad +\quad
		\parbox{2cm}{\centering{	
		 \begin{fmfgraph*}(2,1.25)
			\fmfright{i1}
			\fmfleft{o1}
			\fmf{plain, label=$\dpr_\pp^2$,label.dist=0.05w}{i1,v}
			\fmf{wiggly,label=$\tl{J}_0$ }{o1,v}
			\fmfv{label.angle=-90, label.dist=-0.4w}{v}		
			\SBall{v}			
		\end{fmfgraph*}
		}}  \quad +\quad
		\parbox{2cm}{\centering{	
		 \begin{fmfgraph*}(2,1.25)
			\fmfright{i1}
			\fmfleft{o1}
			\fmf{plain, label=$\dpr_\pp^2$,label.dist=0.05w}{i1,v}
			\fmf{wiggly,label=$\tilde{J}_1$ }{o1,v}
			\fmfv{label.angle=-90, label.dist=-0.4w}{v}		
			\SBall{v}			
		\end{fmfgraph*}
		}} 
	\end{align*}
\end{fmffile}
}{{\bf Local WIs.} The last two diagrams on each line represent the correction terms to the formal WIs coming from the cutoffs.  The meaning of the shaded vertices is pictorically described in fig. \ref{tilde_J0} and \ref{tilde_J1}. 
}{LWI}  
%
%-----------------------------------------------------------  LWI with corrections
%
\subsection{Discussion of the complete local WIs}

We are now ready to discuss the complete local WIs, where ``complete'' refers to the fact that with respect to the formal WIs we consider here the correction terms coming from the cutoff function. The local WIs are shown in fig. \ref{LWI}, with external momentum $p_\n$ taken of order $\g^h$. In the following discussion, using the result of the last sections we will use that: \\

\begin{enumerate}[a)]
\item  the kernels of the one--step potential evaluated in $p_\n=\g^h$ are ``equal'' (at leading order in $h$) to the same kernels evaluated at zero external momentum;

\item  the kernels of the one--step potential and the corresponding kernels of the multiscale potential are ``equal'', since they differ for a term which is subleading both in $\g^h$ and in the small parameter $\e$ in $3d$ and $\l \l_h$ in $2d$. This is due to the fact that the difference between the two effective potentials depends only on the integration over the lowest scale. 
\end{enumerate}

\vskip 0.5cm

{\centering \subsubsection{Local WI for $E_{h}$ } }

The local WI for $E_h$ is obtained by choosing in 
\[
& \hW^{(h)}_{11}(k+p) -\hW^{(h)}_{11}(-k) -3\sqrt{2}\,\hW^{(h)}_{03}(k,p) \non \\[6pt] & \hskip 1cm = 2\,p_{0}\hW^{(h)}_{02;0}(k+p,-k)  + 2\,\TL{W}_{02;0}^{(h)}(k+p,-k) + 2\,\TL{W}_{02;1}^{(h)}(k+p,-k)
%\D^{(h)}_{02;0}(k+p,-k)+\D^{(h)}_{02;1}(k+p,-k)
\]
the external momentum $p=(p_{0},\bz)$ and by developing the identity so obtained at the first order in $p_{0}$. We obtain the equation
\[
\hat{E}_{h}= 2\hat{\m}_{h}^{J_{0}}+2\,\dpr_{0}\TL{W}^{(h)}_{02;0}\big|_{p_0=0} +2\,\dpr_{0}\TL{W}^{(h)}_{02;1}\big|_{p_0=0}
\]
as graphically represented on the first line of fig.~\ref{LWI}.  By the properties of the $C_0(k,p)$ function, see \ref{C_properties}, it follows that 
\[
\dpr_{0}\TL{W}^{(h)}_{02;0}=p_{0}\,\mu_{h}^{\tilde{J}_{0}} + \dpr_{0}\D^{(h)}_{02;0}
\]
with $\D^{(h)}_{02;0}(k,p)$ the lowest scale contribution coming from the contraction  at scale $h$ of at least one of the $\ps$ fields outgoing from the squared vertex representing $C_0(k,p)$. Note that the WIs are derived for the kernels of the one--step potentials at scale $h$, then the lowest scale of the multiscale decomposition is just $h$. 

Regarding the correction term $ \dpr_{0}\TL{W}^{(h)}_{02;1}(k,p) $ it is zero for parity reasons for $p_0=0$ and of order $\g^h$ when $p_0=\g^h$. By using \eqref{m_tilde0} and the estimate of $\dpr_{0}\D^{(h)}_{02;0}$ we finally find
\[
& E_{h}\simeq \e^{-1} \m_h  \bigl(1+ O(\l)\bigr)
\]
both in three and two dimensions.

\feyn{
\begin{fmffile}{feyn-TESI/tilde_J0}
 \unitlength = 0.8cm
\begin{align*}
\parbox{2cm}{\centering{	
		 \begin{fmfgraph*}(2,1.25)
			\fmfright{i1}
			\fmfleft{o1}
			\fmf{plain, label=$\partial_0^2$,label.dist=0.05w}{i1,v}
			\fmf{wiggly,label=$\tilde{J}_0$}{o1,v}
			\fmfv{label.angle=-90, label.dist=-0.4w}{v}		
			\SBall{v}			
		\end{fmfgraph*}
		}} \quad = \quad &
p_0 
\parbox{3cm}{\centering{	
		 \begin{fmfgraph*}(3,1.25)
			\fmfright{i1}
			\fmfleft{o1}
                   \fmf{wiggly,label=$\tilde{J}_0$, tension=1.5}{o1,v}
                   \fmf{plain, left=0.8, tension=0.6}{v,v2}
                   \fmf{plain, right=0.8, tension=0.6}{v,v2}
			\fmf{plain, label=$\dpr_0$,label.dist=-0.2w}{v2,i1}	
			\BBall{v2}
                   \Ball{v}	
                   \fmfv{label=$\m^{\tl{J}_0}_k$,label.angle=-90, label.dist=-0.33w}{v}							\end{fmfgraph*}
		}} \;+ \;  \parbox{3cm}{\centering{	
		 \begin{fmfgraph*}(3,1.25)
			\fmfright{i1}
			\fmfleft{o1}
                   \fmf{wiggly,label=$\tilde{J}_0$, tension=1.5}{o1,v}
                   \fmf{plain, left=0.8, tension=0.6, label=\small{$h$}, label.dist=0.02w}{v,v2}
                   \fmf{plain, right=0.8, tension=0.6,label=\small{$h$}, label.dist=0.03w}{v,v2}
			\fmf{plain, label=$\dpr_0^2$,label.dist=-0.2w}{v2,i1}	
                   \Square{v}
			\BBall{v2}			
		\end{fmfgraph*}
		}}
 % \non \\[12pt] &
+\; \parbox{3cm}{\centering{	
		 \begin{fmfgraph*}(3,1.25)
			\fmfright{i1}
			\fmfleft{o1}
                   \fmf{wiggly,label=$\tilde{J}_0$, tension=1.5}{o1,v}
                   \fmf{plain, left=0.8, tension=0.6, label=\small{$0$}, label.dist=0.02w}{v,v2}
                   \fmf{plain, right=0.8, tension=0.6,label=\small{$h$}, label.dist=0.03w}{v,v2}
			\fmf{plain, label=$\dpr_0^2$,label.dist=-0.2w}{v2,i1}	
                   \Square{v}
			\BBall{v2}			
		\end{fmfgraph*}
		}} % \; +\; \b^{\tl{J}_0, I}_{E}
\end{align*}
\end{fmffile}
}{Diagrams contributing to the beta function of the correction term $\dpr_0^2\TL{W}_{01;0}^{[h^*]}(p_0)$, appearing in the WI for $B_h$. The squared vertex with external $\tl{J}_0$ line represents $C_0(k,p)$. 
%With $\b^{\tl{J}_0, I}_{E}$ we mean the irrelevant diagrams contributing to the flow equation for the correction term, \ie the diagrams with one external plain line containing one of the irrelevant vertices with an outgoing $\tl{J}_0$ field.
}{tilde_J0}

{\centering \subsubsection{ Local WI for $B_{h}$ } }

The identity useful to control the flow of  $B_{h}$ is obtained deriving
\eqref{WI_complete2} with respect to $\f_{y}^{t}$, setting the external fields equal to zero and than choosing as external momentum $p=(p_0, \bz)$: 
\[
2\sqrt{2}\, \lft(\hW_{02}^{(h)}(p_0)- \hW_{02}^{(h)}(0)\rgt)  = -\,p_0 \hW_{01;0}^{(h)}(p_0) - \TL{W}_{01;0}^{(h)}(p_0) -  \TL{W}_{01;1}^{(h)}(p_0)
\]
Developing the previous identity at the second order in $p_{0}$ we get: 
\[ \label{WI_B_complete}
\sqrt{2}\,\hat{B}_{h}= -\hat{E}_{h}^{J_{0}}+\partial_{0}^{2}\TL{W}_{01;0}^{(h)}(p_0)+\partial_{0}^{2}\TL{W}_{01;1}^{(h)}(p_0)
\]
as pictorially showed in the second line of fig. \ref{LWI}, where the shaded vertices  represent the sum of the diagrams contributing to the kernels $\TL{W}^{(h)}_{01;0}$ and $\TL{W}^{(h)}_{01;1}$. The last two terms in eq. \eqref{WI_B_complete}  represent the corrections with respect the formal WI \eqref{4.WI_B}. 
The flow of $E_h^{J_0}$ has been studied in chap.~\ref{flows}; one finds
\[
& E_{h}^{J_0}=-\e^{-1} \lft( 1- E_h  +O(\l\,\e^{\frac{1}{2}}) \rgt)   & d=3 \non \\[6pt]
& E_{h}^{J_0}=-\e^{-1}\lft( 1- E_h +O(\l ) \rgt)   & d=2
\]
For what concerns the first of the two correction terms coming from the cutoff function we have 
\[
\partial_{0}^{2}\TL{W}_{01;0}^{(h)}(p_0) = \hat{E}_h^{\tl{J}_0} + \D_{h}^{E,\tl{J_0}}(p_0) 
\]
with $\D_{h}^{E,\tl{J_0}}$ the last scale contribution coming from the second and third diagram on the r.h.s. of the identity in fig.~\ref{tilde_J0}. Using \eqref{E_tilde0}
and the bounds
\[
& \D_h^{E,\tl{J_0}}=O(\l \e^{-1} E_h\, \l \e^{-\frac{1}{2}} \m_h) & d=3 \non \\
& \D_h^{E,\tl{J_0}}= O(\l \e^{-1} E_h\, \z_* \m_h)  & d=2
\]
we obtain
\[ \label{Bh_complete}
\sqrt{2}\,B_h \simeq \e^{-1} \bigl(1 -E_h +O(\l )\bigr)
\]
both in three and two dimensions. We stress here that in $3d$ the formal WI gives $\sqrt{2}\,B_h \simeq \e^{-1} \bigl(1 -E_h +O(\l \e^{\frac{1}{2}})\bigr)$, \ie the correction term coming from the cutoff change the magnitude in $\l$ of the second non trivial order.
For what concerns the second correction term $\partial_{0}^{2}\TL{W}_{01;1}^{[h]}(p_0)$, it is null for parity reasons, since the kernels contributing to $\TL{W}_{01;1}^{[h]}(p_0)$ are odd in the spatial variable $\kk$.  
%
%For what concern the correction $\partial_{0}^{2}\Delta_{t;1}(p)$ it is null for parity reasons, since the integrand is odd in $\kk$.
%We can see this fact e.g. at the lower order in $\lambda$, when the
%contribution to the flow is the following:
%
%\noindent \begin{center}
%\includegraphics[width= 0.8\textwidth]{\string"C:/Users/Serena/Desktop/PhD-2012.05/immagini/Bh_4\string".pdf}
%\par\end{center}
%
%\[
%\partial_{0}\int\frac{d^{4}k}{(2\pi)^{4}}2\underline{k}\,\chi(k+p)\,\chi(k)\,\frac{\lambda\underline{k}^{2}+k_{0}(k_{0}+p_{0})}{D(k)D(k+p)}=0
%\]

\feyn{
\begin{fmffile}{feyn-TESI/tilde_J1}
 \unitlength = 0.8cm
\begin{align*}
\parbox{2cm}{\centering{	
		 \begin{fmfgraph*}(2,1.25)
			\fmfright{i1}
			\fmfleft{o1}
			\fmf{plain, label=$\dpr_\xx^2$,label.dist=0.05w}{i1,v}
			\fmf{wiggly,label=$\tl{\bf J}_1$}{o1,v}
			\fmfv{label.angle=-90, label.dist=-0.4w}{v}		
			\SBall{v}			
		\end{fmfgraph*}
		}} \quad = \quad &
\pp \cdot  \lft[
\parbox{3cm}{\centering{	
		 \begin{fmfgraph*}(3,1.25)
			\fmfright{i1}
			\fmfleft{o1}
                   \fmf{wiggly,label=$\tl{\bf J}_1$, tension=1.5}{o1,v}
                   \fmf{plain, left=0.8, tension=0.6, label=\small{$\dpr_\xx$}, label.dist=0.005w}{v,v2}
                   \fmf{dashes, right=0.8, tension=0.6}{v,v2}
			\fmf{plain, label=$\dpr_\xx$,label.dist=-0.2w}{v2,i1}	
			\BBall{v2}
                   \Ball{v}	
                   \fmfv{label=$\m^{J_1}_k$,label.angle=-60, label.dist=-0.33w}{v}							\end{fmfgraph*}
		}} \;+ \;  
  \parbox{3cm}{\centering{	
		 \begin{fmfgraph*}(3,1.25)
			\fmfright{i1}
			\fmfleft{o1}
                   \fmf{wiggly,label=$\tl{\bf J}_1$, tension=1.5}{o1,v}
                   \fmf{plain, left=0.8, tension=0.6}{v,v2}
                   \fmf{dashes, right=0.8, tension=0.6, label=\small{$\dpr_\xx$}, label.dist=0.05w}{v,v2}
			\fmf{plain, label=$\dpr_\xx$,label.dist=-0.2w}{v2,i1}	
			\BBall{v2}
                   \Ball{v}	
                   \fmfv{label=$\m^{J_1}_k$,label.angle=-60, label.dist=-0.33w}{v}							\end{fmfgraph*}
		}}  \rgt]  \non \\[12pt] 
& +\; \parbox{3cm}{\centering{	
		 \begin{fmfgraph*}(3,1.25)
			\fmfright{i1}
			\fmfleft{o1}
                   \fmf{wiggly,label=$\tl{\bf J}_1$, tension=1.5}{o1,v}
                   \fmf{plain, left=0.8, tension=0.6, label=\small{$h$}, label.dist=0.02w}{v,v2}
                   \fmf{dashes, right=0.8, tension=0.6,label=\small{$h$}, label.dist=0.03w}{v,v2}
			\fmf{plain, label=$\dpr_\xx^2$,label.dist=-0.2w}{v2,i1}	
                   \Square{v}
			\BBall{v2}			
		\end{fmfgraph*}
		}}
 +\; \parbox{3cm}{\centering{	
		 \begin{fmfgraph*}(3,1.25)
			\fmfright{i1}
			\fmfleft{o1}
                   \fmf{wiggly,label=$\tl{\bf J}_1$, tension=1.5}{o1,v}
                   \fmf{plain, left=0.8, tension=0.6, label=\small{$0$}, label.dist=0.02w}{v,v2}
                   \fmf{dashes, right=0.8, tension=0.6,label=\small{$h$}, label.dist=0.03w}{v,v2}
			\fmf{plain, label=$\dpr_\xx^2$,label.dist=-0.2w}{v2,i1}	
                   \Square{v}
			\BBall{v2}			
		\end{fmfgraph*}
		}} % \; +\; \b^{E,J_1}_{I}
\end{align*}
\end{fmffile}
}{Diagrams contributing to the beta function of the correction term to the WI for $A_h$. The squared vertex with external $\tl{\bf J}_1$ line represents $C_1(k,p)$. %With $\b_{J_1}^{\text{irr}}$ we mean the irrelevant diagrams contributing to the flow equation for the correction term, \ie the diagrams with one external plain line containing one of the irrelevant vertex with an external $\tl{\bf J}_1$ field.
}{tilde_J1}

\vskip 1cm

{\centering \subsubsection{Local WI for $A_{h}$ } }

In order to derive this identity we derive \eqref{WI_complete2} with respect to $\f^t$, setting the external fields equal to zero and than choosing as external momentum $p=(0,\pp)$: 
\[
2\sqrt{2} \lft(\hW_{tt}(\pp)-\hW_{tt}(\bz)\rgt)=
%\pp\cdot\underline{W}_{t;1}(\underline{p})+\Delta_{t;0}(\underline{p})+\Delta_{t;1}(\underline{p})
\pp\cdot\hW_{01;1}^{[h]}(p_0) + \TL{W}_{01;0}^{[h]}(\pp) + \TL{W}_{01;1}^{[h]}(\pp)
\]
By developing the previous identity at the second order in $\pp$ we get: 
\[
\sqrt{2}\,\lft(\hat{A}_{h}-1\rgt)= \hat{E}_{h}^{J_{1}}+ \dpr^2_\pp\TL{W}_{01;0}^{[h]}(\pp) \Big|_{\pp=0} +\, \dpr^2_\pp\TL{W}_{01;1}^{[h]}(\pp)\Big|_{\pp=0}
\]
which is pictorially represented in fig.~\ref{tilde_J1}. In this figure the diagrams on the first line of the r.h.s. represent the leading order to the part of the beta function where the $J_1$ line comes from $\m^{\tl{J}_1}_\bh$ or other irrelevant diagrams at scale $\bh$; the  diagrams on the second line with squared vertex are the last scale contribution. We have
\[
\dpr^2_\pp\TL{W}_{01;0}^{[h]}(\pp) = \hat{E}_h^{\tl{J}_1} +  \D_{h}^{E,\tl{J_1}}(\pp) 
\]  

%so that the contribution coming from the correction term is of higher order with respect to $E_{h}^{J_{1}}$ and does not change the behaviour of $A_{h}$ with respect to what we have found by neglecting the cutoff function.
%
For what regards the third correction term, $\dpr{\pp}^{2}\D^{[h]}_{01;0}(p)\big|_{\pp=0}$, it is null for parity reasons. Finally we get
\[
& A_h-1  \simeq O(\l \e^\frac{1}{2})   & d=3\non \\
& A_h -1 \simeq  O(\l )   & d=2
\]
with the correction term coming from the cutoffs being of order $o(\l^2)$.

\feyn{
\begin{fmffile}{feyn-TESI/fig_prop}
 \unitlength = 0.8cm
\begin{align*}
 \parbox{2.5cm}{\centering{	 \vskip 0.5cm
		\begin{fmfgraph*}(2.2,1.25)
			\fmfleft{i}
			\fmfright{o}
			\fmf{plain,label=$\dpr_0$,label.dist=-0.22w}{i,v}
			\fmf{wiggly,label=$J_0$,label.dist=-0.3w}{v,o}
                    \Ball{v}   
		\end{fmfgraph*} \\
                $E_h^{J_0}$
		}} 
		 & = \quad
		 \parbox{2cm}{\centering{	\vskip 0.5cm
		 \begin{fmfgraph*}(2.2,1.25)
			\fmfright{i1}
			\fmfleft{o1}
			\fmf{wiggly, label=$J_0$,label.dist=0.08w}{i1,v}
			\fmf{wiggly,label=$J_0$}{o1,v}	
                    \Ball{v}		
		\end{fmfgraph*}\\
                $J_h$
		}} 
		\quad +\quad
		\parbox{2cm}{\centering{	\vskip 0.5cm
		 \begin{fmfgraph*}(2.2,1.25)
			\fmfright{i1}
			\fmfleft{o1}
			\fmf{wiggly, label=$J_0$,label.dist=0.08w}{i1,v}
			\fmf{wiggly,label=$\tl{J}_0$ }{o1,v}	
			\SBall{v}			
		\end{fmfgraph*}\\
                $\tl{J}_h$
		}}  
	\end{align*}
\end{fmffile}
}{Local WI useful to control the behavior of the propagator.
}{fig_prop}  

\vskip 0.5cm

{\centering \subsubsection{ Local WI for the propagator } }

The complete local WI relating $E_h^{J_0}$ with $J_h$ is shown in fig.~\ref{WI_prop}, where the last term is the correction with respect the formal WI \eqref{WI_JK}. The flow of the coupling $\tl{J}_h$ is controlled as the one for $J_h$, the only difference being that the vertex with the outgoing $\tl{J}$ line has an extra $\l$ with respect the vertex with outgoing $J$ line and then $\tl{J}_h =  O(\l J_h)$. Using \eqref{WI_B_complete} we have
\[
B_h \simeq -E_h^{J_0} \lft(1+O(\l) \rgt) 
\]
and finally find
\[ \label{WI_J_complete}
B_h \simeq J_h \lft(1+O(\l) \rgt) 
\]
which has been used in chap.~\ref{flows} to prove that
\[
& B_h + \frac{E_h^2}{Z_h} \simeq \e^{-1} (1 +O(\l)) 
\]

%---------------APPENDICI
%\appendix \input{APP-WI} \end{document}

%{\backmatter 
%\input{intro-senza-sapclass} \input{intestazione-sap} \begin{document}

\chapter{Conclusions}  \label{conclusions}
%\addcontentsline{toc}{chapter}{\numberline{}Conclusions}
%\addcontentsline{toc}{chapter}{Conclusions}
%\markboth{\textsc{Conclusions}}{}

\subsection*{Summary}

%The theory of ultracold, dilute Bose gases is the subject of intensive studies, driven by always new experimental applications, which also motivate the study of BEC in low dimensions. From a theoretical point of view there is a single, quite special, model~\cite{hard-sphere-bosons} proving BEC for homogeneous interacting bosons. More recent results exist for different scaling limits. However the  fundamental problem of proving the occurrence of BEC in the thermodynamic limit is still open. 

Let us now summarize the results of the analysis performed in this thesis. 

\vskip 0.2cm

With the aim of studying the condensation problem for a homogeneous three and two dimensional system of bosons interacting with a repulsive short range potential  at zero temperature, we considered a simplified model, obtained by introducing an ultraviolet momentum cutoff. 
Such a model was analyzed  by {\it exact Renormalization Group} methods, which have already been proved effective in the study of several low dimensional condensed matter systems, see the introduction of~\cite{GRanom} for a list of references. 

In the three dimensional case we proved that the renormalized expansion is order by order finite in the running coupling constants. In two dimensions we proved that 
the interacting theory is well defined at all orders in terms of two effective parameters related to the intensity of the three and two particles interactions.  In both cases we have explicit bounds on the coefficient of order $n$.

In both dimensions the correlations do not exhibit anomalous dimensions, i.e. the model is in the same universality class of the exactly soluble Bogoliubov model. 
%In fact the linear spectrum is exactly constrained by Ward Identities. 
We stress that the power series expansion around Bogoliubov model, in the bare couplings,  is plagued by {\it logarithmic divergences} in the three dimensional case and even more ``dangerous'' divergences in the two dimensional case. Then the absence of anomalous dimensions is a quite remarkable result, since we may expect the summations of the divergences to deeply change the qualitative behavior of Bogoliubov propagator.

\vskip 0.2cm

Our results are obtained by implementing local Ward identities (WIs) within an exact RG scheme, thanks to the technique developed in~\cite{BM-luttinger}. These identities  reduce the number of independent effective parameters, this fact being crucial for the control of the two dimensional theory, where the four and three points effective interactions are relevant and there are eleven effective running couplings to be controlled. 

Since the momentum cut--offs, introduced in the momentum regularization scheme we exploited, break the local gauge invariance, the study of the corrections term to the formal local WIs is among the main goals of this thesis. In fact these  terms may a priori may be responsible for anomalous dimensions, since they are dimensionally marginal in $3d$ and also relevant in $2d$. Remarkably, the corrections terms to the formal WIs can be rigorously bounded at all orders in renormalized perturbation theory (see chapter \ref{WI}) and in our model turn out to be of higher order in the small parameter $\l$, describing the intensity of the interaction. This means that the formal WIs are exact at leading order, \ie we may neglect the correction terms coming from the cut--off as far as we are interested in the leading order relations between the effective parameters of the system. 

%The key idea of using WIs to simplify the flow equations was first introduced by Pistolesi et al.~\cite{CaDiC1, CaDiC2} within a scheme based on a dimensional regularization. Thanks to the technique developed in~\cite{BM-luttinger} for the analysis of Luttinger liquids, we were able to implement the same ideas within an exact RG scheme. In this latter case, since the ultraviolet momentum cut--offs introduced in the multiscale decomposition of the propagator break the local gauge invariance, 

%Due to the presence of the ultraviolet cut--off  the WIs have {\it corrections} with respect to the formal ones, which a priori may be responsible for anomalous dimensions. This is the case of other low dimensional condensed matter systems, as Luttinger liquids, were the anomalous dimensions are only found when the corrections to formal WIs are taken into account. 

\vskip 0.2cm

The proof that the renormalized expansion for a three dimensional system of bosons, with an ultraviolet momentum cut--off, is order by order finite was already obtained by Benfatto~\cite{benfatto} some years ago. With respect to Benfatto's our method also applies to the much more subtle $2d$ case. Moreover also in the $3d$ case our method appears more satisfactory from a conceptual point of view, since we exploit  the symmetries of the problem at best\footnote{From a technical point of view our method does not really simplify the treatment of the $3d$ case. Even if we have to study only two flow equations, instead of the six of Benfatto's work, the analysis of the correction terms to WIs is quite hard--working.  }. In fact, as showed in chap.~\ref{flows}, WIs exactly constrain the behavior of the propagator and the unique effect of the summation of the divergences appears in the renormalization of the speed of sound of quasi--particles. Two less significant differences between our work and Benfatto's one are the study of the high momentum transient regime -- where the contribution of the quadratic Bogoliubov potential in the measure is negligible -- which has been discussed in great detail in this thesis and the analysis of the dependence of the perturbative series by the dimensional quantities  characterizing  the problem, \ie the condensate density and the range of the interacting potential.

\vskip 0.2cm
% remade the three dimensional case with an alternative and more satisfactory method, which also allow us to control the two dimensional case. Our method is based on ideas introduced in a non rigorous way by Pistolesi et al, that we have applied within a rigorous scheme, that is the constructive renormalization group approach. \\

%Rivedere bosoni interagenti con gruppo di rinormalizzazione con un approccio costruttivo, implementando local WI che in 3d riducono lo studio del flusso, semplificando la trattazione dal punto di vista concettuale (pi\'u che pratico, visto che lo studio delle correzioni e' complicato!) lo studio delle costanti correnti. In 2d invece le WI sono cruciali e permettono di dimostrare che le costanti rinormalizzate sono finite ordine per ordine, supposto che $\l$ ammetta un punto fisso non banale, come congetturato da Pistolesi et al. ma che con le nostre tecniche non si puo' vedere (forse si puo' con analisi asintotiche numeriche).

%Sottolineare nelle conclusioni la differenza con Benfatto e Pistolesi et al.

With respect to the work by Pistolesi et al., we have implemented local WI's in a exact RG scheme, where {\it exact} means that we have a complete control of all the diagrams arising in the perturbation theory at each order, without neglecting the irrelevant terms. Our methods allow to rigorously prove the relations between the effective couplings stated by Pistolesi et al. The agreement with their result comes from the fact that the correction terms coming from the ultraviolet cut--off are subleading with respect to the terms already present in the formal WIs; however these terms give effects which are in principle observable in the relations between thermodynamical and response functions. 

\vskip 0.2cm

By using the Wilsonian RG scheme, rather than a dimensional regularization, we found that in the $2d$ case -- in addition to the eight running coupling constants which comes from a ``na\"ive'' dimensional analysis
%already indicated by Pistolesi et al.
-- {\it three new effective marginal coupling constants arise}, which have not been recognized before. Thanks to three additional (w.r.t.~the $3d$ case) global WIs these new couplings can be written in terms of the three particle effective coupling $\l_{6}$. The latter coupling was completely neglected by Pistolesi et al. in their paper. The latter authors suggest that a study of this coupling is not necessary, being it irrelevant in the $3d$ case\footnote{
Pistolesi et al. exploited an {\it $\e$ expansion} with $\e=d-3$, with $d$ the spatial dimension of the system; it is not clear within this scheme how to extrapolate the behavior of an effective parameter in $2d$ which was irrelevant in the $3d$ case.}. 
%it cannot be studied with the dimensional regularization method exploited by these authors

On the contrary not only the flow of the coupling $\l_{6,h}$  is {\it not trivial} but it changes the leading order flow equation of the two--particles effective interaction $\l_h$, with respect the one suggested by Pistolesi et al. 
To be more precise the leading order flow equations of the two independent effective parameters $\l_h$ and $\l_{6,h}$ are coupled among them at all orders.  By a leading order calculation one finds that both $\l_h$ and $\l_{6,h}$ admit fixed points of order one. In particular our prediction for the fixed point of $\l_h$ is not significantly different from the one by Pistolesi et al.   Remarkably neither the presence of the new marginal couplings nor the fact that $\l_{6,h}$ has a non trivial flow change the conclusions on the behavior of the propagator, which only depends on the existence of the two fixed points.

%together with the new six--body marginal coupling $\l_{6}$ which was recognized by Pistolesi et al. but not discussed  enters in the leading order flow of the four--body effective parameter $\l_h$.   \\

%In the two dimensional case we have also studied a marginal effective coupling, in the thesis denoted with $\l_{6}$ see sect. \ref{lambda6}, whose study was not discussed in the work by Pistolesi et al. 

%La differenza rispetto al lavoro di Pistolesi et al. e' nel controllo delle correzioni che danno effetti sottodominanti ma possibilmente osservabili e lo studio di una costante corrente marginale che non viene fatta da Pistolesi et al. visto che nel contesto della $\e$ expansion non c'e' modo di estrapolare una costante corrente che da marginale diventa irrilevante. Rispetto al lavoro di Benfatto studio la regione transiente. \\

\vskip 0.2cm

%------------------------- {Correlazioni.}

Coming back to our main result, in this thesis we have proved that the behavior of the renormalized propagator for small momenta is equal to Bogoliubov's except for a renormalized speed of sound. With exactly the same methods -- here applied to the multiscale analysis of the free energy, see sec. \ref{multiscale}, and to the generating functional of density and current correlations, see sec. \ref{sec:gen_fun} --
without much additional efforts, one can calculate the correlation functions and show that the result we have proved for the renormalized propagator also holds for the two--point Schwinger function. We will not belabor the details here, referring to~\cite{GM, THGiuliani} for a description of how the multiscale analysis applies to the generator of Schwinger functions.  \\

%The constructive method we have presented in this thesis allows to derive an expansion for all the Schwinger functions and density and current response functions, in presence of  condensation. 

%Con poco sforzo con gli stessi metodi di questa tesi si possono calcolare le funzioni di correlazione e mostrare che il risultato che abbiamo trovato per propagatore vestito e' valido anche per le funzioni di Schwinger but we will belabor the details here. 

%Derivo un'espansione convergente per tutte le funzioni di Schwinger, per tutte le funzioni di correlazione densita', corrente con in mente il fenomeno della rottura di gauge globale associata alla condensazione. Alcune delle relazioni che troviamo tra le osservabili dipendono dalle simmetrie locali.  \blue{Scrivere tra i rusultati: a) le espressioni dei campi esterni j0 e J1 in termini degli operatori di creazione e distruzione; b) come scalano le correlazioni densita'-densita' e corrente-corrente; c) le identita' di ward tra le funzioni di correlazione, in maniera grafica} \\

%-------------------- OPEN PROBLEMS

\vskip 1cm

\subsection*{Outlook}

%Let us now discuss the open problems we can naturally face with the s. 
Several interesting open problems may be naturally faced with the same methods used in this thesis; we plan to go through them in the immediate future.   \\

{\bf 1.} In this work we were mainly interested in the study of the long--range behavior of correlations, especially in the two dimensional case. For this reason we have introduced a rotational invariant ultraviolet momentum cutoff, which greatly simplify the problem without affecting  the infrared behavior of the system.
%which does not affect, but greatly simplify the study since we have not to deal with the ultraviolet divergences.
%
%which greatly simplify the study of the problem but gives unphysical prediction for the ground state energy and the chemical potential, as stressed in chapter \ref{model}. 
%

However, the study of the renormalizability of the ultraviolet region turns to be crucial to get quantitative predictions on physical quantities of interest, such as the corrections to the ground state energy or to the chemical potential, with respect to Bogoliubov's predictions. Due to the presence of the ultraviolet momentum cut--off, precise values of the subleading corrections to the thermodynamic and correlation functions may be quantitatively different from those in the Hamiltonian model,  as stressed in the section \ref{eff_mod}. 

A Renormalization Group analysis, similar to the one we have performed here, may possibly be effective also for the continuum ultraviolet problem, the only difference lying in the fact that for momenta greater than the inverse range of the potential, the interaction does not appear local anymore. 

\vskip 0.2cm

An alternative direction which may be  followed  is the extension of the results here obtained with the ultraviolet momentum cut--off to a non rotational symmetric theory of bosons on a lattice. A motivation for this model is provided by the recent experiments of condensation in optical traps. However the lattice case seems to be much more subtle than the one considered in this thesis, since  the  symmetries we have used to reduce the flow equations  break down and one has to look for new discrete symmetries which may play the same roles of the other continuous ones that we employed above.

{\bf 2.} In two dimensions the renormalized expansion is order by order finite provided that the effective two--particles interaction $\l_h$ is small with respect to $\l^{-1}$, with $\l$ the intensity of the interaction. A second order calculation shows $\l \l_h$ to be of order one, see sec. \ref{lambda6}, {\it independently of the values of the other parameters of the problem}, which are the condensate density $\r_0$ and  the range $R_0$ of the interacting potential. Then it is audacious to state that the perturbation theory is well defined, even order by order. 
%
%We remark that the possibility of proving that $\l \l_h$ is small is beyond the possibility of the perturbative theory; however, even if not conclusive, our second order calculation makes reasonable the conjecture that the latter condition may be verified. Of course the only hope to get convergence is to have alternating series. %cancellata da alessandro

The flow of the effective coupling $\l_h$ might of course be studied by numerical analysis. However the effective model we considered is clearly not fundamental: a more realistic microscopic model is needed to get trustworthy conclusions.

%Da un conto al secondo ordine trovo che e' minore di costante che non e' conclusivo ma non e' neanche disperato. Uno potrebbe studiare numericamente il flusso di $\l_*$ ma allora serve un modello microscopico di cui ci si puo' fidare di piu'. 

\vskip 0.2cm

%With the notations used in this thesis, the expected regime of validity of Bogoliubov's theory is  $\l^{-1} \gg 1 \gg \sqrt{\l \r_0 R_0^d} \gg \l$. 

The regime $\l \r_0 R_0^d \leq 1$ we have considered in the thesis is not the most general one in which one expects Bogoliubov's theory to be valid\footnote{
%With the notations used in this thesis, the regime $\l \r_0 R_0^d \ll 1$ with $\l \ll 1$ corresponds to the one in which one expects to rigorously prove, in the $3d$ case, the  Lee--Huang--Yang formula~\cite{Lee-Huang-Yang} for the ground state energy of the Bose gas with repulsive interactions. 
With ``regime of validity of  Bogoliubov theory'' we refers here to the regime in which  
Bogoliubov approximation is expected to be valid, \ie in which the truncation of the interacting potential performed in Bogoliubov approximation, see~\eqref{step1} pag.~\pageref{step1}, is justified by the fact that the contributions to the thermodynamic functions coming from the terms of the interacting potential which are cubic or quartic in the bosonic fields with non zero momentum are subleading with respect to those coming from the quadratic terms in the same fields.}. 
%This is pictorially referred to saying that the scattering of particle outside the condensate is negligible.}  
%
One may wonder if there exists a different regime in which the two dimensional theory turns to be stable. In particular in the weak coupling and high density regime $  \r_0 R_0^d \gg \l^{-1} \gg 1$ the choice of an ultraviolet momentum cut--off at the scale of the range of the interacting potential is not justified. On the contrary the effective parameters at scale $R_0^{-1}$ are  renormalized by the integration over an higher region of momenta, where the interaction is not local. The latter regime may possibly shows different features.   \\

{\bf 3.} Last but not least, it would be interesting to extend our analysis to a system of interacting bosons at {\it finite temperature}.  In particular the computation of the critical temperature, where the condensate density is zero but the correlation function has a power low decay, is a debated point (see~\cite{Tc} and references therein), which also has an undoubted experimental interest.   \\

%\vskip 0.2cm
%The behavior of condensates at finite temperatures is also one of the frontiers for the experimental exploration, after being for a long time mainly concentrated on the initial formation of condensates.  \\

The perspective to make the treatment of BEC rigorous, \ie to provide a full {\it non perturbative construction} for the model, is far to be reached. As well known this is an intrinsic problem for bosonic theories. The main missing point, in the context of the Bose gas, is to solve the large field problem. 
This is not expected to be a trivial generalization of known techniques -- as those used to analyze the infrared $\f^4$ problem in $d = 2, 3$ -- since in the Bose gas case one has to use complex Gaussian measures, instead of positive Gaussian measures, and this introduces new technical problems. Some attempts of dealing with complex measures, using stationary phase approximations techniques, have been recently developed by Balaban et al. (see ~\cite{BTFK_ultraviolet} and ref. therein) to control the temporal ultraviolet limit. However the problem remains wide open. 

%\vskip 0.2cm

On the other hand, to date it is not even clear how to recover the estimates by Dyson~\cite{Dyson-upper-bound} and Lieb-Yngvason~\cite{LY-lower-bound} for the ground state energy in a functional integral approach. The latter goal seems achievable and is expected to be  good warm up exercise to a deeper comprehension of the description of interacting bosons in terms of functional integrals.

%A long term goal for my PhD research is to combine renormalization group ideas with the stationary phase approximations techniques, developed by Balaban et al. [22] for analysing the large distance/infrared behaviour of a system of identical bosons, as the temperature tends to zero.

%\end{document}
%}
%

%-------------------------------------------------------------------
\appendix

%\chapter{Bogoliubov model}
%\input{capitoli/APP-Bogoliubov_Fock_space.tex}

%\chapter{Coherent states}
%\input{capitoli/APP-coherent_states.tex}

%\input{intro-senza-sapclass} \input{intestazione-sap} \begin{document}

\chapter{Multiscale analysis and power counting}

In the present appendix we list some definitions, properties and lemmas which constitute main technical points  in chapter \ref{chap_multiscale}.

\section{Some properties of Gaussian measures} \label{A1}

In this section we list some basic properties of the Gaussian integration and the definition of truncated expectation, crucial in deriving the Feynman diagrams expansion. 

Let consider the complex fields $\ps^-_x$ and $\ps^+_x=(\ps^-_x)^*$, with $x \in \L$. Let denote with $P_M(d\ps)$ the Gaussian measure in the fields $\ps^\pm_x$ with covariance matrix $M$:
\[
P_M(d\ps) = \int \Bigl(\prod_{x \in \L}\frac{d\ps^+_x d\ps^-_x}{2\pi i} \Bigr) \lft(\det M\rgt)^{-1} e^{-\sum_{i,j} \ps^+_i M^{-1}_{i j} \ps^-_j } 
\]
where 
\[
\frac{d\ps^+_x d\ps^-_x}{2\pi i} := \frac{d \Re(\ps^-_x) \, d \Im (\ps^-_x) }{\pi}
\]
and $M$ is a positive definite complex $|\L| \times |\L|$ matrix. 
By construction one has 
\[
\int P_M(d\ps)=1, \qquad \int P_M(d\ps) \ps^-_i \ps^+_j = M_{i j}
\]
For each analytic function $F(\ps)$ we can write
\[
\EE(F) = \int P_M(d \ps) F(\ps)
\] 

It is useful to introduce the notion of {\it truncated expectation}, since it appears naturally considering the integration of an exponential with respect to a Gaussian measure.
Given $p$ functions $X_1, \ldots, X_p$ defined on the fields $\ps$ and $p$ positive integer numbers $n_1, \ldots, n_p$, the truncated expectation is defined as
\[ \label{trunc_def}
\EE^T (X_1, \ldots, X_p; n_1, \ldots, n_p)= \frac{\dpr^{n_1, \ldots, n_p}}{\dpr \l_1^{n_1}, \ldots, \dpr \l_p^{n_p}} \log \int P(d\ps) e^{\l_1 X_1(\ps)+\ldots +\l_p X_p(\ps)} \Bigl|_{\l=0}
\]
where $\l = \{\l_1, \ldots, \l_p\}$. It is easy to check that $\EE^T$ satisfies the following property
\[
\EE^T(c_1 X_1 + \ldots + c_p X_p; n)= \hskip -0.2cm \sum_{n_1 + \ldots + n_p = n} \hskip -0.15cm \frac{n!}{n_1 ! \ldots n_p!}\,c_1^{n_1}\ldots c_p^{n_p}  \EE^T( X_1,\ldots, X_p; n_1, \ldots, n_p)
\]
so that the following relations immediately follow:
\begin{enumerate}[(1)]
\item $\EE^T(X,1)= \EE(X)$,
\item $\EE^T(X,0)= 0$
\item $\EE^T(X, \ldots, X;n_1, \ldots, n_p )= \EE^T(X; n_1 +\ldots + n_p)$
\end{enumerate}
Moreover one has 
\[ \label{trunc1}
\EE^T( \underbrace{\;X_1, \ldots, X_1\;}_{n_1}, \ldots, \underbrace{\;X_p, \ldots, X_p\;}_{n_p} \, ; \underbrace{\,1, \ldots, 1\,}_{n_1 + \ldots +n_p} ) = \EE^T (X_1, \ldots, X_p; n_1, \ldots, n_p)
\]
We define also
\[ \label{trunc2}
\EE^T (X_1, \ldots, X_p)= \EE^T (X_1, \ldots, X_p; 1, \ldots, 1)
\]
By \eqref{trunc1} we see that all the truncated expectations can be expressed in terms of \eqref{trunc2}; it easy to see that \eqref{trunc2} is vanishing if $X_j=0$ for at least one $j$.
As a particular case of \eqref{trunc_def} one has:
\[
\EE^T(X;n)= \frac{\dpr^n}{\dpr \l^n} \log \int P(d\ps) e^{\l X(\ps)} \Bigl|_{\l=0}
\]
so that we can rewrite formally
%formally: se la funzione e' analitica e' vero altrimenti nessuno ci garantisce che lo e' 
the integral of an exponential with respect to the Gaussian measure $P(d \ps)$ as:
\[
\log  \int P(d\ps) e^{X(\ps)} & = \sum_{n =0}^{\io} \frac{1}{n!} \frac{\dpr^n}{\dpr \l^n}
\log \int P(d\ps) e^{\l X(\ps)} \Bigl|_{\l=0} \non \\
& = \sum_{n =0}^{\io} \frac{1}{n!} \EE^T(X;n)
\]
The following properties holds:
\begin{enumerate}[\it (1)]
\item {\it Wick rule.} Given two set of labels $\{\a_1, \ldots, \a_n\}$ and $\{\b_1, \ldots, \b_n\}$ one has
\[
 \int P_M(d \ps) \ps^-_{\a_1}\ldots \ps^-_{\a_n} \ps^+_{\b_1}\ldots \ps^+_{\b_n}= \sum_{\p} \prod_{i=1}^{n} M_{\a_{i} \b_{\p(i)}}
\]
where the sum is over all the permutations $\p= \{ \p(1), \ldots, \p(n)\}$ of the indices $\{1, \ldots, n\}$.
\item {\it Addition principle.} Given two integrations $P_{M_1}(d \ps_1)$  and $P_{M_2}(d \ps_2)$, with covariance $M_1$ and $M_2$ respectively, for any function $F=F(\ps)$ with $\ps= \ps_1 + \ps_2$ one has
\be \label{addition_principle}
\int P_{M_1}(d \ps_1) \int P_{M_2}(d \ps_2) F(\ps_1 + \ps_2) = \int P_M(d \ps) F(\ps)
\ee 
where $M \equiv M_1 + M_2$.
\item {\it Invariance of exponentials.} From the definition of truncated expectations, it follows that, if $\f$ is an ``external field'', \ie a not integrated field, then
\[ \label{invariance_exp}
\int P_M(d \ps) e^{X(\ps + \f)} = \exp \lft[ \sum_{n =0}^{\io} \frac{1}{n!} \EE^T \lft(X(\cdot+\f) ;n \rgt) \rgt] \equiv e^{X'(\f)}
\]
This property says that 
% vero se la somma converge ! ! !
integrating an exponential one still gets an exponential, whose argument is expressed by the sum of truncated expectations.
\item {\it Change of integration.} If $P_M(d \ps)$ denotes the integration with covariance $M$, then for any analytic function $F(\ps)$ one has
\[ \label{change_integration}
\frac{1}{\NN_N} \int P_M(d \ps) e^{-\sum_{i,j \in \L} \ps^+_i N^{-1}_{ij}\ps^-_j} F(\ps) = \int P_{\tl{M}}(d \ps) F(\ps)
\]
where
\[
\tl{M}^{-1} = M^{-1} + N^{-1}
\]
and
\[
\NN_N = \int P_M(d \ps) e^{-\sum_{i,j \in \L} \ps^+_i N^{-1}_{ij}\ps^-_j}= \frac{\det M^{-1}}{\det \lft( M^{-1} + N^{-1}\rgt)} = \lft[\det (\unit + N^{-1}M)\rgt]^{-1}
\]
\end{enumerate}

%\blue{Expansion of the truncated expectation in terms of Feynman diagrams?}

\pagina 
\section{Proof of lemma \ref{lemma}} \label{App-lemma}
We want to obtain a bound on the modulo of the propagator $g^{(h)}_{\a \a'}(x)$ defined as follows:
\be 
g_{\a \a'}^{(h)}(x)=\frac{1}{(2\pi)^4}\int d^{d}k \,f_{h}(|k|)\, 
\frac{ p^{h}_{\a \a'}(k)}{|k|^2}\, e^{-ikx}
\ee
where
\[
p_{\a \a'}^{h}(k) & =(\r_0 R_0^2)^{-1} \left(\begin{array}{cc}
\kk^{2}   & -k_{0}\\
k_{0} & \quad \kk^{2}
\end{array}\right)  
&  |k|^2=k^2_0 + \kk^4 \quad  \text{for } h \geq \bh \non \\[6pt]
p_{\a \a'}^{h} (k) & =(\r_0 R_0^2)^{-1} \left(\begin{array}{cc}
 \kk^{2}  & -k_{0}\\
k_{0} &   \e
\end{array}\right)  & |k|^2=k^2_0 + \e \kk^2  \quad \text{for } h < \bh 
\] 
with the first row and column referring to $\a=l$ and the second row and column to $\a=t$. 

{\centering \subsubsection{Case $\bh <h \leq 0$ }}
Let start from the case $\bh <h \leq 0$ . The bound we want to prove is:
\[ \label{bound_above}
 \lft|g_{\a \a'}^{(h)}(x) \rgt| \leq \g^{\frac{d}{2}h}\, \frac{C_N\,(\r_0 R_0^2)^{-1} }{1+\lft[(\g^h x_0)^{2}  + (\g^{\frac{h}{2}} \xx )^{2} \rgt]^N}
\]
Let consider the modulo of the product $(\g^h x_0)^{2}g^{(h)}(x)$:
\[ \label{bound1}
\lft|(\g^h x_0)^{2} g_{\a \a'}^{(h)}(x) \rgt| & = \lft|\frac{1}{(2\pi)^4}\int d^{d}k \,f_{h}(|k|)\, \frac{ p^{h}_{\a \a'}(k)}{|k|^2}\,\g^{2h}\, \dpr^2_{k_0}\,e^{-ikx} \rgt| \non \\
& \leq \frac{1}{(2\pi)^4}\int d^{d}k \lft| \g^{2h}\, \dpr^2_{k_0} \lft[ \,f_{h}(|k|)\, \frac{ p^{h}_{\a \a'}(k)}{|k|^2}\,e^{-ikx} \rgt] \rgt|
\]
where in the second line we have used the integration by parts. Now making a change of variable
\[
& \g^{-h}k_0 \arr k_0 \non \\
& \g^{-\frac{h}{2}} \, \kk \arr \kk
\]
$|k|^2= k_0^2 + \kk^4 \arr \g^{2h}\,|k|^2$ and the integral in \eqref{bound1} becomes equal to an adimensional integral (which can be bounded by a constant independent on $h$) multiplying a dimensional factor
\[
\eqref{bound1} \leq C\,(\r_0 R_0^2)^{-1}\,\g^{\lft(\frac{d}{2}+1\rgt)}\,\g^{2h}\, \g^{-2h}\, \g^{-h} = C\,(\r_0 R_0^2)^{-1}\,\g^{\frac{d}{2}h}
\]
where the factor $\g^{\lft(\frac{d}{2}+1\rgt)}$ comes from the integration over $d^d\kk dk_0$; the factor $\g^{2h}$ was already present in the integrand; the factor $\g^{-2h}$ comes from the second derivative with respect to $k_0$; $\g^{-h}$ comes from $p_{\a \a'}^{(h)}(k)/|k|^2$. 
Following the same strategy we can obtain the general bound
\[  \label{bound2}
\lft|(\g^h x_0)^{2N} g_{\a \a'}^{(h)}(x) \rgt| \leq C_N\,(\r_0 R_0^2)^{-1} \g^{\frac{d}{2}h}
\]
and the analogous result for the $\xx$ variable:
\[ \label{bound3}
\lft|(\g^{\frac{h}{2}} \xx )^{2N} g_{\a \a'}^{(h)}(x) \rgt| \leq C_N\,(\r_0 R_0^2)^{-1}\, \g^{\frac{d}{2}h}
\]
where the fact that the $\xx$ variable in \eqref{bound2} is multiplied for a factor $\g^{\frac{h}{2}}$ instead of $\g^{h}$ comes from the different power of $k_0$ and $\kk$ in the definition of $|k|^2$ for $h \geq \bh$. Putting together the bounds \eqref{bound2} and \eqref{bound3} one obtains:
\[
 \lft|\Bigl \{ 1+\bigl[(\g^h x_0)^{2}  + (\g^{\frac{h}{2}} \xx )^{2} \bigr]^N   \Bigr \} g_{\a \a'}^{(h)}(x) \rgt| \leq C_N \,(\r_0 R_0^2)^{-1}\,\g^{\frac{d}{2}h}
\]
which is equivalent to \eqref{bound_above}. \\

{\centering \subsubsection{Case $h \leq \bh$ }}
In order to prove the bound in lemma \ref{lemma} for the behavior of the propagator for $h < \bh$ we will follow the same strategy used in proving \eqref{bound2}, but with a different change of variable
\[
k'_0  &=\g^{-h}\,k_0\non \\
\kk' & = \g^{-h}\, \sqrt{\e}\,\kk 
\]
since in this region $|k|^2= k_0^2 + \e\,\kk^2$. Besides, $p_{\a \a'}^{(h)}(k)$ has a dimensional scaling depending on $\a$ and $\a'$. We have:
\be \label{matrix_below}
p_{\a \a'}^{(h)}(k') = (\r_0 R_0^2)^{-1}\,
\left(\begin{array}{cc}
	      \e^{-1}\, \g^{2h}\kk'^{2}   & -\g^{h} k'_{0}\\[6pt]
	       \g^{h} k'_{0} &   \e
		\end{array}\right)  
\ee
From \eqref{matrix_below} we see that if one of the labels $\a, \a'$ are equal to $l$ we get a factor $\g^h$; if both are of type $l$ we have $\g^{2h}$; if both are of type $t$ we have no dependence on $h$.  Then we can write the dimensional factor coming from $p_{\a \a'}^{(h)}(k') $ as $\g^{h \lft( \d_{\a l}+ \d_{\a' l}  \rgt)}$. For what regards the order in $\e$ we have a factor $\e^{-d/2}$ coming from the change of variable in $d^d\kk$. Putting all this factors together we get, for example, 
\[ 
\lft|(\g^h x_0)^{2} g_{\a \a'}^{(h)}(x) \rgt| &   \leq \e^{-(\frac{d}{2}-1+\d_{\a l}+\d_{\a' l})}\int \frac{d^{d+1}k'}{(2\pi)^{d+1}} \lft| \g^{2h}\, \dpr^2_{k'_0} \lft[ \,f_{h}(|k'|)\, \frac{ p^{h}_{\a \a'}(k')}{|k'|^2}\,e^{-ik'x} \rgt] \rgt| \non \\[6pt]
% &  \leq C\,\e^{\frac{1}{2}(-d-1+\d_{\a l}+\d_{\a' l})}\,\g^{\lft(d+1\rgt)\,h}\,\g^{2h}\, \g^{-2h}\, \g^{-2h}\, \g^{h \lft( \d_{\a l}+ \d_{\a' l}\rgt)} \non \\
& \leq C\,(\r_0 R_0^2)^{-1}\,\e^{-(\frac{d}{2}-1+\d_{\a l}+\d_{\a' l})}\g^{\lft(d-1 + \d_{\a l}+ \d_{\a' l} \rgt)\,h}
\]
The final bound for  $h \leq \bh$  results to be:
\[ \label{bound_above}
\lft|g_{\a \a'}^{(h)}(x) \rgt| \leq \e^{\frac{d}{2}}(\e^{-1}\g^h)^{\lft(d-1 + \d_{\a l}+ \d_{\a' l} \rgt)} \, \frac{C_N\,(\r_0 R_0^2)^{-1}}{1+\lft[(\g^h x_0)^{2}  + (\g^h \sqrt{\e}\,\xx )^{2} \rgt]^N}
\]

\pagina
%------------------------------------------------------------------------------------------------
\section{Order in $\e$, $\r_0$ and $R_0$ for the Feynman diagrams} \label{order_e}

In this section we will  evaluate the dependence on $\e$, $\r_0$ and $R_0$ of a generic Feynman diagram contributing to $|| V_{n_l^\txe n_t^\txe}^{(h)}||$ in both region $\bh<h \leq 0$ and $h \leq \bh$. The difference between the two regions stays in the fact that in the lower region the presence of Bogoliubov measure in the propagator gives a non trivial dependence on $\e$. In particular  the order in $\e$ in the lower region turns to depend on the number of loops in the Feynman diagrams, instead than on the number of vertices as usual. 
%The leading contribution to $|| V_{n_l^\txe n_t^\txe}^{(h)}||$ for $h \leq \bh$ are then given by the one loop Feynman diagrams without vertices $\n_h$ in the $3d$ case and without vertices $\n_h$ or $\l_{6,h}$ in the $2d$ case. 

\subsection{Dimensional  factor}
First of all, let's consider the contribution coming from the dimensional factors associated to vertices, propagators and integrations. We have:
\begin{enumerate}[a)]
\item for each vertex a factor $\r_0 R_0^{-2}$;
\item for each propagator in the momentum space $(\r_0 R_0^{-2})^{-1}$
\item for each loop a factor $R_0^{-2-d}$
\end{enumerate}
Denoting with $L$ the number of loops one gets
\[ \label{fixing_dim}
\r_0 R_0^{-2} \,(\r_0 R^d_0)^{-L}
\]
being $L= p - m +1$, with $p$ and $m$ respectively the total number of propagator and vertices. In \eqref{fixing_dim} the factor $\r_0 R_0^{-2}$ the exact physical dimension for a generic term of the effective potential, while $\r_0 R^d_0$ is an adimensional factor. 
%Since we have introduced adimensional running coupling constants, their beta function does not contain the dimensional factor $\r_0 R_0^{-2}$. 
By definition of the adimensional running coupling constants, the beta function is also adimensional and it is equal to the sum of the diagrams with at least two vertices contributing to $|| V_{n_l^\txe n_t^\txe}^{(h)}||$, divided by the fixing dimension factor $\r_0 R_0^{-2}$.

\subsection{Region $\bh<h \leq 0$}

Let denote with $\bar{C}(P_v;\e, \r_0, R_0)$ the factor in $|  \int dx_{v_0} \Val(\G)|$  containing the dependence on $\e$, $\r_0$ and $R_0$, with $\G$ a generic diagram contributing to the kernel $|| V_{n_l^\txe n_t^\txe}^{(h)}||$ for $\bh<h \leq 0$. Using \eqref{fixing_dim}, introducing the number of loops $L^>$ 
\[ \label{loop0}
L^> & =  \frac{1}{2}(n_{l}^\txi+n_{t}^\txi)-(m_{v}-1)  =  m_{4}+\frac{1}{2}m_{3}  -\frac{1}{2}n_{l}^\txe-\frac{1}{2}\,n_{t}^\txe+1
\]
 and taking into account the fact that $\e=\l \r_0 R_0^d$, we have:
\[  \label{bar_C}
& \bar{C}(P_v;\e, \r_0, R_0)  
 = \r_0 R_0^{-2}\:(\l \e^{-1})^{1-\frac{1}{2}(n_l^\txe + n_t^\txe)} \non \\[6pt]
& \qquad \quad  \prod_v (\l \e^{-1}\,\bl_{h_v})^{m_{4,v}}  (\sqrt{\l \e^{-1}\,}\bm_{h_v})^{m_{3,v}} (\bn_{h_v})^{m_{2,v}} (\bar{a}_{h_v})^{m_{a,v}} (\bar{e}_{h_v})^{m_{e,v}}
\]
where in three dimensions the vertices of type $\l$ can be present only at scale $h=0$, being irrelevant. Here $m_{4,v}$,  $m_{3,v}$ and $m_{a,v}$ are respectively the sum of vertices of type $\{\bl_{h_v}, \bl'_{h_v}, \bl''_{h_v}\}$, $\{\bm_{h_v}, \bm'_{h_v}\}$  and $\{\bar{a}_{h_v}, \bar{a}'_{h_v}\}$, while $m_{2,v}$ and $m_{e,v}$ denote the number of vertices of type $\bn$ and $\bar{e}$ at scale $h_v$.  If all the vertices in \eqref{bar_C} are at scale $0$ we have
\[ \label{bar_C0}
\bar{C}_0(P_v;\e, \r_0, R_0) & 
 = \r_0 R_0^{-2}\:(\l \e^{-1})^{1-\frac{1}{2}(n_l^\txe + n_t^\txe)} \prod_v (\l)^{m_{4,v}+m_{2,v}}  (\sqrt{\l \e\,})^{m_{3,v}} 
\]
being $\bl=\bm=\e$ and $\n_0=\l$. Assuming that
\[
\bar{\h}=\max_{\bh <h \leq 0}\{\l \e^{-1}\bl_h,\sqrt{\l \e^{-1}\,}\bm_h,\, \bn_h\,,\bar{a}_h,\,\bar{e}_h \} 
\]
is a small constant the factor \eqref{bar_C} may be bounded for each $\bh<h \leq 0$ by 
\[
\bar{C}(P_v;\e, \r_0, R_0) & 
 \leq \r_0 R_0^{-2}\:(\l \e^{-1})^{1-\frac{1}{2}(n_l^\txe + n_t^\txe)} \,\bar{\h}^{m}
\]
with $m$ the total number of vertices. It turns to be $\bar{\h}=\sqrt{\l}$ both in the three and two dimensional case. We see as in the higher momentum region the dominant diagrams -- apart for the factor $(\l \e^{-1})$ that we can consider one for the moment, are the diagrams with minimum number of vertices. \\

A particular estimate can be obtained considering a diagram where all the vertices are at scale $\bh$. In this case $\bl_\bh=\e$, $\bm_\bh=\sqrt{\e}$ while the two legged vertices are $O(\l)$; we get
\[
\bar{C}^{(\bh)}(P_v;\e, \r_0, R_0) & 
 \leq \r_0 R_0^{-2}\: \e^{\frac{1}{2}(n_l^\txe + n_t^\txe)-1} \,\l^{L+m_2+m_a+m_e}
\]
that is the dominant diagrams at scale $\bh$ are the one--loop diagrams without two--legged vertices. This will be the typical situation for $h \leq \bh$, as we will see in the next paragraph.

%The renormalized propagator is not changed by the integration over the momenta in the region $ \bh < h\leq 0$, since $A_\bh$ and $E_\bh$ differ from one for higher order in $\e$, $Z_\bh$ is equal to $\e$ except for higher order corrections and $B_\bh$ gives neglectful contribution to the propagator, since 

%Since $ \g^{2(h-1)} \leq k_{0}^{2}+\e\kk^{2} \leq \g^{2h}$ at each step $h$  the spatial momentum satisfies $ \e^{-1} \g^{2(h-1)} \leq \kk^2 \leq \e^{-1} \g^{2h}$. Then for each integration in the $d$--dimensional momentum space we get a factor $\e^{-\tfrac{d}{2}}$ and each propagator $g^{(h)}_{ll}(k)$ can be bounded by $\e^{-1}$. The estimate for the renormalized propagator in the $x$-space for $h \leq \bh$ is given by the following lemma: 
%{\lemma (Renormalized propagator for $h \leq \bh$)
%\[
%&  \lft|g_{tt}^{(h)}(x)\rgt|  \leq \lft( \r_0 R_0^d\rgt)^{-1}\, \e^{-\tfrac{d}{2} -1}\,\g^{(d-1)h} \,\frac{C_N}{ 1+ \lft[(\g^h x_0 + \g^h \xx)^2 \rgt]^N} \non \\
%&  \lft|g_{tl}^{(h)}(x)\rgt|\leq \lft( \r_0 R_0^d\rgt)^{-1}\, \e^{-\tfrac{d}{2}}\,\g^{d h} \,\frac{C_N}{ 1+ \lft[(\g^h x_0 + \g^h \xx)^2 \rgt]^N}  \non \\
%& \lft|g_{ll}^{(h)}(x)\rgt|\leq \lft( \r_0 R_0^d\rgt)^{-1}\, \e^{-\tfrac{d}{2}+1}\,\g^{(d+1) h} \,\frac{C_N}{ 1+ \lft[(\g^h x_0 + \g^h \xx)^2 \rgt]^N} 
%\]
%}

\subsection{Region $h\leq \bar{h}$} 

\subsubsection{Order in $\e$ coming from integrations and propagators}

In the region $h \leq \bh$, due to the presence of $\e$ in the definition of the cutoff function, the order in $\e$ turns to be not only dependent on the number of vertices. For this reason it is useful to evaluate the power in $\e$, for a generic Feynman diagram $\G$ contributing to a certain kernel $|| V_{n_l^\txe n_t^\txe}^{(h)}||$, coming from the integrations over the connected lines of the diagrams. We denote with $D(P_v,\e)$ this factor. We have that:
\begin{enumerate}[a)]
\item each internal half--line of type $l$ gives a contribution $\e^{\frac{-2-d}{4}}$, while each internal half--line of type $t$ gives a contribution $\e^{\frac{2-d}{4}}$; then each $l$ half--line has got an extra factor $\e^{-1}$ with respect to a $t$ line (see lemma \ref{lemma});
\item each integration in the $x$-space gives a contribution proportional to $\e^{\frac{d}{2}}$; 
%since at each step $h$ we have $ \g^{2(h-1)} \leq k_{0}^{2}+\e\kk^{2} \leq \g^{2h}$;
\item for each derivative $\dpr_\xx$ acting on a propagator $g^{(h)}(x)$ we get an extra contribution $\e^{-1}$ coming from the change of variables $\kk^2 \arr \e^{-1} \kk'^2$. Then we may associate at each contracted half--line bringing a label $\dpr_\xx$ a factor $\e^{\frac{1}{2}}$ 
\end{enumerate}
Using the same notations introduced in section \ref{dimensional} and taking into account only the marginal and relevant couplings without external fields we get:
\[ 
D(P_v,\e) =  \frac{d}{2}\,(m -1) - \frac{2+d}{4} \,n^\txi_{l}  +
\frac{2-d}{4}\,n^\txi_{t}  -\frac{1}{2}\, n^\txi_{\dpr_\xx} 
\]
where $m$ is the total number of vertices.  Using the relations \eqref{prop2} and \eqref{prop3} we get:
\[ \label{delta_e}
D(P_v,\e)  =  &\, -\frac{d}{2}+\lft(2-\frac{d}{2}\rgt)m_{4}+\frac{1}{2}\lft(1-\frac{d}{2}\rgt)m_{3}+m_{2}+ m_6 \,\c(d=2)  \\[6pt] \non 
&\, + \frac{d+2}{4}\, n_{l}^{\txe} + \frac{d-2}{4}\,n_{t}^\txe + \frac{1}{2}\, n_{\dpr_\xx}^\txe
\]
with $\c(d=2)$ different from zero only in the two dimensional case.
Introducing the number of loops $L$ in the low momenta region, which is given by
\[ \label{loop}
L & =  m_{4}+\frac{1}{2}m_{3} + 2m_6\, \c(d=2) -\frac{1}{2}n_{l}^\txe-\frac{1}{2}\,n_{t}^\txe+1
\]
we can conveniently rewrite \eqref{delta_e} as follows 
\[   
D_{3d}(P_v,\e)  &=  -2 + \frac{1}{2}L -\frac{1}{2}m_3 +m_2 +\frac{3}{2}n_l^\txe +\frac{1}{2}n_t^\txe +\frac{1}{2}n_{\dpr_\xx}^\txe  \label{D3d}  \\[6pt]
D_{2d}(P_v,\e)  & =  -2 + L -\frac{1}{2}m_3 +m_2 -m_6 +\frac{3}{2}n_l^\txe +\frac{1}{2}n_t^\txe +\frac{1}{2}n_{\dpr_\xx}^\txe   \label{D2d}
\]
Using  \eqref{fixing_dim}, \eqref{D3d} and \eqref{D2d} we obtain the dependence on $\e$, $\r_0$ and $R_0$ for $|\int dx_{v_0} \Val(\G)|$, with $\G$ a generic diagram contributing to the kernel $|| V_{n_l^\txe n_t^\txe}^{(h)}||$ for $h \leq \bh$. Let denote with $\tl{C}(\e, \r_0, R_0)$ the factor containing this dependence. In the three dimensional case we have
\[ \label{order_e_3d}
\tl{C}_{3d}(P_v;\e, \r_0, R_0) =  &\, \r_0 R_0^{-2} \, \bigl(\l \e^{-\frac{1}{2}}\bigr)^L\, 
 \e^{-2  +\frac{3}{2} n_{l}^\txe +  \frac{1}{2}n_{t}^\txe + \frac{1}{2} n_{\dpr_\xx}^\txe}\, \e^{-\frac{1}{2}m_3+m_{2}} 
\]
where we have split the terms containing the dependence on the loop number, on the vertices contained in the diagram and on the external legs (the latter term being fixed for each kernel $V_{n_l^\txe n_t^\txe}^{(h)}$).
The analogous factor for $d=2$ is
\[ \label{order_e_2d}
\tl{C}_{2d}(P_v;\e, \r_0, R_0) =  &\, \r_0 R_0^{-2} \, \l^{L}\,
 \e^{-2 +\frac{3}{2}  n_{l}^\txe + \frac{1}{2}  n_{t}^\txe + \frac{1}{2} n_{\dpr_\xx}^\txe}\, \e^{-\frac{1}{2}m_3 +m_2 -m_6} 
 \non \\[6pt]
= & \, \r_0 R_0^{-2} \, \l^{1-\frac{1}{2}(n_l^\txe + n_t^\txe)}\,  
 \e^{-2  +\frac{3}{2}  n_{l}^\txe + \frac{1}{2}  n_{t}^\txe + \frac{1}{2} n_{\dpr_\xx}^\txe} \non \\
& \hskip 1.2cm \l^{m_4}\, (\sqrt{\l \e^{-1}\,})^{m_3}\, \e^{m_2} (\l^2 \e^{-1})^{m_6}
\]

In the following we will add to the estimates \eqref{order_e_3d} and \eqref{order_e_2d} the contribution coming from the vertices in two particular situations: the one in which all the vertices are at scale $\bh$ and the one in which they are all at scale $h^* \ll \bh$. 

\subsubsection{Diagrams with all vertices at scale $\bh$}

The values of the running coupling constants at $\bh$ are:
\[
& \begin{array}{lllll}
\l_ \bh= \frac{\e}{16}   &  \m_\bh= \frac{\sqrt{2}}{4}\,\e &  \n_\bh = O(\l \e^{-1/2}) & & d=3\non \\[9pt]
\l_ \bh= \frac{1}{16}  \quad &  \m_\bh= \frac{1}{4}\sqrt{2\,\e} \quad  & \n_\bh = O(\l) &  \quad \l_{6, \bh} = O(\l\e) \qquad&  d=2 
\end{array}  
\]
Adding the contributions of the vertices to \eqref{order_e_3d} and \eqref{order_e_2d}  we get:  
\[ 
\tl{C}^{(\bh)}_{3d}(P_v;\e, \r_0, R_0)=  &\, \r_0 R_0^{-2} \, \bigl(\l \e^{\frac{1}{2}}\bigr)^L\, 
 \e^{-3  +2 n_{l}^\txe + n_{t}^\txe + \frac{1}{2} n_{\dpr_\xx}^\txe}\, \e^{\frac{3}{2}m_{2}}  \label{3d_bh} \\[6pt]
\tl{C}^{(\bh)}_{2d}(P_v;\e, \r_0, R_0) =  &\, \r_0 R_0^{-2} \, \l^{L}\,
 \e^{-2 +\frac{3}{2}  n_{l}^\txe + \frac{1}{2}  n_{t}^\txe + \frac{1}{2} n_{\dpr_\xx}^\txe}\, (\l \e)^{m_2} (\l)^{m_6}
 \non \\[6pt]
= & \, \r_0 R_0^{-2} \, \l^{1-\frac{1}{2}(n_l^\txe + n_t^\txe)}\,  
 \e^{-2  +\frac{3}{2}  n_{l}^\txe + \frac{1}{2}  n_{t}^\txe + \frac{1}{2} n_{\dpr_\xx}^\txe}
\l^{L}\,(\l \e)^{m_2}  \l^{m_6}   \label{2d_bh}
\]
If all the vertices are at scale $\bh$ we see as the order in $\l$ and $\e$ depends on:
\begin{itemize}
\item the number of loops;
\item the structure of the external legs, in particular if we substitute a $t$ external leg with a leg of type $l$ we have an extra $\e$;
\item the number of $2$-legged vertices; 
%which counts as extra loops, \ie a diagram with one loop and a 2--legged vertex has the same order in $\e$ of a two loops diagram;
\item in the two dimensional case the number of $6$--legged vertices, which counts as a loop.
\end{itemize}
Once the structure of the external legs is fixed, the dominant diagrams are the one-loop graphs without $2$-legged in three dimensions and without $2$-legged and $6$-legged vertices in two dimensions. 
%This means, for example, that the relevant running coupling $\n_\bh$ does not appear in the flow equations of $\l_\bh$ and $\m_\bh$ at leading order.

\subsubsection{Diagrams with vertices at scale $h \ll \bh$}

If $h\ll \bh$ we have also to take into account the expression for the renormalized propagator, which is:
\[
g^{(h)}(k)=\frac{\left(\begin{array}{cc}
 Z_{h}^{-1} \lft(k_0^2 + \e^{-1}\kk^2 \rgt )\quad & k_{0}\\
-k_{0} & \e
\end{array}\right)}{k_{0}^{2}+\e \kk^2}
\] 
With respect to the estimate for the unrenormalized propagator we have a factor $\e Z_{h}^{-1}$ for each propagator $g^{(h)}_{ll}(k)$. Then the order in $\e$ for a generic diagram, taking also into account the contribution of the vertices,  is given by
\[
\tl{C}^{\{h_v\}}_{3d}(P_v;\e, \r_0, R_0) =  &\, \r_0 R_0^{-2} \, \bigl(\l \e^{-\frac{1}{2}}\bigr)^L\, 
 \e^{-2  +\frac{3}{2} n_{l}^\txe +  \frac{1}{2}n_{t}^\txe + \frac{1}{2} n_{\dpr_\xx}^\txe} \non \\
& \prod_v (\l_{h_v}) (\e^{-\frac{1}{2}}\m_{h_v}) (\e \n_{h_v})\prod_{n_{ll}}(\e Z_{h_v}^{-1})  \label{3d_h}  \\[6pt]
\tl{C}^{\{h_v\}}_{2d}(P_v;\e, \r_0, R_0) = 
& \, \r_0 R_0^{-2} \, \l^{1-\frac{1}{2}(n_l^\txe + n_t^\txe)}\,  
 \e^{-2  +\frac{3}{2}  n_{l}^\txe + \frac{1}{2}  n_{t}^\txe + \frac{1}{2} n_{\dpr_\xx}^\txe} \non \\
& \prod_v (\l \l_{h_v})\,(\sqrt{\l \e^{-1}\,} \m_{h_v})\, (\e \n_{h_v}) (\l^2 \e^{-1}\l_{6,h_v}) \prod_{n_{ll}}(\e Z_{h_v}^{-1})  \label{2d_h}
\]
where $n_{ll}$ is the number of $g^{(h)}_{ll}(k)$ propagators and may be bounded by  
\[ \label{n_ll}
n_{ll}\leq \Bigl\lfloor \frac{m_{3}-n_{l}^\txe }{2}\Bigr\rfloor  
%= \frac{m_3 - n_l^\txe - \c( \text{odd})}{2}
\]
%with $\c(\text{odd})$ equal to one if $m_3 - n_l^\txe$ is odd and equal to zero otherwise. 
In the particular case in which {\it all the vertices and propagators have same scale $h$} we get the following factors.

{\subsubsection{\it  Asymptotic three dimensional case.} }

Using the fact that - as proved in chapter \ref{WI} - 
\[
16 \l_{h}  = 2\,\sqrt{2}\,\m_{h}= Z_{h}
\]
the dependence on $\e$, $\r_0$ and $R_0$ for a generic diagram $\G$ contributing to the kernel $|| W_{n_l^\txe n_t^\txe}^{(h)}||$ with all the vertices at scale $h$ is given by
\[
\tl{C}^{(h)}_{3d}(P_v;\e, \r_0, R_0) =  &\, \r_0 R_0^{-2} \, \bigl(\l \e^{-\frac{1}{2}}\bigr)^L\, 
 \e^{-2  +\frac{3}{2} n_{l}^\txe +  \frac{1}{2}n_{t}^\txe + \frac{1}{2} n_{\dpr_\xx}^\txe} \non \\
&  (Z_{h})^{L-1 +\frac{1}{2}(n_l^\txe+n_t^\txe)} \,(\e^{-1}Z_h)^{(\frac{1}{2}m_3 -n_{ll})} \,(\e \n_h)^{m_2}  \label{3d_hh}
\]
Asymptotically, being $Z_{h_*} = \e^{\frac{1}{2}}(\l \, |h_* -\bh|)^{-1}$ the factor in \eqref{3d_hh} becomes
\[
\tl{C}^{(h_*)}_{3d}(P_v;\e, \r_0, R_0) =  &\, \r_0 R_0^{-2} \: \l^{1-\frac{1}{2}(n_l^\txe +n_t^\txe) }\:
 \e^{-\frac{5}{2}+\frac{7}{4} n_{l}^\txe +  \frac{3}{4} n_{t}^\txe   + \frac{1}{2} n_{\dpr_\xx}^\txe} \non \\
&  \lft|h_*-\bh \rgt|^{-\lft(L-1 +\frac{1}{2}(n_l^\txe+n_t^\txe)\rgt)} \,\lft(\l\,\e^{\frac{1}{2}} |h_*-\bh| \rgt)^{-(\frac{1}{2}m_3 -n_{ll})}\,(\e \n_h)^{m_2}    \label{3d_hstar}
\]
where 
\begin{itemize}
\item the first line on the r.h.s of  \eqref{3d_hh} is fixed once $\{P_{v_0}\}$ is fixed;
\item the second term on the second line of \eqref{3d_hh} can only improve the estimate, being $m_3/2 -n_{ll} \geq 0$;
\item the addition of two--legged $m_2$ vertices improves the order in $\e$, being $\n_h$ bounded by a constant.  
\end{itemize}
Note that all the terms in the beta functions for $\l_h$, $\m_h$, $Z_h$ and $E_h$ with more than one loop are at least of order $|h - \bh|^{-3}$. For what concerns the wave function renormalization constants $A_h$ and $B_h$ the power in $|h - \bh|$ improves with respect to the estimate \eqref{3d_hh} due to some cancellations, encoded in the WIs and that can also be checked with a one loop calculation (see \ref{appB.WI_B}). \\

%Using \eqref{C3d_1} we see that the one-loop graph contribution to the beta function of $Z_{h}$ is proportional to $\l\,\e^{-\frac{1}{2}} \, Z_h^2$. Then the leading order flow equation is
%\[
%Z_{h-1}-Z_{h} & =  - c\, \l\,\e^{-\frac{1}{2}} Z_{h}^{2} 
%\]
%with $c$ a positive constant independent on $\l$ and $\e$ and we get \eqref{asymptotic_3d}. \\

{\subsubsection{\it Asymptotic two dimensional case.} }

The calculation of the order in $\e$ and $\l$ for the asymptotic two dimensional case is much subtle, since -- as discussed in sec. \ref{effective} -- three new effective marginal couplings arise when $h \ll \bh$, see fig. \ref{effective} pag. \pageref{effective}. The presence of these additional vertices -- in this thesis denoted with $\o_h$, $\l'_h$ and $\m'_h$, see \eqref{L_extra} for a definition -- changes the factor \eqref{order_e_2d}, which gives the order in $\e$ and $\l$ coming from integrations and propagators, as follows:
\[  
& \tl{C}'_{2d}(P_v;\e, \r_0, R_0) = 
  \r_0 R_0^{-2} \, \l^{1-\frac{1}{2}(n_l^\txe + n_t^\txe)}\,  
 \e^{-2  +\frac{3}{2}  n_{l}^\txe + \frac{1}{2}  n_{t}^\txe + \frac{1}{2} n_{\dpr_\xx}^\txe} \non \\[6pt]
& \qquad  \qquad  \l^{m_4}\,  (\l \e^{-1})^{\frac{1}{2}m_3}\, \e ^{m_2}\, (\l^2 \e^{-1})^{m_6}   (\l \e^{-2})^{m'_4}\, (\l \e^{-5})^{\frac{1}{2}m'_3}\,(\l \e^{-1})^{\frac{3}{2}m_5}\,
\]
with $m_5$, $m'_4$ and $m'_3$ respectively the numbers of $\m_h$, $\l'_h$ and $\m'_h$ vertices and the loop number given by
\[ \label{Loop_eff}
L'=m_4 + m'_4 + \frac{1}{2}(m_3+m'_3) +\frac{3}{2}m_5 +2m_6 -\frac{1}{2} (n_l^\txe +n_t^\txe) +1
\]
Taking into account the contribution of the vertices and of the longitudinal propagators $g_{ll}^{(h)}$ and using the following global WIs  
\[   \label{asymptotic_2d}
\m_{h}  & =  4 \sqrt{2} \,\g^{\frac{h}{2}}\,\l_h   & Z_h = \g^{h}\l_h & \non \\[6pt]
 \o_h  & =6 \sqrt{2}\, \g^{\frac{h}{2}}\,\l_{6,h} 
& \l'_h =  24\, \g^{h}\,\l_{6,h} &
& \m'_h  = 16 \sqrt{2}\, \g^{\frac{3}{2}h}\,\l_{6,h} 
\]
we obtain
\[
& \tl{C}^{\{h_v\}}_{2d}(P_v;\e, \r_0, R_0) \leq \cst \r_0 R_0^{-2}\,  \l^{1-\frac{1}{2}(n_l^\txe+n_t^\txe)}\,\e^{-2 +\frac{3}{2}n_l^\txe+ \frac{1}{2}n_t^\txe + \frac{1}{2}n_\dpr^\txe}  \non \\[6pt]
&  \qquad \quad \prod_{v \text{e.p.}}\, \lft(\,\l \l_{h_v} \,\rgt)^{m_{4,v}+ m'_{4,v}+\frac{1}{2}(m_{3,v}+m'_{3,v}) +\frac{3}{2}m_{5,v} +2m_{6,v}}\,(\e \n_{h_v})^{m_{2,v}}\, \non \\[6pt]
&\qquad \lft(\,\frac{\l_{6,{h_v}}}{\e \l_{h_v}^2} \,\rgt)^{m_{6,v}+ m_{5,v} + m'_{4,v} +m'_{3,v}}  \lft(\e^{-1}\g^{h_v} \l_{h_v}\rgt)^{\frac{1}{2}(m_{3,v} + m_{5,v} + 2m'_{4,v} +3m'_{3,v})- n^v_{ll}}
\]
By imposing the renormalization condition one find that in the infrared limit $|\n_h| \leq \cst$, see sec.~\ref{nu}. Let us also assume $\l \l_h$ and $\l_{6,h}/(\l \l_h^2)$ to have fixed points $\l \l_*$ and $c_{6,*}$, hypothesis whose consistency will be proved at leading order, see sec.~\ref{lambda6}. Then the order in $\e$ and $\l$ for a generic diagram $\G$ where all the vertices are at scale $h_* \ll \bh$ turns out to be
\[  \label{2d_hstar}
\tl{C}^{(h_*)}_{2d}(P_v;\e, \r_0, R_0)  \leq &\; \cst \,\r_0 R_0^{-2}  \l^{1-\frac{1}{2}(n_l^\txe+n_t^\txe)}\,\e^{-2 +\frac{3}{2}n_l^\txe+ \frac{1}{2}n_t^\txe + \frac{1}{2}n_\dpr^\txe} 
\non \\[6pt]
&  \, \lft(\,\l \l_{*} \,\rgt)^{L' -1+\frac{1}{2}(n_l^\txe +n_t^\txe )}\, \lft(\e \n_* \rgt)^{m_2} c_{6,*}^{m_{6,v}+ m_{5,v} + m'_{4,v} +m'_{3,v}}\non \\[6pt]
& \lft(\e^{-1}\g^h \l_* \rgt)^{\frac{1}{2}(m_3 + m_5 + 2m'_4 +3m'_3)- n_{ll}} 
\]
where $L'$ is the number of loops for a generic diagram obtained by contracting vertices of type $\l_{6,h}$, $\o_h$, $\l_h$, $\l'_h$, $\m_h$, $\m'_h$ and $\n_h$.
%
%The leading order diagrams are the one loop diagram (remind that the loop number is equal to )
Provided that $c_{6,*}$ is a constant and $\l\l_* \ll 1$, the leading order diagrams are the {\it one loop diagrams  where all the internal dashed legs are contracted among them}, in order to minimize the exponent of $\g^h$, on the last line of \eqref{2d_hstar}. Moreover, the insertion of a two--legged vertex give an extra $\e$ with respect the order of the diagram without $m_2$ vertices. \\

%A second order calculation for $\l_h$ suggests $\l_*= \l^{-1}$; in fact, using \eqref{C2d_1}, we find that the beta function for $\l_h$ at leading order is equal to 
%\[
%\b^\l_h =- c\, \l \, (\l_*)^2
%\]
%with $c$ a positive constant independent on $\l$, $\r_0$ and $R_0$. Then the flow equation
%\[
%\l_{h-1} - \g \l_h = \b^\l_h
%\]
%in the asymptotic regime in which $\l_h \arr \l_*$ gives
%\[
%\l_* = \frac{(\g -1)}{c} \,\l^{-1}
%\]

\subsection{Order in $\e$ coming from external fields}

In this section we will describe how the insertion of the running coupling constants with the external fields $J_\n$ change the order in $\e$ of a generic Feynman diagram in the region $h \leq \bh$. 

First af all we note that the two legged vertices $Z_h^{J_0}$, $E_h^{J_0}$ and $E_h^{J_1}$ do not contribute to the local part of the Feynman graphs since they have only one leg free to be contracted. The unique vertices with external field contributing to the flows of the running coupling constants are $\bm_h^{J_0}$ and $\bm_h^{J_1}$  for $\bh < h \leq 0$ and $\m_h^{J_0}$ for $h \leq 0$.

\vskip 0.2cm

For what concerns the high momentum region, including also $\bm^{J_0}_h$ and $\bm^{J_1}_h$ into \eqref{delta_e} we see that the order in $\e$ due to the integrations and propagator is changed by a factor
\[
\e^{m_{J_0}-\frac{1}{2}m_{J_1}}
\]
with $m_{J_0}$ and $m_{J_1}$ the number of $\m^{J_0}_h$ and $\m^{J_1}_h$ vertices. In order to include the contribution coming from the external fields one has only to multiply the r.h.s. of \eqref{bar_C} by the product
\[ \label{prod_Jbar}
\prod_v \left( \e \bm_h^{J_0} \rgt)^{m_{J_0}} \left( \e^{-\frac{1}{2}} \bm_h^{J_1} \rgt)^{m_{J_1}}
\]
Note that the inclusion of $\bm_h^{J_0}$ and $\bm_h^{J_1}$ in the list of possible vertices does not change the relation \eqref{loop0} for the loop number; that is why in the product \eqref{prod_Jbar} there is not dependence on $\l$, see \eqref{bar_C} for a comparison.  

\vskip 0.2 cm

For what concerns low momentum region, the initial value of $\m^{J_0}_0$ is
\[
 \m_\bh^{J_0} = \begin{cases}
\, 1 + O\bigl(\l \e^{\frac{1}{2}} \bigr)& d=3 \\
\, \e^{-\frac{1}{2}} \bigl(1+ O\lft(\l\bigr) \rgt) & d=2 
\end{cases}
\]
Since the order in $\e$ due to the integrations and propagator is changed by $\e^{m_{J_0}}$, and the inclusion on $\m_h^{J_0}$ does not change the relation \eqref{loop} for the loop numbers,  the factors \eqref{3d_h} and \eqref{2d_h} are changed only by the extra factor
\[
\prod_v \left( \e \m_h^{J_0} \rgt)^{m_{J_0}}
\]
Being $\m_h^{J_0}=\e^{-1}\m_h$ for each $h \leq \bh$, the latter product stays small as far as $\m_h$ is small. We see that the substitution of a vertex of type $\m_h$ with a vertex of type $\m^{J_0}_h$ gives a factor $\e^{-1}$.

\pagina
%------------------------------------------------------------------------------------------------
\section{Counting of trees and Feynman diagrams} \label{App-counting}

This section recollects some very standard lemmas referring to the bounds on the number of Gallavotti--Nicol\`o trees and Feynman diagrams. For most of them we have skipped the proof, which can be found in \cite{GalReview, GM}. In the following we will use the same definitions and conventions introduced in sec. \ref{multiscale}

{\lemma
The number of rooted unlabeled trees with $m$ vertices is bounded by $C_1^m$ for some constant $C_1$. 

The number of rooted unlabeled trees with $n$ endpoints is bounded by $C_2^n$ for some constant $C_2$. 
}\\

Let  $\TT_{h,n}$ the set of the labeled trees with $n$ endpoints and root at frequency $h$. Their number cannot be uniformly estimated in $h$ for $h \arr -\io$. In fact, even if the maximum number of non trivial vertices is fixed by $n$ and is equal to $n-1$, we can add between them the non trivial vertices. Of course in the limit  $h \arr -\io$ the number of possible insertions tends to infinite. However is it possible to proof the following result.

{ \lemma \label{lem:labeled_trees}
Let $\TT_{h,n}$ the set of labeled trees with $n$ endpoints and root at scale $h$. If $\g >1$ and $a>0$ we have
\be
\sum_{\TT_{h,n}} \prod_{v\,\text{not e.p.}} \g^{-a(h_v -h_{v'})} \leq C^n
\ee
for some constant $C$.
}\\

If now consider a tree without root, which corresponds to a ``graph'' (\ie a set of points and a set of lines whose extremes are these points) connected (\ie for each couple of points there exists a set of lines of the graph connecting them) acyclic (\ie for each couple of points there is a unique path connecting them). Then, if we call the points as $P_1, \ldots, P_n$ the following lemma holds, where the $n!$ factor comes from the number of ways of assigning the names to the points.

{ \lemma 
Given $n$  points, the number of trees connecting them is bounded by $C^n n!$
}\\

Let now consider $n$ points $v_1, \ldots, v_n$. From each point $v_j$ in this set we denote with $P_{v_j}$ the set of  lines outgoing from the point. By contracting some of these lines among them, we get a graph. We well call Feynman graphs the subset of connected graphs obtained by connecting some of the lines $P_{v_1}, \ldots, P_{v_n}$. We will call {\it internal lines} the contracted lines of the graph and {\it external lines} the lines of the graph that have been not contracted. Then we have the following lemmas.

{ \lemma \label{lem:Feyn_diagrams}

Let consider the Feynman diagrams with $n$ vertices $v_1, \ldots, v_n$ such that $P_{v_1}, \ldots, P_{v_n}$ are the external lines outgoing from these points and  $n^\txe$ is the number of external lines. Their number is bounded by $C^n (2n)!$ uniformly in the number $n^\txe$.  
}

{ \lemma \label{lem:label_choice}
Let $\{P_v\}$ the set of the labels which, for a given collection of cluster $v$ corresponding to a tree $\t$, describes how many and which lines are outgoing from each cluster $v$. Let $a$ to be a positive constant. Then
\be
\sum_{\{P_v\}} \sum_{v \,\text{not e.p.}} \g^{-a |P_v|} \leq C^n_a
\ee
where $n$ is the number of endpoint of $\t$ and $C^n_a$ is a constant depending on $a$.
}

%\section{Bounds on the  renormalized propagator}

%\blue{inserisco i dettagli dei bounds?}

\pagina

\tik{
\[
\begin{tikzpicture}[scale=0.65]
\foreach \x in {0,...,8}   
    {
        \draw[th] (\x , -3) -- (\x , 2);      
    }
\draw(-1.2, 0) node{$\t=$};
\draw[med] (0,0) node[vertex]{} -- (1,0) node [vertex] {} -- (2,0) node [vertex] {}-- (3,0.5) node [vertex] {} -- (4,1) node[vertex] {};  
\draw[med] (2,0) -- (3,0.15) node[vertex] {} -- (4,0.3) node[vertex] {} -- (5,0.45) node[vertex] {} -- (6,0.6) node[vertex] {} -- (7,1.1) node[vertex] {}; 
\draw[med] (6,0.6)-- (7,0.6) node[vertex] {} -- (8,0.6) node[vertex] {};  
\draw[med] (6,0.6) -- (7,0.1) node[vertex] {} -- (8,-0.4) node[vertex] {};  
\draw[med] (4,-1) node[vertex] {} -- (5,-0.75) node[vertex] {}; 
\draw[med] (2,0) -- (3,-0.5) node[vertex] {}-- (4,-1) node[vertex] {} -- (5,-1.5) node[vertex] {}-- (6,-1.5) node[vertex] {}  -- (7,-1.5) node[vertex] {};  
\draw[med] (5,-1.5) -- (6,-2) node[vertex] {}-- (7,-2.5) node[vertex] {};  
\draw[med] (1,0) -- (3,-2) node[vertex]{};
\draw[med] (2,-1) node[vertex]{} -- (3,-1.4) node[vertex]{}-- (4,-1.8) node[vertex]{};
\draw(0, -3.5)  node{$h^*$};
\draw(2, -3.5)  node{$v_0$};
\draw(4, -3.5)  node{$\bh$};
\fill [color=gray, fill opacity=0.2] (5,-3) rectangle (8,2);
\end{tikzpicture} 
\quad
\begin{tikzpicture}[scale=0.65]
\foreach \x in {0,...,8}   
    {
        \draw[th] (\x , -3) -- (\x , 2);      
    }
%\foreach \y in {-3,-2.5,...,2}   
    %{        \draw[th] (0,\y ) -- (8, \y );         }
\draw(-1.2, 0) node{$\t^<=$};
\draw[med] (0,0) node[vertex]{} --   (1,0) node [vertex] {}  --   (2,0) node [vertex] {} --   (3,0.5) node [vertex] {} -- (4,1) node[vertex] {};  
\draw[med] (2,0) -- (3,0.15) node[vertex]{} -- (4,0.3) node[vertex]{}-- (5,0.45) node[whitevex]{};   
\draw[med] (4,-1)node[vertex] {} -- (5,-0.75) node[vertex] {}; 
\draw[med] (2,0) -- (3,-0.5) node[vertex] {} -- (4,-1) node[vertex] {} -- (5,-1.5) node[whitevex] {};  
\draw[med] (1,0) -- (3,-2) node[vertex]{};
\draw[med] (2,-1) node[vertex]{} -- (3,-1.4) node[vertex]{}-- (4,-1.8) node[vertex]{};
\draw(0, -3.5)  node{$h^*$};
\draw(2, -3.5)  node{$v_0$};
\draw(4, -3.5)  node{$\bh$};
\fill [color=gray, fill opacity=0.3] (5,-3) rectangle (8,2);
%\draw(6.5,-0.5) node{$g_2$};
\end{tikzpicture} \non
\]
}{On the left a generic tree $\t$ with endpoints above and below $\bh+1$; on the right the tree $\t^<$ obtained from $\t$ by substituting the subtrees with root at scale $\bh+1$ with generalized vertices, depicted as white circles. The ``simple'' points at scale $\bh +1$ represent the local term defined by the localization procedure in the high momenta region $\bh < h \leq 0$, represented by the region filled in gray.}{above_below}
%
%---------------------------------------------

\section{Dimensional bounds for diagrams with any vertex at scale $\bh$}  \label{app:hbar}

Let consider the set of renormalized  trees $\TT^<_{h^*,n}$ with root at scale $h^* < \bh$ and $n$ endpoints at scales $ h^*<h\leq \bh$.
For sake of simplicity in chapter \ref{chap_multiscale} we have only considered the subset of these trees with vertices at scale $\bh$ equal to the ones at scale $0$, plus the six--legged vertex in the two dimensional case. However the integration over the first $|\bh|$ scales gives rise to each possible vertex at scale $\bh$, with an arbitrary number and type of external legs (except for some symmetry constraints). In this appendix we will prove that the dimensional bounds for the kernel of the effective potentials are not changed when we consider {\it all the possible vertices at scale $\bh$} instead than only those considered in chapter \ref{chap_multiscale}.  \\

Let consider a tree $\t^< \in \TT^<_{h^*,n}$. The latter can be imagined as obtained 
%Let denote with $\SS(\t)$ the set of the subtrees in $\t$ with root at scale $\bh +1$. It is clear that each $t_j$ belongs to the set $\TT^>_{\bh,n_j}$ with $n_j$ the number of endpoints of $\t_j$. By construction a Feynman diagram compatible with a tree in $\SS(\t)$ will have internal lines contracted on scales greater then $\bh$, \ie 
%\[
%\forall l \in \GG(\t_j),\, \t_j \in \SS(\t) \quad \Rightarrow  \quad h_{l}> \bh 
%\]
%and external legs to be contracted on scales equal or lower than $\bh$. 
by a generic tree $\t$, with endpoints at each possible scale $h^*<h\leq 0$,  substituting to each subtrees $\t_j$ in the region $\bh < h \leq 0$ a generalized vertex $g_j$, see fig. \ref{above_below}. In the following we will denote with $\TT^{\,>}_{\bh,n_j}$ the set of the trees with root at scale $\bh$ and $n_j$ endpoints at scales between $\bh$ and $0$. Three different types of endpoints belong to $\t^<$:
\begin{enumerate}[(1)]
\item ``usual'' endpoints at scale $k \in [h+1,\bh]$ which correspond to one of the terms of $\LL_< \BV_{k -1}$,\ie to one of  the marginal and relevant vertices $r^<_{i,k}$; 
\item  ``generalized''  $g_j$  at scale $\bh +1$, coming from the integration at scales lower than $\bh$.  These may be in turn divided in two categories:
\begin{enumerate}[(2a)]
\item generalized endpoints corresponding to $\LL_> \BV_\bh$, \ie to the sum of all the trees in  $\TT^{\,>}_{\bh,n_j}$ with index $\LL$ on the vertex at scale $\bh +1$; these will be depicted as ``usual'' points at scale  $\bh +1$.
\item generalized endpoints corresponding to the trees  $\t_{j} \in \TT^{\,>}_{\bh,n_j}$ with index $\RR$ on the vertex at scale $\bh+1$. We will denote graphically this second category of endpoints with white circles, see fig. \ref{above_below}. These are just the endpoints we have neglected in the simplified discussion of chapter \ref{multiscale}.
\end{enumerate}
\end{enumerate}
 Referring to a generalized endpoint $g_j$ in the following we will denote
\begin{itemize}
\item  $n_{\a}(g_{j})$ the number of external legs of type $\a$ coming out $g_{j}$;
\item  $n_{\a,v}(g_{j})$ the number of legs in the set $n_{\a}(g_{j})$  which are external to the cluster at scale $h_v$; the latter legs can be contracted at scales $ h_c < h_v$.  
\end{itemize}
Our goal is to recompute the scaling dimension $\d_{v}^{<}$ in the general case in which $\t^<$ contains whatever endpoints at scale $\bh$. In order to make the strategy of the proof clearer we choose to consider a specific case, instead than the generic $d$ dimensional case, \ie the estimate for the free energy in the three dimensional case.  \\

Let consider a tree $\t^< \in \TT^{\,<}_{h,n} $ and a Feynman graph $\G(\t{^<})$ belonging  to the set of the Feynman diagrams compatible with $\t^<$. 
As seen in section \ref{multiscale}, if there are not generalized vertices, the diagram  $\G^<$ (neglecting for the moment the effect of regularization) has scaling dimension 
\[
\delta_{v}^{<}=4-2n^\txe_{l,v}-n^\txe_{t,v} -n^\txe_{\dpr, v}
\]
which comes from
\[
\d^<_{v}=-4(m_{v}-1)+2n_{l,v}^\txi+n_{t,v}^\txi-n_{\dpr, v}^\txi
\]
We want to include in the calculation the generalized vertices $g_{j}$. Let denote with $\{g_{v}\}$ the set of the generalized vertices contained in a cluster with scale equal or lower than $h_v$, \ie with at least one of its outgoing lines contracted on scale $h_{v}$.
We have:
\[
m_{v} & =  m_{4,v}+m_{3,v}+\sum_{g \in \{g_{v}\}} 1 \non \\
n_{l,v}^\txi & =  m_{3,v}+\sum_{g \in \{g_{v}\}} n^\txe_{l}(g)-n^\txe_{l,v} \non \\
n_{t,v}^\txi & =  4m_{4,v}+2m_{3,v}+\sum_{g \in \{g_{v}\}}n^\txe_{t}(g)-n^\txe_{t,v}
\]
The complete dimension is then:
\[ \label{generalized_dim}
\bar{\delta}_{v}^{<}& =\, 4 - 2n^\txe_{l,v} -n^\txe_{t,v} -\sum_{g \in \{g_{v}\}}\lft[4-2n^\txe_{l} (g)-n^\txe_{t} (g)\rgt] \non \\
& =\, \d_{v}^{<} \, -\sum_{g \in \{g_{v}\}}\d^{<}(g)
\]
Putting \eqref{generalized_dim}  together with the estimate for each of the generalized vertices, which is given by the dimensional bound for a tree with root at scale $\bh$, 
we obtain the following {\it unrenormalized bound} for the contribution to $||W^{(h);n}_{n^\txe_l,n^\txe_t}||$ coming from an diagram $\G(t^<)$: 
\[ \label{val_G}
\bigl \|\Val (\G(\t^<))\bigr\|  \leq  & \, C^n \,C(P_v; \e, \r_0, R_0)\, %\prod_{\substack{v \, \text{e.p.}\\ h_v \leq \bh} } (r_{h_v}) 
%\prod_{\substack{v \, \text{e.p.}\\ \bh<h_v \leq 0}}(\bar{r}_{h_v})  \non \\[6pt]
 \e^n\;\Bigl(\g^{h\d_{v_0}^{<}} \prod_{\substack{v\, \text{not e.p.} \\ v>v_0 ,\, h_v \leq \bh}} 
\g^{(h_{v}-h_{v'})\d_{v}^{<}}\Bigr) \non \\[6pt] 
& \Bigl(\g^{-h \sum_{g \in \{g_{v_0}\}}\d^{<}(g)}
\prod_{\substack{v\, \text{not e.p.} \\ v>v_0 ,\, h_v \leq \bh}} \g^{-(h_{v}-h_{v'}) \sum_{g \in \{g_{v}\}}\d^{<}(g)}\Bigr) \non \\[6pt]
& \prod_{g_j \in \{g_{v_0}\}} \Bigl(\g^{h_{r_{j}}\d_{v}^{>}}\prod_{\substack{v\,\text{not e.p.}\\ v>r_{j}}}\g^{(h_{v}-h_{v'})\delta_{v}^{>}}\Bigr)
\]
where 
\begin{enumerate}[i)]
\item the factor $C(P_v; \e, r_0, R_0)$ contains the dependence on $\e$ coming from the integration in the region $h \leq \bh$ and the dependence on $\r_0$ and $R_0$ coming from the number of loops of diagram $\G(\t)$;
\item the factor $\e^n$ comes from the product over the $n$ endpoints, which are all at scale $0$ in the unrenormalized case;
%we have denoted the maximum value of the endpoint set of the running coupling constants in the region $\bh <  h\leq 0$ and $h \leq \bh$;
\item the product on the fist line of \eqref{val_G} is restricted to the ``usual endpoints'', \ie the endpoints with scale $h_v \leq \bh$;
\item the product on the last line is the dimensional estimate for each of the subgraph  $\G(\t_j)$, having scaling dimension  
\be
\delta_{v}^{>} \;=\;  \frac{1}{4}(10-3n_{v}^\txe) -\frac{1}{2} n^\txe_{\dpr_\xx,v} - n^\txe_{\dpr_0,v} 
\ee
\end{enumerate}
We note that the product in the second bracket of \eqref{val_G} can be rearranged as follows:
\[
\prod_{\substack{v\, \text{not e.p.} \\ v>v_0 ,\, h_v < \bh}} 
\prod_{g \in \{g_{v}\}} \g^{-h_{v} \d^{<}(g)} = \prod_{g_j \in \{g_{v_0}\}} \g^{-h_{r'_{j}} \d^{<}_{r_j}}
\]
where $r_j$ is the root of the tree $\t_j$ giving the vertex $g_j$ and the vertex $r_{j}'$ is the vertex immediately preceding the vertex $r_{j}$ along the tree $\t^<$, \ie $h_{r'_j} \leq \bh$ is equal to first the scale $h_v$ where at least one of the outgoing lines of $g_j$ is contracted. Recollecting the different terms together we get:
\[
 \eqref{val_G}   \leq \,  & \, C^n \,C(P_v; \e, \r_0, R_0)\, \e^n \; 
%\prod_{\substack{v \, \text{e.p.}\\ h_v \leq \bh} } (r_{h_v}) 
%\prod_{\substack{v \, \text{e.p.}\\ \bh<h_v \leq 0}}(\bar{r}_{h_v}) \; 
\Bigl(\g^{h\d_{v_0}^{<}} 
\prod_{\substack{v\, \text{not e.p.} \\ v>v_0 ,\, h_v \leq \bh}} 
\g^{(h_{v}-h_{v'})\d_{v}^{<}}\Bigr) \non \\[6pt] 
&  \Bigl(\g^{h\d_{v_0}^{<}} \prod_{\substack{v\, \text{not e.p.} \\ v>v_0 ,\, h_v \leq \bh}} \g^{(h_{v}-h_{v'})\d_{v}^{<}}\Bigr) 
\prod_{g_j \in \{g_{v_0}\}} \Bigl(\g^{-h_{r'_{j}} \d^{<}_{r_j}} \, \g^{h_{r_{j}}\d_{r_j}^{>}} \prod_{v>r_{j}}\g^{(h_{v}-h_{v'})\delta_{v}^{>}}\Bigr)
 \non \\[9pt]
%%%%%
   =\, & \, C^n \,C(P_v; \e, \r_0, R_0)\,\e^n \; 
%\prod_{\substack{v \, \text{e.p.}\\ h_v \leq \bh} } (r_{h_v}) 
%\prod_{\substack{v \, \text{e.p.}\\ \bh<h_v \leq 0}}(\bar{r}_{h_v}) \non \\[6pt]
 \g^{h\d_{v_0}^{<}} \prod_{\substack{v\, \text{not e.p.} \\ v>v_0 ,\, h_v \leq \bh}} \g^{\lft(h_{v}-h_{v'}\rgt)\,\d_{v}^{<}} 
\; \prod_{g_j \in \{g_{v_0}\}} \g^{(h_{r_{j}}-h_{r_{j}'})\,\d_{r_j}^{<}} \non \\
& \hskip 3cm 
\prod_{g_j \in \{g_{v_0}\}}
\biggl(  \g^{\lft(\d_{j}^{>}-\d_{j}^{<}\rgt)\,h_{r_{j}}}\,
\prod_{\substack{v\, \text{not e.p.} \\ v>r_j}} 
\g^{\lft(h_{v}-h_{v'}\rgt) \d_{v}^{>}} \biggr)   
\]
Then, finally
\[
\bigl \|\Val (\G(\t^<))\bigr\|  & \leq \, C^n \,C(P_v; \e, \r_0, R_0) \e^n \; %\prod_{\substack{v \, \text{e.p.}\\ \bh<h_v \leq 0}}(\bar{r}_{h_v}) \non \\[6pt]
%\; \prod_{\substack{v \, \text{e.p.}\\ h_v \leq \bh} } (r_{h_v}) 
 \g^{h\d_{v_0}^{<}} \prod_{\substack{v\, \text{not e.p.}\\ v>v_0}} \g^{\lft(h_{v}-h_{v'}\rgt)\,\d_{v}^{<}}   \non \\[6pt]
& \prod_{g_j \in \{g_{v_0}\}}
\biggl(  \g^{\lft(\d_{j}^{>}-\d_{j}^{<}\rgt)\, \bh}\, 
\prod_{\substack{v\, \text{not e.p.} \\ v>r_j}} 
\g^{\lft(h_{v}-h_{v'}\rgt) \d_{v}^{>}} \biggr)     \label{complete} 
\]
The product on the first line of  \eqref{complete} is the usual dimensional estimates for a tree with vertices at scales $h\leq \bh$. 
%Note that in the product over the vertices of $\t^<$ which are not endpoints are also included the vertices at scale $\bh$. 
Besides of that, for each generalized vertex $g_j$ we must add the term on the second line, which is corresponds to the dimensional estimate for the tree $t_j$ from which generalized vertex comes from, 
%This has the same form \eqref{power_fin_up} of the dimensional bound for a tree in $\TT^{\,>}_{h,n}$, 
apart for the external dimension, which is \be \label{transition}
\g^{\lft(\d_{j}^{>}-\d_{j}^{<}\rgt)\,h_{r_j}}
\ee 
instead of $\g^{\d_{j}^{>}\,h_{r_j}}$ as in \eqref{power_fin_up}. The factor \eqref{transition} is associated to the change of dimensions in the transition from the first to the second region and  we will discuss it in a while. Before of that let's see which is the effect of the renormalization on a the estimate \eqref{complete}.

If we take into account the effect of the renormalization, the dimensions in the products 
$\prod_{\substack{ v>v_0}} \g^{(h_{v}-h_{v'})\d_{v}^{<}}$ and $\prod_{\substack{ v>r_j}} \g^{(h_{v}-h_{v'})\d_{v}^{<}}$  in \ref{complete} become negative for each cluster, provided to have the right improvement of the dimensional bound also if $\RR$ acts on a cluster at scale greater than $\bh$ whose first outgoing contracted line is at scale $h_{r'_j} \leq \bh= h_{r_j}$. This is immediately verified to be true. In fact the derivative coming from the renormalization procedure will fall on a propagator $g^{h_j}$ at scale $h_j > \bh$, giving a factor $\g^{-h_j}$ or $\g^{-\frac{h_j}{2}}$ according as the derivative is with respect to $x_0$ or $\xx$ variable. On the other side the external momenta will be associated to a line contracted  at scale $h_k \leq h_{r'_j} \leq \bh$  and will give a factor $\g^{h_k}$. Putting together the two estimates we could verify that the action of the $\RR$ operator gives the same factor
\[
 \g^{-z^<_j (h_{r_j} - h_{r'_j})} 
\]
we obtain when $\RR$ acts on a vertex at scale lower than $\bh+1$. Then, if the transition factor \eqref{transition} is bounded, we have that all the scaling dimensions in \eqref{complete} are negative, so allowing the sum over the scale indices. \\

Now it's time to discuss the factor \eqref{transition}. Let rewrite the bound \eqref{complete} in the case of the renormalized expansion
\[
\bigl \|\Val (\G)\bigr\|  \leq  & \, C^n \,C(P_v; \e, \r_0, R_0) \non \\[6pt]
& \biggl(\,\prod_{\substack{v \, \text{e.p.} \\ v \neq g_j  }} r^<_{v,h_v-1} \,\biggr)\,
 \g^{h D_{v_0}^{<}} \prod_{\substack{v\, \text{not e.p.} \\ v>v_0}} \g^{\lft(h_{v}-h_{v'}\rgt)\, D_{v}^{<}} \non \\
& 
\prod_{g_j \in \{g_{v_0}\}}
\biggl(  \g^{\lft(\d_{j}^{>}-\d_{j}^{<}\rgt)\,\bh}\, 
\Bigl(\,\prod_{\substack{v\, \text{e.p.} \\ v \in \t_j} } r^>_{j,h_v-1} \,\Bigr)\,
\prod_{\substack{v\, \text{not e.p.} \\ v>r_j}} 
\g^{\lft(h_{v}-h_{v'}\rgt) D_{v}^{>}} \biggr)    \label{complete_ren} 
\]
where $h_{r_j}= \bh$ for each $g_j$, $D_v=\d_v + z_v$ is the renormalized scaling dimension and we have written explicitly the product over the endpoints, denoting with  $\{r^<_{i,h}\}$ and $\{r^>_{i,h}\}$ the running coupling constants for $h \leq \bh$ and $h > \bh$ respectively.  Now, if to the  generalized endpoint $g_j$  is associated the index $\LL$ the factor $\g^{\lft(\d_{j}^{>}-\d_{j}^{<}\rgt)\,\bh} $ takes just into account the different scaling dimension of the running couplings in the two regions, \ie
the fact that at the borderline $h = \bh$ we must have
\[
\g^{\d_{j}^{>} \bh} \,r^>_\bh  = \g^{\d_{j}^{<}\bh} \,r^<_\bh
\]
and then we may write $\g^{\lft(\d_{j}^{>}-\d_{j}^{<}\rgt)\,\bh}\, r^>_\bh = r^<_\bh$.
 For what concerns the generalized point of type (b), with index $\RR$ associated to $\bh +1$, the factor $\g^{\lft(\d_{j}^{>}-\d_{j}^{<}\rgt)\,\bh}$ may be bounded by one if %$\D_j = \d_{j}^{>}-\d_{j}^{<}$
\be
\D_v=\delta_{v}^{>}-\delta_{v}^{<}  \;=\;  \frac{1}{4}(-6+5n_{l,v}^\txe+n_{t,v}^\txe) +\frac{1}{2} n_{\dpr_\xx,v}
\ee
 is positive. However there are a few cases in which the dimension $\D_v$ is not positive, characterized by the condition $5n_{l,r_j}^\txe+n_{t,r_j}^\txe \leq 6$; these are the renormalized vertices with
\[ 
(n_{l,r_j}^\txe,\, n_{t,r_j}^\txe) = (0,2),\, (0,3),\,(0,4),\,(0,6),\,(1,1) \non
\]
and eventually some $k_0$ momenta on the external legs. In these cases we have a large factor $\g^{-|\D_{r_j}| \bh}  = \e^{-|\D_{r_j}|}$, being $\g^\bh=\e$. However this bad factor is over--compensated by the $\e$ factors coming from the vertices contained in $g_j$, still leaving a small constant per vertex. For example, in the worst case, that is $(n_{l,r_j}^\txe,\, n_{t,r_j}^\txe) = (0,2)$ and only two vertices $\bar{\m}_{h_v}$, with $h_v > \bh$ contained in $g_j$, we have a factor $\e^{-1}$ coming from the change of dimension at the transition between the two regions and a factor $\e \g^{-h_v/4} \leq \e \g^{-\bh/4}=\e^{3/4}$ coming from each of the vertex $\bar{\m}_{h_v}$, see sec. \ref{flows}. Then, we get that the product over the vertex $g_j$. 

We can rewrite \eqref{complete_ren} as follows
\[
\bigl\|\Val (\G)\bigr\|  =  \, C^n \, \tl{\h}^n \,\g^{h D_{v_0}^{<}} \prod_{\substack{v\, \text{not e.p.} \\ v>v_0}} \g^{\lft(h_{v}-h_{v'}\rgt)\, D_{v}^{<}}  
\prod_{g_j \in \{g_{v_0}\}}
%\biggl(  \g^{\lft(\d_{j}^{>}-\d_{j}^{<}\rgt)\,\bh}\, 
\prod_{\substack{v\, \text{not e.p.} \\ v>r_j}} \g^{\lft(h_{v}-h_{v'}\rgt) D_{v}^{>}}    \label{complete_ren2} 
\]
where $  \tl{\h} = \max \lft \{ \bigl(\max_{h \leq \bh} r_{i,h}\bigr) , \e^a  \rgt\} $
with $a$ an appropriate constant taking into account of the worst factor between those coming from the generalized vertices with index $\RR$ at scale $\bh +1$. Since the scaling dimensions in  \eqref{complete_ren2} are all negative, assuming $\h$  bounded, we can prove the $n!$ bounds for the $n$--th order contribution to the free energy.  \\

%\end{document}

%\input{intro-senza-sapclass} \input{intestazione-sap} \usepackage{showkeys} \begin{document}  

\chapter{Transient region $\bh<h\leq 0$} \label{app_transient}

\section{Ward Identities in the transient region}

For completeness we report here the expressions for the global and local WIs in the region $\bh< h \leq 0$, which relate the coupling constants $\bl_h$, $\bm_h$, $\bar{Z}_h$, $\bar{A}_h$, $\bar{B}_h$ and $\bar{E}_h$. 
\begin{center}
\[  
 \hline \non \\[-12pt]
 %\bh < h \leq0  \qquad  
\hskip -0.5cm \text{ Global WIs } \qquad
d & =3  & d &=2  \non \\ \hline & \non \\[-9pt]
  & - &      \bm_h  & =4\sqrt{2}\,\g^{\frac{h}{2}}\bl_h     
 \label{GWI_m_low}\\[3pt]
\g^h\,\bar{Z}_h & = 2\,\sqrt{2}\,\g^{\frac{h}{4}}\bm_h + 2\,\g^{h}\bn_h     &  \g^h\,\bar{Z}_h & = 2\,\sqrt{2}\,\g^{\frac{h}{2}} \bm_h + 2\,\g^{h}\bn_h    \label{GWI_Z_low}
\]
\[  
 \hline \non \\[-12pt]
\hskip -1.8cm\text{Formal local WIs } \hskip 1.5cm
 %\bh < h \leq0  \quad \quad
 d & =3  & d &=2  \non \\ \hline & \non \\[-9pt]
\bm_h^{J_0} -1 & =  \bar{E}_h      & \bm_h^{J_0} -1 & =  \bar{E}_h \label{LWI_E_min}   \\[3pt]
\g^{\frac{3}{4}h}\,\bar{E}_h^{J_1}  & = \sqrt{2}\,\bar{A}_h       
&  \g^{\frac{h}{2}}\,\bar{E}_h^{J_1}  & = \sqrt{2}\,\bar{A}_h    \label{LWI_A_min} %\\[3pt]
%\bar{E}_h^{J_0} & = -   \sqrt{2}\,\ber{B}_h     & \sqrt{2} B_h & = -E_h^{J_0}    \label{LWI_B}  \\[3pt]
%\g^{\frac{3}{4}h}\,\bar{Z}_h^{J_0}& = 2\sqrt{2}\lft(\e^{-1}\g^h\,\bar{Z}_h -1\rgt)      %& 
%\g^{\frac{h}{2}}\,\bar{Z}_h^{J_0}& = 2\sqrt{2}\lft(\e^{-1}\g^h\,\bar{Z}_h -1\rgt)     \label{LWI_Z_min}
\]
\end{center}
The main difference between the previous identities and the expression of the corresponding WIs in the lower momentum region $h \leq \bh$ stays in the different external dimension of the kernels $V^{(h)}_{n_l n_t}$ in the two regions. Then the expression of the identities also depends on the initial values of the coupling constants at scale $0$. For example, by definition $\bar{E}_0=0$ while $E_\bh=1$, from which comes the difference between \eqref{LWI_E_min} and \eqref{LWI_E}.

\section{Flow in the transient region $ \bh < h \leq 0 $} \label{transient}

In this section we will study perturbatively the flow of the running coupling constants in the transient region  $\bh < h \leq 0$ in order to find the initial values of the marginal and relevant couplings for $h = \bh$. In order to find bounds on the value of the running coupling constants  we will use the dimensional estimate \eqref{thm_nfactorial0} pag. \pageref{thm_nfactorial0} for the leading order diagrams contributing to the beta function. In the transient region $\bh < h \leq 0$ the leading order diagrams are the diagrams with the minimum number of vertices, see \eqref{bar_C}.  

We remind to the reader the scaling dimensions for the kernels of the effective potential both in the transient region :
\[
 \d^>_{v} & =  \frac{5}{2}-\frac{3}{4}n_{v}^\txe-n_{J_{0},v}^\txe -\frac{1}{2}n_{J_{1},v}^\txe  - n_{\dpr_{0},v}^\txe -\frac{1}{2}n_{\dpr_\xx,v}^\txe
& d=3  \non \\[6pt]
 \d^>_{v} & =  2 -\frac{1}{2}n_{v}^\txe -n_{J_{0},v}^\txe -\frac{1}{2}n_{J_{1},v}^\txe  - n_{\dpr_{0},v}^\txe -\frac{1}{2}n_{\dpr_\xx,v}^\txe & d=2
\]

\vskip 1cm

{\centering  \subsection{Three dimensions}}

\subsubsection{\it Vertices without external fields}

The running coupling constants in the transient region and for $d=3$ are $\bm_h$, $\bar{Z}_h$, $\bar{A}_h$ and $\bar{E}_h$, whose flow may be bounded using the dimensional estimate \ref{bar_C} for the leading order diagrams contributing to the beta functions. However we are also interested in calculating the value at $h=\bh$ of the four--legged diagram $\bl_\bh$ and the two--legged diagram with two external $\dpr_0$ derivatives $\bar{B}_\bh$, since these diagrams become marginal in the lower region. 

Let us start with the study of the running coupling constants. 
There exists a constant $c$ such that:
\[
\lft| \bm_h - \g^{\frac{1}{4}}\bm_{h+1} \rgt|& \leq c \,\bigl(\r_0 R_0^3 \bigr)^{-1} |\bm_h|^3 \non \\[6pt]
\lft| \bar{Z}_h - \g\,\bar{Z}_{h+1} \rgt| & \leq c \,\bigl(\r_0 R_0^3 \bigr)^{-1} |\bm_h|^2 \non \\[6pt]
 \lft| \bar{A}_h - \bar{A}_{h+1} \rgt| & \leq c \,\bigl(\r_0 R_0^3 \bigr)^{-1}  |\bm_h|^2 \non \\[6pt]
 \lft| \bar{E}_h - \bar{E}_{h+1} \rgt| & \leq c \,\bigl(\r_0 R_0^3 \bigr)^{-1}  |\bm_h|^2
\]
Then, assuming $|\bm_h| \leq c\, \g^{-\frac{h}{4}} \e$ for each $\bh<h\leq 0$ one obtain
\[
\bigl| \bm_h - \g^{-\frac{h}{4}}\,\bm_0 \bigr| \leq c \,\bigl(\r_0 R_0^3 \bigr)^{-1} \e^{\frac{5}{2}}\, \g^{-\frac{h}{4}}
\]
that is consistent with the assumption. Using the latter bound for $|\bm_h|$ we find the following values for the running coupling constants at $\bh$
\[
\bigl|\bm_\bh - \tfrac{\sqrt{2}}{2}\,\e^\frac{3}{4}  \bigr| & \leq c\,\bigl(\r_0 R_0^3 \bigr)^{-1} \e^{\frac{9}{4}} = c\,\l\,\e^{\frac{5}{4}}  \non \\
 \lft| \bar{Z}_\bh - 1 \rgt| & \leq c \,\bigl(\r_0 R_0^3 \bigr)^{-1} \e^{\frac{3}{2}}= c\,\l\,\e^{\frac{1}{2}} \non \\[6pt]
 \lft| \bar{A}_\bh - 1 \rgt| & \leq c \,\bigl(\r_0 R_0^3 \bigr)^{-1} \e^{\frac{3}{2}} = c\,\l\,\e^{\frac{1}{2}} \non \\[6pt]
 \lft| \bar{E}_\bh - 1 \rgt| & \leq c \,\bigl(\r_0 R_0^3 \bigr)^{-1} \e^{\frac{3}{2}} = c\,\l\,\e^{\frac{1}{2}} 
\]
Note that one could find the same estimates on $\bar{Z}_h$ by using the global WI \eqref{GWI_Z_low}
\[
\g^{\frac{3}{4}h}\,\bar{Z}_h= 2\,\sqrt{2}\,\bm_h
\]
valid at each order in perturbation theory. Regarding $\bl_\bh$ and $\bar{B}_\bh$ one find the following estimates:
\[
\bl_{h}=\g^{\frac{h}{2}}\l_{0}=\g^{\frac{h}{2}}\, \frac{\e}{16} 
\]
which satisfies  the global WI $\bl_h = \g^{\frac{3}{2}h}\,\bar{Z}_h/8$ and
\[
\lft|\bar{B}_\bh\rgt| \leq c \,\bigl(\r_0 R_0^3 \bigr)^{-1} \e^{\frac{3}{2}}
\]
Denoting with $\bar{\a}_{\bh}$ a generic vertex of the effective theory
at scale $h=\bh$, obtained by the integration on momenta higher than $\bh$, and with $\a_{\bh}$ the corresponding running coupling for the effective theory in the lower region $h\leq\bar{h}$, the following continuity condition must be satisfied 
\[ \label{continuity}
\g^{\bar{\d}\,\bh}\bar{\a}_{\bh}=\g^{\d\,\bh}\a_{\bh}
\]
with $\bar{\d}$ and $\d$ the scaling dimensions of the kernels upper and below $\bh$.  Using \eqref{continuity} we obtain the initial values of the couplings at the beginning of the lower region $h\leq \bh$ which are:  
\[
& \mu_{\bar{h}}=\g^{\frac{\bar{h}}{4}} \bm_\bh= \m_0\, \lft(1+O\bigl(\l\,\e^{\frac{1}{2}}\bigr) \rgt) \non \\
& \l_{\bar{h}}=\g^{-\frac{h}{2}}\bl_\bh= \l_0\, \lft(1+O\bigl(\l\,\e^{\frac{1}{2}}\bigr) \rgt) \non \\
& Z_\bh = \g^{\bh} \bar{Z}_\bh = Z_0 \lft(1+O\bigl(\l\,\e^{\frac{1}{2}}\bigr) \rgt) \non \\
& A_\bh = \bar{A}_\bh =O\bigl(\l\,\e^{\frac{1}{2}}\bigr)\non \\
& E_\bh = \bar{E}_\bh = O\bigl(\l\,\e^{\frac{1}{2}}\bigr) \non \\
& B_\bh = \g^{-\bh}\bar{B}_\bh = O\bigl(\l\,\e^{-\frac{1}{2}}\bigr) 
\]
with $\l\, \e^{-\frac{1}{2}} \ll 1$ if $\r_0 R_0^3 \gg \l$. This condition toghether with $\e \ll 1$ gives $\r_0 R_0^3 = \l^\a$ with $\a \in [-1,1]$. Referring to $\bar{Z}_h$,  its flows start from $\bar{Z}_{0}=\e$ and grows to the value $\bar{Z}_\bh=1$. Since the propagator for $\bh<h\leq 0$ is given by 
\[ \label{3:prop_bh}
%g^{(h)}(k)=\frac{\left(\begin{array}{cc}
%\bar{A}_{h}\,\kk^{2} \quad & \bar{E}_{h}\,k_{0}\\
%-\bar{E}_{h}\,k_{0}\quad & \kk^{2}+\g^{h}\bar{Z}_{h}
%\end{array}\right)}{k_{0}^{2}\,\bar{E}_{h}^{2}+\kk^{2}\,\bar{A}_{h}(\kk^{2}+\g^{h}\bar{Z}_{h})}
g^{(h)}_{\a \a'}(k)= (\r_0 R_0^{-2})^{-1}\,\frac{\left(\begin{array}{cc}
\kk^{2} \quad & k_{0}\\
-k_{0}\quad & \kk^{2}+\g^{h}\bar{Z}_{h}
\end{array}\right)}{k_{0}^{2}+\kk^{2}\,(\kk^{2}+\g^{h}\bar{Z}_{h})}
\]
the bound  $\g^{2(h-1)} \leq k_{0}^{2}+\kk^{4} \leq \g^{2h}$ works
until $\g^{h}\bar{Z}_{h}$ becomes comparable with $\kk^{2} \simeq \g^\bh$. This happens when $\bar{Z}_{\bh}=\gamma^{-\bar{h}}\e$ is equal to one, \ie for $\bh$ such that $\g^\bh=\e$. \\

\subsubsection{\it Vertices with external fields}

The scaling dimensions for the external fields in the region $h>\bh$ are $n_{J_0}=-1$ and $n_{J_1}=\tfrac{1}{2}$, \ie we have chosen for $J_0$ and $J_1$ the same scaling dimension than the derivative with respect to $x_0$ or $\xx$ respectively. With this choice $\bm^{J_0}_h$ and $\bm^{J_1}_h$ are marginal, $\bar{E}^{J_0}_h$ is irrelevant with dimension $-1/4$, and  $\bar{Z}^{J_0}_h$ and $\bar{E}^{J_1}_h$ are relevant with dimension $3/4$. Using the dimensional estimates \eqref{bar_C} and \eqref{prod_Jbar} one proof that:
\[
&  \m_\bh^{J_0}  =  \bm_\bh^{J_0} = 1 +   O(\l\,\e^{\frac{1}{2}}) \non \\[6pt]
& \m_\bh^{J_1}  =   \g^{-2\bh}\,\bm_\bh^{J_1} = \e^2 \left( 1+ O(\l\,\e^{\frac{1}{2}}) \rgt) \non \\[6pt]
&  E_\bh^{J_0}  =  \g^{-\frac{h}{4}}\,\bar{E}_\bh^{J_0} =  O(\l \e^{-\frac{1}{2}}) \non \\[6pt]
&  Z_\bh^{J_0}  =  \g^{\frac{3}{4}h}\,\bar{Z}_\bh^{J_0} =  O(\l) \non \\[6pt]
& E_\bh^{J_1}  = \g^{\frac{3}{4}h}\,\bar{E}_\bh^{J_1} =    O(\l\,\e^{\frac{1}{2}})
\]
%The bounds on $\m_\bh^{J_0}$, $Z_\bh^{J_0}$ and $E_\bh^{J_1} $ may be equivalently found using the local WI's \eqref{LWI_E_min}, \eqref{LWI_Z_min} and \eqref{LWI_A_min}. 

\vskip 0.5cm

{\centering  \subsection{Two dimensions}}

\subsubsection{\it Vertices without external fields}

The running coupling constants in the transient region and for $d=2$ are $\bl_h$, $\bm_h$, $\bar{Z}_h$, $\bar{A}_h$ and $\bar{E}_h$. As in the three dimensional case we are also interest in calculating the value at $h=\bh$ of some irrelevant diagrams which become relevant in the lower region, namely the two--legged diagram with two external $\dpr_0$ derivatives, $\bar{B}_\bh$, and the six--legged diagram $\bl_{6,\bh}$. The dimensional estimate for the diagrams contributing to the running coupling constants are the following
\[
\lft| \bl_h - \bl_{h+1} \rgt| & \leq c \,\bigl(\r_0 R_0^2 \bigr)^{-1} |\bl_h|^2 \non \\[6pt]
\lft| \bm_h - \g^{\frac{1}{2}}\bm_{h+1} \rgt|& \leq c \,\bigl(\r_0 R_0^2 \bigr)^{-1} |\bm_h|\, |\bl_h| \non \\[6pt]
\lft| \bar{Z}_h - \g\,\bar{Z}_{h+1} \rgt| & \leq c \,\bigl(\r_0 R_0^2 \bigr)^{-1} |\bm_h|^2 \non \\[6pt]
 \lft| \bar{E}_h - \bar{E}_{h+1} \rgt| & \leq c \,\bigl(\r_0 R_0^2 \bigr)^{-1}  |\bm_h|^2 \non \\[6pt]
 \lft| \bar{A}_h - \bar{A}_{h+1} \rgt| & \leq c \,\bigl(\r_0 R_0^2 \bigr)^{-1}  |\bm_h|^2 
\]
with $c$ a suitable constants. 
%These estimates give logarithmically divergent beta function. However it is simple to check that the sum if the contributions coming from the leading order diagrams (\ie the diagrams with two vertices) giving such divergence cancel for each of the previous coupling except $\bar{A}_h$. The leading contributions to the beta function f turn to be the one--loop diagrams with three vertices for $\m_h$ and $\l_h$, and the two--loops diagrams with two vertices for $\bar{Z}_h$ and $\bar{E}_h$. Then the behavior is given by the diagrams with more than two vertices.  
%
%\[
%\lft| \bl_h - \bl_{h+1} \rgt| & \leq c \,\bigl(\r_0 R_0^2 \bigr)^{-1} |\bl_h||\bm_h|^2  \non \\[6pt]
%\lft| \bm_h - \g^{\frac{1}{2}}\bm_{h+1} \rgt|& \leq c \,\bigl(\r_0 R_0^2 \bigr)^{-1} |\bm_h|^3 \non \\[6pt]
%\lft| \bar{Z}_h - \g\,\bar{Z}_{h+1} \rgt| & \leq c \,\bigl(\r_0 R_0^2 \bigr)^{-2} |\bm_h|^2 \non \\[6pt]
% \lft| \bar{E}_h - \bar{E}_{h+1} \rgt| & \leq c \,\bigl(\r_0 R_0^2 \bigr)^{-2}  |\bm_h|^2
%\]
Using the global WI  $\m_h = 4\sqrt{2}\,\g^{-\frac{h}{2}}\l_h$ we find the following values for the running coupling constants at scale $h >\bh$  
\[
\lft| \bl_h - \bl_0 \rgt| & \leq c \,\bigl(\r_0 R_0^2 \bigr)^{-1} \e^2 |h|= c\,\l \e |h|  \non \\[6pt]
\bigl|\g^{\frac{h}{2}}\,\bm_h - \bm_0  \bigr| & \leq c\,\bigl(\r_0 R_0^2 \bigr)^{-1} \e^{2}|h| = c\, \l \e |h|  \non \\
 \lft| \g^{h}\,\bar{Z}_h - Z_0 \rgt| & \leq c \,\bigl(\r_0 R_0^2 \bigr)^{-1} \e^2 |h|= c\,\l \e |h|  \non \\[6pt]
 \lft| \bar{E}_h \rgt| & \leq c \,\bigl(\r_0 R_0^2 \bigr)^{-1} \e^2 \g^{-h}= c\,\l \e \g^{-h}   \non \\[6pt]
 \lft| \bar{A}_h  \rgt| & \leq c \,\bigl(\r_0 R_0^2 \bigr)^{-1} \e^2 \g^{-h} = c\,\l \e \g^{-h}
\]
with
\[
& 16\,\bl_0 = 2\sqrt{2}\bm_0 = \bar{Z}_0 = \e \non 
& \bar{A}_0 =\bar{B}_0 =\bar{E}_0 =0
\]
Regarding $\bar{B}_\bh$ and  $\bl_{6,\bh}$ one find the following estimates:
\[
\lft|\bar{B}_\bh\rgt| & \leq c \,\bigl(\r_0 R_0^2 \bigr)^{-1} \e = c\, \l \non \\[6pt]
\lft| \bl_{6,\bh} \rgt| & \leq c\, \bigl(\r_0 R_0^2 \bigr)^{-1} \e^3 =  c\, \l\, \e^2
\] 
Using \eqref{continuity} we obtain the initial values of the couplings at the beginning of the lower region $h\leq \bh$ in the $2d$ case which are:  
\[
& \l_{6,\bar{h}} =  \g^{-\bh}\,\bl_{6,\bar{h}} = O \bigl(\l \e \bigr)    \non \\[6pt]
& \l_{\bar{h}}=\g^{-h}\bl_\bh=\bl_0 \lft(1+O\bigl(\l \e \log\e \bigr) \rgt)\non \\[6pt]
& \m_\bh=\g^{-\frac{\bh}{2}}\,\bm_\bh=\bm_0 \lft(1+O\bigl( \l \e \log\e\bigr) \rgt) \non \\[6pt]
& Z_\bh = \g^{-\bh} \bar{Z}_\bh = \bar{Z}_0 \lft( 1+O\bigl(\l \e \log\e \bigr) \rgt)\non \\[6pt]
& A_\bh = 1+ \bar{A}_\bh =1 + O\bigl(\l\ \bigr)\non \\[6pt]
& E_\bh = 1+ \bar{E}_\bh =1 + O\bigl(\l  \bigr) \non \\[6pt]
& B_\bh = \g^{-\bh}\bar{B}_\bh = O\bigl( \l \e^{-1}  \bigr) 
\]
Referring to the behaviour of the propagator at $\bh$ in the two dimensional case, we get the same expression \eqref{3:prop_bh} of the three dimensional case, with the same wave function renormalization $A_\bh$, $E_\bh$ and $Z_\bh$ at leading order. For $\g^\bh=\e$ the term $\g^{h}\bar{Z}_{h}$ becomes comparable with $\kk^{2} \simeq \g^\bh$ and the bounds on the propagator change, as showed in lemma \ref{lemma}.  \\

\vskip 0.5cm

{\it Coupling constants with external fields} \\

The dimensions of the external field $J_0$ and $J_1$ are the same than  in the three dimensional case, \ie $1$ and $1/2$ respectively. With this choice  $\bm^{J_0}_h$ and $\bm^{J_1}_h$ are marginal, $\bar{E}^{J_0}_h$ is irrelevant with dimension $-1/2$, and  $\bar{Z}^{J_0}_h$ and $\bar{E}^{J_1}_h$ are relevant with dimension $1/2$. The diagrams with two $J_0$ or $J_1$ fields and without other external legs are also marginal. 
All the other diagrams are irrelevant. \\

We first note that the running couplings $\bm_h^{J_0}$ and $\bar{Z}_h^{J_0}$ have the same beta function than the couplings $\m_h$ and $Z_h$, apart for the substitution of the vertices $\m_h^{J_0}$ and $\m'^{J_0}_h$ with $\m_h$ and $\m'_h$. However the combinatorial factors in this case are different and in particular one find that the sum of the contributions to the beta function for $\bm_h^{J_0}$ and $\bar{Z}_h^{J_0}$  which comes from the diagrams with two vertices is equal to zero, see fig.~\ref{tra1} and the explicit computations in the next section.

%These estimates give logarithmically divergent beta function. However it is simple to check that the sum if the contributions coming from the leading order diagrams (\ie the diagrams with two vertices) giving such divergence cancel for each of the previous coupling except $\bar{A}_h$. The leading contributions to the beta function f turn to be the one--loop diagrams with three vertices for $\m_h$ and $\l_h$, and the two--loops diagrams with two vertices for $\bar{Z}_h$ and $\bar{E}_h$. Then the behavior is given by the diagrams with more than two vertices.  

Then, differently from the cases of $\m_h$ and $Z_h$ one find that the leading order contributions to the beta function are summable over $h$:
\[ \label{mu_zeta}
& |\bar{\b}_{h}^{\m,J_0}| \leq c\,\l \e \g^{-h} \non \\
& |\bar{\b}_{h}^{Z,J_0}| \leq c\,\l^2 \e \g^{-\frac{3}{2}h} 
\]
which implies
\[
&   \bm_h^{J_0} = 1 +   O\bigl(\l \e \g^{-h} \bigr) \non \\[6pt] 
&  \g^{\frac{h}{2}}\,\bar{Z}_h^{J_0} =  O\bigl(\l^2 \e \g^{-h}\bigr)
\]
and then
\[
&  \m_\bh^{J_0}  =  \bm_\bh^{J_0} = 1 +   O\bigl(\l  \bigr) \non \\[6pt] 
&  Z_\bh^{J_0}  =  \g^{\frac{\bh}{2}}\,\bar{Z}_\bh^{J_0} =  O\bigl(\l^2 \bigr)
\]
The difference between the two bounds in \eqref{mu_zeta} stays in the fact that the first non trivial diagrams contributing to $\bm_\bh^{J_0}$ are one--loop diagrams of order three, while the first non trivial diagrams contributing to $\bar{Z}_\bh^{J_0}$ are two--loops diagrams of order four.  \\

For what concerns the running couplings $\bm^{J_1}_h$ and $\bar{E}^{J_1}_h$, they have the same beta function, once the $\bl_k$ vertex in $\bm^{J_1}_h$ is substituted by a $\bm_k$ vertex. From a dimensional point of view $\l_h = \g^{\frac{h}{2}} \m_h$ and this explain the different scaling dimension for $\bm^{J_1}_h$ and $\bar{E}^{J_1}_h$. The na\"ive dimensional estimate of the beta function for $\bm^{J_1}_h$ and $\bar{E}^{J_1}_h$ is not summable over $h$. However an explicit  second order calculation (see \ref{leading_transient}) again shows that the contributions coming from the second order diagrams cancel and that the beta function may be estimates as follows:
\[ 
& |\bar{\b}_{h}^{\m,J_1}| \leq c\,\l \e \g^{-h} \non \\
& |\bar{\b}_{h}^{E,J_1}| \leq c\,\l^2 \e \g^{-\frac{3}{2}h} 
\]
which implies
\[
&   \bm_h^{J_1} = 1 +   O\bigl(\l \e \g^{-h} \bigr) \non \\[6pt] 
&  \g^{\frac{h}{2}}\,\bar{E}_h^{J_1} =  O\bigl(\l^2 \e \g^{-h}\bigr)
\]
and then
\[
& \m_\bh^{J_1}  = \g^{\frac{3}{2}h}\,\bm_\bh^{J_1} = \e^{\frac{3}{2}} \lft( 1 + O(\l) \rgt) \non \\[6pt]
& E_\bh^{J_1} = \g^{\frac{\bh}{2}}\bar{E}_\bh^{J_1} = O(\l) 
\]
Finally, regarding $E^{J_0}_\bh$ and  $J^{J_0}_\bh$ one can proof that
\[
 E_\bh^{J_0}  & =  O\bigl( \l^2 \e^{-1}  \bigr) \non \\[6pt]
 J_\bh^{J_0}  &  =  O \bigl(\l^2 \e^{-1} \bigr)
\]

%-------------------------------------------------------TABLE N1
\begin{center}
\begin{table}[H]
\renewcommand{\arraystretch}{1.5}   %distanza tra le righe
\vskip 1cm
\noindent \begin{centering}
\begin{tabular}{|c|c||c|c||c|c|}
\hline
\multicolumn{6}{|c|}{RCC for $\bh < h\leq 0$}\\
\hline 
& $\bar{r}_{0}$ & $\bar{\d}_{3d}$ & $\bar{r}^{3d}_{h}$ & $\bar{\d}_{2d}$  & $\bar{r}^{2d}_{h}$  \\
\hline 
\hline 
$\bl_{h}$ & $\frac{\e}{16}$ & -- & -- & $0$ & $\l_0\,\lft(1+O\bigl(\l\,\e\,|h| \bigr)\rgt)$  \\
\hline 
$\bm_{h}$ & $\frac{\sqrt{2}}{4}\e$ & $1/4$ & $\e \g^{-\frac{h}{4}}$ & $ 1/2 $ & $\m_0\,\g^{-\frac{h}{2}}\lft(1+O\bigl(\l\,\e\,|h|\bigr)\rgt)$\\
%\hline 
%$\bar{\nu}_{h}$ & $\e (1+ o(1))$ & $1$ & $\e^{\frac{1}{2}}$ & $1$ & $\cst$\tabularnewline
\hline 
\multicolumn{6}{|c|}{Quadratic RCC}\\
\hline 
$\bar{A}_{h}$ & $0$ & $0$ & $O(\l \,\e\,\g^{-\frac{h}{2}})$ & 0 & $O\lft( \l \e \g^{-h} \rgt)$ \\
\hline 
$\bar{E}_{h}$ & $0$ & $0$ & $O(\l \,\e\,\g^{-\frac{h}{2}})$ & $0$ & $O\lft( \l \e \g^{-h} \rgt)$\\
\hline 
$\bar{Z}_{h}$ & $\e$ & 1 & $\e\,\g^{-h}$ & 1 & $\g^{-h}Z_0\,\lft( 1+O(\l \e\,|h|) \rgt)$ \\
\hline 
\multicolumn{6}{|c|}{RCC with external fields}\\
\hline 
$\bar{\mu}_{h}^{J_{0}}$ & $1$ & $0$ & $1+O(\l\, \e\, \g^{-\frac{h}{2}})$ & 0 & $1+O(\l \,\e\,\g^{-h}) $  \\
\hline 
$\bar{\mu}_{h}^{J_{1}}$ & $1$ & $0$ & $1+O(\l\, \e\, \g^{-\frac{h}{2}})$ & $0$ & $1+O(\l \,\e\,\g^{-h}) $  \\
\hline 
$\bar{Z}_{h}^{J_{0}}$ & $0$ & $3/4$ & $O(\l\,\g^{-\frac{3}{4}h}$) & $1/2$ & $\g^{-\frac{h}{2}}\,O\lft( \l^2 \,\e \g^{-h}\rgt)$ \\
\hline 
%$\bar{E}_{h}^{J_{0}}$ & $0$ & $-1/4$ & $O(\l\,\g^{-\frac{h}{4}})$ & $-1/2$ & $O\lft(\l\,\g^{-h}\rgt)$ \\
%\hline 
$\bar{E}_{h}^{J_{1}}$ & $0$ & $3/4$ & $\g^{-\frac{3}{4}h} O(\l \,\g^{\frac{h}{2}})$ & $1/2$ & $\g^{-\frac{h}{2}}\,O\lft( \l^2 \,\e \g^{-h}\rgt)$ \\
\hline 
\end{tabular}
\par\end{centering}
\vskip 0.2cm
\noindent \centering{}\caption{Flow of the running coupling constants in the transient region $\bh<h\leq 0$ in three and two dimensions. With $\bar{r}_0$ we have denoted the initial values of the couplings, while $\bar{\d}$ is the scaling dimension of each coupling for $\bh<h\leq 0$. }
\end{table}
\end{center}
%
%-------------------------------------------------------END TABLE

\pagina

\section{Leading order computations} \label{leading_transient}

In this section some of the leading order flow equations for the running coupling constants in the transient region are collected. They show as in two dimensions the behavior of the coupling constants with external fields is better than expected on the basis of their dimensional estimate. 
%the log--divergent contribution to the beta function cancels and the estimates are improved with respect the na\"ive dimensional estimate.

In order to calculate the beta function at scale $h$, $\b_h$, one has to evaluate some Feynman graphs at zero momentum of the external line. Therefore: 
\begin{enumerate}
\item only terms with at least one loop can contribute, since the single scale propagator vanishes at zero momentum. 
\item In the graphs with only one loop, all the internal lines must carry the same momentum. Hence, due to our choice of the cutoff function $f_0(k)$, the internal lines of the loop may only have propagators of scale $h$ or $h+1$; in fact at least one propagator must be of scale $h$ (by definition of $\b_h$) and the supports of the Fourier transform of the propagators at scale $h$ and and $h'\geq h$ are diskoint if $h'>h+1$. 
%\item In the calculation of the graphs with one loop, the subgraph associated with the tree vertex of scale $h+1$ has no loop. Therefore in this tree vertex the $\RR$ operator coincides with the identity.
%\item Since we are interested only in the leading orders, we can neglect in the rescaled propagators \eqref{RENg} the terms proportional to $\g^{2h}$ and the dependence on $k$ of $Z_h$, $A_h$, $B_h$ and $E_h$, see \eqref{wave_funct}. For the same reason we can approximate $Z_{h+1}$ , $A_{h+1}$, $B_{h+1}$ and $E_{h+1}$ by $Z_h$, $A_h$, $B_h$ and $E_h$ in the expression of $g^{(h+1)})_{\a\a'}(x)$. 
\item The leading order diagrams in the small parameters $\l$ and $\e$ in the region $\bh <h \leq 0$ are the diagrams with minimum number of vertices. 
\item Since we are interested only in the leading orders, we can neglect in the rescaled propagators \eqref{3:prop_bh} the factor $\g^h Z_h$. For the same reason we can neglect the difference between $\{\a_{h+1}\}$ and $\{\a_h\}$ in the endpoints of the trees involving a tree vertex on scale $h+1$.  
\end{enumerate}

The previous remarks imply that the leading terms in the beta function can be obtained by the following steps:
\begin{enumerate}[a)]
\item Evaluate the graphs with one loop and minimum number of endpoints and propagator given by the sum of the single scale propagators at scale $h$ and $h+1$, approximated as explained in the remark. 
\item Evaluate the same graphs with propagator of scale $h+1$, again approximated as in the remark 3. 
\item Subtract the values found in b) with the values found in a) and add the trivial graphs without any internal line. 
\item Approximate in the result the cutoff function $f_0(k)$ by the characteristic function of the set $\{k_0^2 +\kk^4  \leq 1 \}$. Note that this approximation is everywhere equivalent to calculating the graphs with all propagators on the single scale $h$, except in the case of the beta function for the coupling constants which involve derivatives with respect to the loop momentum. Hence, except in this case, we will calculate the graphs by using only propagators on scale $h$.
\end{enumerate}
The last remark regards the {\it combinatorial factors} one has to take into account: 
\begin{enumerate}[i.]
\item the coefficient of the truncated expectations, which is $(-1)^{n+1}/n!$ if we are contracting $n$ vertices;
\item the different possibilities of choosing different vertices between those present in the expectations, giving rise to the same graph;
\item for each vertex, the different possibilities of choosing the external lines;
\item the different possibility of contracting the internal lines.
\end{enumerate}

\vskip 0.5cm
Now we are ready to start.

\feyn{
\begin{fmffile}{feyn-TESI/tra1}
 \unitlength = 0.8cm
\def\myl#1{2.5cm}
\begin{align*}
\bar{\b}^{\m,J_0}_h =
 \parbox{\myl}{\centering{ \vskip 0.8cm
	\begin{fmfgraph*}(2.3,1.5)
			\fmfleft{i1}
			\fmfright{o1,o2}
			\fmftop{t}
			\fmfbottom{b}
			\fmf{phantom, tension=1.8}{t,v3}    %---
			\fmf{phantom, tension=1.8}{b,v4}    %---
                   \fmf{wiggly, tension=1.5}{i1,v1}			
			\fmf{plain,left=0.3}{v1,v3}
			\fmf{plain,left=0.3}{v3,v2}			
			\fmf{plain,right=0.3}{v1,v4}
			\fmf{plain,right=0.3}{v4,v2}
			\fmf{plain, tension=1.5}{o1,v2}
                   \fmf{plain, tension=1.5}{v2,o2}
			\Ball{v1,v2}	
		\end{fmfgraph*} \\[3pt]
   $-\frac{1}{2}\,2 \,\binom{4}{2}\, 2$
		}}   +
 \parbox{\myl}{\centering{  \vskip 0.8cm
	\begin{fmfgraph*}(2.3,1.5)
			\fmfleft{i1}
			\fmfright{o1,o2}
			\fmftop{t}
			\fmfbottom{b}
			\fmf{phantom, tension=1.8}{t,v3}    %---
			\fmf{phantom, tension=1.8}{b,v4}    %---
                   \fmf{wiggly, tension=1.5}{i1,v1}			
			\fmf{plain,left=0.3}{v1,v3}
			\fmf{dashes,left=0.3}{v3,v2}			
			\fmf{plain,right=0.3}{v1,v4}
			\fmf{dashes,right=0.3}{v4,v2}
			\fmf{plain, tension=1.5}{o1,v2}
                   \fmf{plain, tension=1.5}{v2,o2}
			\Ball{v1,v2}	
		\end{fmfgraph*}\\[3pt]
   $-\frac{1}{2}\,2 \cdot 2$
		}}  +
 \parbox{\myl}{\centering{ \vskip 0.8cm
	\begin{fmfgraph*}(2.3,1.5)
			\fmfleft{i1}
			\fmfright{o1,o2}
			\fmftop{t}
			\fmfbottom{b}
			\fmf{phantom, tension=1.8}{t,v3}    %---
			\fmf{phantom, tension=1.8}{b,v4}    %---
                   \fmf{wiggly, tension=1.5}{i1,v1}			
			\fmf{dashes,left=0.3}{v1,v3}
			\fmf{plain,left=0.3}{v3,v2}			
			\fmf{dashes,right=0.3}{v1,v4}
			\fmf{plain,right=0.3}{v4,v2}
			\fmf{plain, tension=1.5}{o1,v2}
                   \fmf{plain, tension=1.5}{v2,o2}
			\Ball{v1,v2}	
		\end{fmfgraph*}\\[3pt]
   $-\frac{1}{2}\,2 \,\binom{4}{2}\, 2$
		}}  +
 \parbox{\myl}{\centering{ \vskip 0.8cm
	\begin{fmfgraph*}(2.3,1.5)
			\fmfleft{i1}
			\fmfright{o1,o2}
			\fmftop{t}
			\fmfbottom{b}
			\fmf{phantom, tension=1.8}{t,v3}    %---
			\fmf{phantom, tension=1.8}{b,v4}    %---
                   \fmf{wiggly, tension=1.5}{i1,v1}			
			\fmf{dashes,left=0.3}{v1,v3}
			\fmf{dashes,left=0.3}{v3,v2}			
			\fmf{dashes,right=0.3}{v1,v4}
			\fmf{dashes,right=0.3}{v4,v2}
			\fmf{plain, tension=1.5}{o1,v2}
                   \fmf{plain, tension=1.5}{v2,o2}
			\Ball{v1,v2}	
		\end{fmfgraph*} \\[3pt]
$-\frac{1}{2}\,2 \cdot 2$
		}}  \\
\bar{\b}^{Z,J_0}_h =
 \parbox{\myl}{\centering{ \vskip 0.8cm
	\begin{fmfgraph*}(2.3,1.5)
			\fmfleft{i1}
			\fmfright{o1}
			\fmftop{t}
			\fmfbottom{b}
			\fmf{phantom, tension=1.8}{t,v3}    %---
			\fmf{phantom, tension=1.8}{b,v4}    %---
                   \fmf{wiggly, tension=1.5}{i1,v1}			
			\fmf{plain,left=0.3}{v1,v3}
			\fmf{plain,left=0.3}{v3,v2}			
			\fmf{plain,right=0.3}{v1,v4}
			\fmf{plain,right=0.3}{v4,v2}
			\fmf{dashes, tension=1.5}{o1,v2}
			\Ball{v1,v2}	
		\end{fmfgraph*} \\[3pt]
   $-\frac{1}{2}\,2 \cdot 2$
		}}   +
 \parbox{\myl}{\centering{  \vskip 0.8cm
	\begin{fmfgraph*}(2.3,1.5)
			\fmfleft{i1}
			\fmfright{o1}
			\fmftop{t}
			\fmfbottom{b}
			\fmf{phantom, tension=1.8}{t,v3}    %---
			\fmf{phantom, tension=1.8}{b,v4}    %---
                   \fmf{wiggly, tension=1.5}{i1,v1}			
			\fmf{plain,left=0.3}{v1,v3}
			\fmf{dashes,left=0.3}{v3,v2}			
			\fmf{plain,right=0.3}{v1,v4}
			\fmf{dashes,right=0.3}{v4,v2}
\fmf{dashes, tension=1.5}{o1,v2}
			\Ball{v1,v2}	
		\end{fmfgraph*}\\[3pt]
   $-\frac{1}{2}\,2 \cdot 2\cdot 3$
		}}  +
 \parbox{\myl}{\centering{ \vskip 0.8cm
	\begin{fmfgraph*}(2.3,1.5)
			\fmfleft{i1}
			\fmfright{o1}
			\fmftop{t}
			\fmfbottom{b}
			\fmf{phantom, tension=1.8}{t,v3}    %---
			\fmf{phantom, tension=1.8}{b,v4}    %---
                   \fmf{wiggly, tension=1.5}{i1,v1}			
			\fmf{dashes,left=0.3}{v1,v3}
			\fmf{plain,left=0.3}{v3,v2}			
			\fmf{dashes,right=0.3}{v1,v4}
			\fmf{plain,right=0.3}{v4,v2}
			\fmf{dashes, tension=1.5}{o1,v2}
			\Ball{v1,v2}	
		\end{fmfgraph*}\\[3pt]
   $-\frac{1}{2}\,2 \cdot 2$
		}} +
 \parbox{\myl}{\centering{ \vskip 0.8cm
	\begin{fmfgraph*}(2.3,1.5)
			\fmfleft{i1}
			\fmfright{o1}
			\fmftop{t}
			\fmfbottom{b}
			\fmf{phantom, tension=1.8}{t,v3}    %---
			\fmf{phantom, tension=1.8}{b,v4}    %---
                   \fmf{wiggly, tension=1.5}{i1,v1}			
			\fmf{dashes,left=0.3}{v1,v3}
			\fmf{dashes,left=0.3}{v3,v2}			
			\fmf{dashes,right=0.3}{v1,v4}
			\fmf{dashes,right=0.3}{v4,v2}
			\fmf{dashes, tension=1.5}{o1,v2}
			\Ball{v1,v2}	
		\end{fmfgraph*} \\[3pt]
$-\frac{1}{2}\,2 \cdot 2 \cdot 3$
		}} 
\end{align*}
\end{fmffile}
}{Leading order flow equation for $\bm_h^{J_0}$ and $\bar{Z}_h^{J_0}$. The combinatorial factors are such that in both cases we find the integral over $k$ of the combination $k_0^2 - \kk^4$, which gives zero.}{tra1}

\subsection{Leading order beta function of $\bm_h^{J_0}$ and $\bar{Z}_h^{J_0}$}

The leading order contributions to the beta function of $\bm_h^{J_0}$ are shown in fig. \ref{tra1}, whose sum is equal to:
\[ \label{L1}
\bar{\b}_{h,1}^{\m,J_0}= -\l \e^{-1} \m_h^{J_0}\lft(12\l_h +2\l'_h \rgt) \frac{1}{(2\pi)^3}\int dk_0 d^2\kk \,f_h(\kk^4+k_0^2)\,\frac{\kk^2 (\kk^2 +\g^h \bar{Z}_h)-k_0^2}{\lft(\kk^2(\kk^2 +\g^h \bar{Z}_h) +k_0^2 \rgt)^2}
\]
where $\e \leq \bar{Z}_h \leq 1$. Now, it is sufficient to note that 
\[
I_1=\frac{1}{(2\pi)^3}\int dk_0 d^2\kk\,f_h(\kk^4+k_0^2)\,\frac{\kk^4 -k_0^2}{(\kk^4 +k_0^2)^2} =0
\]
which can be easily seen with the following change of coordinates
\[
\kk^2 = \r \sin\th \qquad k_0 = \r \cos\th
\]

%-------------------------------------------------------------------------

\feyn{
\begin{fmffile}{feyn-TESI/tra2}
 \unitlength = 0.8cm
\def\myl#1{2.5cm}
\begin{align*}
& \bar{\b}^{\l}_h =
 \parbox{\myl}{\centering{ \vskip 0.8cm
	\begin{fmfgraph*}(2.3,1.5)
			\fmfleft{i1,i2}
			\fmfright{o1,o2}
			\fmftop{t}
			\fmfbottom{b}
			\fmf{phantom, tension=1.8}{t,v3}    %---
			\fmf{phantom, tension=1.8}{b,v4}    %---
                   \fmf{plain, tension=1.5}{i1,v1,i2}			
			\fmf{plain,left=0.3}{v1,v3}
			\fmf{plain,left=0.3}{v3,v2}			
			\fmf{plain,right=0.3}{v1,v4}
			\fmf{plain,right=0.3}{v4,v2}
			\fmf{plain, tension=1.5}{o1,v2}
                   \fmf{plain, tension=1.5}{v2,o2}
			\Ball{v1,v2}	
		\end{fmfgraph*} \\[3pt]
   $-\frac{1}{2} \,\binom{4}{2}^2\, 2$
		}}   +
 \parbox{\myl}{\centering{  \vskip 0.8cm
	\begin{fmfgraph*}(2.3,1.5)
			\fmfleft{i1,i2}
			\fmfright{o1,o2}
			\fmftop{t}
			\fmfbottom{b}
			\fmf{phantom, tension=1.8}{t,v3}    %---
			\fmf{phantom, tension=1.8}{b,v4}    %---
                   \fmf{plain, tension=1.5}{i1,v1,i2}			
			\fmf{plain,left=0.3}{v1,v3}
			\fmf{dashes,left=0.3}{v3,v2}			
			\fmf{plain,right=0.3}{v1,v4}
			\fmf{dashes,right=0.3}{v4,v2}
			\fmf{plain, tension=1.5}{o1,v2}
                   \fmf{plain, tension=1.5}{v2,o2}
			\Ball{v1,v2}	
		\end{fmfgraph*}\\[3pt]
   $-\frac{1}{2}\,2 \,\binom{4}{2}\,2$
		}}  +
 \parbox{\myl}{\centering{ \vskip 0.8cm
	\begin{fmfgraph*}(2.3,1.5)
			\fmfleft{i1,i2}
			\fmfright{o1,o2}
			\fmftop{t}
			\fmfbottom{b}
			\fmf{phantom, tension=1.8}{t,v3}    %---
			\fmf{phantom, tension=1.8}{b,v4}    %---
                   \fmf{plain, tension=1.5}{i1,v1,i2}			
			\fmf{dashes,left=0.3}{v1,v3}
			\fmf{dashes,left=0.3}{v3,v2}			
			\fmf{dashes,right=0.3}{v1,v4}
			\fmf{dashes,right=0.3}{v4,v2}
			\fmf{plain, tension=1.5}{o1,v2}
                   \fmf{plain, tension=1.5}{v2,o2}
			\Ball{v1,v2}	
		\end{fmfgraph*} \\[3pt]
$-\frac{1}{2}\,2 $
		}}  \\
& \bar{\b}^{\m}_h =
 \parbox{\myl}{\centering{ \vskip 0.8cm
	\begin{fmfgraph*}(2.3,1.5)
			\fmfleft{i1}
			\fmfright{o1,o2}
			\fmftop{t}
			\fmfbottom{b}
			\fmf{phantom, tension=1.8}{t,v3}    %---
			\fmf{phantom, tension=1.8}{b,v4}    %---
                   \fmf{dashes, tension=1.5}{i1,v1}			
			\fmf{plain,left=0.3}{v1,v3}
			\fmf{plain,left=0.3}{v3,v2}			
			\fmf{plain,right=0.3}{v1,v4}
			\fmf{plain,right=0.3}{v4,v2}
			\fmf{plain, tension=1.5}{o1,v2}
                   \fmf{plain, tension=1.5}{v2,o2}
			\Ball{v1,v2}	
		\end{fmfgraph*} \\[3pt]
   $-\frac{1}{2} \,\binom{4}{2}\, 2$
		}}   +
 \parbox{\myl}{\centering{  \vskip 0.8cm
	\begin{fmfgraph*}(2.3,1.5)
			\fmfleft{i1}
			\fmfright{o1,o2}
			\fmftop{t}
			\fmfbottom{b}
			\fmf{phantom, tension=1.8}{t,v3}    %---
			\fmf{phantom, tension=1.8}{b,v4}    %---
                   \fmf{dashes, tension=1.5}{i1,v1}			
			\fmf{plain,left=0.3}{v1,v3}
			\fmf{dashes,left=0.3}{v3,v2}			
			\fmf{plain,right=0.3}{v1,v4}
			\fmf{dashes,right=0.3}{v4,v2}
			\fmf{plain, tension=1.5}{o1,v2}
                   \fmf{plain, tension=1.5}{v2,o2}
			\Ball{v1,v2}	
		\end{fmfgraph*}\\[3pt]
   $-\frac{1}{2}\,2 \cdot2$
		}}  +
 \parbox{\myl}{\centering{ \vskip 0.8cm
	\begin{fmfgraph*}(2.3,1.5)
			\fmfleft{i1}
			\fmfright{o1,o2}
			\fmftop{t}
			\fmfbottom{b}
			\fmf{phantom, tension=1.8}{t,v3}    %---
			\fmf{phantom, tension=1.8}{b,v4}    %---
                   \fmf{dashes, tension=1.5}{i1,v1}			
			\fmf{dashes,left=0.3}{v1,v3}
			\fmf{plain,left=0.3}{v3,v2}			
			\fmf{dashes,right=0.3}{v1,v4}
			\fmf{plain,right=0.3}{v4,v2}
			\fmf{plain, tension=1.5}{o1,v2}
                   \fmf{plain, tension=1.5}{v2,o2}
			\Ball{v1,v2}	
		\end{fmfgraph*} \\[3pt]
$-\frac{1}{2}\,2\cdot 3 \,\binom{4}{2} \,2 $
		}} +\parbox{\myl}{\centering{ \vskip 0.8cm
	\begin{fmfgraph*}(2.3,1.5)
			\fmfleft{i1}
			\fmfright{o1,o2}
			\fmftop{t}
			\fmfbottom{b}
			\fmf{phantom, tension=1.8}{t,v3}    %---
			\fmf{phantom, tension=1.8}{b,v4}    %---
                   \fmf{dashes, tension=1.5}{i1,v1}			
			\fmf{dashes,left=0.3}{v1,v3}
			\fmf{dashes,left=0.3}{v3,v2}			
			\fmf{dashes,right=0.3}{v1,v4}
			\fmf{dashes,right=0.3}{v4,v2}
			\fmf{plain, tension=1.5}{o1,v2}
                   \fmf{plain, tension=1.5}{v2,o2}
			\Ball{v1,v2}	
		\end{fmfgraph*} \\[3pt]
$-\frac{1}{2}\,2\cdot3\cdot2 $
		}}  \\
& \bar{\b}^{Z}_h =
 \parbox{\myl}{\centering{ \vskip 0.8cm
	\begin{fmfgraph*}(2.3,1.5)
			\fmfleft{i1}
			\fmfright{o1}
			\fmftop{t}
			\fmfbottom{b}
			\fmf{phantom, tension=1.8}{t,v3}    %---
			\fmf{phantom, tension=1.8}{b,v4}    %---
                   \fmf{dashes, tension=1.5}{i1,v1}			
			\fmf{plain,left=0.3}{v1,v3}
			\fmf{plain,left=0.3}{v3,v2}			
			\fmf{plain,right=0.3}{v1,v4}
			\fmf{plain,right=0.3}{v4,v2}
			\fmf{dashes, tension=1.5}{o1,v2}
			\Ball{v1,v2}	
		\end{fmfgraph*} \\[3pt]
   $-\frac{1}{2}\,2$
		}}   +
 \parbox{\myl}{\centering{  \vskip 0.8cm
	\begin{fmfgraph*}(2.3,1.5)
			\fmfleft{i1}
			\fmfright{o1}
			\fmftop{t}
			\fmfbottom{b}
			\fmf{phantom, tension=1.8}{t,v3}    %---
			\fmf{phantom, tension=1.8}{b,v4}    %---
                   \fmf{dashes, tension=1.5}{i1,v1}			
			\fmf{plain,left=0.3}{v1,v3}
			\fmf{dashes,left=0.3}{v3,v2}			
			\fmf{plain,right=0.3}{v1,v4}
			\fmf{dashes,right=0.3}{v4,v2}
\fmf{dashes, tension=1.5}{o1,v2}
			\Ball{v1,v2}	
		\end{fmfgraph*}\\[3pt]
   $-\frac{1}{2}\,2 \cdot 2\cdot 3$
		}}  +
 \parbox{\myl}{\centering{ \vskip 0.8cm
	\begin{fmfgraph*}(2.3,1.5)
			\fmfleft{i1}
			\fmfright{o1}
			\fmftop{t}
			\fmfbottom{b}
			\fmf{phantom, tension=1.8}{t,v3}    %---
			\fmf{phantom, tension=1.8}{b,v4}    %---
                   \fmf{dashes, tension=1.5}{i1,v1}			
			\fmf{dashes,left=0.3}{v1,v3}
			\fmf{dashes,left=0.3}{v3,v2}			
			\fmf{dashes,right=0.3}{v1,v4}
			\fmf{dashes,right=0.3}{v4,v2}
			\fmf{dashes, tension=1.5}{o1,v2}
			\Ball{v1,v2}	
		\end{fmfgraph*} \\[3pt]
$-\frac{1}{2}\,3^2 \cdot 2 $
		}} 
\end{align*}
\end{fmffile}
}{Leading order flow equation for $\bl_h$, $\bm_h$ and $\bar{Z}_h$. The one--loop diagrams are the same appearing in the beta function for $\bm^{J_0}_h$ and $\bar{Z}^{J_0}_h$ but with different combinatorial factors.}{tra2}

The integral becomes
\[
I_1 & = - \frac{1}{8 \pi^2}\int \r \,d\r f_h(\r)\int_{0}^{\pi}  d\th \,\frac{\r^2(\cos^2\th -\sin^2\th)}{\r^4}  %(\r^2 +\g^h Z_h \r \cos \th)^2 
\\ \non
& =  -\frac{1}{8 \pi^2}\int \frac{d\r}{\r} \,f_h(\r)\int_0^\pi d\th \,\cos 2\th  =0
\]
since the angular part is zero. Then $ \eqref{L1}= O(\l \e \g^h)$ which is summable over $h$. This is not the leading order contribution to the beta function (and in fact in the following calculations we will neglect the factor $\g^h Z_h$ in the propagator, as also stressed in the introduction). The leading order contribution to $\bar{\b}_h^{\m,J_0}$ comes from the diagrams with one vertex $\bm_h^{J_0}$ and two vertices $\m_h$. By summing over the latter diagrams one finds
\[
\bar{\b}_{h,2}^{\m,J_0}= 2 \m_h^{J_0} \m^2_h \frac{1}{(2\pi)^3}\int dk_0 d^2\kk\,f_h(\kk^4+k_0^2)\,\frac{2\kk^2 +\g^h \bar{Z}_h }{(\kk^4 +k_0^2)^2} 
\]
By using that $\m_h \leq \g^{-\frac{h}{2}}\e$ we find $|\bar{\b}_{h,2}^{\m,J_0}| \leq \l \e \g^{-h}$, which substituted in
\[
\bm_h^{J_0} = \bm^{J_0}_0 + \sum_{k=h +1}^0 \bar{\b}_{k,1}^{\m,J_0} 
\]
gives 
\[
\bm_h^{J_0} = \bm^{J_0}_0 \lft(1+ O(\l \e \g^{-h}) \rgt)
\]

The same situation occurs when one calculates the leading order beta function for $\bar{Z}_h^{J_0}$. The two vertices diagrams in fig.~\ref{tra1} cancel and we are left with the next order contribution, which cames from the diagrams with one vertex $\m_h^{J_0}$ and three vertices $\m_h$. By using that $\m_h \leq \g^{-\frac{h}{2}}\e$ we find 
\[
|\bar{\b}_{h}^{Z,J_0}| \leq \l^2 \e \g^{-\frac{3}{2}h}
\] 
which finally gives
\[
\g^{\frac{h}{2}}\bar{Z}_h^{J_0} =   O\lft(\l^2 \e \g^{-h}\rgt)
\]
On the contrary, if one calculates the beta functions for $\bl_h$, $\bm_h$ and $\bar{Z}_h$ finds that the diagrams of second order (see fig.~\ref{tra2}) do not cancel, since the combinatorial factors are different. In particular the point is that the vertex $\m'_h$ has an additional combinatorial factor $3$ with respect to $\m'^{J_0}_h$ related to the fact that we can choice the external dashed leg among its three legs. 

\vskip 0.5cm

\feyn{
\begin{fmffile}{feyn-TESI/tra3}
 \unitlength = 0.8cm
\def\myl#1{2.5cm}
\begin{align*}
& \bar{\b}^{\m,J_1}_h =
 \parbox{\myl}{\centering{ \vskip 0.8cm
	\begin{fmfgraph*}(2.3,1.5)
			\fmfleft{i1}
			\fmfright{o1,o2}
			\fmftop{t}
			\fmfbottom{b}
			\fmf{phantom, tension=1.8}{t,v3}    %---
			\fmf{phantom, tension=1.8}{b,v4}    %---
                   \fmf{wiggly, tension=1.5, label=$J_1$, label.dist=-0.5cm}{i1,v1}			
			\fmf{plain,left=0.3, label=$\dpr_\xx$,label.dist=0.05cm}{v1,v3}
			\fmf{plain,left=0.3}{v3,v2}			
			\fmf{dashes,right=0.3}{v1,v4}
			\fmf{dashes,right=0.3}{v4,v2}
			\fmf{dashes, tension=1.4}{o1,v2}
                   \fmf{plain, tension=1.5, label=$\dpr_\pp$, label.dist=0.05cm}{v2,o2}
			\Ball{v1,v2}	
		\end{fmfgraph*} \\[3pt]
   $-\frac{1}{2}\,2 \cdot 2^2 $
		}}   +
 \parbox{\myl}{\centering{ \vskip 0.8cm
	\begin{fmfgraph*}(2.3,1.5)
			\fmfleft{i1}
			\fmfright{o1,o2}
			\fmftop{t}
			\fmfbottom{b}
			\fmf{phantom, tension=1.8}{t,v3}    %---
			\fmf{phantom, tension=1.8}{b,v4}    %---
                   \fmf{wiggly, tension=1.5, label=$J_1$, label.dist=-0.5cm}{i1,v1}			
			\fmf{plain,left=0.3, label=$\dpr_\xx$,label.dist=0.05cm}{v1,v3}
			\fmf{dashes,left=0.3}{v3,v2}			
			\fmf{dashes,right=0.3}{v1,v4}
			\fmf{plain,right=0.3}{v4,v2}
			\fmf{dashes, tension=1.4}{o1,v2}
                   \fmf{plain, tension=1.5, label=$\dpr_\pp$, label.dist=0.05cm}{v2,o2}
			\Ball{v1,v2}	
		\end{fmfgraph*} \\[3pt]
   $-\frac{1}{2}\,2 \cdot 2^2 $
		}}   - \parbox{\myl}{\centering{ \vskip 0.8cm
	\begin{fmfgraph*}(2.3,1.5)
			\fmfleft{i1}
			\fmfright{o1,o2}
			\fmftop{t}
			\fmfbottom{b}
			\fmf{phantom, tension=1.8}{t,v3}    %---
			\fmf{phantom, tension=1.8}{b,v4}    %---
                   \fmf{wiggly, tension=1.5, label=$J_1$}{i1,v1}			
			\fmf{plain,left=0.3}{v1,v3}
			\fmf{plain,left=0.3}{v3,v2}			
			\fmf{dashes,right=0.3, label=$\dpr_\xx$,label.dist=0.05cm}{v1,v4}
			\fmf{dashes,right=0.3}{v4,v2}
			\fmf{dashes, tension=1.4}{o1,v2}
                   \fmf{plain, tension=1.5, label=$\dpr_\pp$, label.dist=0.05cm}{v2,o2}
			\Ball{v1,v2}	
		\end{fmfgraph*} \\[3pt]
   $-\frac{1}{2}\,2 \cdot 2^2 $
		}}   -
 \parbox{\myl}{\centering{ \vskip 0.8cm
	\begin{fmfgraph*}(2.3,1.5)
			\fmfleft{i1}
			\fmfright{o1,o2}
			\fmftop{t}
			\fmfbottom{b}
			\fmf{phantom, tension=1.8}{t,v3}    %---
			\fmf{phantom, tension=1.8}{b,v4}    %---
                   \fmf{wiggly, tension=1.5, label=$J_1$}{i1,v1}			
			\fmf{plain,left=0.3}{v1,v3}
			\fmf{dashes,left=0.3}{v3,v2}			
			\fmf{dashes,right=0.3, label=$\dpr_\xx$,label.dist=0.05cm}{v1,v4}
			\fmf{plain,right=0.3}{v4,v2}
			\fmf{dashes, tension=1.4}{o1,v2}
                   \fmf{plain, tension=1.5, label=$\dpr_\pp$, label.dist=0.05cm}{v2,o2}
			\Ball{v1,v2}	
		\end{fmfgraph*} \\[3pt]
   $-\frac{1}{2}\,2 \cdot 2^2 $
		}}  
  \\
%----------------------------------------------------------------------
& \bar{\b}^{E,J_1}_h =
\parbox{\myl}{\centering{ \vskip 0.8cm
	\begin{fmfgraph*}(2.3,1.5)
			\fmfleft{i1}
			\fmfright{o1}
			\fmftop{t}
			\fmfbottom{b}
			\fmf{phantom, tension=1.8}{t,v3}    %---
			\fmf{phantom, tension=1.8}{b,v4}    %---
                   \fmf{wiggly, tension=1.5, label=$J_1$}{i1,v1}			
			\fmf{plain,left=0.3, label=$\dpr_\xx$,label.dist=0.05cm}{v1,v3}
			\fmf{plain,left=0.3}{v3,v2}			
			\fmf{dashes,right=0.3}{v1,v4}
			\fmf{dashes,right=0.3}{v4,v2}
                   \fmf{plain, tension=1.5, label=$\dpr_\pp$, label.dist=0.05cm}{v2,o1}
			\Ball{v1,v2}	
		\end{fmfgraph*} \\[3pt]
   $-\frac{1}{2}\,2 \cdot 2 $
		}}   +
 \parbox{\myl}{\centering{ \vskip 0.8cm
	\begin{fmfgraph*}(2.3,1.5)
			\fmfleft{i1}
			\fmfright{o1}
			\fmftop{t}
			\fmfbottom{b}
			\fmf{phantom, tension=1.8}{t,v3}    %---
			\fmf{phantom, tension=1.8}{b,v4}    %---
                   \fmf{wiggly, tension=1.5, label=$J_1$}{i1,v1}			
			\fmf{plain,left=0.3, label=$\dpr_\xx$,label.dist=0.05cm}{v1,v3}
			\fmf{dashes,left=0.3}{v3,v2}			
			\fmf{dashes,right=0.3}{v1,v4}
			\fmf{plain,right=0.3}{v4,v2}
                   \fmf{plain, tension=1.5, label=$\dpr_\pp$, label.dist=0.05cm}{v2,o1}
			\Ball{v1,v2}	
		\end{fmfgraph*} \\[3pt]
   $-\frac{1}{2}\,2 \cdot 2 $
		}}   - \parbox{\myl}{\centering{ \vskip 0.8cm
	\begin{fmfgraph*}(2.3,1.5)
			\fmfleft{i1}
			\fmfright{o1}
			\fmftop{t}
			\fmfbottom{b}
			\fmf{phantom, tension=1.8}{t,v3}    %---
			\fmf{phantom, tension=1.8}{b,v4}    %---
                   \fmf{wiggly, tension=1.5, label=$J_1$, label.dist=-0.5cm}{i1,v1}			
			\fmf{plain,left=0.3}{v1,v3}
			\fmf{plain,left=0.3}{v3,v2}			
			\fmf{dashes,right=0.3, label=$\dpr_\xx$,label.dist=0.05cm}{v1,v4}
			\fmf{dashes,right=0.3}{v4,v2}
                   \fmf{plain, tension=1.5, label=$\dpr_\pp$, label.dist=0.05cm}{v2,o1}
			\Ball{v1,v2}	
		\end{fmfgraph*} \\[3pt]
   $-\frac{1}{2}\,2 \cdot 2 $
		}}   -
 \parbox{\myl}{\centering{ \vskip 0.8cm
	\begin{fmfgraph*}(2.3,1.5)
			\fmfleft{i1}
			\fmfright{o1}
			\fmftop{t}
			\fmfbottom{b}
			\fmf{phantom, tension=1.8}{t,v3}    %---
			\fmf{phantom, tension=1.8}{b,v4}    %---
                   \fmf{wiggly, tension=1.5, label=$J_1$, label.dist=-0.5cm}{i1,v1}			
			\fmf{plain,left=0.3}{v1,v3}
			\fmf{dashes,left=0.3}{v3,v2}			
			\fmf{dashes,right=0.3, label=$\dpr_\xx$,label.dist=0.05cm}{v1,v4}
			\fmf{plain,right=0.3}{v4,v2}
                   \fmf{plain, tension=1.5, label=$\dpr_\pp$, label.dist=0.05cm}{v2,o1}
			\Ball{v1,v2}	
		\end{fmfgraph*} \\[3pt]
   $-\frac{1}{2}\,2 \cdot 2 $
		}}  
\end{align*}
\end{fmffile}
}{Leading order flow equation for $\bm_h^{J_1}$ and $\bar{E}_h^{J_1}$. The symbol $\dpr_\pp$ on the external leg means the derivative with respect the external momentum $p=(p_0,\pp)$, which at the end of the computation must be taken equal to zero, since we are interested in calculating the local part of the diagrams. }{tra3}

\subsection{Leading order beta function of $\bm_h^{J_1}$ and $\bar{E}_h^{J_1}$}  \label{low_ord_J}

The diagrams contributing to the beta function for $\bm_h^{J_1}$ and $\bar{E}_h^{J_1}$ at leading order are shown in fig.~\ref{tra3}. Both the beta function are proportional to the integral
\[
I_1=  \lim_{\pp\arr 0} \dpr_\pp \int dk_0 d^2\kk \,f_0(k)\, (2\kk + \pp) \, \frac{(\kk+\pp)^2\,\kk^2 +k_0^2}{\bar{\DD}_0(k+p)\,\bar{\DD}_0(k)}
\]
where $p = (p_0, \pp)$
\[
\bar{\DD}_0(k) = k_0^2 + \kk^4
\]
Using that $\dpr_\pp (\kk + \pp)=2$,  $\dpr_\pp (\kk + \pp)^2=2(\kk +\pp)$ and $\dpr_\pp\bar{\DD}_0(k+p) =4(\kk + \pp)^3$ we obtain
\[
I_1=  2 \int dk_0 d^2\kk \,f_0(k)\, \frac{k_0^2- \kk^4}{\bar{\DD_0^2(k)}}
\]
which again is equal to zero. The first non trivial contribution to the beta function of $\m_h^{J_1}$ is given by the diagrams with one vertex $\m_h^{J_1}$ and two vertices $\m_h^2$, \ie
\[
|\bar{\b}_{h}^{\m,J_1}| \leq \l \e \,\g^{-\frac{h}{2}}
\] 
which gives 
\[
\bm_h^{J_1} = \bm^{J_1}_0 \lft(1+ O(\l \e \g^{-h}) \rgt)
\]
Regarding the first non trivial contribution to the beta function of $\bar{E}_h^{J_1}$, this is given by the diagrams with one vertex $\m_h^{J_1}$ and three vertices $\m_h^2$, \ie
\[
|\bar{\b}_{h}^{E,J_1}| \leq \l^2 \e \,\g^{-\frac{3}{2}h}
\] 
which gives
\[
\g^{\frac{h}{2}}\bar{E}_h^{J_1} =   O\lft(\l^2 \e \g^{-h}\rgt)
\]

%\end{document}

%\input{intro-senza-sapclass} \input{intestazione-sap} \begin{document}% \tableofcontents

\chapter{Leading order computations}  \label{lowest_order}

In order to calculate the beta function at scale $h$, one has to evaluate some Feynman graphs at zero momentum of the external line, as already described in appendix~\ref{leading_transient}. We repeat here the rules we will follow in the computations, since there are some differences with respect to the transient region.
\begin{enumerate}
\item only terms with at least one loop can contribute, since the single scale propagator vanishes at zero momentum. 
\item In the graphs with only one loop, all the internal lines must carry the same momentum. Hence, due to our choice of the cutoff function $f_0(k)$, the internal lines of the loop may only have propagators of scale $h$ or $h+1$; in fact at least one propagator must be of scale $h$ (by definition of $\b_h$) and the supports of the Fourier transform of the propagators at scale $h$ and and $h'\geq h$ are diskoint if $h'>h+1$. 
%\item In the calculation of the graphs with one loop, the subgraph associated with the tree vertex of scale $h+1$ has no loop. Therefore in this tree vertex the $\RR$ operator coincides with the identity.
\item Since we are interested only in the leading orders, we can neglect in the rescaled propagators \eqref{RENg} the terms proportional to $\g^{2h}$ and the dependence on $k$ of $Z_h$, $A_h$, $B_h$ and $E_h$, see \eqref{wave_funct}. For the same reason we can approximate $Z_{h+1}$ , $A_{h+1}$, $B_{h+1}$ and $E_{h+1}$ by $Z_h$, $A_h$, $B_h$ and $E_h$ in the expression of $g^{(h+1)})_{\a\a'}(x)$. 

\end{enumerate}

Using these remarks and the results of appendix \ref{order_e} one finds that the leading terms in the beta function for $h \leq \bh$ are obtained by the following steps:
\begin{enumerate}[a)]
\item Evaluate the graphs with one loop and propagator given by the sum of the single scale propagators at scale $h$ and $h+1$, approximated as explained in the remark. 
\item Evaluate the same graphs with propagator of scale $h+1$, again approximated as in the remark 3. 
\item Subtract the values found in b) with the values found in a) and add the trivial graphs without any internal line. 
\item Approximate in the result the cutoff function $f_0(k)$ by the characteristic function of the set $\{k_0^2 +\kk^4  \leq 1 \}$. Note that this approximation is everywhere equivalent to calculating the graphs with all propagators on the single scale $h$, except in the case of the beta function for the coupling constants which involve derivatives with respect to the loop momentum. Hence, except in this case, we will calculate the graphs by using only propagators on scale $h$.
\end{enumerate}
The last remark regards the {\it combinatorial factors} one has to take into account: 
\begin{enumerate}[i.]
\item the coefficient of the truncated expectations, which is $(-1)^{n+1}/n!$ if we are contracting $n$ vertices;
\item the different possibilities of choosing different vertices between those present in the expectations, giving rise to the same graph;
\item for each vertex, the different possibilities of choosing the external lines;
\item the different possibility of contracting the internal lines.
\end{enumerate}

\vskip 0.5cm

\section{Global WIs and flow equations in $3d$} \label{B.flow}

As shown in appendix \ref{order_e}  in order to keep in each flow equation only the leading terms in the small parameter $\e$ in three dimensions it is sufficient to consider the one loop graphs without $\n_h$ vertices. This property holds both at the beginning of the region $h \leq \bh$ and in the asymptotic region $h \arr -\io$, but in the latter case the dominant diagrams in $|h|^{-1}$ are those with all the internal dashed lines contracted among them, see \eqref{3d_hstar}. We shall study the flow equations, by keeping only these contributions; the properties of the corresponding solutions will be used to justify the approximation. 

In this section we will report the leading order flow equations for the running coupling constants $\l_h$ and $\m_h$ and the wave function renormalization constant $Z_h$, in the region $h \leq \bh$. These computations are useful to understand the perturbative interpretation of the global WIs \ref{GW_flow_3d} at the first non trivial order.

\vskip -0.5cm

\feyn{
\begin{fmffile}{feyn-TESI/conti3d}
 \unitlength = 0.8cm
\def\myl#1{2.5cm}
\def\myll#1{2.7cm}
\begin{align*}
	 \b_h^\m & = \;
	\parbox{\myll}{\centering{ \vskip 0.5cm
			\begin{fmfgraph*}(2.8,2) 
			\fmfleft{i1}
			\fmfright{o1,o2}
			\fmftop{t}
			\fmfbottom{b}
			\fmf{phantom, tension=1.8}{t,v3}    %---
			\fmf{phantom, tension=1.8}{b,v4}    %---
            \fmf{dashes, tension=1.5}{i1,v1}			
			\fmf{plain,left=0.3}{v1,v3}
			\fmf{plain,left=0.3}{v3,v2}			
			\fmf{plain,right=0.3}{v1,v4}
			\fmf{plain,right=0.3}{v4,v2}
			\fmf{plain, tension=1.5}{o1,v2,o2}
		\Ball{v1,v2}
		\end{fmfgraph*} \\
$-\frac{1}{2}\,2\,\binom{4}{2}\,2$
			}} 
			+\;
	\parbox{\myll}{\centering{  \vskip 0.5cm
			\begin{fmfgraph*}(2.8,2) 
			\fmfleft{i1}
			\fmfright{o1,R,o2}
                   \fmftop{T}
                   \fmfbottom{B}
			\fmf{phantom, tension=1.8}{B,vB}  %-------
                    \fmf{phantom, tension=1.8}{T,vT}  %-------
			\fmf{dashes,tension=1.5}{i1,v1}
			\fmf{plain, left=0.3, tension=1.2}{v1,vT}
                   \fmf{plain, left=0.2, tension=0.4}{vT,v3}
                    \fmf{phantom, tension=1}{R,vR}  %------
			\fmf{dashes, left=0.3, tension=1}{v3,vR}
                   \fmf{dashes, left=0.3, tension=0.4}{vR,v2}			
                  \fmf{plain, right=0.3, tension =1.4}{v1,vB}
	            \fmf{plain, right=0.2, tension=0.4}{vB,v2}           
		     \fmf{plain, tension=0.8}{o1,v2}
			\fmf{plain,tension=0.8}{o2,v3}
                   \Ball{v1,v2,v3}
			\end{fmfgraph*}  \\
$\frac{1}{3!}\,3\,2^3$
			}}  
+\;
	\parbox{\myll}{\centering{ \vskip 0.5cm
			\begin{fmfgraph*}(2.8,2) 
			\fmfleft{i1}
			\fmfright{o1,R,o2}
                   \fmftop{T}
                   \fmfbottom{B}
			\fmf{phantom, tension=1.8}{B,vB}  %-------
                    \fmf{phantom, tension=1.8}{T,vT}  %-------
			\fmf{dashes,tension=1.5}{i1,v1}
			\fmf{plain, left=0.3, tension=1.2}{v1,vT}
                   \fmf{dashes, left=0.2, tension=0.4}{vT,v3}
                    \fmf{phantom, tension=1}{R,vR}  %------
			\fmf{plain, left=0.3, tension=1}{v3,vR}
                   \fmf{plain, left=0.3, tension=0.4}{vR,v2}		
                  \fmf{plain, right=0.3, tension =1.4}{v1,vB}
	            \fmf{dashes, right=0.2, tension=0.4}{vB,v2}           
	             \fmf{plain, tension=0.8}{o1,v2}
			\fmf{plain,tension=0.8}{o2,v3}
                   \Ball{v1,v2,v3}
			\end{fmfgraph*}    \\
$\frac{1}{3!}\,3\,2^3$
			}}  +\;
	\parbox{\myll}{\centering{  \vskip 0.5cm
			\begin{fmfgraph*}(2.8,2) 
			\fmfleft{i1}
			\fmfright{o1,R,o2}
                   \fmftop{T}
                   \fmfbottom{B}
			\fmf{phantom, tension=1.8}{B,vB}  %-------
                    \fmf{phantom, tension=1.8}{T,vT}  %-------
			\fmf{dashes,tension=1.5}{i1,v1}
			\fmf{plain, left=0.3, tension=1.2}{v1,vT}
                   \fmf{dashes, left=0.2, tension=0.4}{vT,v3}
                    \fmf{phantom, tension=1}{R,vR}  %------
			\fmf{plain, left=0.3, tension=1}{v3,vR}
                   \fmf{dashes, left=0.3, tension=0.4}{vR,v2}			
                  \fmf{plain, right=0.3, tension =1.4}{v1,vB}
	            \fmf{plain, right=0.2, tension=0.4}{vB,v2}           
			\fmf{plain, tension=0.8}{o1,v2}
			\fmf{plain,tension=0.8}{o2,v3}
                   \Ball{v1,v2,v3}
			\end{fmfgraph*}    \\
$\frac{1}{3!}\,3\,2^4$
			}}  \\
& \hskip 5cm \tfrac{1}{2}\,\b_h^Z  = \;
	\parbox{\myll}{\centering{  \vskip 0.5cm
			\begin{fmfgraph*}(2.8,2) 
			\fmfleft{i1}
			\fmfright{o1}
			\fmftop{t}
			\fmfbottom{b}
			\fmf{phantom, tension=1.8}{t,v3}    %---
			\fmf{phantom, tension=1.8}{b,v4}    %---
            \fmf{dashes, tension=1.5}{i1,v1}			
			\fmf{plain,left=0.3}{v1,v3}
			\fmf{plain,left=0.3}{v3,v2}			
			\fmf{plain,right=0.3}{v1,v4}
			\fmf{plain,right=0.3}{v4,v2}
			\fmf{dashes, tension=1.5}{o1,v2}
		\Ball{v1,v2}
		\end{fmfgraph*}   \\
$-\frac{1}{2!}\,2$
			}} 
\end{align*}
\end{fmffile}	
}{ One--loop beta functions for $\m_h$ and $Z_h$, d=3. }{conti3d}

%--------------------------------------------------

\feyn{
\begin{fmffile}{feyn-TESI/trick}
 \unitlength = 0.8cm
\def\myl#1{2.5cm}
\def\myll#1{1cm}
\begin{align*} \hskip -2cm
	\parbox{1.5cm}{\centering{
			\begin{fmfgraph*}(2.5,2) 
			\fmfright{o1,o2}
			\fmftop{t}
			\fmfbottom{b}
			\fmf{phantom}{t,v3}    %---
			\fmf{phantom}{b,v4}    %---			
			\fmf{plain,left=0.3, tension=0.6}{v3,v2}			
			\fmf{plain,right=0.3, tension=0.6}{v4,v2}
			\fmf{plain}{o1,v2,o2}
		\Ball{v2}
		\end{fmfgraph*} 
			}} 
			\qquad \arr 
	\parbox{\myll}{\centering{  
			\begin{fmfgraph*}(2.5,2) 
			\fmfright{o1,R,o2}
                   \fmftop{T}
                   \fmfbottom{B}
			\fmf{phantom, tension=1}{B,vB}  %-------
                    \fmf{phantom, tension=1}{T,vT}  %-------
                   \fmf{plain, left=0.2, tension=0.4}{vT,v3}
                    \fmf{phantom, tension=1}{R,vR}  %------
			\fmf{dashes, left=0.2, tension=1}{v3,vR}
                   \fmf{dashes, left=0.2, tension=0.8}{vR,v2}			
	            \fmf{plain, right=0.2, tension=0.4}{vB,v2}           
		     \fmf{plain, tension=0.8}{o1,v2}
			\fmf{plain,tension=0.8}{o2,v3}
                   \Ball{v2,v3}
			\end{fmfgraph*} 
			}}  
\hskip 1.5cm + \hskip -0.5cm
	\parbox{\myll}{\centering{  
			\begin{fmfgraph*}(2.5,2) 
			\fmfright{o1,R,o2}
                   \fmftop{T}
                   \fmfbottom{B}
			\fmf{phantom, tension=1}{B,vB}  %-------
                    \fmf{phantom, tension=1}{T,vT}  %-------
                   \fmf{dashes, left=0.2, tension=0.4}{vT,v3}
                    \fmf{phantom, tension=1}{R,vR}  %------
			\fmf{plain, left=0.2, tension=1}{v3,vR}
                   \fmf{plain, left=0.2, tension=0.8}{vR,v2}			
	            \fmf{dashes, right=0.2, tension=0.4}{vB,v2}           
		     \fmf{plain, tension=0.8}{o1,v2}
			\fmf{plain,tension=0.8}{o2,v3}
                   \Ball{v2,v3}
			\end{fmfgraph*} 
			}}  \hskip 1.5cm + \hskip -0.5cm
	\parbox{\myll}{\centering{  
			\begin{fmfgraph*}(2.5,2) 
			\fmfright{o1,R,o2}
                   \fmftop{T}
                   \fmfbottom{B}
			\fmf{phantom, tension=1}{B,vB}  %-------
                    \fmf{phantom, tension=1}{T,vT}  %-------
                   \fmf{plain, left=0.2, tension=0.4}{vT,v3}
                    \fmf{phantom, tension=1}{R,vR}  %------
			\fmf{dashes, left=0.2, tension=0.6}{v3,vR} 
                   \fmf{plain, left=0.2, tension=1.2}{vR,v2}			
	            \fmf{dashes, right=0.2, tension=0.4}{vB,v2}           
		     \fmf{plain, tension=0.8}{o1,v2}
			\fmf{plain,tension=0.8}{o2,v3}
                   \Ball{v2,v3}
			\end{fmfgraph*} 
			}} 
\end{align*}
\end{fmffile}	
}{The substitution of a $\l_h$ vertex with two $\m_h$ vertices. }{trick}

Let us start from the computation of $\b^Z_h$ and $\b^\m_h$, see fig.~\ref{conti3d}. We want to prove that assuming the global WIs
\[
Z_h=2\sqrt{2}\m_h  \hskip 2cm 4\sqrt{2}\l_h =\m_h
\]
to be valid, one can proove the same identities between the one--loop contribution to the beta function. The computation of $\b^Z_h$ is trivial and gives:
\[
\b^Z_h = -2\,\l \e^{-\frac{1}{2}}\m_h^2\,\b^{3d}_2
\]
with
\[  
\b^{3d}_n = \frac{1}{(2\pi)^4}\int dk_0 d^3\kk \,\frac{f_0(k_0^2 +\kk^2)}{(k_0^2 +\kk^2)^n}  
\]
Regarding the beta function of $\m_h$ the ``trick'' is to note that the sum of the three third order diagrams gives 
\[
\b_h^{\m, \text{3rd}} & = 4\,\l \e^{-1} \m_h^3 \,\frac{1}{(2\pi)^4}\int d^4k\,
g^{(h)}_{tt}(k)\,\lft[g^{(h)}_{tt}(k) g^{(h)}_{tt}(k) + g^{(h)}_{tl}(k) g^{(h)}_{lt}(k) +2 \bigl(g^{(h)}_{tl}(k)\bigr)^2  \rgt] \non \\
& = 4\,\l \e^{-1} \m_h^3 \,\frac{1}{(2\pi)^4}\int d^4k\,
g^{(h)}_{tt}(k)\,\lft[g^{(h)}_{tt}(k) g^{(h)}_{tt}(k) + \bigl(g^{(h)}_{tl}(k)\bigr)^2  \rgt] \non \\
& =  4\,\l \e^{-1}\m_h^3 \,\frac{1}{(2\pi)^4}\int d^4k\,\frac{g^{(h)}_{tt}(k)}{\DD_h(k)}
\]
Using that $\DD_h(k)\simeq \e^{-1}Z_h (k_0^2 +\e \kk^2)$  one obtains
\[
\b_h^{\m, \text{3rd}} & =\sqrt{2}\,\l \e^{-\frac{1}{2}}\m_h^2 \,\b^{3d}_2
\]
The contribution to the beta function of $\m_h$ coming from the diagram of the second order is
\[
\b_h^{\m, \text{2nd}} & = -12\,\l \e^{-\frac{1}{2}}\, \l_h \m_h \,\b^{3d}_2 = 
\]
Then  
\[
2\sqrt{2}\b^\m_h &= -2\l \e^{-\frac{1}{2}}\,\b^{3d}_2  \lft(12 \sqrt{2} \l_h \m_h  -2 \m_h^2 \rgt) \non \\
\b^Z_h &= -2\l \e^{-\frac{1}{2}}\,\b^{3d}_2\, \m_h^2
\]
which are equal if $4 \sqrt{2} \l_h=\m_h$. One can also note that the substitution of a vertex of type $\l_h$ with two vertices of type $\m_h$ see fig.~\ref{trick} corresponds to the following substitutions
\[
6 \l_h \bigl(g_{tt}^{(h)}(k)\bigr)^2 \quad &\arr \qquad 4\m_h^2 \frac{g_{tt}^{(h)}(k) }{\DD_h(k)} = 2^3 \l_h \, \bigl(g_{tt}^{(h)}(k)\bigr)^2 \non \\
6 \l_h g_{lt}^{(h)}(k) g_{tl}^{(h)}(k)\quad &\arr \qquad 4\m_h^2 \frac{g_{ll}^{(h)}(k) }{\DD_h(k)} =2^3 \l_h  \, g_{ll}^{(h)}(k) g_{tt}^{(h)}(k)\non \\
6 \l_h g_{tt}^{(h)}(k) g_{tl}^{(h)}(k)\quad &\arr \qquad 4\m_h^2 \frac{g_{tl}^{(h)}(k) }{\DD_h(k)} =2^3 \l_h  \,g_{tl}^{(h)}(k) g_{tt}^{(h)}(k)
\]
By applying the substitution on the first line we can calculate quickly the beta function of $\m_h$ which turns to be
\[
\b_h^\m & = 2\l \e^{-\frac{1}{2}} \lft[ -\frac{1}{2}\,2\,6 +\frac{1}{3!}\,3\,2^3 \rgt] \l_h \m_h\,\b_2^{3d} \non \\
& = -4\,\l \e^{-\frac{1}{2}} \l_h \m_h\,\b_2^{3d}
\]
The beta function for $\l_h$ can be calculated with the same ideas. Denoting with $\b^{\l, \text{2nd}}_h$, $\b^{\l, \text{3rd}}_h$ and $\b^{\l, \text{4th}}_h$ the contribution to the beta function of $\l_h$ coming from the diagrams of second, third and fourth order respectively, see fig.~\ref{conti3d_2}, one finds:
\[
\b^{\l, \text{2nd}}_h &= -36 \l \e^{-\frac{1}{2}} \l^2_h \b_2^{3d} \non \\
\b^{\l, \text{3rd}}_h &=  +48 \l \e^{-\frac{1}{2}} \l^2_h \b_2^{3d} \non \\
\b^{\l, \text{4th}}_h &= - 16 \l \e^{-\frac{1}{2}} \l^2_h \b_2^{3d}
\]
so that 
\[
\b_h^\l = - 4 \l \e^{-\frac{1}{2}} \l^2_h \b_2^{3d}
\]
Then $\b^Z_h = 16 \b^\l_h$ and also the global WI $Z_h=16\l_h$ is proved at the one--loop level.

\feyn{
\begin{fmffile}{feyn-TESI/conti3d_2}
 \unitlength = 0.8cm
\def\myl#1{2.7cm}
\def\myll#1{2.7cm}
\begin{align*}
	 \b_h^\l & = \;
	\parbox{\myll}{\centering{ \vskip 0.5cm
			\begin{fmfgraph*}(2.8,2) 
			\fmfleft{i1,i2}
			\fmfright{o1,o2}
			\fmftop{t}
			\fmfbottom{b}
			\fmf{phantom, tension=1.8}{t,v3}    %---
			\fmf{phantom, tension=1.8}{b,v4}    %---
            \fmf{plain, tension=1.5}{i1,v1,i2}			
			\fmf{plain,left=0.3}{v1,v3}
			\fmf{plain,left=0.3}{v3,v2}			
			\fmf{plain,right=0.3}{v1,v4}
			\fmf{plain,right=0.3}{v4,v2}
			\fmf{plain, tension=1.5}{o1,v2,o2}
		\Ball{v1,v2}
		\end{fmfgraph*} \\
$-\frac{1}{2}\,\binom{4}{2}^2\,2$
			}} 
			+\;
	\parbox{\myll}{\centering{  \vskip 0.5cm
			\begin{fmfgraph*}(2.8,2) 
			\fmfleft{i1,i2}
			\fmfright{o1,R,o2}
                   \fmftop{T}
                   \fmfbottom{B}
			\fmf{phantom, tension=1.8}{B,vB}  %-------
                    \fmf{phantom, tension=1.8}{T,vT}  %-------
			\fmf{plain,tension=1.5}{i1,v1,i2}
			\fmf{plain, left=0.3, tension=1.2}{v1,vT}
                   \fmf{plain, left=0.2, tension=0.4}{vT,v3}
                    \fmf{phantom, tension=1}{R,vR}  %------
			\fmf{dashes, left=0.3, tension=1}{v3,vR}
                   \fmf{dashes, left=0.3, tension=0.4}{vR,v2}			
                  \fmf{plain, right=0.3, tension =1.4}{v1,vB}
	            \fmf{plain, right=0.2, tension=0.4}{vB,v2}           
		     \fmf{plain, tension=0.8}{o1,v2}
			\fmf{plain,tension=0.8}{o2,v3}
                   \Ball{v1,v2,v3}
			\end{fmfgraph*}  \\
$\frac{1}{3!}\,3\,\binom{4}{2}\,2^3$
			}}  
+\;
	\parbox{\myll}{\centering{ \vskip 0.5cm
			\begin{fmfgraph*}(2.8,2) 
			\fmfleft{i1,i2}
			\fmfright{o1,R,o2}
                   \fmftop{T}
                   \fmfbottom{B}
			\fmf{phantom, tension=1.8}{B,vB}  %-------
                    \fmf{phantom, tension=1.8}{T,vT}  %-------
			\fmf{plain,tension=1.5}{i1,v1,i2}
			\fmf{plain, left=0.3, tension=1.2}{v1,vT}
                   \fmf{dashes, left=0.2, tension=0.4}{vT,v3}
                    \fmf{phantom, tension=1}{R,vR}  %------
			\fmf{plain, left=0.3, tension=1}{v3,vR}
                   \fmf{plain, left=0.3, tension=0.4}{vR,v2}		
                  \fmf{plain, right=0.3, tension =1.4}{v1,vB}
	            \fmf{dashes, right=0.2, tension=0.4}{vB,v2}           
	             \fmf{plain, tension=0.8}{o1,v2}
			\fmf{plain,tension=0.8}{o2,v3}
                   \Ball{v1,v2,v3}
			\end{fmfgraph*}    \\
$\frac{1}{3!}\,3\,\binom{4}{2}\,2^3$
			}}  +\;
	\parbox{\myll}{\centering{  \vskip 0.5cm
			\begin{fmfgraph*}(2.8,2) 
			\fmfleft{i1,i2}
			\fmfright{o1,R,o2}
                   \fmftop{T}
                   \fmfbottom{B}
			\fmf{phantom, tension=1.8}{B,vB}  %-------
                    \fmf{phantom, tension=1.8}{T,vT}  %-------
			\fmf{plain,tension=1.5}{i1,v1,i2}
			\fmf{plain, left=0.3, tension=1.2}{v1,vT}
                   \fmf{dashes, left=0.2, tension=0.4}{vT,v3}
                    \fmf{phantom, tension=1}{R,vR}  %------
			\fmf{plain, left=0.3, tension=1}{v3,vR}
                   \fmf{dashes, left=0.3, tension=0.4}{vR,v2}			
                  \fmf{plain, right=0.3, tension =1.4}{v1,vB}
	            \fmf{plain, right=0.2, tension=0.4}{vB,v2}           
			\fmf{plain, tension=0.8}{o1,v2}
			\fmf{plain,tension=0.8}{o2,v3}
                   \Ball{v1,v2,v3}
			\end{fmfgraph*}    \\
$\frac{1}{3!}\,3\,\binom{4}{2}\,2^4$
			}}  \\[3pt]
%----------------------------------------------------------------------quarto ordine
&  + \;
	\parbox{\myl}{\centering{ \vskip 0.5cm
			\begin{fmfgraph*}(2.8,2) 
			\fmfleft{i1,i2}
			\fmfright{o1,o2}
			\fmf{plain}{i1,v1}
			\fmf{plain}{i2,v2}
                   \fmf{plain}{v3,o1}
			\fmf{plain}{v4,o2}
                   %linee centrali
                 \fmf{plain, tension=0.6}{v1,vb}
                   \fmf{plain, tension=0.6}{vb,v3}
	             \fmf{plain, tension=0.6}{v2,vt}
                    \fmf{plain, tension=0.6}{vt,v4}
			\fmf{dashes, tension=0.4}{v1,v2}
			\fmf{dashes, tension=0.4}{v3,v4}
			\Ball{v1,v2,v3,v4}
			\end{fmfgraph*}  \\
$-\frac{1}{4!}\,\binom{4}{2}\,2^5$
			}} 
+\; \parbox{\myl}{\centering{
			\begin{fmfgraph*}(2.8,2) 
			\fmfleft{i1,i2}
			\fmfright{o1,o2}
			\fmf{plain}{i1,v1}
			\fmf{plain}{i2,v2}
                   \fmf{plain}{v3,o1}
			\fmf{plain}{v4,o2}
                   %linee centrali
                   \fmf{dashes, tension=0.6}{v1,vb}
                   \fmf{plain, tension=0.6}{vb,v3}
	             \fmf{dashes, tension=0.6}{v2,vt}
                    \fmf{plain, tension=0.6}{vt,v4}
			\fmf{plain, tension=0.4}{v1,v2}
			\fmf{dashes, tension=0.4}{v3,v4}
			\Ball{v1,v2,v3,v4}
			\end{fmfgraph*}  
			}} +
\; \parbox{\myl}{\centering{
			\begin{fmfgraph*}(2.8,2) 
			\fmfleft{i1,i2}
			\fmfright{o1,o2}
			\fmf{plain}{i1,v1}
			\fmf{plain}{i2,v2}
                   \fmf{plain}{v3,o1}
			\fmf{plain}{v4,o2}
                   %linee centrali
                   \fmf{dashes, tension=0.6}{v1,vb}
                   \fmf{plain, tension=0.6}{vb,v3}
	             \fmf{plain, tension=0.6}{v2,vt}
                    \fmf{plain, tension=0.6}{vt,v4}
			\fmf{plain, tension=0.8}{v1,vl}
                   \fmf{dashes, tension=0.8}{vl,v2}
			\fmf{dashes, tension=0.4}{v3,v4}
			\Ball{v1,v2,v3,v4}
			\end{fmfgraph*}  
			}} + \quad \ldots
\end{align*}
\end{fmffile}	
}{ One--loop beta functions for $\l_h$, d=3. There are six different diagrams of fourth order, obtained by substituting in the third order diagrams the $\l_h$ vertex with two $\m_h$ vertices.}{conti3d_2}

\pagina

%\section{Global WIs in the asymptotic region}

\feyn{
\begin{fmffile}{feyn-TESI/GWI_2d}
 \unitlength = 0.8cm
\def\myl#1{2.3cm}
\begin{align*}
\parbox{\myl}{\centering{ \vskip 0.8cm
			\begin{fmfgraph*}(2.2,1.3) 
			\fmfleft{i1}
			\fmfright{o1}
			\fmf{dashes, tension=1.2}{i1,v1}
			\fmf{plain,left=0.7, tension=0.4}{v1,v2}
			\fmf{plain,right=0.7, tension=0.4}{v1,v2}
			\fmf{dashes, tension=1.2}{o1,v2}
			\Ball{v1,v2}
			\end{fmfgraph*}    \\[3pt]
                   $ -\frac{1}{2!} \cdot 2 $
			}} 
& = \quad 8 \parbox{\myl}{\centering{ \vskip 0.8cm
			\begin{fmfgraph*}(2.2,1.3)
			\fmfleft{i1,i2}
			\fmfright{o1,o2}
			\fmf{plain, tension=1.2}{i1,v1,i2}
			\fmf{plain,left=0.7, tension=0.4}{v1,v2}
			\fmf{plain,right=0.7, tension=0.4}{v1,v2}
			\fmf{plain, tension=1.2}{o1,v2,o2}
			\Ball{v1,v2}
			\end{fmfgraph*}  
 			  \\[3pt]    $ -\frac{1}{2!}\,\binom{4}{2}^2 \, 2 $
			}} +\; 8
 \parbox{\myl}{\centering{ \vskip 0.8cm
		\begin{fmfgraph*}(2.2,1.3) 
			\fmfleft{i1,i2}
			\fmfright{o1,o2}
			\fmf{plain, tension=1.5}{i1,v1,i2}
                   \fmf{plain}{o1,v2}
                   \fmf{plain}{o2,v3}
			\fmf{plain, right=0.4, tension=0.6}{v1,v2}
                    \fmf{plain, left=0.4, tension=0.6}{v1,v3}
                   \fmf{dashes, right=0.4, tension=0.2}{v2,v3}
                    \Ball{v1,v2,v3}
		\end{fmfgraph*}
            \\[3pt] $ \frac{1}{3!}\,3\,\binom{4}{2}\,\binom{2}{1}^2\,2 $
	}} +\; 8
\parbox{\myl}{\centering{  \vskip 0.8cm
			\begin{fmfgraph*}(2.2,1.3) 
			\fmfleft{i1,i2}
			\fmfright{o1,o2}
			\fmf{plain, tension=1.8}{i1,v1}
                    \fmf{plain, tension=1.8}{i2,v2}
                    \fmf{plain}{v1,v2}
                    \fmf{dashes}{v2,v4}
                    \fmf{dashes}{v1,v3}
                    \fmf{plain}{v4,v3}
			%\fmf{dashes,left=0.7, tension=0.4}{v1,v2}
			%\fmf{dashes,right=0.7, tension=0.4}{v1,v2}
			\fmf{plain, tension=1.8}{o1,v3}
	             \fmf{plain, tension=1.8}{v4,o2}
			\Ball{v1,v2,v3,v4}
			\end{fmfgraph*}  
                    \\[3pt] $ -\frac{1}{4!}\,\binom{4}{2}\,2 $
			}} \\[12pt]
%------------------------------------------------------------------------
\parbox{\myl}{\centering{ \vskip 0.4cm
		\begin{fmfgraph*}(2,2)
			\fmfleft{i1}
			\fmfright{o1}
			\fmf{plain}{v1,v1}
                    \fmf{dashes}{i1,v1,o1}
                   \Ball{v1}
		\end{fmfgraph*}
             \\[-3pt] $1$
		}}
		& = \quad 8
		\parbox{\myl}{\centering{  \vskip 0.8cm
             \vskip -0.5cm
		\begin{fmfgraph*}(2,2)
                   \fmfright{i1}
                   \fmfleft{i2}
			\fmfbottom{o1,o2}
			\fmf{plain, tension=1.5}{i1,v1,i2}
                   \fmf{plain}{o1,v1,o2}
			\fmf{plain}{v1,v1}
                   \Ball{v1}
		\end{fmfgraph*}
 		\\[3pt] $  \binom{6}{2}$
	}}		
		+\; 8
	\parbox{\myl}{\centering{ \vskip 0.8cm
		\begin{fmfgraph*}(2.2,1.3) 
			\fmfleft{i1,i2,i3}
			\fmfright{o1}
			\fmf{plain, tension=1.2}{i1,v1,i2}
                   \fmf{plain, tension=1.2}{i3,v1}
                   \fmf{plain, tension=1.8}{v2,o1}
			\fmf{dashes, right=0.7, tension=0.6}{v1,v2}
                    \fmf{plain, left=0.7, tension=0.6}{v1,v2}
                    \Ball{v1,v2}
		\end{fmfgraph*}
             \\[3pt] $  -\frac{1}{2!}\,2\,\binom{5}{1}\,\binom{2}{1}\,2$
	}}		+\; 8\;\parbox{\myl}{\centering{ \vskip 0.8cm
		\begin{fmfgraph*}(2.2,1.3) 
			\fmfleft{i1,i2}
			\fmfright{o1,o2}
			\fmf{plain, tension=1.5}{i1,v1,i2}
                   \fmf{plain}{o1,v2}
                   \fmf{plain}{o2,v3}
			\fmf{dashes, right=0.4, tension=0.6}{v1,v2}
                    \fmf{dashes, left=0.4, tension=0.6}{v1,v3}
                   \fmf{plain, right=0.4, tension=0.2}{v2,v3}
                    \Ball{v1,v2,v3}
		\end{fmfgraph*}
                \\[3pt] $  \frac{1}{3!}\,3\,\binom{2}{1}^2\,2$
	}}
	  \\[12pt]
%---------------------------------------------------------------------------
\parbox{\myl}{\centering{ \vskip 0.8cm
			\begin{fmfgraph*}(2.2,1.3) 
			\fmfleft{i1}
			\fmfright{o1}
			\fmf{dashes, tension=1.2}{i1,v1}
			\fmf{dashes,left=0.7, tension=0.4}{v1,v2}
			\fmf{dashes,right=0.7, tension=0.4}{v1,v2}
			\fmf{dashes, tension=1.2}{o1,v2}
			\Ball{v1,v2}
			\end{fmfgraph*}  
  			\\[3pt] $  -\frac{1}{2!}\,\binom{3}{2}^2\,2$
			}}  & = \quad 8
	\parbox{\myl}{\centering{  \vskip 0.8cm
			\begin{fmfgraph*}(2.2,1.3) 
			\fmfleft{i1,i2}
			\fmfright{o1,o2}
			\fmf{plain, tension=1.5}{i1,v1,i2}
			\fmf{dashes,left=0.7, tension=0.4}{v1,v2}
			\fmf{dashes,right=0.7, tension=0.4}{v1,v2}
			\fmf{plain}{o1,v2,o2}
			\Ball{v1,v2}
			\end{fmfgraph*}  
			\\[3pt] $  -\frac{1}{2!}\,2$
			}} 
%------------------------------------------------------------------------
\end{align*}
\end{fmffile}	
}{{\bf Asymptotic global WI $Z_h=16\l_h$, d=2.} The sum of the diagrams on the l.h.s. of the figure represent the leading order beta function of $Z_h/2$ in the asymptotic region $h \arr \io$; the sum of the diagrams on the r.h.s. side is the leading order beta function of $\l_h$. Using inductively that identities \eqref{GWI_m}, \eqref{GWI_Z} and \eqref{new_vertices} one can check that the relation $Z_h=16\l_h$ is satisfied; in particular it is separately verified among the diagrams with the same number of plain (\ie transverse) propagators. }{GWI_2d}

\section{The leading order beta function in $2d$}  \label{leading_2d}

\subsection{Role of the global WIs}

In the two dimensional case three extra global WIs are needed, in order to relate the new ``effectively marginal'' couplings $\m'_h$, $\l'_h$ and $\o_h$ to $\l_{6,h}$. These are:
\[
& \o_h = 6 \sqrt{2}\, \g^{\frac{h}{2}}\,\l_{6,h}  \\[6pt]
& 2 \sqrt{2}\, \g^{-\frac{h}{2}}\o_h -\g^{-h}\l'_h = -2\g^h \l_h  \\[6pt]
& 3\, \g^{-\frac{3}{2}h}\m'_h -2\sqrt{2} \g^{-h}\l'_h = 2 \g^{\frac{h}{2}}\m_h
\]
Assuming $\l_h$ to have a fixes point $\l_*$, in the asymptotic region \mbox{$h \arr -\io$}, since $\g^h\l_h$ and $ \g^{\frac{h}{2}}\m_h$ are going to zero,  the previous identities state that:
\[ \label{new_vertices}
& \o_h = 6 \sqrt{2}\, \g^{\frac{h}{2}}\,\l_{6,h} \non \\[6pt]
& \l'_h = 24\,\g^h \l_{6,h} \non \\[6pt]
& \m'_h = 16 \sqrt{2}\, \g^{\frac{3}{2}h} \l_{6,h}
\]
We are now ready to study the leading order flow equation for $\l_h$ and $\l_{6,h}$. 
%and prove the assumption on their behavior. 
Under the previous assumption on $\l_h$ the order in the small parameters $\l$ and $\e=\l \r_0 R_0^2$ in the asymptotic region is given by:
\[ \label{asym_2d}
\tl{C}_{2d}(P_v; \l, \e)  =   \r_0 R_0^{-2} &\, \l^{1-\frac{1}{2}(n_l^\txe+n_t^\txe)}\,\e^{-2 +\frac{3}{2}n_l^\txe+ \frac{1}{2}n_t^\txe + \frac{1}{2}n_\dpr^\txe} 
\non \\[6pt]
&  \, (\l \l_*)^{m_{4}+\frac{1}{2}m_{3}}\,(\e \n_*)^{m_{2}}\,  (\l \e^{-1} \,\l \l_{6})^{m_6}\non \\[6pt]
& \lft(\l \e^{-1} \,\frac{\l_{6}}{\l_*}\rgt)^{m'_4}\,\lft( \l \e^{-2} \, \frac{\l^2_{6}}{\l_*^3}\rgt)^{\frac{1}{2}m'_3} \,\lft( \e \, \frac{\l^2_{6}}{\l_*}\rgt)^{\frac{1}{2}m'_5}
\non \\[6pt]
& \lft(\e^{-1}\g^h \l_* \rgt)^{\frac{1}{2}(m_3 + m_5 + 2m'_4 +3m'_3)- n_{ll}}
\]
The leading order diagrams are the one loop diagram (remind that the loop number is equal to $L=m_4 + m'_4 + (m_3+m'_3)/2 +3m_5/2 +2m_6 -n^\txe/2 +1$) where all the internal dashed legs are contracted among them, in order to minimize the exponent of $\g^h$, on the last line of \eqref{asym_2d}.

\feyn{
\begin{fmffile}{feyn-TESI/zeta2d}
 \unitlength = 0.8cm
\def\myl#1{2.5cm}
\begin{align*}
\parbox{\myl}{\centering{
		\begin{fmfgraph*}(2,1.1)
			\fmfleft{i1}
			\fmfright{o1}
			\fmf{dashes}{i1,v1,o1}
			\fmfv{label={$Z_{h-1}/2$},label.angle=90,label.dist=-0.4w}{v1} 
                   \bBall{v1}
		\end{fmfgraph*}
		}}
		& =
		\parbox{\myl}{\centering{
		\begin{fmfgraph*}(2,1)
			\fmfleft{i1}
			\fmfright{o1}
			\fmf{dashes}{i1,v1,o1}
			\fmfv{label={$Z_h/2$},label.angle=-90}{v1} 
                   \Ball{v1}
		\end{fmfgraph*}
	}}			-\;
\parbox{\myl}{\centering{
		\begin{fmfgraph*}(2.2,1.5) \fmfkeep{z1}
			\fmfleft{i1}
			\fmfright{o1}
			\fmf{dashes}{i1,v1,o1}
                    \fmf{plain}{v1,v1}
                    \Ball{v1}
		\end{fmfgraph*}
	}}			+\;
	\parbox{\myl}{\centering{
			\begin{fmfgraph*}(2.5,1.4)  \fmfkeep{z2}
			\fmfleft{i1}
			\fmfright{o1}
			\fmf{dashes, tension=1.5}{i1,v1}
			\fmf{plain,left=0.7, tension=0.4}{v1,v2}
			\fmf{plain,right=0.7, tension=0.4}{v1,v2}
			\fmf{dashes}{o1,v2}
			\Ball{v1,v2}
			\end{fmfgraph*}  
			}}   
	+\;
	\parbox{\myl}{\centering{
			\begin{fmfgraph*}(2.5,1.4) \fmfkeep{z3}
			\fmfleft{i1}
			\fmfright{o1}
			\fmf{dashes, tension=1.5}{i1,v1}
			\fmf{dashes,left=0.7, tension=0.4}{v1,v2}
			\fmf{dashes,right=0.7, tension=0.4}{v1,v2}
			\fmf{dashes}{o1,v2}
			\Ball{v1,v2}
			\end{fmfgraph*}  
			}}  
\end{align*}
\end{fmffile}	
}{{\bf Beta function for $Z_h$, d=2.} }{zeta2d}

\subsection{Beta function for $\l_h$ }

In spite of studying directly the leading order beta function $\b_h^\l$ for $\l_h$ it results more convenient to look at the beta function $\b_h^Z$ of $Z_h$, which a smaller number of diagram  contribute to, and then use the global WI
\[
Z_h = 16 \g^h \l_h
\]
(represented in fig.~\ref{GWI_2d}) to obtain the flow equation for $\l_h$: 
\[
\g^{h-1}\l_h-1 = \g^h\l_h + \g^h\b_h^\l
\]
with $\g^h\b^\l_h=\b^Z_h/16$. The leading order diagrams contributing to the flow of $Z_h$ according to \eqref{asym_2d} are showed in fig. \ref{zeta2d}. In the following we will denote as:
\[  \label{beta_def}
\b^{2d}_n = \frac{1}{(2\pi)^3}\int dk_0 d^2\kk \,\frac{f_0(k_0^2 +\kk^2)}{(k_0^2 +\kk^2)^n}  
\]
The behavior of the asymptotic propagator is
\[
g^{(h)}_{\a \a'}(x) & = \frac{1}{(2\pi)^3}\int dk_0 d^2\kk\, f_h(k_0^2 +\e\kk^2)\, g^{(h)}_{\a \a'}(k) \non \\[6pt]
g^{(h)}_{\a \a'}(k) & = (\r_0 R_0^2)^{-1} \,
\left(\begin{array}{cc}
Z_h   & - \frac{k_{0}}{k_0^2 +\e\kk^2}\\
\frac{k_{0}}{k_0^2 +\e\kk^2} & \quad \frac{\e}{k_0^2 +\e\kk^2}
\end{array}\right)
\]
where the first line and row correspond to $a=l$ and the second line and row correspond to $\a'=t$. The values of the diagrams of picture \ref{zeta2d} are calculated below. 
%The combinatorial factors take into account: 
%\begin{enumerate}[a.]
%\item the coefficient of the truncated expectations, which is $(-1)^{n+1}/n!$ if we are contracting $n$ vertices;
%\item the different possibilities of choosing different vertices between those present in the expectations, giving rise to the same graph;
%\item for each vertex, the different possibilities of choosing the external lines;
%\item the different possibility of contracting the internal lines.
%\end{enumerate}
%\vskip 0.5cm
%Below we report their explicit calculation:
\feynH{
\vskip -0.5cm
 \unitlength = 0.8cm
\def\myl#1{2.5cm}
\begin{align*}
\parbox{\myl}{\centering{\fmfreuse{z1}}}
 & = \; \l \e^{-1}\,\b^{2d}_1\,\l'_h = \; 24\, \l \e^{-1}\,\b_1^{2d}\g^h\,\l_{6,h}   \non \\[6pt]
\parbox{\myl}{\centering{\fmfreuse{z2}}} &= \; -\frac{1}{2!}\,2\,\l\,\b^{2d}_2\, \m_h^2 = \; -32\,\l\,\b^{2d}_2 \g^h\,\l_h^2  \non \\[6pt]
 \parbox{\myl}{\centering{\fmfreuse{z3}}} &=\; -\frac{1}{2!}\,2\cdot 3^2\,\l\e^{-2}\,\b^{2d}_0\, \lft(\frac{\m'_h}{Z_h}\rgt)^2 =\; -18\,\l\e^{-2}\,\b^{2d}_0\,\g^h\,\lft(\frac{\l_{6,h}}{\l_h}\rgt)^2
\end{align*}
}
Collecting all the terms together we obtain
\[ \label{lambda_1}
\g^{-1}\l_{h-1} - \l_h= - \l \,(1-\g^{-1})\,\lft( 4\,\tl{\b}^{2d}_2\,\l_h^2 \;-\; 3\,\tl{\b}_1^{2d}\,\e^{-1}\l_{6,h} \;+\; \frac{9}{4}\,\tl{\b}^{2d}_0 \lft(\frac{\l_{6,h}}{\e \l_h}\rgt)^2 \rgt)
\]
with $(1-\g^{-1}) \tl{\b}_n^{2d}=\b_n^{2d}$.  By substituting to $f_0(k)$  the characteristic function of the set $$\lft[\sqrt{\tfrac{2}{1+\g^2}}\,\g^{-1}, \sqrt{\tfrac{2}{1+\g^2}}\,\g \rgt]$$ we obtain
%\[ 
%&\b^{2d}_0 =\frac{1}{2\pi^2}\,\sqrt{\tfrac{2}{\g^2+1}} \,\tfrac{2(\g^3-\g^{-3})}{3(\g^2+1)}
%\non \\[6pt]
%
%& \b^{2d}_1= \frac{1}{2\pi^2}\,\sqrt{\tfrac{2}{\g^2+1}} \,\g^{-1}(\g^2-1)
%\non \\[6pt]
%
%&\b^{2d}_2 = \frac{1}{2\pi^2}\,\sqrt{\tfrac{2}{\g^2+1}} \,\tfrac{\g^{2}+1}{2}(\g-\g^{-1})
%\non \\
%\]  
\[ 
&\tl{\b}^{2d}_0 = \frac{1}{2\pi^2}\,\sqrt{\tfrac{2}{\g^2+1}} \,\tfrac{2}{3} \tfrac{(\g^2+\g+1)(\g^3+1)}{\g^2(\g^2+1)} \non \\[6pt] 
& \tl{\b}^{2d}_1 =  \frac{1}{2\pi^2}\,\sqrt{\tfrac{2}{\g^2+1}} \,(\g+1)
\non \\[6pt]
&\tl{\b}^{2d}_2  = \frac{1}{2\pi^2}\,\sqrt{\tfrac{2}{\g^2+1}} \,\tfrac{\g^{2}+1}{2}\,(\g+1)
\] 
where $\lim_{\g \arr 1}\tl{\b}_n^{2d}=1/\pi^2$ for each $n=0,1,2$.

\feyn{
\begin{fmffile}{feyn-TESI/muprimo2d}
 \unitlength = 0.8cm
\def\myl#1{2.3cm}
\begin{align*}
\parbox{\myl}{\centering{
		\begin{fmfgraph*}(2,1.5)
			\fmfleft{i1}
			\fmfright{o1,o2}
			\fmf{dashes, tension=1.5}{i1,v1}
                    \fmf{dashes}{o1,v1,o2}
			\fmfv{label={$\m'_{h-1}$},label.angle=90,label.dist=-0.4w}{v1} 
                   \bBall{v1}
		\end{fmfgraph*}
		}}
		& =
		\parbox{\myl}{\centering{
		\begin{fmfgraph*}(2,1.5)
                   \fmfleft{i1}
			\fmfright{o1,o2}
			\fmf{dashes, tension=1.5}{i1,v1}
                   \fmf{dashes}{o1,v1,o2}
			\fmfv{label={$\m'_h$},label.angle=-90}{v1} 
                   \Ball{v1}
		\end{fmfgraph*}
	}}			-\;
\parbox{\myl}{\centering{
		\begin{fmfgraph*}(2.2,1.5) \fmfkeep{m1}
			\fmfleft{i1}
			\fmfright{o1,o2}
			\fmf{dashes, tension=1.5}{i1,v1}
                   \fmf{dashes}{o1,v2}
                   \fmf{dashes}{o2,v3}
			\fmf{plain, right=0.4, tension=0.6}{v1,v2}
                    \fmf{plain, left=0.4, tension=0.6}{v1,v3}
                   \fmf{plain, right=0.4, tension=0.2}{v2,v3}
                    \Ball{v1,v2,v3}
		\end{fmfgraph*}
	}}			+\;
	\parbox{\myl}{\centering{
		\begin{fmfgraph*}(2.2,1.5) \fmfkeep{m2}
			\fmfleft{i1}
			\fmfright{o1,o2}
			\fmf{dashes, tension=1.5}{i1,v1}
                   \fmf{dashes}{o1,v2}
                   \fmf{dashes}{o2,v3}
			\fmf{dashes, right=0.4, tension=0.6}{v1,v2}
                    \fmf{dashes, left=0.4, tension=0.6}{v1,v3}
                   \fmf{dashes, right=0.4, tension=0.2}{v2,v3}
                    \Ball{v1,v2,v3}
		\end{fmfgraph*}
	}}			+\;
	\parbox{\myl}{\centering{
			\begin{fmfgraph*}(2.2,1.5) \fmfkeep{m3}
			\fmfleft{i1}
			\fmfright{o1,o2}
			\fmf{dashes, tension=1.5}{i1,v1}
			\fmf{plain,left=0.7, tension=0.4}{v1,v2}
			\fmf{plain,right=0.7, tension=0.4}{v1,v2}
			\fmf{dashes}{o1,v2,o2}
			\Ball{v1,v2}
			\end{fmfgraph*}  
			}}  
\end{align*}
\end{fmffile}	
}{{\bf Beta function for $\m'_h$, d=2.} }{muprimo2d}

\pagina

\subsection{Beta function for $\l_{6,h}$}

Similarly at what happens for $\l_h$, the beta function for $\l_{6,h}$ may be rewritten in term of the beta function of $\m'_h$, using the last of \eqref{new_vertices}: 
\[
\l_{6,h-1} = \l_{6,h} +\b_h^{\l_6}
\]
with $\b^{\l_6}_h=\g^{-\frac{3}{2}h} \b^{\m'}_h/(16 \sqrt{2})$. The leading order diagrams contributing to the flow of $\m'_h$ are showed in fig. \ref{muprimo2d}. Their explicit computation is shown below:
\feynH{
\vskip -0.8cm
 \unitlength = 0.8cm
\def\myl#1{2.5cm}
\begin{align*}
\parbox{\myl}{\centering{\fmfreuse{m1}}}
 & = \; +\frac{1}{3!}\cdot 6  \, \l \e \,\b^{2d}_3\,\m^3_h =   
 2^7\sqrt{2}\, \g^{\frac{3}{2}h}\,\b^{2d}_3\,\l \e\,\l^3_h 
\non \\[6pt]
\parbox{\myl}{\centering{\fmfreuse{m2}}} &= \; +\frac{1}{3!}\,3^3\cdot 6\,\l \e^{-2}\,\b^{2d}_0\, \lft(\frac{\m'_h}{Z_h}\rgt)^3 
= \; 3^3 \cdot 2\sqrt{2} \,\g^{\frac{3}{2}h}\,\b^{2d}_0\,\l \e\, \lft(\frac{\l_{6,h}}{\e \l_h}\rgt)^3 
 \non \\[6pt]
\parbox{\myl}{\centering{\fmfreuse{m3}}} &=\; -\frac{1}{2!}\cdot 2 \cdot 2 \, \l \,\b^{2d}_2\,\l'_h\,\m_h =   
-3\cdot 2^6\sqrt{2}\, \g^{\frac{3}{2}h}\,\b^{2d}_2\,\l\,\l_h\,\l_{6,h} 
\end{align*}
}
where again $ \b^{2d}_3= (1-\g^{-1})\,\tl{\b}^{2d}_3$ and 
\[
& \tl{\b}^{2d}_3 = \frac{1}{2\pi^2}\,\sqrt{\tfrac{2}{\g^2+1}} \,\tfrac{(\g^{2}+1)^2}{12}\tfrac{(\g^2+\g+1)(\g^3+1)}{\g^2}
\]
%\[
%&\b^{2d}_3  = \frac{1}{2\pi^2}\,\sqrt{\tfrac{2}{\g^2+1}} \,\tfrac{(\g^{2}+1)^2}{12}(\g^3-\g^{-3})
%\]
The flow equation for $\l_{6,h}$ at leading order turns to be:
\[ \label{l6_1}
\l_{6,h-1} - \l_{6,h} &=  8\,\b^{2d}_3 \l \e\,\l^3_h -  12\,\b^{2d}_2\,\l \l_h\,\l_{6,h} 
+\frac{27}{8}\,\b^{2d}_0\,\l \e\, \lft(\frac{\l_{6,h}}{\e \l_h}\rgt)^3  \non \\[6pt]
& = \l \l_h \biggl[ \, 8\b^{2d}_3\,\e \l_h^2 - \l_{6,h} \lft(12\,\b^{2d}_2 - \frac{27}{8}\,\b^{2d}_0\, \Bigl(\frac{\l_{6,h}}{\e \l^2_h}\Bigr)^2 \rgt) \biggr] 
\]

\pagina

\section{Local Ward Identities}   \label{B.WI-leading}

In this section we will verify at leading order in perturbation theory the local WI's  that we have proved at all orders in chapter \ref{WI}. The computation of WI's at leading order is useful to understand on which symmetries the WI's are based on. Of course the leading order computations we are describing in the following (which correspond to the one--loop computations in the region $h \leq \bh$ we are interested in) are not sufficient. If we could not exclude that similar equalities or cancellations take place at all orders in perturbation theory there would always be the possibility that higher order produce a completely different behavior. That's why the derivation and study of WI's -- which are relations valid at all orders -- turns to be crucial in our work.

%-----------------------------------------------------------------------------------E_h/Z_h leading order
\feyn{
\begin{fmffile}{feyn-TESI/WI_1}
\unitlength = 1 cm  
\def\myl#1{2.5cm}
\[
   \beta^{Z}_h  = \; 
 \parbox{\myl}{\centering{
	\begin{fmfgraph*}(2.8,2)
			\fmfleft{i1}
			\fmfright{o1}
			\fmftop{t}
			\fmfbottom{b}
			\fmf{phantom, tension=1.8}{t,v3}    %---
			\fmf{phantom, tension=1.8}{b,v4}    %---
                   \fmf{dashes, tension=1.5}{i1,v1}			
			\fmf{plain,left=0.3}{v1,v3}
			\fmf{plain,left=0.3}{v3,v2}			
			\fmf{plain,right=0.3}{v1,v4}
			\fmf{plain,right=0.3}{v4,v2}
		       \fmf{dashes, tension=1.5}{o1,v2}
			\fmfdot{v2}
			\Ball{v1,v2}	%, 
		\end{fmfgraph*}
		}}  \qquad \quad
   \beta^{E}_h  = \;
 \parbox{\myl}{\centering{
	\begin{fmfgraph*}(2.8,2)
			\fmfleft{i1}
			\fmfright{o1}
			\fmftop{t}
			\fmfbottom{b}
			\fmf{phantom, tension=1.8}{t,v3}    %---
			\fmf{phantom, tension=1.8}{b,v4}    %---
                   \fmf{plain, tension=1.5}{i1,v1}			
			\fmf{plain,left=0.3}{v1,v3}
			\fmf{plain,left=0.3}{v3,v2}			
			\fmf{dashes,right=0.3}{v1,v4}
			\fmf{plain,right=0.3}{v4,v2}
		       \fmf{dashes, tension=1.5, label= $\dpr_0$}{o1,v2}
			\Ball{v1,v2}
		\end{fmfgraph*}
		}} \non
\]
\end{fmffile}
}{The beta functions for $Z_h$ and $E_h$ at leading order. 
}{WI_1}
%------------------------------------------------------------------------------------------------

\subsection{Local WI for $E_h/ Z_h$}  

The beta functions for $Z_h$ and $E_h$ at leading order are shown in fig. \ref{WI_1}. We see that they differ only for the change of a $g_{tt}^{(h)}$ propagator with a $g_{lt}^{(h)}$ propagator:
\[
\b^Z_h & = \m^2_h\, \frac{1}{(2\pi)^d}\int d^{d+1}k \,\lft(g_{tt}^{(h)}(k)\rgt)^2 \non \\ 
\b^E_h & = \m^2_h\, \frac{1}{(2\pi)^d}\int d^{d+1}k \,g_{tt}^{(h)}(k)\,\dpr_{p_0}\lft[ g_{lt}^{(h)}(k+p_0)\rgt]_{p_0=0}
\]
Since
\[
g_{lt}^{(h)}(k) & =g_{tt}^{(h)}(k) \,\frac{E_h\,k_0}{Z_h}
\]
we have
\[
\b^E_h & = \m^2_h\, \frac{1}{(2\pi)^d}\int d^{d+1}k \,g_{tt}^{(h)}(k)\,\lft[ \frac{E_h}{Z_h}g_{tt}^{(h)}(k)  - \frac{2k_0}{D^2_h(k)}\rgt]
\]
where the second term in the r.h.s. of the latter equation is zero for parity reasons. We finally get
\[
\b^E_h & =\frac{E_h}{Z_h}\,\b^Z_h
\]
from which
\[
\frac{E_{h-1} -E_h}{E_h} =\frac{Z_{h-1} -Z_h}{Z_h}
\]
As a consequence
\[
 \frac{E_h}{Z_h}=\frac{E_\bh}{Z_\bh}=\e^{-1}
\]

%\pagina

%\subsection{Leading order computation for $ A_h $}   \label{appB.WI_A}

%\pagina

%----------------------------------------------------------------------------------- WI Bh 
\feyn{
\begin{fmffile}{feyn-TESI/WI_Bh}
\unitlength = 1 cm  
\def\myl#1{3.5cm}
\[
   \beta^{B}_h  = \; 
 \parbox{\myl}{\centering{
	\begin{fmfgraph*}(2.8,2)
			\fmfleft{i1}
			\fmfright{o1}
			\fmftop{t}
			\fmfbottom{b}
			\fmf{phantom, tension=1.8}{t,v3}    %---
			\fmf{phantom, tension=1.8}{b,v4}    %---
                   \fmf{plain, tension=1.5,label=$\dpr_0$, label.dist=-0.4cm}{i1,v1}			
			\fmf{plain,left=0.3}{v1,v3}
			\fmf{plain,left=0.3}{v3,v2}			
			\fmf{dashes,right=0.3}{v1,v4}
			\fmf{dashes,right=0.3}{v4,v2}
		       \fmf{plain, tension=1.5, label=$\dpr_0$, label.dist=0.05cm}{o1,v2}
			\fmfdot{v2}
			\Ball{v1,v2}	%, 
		\end{fmfgraph*}
		}}  +
 \parbox{\myl}{\centering{
	\begin{fmfgraph*}(2.8,2)
			\fmfleft{i1}
			\fmfright{o1}
			\fmftop{t}
			\fmfbottom{b}
			\fmf{phantom, tension=1.8}{t,v3}    %---
			\fmf{phantom, tension=1.8}{b,v4}    %---
                   \fmf{plain, tension=1.5, label= $\dpr_0$, label.dist=-0.4cm}{i1,v1}			
			\fmf{plain,left=0.3}{v1,v3}
			\fmf{dashes,left=0.3}{v3,v2}			
			\fmf{dashes,right=0.3}{v1,v4}
			\fmf{plain,right=0.3}{v4,v2}
		       \fmf{plain, tension=1.5, label= $\dpr_0$, label.dist=0.05cm}{o1,v2}
			\Ball{v1,v2}
		\end{fmfgraph*}
		}} \non
\]
\end{fmffile}
}{The beta functions for $B_h$ at leading order. }{WI_Bh}

\subsection{Leading order computation for $ B_h$}   \label{appB.WI_B}

At the main order in $\lambda$ the beta function for $B_{h}$ is represented in fig.~\ref{WI_Bh}. If we denote with $\beta_{h}^{B,1}$ and $\beta_{h}^{B,2}$ the two contribution in fig.~\ref{WI_Bh} starting from the left, we see that $\beta_{h}^{B,1}$ is apparently not summable over $h$ for $d=3$:
\[
|\beta_{h}^{B,1}| & \leq  c\,|\mu_{h}^{2}|\,\frac{1}{|Z_{h}|}= O \lft( (1+\l \e^{\frac{1}{2}}|h-\bh|)^{-1} \rgt)   \non \\[6pt]
|\beta_{h}^{B,2} |& \leq  c\,|\mu_{h}^{2}|=  O \lft( (1+\l \e^{\frac{1}{2}}|h-\bh|)^{-2} \rgt)
\]
The key to solve the apparent problem is in the fact that to obtain the flow for $B_{h}$ we have first to derive the kernels with respect the external momentum $p_{0}$ and then we can take the limit $p_{0}\rightarrow0$. This procedure gives some ``interferences'' from the boundary which one can not see, of course, if the external momentum is taken equal to zero. When the calculation is performed one discover that the ``bad'' contribution erase, so that the beta function is summable. 

To make the calculation we remind that, by definition, the beta function on scale $h$ has to contain at least one propagator on scale $h$. Since $p_{0}$is small the second propagator - in the case of the graphs in the picture - can be or on scale $h$ or on scale $h+1$,
otherwise the cutoff function do not have common supports: 
\[
f_{h}(k+p_{0})f_{j}(k)=0\quad\text{if}\; j>h+1
\]
So we have three possible cases: both the propagator on scale $h$,
that is $g_{h}(k+p_{0})g_{h}(k)$ or one propagator on scale $h$
and the second on scale $h+1$, that is $g_{h+1}(k+p_{0})g_{h}(k)$
or $g_{h}(k+p_{0})g_{h+1}(k)$. This can be written as:
\[
\left[g_{h}(k+p)+g_{h+1}(k+p)\right]\left[g_{h}(k)+g_{h+1}(k)\right]-g_{h+1}(k+p)g_{h+1}(k)
\]
In the following we will indicate with $T_{1}$ and $T_{2}$
the cutoff functions respectively on scale $h+(h+1)$ and $h$ 
\[
T_{1}(k) & = f_{h+1}(k)+f_{h}(k)=\chi_{h+1}(k)-\chi_{h-1}(k) \non \\[6pt]
T_{2}(k) & =  f_{h}(k)=\chi_{h}(k)-\chi_{h-1}(k)
\]
where 
\[
\chi_{h}(k)=\chi(\gamma^{-2h}(k_{0}^{2}+\kk^{2}))
\]
%
%\noindent \begin{center}
%\includegraphics[height=4cm,bb = 0 0 200 100, draft, type=eps]{cutoffs_Bh.emf}
%\par\end{center}
%
Since we are interested in the lowest order computation we can neglect the dependence on $k$ of the rescaled propagators. Then the expression of the beta function with cutoff $T_{i}$ is the following:
\[
\beta_{h,i}^{B} & =  \mu_{h}^{2}\int\frac{d^{4}k}{(2\pi)^{4}}\partial_{0}^{2}\left[\frac{E_{h}^{2}(k_{0}+p_{0})k_{0}+Z_{h}\left((1+A_{h})\kk^{2}+B_{h}k_{0}^{2}\right)}{\DD_{h}(k)D_{h}(k+p_{0})} \,T_{i}(k+p_{0})T_{i}(k)\right] \non \\[6pt]
 & = \m_{h}^{2} \int\frac{d^{4}k}{(2\pi)^{4}} \dpr_{0}^{2} \left[\frac{E_{h}^{2}p_{0}k_{0}+\DD_{h}(k)}{\DD_{h}(k)D_{h}(k+p_{0})}\, T_{i}(k+p_{0})T_{i}(k)\right] \non \\[6pt]
 & =  \mu_{h}^{2}\int\frac{d^{4}k}{(2\pi)^{4}}\dpr_{0}\left[\frac{E_{h}^{2}k_{0}T_{i}(k+p_{0})T_{i}(k)}{\DD_{h}(k)\,\DD_{h}(k+p_{0})}+\left(E_{h}^{2}p_{0}k_{0}+\DD_{h}(k)\right)\frac{T_{i}(k)}{\DD_{h}(k)}
\dpr_{0} \left(\frac{T_{i}(k+p_{0})}{\DD_{h}(k+p_{0})}\right)\right] \non \\[6pt]
 & =  \m_{h}^{2}\int\frac{d^{4}k}{(2\pi)^{4}} \left[2E_{h}^{2}k_{0}\frac{T_{i}(k)}{\DD_{h}(k)}\dpr{0}\left(\frac{T_{i}(k+p_{0})}{\DD_{h}(k+p_{0})}\right)+T_{i}(k)\partial_{0}^{2}\left(\frac{T_{i}(k+p_{0})}{\DD_{h}(k+p_{0})}\right)\right]
\]
Integrating by part we get 
\[
& 2E_{h}^{2}k_{0}\frac{T_{i}}{\DD_{h}}\partial_{0}\left(\frac{T_{i}}{\DD_{h}}\right)  =  E_{h}^{2}k_{0}\partial_{0}\left[\left(\frac{T_{i}}{\DD_{h}}\right)^{2}\right]=\partial_{0}\left[E_{h}k_{0}\left(\frac{T_{i}}{\DD_{h}}\right)^{2}\right]-E_{h}^{2}\left(\frac{T_{i}}{D_{h}}\right)^{2} \non \\[6pt]
& T_{i}\partial_{0}^{2}\left[\frac{T_{i}}{\DD_{h}}\right]  =  \partial_{0}\left(T_{i}\partial_{0}\frac{T_{i}}{\DD_{h}}\right)-\left(\partial_{0}T_{i}\right)\partial_{0}\left(\frac{T_{i}}{\DD_{h}}\right)=-\left(\partial_{0}T_{i}\right)\partial_{0}\left(\frac{T_{i}}{\DD_{h}}\right)
\]
so that 
\[
\beta_{h,i}^{B}=-\mu_{h}^{2}\int\frac{d^{4}k}{(2\pi)^{4}}\left[E_{h}^{2}\left(\frac{T_{i}}{\DD_{h}}\right)^{2}+\left(\partial_{0}T_{i}\right)\partial_{0}\left(\frac{T_{i}}{\DD_{h}}\right)\right]
\]
The first term in the beta function $\beta_{h,i}^{B}$ is summable over $h$ since 
\[
\frac{E_{h}^{2}}{\DD_{h}^{2}}=\left(\frac{E_{h}}{B_{h}Z_{h}k_{0}^{2}+A_{h}Z_{h}\kk^{2}}\right)^{2}\leq\left(\frac{\e E_{h}}{Z_{h}}\gamma^{-2h}\right)^{2}=\gamma^{-4h}
\]
and $\mu_{h}^{2}$ with $\mu_{h}=\e \left(1+\e \l^{\frac{1}{2}}|h-\bh|\right)^{-1}$
is summable for $d+1=4$. The second term is not summable since contains
\[
\frac{\mu_{h}^{2}}{Z_{h}}=\mu_{h}=\frac{\lambda}{1+\lambda^{\frac{3}{2}}|h|}
\]
However the non summable contribution to the beta function $\beta_{h}^{B}=\beta_{h,1}^{B}-\beta_{h,2}^{B}$,
that we call $B=B_{1}-B_{2}$, is null because the terms containing
the derivatives of $T_{i}$ give the same contribution to $\beta_{h,1}^{B}$
and $\beta_{h,2}^{B}$. First of all we can easily see that $B_{i}$
receives contribution only from the region in the momentum space such
that $T'_{i}\neq0$, that is from its boundaries: 
\[
B_{i} & =  \int\frac{d^{4}k}{(2\pi)^{4}}\,\partial_{0}\left(\frac{T_{i}}{\DD_{h}}\right)\partial_{0}T_{i} \non \\[6pt]
 & = \int\frac{d^{4}k}{(2\pi)^{4}}\left[\frac{\left(\partial_{0}T_{i}\right)^{2}}{\DD_{h}}-\frac{\left(\partial_{0}T_{i}\right) \left(\partial_{0}\DD_{h}\right)}{\DD_{h}^{2}}\right]
\]
where $\mathcal{D}_{h}(k)=b_{h}k_{0}^{2}+a_{h}\kk^{2}$. Since $B_{1}$and $B_{2}$ have the same right boundaries, the contributions coming from $\chi_{h+1}(k)$ cancel. For what concern the left sides $B_{1}^{\text{lx}}$ and $B_{2}^{\text{lx}}$ we have 
\[
A_{1}^{\text{lx}} & =  \int\frac{d^{4}k}{(2\pi)^{4}}\left[\frac{\left(\partial_{0}\chi_{h-1}\right)^{2}}{\mathcal{D}_{h}}-\frac{\left(\partial_{0}\chi_{h-1}\right)\left(\partial_{0}\mathcal{D}_{h}\right)}{\mathcal{D}_{h}^{2}}\right] \non \\[6pt]
 & \qquad =  \int\frac{d^{4}k}{(2\pi)^{4}}\left[\frac{4k_{0}^{2}\left(\chi'\right)^{2}\gamma^{-4h}\gamma^{4}}{\mathcal{D}_{h}}-\frac{4k_{0}^{2}b_{h}\gamma^{-2h}\gamma^{2}\chi^{'}}{\mathcal{D}_{h}^{2}}\right] \non \\[6pt]
A_{2}^{\text{lx}} & =  \int\frac{d^{4}k}{(2\pi)^{4}}\left[\frac{\left(\partial_{0}\chi_{h}\right)^{2}}{\mathcal{D}_{h}}-\frac{\left(\partial_{0}\chi_{h}\right)\left(\partial_{0}\mathcal{D}_{h}\right)}{\mathcal{D}_{h}^{2}}\right] \non \\[6pt]
 & \qquad =  \int\frac{d^{4}k}{(2\pi)^{4}}\left[\frac{4k_{0}^{2}\left(\chi^{'}\right)^{2}\gamma^{-4h}}{\mathcal{D}_{h}}-\frac{4k_{0}^{2}b_{h}\gamma^{-2h}\chi^{'}}{\mathcal{D}_{h}^{2}}\right]
\]
where we have used 
\[
\partial_{0}\chi_{h}(k) & =  2k_{0}\gamma^{-2h}\chi'  \non \\[6pt]
\partial_{0}\mathcal{D}_{h} & =  2k_{0}b_{h}
\]
If we now consider the scaling $k\rightarrow\g^{-4}k$ in $B_{1}$
we find:
\[
B_{1}^{\text{lx}}=\int\frac{d^{4}k' \g^{-4}}{(2\pi)^{4}}4k_{0}^{2}\g^{-2}\left[\frac{\bigl(\chi^{'}\bigr)^{2}\g^{-4h}\g^{4}}{\DD_{h}\g^{-2}}-\frac{b_{h}\gamma^{-2h}\gamma^{2}\chi^{'}}{\mathcal{D}_{h}^{2}\gamma^{-4}}\right]=B_{2}^{\text{lx}}
\]
so that also the contributions coming from the left boundaries of $T_{1}$ and $T_{2}$ cancel. The motivation is that the integral giving $B_{1}$ and $B_{2}$ is logarithmic, so we can always rescale the momenta with respect to a $\g^{\a}$ without change the
result.

%\end{document}

\chapter{Technical tools for the WIs analysis} \label{tools}

\section{One--step potential vs multiscale potential}  \label{one-step}

In this section we derive the relation between the kernels of the effective potential $\VV_{h}(\ps)$ defined in chap.~\ref{multiscale}  and the kernels of the ``one--step'' potentials $\WW_{h}(\ps)$ introduced in chap.~\ref{WI} to derive the WIs. The effective potential at scale $h$ is defined by
\[ \label{eff_pot}
e^{-\VV_{h}(\PS{\leq h}) -\tl{\EE}_{h} }=\int P_{Q_{h+1}, f_{h+1} }(d \PS{h+1})\, e^{-\VV_h(\PS{h+1}+\PS{\leq h})}
\]
with $P_{Q_h, f_h }(d \PS{h})$ the measure with covariance
\[   \label{APP-g}
 g^{\,(h)}_{\a \a'}(x) &= \int\frac{d^{4}k}{(2\pi)^{4}}e^{-ikx}f_h(k)\,g^{\,(h)}_{\a \a'}(k) \non  \\
\lft( g^{\,(h)}_{\a \a'}(k) \rgt)^{-1} & = \lft(g^{\,(0)}_{\a \a'}(k)\rgt)^{-1}+\,\sum_{j=h}^{-1}\chi_{[h^*,j]}(k)\,Q^{(h)}_{\a \a'} 
\]
with $Q^{(h)}_{\a \a'} $ the meatrix of the local quadratic terms which renormalize the measure at each step of the multiscale integration. The ``one-step'' effective potential at scale $h$ is defined by 
\[ \label{one-step}
e^{-\WW_{h}(\PS{\leq h})}=\int \hat{P}_{Q_0, \c_{[h,0]}}(d\PS{\leq 0})e^{-\VV_{0}(\PS{\leq 0})}
\]
with $\hat{P}_{Q_0, \c_{[h,0]} }(d \PS{0})$ the measure with covariance 
\[   
 \hat{g}^{\,(h)}_{\a \a'}(x) &= \int\frac{d^{4}k}{(2\pi)^{4}}\,e^{-ikx} \c_{[h,0]}(k)\,\hat{g}^{\,(0)}_{\a \a'}(k) 
\non  \\
\lft( \hat{g}^{\,(h)}_{\a \a'}(k) \rgt)^{-1} & = \lft(g^{\,(0)}_{\a \a'}(k)\rgt)^{-1}+\,\chi_{[h,0]}(k)\,Q^{(0)}_{\a \a'} 
\]
To derive a relation between \eqref{eff_pot} and \eqref{one-step} we fix $h$ and  perform successively the integrations on the momentum slices with indices $j=0,1,\ldots,h$ in the definition \eqref{one-step}, by including each time the local quadratic part $Q_{j}$ into the measure. We have: 
\[
e^{-\WW_{h}(\PS{\leq h})}  & =  \int \hat{P}_{Q_{0}, \c_{[h,-1]}}(\PS{\leq -1})  \hat{P}_{Q_{0}, f_0}(\PS{\leq 0})\,e^{-\VV_{0}(\PS{\leq -1}+ \PS{0})} \non \\
& =  e^{-\tl{\EE}_{-1}} \int \hat{P}_{Q_{0}, \c_{[j,-1]}}(\PS{\leq -1}) \,e^{-\HV_{-1}(\PS{\leq -1})} \non \\
& = e^{-\tl{\EE}_{-1} -t_{-1}}   \int \hat{P}_{Q_{-1}, \c_{[h,-1]}}(\PS{\leq -1}) \,e^{-\VV_{-1}(\PS{\leq -1})} \non \\
& = e^{-\EE_{-2}}   \int \hat{P}_{Q_{-2}, \c_{[h,-2]}}(\PS{\leq -2}) \,e^{-\VV_{-2}(\PS{\leq -2})} \non \\
& =  \ldots \non \\
&  = e^{-\EE_{h}} \int \hat{P}_{Q_{h}, \c_{[h,h]}}(d\PS{h}) \,e^{-\VV_{h}(\PS{\leq h})}
\]
with $\c_{[h,h]}(k) = f_h(k)$ and $\hat{P}_{Q_h, f_h }(d \PS{h})$ the measure with covariance
\[ \label{APP-gbar}
& \hat{g}^{\,(h)}_{\a \a'}(x)= \int\frac{d^{4}k}{(2\pi)^{4}}e^{-ikx} f_h(k) \,\hat{g}^{\,(h)}_{\a \a'}(k) \\
& \lft( \hat{g}^{\,(h)}_{\a \a'}(k) \rgt)^{-1} = \lft(g^{\,(0)}_{\a \a'}(k)\rgt)^{-1}+\,\sum_{j=h}^{-1}\chi_{[h,j]}(k)Q_{j}
\]
The difference between the measures $P_{Q_h,f_h}$ and $\hat{P}_{Q_h,f_h}$ constists in the different lower scale of the cutoff function in their propagators, see \eqref{APP-g} and \eqref{APP-gbar}. If $h^*<h$ 
\[
f_{h}(k)\,\chi_{[h^*,\, j]}(k)=f_{h}(k)\,\chi_{[h,\, j]}(k)\quad\forall j>h
\]
so that 
\[ \label{g_vs_ghat}
f_h(k) \,\lft(\hat{g}^{\,(h)}_{\a \a'}(k) \rgt)^{-1} & = f_h(k)\, \lft[ \lft(g^{\,(0)}_{\a \a'}(k)\rgt)^{-1}+\,\sum_{j=h+1}^{-1}\chi_{[h,j]}(k)q_{\a \a'}^{(j)} +f_h(k)\,q_{\a \a'}^{(h)}  \rgt] \non \\
 &  = f_h(k)\, \lft[ \lft(g^{\,(0)}_{\a \a'}(k)\rgt)^{-1} +\,\sum_{j=h}^{-1}\chi_{[h^*,j]}(k)q_{\a \a'}^{(j)} +(f_h(k)-\c_{[h^*,h]})\,q_{\a \a'}^{(h)}   \rgt] \non \\
& = f_h(k)\, \lft[ \lft(g^{\,(h)}_{\a \a'}(k)\rgt)^{-1}  -\c_{[h^*,h-1]} \,q_{\a \a'}^{(h)} \rgt]
\]
Note that $f_h(k) \,\hat{g}^{\,(h)}_{\a \a'}(k)$ and $f_h(k) \,g^{\,(h)}_{\a \a'}(k)$ would be equal if the functions $\chi_{[h,j]}(k)$ were the characteristic functions of their support, since in that case $$ f_h(k)\chi_{[h^*,\, h-1]}(k)=0$$ 
However, as we shall prove below, $q^{(h)}_{\a \a'}$ is a small quantity, so that the potentials $\VV_{h}(\ps)$ and $\WW_{h}(\ps)$ are equal at leading order.  

\feyn{
\begin{fmffile}{feyn-TESI/last_scale}
 \unitlength = 1cm
\def\myl#1{3cm}
\begin{align*}
& q^{(h)}_{ll} =\parbox{\myl}{\centering{
			\begin{fmfgraph*}(2.5,1.4) 
			\fmfleft{i1}
			\fmfright{o1}
			\fmf{dashes, tension=1.5}{i1,v1}
			\fmf{plain,left=0.8, tension=0.4, label=$f_h(k)$}{v1,v2}
			\fmf{plain,right=0.8, tension=0.4,label=$f_j(k)$}{v1,v2}
			\fmf{dashes}{o1,v2}
			\Ball{v1,v2}
			\end{fmfgraph*}  
			}} & d=3\\[12pt]
& q^{(h)}_{ll} =\parbox{\myl}{\centering{
			\begin{fmfgraph*}(2.5,1.4) 
			\fmfleft{i1}
			\fmfright{o1}
			\fmf{dashes, tension=1.5}{i1,v1}
			\fmf{plain,left=0.8, tension=0.4}{v1,v2}
			\fmf{plain,right=0.8, tension=0.4}{v1,v2}
			\fmf{dashes}{o1,v2}
			\Ball{v1,v2}
			\end{fmfgraph*}  
			}} 
+\parbox{\myl}{\centering{
			\begin{fmfgraph*}(2.5,1.4) 
			\fmfleft{i1}
			\fmfright{o1}
			\fmf{dashes, tension=1.5}{i1,v1}
			\fmf{dashes,left=0.8, tension=0.4}{v1,v2}
			\fmf{dashes,right=0.8, tension=0.4}{v1,v2}
			\fmf{dashes}{o1,v2}
			\Ball{v1,v2}
			\end{fmfgraph*}  
			}}
+\parbox{\myl}{\centering{
			\begin{fmfgraph*}(2,1.2) 
			\fmfleft{i1}
			\fmfright{o1}
			\fmf{dashes}{i1,v1,o1}
	             \fmf{plain}{v1,v1}
			\Ball{v1}
			\end{fmfgraph*}  
			}} & d=2
\end{align*}
\end{fmffile}	
}{{\bf Local quadratic terms, $h \leq \bh$.} By definition of $f_h(k)$, the diagrams in the figure are different from zero only if the cutoff function $f_j(k)$ is at scale $j=h$ or $j=h+1$. }{last_scale}

\pagina
As an example let us calculate $q_{ll}^{(h)} \equiv z_h$, whose leading order contribution is represented in fig.~\ref{last_scale}. An explicit computation shows that
\[
& z_h = c_1 \, \l \e^{-\frac{1}{2}} \m_h^2 = c'_1\, Z_h \, \frac{\l \e^{\frac{1}{2}} }{(1+c \l \e^{\frac{1}{2}}|h -\bh|)}  & d=3 \non \\[6pt]
& z_h = c_2\, \l \m_h^2 = \g^h \l_h\, \l \l_h \bigl(c'_2  + c''_2 \,\frac{\l_{6,h}}{\e \l_h^2} \bigr)   & d=2
\]
with $c_1$, $c'_1$, $c$, $c_2$, $c'_2$ and $c''_2$ explicitly computable constants. By substituting these expression in \eqref{g_vs_ghat} we obtain
\[
f_h(k) \,\lft(\hat{g}^{\,(h)}_{ll}(k) \rgt)^{-1} =  f_h(k)\,Z_h(k) \lft( 1+ {\cal{Z}}(\l) \rgt) 
\]
with 
\[
{\cal{Z}}(\l) = \begin{cases}
O\lft( \l \e^{\frac{1}{2}} \bigl(1+c \l \e^{\frac{1}{2}}|h -\bh|\bigr)^{-1}  \rgt) & d=3 \\[6pt]
O(\l \l_h) & d=2
\end{cases}
\]
In three dimensions ${\cal{Z}}(\l)$ is subdominant in $\e$ at the beginning of the second region and goes to zero as $h \arr -\io$; in two dimensions the condition $\l \l_h$ smallest than one is the condition making the perturbative theory meaningful. A similar discussion can be done for the other local quadratic terms, obtaining similar conclusions.
\pagina 

\section{Properties of the correction term $C_\n(k,p)$}   \label{C_properties}

This section is devoted to the prove of the properties of the functions $C_\n(k,p)$ 
\[ 
C_{0}(k,p) & =  \frac{1}{2} \left[(k_{0}+p_{0})(\chi_{[h^*,0]}^{-1}(k,p)-1)-k_{0}(\chi_{[h^*,0]}^{-1}(k)-1)\right] \label{Cnu0}\\[6pt]
C_{1}(k,p) & =  (\kk+\pp)^{2}(\chi_{[h^*,0]}^{-1}(k+p)-1)-\kk^{2}(\chi_{[h^*,0]}^{-1}(k)-1) \label{Cnu1}
\]
which are crucial in order to study the functional integral \eqref{W_tilde} by RG methods. \\

The first property is that, due to the presence of the cutoff function $\c_{[h^*,0]}(k+p)$ in the definitions \eqref{Cnu0} and \eqref{Cnu1}, the contraction of the outgoing lines of the kernels 
$$
\tl{J}_0C_0(k,p) \ps^l_{k+p}\ps^l_{-k} \quad \tl{J}_0C_0(k,p) \ps^t_{k+p}\ps^t_{-k} \quad \tl{J}_1C_1(k,p) \ps^l_{k+p}\ps^t_{-k}
$$
is {\it different from zero only if at least one of the bosonic  $\ps$ fields are contracted at scale $h=0$ or $h=h^*$}. The proof of this property is trivial. Let us consider the contraction of the function $C_\n(k,p)$ with the cutoff functions of two propagators respectively at scale $i$ and $j$; then
\[
\D_{\n}^{(j,l)}(k,p) =  f_{j}(k+p)C_{\n}(k,p)f_{l}(k)  =0 \qquad \text{for }0<j,l<h
\]
If $0<j<h$ the cutoff function $\c_{[h^*,0]}(k+p)$ is equal to the unit operator on the support of $f_j(k+p)$; moreover, if $0<l<h$ also  $\c_{[h^*,0]}(k)=\unit$ on the support of $f_l(k)$ . Then \eqref{Cnu0} and \eqref{Cnu1} are identically equal to zero. 

\vskip 0.5cm

In the cases in which $\D_{\n}^{(j,l)}(k,p)$ is not identically equal to zero, since $\D_0^{(j,l)}(k,p)= \D_0^{(l,j)}(-(k+p),p)$ and $\D_i^{(j,l)}(k,p)= -\D_i^{(l,j)}(k+p,p)$, we can restrict the analysis to the case $j \geq l$. The different cases are studied in the following.  We remark that what {\it we need} in order to control the contraction of the corrections term coming from the cutoff is that
\begin{description}
\item[$(i,j)=(0,0)$]: the function $\D^{(0,0)}_\n(k,p)$ must be bounded by $|p_n|$ times a cutoff function which has the same support of $f_0(k)$ (but can also have a slight different behavior); we will prove this property in subsection I below.
\item[$(i,j)= (0,h^*)$]:  the function $\D^{(0,h^*)}_\n(k,p)$ must be bounded by $|p_n|$ times a cutoff function which has the same support of $f_0(k)\,f_h(k)$; we will prove this property in subsection II.  
\item[$(i,j)= (h^*,h^*)$]: $\D^{(h^*,h^*)}_\n(k,p)$ must bounded by $|p_n|$ times a cutoff function which has the same support of $f_{h^*}(k)$. This requirement is sufficient provided that we have chosen as localization point $|p_\n|=\g^h$ and it is proven in subsection III. We stress that if one had taken the external momentum is taken equal to zero it would not have been sufficient to have a single support function $f_{h^*}(k)$, but we need one support function for each propagator, to prove that $\D^{(h^*,h^*)}_\n(k,p)$ is well defined. 
\end{description}
%for the kernels with external $\tl{J}$ fi

\vskip 0.5cm

\subsection*{ I. Contractions of both the $\ps$ fields at scale $h=0$}

\vskip 0.2cm

{\centering \subsubsection{Vertices with $C_0(k,p)$}}

Let us analyze what happens  when  both the external legs of the correction term $\tilde{J}_{0} C_0(k,p) (\psi_{k+p}^{t}\psi_{-k}^{t}+\psi_{k+p}^{l}\psi_{-k}^{l})$ are contracted at scale $h=0$. Here $p$ is the external momentum associated to the fields $\tl{J}_0$, which we will choose at scale $\g^{h^*}$.

We denote $\D_{0}^{(0,0)}(k,p) =  f_{0}(k+p)C_{0}(k,p)f_{0}(k) $ the contraction of the squared vertex $C_{0}(k,p)$ with the cutoff functions of the two propagators. At scale $0$ we have
\[
\Delta_{0}^{(0,0)}(k,p)  & =  (k_{0}+p_{0})u_{0}(k+p)f_{0}(k)-k_{0}u_{0}(k)f_{0}(k+p)
\]
with 
\[ \label{u0}
u_0(k)= f_0(k)\, \lft(\c^{-1}_{[h^*,0]}(k) -1 \rgt)=
\begin{cases}
1-f_0(k)  &  |k|_{>}^2= k_0^2 + \kk^4 \geq \frac{2}{\g^2 +1} \\
0 & \text{otherwise } \\
\end{cases}
\]
In the following for simplicity of notation, we will forget about the factor $2/(\g^2 +1)$ and consider $u_0(k)$ defined in \eqref{u0} different from zero for $|k|^2_> \geq1$.    By developing $f_0(k,p)$ and $u_0(k,p)$ around small values of $p$ and taking into account the fact that $f_0(k)=1-u_0(k)$ on the support of $\dpr_\n u_0$ and $\dpr_\n f_0 = -\dpr_\n u_0$ on the support of $u_0(k)$ we obtain
\[ \label{D0}
\Delta_{0}^{(0,0)}(k,p) & =  p_{0}u_{0}(k)f_{0}(k)+(k_{0}+p_{0})\, p_{\nu}\partial_{\nu}u_{0}(k^{*})\, f_{0}(k)-k_{0}u_{0}(k)p_{\nu}\partial_{\nu}f_{0}(k^{*})  \non \\[6pt]
 & =  p_{0}\big[u_{0}(k)f_{0}(k)+ p_{\nu}\,\partial_{\nu}u_{0}(k^{*})\,f_{0}(k)\big]+k_{0}p_{\nu}\partial_{\nu}u_{0}(k^{*})
\]
If we choose $p=(p_{0},\bz)$
\[ \label{D0_0}
\Delta_{0}^{(0,0)}(k,p_{0})=p_{0}\Big[u_{0}(k)f_{0}(k)+f_{0}(k)\, p_{0}\partial_{0}u_{0}(k^{*})+k_{0}\partial_{0}u_{0}(k^{*})\Big]
\]
with $|k|<|k^*|<|k+p|$.  The brackets  in \eqref{D0_0} contains the sum of cutoff functions which are supported on the same support than $f_0(k)$. In fact $\dpr_0 u_0(k^*)$ is different from zero for each $\g^{-1}\leq |k^*|\leq \g$ and $|k^*|\simeq |k|$ being $|p|= \g^h$. 

With respect to the contraction of two lines  outgoing from a simple vertex (\ie without $C_0(k,p)$), here we get $p_0$ which multiplies a sum of cutoff functions on a similar support. From a dimensional point of view the factors in the square brackets in \eqref{D0_0} are bounded by a constant; in fact $p_{0}\dpr_{0}$ and  $k_{0}\dpr_{0}$ are equal or lower than one. Then $\D_{0}^{(0,0)}(k,p_0)\ps^t_{k+p}\ps^t_{-k}$  has the same scaling dimension of $p_{0}$ multiplied for a vertex $\bar{\mu}_{0}^{J_{0}}(k,p)$ whose external legs are contracted at scale $0$. Similarly $\D_{0}^{(0,0)}(k,p_0)\ps^l_{k+p}\ps^l_{-k}$  has the same scaling dimension of $p_{0}$ multiplied for the irrelevant kernel $J_{0}\ps^l_{k+p}\ps^l_{-k}$.   

The only difference with the kernels $J_{0}C_0(k,p)\ps^t_{k+p}\ps^t_{-k}$ and $J_{0}\ps^l_{k+p}\ps^l_{-k}$ is the fact that in the contraction with the squared vertex one of the cutoff function associated to the propagators is ``lost''; however the presence of a single cutoff function is sufficient to guarantee that the behavior of the kernels with external fields $J_0$ and $\tl{J}_0$ is the same if their legs are contracted at scale $h=0$. \\

If we choose $p_\n=0$ for each $\n \neq i$ and $p_i=\g^h$  \eqref{D0} becomes
\[ \label{D0_1}
\Delta_{0}^{(0,0)}(k,p_i) & =  p_i \, k_{0}\dpr_i u_{0}(k^{*})
\]
where $ p_i \, k_0 = \g^{h^*-\frac{j}{2}}\leq \g^{\frac{h^*}{2}} $. Then $\tl{J}_0\,C_0(k,p_i) \ps^{t,(0)}_{k+p} \ps^{t,(0)}_{-k}$ has the same dimensional scaling of $p_i \,\bar{\mu}_{0}^{J_{0}}(k,p_i)$, the only difference being the absence of one of the two cutoff functions associated to the contractions of the external $\ps$ fields. Similarly $\tl{J}_0 C_0(k,p_i) \ps^{l,(0)}_{k+p} \ps^{l,(0)}_{-k}$ has the same dimensional scaling of $p_i \,\bar{\mu}'^{J_{0}}_{0}(k,p_i)$.

{\centering \subsubsection{Vertices with $C_{1}(k,p)$}}

Let's now analyze what happens  when  both the external legs of the correction term $\tilde{J}_{1}\psi_{k+p}^{l}\psi_{-k}^{t}$ are contracted at scale $h=0$. We first remind that
\[
C_{1}(k,p)=(\kk+\pp)^{2}\big(\chi_{[h^*,0]}^{-1}(k+p)-1\big)-\kk^{2}\big(\chi_{[h^*,0]}^{-1}(k)-1\big)
\]
Then, defining $\D_{1}^{(0,0)}(k,p) =  f_{0}(k+p)C_{1}(k,p)f_{0}(k)  $ the contraction of the squared vertex $C_{1}(k,p)$ with the cutoff functions of the two propagators at scale $0$ we have
\[
\D_{1}^{(0,0)}(k,p) &  =  (\kk+\pp)^{2}u_{0}(k+p)f_{0}(k)-\kk^{2}u_{0}(k)f_{0}(k+p)
\]
with $u_0(k)$ defined in \eqref{u0}. By developing $f_0(k,p)$ around small values of $p$ we obtain:
\[
\D_{1}^{(0,0)}(k,p) & =  \pp\cdot(2\kk+\pp)u_{0}(k)f_{0}(k)-p_{\nu} \big[\pp\cdot(2\kk+\pp)f_{0}(k)-\kk^{2}\big]\dpr_{\nu}u_{0}(k^{*})
\]
Choosing as external momentum $p=(p_{0},\bz)$ we get
\[
\Delta_{1}^{(0,0)}(k,p_0)=p_{0}\,\kk^{2}\dpr_{0}u_{0}(k^{*})
\]
Each of the $\kk$'s corresponds to a derivative $\dpr_\xx$ over the contracted line carrying momenta $k+p_0$, with $p_0=\g^{h^*}$ and $k$ order one, since $k$ belongs to the support of $f_0(k)$. Then $\tl{J}_1 \, C_{1}^{(0,0)}(k,p_{0})\ps^l_{k+p}\ps^t_{-k}$ has the same scaling dimension of $p_0\,J_1 \ps^{l}_{k+p_0} \ps^{t}_{-k}$, with both the $\ps$ fields contracted at scale $0$. \\

If we choose $p_\n=0$ for each $\n \neq i$ and $p_i=\g^h$
\[
\D_{1}^{(0,0)}(k,p_i)=p_i \lft[\,(2k_i+p_i)\Big(u_{0}(k)f_{0}(k)+\pp\cdot\dpr_\xx f_{0}(k^{*})\Big)- \dpr_i f_{0}(k^{*})\kk^{2} \rgt]
\]
that is $\tl{J}_1 \, C_{1}(k,p_i)\ps^{l,(h)}_{k+p}\ps^{t,(h)}_{-k}$ has the same dimensional estimate of $p_i\, \bm_{0,i}^{J_1}$.
 
%\vskip 1cm

\pagina

\subsection*{II. Contraction of one of the $\ps$ fields on scale $0$}
If one of the $\ps$ fields emerging from the correction term $C_0(k,p)$ is contracted at scale $0$ and the other one at scale $h^*\leq j<0$ we get
\[
\D_0^{(0,j)}(k,p)  & =  f_0(k+p)C_0(k,p)f_j(k)
\]
The key observation here is that 
\[
\lft(\c^{-1}_{[h^*,0]}(k) -1 \rgt) f_j(k) = \tl{u}_{h^*}(k) \d_{jh^*}
\]
with 
\[ \label{tl_u0}
\tl{u}_{h^*}(k)= f_{h^*}(k)\, \lft(\c^{-1}_{[h^*,0]}(k) -1 \rgt)=
\begin{cases}
1-f_{h^*}(k)  &  |k|^2\leq \frac{2}{\g^2 +1}\,\g^{2h} \\
0 & \text{otherwise} \\
\end{cases}
\]
where $|k|^2$ in \eqref{tl_u0} stays for $|k|^2_{<}=k_0^2 + \kk^4 $ for $\bh<h^*\leq 0$ or $|k|_>^2=k_0^2 + \e \kk^2 $ for $h^* \leq \bh$. Since we are interested in using the WIs only in the lower region $h \leq \bh$ in the following we will consider only the case $h^* \leq \bh$. 
Then
\[ \label{D0_j}
\D_0^{(0,j)}(k,p)  & =  (k_0+p_0)u_0(k+p)f_j(k)- k_0 f_0(k+p) \tl{u}_{h^*}(k) \d_{jh^*}
\]
Note that if $j < -1$ the first term in the r.h.s. of  \eqref{D0_j} vanishes for $|p|\leq 1-\g^{-1}$~\footnote{In fact $u_0(k+p) \neq 0$ implies $|k+p|\geq 1$. If $|p| \leq 1-\g^{-1}$ we have $|k| \geq \g^{-1}$ and as a consequence $f_j(k) = 0$, being  $j <{-1}$.}.  Analogously the second term in the r.h.s. of \eqref{D0_j}  vanishes for $|p| \leq 1-\g^{-1}-\g^{h^*}$~\footnote{In fact $f_0(k+p) \neq 0$ implies $|k+p| > 1-\g^{-1}$; for $|p| \leq 1-\g^{-1}-\g^{h^*}$ it holds $|k| > \g^{h^*}$ and as a consequence $\tl{u}_{h^*}(k)=0$.}. Since we are choosing $|p| = \g^h$ for $j<-1$ the function $\D_0^{(0,j)}(k,p) $ is zero. If $j=-1$ we have
\[ 
\D_0^{(0,-1)}(k,p)  & =  (k_0+p_0)u_0(k+p)f_{-1}(k)
\]
with $u_0(k)f_{-1}(k)=0$. Then %If $j \neq h^*$ we have
 \[ 
\D_0^{(0,-1)}(k,p)  & =  (k_0+p_0)\,p_\n \dpr_\n u_0(k^*)\,f_{-1}(k)
\]
which is dimensionally equal to $ \cst\,|p_\n|f_{-1}(k)$. 
%The contraction of $f_0(k+p)f_j(k)$ with the kernel $C_0(k,p)$ in this case does not ``cancel'' one of the cutoff functions, but also after the contraction we have two cutoff functions, one at scale $0$ and the second one at scale $j$. 
Finally, if $j=h^*$ we find
\[  \label{D0_0h}
\D_0^{(0,h^*)}(k,p)  & = p_\n \lft[(k_0+p_0)\,\dpr_\n u_0(k^*)\,f_{h^*}(k) - 
k_0\, \dpr_\n f_0(k^*)\tl{u}_{h^*}(k)\rgt]
\]
which again can be bounded by $\cst |p_\n|$ times the multiplication of cutoff functions. However in the second term in the r.h.s. of \eqref{D0_0h} $\tl{u}_{h^*}(k)$ is a function which does not goes to zero for $k \arr 0$, differently from $f_{h^*}(k)$. This does not allow, as we will see, to choose $0<|p_\n|\ll \g^{h^*}$. 

A similar discussion holds for the function $C_1(k,p)$; we will not belabor the details here.

\pagina

\subsection*{III. Contractions of both the $\ps$ fields on scale $h^*$}

%Since this expression can only appear at the last integration step, it is not involved in any regularization procedure. Hence we only need to estimate its size for values of the external momentum $p$ of order $\g^{h^*}$. \blue{The reason why we cannot choose $0\leq |p| \ll \g^{h^*}$ is stressed in section..}

\vskip 0.2cm

{\centering \subsubsection{Vertices with $C_{0}(k,p)$}}

In the following we will analyze what happens  when both the external legs of the correction term $\tilde{J}_{0}(\psi_{k+p}^{t}\psi_{-k}^{t}+\psi_{k+p}^{l}\psi_{-k}^{l})$ are contracted at the lowest scale $h^*$ of the cutoff function $\c_{[h^*,0]}(k)$. We define $\D_0^{(h^*,h^*)}(k,p) =  f_{h^*}(k+p)C_0(k,p)f_{h^*}(k) $ the contraction of the squared vertex $C_{0}(k,p)$ with the cutoff functions of two propagators contracted at scale $h$. It turns to be: 
\[
\D_0^{(h^*,h^*)}(k,p)  & =  (k_0+p_0)\tl{u}_{h^*}(k+p)f_{h^*}(k)-k_0\tl{u}_{h^*}(k)f_{h^*}(k+p)
\]
with $\tl{u}_{h^*}$ defined in \eqref{tl_u0}. By developing $f_0(k,p)$ around small values of $p$ we obtain
\[  \label{Dh}
\D_{0}^{(h^*,h^*)}(k,p) & =   p_0\,f_{h^*}(k)\big[\tl{u}_{h^*}(k)-\, p_{\nu}\dpr_{\nu}\tl{u}_{h^*}(k^{*})\big]+k_{0}p_{\nu}\dpr_{\n} \tl{u}_{h^*}(k^*)
\]
where we have used the fact that on the support of $\tl{u}_{h^*}(k)$ we have $\dpr_\n f_{h^*}(k)=-\dpr_\n u_{h^*}(k)$ and that on the support of $\dpr_\n u_{h^*}(k)$ it turns $f_{h^*}(k)=1-\tl{u}_{h^*}(k)$. 

If we choose as external momentum $p=(p_0, \bz)$ \eqref{Dh} becomes:
\[  \label{Dh_0}
\D_{0}^{(h^*,h^*)}(k,p_0) & =   p_0\,\big[\tl{u}_{h^*}(k)f_{h^*}(k) -\, p_{0}\dpr_{0}\tl{u}_{h^*}(k^{*})f_{h^*}(k) +k_{0}\dpr_{0} \tl{u}_{h^*}(k^*) \big]
\]
which from a dimensional point of view is bounded by $\cst \, \g^{h^*}$. While the first two terms in the r.h.s. side of \eqref{Dh_0} are still the product of two cutoff functions, even if on a smaller support with respect $f^2_{h^*}(k)$, the last term in the r.h.s. side of \eqref{Dh_0} has a single cutoff function, fact that won't allow to choose for $|p|$ values smaller than $\g^{h^*}$.

For a choice of the external momentum such that $p_\n=0$ for each $\n \neq i$ and $p_i=\g^{h^*}$ \eqref{Dh} becomes:
\[  \label{Dh_1}
\D_{0}^{(h^*,h^*)}(k,p_i) & =  p_i \,k_0\,\dpr_{i} \tl{u}_{h^*}(k^*)
\]
From a dimensional point of view, $k_0$ is the loop variable associated to the propagator at scale $h^*$. The derivative $\dpr_i$ also falls on one of the internal propagator, on scale equal or greater than $h^*$. Then the dimensional estimate for \eqref{Dh_1} is $\cst\,\g^{h^*}$. 

\vskip 0.5cm

{\centering \subsubsection{Vertices with $C_{1}(k,p)$}}

The discussion for the squared vertex representing $C_{1}(k,p)$ follows the same ideas just presented: 
\[
& \Delta_{1}^{(h^*,h^*)}(k,p)  =  f_{h^*}(k+p)C_{1}(k,p)f_{h^*}(k) \non \\
 &  \quad = (\kk+\pp)^{2}f_{h^*}(k)\,\tl{u}_{h^*}(k+p)-\kk^{2}f_{h^*}(k+p)\,\tl{u}_{h^*}(k) \non \\
&  \quad =  \pp\cdot(\pp+2\kk)\left[f_{h^*}(k)\tl{u}_{h^*}(k) - p_\n \dpr_\n \tl{u}_{h^*}(k^*)\, \tl{u}_{h^*}(k) \right]
+\lft(\kk + \pp \rgt)^2 \,p_\n \,\partial_{\nu}\tl{u}_{h^*}(k^*)
\]
If $p=(p_{0},\bz)$ we get
\[
\Delta_{1}^{(h^*,h^*)}(k,p_{0})=p_0\,\kk^2 \partial_0\,\tl{u}_{h^*}(k^*)
\]
If $p_\n=0$ for each $\n\neq i$ and $p_i=\g^h$ we get
\[
\Delta_{1}^{(h^*,h^*)}(k,p_i)=&\, p_i \, (p_i+2k_i) \left[f_{h^*}(k)\tl{u}_{h^*}(k) - p_i \dpr_i \tl{u}_{h^*}(k^*)\, \tl{u}_{h^*}(k) \right] \non \\
& +\lft(\kk + \pp \rgt)^2 \,p_i \,\partial_i \tl{u}_{h^*}(k^*)
\]
We can conclude that $\tl{J}_1\,C_{1}(k,p_\n)\ps^{l,(h^*)}_{k+p}\ps^{t,(h^*)}_{-k}$ has the same dimensional estimate then $|p_\n|\,\m^{J_1}_{h,i}$, with $i=1,2,3$. The difference between the vertex $\m_h^{J_1}$ and $\m_h^{\tl{J}_1}$ is the fact that in the last case in the contraction of the squared vertex with the two cutoff functions $f_{h^*}(k)$ coming from the propagators, one of the function is ``lost''. This fact does not allow to choose $p_i=0$ for each $i$.

\pagina

\section{Localization in $p=0$ vs $p=\g^h$} \label{APPzero} 

In this section we list which are the differences arising when we evaluate the local WIs at external momentum $|p|=\g^{h^*}$ rather than in $|p|=0$ and how to prove that they are all subdominant in $h$ as $h \arr -\io$. 

\vskip 0.5cm

{\centering { \bf i) Quadratic local terms.}}  We compare the values of the quadratic local terms 
$$\uQ_{\,h}(p)=\{A_{\,h^*}(p),B_{\,h^*}(p), E_{\,h^*}(p),Z_{\,h^*}(p)\}$$ 
localized in $|p|=\g^h$ with the local quadratic terms of the one--step potential appearing in the local WIs. We remind that the $\uQ_{\,h}(p)$ are obtained by inserting at each step of the multiscale integration the local quadratic part in the measure, as described in section \ref{effective}. Due to this procedure each of the $\uQ_{\,h}(p)$ contains a dependence on  the cutoff functions $\chi_{[h^*,j]}(p)$ with $ h^* \leq j \leq \bh$. We will detail the discussion only for $Z_h$, the discussion for the other renormalization function being similar. We have:
\[ 
Z_{h^*}(p)=Z_{\bh}+ \sum_{j=h^*}^{\bh} z_{j}\,\chi_{[h^*,j]}(p)
\]
with $z_j=Z_{j-1} - Z_{j}$. For $k=\g^{h^*}$ each cutoff function is equal to one and we get
\[
Z_{h^*} := Z_{h^*}(\g^{h^*})=Z_{\bh}+ \sum_{j=h^*}^{\bh} z_{j}
\]
We want to compare $Z_{h^*}$ with the local quadratic terms $\hat{Z}_{\,h^*}$ of the one--step potential for which the WIs have been derived. 
%For a definition
%\[
%\hat{Z}_{h^*}=\frac{\d^{2}}{\d \phi_{k}^{l}\phi_{-k}^{l}}\,\WW_{h^*}(\f)\Big|_{\f=0}
%\]
%with
%\[ 
%\WW_{h^*}(\f)= \frac{1}{|\L|} \log \int P_{\c_{[h^*,0]}} (d\ps)\,
% e^{-\VV_I(\ps+\f) }  
% ---- effective potential con campi J
% & \WW_{h^*}(\f,J)= \non \\
%& \quad \frac{1}{|\L|} \log \int P_{\c_{[h^*,0]}} (d\ps) e^{-\VV_I(\ps+\f) + \int_\L dx \left[\,J_{x}^{0} \,\psi_x^{+}\psi_x^{-} +{\bf J}_{x}^{1} \cdot \left(\ps_x^- \dpr_\xx \ps_x^{+} - \psi_x^+ \dpr_\xx \ps_x^- \right)\right] }  
%\]
The iterative relation $\hat{Z}_{h-1}=\hat{Z}_{h}+\hat{z}_{h-1}$
holds and one gets $\hat{Z}_{h^*}=Z_{\bh}+\sum_{j=h^*}^{\bh}\hat{z}_{j}$. 
Since $\hat{z}_{j}=z_j$ for each $j>h^*$ and $\hat{z}_{h^*}-z_{h^*}=o(Z_h)$, as shown in appendix \ref{one-step}, we get $\hat{Z}_{h^*}=Z_{h^*}$ at leading order.  

\vskip 1cm

{\centering {  \bf ii) $\hW_{12}^{(h)}(p)$ and kernels with one external field.}} 
In this section we describe how to prove that the difference between the kernels one--step potential appearing in the local WIs can be equivalently localized in $p_0=\g^h$ or in zero, the difference being subdominant in the small parameter of the perturbation theory. Let first consider the kernel $\hW_{12}^{(h)}(p_0)$ appearing in the WI for $E_h$. In this case we are interested in estimating the difference
\[
\hW^{(h)}_{n_l\,n_t;\n}(p_0)-\hW^{(h)}_{n_l\,n_t\n}(0)
\]
with $p_0=\g^{h^*}$ and $p_i=0$ for all $i$'s. In order to do this it is sufficient to note that we can express $\hW^{(h)}_{n_l\,n_t}(0)$ as a sum over GN trees with external momentum $p_0\neq 0$:
\[ \label{diff_ph}
\hW^{(h)}_{n_l\,n_t}(0) =\hW^{(h)}_{n_l\,n_t}(p_0) + p_0\dpr_0 \hW^{(h)}_{n_l\,n_t}(p^{*})
\]
%where $\hW^{[h]}_{12}(p)$ and $\hW^{[h]}_{12}(0)$ are both expressed as sums over GN trees. 
First of all we note that the trees contributing to $\dpr_0 \hW^{(h)}_{12}(p^{*})$ have a short memory factor  $\g^{(h-k)}$ due to the fact that the derivative $\dpr_p$ acts on one of the internal propagators on scale $k\geq h$, while the external $p$ lives on scale $h$.  Then the dominant contribution to the r.h.s in \eqref{diff_ph} comes from the diagrams where there is at least a propagator at scale $h$. At this point it is sufficient to note that the trees contributing to  $p_0 \dpr_0 \hW^{(h)}_{12}(p^{*})$ must have at least two vertices, due to the presence of the derivative. Then, with respect to the trees giving $\hW^{(h)}_{12}(p)$ or $\hW^{(h)}_{12}(0)$ we always have an extra vertex $\m_h$ or $\l_h$, which means a small factor. In fact for each extra $\l_h$ or $\m_h$ vertex we have a factor $(\l \sqrt{\e} |h|)^{-1}$ in the three dimensional case and a factor $\l\,\l_*$ or $\g^{\frac{h}{2}}\l\,\l_*$ in the two dimensional case. 

The same result also holds for the kernels with external fields appearing in the WIs, \eg $\m_h^{J_0}$ and $\m_h^{\tl{J_0}}$, since their flow is controlled by comparison with the flow of $\m_h$.
%In fact also in the latter case the difference between the kernels at external momentum $|p|=\g^{h}$ or zero is equal to the sum of diagrams with one extra vertex $\l_h$ or $\m_h$ vertex.

\vskip 1cm

{\centering { \bf iii) Kernels vanishing at $|p|=0$.} } The WIs calculated at zero external momentum differs with respect to the ones calculated at non zero momentum also because there are some terms which are zero if $|p|=0$.  
%whose local part is null and so does not appear in the first case.
For example, the (formal) local WI for $E_{h}$ is obtained by deriving the identity
\[
\hW_{03}^{(h)}(k,\, p_{0})+\hW_{11}^{(h)}(-k)-\hW_{11}^{(h)}(k+p_{0})=p_{0}\hW_{02;0}^{(h)}(k,-k-p_{0})
\]
with respect to the external momentum $p_0$ in $J_0$. For $k=(0, \bz)$ and $p_0=0$  the kernel $W^{(h)}_{03}(0,0)=0 $ is zero by parity reasons. This can be seen immediately at level of one--loop  calculations. In fact, denoting with $q$ the loop integral, we have 
\[
W^{(h)}_{03}(0,p_0)=\int dq_0 d^{d}q\,\frac{(p_0 +q_0)\,\qq^{2}}{\DD_{h}(q)\DD_{h}(p_0+q)}
\]
where the term proportional to $q_0$ is zero for parity reasons; then the integral is equal to zero as soon as $p_0=0$. If $p_0\neq 0$ the kernel $W^{(h)}_{03}(0,p_0) \neq 0$ but since $p_0=\g^{h^*}$ it is dimensionally smaller than the dominant terms appearing in the WI. 

%For what concern the derivative with respect to $p_0$ of $W_{ttt}(0)$ we get:
%\[
%\dpr_0 W_{ttt}(0,p_0)= \int d^{d+1}q \lft\{ \frac{\qq^2}{D^2(q)}\frac{\qq^2 - q^2_0 }{D^2(q+p_0)} +\frac{p^2_0\, \qq^2}{D(q)\,D^2(q+p_0)} \rgt\}
%\]

\vskip 1cm

{\centering { \bf iv) Discrete derivatives.} }  If $p_\n=\g^{h^*}$ all the derivative with respect to $p_\n$ we have performed to obtain the formal local WIs are discrete derivatives. However, since we are only considering the dominant behavior in $\g^{h^*}$ as $h^* \arr -\io$ the difference between the discrete derivative and the derivative taken in $p_\n=0$ is subdominant. As an example consider the following one--loop computation:
\[
\frac{1}{p^2_{0}}\left(W_{02}^{(h)}(0)-W_{02}^{(h)}(p_{0})\right)= b_h\,\int d^{d+1}q\,\frac{q_0^2}{\DD^2_h(q)\,\DD_h(p_0+q)} 
\]
We see that the difference between the discrete derivative and $\dpr_0\,W^{(h)}_{22}(p)$ stays in the denominator, which is $\DD_h(q+p_0)=\DD_h(q)+b_h\,p_0(p_0+2)$ instead of $\DD^2_h(q)$. However being $|p_0|=\g^{h^*}$ with $h \geq h^*$ this difference is subdominant.

\pagina

\section{Explicit computations}

\subsection{Initial values of the RCC with external field $\tl{J}$}  \label{appB.initial_val}

\feyn{
\begin{fmffile}{feyn-TESI/mu0_tilde}
 \unitlength = 0.8cm
\def\myl#1{2.5cm}
\begin{align*}
p_0\, \parbox{1.8cm}{\centering{
			\begin{fmfgraph*}(1.8,1.4) 
			\fmfleft{i1}
			\fmfright{o1,o2}
			\fmf{wiggly, tension=1.5, label=$\tl{J}_0$}{i1,v1}
			\fmf{plain}{o1,v1,o2}
			\Ball{v1}
			\end{fmfgraph*}  
			}}=
	\parbox{\myl}{\centering{
			\begin{fmfgraph*}(2.5,1.4) 
			\fmfleft{i1}
			\fmfright{o1,o2}
			\fmf{wiggly, tension=1.2, label=$\tl{J}_0$}{i1,v1}
			\fmf{plain,left=0.8, tension=0.4}{v1,v2}
			\fmf{plain,right=0.8, tension=0.4}{v1,v2}
			\fmf{plain}{o1,v2,o2}
                   \fmfdot{v2}
			\Square{v1}
                   \fmfv{\label=\bl_0}{v2}
			\end{fmfgraph*}  
			}} +
\parbox{\myl}{\centering{
			\begin{fmfgraph*}(2.5,1.4) 
            \fmfleft{i}
			\fmfright{o1,o2}
			\fmftop{t}
			\fmfbottom{b}
			\fmf{phantom, tension=4}{t,v3}    
			\fmf{phantom, tension=4}{b,v4}   
            \fmf{wiggly, tension=1.2, label=$\tl{J}_0$, label.dist=-0.3w}{i,v1}			
			\fmf{plain,left=0.4, tension=1}{v1,v3}
			\fmf{dashes,left=0.4, tension=0.8}{v3,v2}			
			\fmf{plain,right=0.4, tension=1.2}{v1,v4}
			\fmf{dashes,right=0.4, tension=0.8}{v4,v2}
			\fmf{plain, tension=1.2}{o1,v2,o2}
                    \fmfdot{v2}
                    \Square{v1}
			\fmfv{label=$\bl'_0$}{v2}					
\end{fmfgraph*}  
			}}  
			+
\parbox{\myl}{\centering{
			\begin{fmfgraph*}(2.5,1.4) 
			\fmfleft{i1}
			\fmfright{o1,o2}
			\fmf{wiggly, tension=1.2, label=$\tl{J}_0$}{i1,v1}
			\fmf{dashes,left=0.8, tension=0.4}{v1,v2}
			\fmf{dashes,right=0.8, tension=0.4}{v1,v2}
			\fmf{plain}{o1,v2,o2}
			\fmfv{label=$\bl'_0$}{v2}
                    \fmfdot{v2}
                    \Square{v1}
			\end{fmfgraph*}  
			}} 
			+
\parbox{2.5cm}{\centering{
			\begin{fmfgraph*}(2.5,1.4) 
                   \fmfleft{i}
			\fmfright{o1,o2}
			\fmftop{t}
			\fmfbottom{b}
			\fmf{phantom, tension=4}{t,v3}    
			\fmf{phantom, tension=4}{b,v4}   
            \fmf{wiggly, tension=1.2, label=$\tl{J}_0$, label.dist=-0.3w}{i,v1}			
			\fmf{dashes,left=0.4, tension=1}{v1,v3}
			\fmf{plain,left=0.4, tension=0.8}{v3,v2}			
			\fmf{dashes,right=0.4, tension=1.2}{v1,v4}
			\fmf{plain,right=0.4, tension=0.8}{v4,v2}
			\fmf{plain, tension=1.2}{o1,v2,o2}
			\fmfv{label=$\bl_0$}{v2}
                    \fmfdot{v2}
                    \Square{v1}					
			\end{fmfgraph*}  
			}}
\end{align*} 
\end{fmffile}}
{Leading order diagrams contributing to $\m^{\tl{J}_0}_{-1}$.}{mu0_tilde}

{\centering \subsubsection{Lowest order computation of $\m_{-1}^{\tl{J}_{0}}$} }

The running coupling constant $\mu_{-1}^{\tilde{J}_{0}}$ is given by the contraction of the correction term $C_0(k,p)\,\bigl(\psi^{t}_{k+p}\psi^{t}_{-k} +\psi^{l}_{k+p}\psi^{l}_{-k}\bigr)$ at scale $h=0$. The leading order diagrams contributing to $\mu_{-1}^{\tilde{J}_{0}}$ are shown in fig. \ref{mu0_tilde}, with the squared vertex representing the kernel $C_0(k,p)$:
\[
C_{0}(k,p) & =  \frac{1}{2} \lft[(k_0+p_0)\bigl(\chi_{[h^*,0]}^{-1}(k+p)-1\bigr)-k_{0}\bigl(\chi_{[h^*,0]}^{-1}(k)-1\bigr)\rgt]
\]
Here $\chi_{[h^*,0]}(k)=\chi_0(k) - \chi_{h^*-1}(k)$ with
\[
\c_h(k)=\begin{cases}
1 & k_0^2 + |\kk|^4  \leq\frac{2}{\g^{2}+1}\,\g^{2h}\\
0 & k_0^2 + |\kk|^4  \geq\frac{2}{\g^{2}+1}\,\g^{2(h+1)}
\end{cases}
\]
with $\gamma$ a fixed number grater then 1. With these definitions $\chi_{[h^*,0]}(k)$ is different from zero in the interval $[\frac{2}{\g^{2}+1}\g^{2(h^*-1)},\frac{2}{\g^{2}+1}]$. Summing the four diagrams in fig. \ref{mu0_tilde}, taking into account the combinatorial factors associated to them, we get:
\[ \label{app-mu}
\mu_{0}^{\tilde{J}_{0}} & = -\lim_{p_{0}\arr0}\,\frac{1}{p_0}\,16 (\l \e^{-1}) \bl_0\int\frac{d^{4}k}{(2\pi)^{4}}f_{0}(k+p)f_{0}(k)C_{0}(k,p)\frac{|\kk|^{4}-k_{0}(k_0 +p_0)}{D_0(k+p)D_0(k)}\]
with $\bl_0=\e/16$, 
$f_{0}(k)  =  \chi_0(k)-\chi_{-1}(k)$ and $ D_0(k)  =  k_{0}^{2}+ |\kk|^{4}$. Let's define the function:
\[
u_{0}(k)=f_{0}(k)(\chi^{-1}_{[h^*,0]}(k)-1)=\begin{cases}
1-f_{0}(k) & \quad\text{if } \frac{2\g^{-2}}{\g^2 +1}\leq k_{0}^{2}+ |\kk|^{4} \leq\frac{2\g^2}{\g^{2}+1}\\
0 & \quad\text{otherwise}
\end{cases}
\]
We can easily extraxt a $p_{0}$ from the integral in \eqref{app-mu} using \eqref{D0_0}:
\[
 f_{0}(k+p)f_{0}(k)C_{0}(k,p)  & =  p_{0}\big[f_{0}(k)u_{0}(k+p)+k_{0}\partial_{0}u_{0}(k)\big] + p_0^2 f_0(k) \dpr_0 f_0(k)\non \\[6pt]
 & =  p_{0}\big[f_{0}(k)u_{0}(k+p)+2k_{0}^{2}u_{0}'\big] + p_0^2 f_0(k) \dpr_0 f_0(k)
\]
Here we have used  that $\dpr_0 u_0(k)=2k_0u'_0$, being $u_{0}(k)\equiv u_{0}(k_{0}^{2}+\kk^4)$. Taking the limit $p_{0}\arr0$ we get: 
\[
\mu_{0}^{\tilde{J}_{0}} & =  -\l \int\frac{d^{d+1}k}{(2\pi)^{d+1}} \left(f_{0}(k)u_{0}(k)+2k_{0}^{2}u'_0(k) \right) \frac{|\kk|^{4}-k_{0}^{2}}{\big(|\kk|^{4}+k_{0}^{2}\big)^{2}}
\]
In the following we will denote with $t_{0}(k)=f_{0}(k)u_{0}(k)$. 

{\centering \subsubsection{\it 1.Three dimensions}}
Using spherical coordinates for the variable $\kk$ we get:
\[
\mu_{0}^{\tilde{J}_{0}}=- \l \,\frac{4\pi}{(2\pi)^{4}}\int_{-\infty}^{+\infty}dk_{0}\int_{0}^{+\infty}|\kk|^{2}d|\kk|\left(t_{0}(k)+2k_{0}^{2}u'_{0}\right)\frac{|\kk|^{4}-k_{0}^{2}}{\big(|\kk|^{4}+k_{0}^{2}\big)^{2}}
\]
Since the cutoff functions $t_{0}(k)$ and $u'_0(k)$ are functions of  $\{\kk^{4}+k_{0}^{2}\}$ it is convenient to make the change of variables $|\kk|^{2}=y$
\[
\mu_{0}^{\tilde{J}_{0}}=-\frac{\l}{4\pi^{3}} \int_{0}^{+\infty}dk_{0}\int_{0}^{+\infty}dy\,\sqrt{y}\,\frac{y^{2}-k_{0}^{2}}{\big(y^{2}+k_{0}^{2}\big)^{2}}\,\left(t_{0}(y^{2}+k_{0}^{2})+2k_{0}^{2}u'_{0}\right)
\]
and then pass to polar coordinates, that is
\[ \label{polar}
y & =  \rho\sin\a; \quad k_{0} =  \rho\cos\a \qquad\a \in[0,\frac{\pi}{2}]
\]
\[
\frac{\mu_{0}^{\tilde{J}_{0}}}{\l} & =-  \frac{1}{4\,\pi^{3}}\int_{0}^{+\infty}\rho d\rho\int_{0}^{\nicefrac{\pi}{2}}d\a \sqrt{\r\sin\a}\,\frac{\r^{2}(\sin^{2}\a-\cos^{2}\a)}{\r^{4}}\,\left(t_{0}(\r)+2\r^{2}u'_{0} \cos^{2}\a\,\right)  \non \\
 & =  \frac{1}{4\pi^{3}}\int_{0}^{+\infty}d\r\, \r^{-\nicefrac{1}{2}}t_{0}(\r)\int_{0}^{\nicefrac{\pi}{2}}d\a \sqrt{\sin\a}\,(\cos^{2}\a-\sin^{2}\a)  \non \\
 &   \qquad +\frac{1}{4\pi^{3}}\int_{0}^{+\infty}d\r \,\r^{\frac{3}{2}}u'_{0}(\rho)\int_{0}^{\frac{\pi}{2}} d\a \sqrt{\sin\a}\,(\cos^{2}\alpha-\sin^{2}\a)2\cos^{2}\a \non \\[6pt]
 & =  -\frac{1}{8\pi^{3}}\frac{\G^{2}(-\frac{1}{4})}{20\sqrt{2\pi}}\,\int_{0}^{+\infty} d\rho\rho^{-\nicefrac{1}{2}} t_{0}(\rho)+\frac{1}{8\pi^{3}} \frac{\G^{2}(-\frac{1}{4})}{15\sqrt{2\pi}}\,\int_{0}^{+\infty}d\rho\rho^{\frac{3}{2}}u'_{0}(\r) \non\\[6pt]
 & = \frac{1}{8\pi^{3}}\,\frac{\G^{2}(-\frac{1}{4})}{60\sqrt{2\pi}}\int_{0}^{+\infty}d\rho\left(-3\rho^{-\frac{1}{2}}t_{0}(\rho)+4\rho^{\frac{3}{2}}u'_{0}(\rho)\right)
\]
Approximating in the result the cutoff functions $t_{0}(\r)$ and $u'_{0}(\r)$
by the characteristic function of the set $\lft \{\frac{2}{\g^{2}+1}\leq\kk^{4}+k_{0}^{2}\leq\frac{2\g^{2}}{\g^{2}+1}\rgt\}$
we get
\[
\frac{\mu_{0}^{\tilde{J}_{0}}}{\l} & =  
\frac{1}{8\pi^{3}} \frac{\G^{2}(-\frac{1}{4})}{30\sqrt{2\pi}}\,\left(-3\r^{\nicefrac{1}{2}} +\frac{4}{3}\rho^{\frac{5}{2}}\right)
\Big|_{\big(\frac{2}{\g^{2}+1}\big)^{\frac{1}{2}}}^{\big(\frac{2\g^{2}}{\g^{2}+1}\big)^{\frac{1}{2}}} \non\\
 & =  \frac{1}{8\pi^{3}} \frac{\G^{2}(-\frac{1}{4})}{30\sqrt{2\pi}} \,\frac{2^{\frac{1}{4}}}{(\g^{2}+1)^{\frac{5}{4}}}\,
\lft( \g^{\frac{1}{2}} (5\g^{2}+1)+3\g^{2}-5 \rgt)
\]
which is positive for all $\g>1$. \\  

{\centering \subsubsection{\it 2.Two dimensions}}
Using spherical coordinates for the variable $\kk$ we get:
\[
\mu_{0}^{\tilde{J}_{0}}=- \l \,\frac{2\pi}{(2\pi)^{3}}\int_{-\infty}^{+\infty}dk_{0}\int_{0}^{+\infty}|\kk|d|\kk|\left(t_{0}(k)+2k_{0}^{2}u'_{0}\right)\frac{|\kk|^{4}-k_{0}^{2}}{\big(|\kk|^{4}+k_{0}^{2}\big)^{2}}
\]
With the changes of variables $|\kk|^{2}=y$ 
\[
\mu_{0}^{\tilde{J}_{0}}=-\frac{\l}{4\pi^{2}} \int_{0}^{+\infty}dk_{0}\int_{0}^{+\infty}dy\,\frac{y^{2}-k_{0}^{2}}{\big(y^{2}+k_{0}^{2}\big)^{2}}\,\left(t_{0}(y^{2}+k_{0}^{2})+2k_{0}^{2}u'_{0}\right)
\]
and then \eqref{polar} we get
\[
\frac{\m_{0}^{\tl{J}_{0}}}{\l} & =  \frac{1}{4\pi^{2}}\int_{0}^{+\infty} \frac{d\r}{\r}\,t_{0}(\r) \int_{0}^{\nicefrac{\pi}{2}}d\a (\cos^{2}\a - \sin^{2}\a) \non \\
& \hskip 1cm +\frac{1}{4\pi^{2}} \int_{0}^{+\infty} \r\,d\r\,u'_{0}(\r) 
\int_{0}^{\nicefrac{\pi}{2}}d\a \cos^2\a \,(\cos^2\a - \sin^2\a) \non \\
& = \frac{1}{32 \pi} \r^2 \Big|_{\big(\frac{2}{\g^{2}+1}\big)^{\frac{1}{2}}}^{\big(\frac{2\g^{2}}{\g^{2}+1}\big)^{\frac{1}{2}}} = \frac{1}{16 \p} \frac{\g^2-1}{\g^2+1}
\]
which is positive for each $\g > 1$.

\vskip 1cm
{\centering \subsubsection{Lowest order computation of of $\m_{-1}^{\tl{J}_1}$} }

The lowest order computation of the beta function of $\bm_{-1}^{\tl{J}_0}$ is similar to the one done in sec.~\ref{low_ord_J} for $\bm_h^{J_1}$. One finds that the second order diagrams contributing to $\bm_{-1}^{\tl{J}_1}$ cancel among them and that the first non trivial contribution is given by the third order diagrams, which give
\[
& \bm_{-1}^{\tl{J}_1} = O(\l \e) & d=2,3
\]
By studying the flow equation for $\bm_h^{\tl{J}_1}$ one finds
\[
& \bm_\bh^{J_1}= \bm_{-1}^{\tl{J}_1} \lft(1 + O(\l \e^{\frac{1}{2}}) \rgt) & d=3 \non \\
& \bm_\bh^{J_1}= \bm_{-1}^{\tl{J}_1} \lft(1 + O(\l ) \rgt) & d=2
\]
with   $ \m_\bh^{J_1} = \e^{2}\bm_\bh^{J_1} $ in $3d$ and $ \m_\bh^{J_1} = \e^{\frac{3}{2}}\bm_\bh^{J_1}$ in $2d$.

%\blue{ Il conto per la funzione beta e' identico a quello gia' fatto per $\m_h^{J_1}$ in section \ref{low_ord_J}; la differenza e' che $\m_h^{J_1}$ parte da 1 per cui la funzione beta da' correzioni di ordine $\l^2$ ad 1, mentre $\m_h^{\tl{J}_h}$ parte da zero, per cui diventa poi $O(\l^2)$: il contributo alla WI per $A_h$ e' $O(\l^2)$ in piu' rispetto a quello di $\m_h^{J_1}$!  }

%\blue{Inserire anche $E_{0}^{\tl{J}_{1}}$? Si, serve sapere come fluisce per l'identita' di ward per $B_h$. D'altra parte $E_{0}^{\tl{J}_{\bh}}$ sara' identico a $E_{0}^{J_{\bh}}$ perche' entrambi partono da zero.}

\pagina

\feyn{
\begin{fmffile}{feyn-TESI/APP_TT}
 \unitlength = 0.8cm
\def\myl#1{2.2cm}
\begin{align*}
 \parbox{\myl}{\centering{	\vskip 0.5cm
		 \begin{fmfgraph*}(2,1.25)
			\fmfright{i1}
			\fmfleft{o1}
			\fmf{plain, label= $p$, label.dist=0.05w}{i1,v}
			\fmf{plain}{o1,v}
                    \Ball{v}		
		\end{fmfgraph*} \\
    $\hW_{02}(p)$
		}} -
 \parbox{\myl}{\centering{	\vskip 0.5cm
		 \begin{fmfgraph*}(2,1.25)
			\fmfright{i1}
			\fmfleft{o1}
			\fmf{plain, label= $0$, label.dist=0.05w}{i1,v}
			\fmf{plain}{o1,v}
                    \Ball{v}		
		\end{fmfgraph*} \\
    $\hW_{02}(0)$
		}} 
		 & = \quad
		 \parbox{\myl}{\centering{	\vskip 0.5cm
		 \begin{fmfgraph*}(2,1.25)
			\fmfright{i1}
			\fmfleft{o1}
			\fmf{plain, label= $p$, label.dist=0.05w}{i1,v}
			\fmf{wiggly,label=$J_0$}{o1,v}
                    \sSquare{v}		
		\end{fmfgraph*} \\
    $\hW_{02;0}(p)$
		}} 
		\;+\;
		\parbox{\myl}{\centering{	\vskip 0.5cm
		 \begin{fmfgraph*}(2,1.25)
			\fmfright{i1}
			\fmfleft{o1}
			\fmf{plain, label=$p$ , label.dist=0.05w}{i1,v}
			\fmf{wiggly,label=$J_1$}{o1,v}
                    \sSquare{v}		
		\end{fmfgraph*} \\
    $\hW_{02;1}(p)$
		}} 
	\end{align*}
\end{fmffile}
}{Local WI for $A_h$ }{APP_TT}  

\subsection{(Non formal) Local WI for $ A_h $}   \label{appB.WI_A2}

The local WI useful to establish that $A_h = 1 + o(1)$ is 
\[ \label{eq:WI_AH}
\sqrt{2}\,[\hW^{(h)}_{02}(k)-\hW^{(h)}_{02}(0)]=\hW^{(h)}_{01;0}(k) - \hW^{(h)}_{01;1}(k)
\]
where $\hW^{(h)}_{01;0}(k) $ and $\hW^{(h)}_{01;1}(k) $ defined at pag.~\pageref{W_tilde}. The identity \eqref{eq:WI_AH} is pictorically represented in fig.~\ref{APP_TT}, where  the shaded squared vertices attached to the fields $J_{0}$ and $J_1$ represent  $T_{0}(k,p)$ and $T_{1}(k,p)$ respectively: 
\[ \label{appT}
T_{0}(k,p) & =  \frac{1}{2}\left[(k_{0}+p_{0})\chi_h^{-1}(k+p)-k_{0}\chi_h^{-1}(k)\right]  \non \\[6pt]
T_{1}(k,p) & =   \lft[(\kk+\pp)^{2}\chi_h^{-1}(k+p)-\kk^{2}\chi_h^{-1}(k) \rgt]
\]
At the main order in $\e$ the diagrams contributing to $\hW_{02}^{(h)}$, $\hW_{02;0}^{(h)}$ and $\hW_{02;1}^{(h)}$ are shown in fig.~\ref{f_tt}. An explicit computation gives:
\[
& \hW^{(h)}_{02}(p)  -W^{(h),t}_{02}= \non \\
& \qquad -2\l \e^{-\frac{3}{2}}\m_h \int\frac{d^{4}k}{(2\pi)^{4}}\,\c_h(k+p) \c_h(k) \left[\frac{\left((\kk+\pp)^{2}+ \chi_h(k+p)\right) \kk^{2} +(k_{0}+p_{0})k_{0}}{\DD_h(k+p)\,\DD_h(k)}\right]  \non \\[6pt]
& \hW^{(h)}_{01;0}(p)  =  -4\,\l \e^{-\frac{3}{2}}\m_h\frac{1}{2}\int\frac{d^{4}k}{(2\pi)^{4}}\,\left[(k_{0}+p_{0})\chi_h(k) - k_{0}\chi_h(k+p)\right] \non \\
& \hskip 5cm 
\left[\frac{\left((\kk+\pp)^{2}+\chi_h(k+p)\right)k_{0} -(\kk+\pp)^{2}k_{0}}{\DD_h(k+p)\,\DD_h(k)}\right] \non \\[6pt]
& \hW^{(h)}_{01;1}(p)  =  -2\,\l \e^{-\frac{3}{2}}\m_h\,\int\frac{d^{4}k}{(2\pi)^{4}} \left[(\kk+\pp)^{2}\chi_h(k)-\kk^{2}\chi_h(k+p)\right]
\non \\ & \hskip 5cm
\left[\frac{\left((\kk+\pp)^{2}+\chi_h(k+p)\right)\kk^{2}+(k_{0}+p_{0})k_{0}}{\DD_h(k+p)\,\DD_h(k)}\right] 
\]
with $W^{(h),t}_{02}$ the contribution to $\hW^{(h)}_{02}(p)$ coming from the tadpole and
\[
\DD_h(k) & =  \big(\kk^{2}+\chi_h(k)\big)\,\kk^{2}+k_{0}^{2}
\]
Using the symmetry with respect to the change of variables:
\begin{eqnarray*}
k+p & \rightarrow & -k\\
-k & \rightarrow & k+p
\end{eqnarray*}
the validity of  \eqref{eq:WI_AH} at the one--loop level is immediately proven:
\[
W^{(h)}_{02}(p)-W^{(h)}_{02}(0) & =  -\l \e^{-\frac{3}{2}}\m_h\,\int\frac{d^{4}k}{(2\pi)^{4}}\frac{1}{\DD_h(k)\,\DD_h(k+p)} \non \\
& \qquad [\,k_{0}^{2}\,\lft(\chi_h(k+p)-\chi_h(k) \rgt)^{2} + \lft(\,\chi_h(k+p)\kk^{2}-\chi_h(k)(\kk+\pp)^{2} \,\rgt)^{2}\,] \non \\[6pt]
W^{(h)}_{01;1}(p) & =   -\l \e^{-\frac{3}{2}}\m_h\,\int\frac{d^{4}k}{(2\pi)^{4}}\frac{1}{\DD_h(k)\DD_h(k+p)}(\chi_h(k+p)\kk^{2}-\chi_h(k)(\kk+\pp)^{2})^{2} \non \\[6pt]
W^{(h)}_{01;0}(p) & =   -\l \e^{-\frac{3}{2}}\m_h\,\int\frac{d^{4}k}{(2\pi)^{4}}\frac{1}{\DD_h(k)\DD_h(k+p)}\,k_{0}^{2}(\chi_h(k+p)-\chi_h(k))^{2}
\]

%--------NOTA IMPORTANTE
%\blue{Nota: qual e' il ruolo della correzione? l'identita' per Ah torna formalmente anche senza correzione? }  Dopo aver verificato la cancellazione nella forma sopra, riesprimi il risultato finale nella forma $A_h$=(pezzo dell'identita' di Ward ingenua)+(correzioni), da cui dovrebbe risultare evidente sia la cancellazione che migliora la stima dimensionale di $A_{h}$ , sia il ruolo della correzione. \\

\feyn{
\begin{fmffile}{feyn-TESI/f_tt}
 \unitlength = 0.8cm
\def\myl#1{2.2cm}
\def\myll#1{3cm}
\begin{align*}
\b^h_{02}
		& =
	\parbox{\myll}{\centering{  \vskip 0.5cm
			\begin{fmfgraph*}(2.6,1.4) 
			\fmfleft{i1}
			\fmfright{o1}
			\fmf{plain, tension=1.5}{i1,v1}
			\fmf{plain,left=0.7, tension=0.4}{v1,v2}
			\fmf{dashes,right=0.7, tension=0.4}{v1,v2}
			\fmf{plain, tension=1.5}{o1,v2}
			\Ball{v1,v2}
			\end{fmfgraph*}   \\
                    $-\frac{1}{2}\,2\,2$
			}}  	\;+\;
	\parbox{\myll}{\centering{ \vskip 0.5cm
			\begin{fmfgraph*}(2.6,1.4) 
            \fmfleft{i}
			\fmfright{o}
			\fmftop{t}
			\fmfbottom{b}
			\fmf{phantom, tension=4}{t,v3}    
			\fmf{phantom, tension=4}{b,v4}   
            \fmf{plain, tension=1.5}{i,v1}			
			\fmf{plain,left=0.4, tension=1.2}{v1,v3}
			\fmf{dashes,left=0.4}{v3,v2}			
			\fmf{dashes,right=0.4}{v1,v4}
			\fmf{plain,right=0.4}{v4,v2}
			\fmf{plain, tension=1.5}{o,v2}
			\Ball{v1,v2}			
			\end{fmfgraph*}  \\
                    $-\frac{1}{2}\,2\,2$
			}} +\;  \parbox{\myl}{\centering{ \vskip 0.4cm
		\begin{fmfgraph*}(2,2)
			\fmfleft{i1}
			\fmfright{o1}
			\fmf{plain, tension=0.8}{i1,v}
			\fmf{plain}{v,v}
			\fmf{plain, tension=0.8}{v,o1} 
			\Ball{v}
		\end{fmfgraph*} \\[-6pt]
                    $\binom{4}{2}$
	}}		\non \\[6pt]
%-----------------------------------------------------------------------------
\b^h_{02;0}
		& =	\parbox{\myll}{\centering{  \vskip 0.5cm
			\begin{fmfgraph*}(2.6,1.4) 
            \fmfleft{i}
			\fmfright{o}
			\fmftop{t}
			\fmfbottom{b}
			\fmf{phantom, tension=4}{t,v3}    
			\fmf{phantom, tension=4}{b,v4}   
            \fmf{wiggly, label=$J_0$, tension=1.5}{i,v1}			
			\fmf{plain,left=0.4, tension=1.2}{v1,v3}
			\fmf{dashes,left=0.4}{v3,v2}			
			\fmf{plain,right=0.4}{v1,v4}
			\fmf{plain,right=0.4}{v4,v2}
			\fmf{plain, tension=1.5}{o,v2}
			\Ball{v2}
                   \sSquare{v1}				
			\end{fmfgraph*}  \\
                   $-\frac{1}{2}\,2\,2\,2$
			}}  	\;+\;
	\parbox{\myll}{\centering{ \vskip 0.5cm
			\begin{fmfgraph*}(2.6,1.4) 
            \fmfleft{i}
			\fmfright{o}
			\fmftop{t}
			\fmfbottom{b}
			\fmf{phantom, tension=4}{t,v3}    
			\fmf{phantom, tension=4}{b,v4}   
            \fmf{wiggly, label=$J_0$, tension=1.5}{i,v1}			
			\fmf{dashes,left=0.4, tension=1.2}{v1,v3}
			\fmf{dashes,left=0.4}{v3,v2}			
			\fmf{dashes,right=0.4}{v1,v4}
			\fmf{plain,right=0.4}{v4,v2}
			\fmf{plain, tension=1.5}{o,v2}
			\Ball{v2}
                   \sSquare{v1}			
			\end{fmfgraph*}  \\
                   $-\frac{1}{2}\,2\,2\,2$
			}}  \non \\[6pt]
%-----------------------------------------------------------------------------------
\b^h_{02;1}
		& =	\parbox{\myll}{\centering{ \vskip 0.5cm
			\begin{fmfgraph*}(2.6,1.4) 
            \fmfleft{i}
			\fmfright{o}
			\fmftop{t}
			\fmfbottom{b}
			\fmf{phantom, tension=4}{t,v3}    
			\fmf{phantom, tension=4}{b,v4}   
            \fmf{wiggly, label=$J_1$, label.dist=-0.5cm, tension=1.5}{i,v1}			
			\fmf{plain, label=$\dpr_\xx$, label.dist=-0.01cm, left=0.4, tension=1.2}{v1,v3}
			\fmf{dashes,right=0.4}{v1,v4}
			\fmf{plain,left=0.4}{v3,v2}			
			\fmf{dashes,right=0.4}{v4,v2}
			\fmf{plain, tension=1.5}{o,v2}
			\Ball{v2}
                   \sSquare{v1}				
			\end{fmfgraph*}  \\
                   $-\frac{1}{2}\,2\,2$
			}}  	\;+\;
\parbox{\myll}{\centering{ \vskip 0.5cm
			\begin{fmfgraph*}(2.6,1.4) 
            \fmfleft{i}
			\fmfright{o}
			\fmftop{t}
			\fmfbottom{b}
			\fmf{phantom, tension=4}{t,v3}    
			\fmf{phantom, tension=4}{b,v4}   
            \fmf{wiggly, label=$J_1$, label.dist=-0.5cm, tension=1.5}{i,v1}			
			\fmf{plain, label=$\dpr_\xx$, label.dist=-0.01cm, left=0.4, tension=1.2}{v1,v3}
			\fmf{dashes,right=0.4}{v1,v4}
			\fmf{dashes,left=0.4}{v3,v2}			
			\fmf{plain,right=0.4}{v4,v2}
			\fmf{dashes, tension=1.5}{o,v2}
			\Ball{v2}
                   \sSquare{v1}					
			\end{fmfgraph*}   \\
                   $-\frac{1}{2}\,2\,2$
			}} 
\end{align*}
\end{fmffile}	
}{ Beta function for $\hW_{02}^{(h)}$, $\hW_{02;0}^{(h)}$ and $\hW_{02;1}^{(h)}$ $h \leq \bh$, at leading order in $\e$. The shaded vertices represent $T_0(k,p)$ and $T_1(k,p)$, see \eqref{appT} for a definition. }{f_tt}

%\blue{WI con derivate nel file 8.Bozze-Ah}

%--------------------------------------------------------------

\bibliographystyle{unsrt}  
\bibliography{biblioBOSE}

\begin{thebibliography}{10}

\bibitem{Anderson1995}
M.~H. Anderson, J.~R. Ensher, M.~R. Matthews, C.~E. Wieman, and E.~A. Cornell.
\newblock Observation of {B}ose--{E}instein condensation in a dilute atomic
  vapor.
\newblock {\em Science}, 269(5221), 1995.

\bibitem{Bradley1995}
C.~C. Bradley, C.~A. Sackett, J.~J. Tollett, and R.~G. Hulet.
\newblock Evidence of {B}ose--{E}instein condensation in an atomic gas with
  attractive interactions.
\newblock {\em Phys. Rev. Lett.}, 75(9), 1995.

\bibitem{Davis1995}
K.~B. Davis, M.~O. Mewes, M.~R. Andrews, N.~J. van Druten, D.~S. Durfee, D.~M.
  Kurn, and W.~Ketterle.
\newblock {B}ose--{E}instein condensation in a gas of sodium atoms.
\newblock {\em Phys. Rev. Lett.}, 75(22), 1995.

\bibitem{Einstein}
A.~Einstein.
\newblock {\em Quantentheorie des einatomigen idealen Gases}.
\newblock Sber. Preuss. Akad. Wiss., Part {I}: 1924. Part {II}: 1925.

\bibitem{vortices}
J.~R. Abo-Shaeer, C.~Raman, J.~M. Vogels, and W.~Ketterle.
\newblock Observation of vortex lattices in bose-einstein condensates.
\newblock {\em Science}, 292(5516):476--479, 2001.

\bibitem{Atom-chip1}
H.~Ott, J.~Fortagh, G.~Schlotterbeck, A.~Grossmann, and C.~Zimmermann.
\newblock Bose-{E}instein {C}ondensation in a {S}urface {M}icrotrap.
\newblock {\em Phys. Rev. Lett.}, 87, 2001.

\bibitem{Atom-chip2}
W.~Hansel, P.~Hommelhoff, T.~W. Hansch, and J.~Reichel.
\newblock Bose-einstein condensation on a microelectronic chip.
\newblock {\em Nature}, 413, 2001.

\bibitem{BEC_optical}
I.~Bloch.
\newblock Ultracold quantum gases in optical lattices.
\newblock {\em Nat. Phys.}, 2005.

\bibitem{AtomLaser}
J.~Billy, V.~Josse, Z.~Zuo, W.~Guerin, A.~Aspect, and P.~Bouyer.
\newblock Guided atom laser: a new tool for guided atom optics.
\newblock {\em Ann. Phys. Fr.}, 32(2), 2007.

\bibitem{BEC_phonons}
J.~Klaers, J.~Schmitt, F.~Vewinger, and M.~Weitz.
\newblock Bose-einstein condensation of photons in an optical microcavity.
\newblock {\em Nature}, 2010.

\bibitem{He_films}
D.~J. Bishop and J.~D. Reppy.
\newblock Study of the superfluid transition in two-dimensional
  $^{4}\mathrm{He}$ films.
\newblock {\em Phys. Rev. Lett.}, 40(26), 1978.

\bibitem{BEC_2d}
H.~Hadzibabic, P.~Kr\"uger, M.~Cheneau, B.~Battelier, and J~Dalibard.
\newblock Berezinskii--{K}osterlitz--{T}houless crossover in a trapped atomic
  gas.
\newblock {\em Nat. Lett.}, 2006.

\bibitem{BEC_low_dim}
A.~G\"orlitz, J.~M. Vogels, A.~E. Leanhardt, C.~Raman, T.~L. Gustavson, J.~R.
  Abo-Shaeer, A.~P. Chikkatur, S.~Gupta, S.~Inouye, T.~Rosenband, and
  W.~Ketterle.
\newblock Realization of bose--einstein condensates in lower dimensions.
\newblock {\em Phys. Rev. Lett.}, 87(13), 2001.

\bibitem{hard-sphere-bosons}
F.~J. Dyson, E.~H. Lieb, and B.~Simon.
\newblock Phase transitions in quantum spin systems with isotropic and
  nonisotropic interactions.
\newblock {\em J. Stat. Phys.}, 1978.

\bibitem{Lse}
E.~H. Lieb and R.~Seiringer.
\newblock Proof of {Bose-Einstein} condensation for dilute trapped gases.
\newblock {\em Phys. Rev. Lett.}, 88(17), 2002.

\bibitem{LSeY5}
E.~H. Lieb, R.~Seiringer, and J.~Yngvason.
\newblock Superfluidity in dilute trapped {B}ose gases.
\newblock {\em Phys. Rev. B}, 66(13), 2002.

\bibitem{Bogoliubov}
N.N. Bogoliubov.
\newblock On the theory of superfluidity.
\newblock {\em Eng. Trans. J. Phis. (USSR)}, 1947.

\bibitem{Landau}
L.D. Landau and E.~M. Lifshits.
\newblock {\em Stat. Phys.}
\newblock Pergamon Press, 1968.

\bibitem{Beliaev}
S.T. Beliaev.
\newblock Application of the methods of quantum field theory to a system of
  bosons.
\newblock {\em Sov. Phys. JETP}, 1958.

\bibitem{Pines}
N.~M. Hugenholtz and D.~Pines.
\newblock Ground-state energy and excitation spectrum of a system of
  interacting bosons.
\newblock {\em Phys. Rev.}, 116, 1959.

\bibitem{LeeYang5}
T.~D. Lee and C.~N. Yang.
\newblock Many-body problem in quantum statistical mechanics. {V}. {D}egenerate
  phase in {B}ose-{E}instein condensation.
\newblock {\em Phys. Rev.}, 117(4), 1960.

\bibitem{Lee-Huang-Yang}
T.~D. Lee, K.~Huang, and C.~N. Yang.
\newblock Eigenvalues and eigenfunctions of a {B}ose system of hard spheres and
  its low-temperature properties.
\newblock {\em Phys. Rev.}, 106(6), 1957.

\bibitem{Gavoret}
J.~Gavoret and P.~Nozi\'eres.
\newblock Structure of the perturbation expansion for the bose liquid at zero
  temperature.
\newblock {\em Annals of Physics}, 28(3), 1964.

\bibitem{Nepom}
Yu.~A. Nepomnyashchii and A.A. Nepomnyashchii.
\newblock Infrared divergence in field theory of a bose system with a
  condensate.
\newblock {\em Sov. Phys. - JETP}, 48(3), 1978.

\bibitem{Popov}
V.N. Popov and A.V. Seredniakov.
\newblock Low-frequency asymptotic form of the self-energy parts of a
  superfluid bose system at $t=0$.
\newblock {\em Sov. Phys. - JETP}, 1979.

\bibitem{benfatto}
G.Benfatto.
\newblock Renormalization group approach to zero temperature {B}ose
  condensation.
\newblock Palaiseau, July 25-27, 1994.
\newblock Proceedings of the workshop ``Constructive results in Field Theory,
  Statistical Mechanics and Condensed Matter Physics''.

\bibitem{CaDiC1}
C.~Castellani, C.~Di~Castro, F.~Pistolesi, and G.~C. Strinati.
\newblock Infrared behavior of interacting bosons at zero temperature.
\newblock {\em Phys. Rev. Lett.}, 78(9), 1997.

\bibitem{CaDiC2}
F.~Pistolesi, C.~Castellani, C.~Di~Castro, and G.~C. Strinati.
\newblock Renormalization-group approach to the infrared behavior of a
  zero-temperature bose system.
\newblock {\em Phys. Rev. B}, 69(2), 2004.

\bibitem{Wilson}
K.G. Wilson.
\newblock The {R}enormalization {G}roup and {C}ritical {P}henomena.
\newblock In {\em Nobel Lectures, Physics 1981--1990}. World Scientific, 1993.

\bibitem{GalReview}
G.~Gallavotti.
\newblock Renormalization theory and ultraviolet stability for scalar fields
  via renormalization group methods.
\newblock {\em Rev. Mod. Phys.}, 57(2), 1985.

\bibitem{BM-luttinger}
G.~Benfatto and V.~Mastropietro.
\newblock Ward identities and chiral anomaly in the {L}uttinger liquid.
\newblock {\em Comm. Math. Phys.}, 258(3), 2005.

\bibitem{BGM-Hubbard}
G.~Benfatto, A.~Giuliani, and V.~Mastropietro.
\newblock Fermi liquid behavior in the 2d {H}ubbard model at low temperatures.
\newblock {\em Ann. Henri Poincar\'e}, 7(5), 2006.

\bibitem{SRGraph1}
A.~Giuliani and V.~Mastropietro.
\newblock Rigorous construction of ground state correlations in graphene:
  Renormalization of the velocities and ward identities.
\newblock {\em Phys. Rev. B}, 79(20), 2009.

\bibitem{SRGraph2}
A.~Giuliani and V.~Mastropietro.
\newblock The two-dimensional hubbard model on the honeycomb lattice.
\newblock {\em Comm. {M}ath. {P}hys.}, 293(2), 2010.

\bibitem{RV}
V.~Rivasseau.
\newblock {\em From perturbative to constructive renormalization}.
\newblock Princeton University Press, 1991.

\bibitem{Brydges}
D.~Brydges.
\newblock A short course on cluster expansions.
\newblock In {\em Les Houches summer school on ``Critical phenomena, random
  systems, Gauge theories''}. K. Osterwalder and R. Stora, 1986.

\bibitem{GK}
K.~Gawedzki and A.~Kupiainen.
\newblock Asymptotic freedom beyond perturbation theory.
\newblock In {\em Les Houches summer school on ``Critical phenomena, random
  systems, Gauge theories''}. K. Osterwalder and R. Stora, 1986.

\bibitem{BTFK_ultraviolet}
T.~Balaban, J.~Feldman, H.~Kn\"orrer, and E.~Trubowitz.
\newblock The temporal ultraviolet limit for complex bosonic many-body models.
\newblock {\em Annales Henri Poincare}, 2010.

\bibitem{NO}
J.W. Negele and H.~Orland.
\newblock {\em Quantum many--particle systems}.
\newblock Addison--Wesley, 1987.

\bibitem{Book-Lieb}
E.H. Lieb, R.~Seiringer, J.P. Solovej, and J.~Yngvason.
\newblock {\em The {M}athematics of the {B}ose {G}as and its {C}ondensation.}
\newblock Birkh\"auser Basel, 2005.

\bibitem{Penrose-Onsager}
O.~Penrose and L.~Onsager.
\newblock {B}ose-{E}instein condensation and liquid helium.
\newblock {\em Phys. Rev. 104}, 1956.

\bibitem{Yang-ODLRO}
C.~N. Yang.
\newblock {C}oncept of {O}ff-{D}iagonal {L}ong-{R}ange {O}rder and the
  {Q}uantum {P}hases of {L}iquid {H}e and of {S}uperconductors.
\newblock {\em Rev. Mod. Phys.}, 34, 1962.

\bibitem{LeeYangGROUP}
T.~D. Lee and C.~N. Yang.
\newblock {\it Phys. Rev.}, 113(5), 1959; {\it Phys. Rev.}, 116(1), 1959; {\it
  Phys. Rev.}, 117(1), 1960 12--21, {\it Phys. Rev.}, 117(1), 1960 22--36; {\it
  Phys. Rev.}, 117(4), 1960.

\bibitem{Dyson-upper-bound}
F.~J. Dyson.
\newblock Ground-state energy of a hard-sphere gas.
\newblock {\em Phys. Rev.}, 106(1), 1957.

\bibitem{LY-lower-bound}
E.~H. Lieb and J.~Yngvason.
\newblock Ground state energy of the low density {B}ose gas.
\newblock {\em Phys. Rev. Lett.}, 80(12), 1998.

\bibitem{LeeYang0}
T.~D. Lee and C.~N. Yang.
\newblock Many-body problem in quantum mechanics and quantum statistical
  mechanics.
\newblock {\em Phys. Rev.}, 105, 1957.

\bibitem{Erdos-Schlein-Yau}
L.~Erd\"os, B.~Schlein, and H.~Yau.
\newblock Ground-state energy of a low-density {B}ose gas: a second-order upper
  bound.
\newblock {\em Phys. Rev. A}, 78(5), 2008.

\bibitem{YauYin}
H.~Yau and J.~Yin.
\newblock The second order upper bound for the ground energy of a bose gas.
\newblock {\em J. Stat. Phys.}, 136(3), 2009.

\bibitem{Giuliani-Seiringer}
A.~Giuliani and R.~Seiringer.
\newblock The ground state energy of the weakly interacting {B}ose gas at high
  density.
\newblock {\em J. Stat. Phys.}, 135(5), 2009.

\bibitem{Seiringer_mean_field}
R.~Seiringer.
\newblock The excitation spectrum for weakly interacting bosons.
\newblock {\em Comm. Math. Phys.}, 306(2), 2011.

\bibitem{Schick}
M.~Schick.
\newblock Two-dimensional system of hard-core bosons.
\newblock {\em Phys. Rev. A}, 3(3), 1971.

\bibitem{LY2D}
E.~H. Lieb and Yngvason J.
\newblock The ground state energy of a dilute two-dimensional {B}ose gas.
\newblock {\em J. Stat. Phys.}, 103(3), 2001.

\bibitem{Hines}
D.F. Hines, N.E. Frankel, and D.J. Mitchell.
\newblock Hard-disc bose gas.
\newblock {\em Phys. Lett. A}, 68(1), 1978.

\bibitem{Hohenberg}
P.~C. Hohenberg.
\newblock Existence of long-range order in one and two dimensions.
\newblock {\em Phys. Rev.}, 1967.

\bibitem{Mermin-Wagner}
N.~D. Mermin and H.~Wagner.
\newblock Absence of ferromagnetism or antiferromagnetism in one- or
  two-dimensional isotropic heisenberg models.
\newblock {\em Phys. Rev. Lett.}, 17, 1966.

\bibitem{Mastropietro}
V.~Mastropietro.
\newblock {\em Non-perturbative {R}enormalization.}
\newblock World Scientific, 2009.

\bibitem{EMGraph1}
A.~Giuliani, V.~Mastropietro, and M.~Porta.
\newblock Lattice quantum electrodynamics for graphene.
\newblock {\em Annals of Physics}, 327(2), 2012.

\bibitem{EMGraph2}
A.~Giuliani, V.~Mastropietro, and M.~Porta.
\newblock Lattice gauge theory model for graphene.
\newblock {\em Phys. Rev. B}, 82(12), 2010.

\bibitem{BM}
G.~Benfatto and G.~Gallavotti.
\newblock {\em Renormalization Group}.
\newblock Princeton University Press, 1995.

\bibitem{GM}
G.~Gentile and V.~Mastropietro.
\newblock Renormalization group for fermions: a review on mathematical results.
\newblock {\em Phys. Rep.}, 352, 2001.

\bibitem{XYZ}
G.~Benfatto and V~Mastropietro.
\newblock Renormalization group, hidden symmetries and approximate ward
  identities in the {XYZ} model, {I}.
\newblock {\em Rev. Math. Phys.}, 13, 2001.

\bibitem{THGiuliani}
A.~Giuliani.
\newblock Gruppo di rinormalizzazione per un sistema di fermioni interagenti in
  due dimensioni.
\newblock Master's thesis, Universit\`a degli studi di Roma ``{L}a
  {S}apienza'', 2000/2001.

\bibitem{AshkinTeller}
A.~Giuliani and V.~Mastropietro.
\newblock Anomalous universality in the anisotropic {A}shkin--{T}eller model.
\newblock {\em Comm. Math. Phys.}, 256, 2005.

\bibitem{GRanom}
A.~Giuliani, V.~Mastropietro, and M.~Porta.
\newblock Anomalous behavior in an effective model of graphene with {C}oulomb
  interactions.
\newblock {\em Annales Henri Poincar\`e}, 11(8), 2010.

\bibitem{Tc}
R.~Seiringer and D.~Ueltschi.
\newblock Rigorous upper bound on the critical temperature of dilute bose
  gases.
\newblock {\em Phys. Rev. B}, 80(1), 2009.

\end{thebibliography}
\addcontentsline{toc}{chapter}{Bibliography}
\markboth{\textsc{Bibliography}}{}

%\addcontentsline{toc}{chapter}{References}
%\markboth{\textsc{References}}{}

% bibliography
% \cleardoublepage
%\phantomsection
%\bibliography{bibliography} % BibTeX database without .bib extension

\end{document}